\documentclass{book}[11pt,a5paper,oneside,notitlepage]

\usepackage{tikz}
   \usetikzlibrary{arrows,calc,patterns,positioning,shapes}
   \usetikzlibrary{decorations.pathmorphing}
   \tikzset{
       modal/.style={>=stealth',shorten >=1pt,shorten <=1pt,auto,
                     node distance=1.5cm,semithick},
       world/.style={circle,draw,minimum size=1cm,fill=gray!15},
       point/.style={circle,draw,fill=black,inner sep=0.5mm},
       reflexive/.style={->,in=120,out=60,loop,looseness=#1},
       reflexive/.default={5},
       reflexive point/.style={->,in=135,out=45,loop,looseness=#1},
       reflexive point/.default={25},
       }

 \tikzset{
       reflexive above/.style={->,loop,in=120,out=60,looseness=#1},
       reflexive above/.default={7},
       reflexive below/.style={->,loop,in=240,out=300,looseness=#1},
       reflexive below/.default={7},
       reflexive left/.style={->,loop,in=150,out=210,looseness=#1},
       reflexive left/.default={7},
       reflexive right/.style={->,loop,in=30,out=330,looseness=#1},
       reflexive right/.default={7}
}

\usepackage{lmodern}
\usepackage{stmaryrd}
\usetikzlibrary{arrows.meta}
\usetikzlibrary{calc}
\usepackage{tikz-cd}
\usepackage{xfrac}
\usepackage{graphicx}
\graphicspath{ {./graphics/} }
\usepackage{amsmath}
\usepackage{amssymb}
\usepackage{relsize}
\usepackage{mathtools}
\usepackage{bussproofs}
\usepackage{proof}
\usepackage{enumitem}
\usepackage{comment}
\usepackage{float}
\usepackage{amsthm}
\usepackage{mathtools}
\usepackage[backend=biber,style=numeric,maxnames=99,minnames=99]{biblatex}
\usepackage[margin=1.5in]{geometry}

\usepackage{hyperref}
\hypersetup{
    colorlinks=true,
    linkcolor=black,
    filecolor=black,      
    urlcolor=black,
    citecolor = black,
    pdftitle={Overleaf Example},
    pdfpagemode=FullScreen,
    }

\DeclarePairedDelimiter\floor{\lfloor}{\rfloor}

\newenvironment{scprooftree}[1]%
  {\gdef\scalefactor{#1}\begin{center}\proofSkipAmount \leavevmode}%
  {\scalebox{\scalefactor}{\DisplayProof}\proofSkipAmount \end{center} }

\newcommand\C {\mathsf{C}}
\newcommand\IM{\mathsf{IM}}
\newcommand{\ICK}{\mathsf{ICK}}
\newcommand\Cl {\mathsf{Cl}}
\newcommand\Prop {\mathsf{Prop}}

\newcommand{\X} {\mathbin{{\bigcirc}}}
\newcommand\G {\mathbin{\Box}}
\newcommand\NWIM{\mathrm{nIM}}
\newcommand\CIM{\mathrm{cIM}}
\newcommand\Ct{\mathsf{C_t}}
\newcommand\Cf{\mathsf{C_f}}
\newcommand\R{\mathsf{R}}

\newcommand\A{\mathsf{A}}
\newcommand\iLTL{\mathsf{iLTL}}
\newcommand{\K}{\mathsf{K}}
\newcommand{\LIM}{\mathcal{L}_\mathrm{IM}}
\newcommand{\Ek}{\mathsf{E}}
\newcommand{\CICK}{\mathrm{cICK}}

\newcommand{\LICK}{\mathcal{L}_\mathsf{ICK}}
\newcommand{\LIPL}{\mathcal{L}_\mathsf{IPL}}
\newcommand{\Rel}{\mathrel{R}}

\newcommand{\f}{\mathsf{f}}
\renewcommand{\u}{\mathsf{u}}

\newcommand{\ld}{\mbox{\larger[-1.5]$\Diamond$}}

\newcommand{\lb}{\mbox{\larger[-1.5]$\square$}}
\newcommand{\lbm}{\ooalign{\lb \cr \hidewidth\raise.035ex\hbox{$* \mkern1.12mu$}\cr}}
\newtheorem{theorem}{Theorem}[chapter]
\newtheorem{definition}[theorem]{Definition}
\newtheorem{lemma}[theorem]{Lemma}
\newtheorem{proposition}[theorem]{Proposition}
\newtheorem{corollary}[theorem]{Corollary}
\newtheorem{example}[theorem]{Example}

\newtheorem{question}[]{Question}
\numberwithin{theorem}{chapter}
\newtheorem{convention}[theorem]{Convention}

\newcommand\LILTL{\mathcal{L}_\mathsf{iLTL}}

\newcommand\Lbi{\mathcal{L}_{\mathsf{bIM}}}
\newcommand\type{\mathrm{T}}
\newcommand\lanfull{\mathcal{L}_\mathsf{bIM}}
\newcommand\increasing{\mathop \uparrow}
\newcommand\decreasing{\mathop \downarrow}
\newcommand\moment{\mathbb M}
\newcommand\fw{\mathcal{M}}
\newcommand\fv{\mathcal{N}}
\newcommand{\redu}{\mathrel\unlhd}
\newcommand{\simm}{\mathrel\triangleq}
\newcommand{\irr}[1]{\mathrm I_{ #1 }}
\newcommand{\niltle}{\mathrm{niLTL_e}}
\newcommand{\niltlp}{\mathrm{niLTL_p}}

\newcommand{\defined}[1]{\exists{#1}}
\newcommand{\undefined}[1]{\nexists{#1}}

\newcommand\cardinality[1]{\left|#1\right|}

\renewcommand{\u}{\mathsf{u}}
\newcommand{\dimp}{\mathbin{\tikz[baseline=-.5ex] \draw[-to reversed] (0,0) -- (1em,0);}}

\newcommand\U {\mathbin{\mathsf{U}}}
\newcommand\E{\Diamond}

\newcommand{\Lbiltl}{\mathcal{L}_\mathsf{biLTL}}
\newcommand{\Lbiltlnext}{\mathcal{L}_{\X}}
\newcommand{\biltl}{\mathsf{biLTL}}
\newcommand{\biltlH}{\mathrm{biLTL_H}}
\newcommand{\biltlHX}{\mathrm{biLTL_H^{\X}}}
\newcommand{\final}{\mathbf{f}}
\newcommand{\cqm}{\sfrac{\mathcal M_\mathrm{c}}\Sigma}

\renewcommand{\u}{\mathsf{u}}

\numberwithin{theorem}{chapter}
\newtheorem{remark}[theorem]{Remark}

\renewcommand{\u}{\mathsf{u}}

\numberwithin{theorem}{chapter}

\usepackage{makeidx}
\makeindex

\usepackage[left, pagewise]{lineno} 

\title{Intuitionistic Dynamic Logic}

\author{Lukas Zenger}
\date{}

\begin{document}
\frontmatter

\nolinenumbers



\begin{titlepage}

\setlength{\hoffset}{-0.6in}
\setlength{\voffset}{-1in}
\setlength{\topmargin}{1.5cm}
\setlength{\headheight}{0.5cm}
\setlength{\headsep}{1cm}
\setlength{\oddsidemargin}{3cm}
\setlength{\evensidemargin}{3cm}
\setlength{\footskip}{1.5cm}
\enlargethispage{1cm}

\fontsize{12pt}{14pt}
\selectfont

\begin{center}

\includegraphics[height=2cm]{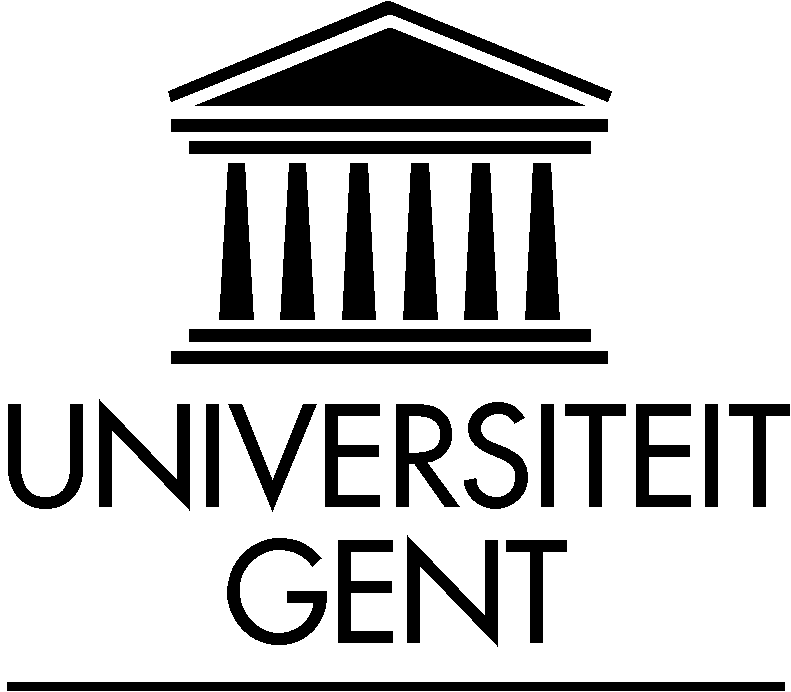}

\vspace{0.5cm}

Faculty of Sciences\\
Department of Mathematics: Analysis, Logic and Discrete Mathematics\\

\vspace{3.5cm}

\fontseries{bx}
\fontsize{17.28pt}{21pt}
\selectfont

Intuitionistic Dynamic Logic

\fontseries{m}
\fontsize{12pt}{14pt}
\selectfont

\vspace{.6cm}

\vspace{.4cm}

Lukas Zenger

\vspace{3.5cm}

Supervisor: Prof.~Dr.~Andreas~Weiermann\\
Co-Supervisor: Prof.~Dr.~Thomas Studer\\
Co-Supervisor: Prof.~Dr.~David Fern\'andez-Duque\\

\vspace{2cm}

Dissertation submitted in fulfilment of the requirements for the degree of\\
Doctor of Science: Mathematics 

\vspace{1cm}

Academic year 2025--2026

\end{center}
\end{titlepage}

\addcontentsline{toc}{chapter}{Preface}
\chapter*{Preface}

This thesis is the culmination of my doctoral studies in computer science at University of Bern between March 2021 and May 2024 and in mathematics at Ghent University between October 2024 and August 2025. My research was conducted under the supervision of Prof. Dr. Andreas Weiermann, Prof. Dr. Thomas Studer and Prof. Dr. David Fern\'andez-Duque.

This thesis investigates a family of intuitionistic modal logics extended with fixed point operators. Such logics are used to reason about dynamic phenomena, such as the change within a mathematical system over time, the change of state in an automaton after a program step or the change of an agent's knowledge after an information update. For this reason I refer to them as \emph{intuitionistic dynamic logics}, where the term `dynamic' highlights their role in formalizing processes of change. At the origin of my doctoral studies stood the goal to develop a coherent mathematical theory of intuitionistic temporal logics, whose modal and fixed point operators capture temporal properties such as `in the next time step' or `eventually in the future'. As it often happens in theoretical research, my interests expanded over the course of my PhD, leading me to study related systems such as intuitionistic epistemic logic, bi-intuitionistic modal logic and intuitionistic modal logic with the master modality. All of these logics are discussed in detail in this thesis.

Although logic lies at the intersection of mathematics, philosophy and computer science, the focus here is primarily mathematical. A significant part of the thesis is devoted to the proof theory of intuitionistic dynamic logics, which is studied using Hilbert-style axiomatizations as well as non-wellfounded and cyclic sequent calculi. The latter two are proof systems that allow formal proofs to be infinitely long. Such infinite proofs are particularly well-suited to handle the infinite nature of fixed point operators, but they present considerable technical challenges. The remaining part of the thesis addresses model theoretic and computational aspects of intuitionistic dynamic logics, including expressivity, finite model property, decidability and complexity results. 

Intuitionistic modal and dynamic logics are also relevant from a philosophical point of view. While I am interested in the philosophy of intuitionism and questions regarding intuitionistic justifications for modalities, such considerations are not included in the thesis. In fact, the intuitionist shall be warned: I use a classical metatheory when reasoning about intuitionistic dynamic logics and I do not attempt to provide an intuitionistic reading of the studied modalities. \smallskip

\noindent \textbf{Acknowledgments} I would like to express my gratitude to the many people who have contributed to my thesis and my life in general during the past four years. First of all, I would like to thank my supervisors, Andreas Weiermann, Thomas Studer and David Fern\'andez-Duque, for their guidance and help throughout my doctoral studies. Thomas, who was my main supervisor during the three years spent at the University of Bern, has helped me greatly to improve as a logician. I am particularly grateful for his course on proof theory, that he held on my request, which helped me learn about areas of proof theory that I had missed in my previous education. As a supervisor, he always provided me with the help I needed while at the same time giving me the academic freedom to pursue my own interests and learn to work independently. I am also grateful for his wise council regarding the many difficulties that I encountered during my doctoral studies. To Andreas, who was my main supervisor at Ghent University, I am particularly grateful that he provided to me the opportunity to continue my doctoral studies for an additional year after my funding ran out in Bern. This allowed me to bring my PhD to a satisfying conclusion. I would also like to thank him for his efforts organizing my defense and ensuring a smooth transition from Bern to Ghent. To David I would like to express my gratitude for his guidance and help in writing some of the papers that lead to this thesis. His impressive mathematical abilities and great ideas have contributed significantly to many of the results presented in the thesis and his explanations to my understanding of the topic. I also sincerely appreciate his council regarding the writing of this thesis and the suggestion for the title. 

I thank the members of the examination committee for reading my thesis and the many helpful comments to improve it: Iris van der Giessen, Bahareh Afshari, Mart\'in Di\'eguez Lodeiro, Seyedmojtaba Mojtahedi and Giovanni Solda.

I am deeply obliged to my co-authors for their contribution to my thesis: To Jan Rooduijn, Lide Grotenhuis, Brett McLean, David Fern\'andez-Duque, Graham E. Leigh, Borja Sierra Miranda, Bahareh Afshari and Thomas Studer. I am very lucky to have had the opportunity to work with so many talented logicians and hope to continue our collaboration in the future. At this place a special thanks goes to Bahareh, who has guided my academic journey ever since I started my master's degree in Amsterdam. As my master's thesis supervisor, she introduced me to the topic of modal fixed point logics, helped me find a PhD position and continued her support throughout my doctoral studies. I also thank the logic groups in Bern and Ghent for the good working atmosphere, the interesting seminars and the social activities.

The last four years have by no means been easy for me. Apart from the pressure and the relative solitude that a PhD life brings, I have also struggled with my mental health. It is important in such situations to have a good social network. Luckily, I have been blessed with good friends and a supportive family. I would like to thank my parents Christoph and Beatrice for their continued support throughout my entire life and in particular in the recent weeks and months. I thank my brother Michael and his (soon to be) wife Luza for their care and visits after the submission of my thesis. A special thanks goes to my friends who have always supported me despite my struggles. Thank you  Bas, Wijnand, Hugh Mee, Simon, Gabriel, Jiajia, Alessandro, Raphael, Tianwei, Irina, Massimo, Tong and Freddy. \\    

\noindent Lukas Zenger \\
\noindent Bern, October 2025

\tableofcontents

\mainmatter

\chapter{Introduction}\label{c: Introduction}

The subject of this thesis is \emph{intuitionistic dynamic logic}: a family of logical systems characterized by two key features. First, they extend intuitionistic propositional logic with modalities and with fixed point operators. As such they are intuitionistic versions of modal fixed point logics. Second, they provide a formal framework for reasoning about change, whether in mathematical structures evolving over time or in the knowledge state of an agent after an information update. This chapter offers an informal overview of intuitionistic dynamic logics, outlines the motivation for their study, and summarizes the main contributions of the thesis. The formal definitions and results are developed in the subsequent chapters.

\section{What Are Intuitionistic Dynamic Logics?}

Intuitionistic dynamic logics build on two foundations: intuitionistic logic and intuitionistic modal logic, further enriched with fixed point operators.  \smallskip

\noindent \textbf{Intuitionistic propositional logic}\index{logic!intuitionistic propositional} ($\mathsf{IPL}$) shares the language of classical propositional logic\index{logic!classical propositional} ($\mathsf{CPL}$) but deviates from the latter in one crucial aspect: classical logic assumes the \emph{law of excluded middle}\index{law of excluded middle}, which states that for any proposition $p$, the formula $p \vee \neg p$ is true. As a consequence, every formula in classical logic is considered to be either true or false. Intuitionistic logic, on the other hand, rejects the law of excluded middle, which makes it a proper subsystem of $\mathsf{CPL}$. Intuitionistic logic was developed by Heyting~\cite{Heyting1930a, Heyting1930b, Heyting1930c} with regard to the foundations of mathematics and has emerged as the logical basis of constructive mathematics~\cite{Moschovakis_2024}. $\mathsf{IPL}$ has also found other interpretations based on topology, computation and, most relevant here, \emph{information}.

In the information based interpretation of $\mathsf{IPL}$ due to Kripke \cite{Kripke_1965}, an \emph{information state}\index{information!state} is a set $w$ of propositions, each  representing a piece of information available at $w$. Propositions absent from $w$ are not considered false but rather unsupported by the available data. The law of excluded middle is therefore not expected to hold, which motivates the use of intuitionistic instead of classical logic to reason about information. Kripke introduced formal semantics for $\mathsf{IPL}$ based on \emph{information orderings}\index{information!ordering}: tuples $(W, \leq)$ where $W$ is a set of information states and $w \leq v$ means that the information state $v$ contains all information of $w$ and possibly more. In this semantics truth is monotone in $\leq$: if $w \models \varphi$ and $w \leq v$, then $v \models \varphi$. The evaluation of implications is given by $w \models \varphi \rightarrow \psi$ if for all $v \geq w$ if $v \models \varphi$, then $v \models \psi$. An implication $\varphi \rightarrow \psi$ thus expresses the consequences of gaining information: ${w \models \varphi \rightarrow \psi}$ expresses that whenever $w$ is extended by information making $\varphi$ true, the information also makes $\psi$ true.

 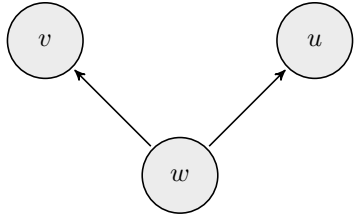
\begin{figure}[t]
    \centering
    \begin{tikzpicture}[modal]
        \node[world](w){$w$};
        \node[world](v)[above left=of w]{$v$};
         \node[world](u)[above right=of w]{$u$};
        \path[->] (w) edge  (v);
        \path[->] (w) edge  (u);
    \end{tikzpicture}
     \caption{A simple information ordering representing the situation about P vs NP outlined in Example \ref{e: information ordering1}. The arrows represent the relation $\leq$.}
    \label{f: information ordering1}
\end{figure}

\begin{example}\label{e: information ordering1}
Let $w$ be an information state which contains all current mathematical knowledge and consider the P vs. NP problem. Clearly, the statements $P = NP$ and $P \not = NP$ are absent in $w$, since the conjecture is unresolved. Two extensions of $w$ are therefore $v$ (containing $P = NP$) and $u$ (containing $P \not = NP$). The information ordering is depicted in Figure \ref{f: information ordering1}. A well-known result in complexity theory states that if $P=NP$, then also $NP = coNP$, implying that $NP=coNP$ is contained in $v$ as well. Hence $w \models (P= NP) \rightarrow (NP = coNP)$. This formula therefore expresses a true statement about the consequences of gaining the information $P=NP$ at $w$.
 \end{example} \smallskip

\noindent \textbf{Intuitionistic modal logic}\index{logic!intuitionistic modal} ($\mathsf{IML}$) extends $\mathsf{IPL}$ by modal operators, typically $\lb$ and $\ld$. Traditionally, $\lb \varphi$ expresses that $\varphi$ is necessarily true and $\ld \varphi$ that $\varphi$ is possibly true. Other interpretations include $\lb \varphi$ as $\varphi$ is known in epistemic logic or $\varphi$ is true in the next time step in temporal logic. Semantically, $\mathsf{IML}$ employs \emph{dynamic models} which are based on information orderings $(W, \leq)$ extended with a binary relation $R \subseteq W \times W$. In difference to classical logic, where truth is static, the evaluation of formulas on dynamic models takes into account higher worlds relative to $\leq$ as well as the worlds accessible over $R$. Truth in $\mathsf{IML}$ is therefore \emph{dynamic}. 

The information theoretic interpretation of $\mathsf{IPL}$ naturally extends to $\mathsf{IML}$, where the relation $R$ is used to model non-monotonic information updates (e.g. updates that revise or discard information). The formula $\ld \varphi$ expresses that there exists an update after which $\varphi$ is true, while $\lb \varphi$ expresses that $\varphi$ is true after every update. This yields the interpretation of $\mathsf{IML}$ as logics to reason about monotone and non-monotonic information updates.

The study of intuitionistic modal logic originated in the seminal work of Fitch~\cite{Fitch_1948} in 1948. Since then $\mathsf{IML}$ has grown into an active research area within logic, motivated by philosophical reasons~\cite{Simpson_1994, Artemov_2016, proietti_2012} and by applications to computer science~\cite{Hirai_2010, kavvos_2016, depaiva_2021}. \smallskip

\noindent \textbf{Intuitionistic dynamic logics}\index{logic!intuitionistic dynamic} further extend $\mathsf{IML}$ with the capacity to express infinitary properties. To that end the language is extended by operators whose truth conditions are characterized as fixed points of specific functions, yielding the name \emph{fixed point operators}\index{fixed point operator}. Such operators are not definable in the language of $\mathsf{IML}$ and significantly increase its expressive power.

\begin{example}
    In the information theoretic interpretation of $\mathsf{IML}$, $w \models \lb \varphi$ expresses that $\varphi$ is true after any information update of $w$. Moreover, $w \models \lb^n \varphi$ where
    \begin{equation*}
        \lb^n \varphi = \underbrace{\lb \ldots \lb}_{n \text{ times }} \varphi
    \end{equation*}
    expresses that $\varphi$ is true after any $n$ updates. The statement that $\varphi$ is true after any finite number of updates is, however, not expressible, since such a statement corresponds to the infinite formula
    \begin{equation*}\label{e: introduction1}
    \bigwedge_{n < \omega} \lb^n \varphi = \varphi \wedge \lb \varphi \wedge \lb^2 \varphi \wedge \ldots
\end{equation*}
   which is not well-formed. A fixed point operator $\lbm$ called \emph{master modality} compactly expresses this infinite statement by equating $\lbm \varphi$ to the infinite conjunction above.
\end{example}

Apart from the master modality, many other fixed point operators will be encountered in this thesis, most notably in the context of \emph{temporal logics}. Here, fixed point operators are used to express statements about the future, such as `eventually in the future, something will be the case' or `henceforth in the future, something will always be the case'. Similar to the study of classical modal fixed point logics, most notably the modal $\mu$-calculus (see e.g.~\cite{Stirling_2001}), the addition of fixed point operators to $\mathsf{IML}$ does not only increase the expressivity of the resulting logics, but also the difficulty of their mathematical theory, mostly due to the infinite nature of these operators. An overview of the state of the art of the study of intuitionistic dynamic logic is provided in the following subsection.

\section{State of the Art}

In difference to the study of intuitionistic modal logic, the study of intuitionistic dynamic logic is arguably still in its infancy. However, in recent years several intuitionistic dynamic logics have been introduced and studied in the literature. Among those are intuitionistic versions of temporal logic, which have been mainly studied in two different contexts: \emph{metaprogramming} and \emph{topological dynamics}. The former involves the addition of temporal operators to $\lambda$-calculi to model aspects of metaprogramming such as staged computation (see e.g.~\cite{Davies_2001, Taha_2003, Yuse_2006}). The latter is of particular interest to this thesis and dates back to the \emph{topological logic project}. This project aims to develop computationally well-behaved logical systems to reason about \emph{topological dynamic systems}, i.e. topological spaces equipped with a continuous function, with potential applications in automated theorem proving in mind. Topological dynamic systems are applied in diverse fields ranging from biology to theoretical computer science. The origin of the project dates back to the seminal work of Artemov, Davoren and Nerode~\cite{Artemov_1999} in the late 1990's, who introduced a classical modal logic and showed how to interpret formulas on topological dynamic systems. In particular, the logic features a standard modal box operator interpreted as the interior operator on topological spaces in the sense of Tarski~\cite{Tarski_1938} as well as a modal operator `next' to reason about the continuous function $f$. Their logic was extended by Kremer and Mints~\cite{Kremer_2005} with the fixed point point `henceforth' to reason about the asymptotic behaviour of $f$, resulting in a logic called dynamic topological logic ($\mathsf{DTL}$). Unfortunately, $\mathsf{DTL}$ was later shown to be undecidable~\cite{Konev_2006}, with which interest in $\mathsf{DTL}$ waned. In 2018, Fern\'andez-Duque introduced an intuitionistic version of linear temporal logic, called $\iLTL$ in this thesis, and showed that it enjoys a natural interpretation both over topological dynamic systems as well as dynamic models based on information orderings~\cite{fernandez-duque_2018}. Furthermore, he established the decidability of $\iLTL$, thus providing a decidable logic capable of expressing interesting properties of topological dynamic systems. The logic $\iLTL$ is an extension of intuitionistic propositional logic with a modality `next' as well as a fixed point operator `eventually' and thus an intuitionistic dynamic logic in our sense. It was later shown that $\iLTL$ admits a sound and complete axiomatization~\cite{Boudou_2022} and that $\iLTL$ extended with additional fixed point operators remains decidable~\cite{Boudou_2017}. The proof theory of $\iLTL$, however, remains largely unexplored, with the exception of a cyclic sequent calculus introduced recently by Men\'endez Turata in his PhD thesis~\cite{menendez_2024}.

Apart from intuitionistic temporal logics, intuitionistic dynamic logics have also been studied in an epistemic setting. Jäger and Marti introduced an intuitionistic version of common knowledge logic called $\mathsf{ICK}$~\cite{jager-intuitionistic_2016, marti_2017}, which extends intuitionistic propositional logic by modal operators $\K_i$, where $\K_i \varphi$ expresses that an agent $i$ \emph{knows} $\varphi$, as well as a fixed point operator expressing a powerful notion of group knowledge called \emph{common knowledge}. Furthermore, variants of $\ICK$ with operators for \emph{distributed knowledge} were also studied in~\cite{jager-distributed_2016, murai_2022, murai_2024}. In difference to $\iLTL$, the work in $\ICK$ is primarily proof theoretic, highlighted by the development and study of axiomatizations and sequent calculi. Artemov and Protopopescu introduced a different version of intuitionistic epistemic logic called $\mathsf{IEL}$~\cite{Artemov_2016}. In difference to $\ICK$, which treats knowledge as in the classical case, Artemov and Protopopescu aim to give an intuitionistic reading to knowledge, resulting in a substantially different logical systems.

Last but not least, intuitionistic versions of the modal $\mu$-calculus have recently been studied by Pacheco~\cite{Pacheco_2024} who focuses on game semantics and by Afshari and Grotenhuis~\cite{afshari2025intuitionistic} who develop non-wellfounded and cyclic proof systems for intuitionistic $\mu$-calculus with the Lewis arrow.

\section{Motivation}

The original motivation for the research presented in this thesis is related to the dynamic topological logic project and the logic $\iLTL$. As mentioned in the previous subsection, the proof theory of $\iLTL$ is largely unexplored, a gap which needs to be filled when it comes to application in automated theorem proving. Moreover, while a sound and complete axiomatization for $\iLTL$ was provided in~\cite{Boudou_2022}, the problem of finding an axiomatization for the extended language with `henceforth' has remained open for several years. Importantly, while the set of validities of $\iLTL$ evaluated over topological models coincides with the set of validities over dynamic models based on information orderings, the same is not true in the extended language (see Chapter \ref{c: biLTL}). The main goal of this thesis was therefore to develop the proof theory of $\iLTL$ and a sound and complete axiomatization for the extended language of $\iLTL$ with `henceforth', in the setting of dynamic models based on information orderings. The presence of fixed point operators complicates this venture, due to the fact that proof systems require a formal counterpart to induction to reason about fixed point operators. Luckily, modal fixed point logics have been studied extensively in the classical realm and it is the methods developed there that come to the rescue. In the classical realm, automata, games, non-wellfounded and cyclic proofs as well as semantic techniques such as filtrations and simulations have proven suitable for the study of fixed points in modal logic~\cite{graedel_2001, Demri_2016, Rowe_?}. Little is known, however, whether these techniques are applicable to the intuitionistic case. In fact, it can be expected that the application of such methods is significantly complicated, due to the underlying intuitionistic logic and the more complicated mathematical structures used to evaluate formulas (i.e. dynamic models). A second motivation, based in mathematical curiosity, is therefore to investigate how to adapt such techniques, in particular non-wellfounded and cyclic proofs, to the intuitionistic case, with the hope that our findings will not only provide a rigorous mathematical theory for $\iLTL$, but also prove useful for the study of other, related logical systems. Thus the second motivation naturally leads to a third one, which is to explore related logics and to adapt the techniques developed for $\iLTL$ to these systems.

From a more general perspective, intuitionistic modal and dynamic logics have found tangible applications in computer science, for example to model aspects of metaprogramming such as staged computation~\cite{Davies_2001, depaiva_2021, jeltsch_2013, kamide_2010, Taha_2003, Maier_2004} and to philosophy, for example to give an intuitionistic account of knowledge~\cite{Artemov_2016, Williamson_1992, proietti_2012}. Furthermore, the information theoretic interpretation of intuitionistic dynamic logics opens the possibility for applications to artificial intelligence, such as to knowledge representation, (epistemic) planning or logic programming. These applications in turn motivate the development of a rigorous mathematical theory, and in particular well-behaved proof systems that facilitate proof-search algorithms. Moreover, applications to computer science lead to the study of the computational properties of intuitionistic dynamic logics.

\section{Contributions}

This thesis studies five intuitionistic dynamic logics:
\begin{enumerate}
    \item \emph{Intuitionistic master modality} ($\IM$): an intuitionistic dynamic logic which extends $\mathsf{IPL}$ by a single modality $\lb$ and the aforementioned master modality $\lbm$.
    
    \item \emph{Intuitionistic common knowledge logic} ($\ICK$): an intuitionistic version of epistemic logic with \emph{common knowledge} introduced by Jäger and Marti~\cite{jager-intuitionistic_2016}.
    
    \item \emph{Intuitionistic linear temporal logic} ($\iLTL$): an intuitionistic version of linear temporal logic introduced by Fern\'andez-Duque~\cite{fernandez-duque_2018}.
    
    \item \emph{Bi-intuitionistic modal logic} ($\mathsf{biML}$): an extension of $\mathsf{IPL}$ with modalities $\lb$ and $\ld$ as well as the co-implication connective from bi-intuitionistic logic. 
    
    \item \emph{Bi-intuitionistic linear temporal logic} ($\biltl$): a bi-intuitionistic version of linear temporal logic.
\end{enumerate}

 Apart from introducing new logics, namely $\mathsf{biML}$ and $\mathsf{biLTL}$, the main contributions of this thesis are threefold: semantic, syntactic and computational. \smallskip

\noindent On the semantic side we study various classes of dynamic models which satisfy different \emph{confluence conditions} and \emph{frame conditions}. The former are interaction principles between $\leq$ and $R$, which play an important role in the study of intuitionistic modal logics. The latter are conditions placed on $R$, such as transitivity. We establish expressivity results, and show that several logics satisfy the finite model property. The most important contribution is the proof of the finite model property for $\mathsf{biML}$ and $\biltl$, which is shown via an intricate combinatorial analysis of the semantics by employing \emph{labelled} and \emph{quasi models}. \smallskip

\noindent On the syntactic side, we develop several proof systems in form of Hilbert-style axiomatizations as well as cyclic and non-wellfounded sequent calculi. We show how to adapt classical techniques such as canonical model constructions and proof search arguments to obtain soundness and completeness results. In some cases, such as for $\IM$ or for $\ICK$, an adaption is relatively straightforward. For other cases, however, such as $\biltl$, adapting the classical techniques to obtain completeness is not possible and we are forced to develop different methods. We highlight the development of a sound and complete axiomatization for an extension of the language of $\iLTL$ with the fixed point operator `henceforth'; a problem that has been open for several years.  \smallskip


\noindent Finally, on the computational side, we establish decidability and complexity results for the problem of checking validity for several intuitionistic dynamic logics. The most important results are, on the one hand, a precise complexity bound for $\ICK$ over S5 models, and on the other hand, the proofs that $\mathsf{biML}$ and $\mathsf{biLTL}$ are decidable.  



\section{Structure of the Thesis and Sources of the Material}

Chapter \ref{c: preliminaries} serves as an introduction to the mathematical formalisms that are used throughout the thesis. In particular we formally introduce $\mathsf{IPL}$ and $\mathsf{IML}$, provide a brief introduction to proof theory and discuss fixed point operators. Furthermore the history of $\mathsf{IPL}$ and $\mathsf{IML}$ is discussed. Chapters \ref{c: IM} - \ref{c: biLTL} each study one of the five intuitionistic dynamic logics mentioned before. Chapters \ref{c: IM} - \ref{c: iLTL} are based on~\cite{afshari_intuitionistic_2024, afshari_ill-founded_2023,rooduijn_analytic_2022}; the first two are joint publications with Bahareh Afshari, Lide Grotenhuis and Graham E. Leigh and the last is a joint publication with Jan Rooduijn. Chapters \ref{c: bi-int ml new} and \ref{c: biLTL} are based on~\cite{fernandez-duque_family_2023, fernandez-duque_sound_2024}, both of which are joint publications with David Fern\'andez-Duque and Brett McLean. A joint publication with Borja Sierra-Miranda and Thomas Studer~\cite{sierra-miranda_coalgebraic_2024} about non-wellfounded proof theory is not discussed in this thesis.

\chapter{Preliminaries}\label{c: preliminaries}

This chapter introduces the logical systems upon which the thesis is built. We begin by setting up notational conventions and basic definitions. Then we formally introduce intuitionistic propositional logic, in particular its syntax and its semantics, which is defined over intuitionistic Kripke models. Moreover, we give a general definition of proof system and introduce a Hilbert-style axiomatization and a Gentzen-style sequent calculus for intuitionistic propositional logic; the axiomatizations and sequent calculi presented later on are based on these two systems. Afterwards we formally introduce intuitionistic modal logic, namely its syntax and its semantics, which is defined over dynamic models: extensions of intuitionistic Kripke models by a modal accessibility relation. Furthermore, we discuss a range of confluence conditions and frame conditions. Finally, we discuss fixed point operators and then finish the chapter with a historical overview of intuitionistic (modal) logic.

\section{Basic Definitions}\label{c: preliminaries, section basic definitions}

 This section introduces the notation and the basic mathematical objects, such as partial functions or trees, used throughout the rest of the thesis.
 
 We denote the set of \emph{natural numbers} by $\omega$ and write $n < \omega$ to denote that $n$ is a natural number. Given a set $X$, the \emph{power set} of $X$ is denoted by $\mathcal{P}(X)$ and the \emph{cardinality} by $\cardinality{X}$.

 \subsection{Binary Relations}
 
 Given sets $X$ and $Y$ a \emph{binary relation}\index{binary relation} is a subset $R \subseteq X \times Y$ of the cartesian product of $X$ and $Y$. Given $x \in X$ and $y \in Y$ we write both $(x,y) \in R$ and  $x \Rel y$ to denote that $x$ and $y$ are related by $R$. Moreover we denote by $dom(R)$ the \emph{domain}\index{binary relation!domain} and by $ran(R)$ the \emph{range}\index{binary relation!range} of $R$. Formally,
\begin{equation*}
    \begin{split}
        dom(R) &\coloneqq \{ x \in X \mid \text{ there exists } y \in Y \text{ with } x \mathrel{R} y \};\\
        ran(R) & \coloneqq \{ y \in Y \mid \text{ there exists } x \in X \text{ with } x \mathrel{R} y \}.\\
    \end{split}
\end{equation*}
If $x \in dom(R)$ we write $\defined{R(x)}$ and otherwise $ \undefined{R(x)}$. Given $U \subseteq X$ and $V \subseteq Y$, we write
\begin{equation*}
    R{\upharpoonright}_{U \times V} \coloneqq \{(x,y) \in U \times V \mid (x,y) \in R\}
\end{equation*}
for the restriction of $R$ to $U$ and $V$. Given two binary relations $R_1 \subseteq X \times Y$ and $R_2 \subseteq Y \times Z$, the \emph{composition}\index{binary relation!composition} of $R_1$ with $R_2$ is denoted by $R_2 \circ R_1$, formally
\begin{equation*}
    R_2 \circ R_1 \coloneqq \{(x,z) \in X \times Z \mid \text{ there exists } y \in Y \text{ with } (x,y) \in R_1 \text{ and } (y,z) \in R_2\}.
\end{equation*}

We are interested in specific binary relations, which are introduced below.

\begin{definition}
    Let $X$ and $Y$ be sets.
    \begin{enumerate}
        \item A \emph{partial function}\index{binary relation!partial function} from $X$ to $Y$ is a binary relation $f \subseteq X \times Y$ which is \emph{functional}: for all $x \in X$ and $y,y' \in Y$ if $(x,y) \in f$  and $(x,y') \in f$, then $y = y'$.
        \item A \emph{function}\index{binary relation!function} is a partial function $f\subseteq X \times Y$ such that $dom(f) = X$.
    \end{enumerate}
\end{definition}

We denote (partial) functions by $f: X \longrightarrow Y$ and write $f(x) = y$ instead of $x \mathrel{f} y$. Functions are always assumed to be total; whenever a function is partial, we will explicitly say so.

\begin{definition}
    Let $X$ be a set and $R \subseteq X \times X$.
    \begin{enumerate}
        \item $R$ is a \emph{partial order}\index{binary relation!partial order} on $X$ if
        \begin{enumerate}
            \item $R$ is reflexive: for all $x \in X$: $x \Rel x$;
            \item $R$ is transitive: for all $x,y,z \in X$: $x \Rel y$ and $y \Rel z$ imply $x \Rel z$;
            \item $R$ is antisymmetric: for all $x,y \in X$: $x \Rel y$ and $y \Rel x$ imply $x=y$.
        \end{enumerate}
        \item $R$ is a \emph{well-order}\index{binary relation!well-order} on $X$ if
        \begin{enumerate}
            \item $R$ is a partial order on $X$;
            \item $R$ is total: for all $x,y \in X$: $x \Rel y$ or $y \Rel x$;
            \item Every non-empty $U \subseteq X$ has a $R$-least element: there exists $x \in U$ such that for all $y \in U$: $x \Rel y$.
        \end{enumerate}
        \item $R$ is an \emph{equivalence relation}\index{binary relation!equivalence relation} on $X$ if
        \begin{enumerate}
            \item $R$ is reflexive;
            \item $R$ is transitive;
            \item $R$ is symmetric: for all $x,y \in X$: $x \Rel y$ implies $y \Rel x$.
        \end{enumerate}
    \end{enumerate}
\end{definition}

Partial orders and well-orders are denoted by $\leq$.\footnote{We also use $\leq$ for the standard smaller-equal relation on the set of natural numbers. Note that $\leq$ is a well-order on $\omega$.} As usual, we write $x < y$ if $x \leq y$ and $x \not = y$. Given a partial order $\leq$ on $X$, the tuple $(X, \leq)$ is called a \emph{poset}. If $\leq$ is a well-order on $X$, then $(X, \leq)$ is called a \emph{well-ordered set}. Given a poset $(X, \leq)$ and $x \in X$, let
\begin{align*}
    x^\uparrow &\coloneqq \{ y \in X \mid x \leq y\}\\
    x^\downarrow &\coloneqq \{ y \in X \mid y \leq x\}.
\end{align*}
We call $x^\uparrow$ the \emph{upset} of $x$ and $x^\downarrow$ the \emph{downset} of $x$.  

Equivalence relations are denoted by $\sim$. If $\sim$ is an equivalence relation on $X$ and $x \in X$, the \emph{equivalence class} of $x$ is given by
\begin{equation*}
    [x] \coloneqq \{ y \in X \mid x \sim y\}.
\end{equation*}

The \emph{quotient} of $X$ under $\sim$ is given by
\begin{equation*}
    \sfrac{X}{\sim} \coloneqq \{[x] \mid x \in X\}.
\end{equation*}

\subsection{Trees}

\emph{Trees} will be used primarily for studying formal proofs. We will usually be quite informal about the precise structure of a tree. Here, the definition of trees and related concepts are introduced.

\begin{definition}
    A \emph{tree}\index{tree} is a tuple $\mathcal{T}=(T, \leq, r)$ where
    \begin{enumerate}
        \item  $(T, \leq)$ is a poset;
        \item   $r \in T$ and for all $t \in T$: $r \leq t$;
        \item for all $t \in T$, $(t^\downarrow, \leq\upharpoonright_{t^\downarrow \times t^\downarrow})$ is a well-ordered set.
    \end{enumerate}
\end{definition}

Elements of $T$ are called \emph{nodes} and $r$ is called the \emph{root}. If $s,t \in T$ with $s < t$ and there does not exist an $s' \in T$ with $s < s' < t$, then $s$ is called the \emph{parent} of $t$ and $t$ is called a \emph{child} of $s$. If $s < t$, then $s$ is called an \emph{ancestor} of $t$ and $t$ is called a \emph{descendant} of $s$. Nodes without children are called \emph{leafs}. A \emph{path}\index{tree!path} is a sequence of nodes $(t_i)_i$ such that for all $i$ the node $t_{i+1}$ is a child of $t_i$. Given a finite path $(t_i)_{i \leq n}$ for $n < \omega$, the \emph{length} $l((t_i)_{i \leq n})$ of $(t_i)_{i \leq n}$ is defined to be $l((t_i)_{i \leq n}) \coloneqq n$.  A \emph{branch}\index{tree!branch} is a path starting at the root which is either infinite or ends in a leaf. A tree $\mathcal{T}=(T, \leq, r)$ is finite if $T$ is finite and infinite otherwise. Given a finite tree $\mathcal{T}$, the \emph{height} $h(\mathcal{T})$ of $\mathcal{T}$ is defined to be $h(\mathcal{T}) \coloneqq max\{l(\rho) \mid \rho \text{ is a branch of } \mathcal{T}\}$. A tree $\mathcal{T}$ is \emph{finitely branching}\index{tree!finite branching} if every node of $\mathcal{T}$ has only finitely many children. The following is a well-known result about finitely branching trees.

\begin{theorem}[K\H{o}nig's Lemma]
    Let $\mathcal{T}$ be a finitely branching tree. If $\mathcal{T}$ is infinite, then $\mathcal{T}$ contains an infinite branch.
\end{theorem}

A \emph{labelled tree} is a tree together with a labelling function which assigns to each node a label.

\begin{definition}
    Let $\mathcal{S}$ be a set. A \emph{$\mathcal{S}$-labelled tree}\index{tree!labelled} is a tuple $\pi = (T, \leq, r, \ell)$ where
    \begin{enumerate}
        \item $(T, \leq, r)$ is a tree;
        \item $\ell: T \longrightarrow \mathcal{S}$ is a function.
    \end{enumerate}
\end{definition}

The \emph{height} of a finite labelled tree $(T, \leq, r, \ell)$ is the height of $(T, \leq, r)$. Finally, the following notion of subtree will be used repeatedly.

\begin{definition}
    Let $\mathcal{T}=(T, \leq, r)$ be a tree and $u \in T$ a node. The \emph{subtree of $\mathcal{T}$ rooted at $u$}\index{tree!rooted subtree} is the tree $\mathcal{T}_u = (u^\uparrow, \leq\upharpoonright_{u^\uparrow \times u^\uparrow}, u)$.
\end{definition}

\section{Intuitionistic Logic}\label{c: preliminaries, Section IPL}\index{logic!intuitionistic propositional}

 Every intuitionistic dynamic logic studied in this thesis is an extension of intuitionistic propositional logic with modalities and fixed point operators. We only study propositional logics, therefore we usually call intuitionistic propositional logic simply intuitionistic logic. This section introduces the syntax and semantics of intuitionistic logic ($\mathsf{IPL}$). An overview of the history of intuitionistic logic and its many interpretations is deferred to Section \ref{c: preliminaries, section historical remarks}. \smallskip

\noindent The \emph{language} $\LIPL$ of $\mathsf{IPL}$ consists of a countable set of \emph{atomic propositions} $\Prop$, the constant \emph{falsum} $\bot$, the connectives \emph{conjunction} $\wedge$, \emph{disjunction} $\vee$ and \emph{implication} $\rightarrow$ as well as \emph{brackets} $($ and $)$. \emph{Formulas} of $\LIPL$ are given by the following grammar in Backus--Naur form:
\begin{equation*}
    \varphi ::= \bot \, \lvert \, p \, \lvert \, (\varphi \wedge \varphi) \, \lvert \, (\varphi \vee \varphi) \, \lvert \, (\varphi \rightarrow \varphi)
\end{equation*}
where $p \in \Prop$. Formulas (in every system studied subsequently) are always denoted by Greek letters $\varphi, \psi, \chi$ and $\gamma$, while propositions are denoted by Latin letters $p,q,r$. Subscripts are also used such as $\varphi_1, \varphi_2$ and so on. We write $\varphi \in \LIPL$ to denote that $\varphi$ is a formula of $\LIPL$. We omit outermost brackets when displaying formulas and simply write e.g. $\varphi \wedge \psi$ instead of $(\varphi \wedge \psi)$. In fact, brackets will only be used to give unique precedence of the connectives in longer formulas. The connectives \emph{negation} $\neg$ and \emph{if and only if} $\leftrightarrow$, as well as the constant \emph{verum} $\top$ are treated as defined connectives, where $\top \coloneqq \bot \rightarrow \bot$, $\neg \varphi \coloneqq \varphi \rightarrow \bot$ and  $\varphi \leftrightarrow \psi  \coloneqq (\varphi \rightarrow \psi) \wedge (\psi \rightarrow \varphi)$.

Formulas of $\mathsf{IPL}$ are evaluated on \emph{intuitionistic Kripke models}.

\begin{definition}
    An \emph{intuitionistic Kripke model}\index{model!intuitionistic Kripke} is a tuple $\fw =(W, \leq, V)$ where 
    \begin{enumerate}
        \item $W$ is a non-empty set;
        \item $\leq$ is a partial order on $W$;
        \item $V: W \longrightarrow \mathcal{P}(\Prop)$ is monotone in $\leq$, i.e. for all $w, v \in W$:
    \begin{equation*}
        w \leq v \text{ implies } V(w) \subseteq V(v).
    \end{equation*}
    \end{enumerate}
\end{definition}

Intuitionistic Kripke models are usually simply called \emph{Kripke models} and denoted by $\fw$ and $\fv$. Elements of $W$ are called \emph{worlds} or \emph{(information) states} and are denoted by $w,v,u$ and so on. The relation $\leq$ is called the \emph{intuitionistic order} and if $w \leq v$, then we call $v$ an \emph{intuitionistic successor} of $w$. The function $V$ is called a \emph{valuation}. A tuple $(\fw, w)$ where $\fw =(W, \leq, V)$ is a Kripke model and $w \in W$ is called a \emph{pointed Kripke model}.

\begin{remark}
    Intuitionistic Kripke models are the mathematical structures based on information orderings mentioned in Chapter \ref{c: Introduction}. The set $W$ is considered to be a set of information states, where for each state $w \in W$ the valuation $V$ assigns to $w$ the information it contains. The intuitionistic order $\leq$ models extensions of the information state, where the monotonicity of $V$ in $\leq$ guarantees that extensions of $w$ contain at least as much information as $w$ itself.
\end{remark}

\begin{definition}\label{d: truth relation for intuitionistic Kripke models}
    The \emph{truth relation} $\models$ between worlds of a Kripke model $\fw= (W, \leq, V)$ and formulas is defined inductively as follows, where $p \in \Prop$ and $w \in W$.
    \begin{center}
    \begin{tabular}{l l l}
    $\fw,w \not \models \bot$ & & \\
    $\fw,w \models p$ & iff & $p \in V(w)$\\
    $\fw,w \models \varphi \wedge \psi$ & iff & $\fw, w \models \varphi$ and $\fw,w \models \psi$\\
    $\fw,w \models \varphi \vee \psi$ & iff & $\fw, w \models \varphi$ or $\fw,w \models \psi$\\
    $\fw,w \models \varphi \rightarrow \psi$ & iff & for all $v \geq w$ if $\fw,v \models \varphi$, then $\fw,v \models \psi$\\
    \end{tabular}
\end{center}
If $\fw, w \models \varphi$, then $\varphi$ is called \emph{true} at $w$.
\end{definition}

 A formula $\varphi$ is called \emph{satisfiable} over the class of Kripke models if there exists a Kripke model $\fw=(W, \leq, V)$ and a world $w \in W$ with $\fw,w \models \varphi$ and \emph{unsatisfiable} otherwise. Furthermore, $\varphi$ is called \emph{valid} over the class of Kripke models if $\fw, w \models \varphi$ for all Kripke models $\fw=(W, \leq, R)$ and all worlds $w \in W$ and \emph{falsifiable} otherwise. 

\begin{definition}
    The set of valid $\LIPL$-formulas over the class of intuitionistic Kripke models is denoted by $\mathbf{IPL}$.
\end{definition}

For the defined connectives a simple computation shows that the following truth conditions hold. The proof is omitted.

\begin{lemma}
    Let $(\fw,w)$ be a pointed Kripke model and $\varphi, \psi \in \LIPL$. The following hold.
    \begin{center}
    \begin{tabular}{l l l}
    $\fw, w \models \top$ & & \\
      $\fw, w \models \neg \varphi$ & iff & for all $v \geq w$, $\fw, v \not \models \varphi$ \\
      $\fw, w \models \varphi \leftrightarrow \psi$ & iff & for all $v \geq w$, $\fw, v \models \varphi$ if and only if $\fw, v \models \psi$
    \end{tabular}
\end{center}
\end{lemma}

A defining property of Kripke semantics for intuitionistic logic is the Monotonicity Lemma, which states that the truth relation is monotone in $\leq$. 

\begin{lemma}[Monotonicity]\label{l: monotonicity for intuitionistic Kripke models}
    Let $\fw =(W, \leq, V)$ be a Kripke model, $w,v \in W$ worlds and $\varphi$ a formula. If $w \leq v$ and $\fw, w \models \varphi$, then $\fw, v \models \varphi$.
\end{lemma}
\begin{proof}
    The proof is by induction on the structure of $\varphi$. The case for $\varphi= \bot$ trivially holds. For $\varphi = p \in \Prop$ if $\fw, w \models p$, then $p \in V(w)$. Since $w \leq v$, the monotonicity of $V$ implies that $p \in V(v)$. Hence $\fw, v \models p$. The cases for $\varphi = \psi \wedge \chi$ and $\varphi = \psi \vee \chi$ follow directly from the truth conditions for $\wedge$ and $\vee$ and from the induction hypothesis.
   Suppose $\varphi = \psi \rightarrow \chi$ and $\fw,w \models \psi \rightarrow \chi$. By definition, for all $u \geq w$ if $\fw,u \models \psi$, then $\fw,u \models \chi$. If $u \geq v$, then the transitivity of $\leq$ implies that $u \geq w$ and so if $\fw, u \models \psi$, then $\fw, u \models \chi$. Hence $\fw, v \models \psi \rightarrow \chi$.
\end{proof}

We finish the section by stating two properties of the Kripke semantics for intuitionistic logic. Recall that the \emph{law of excluded middle} states that for any proposition $p$ the formula $p \vee \neg p$ is valid. 

\begin{figure}[t]
    \centering
    \begin{tikzpicture}[modal]
        \node[world](w){$w$} ;
        \node[world](v)[above=of w]{$v$};
        \path[->] (w) edge node[left] {$\leq$} (v);
    \end{tikzpicture}
    \caption{A simple model witnessing the failure of the law of excluded middle. The reflexive edges are not displayed.}
    \label{f: example lem fails}
\end{figure}
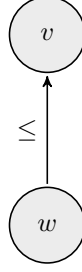

\begin{lemma}
   The law of excluded middle\index{law of excluded middle} does not hold for intuitionistic logic: $p \vee \neg p$ is not valid over the class of intuitionistic Kripke models.
\end{lemma}
 \begin{proof}
     Consider the intuitionistic Kripke model $\fw =(W, \leq, R)$ where $W =\{ w, v\}$, $\leq = \{(w,w), (w,v), (v,v)\}$ and $V(w) = \emptyset$ and $V(v) = \{p\}$ (see Figure \ref{f: example lem fails}).\footnote{Technically speaking, the displayed graph is a \emph{frame} and not a model, since the valuation is not displayed. Since whenever a frame is depicted, the valuation is provided in the example, we do not make this distinction here.} It is clear that $\fw$ is a Kripke model. By definition $\fw, w \not \models p$. Moreover, $\fw, v \models p$, implying that $\fw, w \not \models \neg p$. Hence $\fw, w \not \models p \vee \neg p$.
 \end{proof}

 A Kripke model $\fw=(W, \leq, V)$ is \emph{finite} if $W$ is a finite set. Intuitionistic logic satisfies the \emph{finite model property}\index{finite model property! intuitionistic propositional logic} if every falsifiable formula can be falsified in a finite Kripke model. The following is for example proven in~\cite{VanDalen_1986}.
 
 \begin{lemma}
     Intuitionistic logic satisfies the finite model property.
 \end{lemma}

\section{Proof systems}\label{c: preliminaries, Section proof systems}

An important tool used in our study of intuitionistic dynamic logics are \emph{proof systems}. Intuitively, a proof system consists of a set of axioms - formulas assumed to be true without proof - and a set of inference rules - rules that govern how to prove new formulas from axioms and formulas which have already been proven. The rules of a proof system do not refer to the semantics of the logic. Instead they are purely mechanical and dictate how to manipulate the syntactic structure of formulas to obtain new ones. Accordingly, just as the formal semantics of a logic determine the set of valid formulas, a proof system determines the set of \emph{provable} formulas. The bridge between the semantic and syntactic realm are \emph{soundness}\index{soundness} and \emph{completeness}\index{completeness} proofs: soundness establishes that every provable formula is valid while completeness establishes that every valid formula is provable. 

We will study two types of proof systems: (Hilbert-style) axiomatizations and (Gentzen-style) sequent calculi. Regarding the sequent calculi, we will in particular study \emph{non-wellfounded} and \emph{cyclic proofs}. This section introduces the basic definitions shared by all systems encountered in the thesis. Afterwards we introduce an axiomatization and a sequent calculus for $\mathsf{IPL}$. A discussion of non-wellfounded and cyclic proofs is deferred to Chapter~\ref{c: IM}.

\subsection{General Definitions}

Some proof systems manipulate formulas while others manipulate more complicated mathematical structures. Therefore we give a general definition of inference rules and proof systems relative to an arbitrary set $\mathcal{S}$.

\begin{definition}\label{d: rule instances and rules}
    Let $\mathcal{S}$ be a set. 
    \begin{enumerate}
    \item A \emph{rule instance}\index{proof system!rule instance} (w.r.t. $\mathcal{S}$) is a finite tuple $\langle s, \langle s_1, \ldots, s_n\rangle \rangle$ where $s, s_1, \ldots s_n \in \mathcal{S}$ and $0 \leq n < \omega$. The first element $s$ is called the \emph{conclusion} and the elements $s_i$  for $1 \leq i \leq n$ in the second component are called \emph{premises}.
    \item  An \emph{(inference) rule}\index{proof system!inference rule} $\mathsf{r}$ (w.r.t. $\mathcal{S}$) is a set of rule instances.
    \item A \emph{proof system}\index{proof system} $\mathrm{P}$ (w.r.t. $\mathcal{S}$) is a finite set of inference rules.
\end{enumerate}
\end{definition}

Given a rule instance  $\langle s, \langle s_1, \ldots, s_n\rangle \rangle$, its \emph{arity} is $n$. Rule instances with arity $0$ are called \emph{axioms}\index{proof system!axiom schema}. A rule instance $\langle s, \langle s_1, \ldots, s_n \rangle \rangle \in \mathsf{r}$ is depicted as
\begin{equation*}
    \infer[\mathsf{r}]{s}{s_1 & \ldots & s_n}
\end{equation*}
with the intended reading that if the premises $s_1, \ldots, s_n$ are obtained, then we can derive the conclusion $s$. Inference rules where each rule instance is an axiom are called \emph{axiom schemes}. Inference rules that are not axiom schemes are called \emph{proper}. We will mostly consider inference rules where each rule instance has the same arity.\footnote{However, there are some exceptions, for example the \emph{choice rules} employed in Chapter \ref{c: IM}.} Such inference rules will be depicted schematically, i.e. we depict the rule by providing one rule instance from which all other instances can be obtained via uniform substitution.

\begin{definition}
    Let $\mathrm{P}$ be a proof system (w.r.t. $\mathcal{S}$) and $\pi =(T, \leq, r, \ell)$ a $\mathcal{S}$-labelled tree. Then $\pi$ is \emph{labelled according to the rules of} $\mathrm{P}$ if  whenever $t,t_1, \ldots, t_n \in T$ with $t$ being the parent and $t_1, \ldots, t_n$ the children of $t$ in $\pi$, $\langle \ell(t), \langle \ell(t_1), \ldots, \ell(t_n)\rangle \rangle$ is a rule instance of a rule in $\mathrm{P}$. 
\end{definition}

The two types of proof systems studied in this thesis are axiomatizations and sequent calculi. These two types of systems differ with respect to the set of elements they manipulate.  Axiomatizations manipulate formulas of a logical language (e.g. $\LIPL$). Sequent calculi, on the other hand, manipulate so called \emph{sequents}, which have more structure.

\subsection{An Axiomatization for Intuitionistic Logic}

This subsection introduces an axiomatization for intuitionistic logic. The axiomatization manipulates $\LIPL$-formulas. An axiomatization typically consist of a list of axiom schemes together with a few proper inference rules. The presented system is taken from~\cite{Bezhanishvili_?}.

\begin{definition}
    The Hilbert-style axiomatization\index{axiomatization!$\mathrm{IPL_H}$} $\mathrm{IPL_H}$ consists of the inference rules depicted in Table \ref{d: axiomatization for IPL}.
\end{definition}

\begin{table}
\centering
        \begin{tabular}{|l l|}
        \hline
         $\mathsf{1:}$ & $\varphi \rightarrow (\psi \rightarrow \varphi)$\\
         $\mathsf{2:}$ & $(\varphi \rightarrow (\psi \rightarrow \chi)) \rightarrow ((\varphi \rightarrow \psi) \rightarrow (\varphi \rightarrow \chi))$\\
         $\mathsf{3:}$ & $(\varphi \wedge \psi) \rightarrow  \varphi$\\
         $\mathsf{4:}$ & $(\varphi \wedge \psi) \rightarrow  \psi$\\
         $\mathsf{5:}$ & $\varphi \rightarrow (\varphi \vee \psi)$\\
         $\mathsf{6:}$ & $\psi \rightarrow (\varphi \vee \psi)$\\
         $\mathsf{7:}$ & $(\varphi \rightarrow \chi) \rightarrow ((\psi \rightarrow \chi) \rightarrow ((\varphi \vee \psi) \rightarrow \chi))$\\
         $\mathsf{8:}$ & $\bot \rightarrow \varphi$\\
         $\mathsf{MP:}$ & $\infer{\psi}{\varphi & \varphi \rightarrow \psi}$ \\
         \hline
        \end{tabular}
     \caption{The axiomatization $\mathrm{IPL_H}$}
     \label{d: axiomatization for IPL}
\end{table}

The rules $\mathsf{1}$ -- $\mathsf{8}$ are axiom schemes. Instead of writing for example
\begin{equation*}
    \infer[\mathsf{1}]{\varphi \rightarrow (\psi \rightarrow \varphi)}{}
\end{equation*}
for an axiom, we prefer to simply write $\varphi \rightarrow (\psi \rightarrow \varphi)$ when we list the axiomatization to clearly distinguish axiom schemes from proper inference rules. The rule $\mathsf{MP}$ is the \emph{modus ponens} rule, which will be featured in all axiomatizations considered in this thesis. It states that from $\varphi$ and $\varphi \rightarrow \psi$ we can derive $\psi$. We implicitly assume that every inference rule is closed under uniform substitution. We will therefore not explicitly add a substitution rule to our systems. In other words, each axiom scheme and inference rule can be instantiated by uniformly substituting the formulas displayed in Table \ref{d: axiomatization for IPL} by other formulas. Formal proofs in an axiomatization are called \emph{derivations}.

\begin{definition}
    A \emph{derivation}\index{derivation!$\mathrm{IPL_H}$-derivation} in $\mathrm{IPL_H}$ of a formula $\varphi$ is a finite tree $\pi$ labelled by formulas according to the rules of $\mathrm{IPL_H}$ such that the root of $\pi$ is labelled by $\varphi$ and every leaf of $\pi$ is labelled by an axiom.
\end{definition}

 If there exists a derivation of $\varphi$, then we write $\vdash_\mathrm{IPL_H} \varphi$ and say that $\varphi$ is \emph{$\mathrm{IPL_H}$-derivable}. If the proof system is clear from context, we omit the mention of $\mathrm{IPL_H}$.

\begin{example}\label{e: p implies p derivable}
    The formula $\varphi \rightarrow \varphi$ is derivable in $\mathrm{IPL_H}$. First, recall that $\top = \bot \rightarrow \bot$ which is derivable, since it is an instance of $\mathsf{8}$. Therefore we obtain the following derivation:
    \begin{scprooftree}{0.9}
    \small
        \AxiomC{}
        \RightLabel{$\mathsf{8}$}
        \UnaryInfC{$\top$}
        \AxiomC{}
        \RightLabel{$\mathsf{1}$}
        \UnaryInfC{$\top \rightarrow (\varphi \rightarrow \top)$}
        \RightLabel{$\mathsf{MP}$}
        \BinaryInfC{$\varphi \rightarrow \top$}
        \AxiomC{}
        \RightLabel{$\mathsf{1}$}
        \UnaryInfC{$\varphi \rightarrow (\top \rightarrow \varphi)$}
        \AxiomC{}
        \RightLabel{$\mathsf{2}$}
        \UnaryInfC{$(\varphi \rightarrow (\top \rightarrow \varphi)) \rightarrow ((\varphi \rightarrow \top) \rightarrow (\varphi \rightarrow \varphi))$}
        \RightLabel{$\mathsf{MP}$}
        \BinaryInfC{$(\varphi \rightarrow \top) \rightarrow (\varphi \rightarrow \varphi)$}
        \RightLabel{$\mathsf{MP}$}
        \BinaryInfC{$\varphi \rightarrow \varphi$}
    \end{scprooftree}
   It is easy to see that the above structure is a tree labelled by formulas according to the rules of $\mathrm{IPL_H}$, where the root is labelled by $\varphi \rightarrow \varphi$ and each leaf by an axiom. 
 \end{example}

  We will usually not display derivations as trees and instead simply explain which axioms to choose and what rules to apply in order to get the desired result. The following example illustrates this.

  \begin{example}\label{e: informal derivation}
     We claim that if $\varphi \rightarrow \psi$ and $\psi \rightarrow \chi$ are derivable, then so is $\varphi \rightarrow \chi$. First note that 
     \begin{equation*}
         (\psi \rightarrow \chi) \rightarrow (\varphi \rightarrow (\psi \rightarrow \chi))
     \end{equation*}
     is an instance of the axiom scheme $\mathsf{1}$. Thus since $\psi \rightarrow \chi$ is derivable we obtain 
     \begin{equation*}
         \vdash \varphi \rightarrow (\psi \rightarrow \chi)
     \end{equation*}
     by $\mathsf{MP}$. Now the following is an instance of the axiom scheme $\mathsf{2}$:
     \begin{equation*}
         (\varphi \rightarrow (\psi \rightarrow \chi)) \rightarrow ((\varphi \rightarrow \psi) \rightarrow (\varphi \rightarrow \chi)).
     \end{equation*}
     Thus applying $\mathsf{MP}$ twice yields $\vdash \varphi \rightarrow \chi$ as claimed.
  \end{example}

 The axiomatization $\mathrm{IPL_H}$ is sound and complete with respect to the class of intuitionistic Kripke models.

  \begin{theorem}[Soundness and completeness of $\mathrm{IPL_H}$]\label{t: soundness and completeness of ipl}
      Let $\varphi \in \LIPL$. Then $\varphi$ is valid over the class of intuitionistic Kripke models if and only if $\varphi$ is derivable in $\mathrm{IPL_H}$.
  \end{theorem}

  Soundness is established by a standard induction on the height of derivations. Completeness is established by a canonical model construction. For a proof, see \cite{Bezhanishvili_?}.

  \begin{definition}\label{d: intuitionistic tautologies}
      Let $\mathcal{L}$ be any language extending $\LIPL$. An $\mathcal{L}$-formula $\varphi$ is an \emph{intuitionistic tautology}\index{intuitionistic tautology} if there exists an $\mathrm{IPL_H}$-derivable $\LIPL$-formula $\varphi'$, such that $\varphi$ is obtained from $\varphi'$ via uniform substitution.
  \end{definition}

  For example, in the language extending $\LIPL$ with $\lb$ we have that $\lb \varphi \rightarrow (\lb \psi \rightarrow \lb \varphi)$ is an intuitionistic tautology, since $p \rightarrow (q \rightarrow p)$ is an axiom and we obtain the formula by substituting $p$ by $\lb \varphi$ and $q$ by $\lb \psi$.

  \subsection{A Sequent Calculus for Intuitionistic Logic}

  Hilbert-style axiomatizations provide a neat way to capture the validites of a logic. However, Examples \ref{e: p implies p derivable} and \ref{e: informal derivation} hint at a general problem of such a formalism: the presence of the modus ponens rule makes it difficult to construct a proof for a given formula, since in order to prove a formula, we have to guess the correct instances of axiom schemes and then apply modus ponens. To circumvent this issue, we will consider sequent calculi, which manipulate sequents and are generally much nicer to work with. Here, we introduce a sequent calculus for $\mathsf{IPL}$ which will be the foundation for all sequent calculi studied later on. The system is taken from~\cite{Negri_2001}.

  \begin{definition}
      A \emph{sequent}\index{sequent!for $\mathsf{IPL}$} (for $\mathsf{IPL}$) is a structure $\Gamma \Rightarrow \Delta$ where $\Gamma, \Delta$ are finite sets of $\LIPL$-formulas.
  \end{definition}

  We denote sequents by $\sigma$ and use $\Gamma_\sigma$ to refer to the left side of $\sigma$ and $\Delta_\sigma$ to refer to the right side. As usual $\Gamma, \varphi$ denotes $\Gamma \cup \{\varphi\}$. 

  \begin{convention}
For a non-empty and finite set of formulas $\Gamma = \{\varphi_1, \ldots, \varphi_n\}$, define
\begin{align*}
    \bigwedge \Gamma &\coloneqq (\varphi_1 \wedge (\varphi_2 \wedge (\ldots \wedge \varphi_n)\ldots ) \\
    \bigvee \Gamma &\coloneqq (\varphi_1 \vee (\varphi_2 \vee (\ldots \vee \varphi_n) \ldots )
\end{align*}
Furthermore, we set
    \begin{align*}
        \bigwedge \emptyset &\coloneqq \top \\
        \bigvee \emptyset & \coloneqq \bot.
    \end{align*}
\end{convention}

  The \emph{interpretation} of a sequent $\sigma$ is the formula
  \begin{equation*}
      \sigma^I \coloneqq \bigwedge \Gamma_\sigma \rightarrow \bigvee \Delta_\sigma
  \end{equation*}

 Given a pointed model $(\fw, w)$, we write $\fw, w \models \sigma$ if and only if $\fw, w \models \sigma^I$. Note that sequents are \emph{multi-conclusion}, i.e. $\Delta$ might contain more than one formula. The more standard calculus for $\mathsf{IPL}$ employs \emph{single-conclusion} sequents (see e.g. \cite{Negri_2001}), however as we will later explain, for our purposes multi-conclusion sequents are better suited.

\begin{definition}\label{d: sequent calculus for IPL}
    The sequent calculus $\mathrm{IPL_G}$\index{sequent calculus! $\mathrm{IPL_G}$} consists of the rules depicted in Table \ref{d: sequent rules for IPL}. 
\end{definition}

\begin{table}[t]
    \centering
    \begin{tabular}{|c c|}
    \hline
    & \\
      $\infer[\mathsf{id}]{\Gamma, \varphi \Rightarrow \varphi, \Delta}{}$ & $\infer[\bot]{\Gamma, \bot \Rightarrow \Delta}{}$\\
      & \\
        $\infer[\wedge \mathsf{L}]{\Gamma, \varphi \wedge \psi \Rightarrow \Delta}{\Gamma, \varphi, \psi \Rightarrow \Delta}$ 
         & $\infer[\wedge \mathsf{R}]{\Gamma\Rightarrow \varphi \wedge \psi ,\Delta}{\Gamma \Rightarrow \varphi, \Delta & \Gamma \Rightarrow \psi, \Delta}$\\
        & \\
         $\infer[\vee \mathsf{L}]{\Gamma, \varphi \vee \psi \Rightarrow \Delta}{\Gamma, \varphi \Rightarrow \Delta & \Gamma, \psi \Rightarrow \Delta}$ & $\infer[\vee \mathsf{R}]{\Gamma \Rightarrow \varphi \vee \psi, \Delta}{\Gamma \Rightarrow \varphi, \psi, \Delta}$\\
        & \\
         $\infer[{\to} \mathsf{L}]{\Gamma, \varphi \rightarrow \psi\Rightarrow \Delta}{\Gamma, \varphi \rightarrow \psi \Rightarrow \varphi, \Delta & \Gamma, \psi \Rightarrow \Delta}$ 
         & $\infer[{\to} \mathsf{R}]{\Gamma \Rightarrow \varphi \rightarrow \psi, \Delta}{\Gamma, \varphi \Rightarrow \psi}$\\
         & \\
       \hline
    \end{tabular}
    \caption{The sequent calculus $\mathrm{IPL_G}$}
    \label{d: sequent rules for IPL}
\end{table}

The rules $\mathsf{id}$ and $\bot$ are axioms. All other rules are standard inference rules from classical propositional logic, with the exception of ${\rightarrow} \mathsf{R}$, which has a single-conclusion premise. This restriction ensures that the law of excluded middle is not derivable. Note that commas in sequents are interpreted as conjunctions on the left side of the sequent arrow and as disjunctions on the right side, which explains the rules for conjunction and disjunction. In each inference rule, the distinguished formula in the conclusion is called \emph{principal} and the distinguished formulas in the premises are called \emph{residual}. For example, in the rule ${\wedge}\mathsf{L}$ the principal formula is $\varphi \wedge \psi$ and the residual formulas are $\varphi$ and $\psi$. All other formulas are called \emph{side formulas}.

\begin{definition}
    A \emph{proof}\index{proof!$\mathrm{IPL_G}$-proof} of a sequent $\sigma$ in $\mathrm{IPL_G}$ is a finite tree $\pi$ labelled by sequents according to the rules of $\mathrm{IPL_G}$ such that the root is labelled by $\sigma$ and each leaf is labelled by an axiom.
\end{definition}

If there is a proof of $\sigma$, we write $\vdash_\mathrm{IPL_G} \sigma$ and say that $\sigma$ is $\mathrm{IPL_G}$-\emph{provable}. We omit the mention of $\mathrm{IPL_G}$ if the proof system is clear from context. If $\varphi$ is a formula, then $\varphi$ is provable if and only if $\Rightarrow \varphi$ is provable.

\begin{example}
    The axiom $\mathsf{7}$ is provable in $\mathrm{IPL_G}$.
    \begin{scprooftree}{0.75}
    \AxiomC{}
    \RightLabel{$\mathsf{id}$}
    \UnaryInfC{$\varphi \rightarrow \chi, \psi \rightarrow \chi, \varphi \Rightarrow \varphi, \chi$}
    \AxiomC{}
    \RightLabel{$\mathsf{id}$}
    \UnaryInfC{$\psi \rightarrow \chi, \varphi, \chi \Rightarrow \chi$}
    \RightLabel{${\rightarrow}\mathsf{L}$}
     \BinaryInfC{$\varphi \rightarrow \chi, \psi \rightarrow \chi, \varphi \Rightarrow \chi$}
    \AxiomC{}
    \RightLabel{$\mathsf{id}$}
    \UnaryInfC{$\varphi \rightarrow \chi, \psi \rightarrow \chi, \psi \Rightarrow \psi, \chi$}
     \AxiomC{}
    \RightLabel{$\mathsf{id}$}
    \UnaryInfC{$\varphi \rightarrow \chi, \psi, \chi \Rightarrow \chi$}
    \RightLabel{${\rightarrow}\mathsf{L}$}
     \BinaryInfC{$\varphi \rightarrow \chi, \psi \rightarrow \chi, \psi \Rightarrow \chi$}
     \RightLabel{${\vee}\mathsf{L}$}
    \BinaryInfC{$\varphi \rightarrow \chi, \psi \rightarrow \chi, \varphi \vee \psi \Rightarrow \chi$}
     \RightLabel{${\rightarrow}\mathsf{R}$}
    \UnaryInfC{$\varphi \rightarrow \chi, \psi \rightarrow \chi \Rightarrow (\varphi \vee \psi) \rightarrow \chi$}
     \RightLabel{${\rightarrow}\mathsf{R}$}
    \UnaryInfC{$\varphi \rightarrow \chi \Rightarrow (\psi \rightarrow \chi) \rightarrow ((\varphi \vee \psi) \rightarrow \chi)$}
    \RightLabel{${\rightarrow}\mathsf{R}$}
    \UnaryInfC{$\Rightarrow (\varphi \rightarrow \chi) \rightarrow ((\psi \rightarrow \chi) \rightarrow ((\varphi \vee \psi) \rightarrow \chi))$}
    \end{scprooftree}
\end{example}

One of the advantages of using the sequent calculus $\mathrm{IPL_G}$ instead of the axiomatization $\mathrm{IPL_H}$ stems from the fact that the former allows for upwards proof search\index{proof search}: given a sequent $\sigma$ we find a proof by writing $\sigma$ at the root and then apply rules upwards until axioms are encountered. The rules of $\mathrm{IPL_G}$ are \emph{analytic}: all formulas occurring in the premises of a rule are subformulas of formulas in the conclusion. Hence, in a proof of $\sigma$, only finitely many formulas and thus finitely many different sequents can occur, which implies that there are only finitely many sensible candidates for a proof to check.\footnote{Technically, since sequents are ordered pairs of \emph{sets} of formulas, formulas can serve simultaneously as principal and as side formulas in a rule instance. In such an instance the principal formula also occurs in the premises. Thus it is possible to always continue the proof search even if only finitely many sequents occur; however, such strategy for proof search is hardly reasonable.} Note that this is not possible for a rule like modus ponens: in order to derive $\psi$, we have to guess a formula $\varphi$ so that we can apply modus ponens to $\varphi$ and $\varphi \rightarrow \psi$ and there are infinitely many candidates for $\varphi$. There are thus two ways to read the rules of a sequent calculus: downwards (i.e. going from premises to the conclusion) and upwards (i.e. going from the conclusion to the premises). The first way is consistent with the idea that rules preserve validity: if the premises are valid, then the conclusion is valid too. When thinking of mathematical proofs, the downwards reading is correct, since proofs are arguments starting with axioms and ending with the conclusion that is supposed to be proven. However, when performing proof search, we will read rules upwards instead. In this reading, we think of rules as preserving falsifiability: if the conclusion is falsifiable, then so is at least one of the premises. Since proofs in a sequent calculus are generally constructed by reading rules upwards and this reading makes it also easier to understand a proof, we will usually consider the upwards reading interpretation. However, we will always make it clear when applying rules whether we think of them as upwards or downwards.

\begin{theorem}[Soundness and completeness of $\mathrm{IPL_G}$]
    For any sequent $\sigma$, $\sigma^I$ is valid over the class of intuitionistic Kripke models if and only if $\sigma$ is provable in $\mathrm{IPL_G}$.
\end{theorem}

Soundness is established by a standard induction on the height of proofs. Completeness can be established in two ways: either by a canonical model construction or by embedding the axiomatization $\mathrm{IPL_H}$ into $\mathrm{IPL_G}$. For the latter we are required to extend the sequent calculus with the \emph{cut rule}
\begin{equation*}
    \infer[\mathsf{cut}]{\Gamma \Rightarrow \Delta}{\Gamma, \varphi \Rightarrow \Delta & \Gamma \Rightarrow \varphi, \Delta}
\end{equation*}
where $\varphi$ is any formula, in order to derive $\mathsf{MP}$. Note that the cut rule is not analytic, since $\varphi$ does not need to be a subformula of a formula in the conclusion. Thus to obtain completeness for $\mathrm{IPL_G}$ and to keep the property of analyticity, after embedding $\mathrm{IPL_H}$ into $\mathrm{IPL_G}$ plus the cut rule, we must show that the cut rule can be eliminated without altering the set of provable sequents. This elimination process is called \emph{cut elimination} and can be done for $\mathrm{IPL_G}$, see e.g.~\cite{Negri_2001}. For more details about sequent calculi and the structural properties of $\mathrm{IPL_G}$, the reader is referred to~\cite{Negri_2001}.

\section{Intuitionistic Modal Logic}\label{c: preliminaries, Section IML}

Intuitionistic modal logic\index{logic!intuitionistic modal} ($\mathsf{IML}$) refers to the extension of $\mathsf{IPL}$ by modalities from modal logic. Traditionally, one considers the modalities $\lb$ and $\ld$, whose interpretation is as follows: a formula $\lb \varphi$ is read as \emph{$\varphi$ is necessarily true} and a formula $\ld \varphi$ is read as \emph{$\varphi$ is possibly true}. Many other interpretations for $\lb$ and $\ld$ have been proposed and studied in the literature and some will be encountered in this thesis. In particular, in epistemic logic the modality $\lb$ is read as a knowledge operator: $\lb \varphi$ means that \emph{$\varphi$ is known}. In linear temporal logic, both $\lb$ and $\ld$ correspond to the temporal modality $\X$, where $\X \varphi$ is read as \emph{$\varphi$ is true in the next time step}. In contrast to classical modal logic, the equivalence
\begin{equation*}
   \lb \varphi \leftrightarrow \neg \ld \neg \varphi
\end{equation*}
is usually not taken to be valid in intuitionistic modal logics (e.g. Simpson's fifth criterion for an intuitionistic modal logic is that $\lb$ and $\ld$ are independent~\cite{Simpson_1994}). Nevertheless, intuitionistic modal logics satisfying the equivalence above have been proposed, for example by Bull~\cite{Bull_1965}. This section introduces the syntax and semantics of intuitionistic modal logic. Furthermore, we discuss confluence conditions and frame conditions for the models of $\mathsf{IML}$. A brief overview of the history of intuitionistic modal logic is deferred to Section \ref{c: preliminaries, section historical remarks}. \smallskip

\noindent The \emph{language} $\mathcal{L}_\mathsf{IML}$ of intuitionistic modal logic extends $\LIPL$ by the modal operators \emph{box} $\lb$ and \emph{diamond} $\ld$. \emph{Formulas} of $\mathcal{L}_\mathsf{IML}$ are given by the following grammar in Backus--Naur form:
\begin{equation*}
    \varphi ::= \bot \, \lvert \, p \, \lvert \, \varphi \wedge \varphi \, \lvert \, \varphi \vee \varphi \, \lvert \, \varphi \rightarrow \varphi \, \lvert \, \lb \varphi \, \lvert \, \ld \varphi.
\end{equation*}
where $p \in \Prop$. The modal operator $\lb^k$ for $k < \omega$ is defined inductively by $\lb^0 \varphi \coloneqq \varphi$ and $\lb^{k+1} \varphi \coloneqq \lb \lb^k \varphi$. Similarly, the modal operator $\ld^k$ is defined by $\ld^0 \varphi \coloneqq \varphi$ and $\ld^{k+1} \varphi \coloneqq \ld \ld^k \varphi$.

Formulas of intuitionistic modal logic are evaluated on intuitionistic Kripke models extended by an additional binary relation to evaluate the modalities. The resulting models are called \emph{dynamic models}.

\begin{definition}\label{d: dynamic model}
    A \emph{dynamic model}\index{model!dynamic} is a tuple $\fw =(W, \leq, R, V)$ where
    \begin{enumerate}
        \item  $(W, \leq, V)$ is an intuitionistic Kripke model;
        \item $R \subseteq W \times W$ is a binary relation.
    \end{enumerate}
\end{definition}

The relation $R$ is called the \emph{modal (accessibility) relation}. If $w \Rel v$, then $v$ is called a \emph{modal successor} of $w$. We define the relation $R^k \subseteq W \times W$ for $k < \omega$ inductively as follows. For all $w,v \in W$, $w \Rel^0 v$ if $w=v$. Moreover, $w \Rel^{k+1} v$ if there exists $u \in W$ with $w \Rel u$ and $u \Rel^k v$.

\begin{definition}
    The \emph{truth relation} $\models$ between worlds of a dynamic model ${\fw =(W, \leq, R, V)}$ and formulas extends the truth relation defined in Definition \ref{d: truth relation for intuitionistic Kripke models} with the following two clauses, where $w \in W$.
    \begin{center}
        \begin{tabular}{l l l}
        $\fw, w \models \lb \varphi$ & iff & for all $u,v \in W$ if $w \leq v$ and $v \Rel u$, then $\fw, u \models \varphi$ \\
        $\fw, w \models \ld \varphi$ & iff & for all $v \in W$ if $w \leq v$, then there exists $u \in W$ with $v \Rel u$\\
        & & and $\fw, u \models \varphi$\\
        \end{tabular}
    \end{center}
    If $\fw, w \models \varphi$, then $\varphi$ is called \emph{true} at $w$.
\end{definition}

Let $\mathcal{C}$ be a class of dynamic models and $\varphi$ a formula. Then $\varphi$ is \emph{satisfiable} over $\mathcal{C}$ if there exists a dynamic model $\fw \in \mathcal{C}$ and a world $w$ such that $\fw, w \models \varphi$, and \emph{unsatisfiable} otherwise. Moreover, $\varphi$ is \emph{valid} over $\mathcal{C}$ if for any dynamic model $\fw \in \mathcal{C}$ and any world $w$, $\fw, w \models \varphi$, and \emph{falsifiable} otherwise.

\begin{figure}[t]
    \centering
    \begin{tikzpicture}[modal]
        \node[world](w){$w$} ;
        \node[world](v)[above=of w]{$v$};
        \path[->] (w) edge (v);
        \node[world](w1)[right=of w]{$f(w)$} ;
        \node[world](v1)[above right=of w1]{$f(v)$};
        \node[world](u1)[above=of w1]{$u$};
        \node[world](s)[above=of v1]{$s$};
        \path[->] (w1) edge (v1);
        \path[->] (v1) edge (s);
        \path[->] (w1) edge (u1);
        \path[->] (w) edge[dashed] (w1);
        \path[->] (v) edge[bend right, dashed] (v1);
         \path[->] (w1) edge[reflexive right, dashed] (w1);
         \path[->] (v1) edge[reflexive right, dashed] (v1);
         \path[->] (s) edge[reflexive right, dashed] (s);
         \path[->] (u1) edge[reflexive above, dashed] (u1);
    \end{tikzpicture}
    \caption{A dynamic model. Solid arrows show the intuitionistic order and dashed arrows the modal accessibility relation. The reflexive arrows of the intuitionistic order are omitted.}
    \label{f: example dynamic model}
\end{figure}
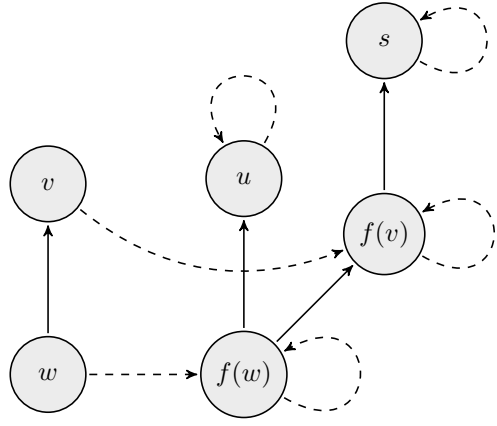

\begin{example}
    Figure \ref{f: example dynamic model} shows a dynamic model $\fw$, where the modal accessibility relation is a function $f$. If we disregard the modal relation, $\fw$ is the disjoint union of two intuitionistic Kripke models: a model $\fv_0$ rooted at $w$ and a model $\fv_1$ rooted at $f(w)$. The modal relation $f$ then models the transformation of $\fv_0$ into $\fv_1$.

    Suppose $p \in \Prop$ and $V$ is the valuation of $\fw$ such that $p \not \in V(s)$. We claim that $\fw, v \models \lb \neg p$. Note that since $p \not \in V(s)$ and $V$ is monotone in $\leq$, also $p \not \in V(f(v))$. Therefore, $\fw, f(v)  \models \neg p$. Since $f(v)$ is the only modal successor of $v$ and there are no (proper) intuitionistic successors of $v$, $\fw, v \models \lb\neg p $ as claimed. Now suppose that additionally ${p \in V(u)}$. Then $\fw, w \models \lb \neg p$ if and only $\fw, f(w) \models \neg p$ and $\fw, f(v) \models \neg p $. The latter has already been confirmed. For the former, however, note that $p \in V(u)$, implying that $\fw, f(w) \not \models \neg p $ which in turn implies that $\fw, w \not \models \lb \neg p$.
\end{example}

Note that the modalities are evaluated using both $\leq$ and $R$. This deviates from the evaluation of modalities in classical modal logic where $w \models \lb \varphi$ if and only if for all $v$ if $w \Rel v$, then $v \models \varphi$, and $w \models \ld \varphi$ if and only if there exists $v$ with $w \Rel v$ and $v \models \varphi$. This deviation is necessary since the classical truth conditions do not preserve the monotonicity property of intuitionistic logic: if for example $w \models \lb \varphi$ and $w \leq v$, then there is no reason that $v \models \lb \varphi$, since the modal successors of $w$ and $v$ might be unrelated to one another. To understand the importance of this, observe that a semantics without the monotonicity property will not yield a `logic' with the most basic property: closure under substitutions. 

\begin{example}
    Consider the formula $p \rightarrow \neg \neg p$ for $p \in \Prop$. It is easy to check that $p \rightarrow \neg \neg p$ is valid over the class of dynamic models. Now substitute $p$ for $\lb q$ where ${q \in \Prop}$. By using the classical truth conditions, the formula $\lb q \rightarrow \neg \neg \lb q$ obtained by the substitution is no longer valid. For a counterexample consider a dynamic model consisting of three worlds, say $w,v$ and $u$, such that $w \leq v$ and $v \Rel u$. The valuation is given by $V(w) = V(v) = V(u) = \emptyset$. Then it easily checked that $w \models \lb q$ (since $w$ does not have any modal successors) while $v \not \models \lb q$, since $v \Rel u$ and $u \not \models q$. Since $v$ has no intuitionistic successor, $v \models \neg \lb q$, implying that $w \not \models \neg \neg \lb q$. Consequently $w \not \models \lb q \rightarrow \neg \neg \lb q$.
\end{example}
By using the intuitionistic truth conditions for the modalities instead, monotonicity of the semantics is preserved, as illustrated in the following lemma.

\begin{lemma}[Monotonicity]\label{l: monotonicity for dynamic models}
    Let $\fw=(W, \leq, R, V)$ be a dynamic model, $w,v \in W$ and $\varphi$ a formula. If $w \leq v$ and $\fw, w \models \varphi$, then $\fw,v \models \varphi$.
\end{lemma}
\begin{proof}
    The proof is by induction on the structure of $\varphi$. The base cases as well as the cases where $\varphi = \psi \ast \chi$ for $\ast \in \{\wedge, \vee, \rightarrow\}$ are covered in Lemma \ref{l: monotonicity for intuitionistic Kripke models}. Suppose $\varphi = \lb \psi$ and $\fw,w \models \lb \psi$. Then for all $u_0, u_1 \in W$ if $w \leq u_0 \Rel u_1$, then $\fw,u_1 \models \psi$. Suppose $v \leq v' \Rel u$ for $v',u \in W$. Since $v \geq w$ also $v' \geq w$ and hence $w \leq v' \Rel u$. Hence $\fw,u \models \psi$ and therefore $\fw,v \models \lb \psi$. The case for $\varphi = \ld \psi$ is similar.
\end{proof}

 In the definition of a dynamic model the intuitionstic order $\leq$ and the modal relation $R$ are unrelated to one another. We now turn our attention to \emph{confluence conditions}: interaction principles between $\leq$ and $R$ that will be of recurring importance in this thesis. We give a general definition.

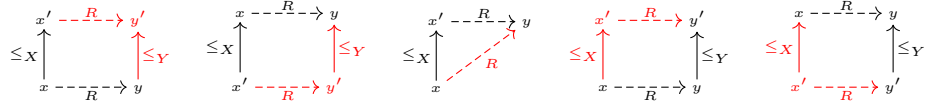
\begin{figure}[t]
\tiny
    \centering
      \begin{tikzcd}[ampersand replacement=\&,row sep=large,column sep=large]
        x' \ar[r,dashed,"R", color=red] \& {\color{red}y'} \\
        x \ar[u,"\le_X"]\ar[r,dashed, "R"'] \& y \ar[u,"\le_Y"', color=red]
    \end{tikzcd}
    \hspace{0.2cm}
    \begin{tikzcd}[ampersand replacement=\&,row sep=large,column sep=large]
        x \ar[r,dashed, "R"] \& y \\
        x' \ar[u,"\le_X"]\ar[r,dashed,"R"', color=red] \& {\color{red}y'} \ar[u,"\le_Y"', color=red]
    \end{tikzcd}
    \hspace{0.2cm}
    \begin{tikzcd}[ampersand replacement=\&,row sep=large,column sep=large]
        x' \ar[r, dashed, "R"] \& y\\
        x \ar[u,"\le_X"]\ar[ur,dashed,"R"', color=red] 
    \end{tikzcd}
    \hspace{0.2cm}
      \begin{tikzcd}[ampersand replacement=\&,row sep=large,column sep=large]
        {\color{red}x'} \ar[r,dashed,"R", color=red] \& y' \\
        x \ar[u,"\le_X", color=red]\ar[r, dashed,"R"'] \& y \ar[u,"\le_Y"']
    \end{tikzcd}
     \hspace{0.2cm}
      \begin{tikzcd}[ampersand replacement=\&,row sep=large,column sep=large]
        x \ar[r, dashed,"R"] \& y \\
        {\color{red}x'} \ar[u,"\le_X", color=red]\ar[r, dashed, "R"', color=red] \& y' \ar[u,"\le_Y"']
    \end{tikzcd}
    \caption{From left to right: Forth-up confluence, forth-down confluence, triangle confluence, back-up confluence and back-down confluence. In each diagram the partial orders $\leq_\mathcal{X}$ and $\leq_\mathcal{Y}$ are depicted by solid arrows, the relation $R$ by dashed arrows and the part stipulated by the confluence condition is marked in red.}
    \label{f: confluence conditions}
\end{figure}

\begin{definition}\label{d: confluence conditions}\index{confluence}
    Let $(X, \leq_X)$ and $(Y, \leq_Y)$ be partially ordered sets and $R \subseteq X \times Y$ a binary relation between $X$ and $Y$. Then 
    \begin{enumerate}
        \item $R$ is \emph{forth-up confluent}\index{confluence!forth-up} (for $(\leq_X, \leq_Y)$) if for all $x,x' \in X$ and all $y \in Y$ it holds that whenever $x \leq_X x'$ and $x \Rel y$, there exists $y' \in Y$ with $y \leq_Y y'$ and $x' \Rel y'$.
        \item $R$ is \emph{forth-down confluent}\index{confluence!forth-down} (for $(\leq_X, \leq_Y)$) if for all $x,x' \in X$ and $y \in Y$ it holds that whenever $x' \leq_X x$ and $x \Rel y$, there exists $y' \in Y$ with $y' \leq_Y y$ and $x' \Rel y'$.
        \item $R$ is \emph{triangle confluent}\index{confluence!triangle} (for $(\leq_X, \leq_Y)$) if for all $x,x' \in X$ and $y \in Y$ it holds that whenever $x \leq_X x'$ and $x' \Rel y$, we have $x \Rel y$.
        \item $R$ is \emph{back-up  confluent}\index{confluence!back-up} (for $(\leq_X, \leq_Y)$) if for all $x \in X$ and $y,y' \in Y$ it holds that whenever $x \Rel y$ and $y \leq_Y y'$, there exists $x' \in X$ with $x \leq_X x'$ and $x' \Rel y'$.
        \item $R$ is \emph{back-down confluent}\index{confluence!back-down} (for $(\leq_X, \leq_Y)$) if for all $x \in X$ and $y,y' \in Y$ it holds that whenever $x \Rel y$ and $y' \leq_{\mathcal{Y}} y$,  there exists $x' \in X$ with $x' \leq_{\mathcal{X}} x$ and $x' \Rel y'$.
    \end{enumerate}
    See Figure \ref{f: confluence conditions} for a depiction of the different confluence conditions.
\end{definition}

We will usually consider dynamic models $\fw=(W, \leq, R, V)$ where $R$ satisfies some confluence condition C for $(\leq, \leq)$. In this case, we simply call $R$ C-confluent. However, we will also study confluent relations between different posets in Chapter \ref{c: bi-int ml new} and Chapter \ref{c: biLTL}. 

One reason why we are interested in confluence conditions is that in forth-up and forth-down confluent dynamic models, the intuitionistic truth conditions for the modalities coincide with the classical truth conditions; the latter can thus be used for convenience. Other reasons will be presented in later chapters. The following is proven in~\cite{Balbiani_2021}.

\begin{lemma}\label{l: confluence classical truth conditions}
    Let $\fw=(W, \leq, R, V)$ be a dynamic model, $w \in W$ and $\varphi$ a formula.
    \begin{enumerate}
        \item If $R$ is forth-up confluent, then $\fw, w \models \ld \varphi$ if and only if there exists $v \in W$ with $w \Rel v$ and $\fw, v \models \varphi$.
        \item If $R$ is forth-down confluent, then $\fw, w \models \lb \varphi$ if and only if for all $v \in W$ if $w \Rel v$, then $\fw, v \models \varphi$.
    \end{enumerate}
\end{lemma}
\begin{proof}[Proof sketch.]
     1. Suppose $R$ is forth-up confluent. The direction from left-to-right is trivial. For the direction from right-to-left, suppose there exists $v \in W$ with $w \Rel v$ and $\fw, v \models \varphi$. If $w \leq u$, then by forth-up confluence there exists $v'$ with $v \leq v'$ and $u \Rel v'$. By monotonicity $\fw,v' \models \varphi$. Hence $\fw, w \models \ld \varphi$. The reasoning for 2. is similar.
\end{proof}

\begin{corollary}\label{cor: triangle model classical truth conditions}
    If $\fw=(W, \leq, R,V)$ is a dynamic model, $w \in W$, $\varphi$ a formula and $R$ is triangle confluent, then $\fw, w \models \lb \varphi$ if and only if for all $v \in W$ if $w \Rel v$, then $\fw, v \models \varphi$.
\end{corollary}
\begin{proof}
    Note that $R$ being triangle confluent implies that $R$ is forth-down confluent.
\end{proof}

We are also interested in dynamic models which satisfy certain frame conditions.

\begin{definition}\label{d: frame conditions}
    A dynamic model $\fw=(W, \leq, R, V)$ is called
    \begin{enumerate}
        \item \emph{reflexive} if $w \Rel w$ holds for all $w \in W$.
        \item \emph{transitive} if $w \Rel v$ and $v \Rel u$ imply $w \Rel u$ for all $w,v, u \in W$.
        \item \emph{symmetric} if $w \Rel v$ implies $v \Rel w$ for all $w,v \in W$.
        \item \emph{serial} if $dom(R) = W$.
        \item an \emph{S4 model}\index{model!S4} if $R$ is reflexive and transitive.
        \item an \emph{S5 model}\index{model!S5} if $R$ is an equivalence relation.
        \item \emph{(total) functional}\index{model!functional} if $R$ is a (total) partial function.
    \end{enumerate}
\end{definition}

In Chapter \ref{c: iLTL} and Chapter \ref{c: biLTL} we will study intuitionistic linear temporal logic, where formulas are evaluated over dynamic models which are total functional. The following lemma is straightforward to check, since in total functional models every world has a unique modal successor. The proof is omitted.

\begin{lemma}\label{l: functional models confluence conditions}
    Let $\fw =(W, \leq, f, V)$ be a total functional model. Then the following are equivalent.
    \begin{enumerate}
        \item $f$ is forth-up confluent.
        \item $f$ is forth-down confluent.
        \item $f$ is monotone in $\leq$, i.e. for all $w,v \in W$: $w \leq v \text{ implies } f(w) \leq f(v)$.
    \end{enumerate} 
\end{lemma}

Given this lemma, we will call total functional models which are forth-up or forth-down confluent simply \emph{forward confluent}. 

\section{Fixed Point Operators}\label{c: preliminaries, Section fixed points}\index{fixed point operator}

Intuitionistic dynamic logics are extensions of intuitionistic modal logics with fixed point operators. Intuitively, fixed point operators express infinite statements, such as `in all future time steps, $\varphi$ is true'. In this section we briefly explain why such operators are called fixed point operators. Strictly speaking, the theory explained in this section is not needed for understanding the rest of the thesis. However, we believe that it is informative to understand the concept of a fixed point operator, as it relates the presented logics to one another and also to other modal fixed point logics. We do not give a general definition, but instead explain the concept with an example.

Consider a dynamic model of the form $\fw =(W, \leq, R, V)$, where $R$ is triangle confluent. Consider the language extending $\LIPL$ by the modality $\lb$ as well as a unary operator $\lbm$ which is characterized semantically as $\fw, w \models \lbm \varphi$ if and only if for all $v \in W$ if $w \Rel^* v$, then $\fw, v \models \varphi$ where $R^*$ is the reflexive transitive closure of $R$. In other words, $\lbm \varphi$ is true at $w$ if $\varphi$ is true in any world $v$ reachable from $w$ over finitely many modal steps. Equivalently, $\lbm \varphi$ is equivalent to the infinite conjunction
\begin{equation*}
    \bigwedge_{n < \omega} \lb^n \varphi.
\end{equation*}
The operator $\lbm$ is called a \emph{fixed point operator}, namely $\lbm \varphi$ is the greatest fixed point of the propositional function $x \mapsto \varphi \wedge \lb x$. Let us make precise what is meant by this.

 Given a formula $\varphi$, let 
 \begin{equation*}
     \llbracket \varphi \rrbracket_\fw \coloneqq \{ w \in W \mid \fw, w \models \varphi\}
 \end{equation*}
 be the \emph{truth set} of $\varphi$ in $\fw$. Let $U \subseteq W$ and $p \in \Prop$ be a proposition not occurring in $\varphi$. Let $U^\uparrow$ be the least upwards closed set containing U. Define $V_U^p: W \longrightarrow \mathcal{P}(\Prop)$ as follows:
\begin{equation*}
    V_U^p(w) \coloneqq \begin{cases}
        V(w) \cup \{p\} & \text{ if } w\in U^\uparrow \\
        V(w) \setminus \{p\} & \text{ otherwise.}
    \end{cases}
\end{equation*}
In other words, the valuation $V_U^p$ changes the valuation $V$ by making $p$ true at the worlds in $U^\uparrow$ and nowhere else. For any other proposition $q$, $q \in V(w)$ if and only if $q \in V_U^P(w)$. Let $\fw_U^p$ denote the dynamic model $(W, \leq, R, V_U^p)$. Now consider the formula $\psi = \varphi \wedge \lb p$. This formula induces a function $f_\psi: \mathcal{P}(W) \longrightarrow \mathcal{P}(W)$ given by
\begin{equation*}
   U \mapsto \llbracket \varphi \wedge \lb p\rrbracket_{\fw_U^p}.
\end{equation*}

Given a function $S: \mathcal{P}(W) \longrightarrow \mathcal{P}(W)$, a \emph{fixed point} of $S$ is a subset $U \subseteq W$ such that $S(U) = U$. Moreover, $U$ is the \emph{greatest} fixed point of $S$ if for any other fixed point $U'$ of $S$ holds that $U' \subseteq U$ and $U$ is the \emph{least} fixed point of $S$ if for any other fixed point $U'$ of $S$ holds that $U \subseteq U'$. If $S$ is monotone on $\mathcal{P}(W)$, meaning that whenever $U \subseteq U'$ we have $S(U) \subseteq S(U')$, then the Knaster--Tarski Theorem~\cite{Tarski_1955} guarantees that $S$ has a least and a greatest fixed point. 

It is not hard to check that the function $f_\psi$ is monotone on $\mathcal{P}(W)$: we have $\llbracket \varphi \wedge \lb p \rrbracket_{\fw} = \llbracket \varphi \rrbracket_\fw \cap \llbracket {\lb p} \rrbracket_\fw$. Since $p$ does not occur in $\varphi$, for any $U \subseteq U'$, $\llbracket \varphi \rrbracket_{\fw_U^p} = \llbracket \varphi \rrbracket_{\fw_{U'}^p}$. Furthermore, clearly $\llbracket \lb p \rrbracket_{\fw_U^p} \subseteq \llbracket \lb p \rrbracket_{\fw_{U'}^p}$. Hence $f_\psi(U) \subseteq f_\psi(U')$. By the Knaster--Tarski Theorem $f_\psi$ has a greatest fixed point $U$. One can now show that the greatest fixed point of $f_\psi$ is exactly the set $\llbracket \lbm \varphi \rrbracket_\fw$; for a proof, see~\cite[Chapter 3]{marti_2017}. This is the precise meaning behind the statement that $\lbm \varphi$ is the greatest fixed point of the function $x \mapsto \varphi \wedge \lb x$.

This construction is not dependent on $\psi$. For any formula $\gamma$ we can consider the function $f_\gamma$ induced by $\gamma$ and if $f_\gamma$ is monotone, then there exist the least and greatest fixed points. A defining property of all intuitionistic dynamic logics studied in this thesis is that they are extensions with fixed point operators for \emph{specific} functions. This is not a necessary restriction; instead we could extend the language of $\mathsf{IML}$ by operators $\mu$ and $\nu$ which are applicable to any formula and define the truth set of a formula $\mu x. \varphi$ as the least fixed point of the function $f_\varphi$ induced by $\varphi$ and similarly, $\nu x. \varphi$ as the greatest fixed point.\footnote{In that case one needs to be careful with the inductive definition of the language to guarantee that every formula induces a \emph{monotone} function.} This results in an expressive logical system which is capable of expressing least and greatest fixed points of \emph{every} monotone function, as opposed to only \emph{some}. This logic is called \emph{intuitionistic modal $\mu$-calculus}, see Pacheco~\cite{Pacheco_2024}. For an introduction to modal fixed point logics in general, we refer the reader to the excellent article about the classical modal $\mu$-calculus by Stirling~\cite{Stirling_2001}.

\section{Historical Remarks}\label{c: preliminaries, section historical remarks}

We finish this chapter by providing a brief historical overview of intuitionistic logic and intuitionistic modal logic.

\subsection{Intuitionism and Intuitionistic Logic}

The origin of intuitionistic logic goes back to \emph{intuitionism}\index{intuitionism}, one of the main positions in philosophy of mathematics developed by Dutch mathematician Brouwer~\cite{Brouwer_1907}. Classical mathematics (at least implicitly) assumes an objective notion of \emph{truth}: every well-formed mathematical statement such as `$2+2= 4$' is either true or false. A common interpretation in philosophy is that truth is a relationship between language and reality~\cite{Priest_2008}. For example, the sentence `Young Boys is the best football club in Bern' is true due to the objective reality regarding football clubs in Bern. But what is the objective reality of the statement `$2+2 = 4$' which is regarded to be true in mathematics? The realist position, which originates in the work of Plato, claims that there are objectively existing mathematical objects, such as natural numbers. In other words, there exists a universe of mathematical objects and the statement `$2+2 = 4$' expresses the objective reality in this universe and is thus true. Intuitionism rejects the realist position and instead claims that mathematics is a mental activity and mathematical objects exist insofar they have been constructed in the mind of the mathematician. Similarly, proofs are mental constructions and a mathematical statement is true insofar it has been proven. As a consequence, intuitionism rejects indirect proofs which are often used in classical mathematics. For example, an existential statement $\exists x. \varphi(x)$ is typically proven in classical mathematics by deriving a contradiction from $\forall x. \neg \varphi(x)$. But such an indirect proof does not result in the construction of an object $c$ which satisfies $\varphi(c)$. Instead the indirect proof assumes that the statement $\exists x. \varphi(x)$ is either true or false, thus deriving a contradiction from its negation $\forall x. \neg \varphi(x)$ means that $\exists x. \varphi(x)$ must be true. In intuitionistic mathematics, such an existential statement is proven constructively by constructing the object $c$ and providing a constructive proof of $\varphi(c)$. Negations of statements are also treated differently in intuitionistic mathematics. A proof of $\neg \varphi$ is a proof that $\varphi$ cannot be proven. Under this interpretation the law of excluded middle does not hold: in classical mathematics $p \vee \neg p$ is always true since either $p$ is true or $p$ is false, in which case $\neg p$ is true; in intuitionistic mathematics any statement $p$ is either proven or not proven, but the lack of a proof for $p$ does not imply that $p$ \emph{cannot} be proven, but rather that no proof for $p$ has been constructed so far. Thus if $p$ formalizes any unresolved mathematical conjecture, such as the Goldbach conjecture, there is no proof of $p \vee \neg p$ and so the law of excluded middle fails. This example shows that intuitionistic mathematics uses different logical reasoning principles than classical mathematics. In 1930, Brouwer's student Heyting isolated the acceptable reasoning principles of intuitionistic mathematics in a formal logical system, leading to the development of intuitionistic logic~\cite{Heyting1930a, Heyting1930b, Heyting1930c}. Since then, intuitionistic logic has developed into arguably one of the most successful logical systems, partly due to the following two reasons. First, it is generally accepted that intuitionistic logic succeeded in capturing intuitionistic reasoning and has thus fulfilled its philosophical purpose. Second, alternative interpretations of intuitionistic logic have emerged, which deviate from the original interpretation regarding intuitionism. Among those are interpretations based on computation, topology or information. For example, the Curry--Howard isomorphism relates, roughly speaking, proofs (in some formal proof system for intuitionistic logic) with programs (in some formal model of computation), illustrating a close connection between intuitionistic proofs and computation~\cite{Srensen_2006}. This close relation is also illustrated by the fact that many foundational systems for computer science are based on intuitionistic logic~\cite{Bezhanishvili_?}. The interpretation of intuitionistic logic guiding this thesis is based on information. As outlined in Chapter \ref{c: Introduction}, we consider intuitionistic logic as a logic to reason about information and information change. This interpretation goes back to Kripke~\cite{Kripke_1965}, who introduced intuitionistic Kripke models for intuitionistic logic in 1965. As we have seen, these models can be regarded as information orderings and the evaluation of implications as expressing consequences of gaining information. Originally, Kripke provided an interpretation of his semantics as a way to capture the intuitionistic idea that mathematical statements are true if proven and false if proven to be unprovable. The worlds $w$ of an intuitionistic Kripke model are thought to be information states containing all mathematical theorems which have been proven up to a certain time point. If $w \leq v$, then $v$ is thought of as an information state at a later time point obtained by adding more theorems that have been proven in the meantime. If $w \not \models \varphi$, then $\varphi$ is not thought to be refuted, but simply to not (yet) been proven. The truth relation should thus be understood as expressing whether a formula has been proven at this point or not. If $w \models \neg \varphi$, then $\varphi$ has been proven to be unprovable, implying that for any extension of $w$, $\varphi$ cannot be true. Kripke semantics thus provide an intuitive interpretation of intuitionistic logic and are furthermore natural to work with mathematically. However, it should be noted that they are not entirely satisfactory from the point of view of intuitionism: the interpretation of formulas on Kripke models does not refer to the notion of proof directly nor to the idea of constructions; moreover while we have seen that the proof systems of intuitionistic logic presented in Section \ref{c: preliminaries, Section proof systems} are sound and complete with respect to the class of Kripke models, the completeness proofs require a classical meta theory and are thus not intuitionistically acceptable~\cite{Simpson_1994}.

For an introduction to intuitionism, we refer the reader to~\cite{Heyting_1956}. For a detailed account of the history of intuitionism and constructivism, see~\cite{Troelstra_2011}. For a detailed introduction to the mathematical theory of intuitionistic logic, see~\cite{VanDalen_1986, Bezhanishvili_?}.

\subsection{Intuitionistic Modal Logic}

The modern origin of modal logic arguably goes back to the seminal work of Lewis~\cite{Lewis_1918} in 1918, who studied logics of strict implications. After the development of intuitionistic logic by Heyting, both logics were studied independently for several decades. The earliest combination of intuitionistic logic and modal logic is found in the work of Fitch~\cite{Fitch_1948} in 1948; however, it was the invention of Kripke semantics for modal logic~\cite{Kripke_1959} and for intuitionistic logic~\cite{Kripke_1965} by Kripke which laid the foundation for studying intuitionistic modal logic. In the 1980's, Plotkin and Stirling~\cite{Plotkin_1988}, Ewald~\cite{Ewald_1986} and Fischer-Servi~\cite{Servi_1981} independently introduced birelational semantics for intuitionistic modal logic by combining the two types of Kripke models, resulting in the development of the structures that we call dynamic models in this thesis. What followed was the development of a sheer bewildering amount of different intuitionistic modal logics, which differ semantically by the confluence conditions imposed on the dynamic models used for their evaluation. Roughly, these logics fall into two camps: \emph{constructive modal logics} such as the systems introduced by Fitch~\cite{Fitch_1948} and Wijesekera~\cite{wijesekera_1990} built for modelling computational aspects such as staged computation~\cite{kavvos_2016}, and \emph{intuitionistic modal logics} such as the systems by Fischer-Servi~\cite{Servi_1981} and Plotkin and Stirling~\cite{Plotkin_1988} aimed at capturing an intuitionistic meta reading of modalities and Kripke semantics. In his influential PhD thesis, Simpson~\cite{Simpson_1994} set up six criteria that an intuitionistic modal logic should satisfy to be classified as `intuitionistic'; one criteria is that there should be an intuitionistically comprehensible explanation of the modalities~\cite[Page 40]{Simpson_1994}. Simpson achieves this by providing a translation of intuitionistic modal logic into intuitionistic first-order logic - akin to the translation of modal logic into classical first-order logic - which can then be used as a formalized meta theory. The resulting logic turns out to be $\mathsf{IK}$, which is semantically obtained by considering all validities over dynamic models satisfying forth-up and back-up confluence.

The restriction to dynamic models satisfying confluence conditions substantially increases the mathematical difficulty of the resulting logics. For example, the logic $\mathsf{IS4}$ - an intuitionistic version of the modal logic $\mathsf{S4}$ which satisfies the six criteria posed by Simpson - has only recently been proven to be decidable~\cite{Girlando_2023} and the presented proof is highly non-trivial. On the other hand, weaker logics obtained by either not considering any confluence conditions or by considering a restricted language behave more similar to classical modal logics and are generally easier to deal with, see e.g.~\cite{jager-intuitionistic_2016}.

\chapter{Intuitionistic Master Modality}\label{c: IM}

\section{Introduction}\label{c: IM, section Introduction}

This chapter introduces a simple intuitionistic dynamic logic called \emph{intuitionistic master modality} ($\IM$). The language of $\IM$ extends $\LIPL$ by the modality $\lb$  and the fixed point operator $\lbm$ called the \emph{master modality}, which we encountered in Section \ref{c: preliminaries, Section fixed points}. The modality $\ld$ is not present in the language. Intuitively, $\lbm \varphi$ expresses that $\varphi$ is true in any world of a dynamic model reachable from the current world over a finite (but arbitrary) number of modal steps. Thus $\lbm \varphi$ is true if $\lb^n \varphi$ is true for any natural number $n$ and it is characterized as the greatest fixed point of the propositional function $x \mapsto \varphi \wedge \lb x$. Given the interpretation that intuitionistic modal logic reasons about information updates, a formula $\lbm \varphi$ expresses that $\varphi$ remains true under any number of information updates. The goal of this chapter is to study the mathematical theory of $\IM$; in particular the expressivity of its language and its proof theory.

To the best of our knowledge, this work constitutes the first mathematical investigation of $\IM$. It is worth noting, however, that a multi-modal variant of $\IM$, known as intuitionistic common knowledge logic ($\ICK$), was previously introduced and studied by Jäger and Marti.~\cite{jager-intuitionistic_2016, jager-distributed_2016}. Classical versions of $\IM$ have also been studied extensively, see e.g.~\cite{rooduijn_2024}. Originally, one of our main motivations to study $\IM$ was the interpretation of $\IM$ as an intuitionistic linear temporal logic. To that end formulas are evaluated on dynamic models where the modal relation is a function modeling time. Such models are called functional models. In difference to stronger temporal logics such as $\iLTL$, which will be studied in Chapter \ref{c: iLTL}, the functional models of $\IM$ do not satisfy any confluence conditions. The modality $\lb$ is then interpreted as a temporal `next' operator and the master modality as a temporal `henceforth', i.e. $\lbm \varphi$ is true if $\varphi$ is true in any future time step. As it turns out, the language of $\IM$ is not expressive enough for such an interpretation. Namely, the set of valid formulas over functional models coincides with the set of valid formulas of arbitrary dynamic models, implying that the restriction to functions does not have an impact on the logic. This discourages an interpretation of $\IM$ as temporal logic. Nevertheless, $\IM$ remains a relevant logical system in its own right: $\IM$ is an interesting case study for applying techniques from the study of classical modal fixed point logics to the intuitionistic realm. This is due to the relatively simple mathematical theory of $\IM$, caused by the lack of meaningful confluence conditions imposed on its dynamic models. Moreover, $\IM$ serves as the stepping stone to investigate other, more expressive, intuitionistic dynamic logics, such as intuitionistic common knowledge logic which will be studied in Chapter \ref{c: ICK} or intuitionistic versions of propositional dynamic logic ($\mathsf{PDL}$). The techniques developed to investigate $\IM$ can be applied or adapted for more complex logics, as  illustrated in Chapter \ref{c: ICK} where much of the proof theoretic work for $\IM$ is applied successfully to $\ICK$.

The next section introduces the syntax and semantics of $\IM$. The semantics is given in terms of dynamic models, where we consider three classes: the class of all dynamic models, the class of functional dynamic models and the class of dynamic models which are triangle confluent. Regarding expressivity, we prove that the language of $\IM$ cannot distinguish between all three classes of models, in the sense that they all produce the same set of validites (see Theorem \ref{t: three classes one logic}). Afterwards we turn to proof theory and introduce a sound and complete axiomatization for $\IM$. The completeness proof is by a standard canonical model construction. Our proof is similar to the completeness proof for classical common knowledge logic presented in~\cite[Chapter 7]{ditmarsch_2017}, highlighting the straightforward adaptation of classical techniques to $\IM$. Completeness for a multi-modal version of $\IM$ was already established for a different axiomatization by Marti in his PhD thesis~\cite{marti_2017}, and the presented proof closely follows Marti's proof. The remaining sections study a sequent calculus for $\IM$. Due to the presence of the master modality, our proof systems require a formal counterpart to induction. Instead of using an induction rule as is done in~\cite{jager-intuitionistic_2016}, we instead employ \emph{non-wellfounded} and \emph{cyclic proofs}. First, we give a brief introduction to such proof formalisms and then introduce the cyclic calculus $\CIM$ and the non-wellfounded calculus $\NWIM$. The calculus $\NWIM$ is not studied in its own right but used as a tool to establish completeness for $\CIM$. We prove soundness of $\CIM$ by an indirect argument and completeness for $\NWIM$ via a proof search argument. It is then shown how to translate non-wellfounded proofs into cyclic proofs, whence establishing completeness for $\CIM$ as well. The presented proof search method is quite robust, which will be illustrated by showing how to obtain dynamic, triangle and functional countermodels from a failed proof search, which requires only minor adjustments to the overall argument. This will also provide an alternative proof of Theorem~\ref{t: three classes one logic} using proof theoretic means.

\section{Syntax and Semantics}\label{c: IM, section syntax}\index{logic!intuitionistic master modality}

The \emph{language} $\LIM$ of $\IM$ extends $\LIPL$ by the modal operator $\lb$ and the fixed point operator $\lbm$ called the \emph{master modality}\index{fixed point operator!master modality}. \emph{Formulas} of $\LIM$ are given by the following grammar in Backus--Naur form:
\begin{equation*}
    \varphi ::= \bot \, \lvert \, p \, \lvert \, \varphi \wedge \varphi \, \lvert \, \varphi \vee \varphi \, \lvert \, \varphi \rightarrow \varphi \, \lvert \, \lb \varphi \, \lvert \, \lbm \varphi.
\end{equation*}
where $p \in \Prop$. As before we write $\varphi \in \LIM$ to denote that $\varphi$ is a $\LIM$-formula and we use the same conventions for the brackets and the notation as introduced for $\mathcal{L}_\mathsf{IPL}$.

\begin{definition}\label{d: closure IM formula}
    The \emph{closure}\index{closure!for $\LIM$} $\Cl(\varphi)$ of a formula $\varphi$ is defined by induction on $\varphi$ as follows.
     \begin{itemize}
        \item $\Cl(\bot) = \{\bot\}$
        \item $\Cl(p) = \{ p \}$ for $p \in \Prop$
        \item $\Cl(\varphi \ast \psi) = \Cl(\varphi) \cup \Cl(\psi) \cup \{ \varphi \ast \psi\}$ for $\ast \in \{\wedge, \vee, \rightarrow\}$
        \item $\Cl(\lb \varphi) = \Cl(\varphi) \cup \{\lb \varphi\}$
        \item $\Cl(\lbm \varphi) = \Cl(\varphi) \cup \{\lbm \varphi, \lb \lbm \varphi\}$
    \end{itemize}
    Given a set $\Gamma$ of formulas, the \emph{closure} $\Cl(\Gamma)$ of $\Gamma$ is defined by
    \begin{equation*}
        \Cl(\Gamma):= \bigcup_{\varphi\in\Gamma}\Cl(\varphi).
    \end{equation*}
A set of formulas $\Gamma$ is \emph{closed} if $\Gamma = \Cl(\Gamma)$.
\end{definition}
 The following lemma is established by a straightforward induction on the structure of $\varphi$. The proof is omitted.

\begin{lemma}\label{l: closure finite}
    For any formula $\varphi$, the closure $\Cl(\varphi)$ is finite.
\end{lemma}

The \emph{complexity} of a formula is defined as follows.

\begin{definition}\label{d: complexity of IM formula}
    The \emph{complexity}\index{complexity!of $\LIM$-formula} $c(\varphi)$ of a formula $\varphi$ is defined by induction on $\varphi$ as follows.
    \begin{itemize}
        \item $c(\bot) = c(p) = 0$
        \item $c(\varphi \ast \psi) = c(\varphi) + c(\psi) +1$ for $\ast \in \{\wedge, \vee, \rightarrow\}$
        \item $c(\lb \varphi) = c(\lbm \varphi) = c(\varphi) + 1$
    \end{itemize}
    Given a finite set of formulas $\Gamma$, the \emph{complexity} $c(\Gamma)$ of $\Gamma$ is defined as 
    \begin{equation*}
       c(\Gamma) = \sum_{\varphi \in \Gamma}c(\varphi).
    \end{equation*}
\end{definition}

We will evaluate formulas of $\mathcal{L}_\mathsf{IM}$ over three classes of dynamic models: the class of all dynamic models (see Definition \ref{d: dynamic model}), the class of functional models (see Definition \ref{d: frame conditions}), and the class of \emph{triangle models}, which are defined as follows. 

\begin{definition}
    A \emph{triangle model}\index{model!triangle} is a dynamic model $\fw=(W, \leq, R, V)$ where $R$ is triangle confluent.\footnote{See Definition \ref{d: confluence conditions}.}
\end{definition}





By Corollary \ref{cor: triangle model classical truth conditions} we can use the classical truth conditions for the modalities when evaluating formulas over triangle models, but in order to obtain monotonicity we shall use the intuitionistic truth conditions for modalities when evaluating formulas over dynamic and functional models. Thus we introduce two truth relations $\models$ and $\models_t$. 

For any binary relation $S$, let $S^*$ denote the reflexive transitive closure of $S$. Given a dynamic model $\fw=(W,\leq,R,V)$, let $\Tilde{R}$ denote the composition $R \circ {\leq}$, i.e. $w \mathrel{\Tilde{R}} v$ holds if and only if  there exists $u \in W$ with $w \leq u$ and $u \Rel v$. 
Note that, since $\leq$ is reflexive,  $ w \mathrel{(\Tilde{R})^*} v$ holds if and only if there exist a natural number $n$ and worlds $u_0, \ldots, u_{2n}$ such that $u_0 = w$, $u_{2n} = v$ and for all $0 \leq i < n$ both $u_{2i} \leq u_{2i+1}$ and $u_{2i+1} \Rel u_{2(i+1)}$ hold.

\begin{definition}\label{d: truth relation for birelational models}
The \emph{truth relation} $\models$ between worlds of a dynamic (functional) model $\fw=(W, \leq, R, V)$ and formulas is defined by extending Definition  \ref{d: truth relation for intuitionistic Kripke models} with the following clauses, where $w \in W$.
\begin{center}
    \begin{tabular}{l l l}
    $\fw,w \models \lb \varphi$ & iff & for all $v \in W$ if $w \mathrel{\Tilde{R}} v$, then $\fw,v \models \varphi$,\\
    $\fw,w \models \lbm \varphi$ & iff & for all $v \in W$ if $w \mathrel{(\Tilde{R})^*} v$, then $\fw,v \models \varphi$.\\
    \end{tabular}
\end{center}
If $\fw, w \models \varphi$, then $\varphi$ is called \emph{true} at $w$.
\end{definition}

A formula $\varphi$ is \emph{satisfiable} over the class of dynamic (functional) models if there exists a dynamic (functional) model $\fw=(W, \leq, R, V)$ and a world $w \in W$ such that ${\fw,w \models \varphi}$, and \emph{unsatisfiable} otherwise. The formula $\varphi$ is \emph{valid} over the class of dynamic (functional) models if for any dynamic (functional) model $\fw=(W, \leq, R, V)$ and any world $w \in W$ holds that $\fw,w \models \varphi$, and \emph{falsifiable} otherwise. \smallskip

\begin{lemma}[Monotonicity of $\models$]\label{l: monotonicity intuitionistic Kripke models}
    Let $\varphi\in \LIM$ and let ${\fw = (W, \leq, R, V)}$ be a dynamic (functional) model with $w,v \in W$. If $w \leq v$ and $\fw,w \models \varphi$, then $\fw,v \models \varphi$.
\end{lemma}
\begin{proof}
   Recall that every functional model is a dynamic model. Therefore it suffices to prove the lemma for dynamic models. Let ${\fw = (W, \leq, R, V)}$ be a dynamic model with $w,v \in W$ and suppose that $w \leq v$. We proceed by induction on the structure of $\varphi$. Each case apart from $\varphi = \lbm \psi$ is covered in Lemma \ref{l: monotonicity for intuitionistic Kripke models} and Lemma \ref{l: monotonicity for dynamic models}. \smallskip
   
   Suppose $\fw,w \models \lbm \psi$. Then for all $u \in W$ with $w \mathrel{(\Tilde{R})^*} u$ holds that $\fw,u \models \psi$. Suppose $v \mathrel{(\Tilde{R})^*} u$ for some $u \in W$. Then there exist $u_0, \ldots, u_{2n} \in W$ with $u_0 = v$, $u_{2n} = u$ and for all $0 \leq i < n$, $u_{2i} \leq u_{2i+1}$ and $u_{2i+1} \Rel u_{2(i+1)}$. Therefore, $v \leq u_1$ and since $w \leq v$, also $w \leq u_1$. Hence $w \mathrel{(\Tilde{R})^*} u$ and so $\fw,u \models \psi$. We conclude that $\fw,v \models \lbm \psi$.\qedhere
\end{proof}

Next, we introduce the truth relation $\models_t$ for evaluating formulas on triangle models.

\begin{definition}
The definition of the truth relation $\models_t$ between worlds of a triangle model $\fw=(W, \leq, R,V)$ and formulas is defined by extending Definition \ref{d: truth relation for intuitionistic Kripke models} with the following clauses, where ${w \in W}$.
\begin{center}
    \begin{tabular}{l l l}
        $\fw,w \models_t \lb \varphi$ & iff & for all $v \in W$ if $w \mathrel{R} v$, then $\fw,v \models_t \varphi$,
        \\
        $\fw,w \models_t \lbm \varphi$ & iff & for all $v \in W$ if $w \mathrel{R^*} v$, then $\fw,v \models_t \varphi$.
    \end{tabular}
\end{center}
\end{definition}

Note that every triangle model $\fw=(W, \leq, R, V)$ satisfies $R = \Tilde{R}$. The monotonicity property for $\models_t$ follows immediately from Corollary \ref{cor: triangle model classical truth conditions} and Lemma \ref{l: monotonicity intuitionistic Kripke models}.

\begin{lemma}[Monotonicity of $\models_t$]
     Let $\varphi\in \LIM$ and let ${\fw = (W, \leq, R, V)}$ be a triangle model with $w,v \in W$. If $w \leq v$ and $\fw,w \models_t \varphi$, then $\fw,v \models_t \varphi$.
\end{lemma}




\begin{definition}\label{d: logics for master modality}
Let
 \begin{enumerate}
     \item $\mathbf{IM}$ be the set of valid $\LIM$-formulas over the class of dynamic models.
     \item $\mathbf{IM_f}$ be the set of valid $\LIM$-formulas over the class of functional models.
     \item $\mathbf{IM_t}$ be the set of valid $\LIM$-formulas over the class of triangle models.
 \end{enumerate}
\end{definition}

The logic $\mathbf{IM_f}$ can be interpreted as an intuitionistic version of linear temporal logic, where $f$ is the function mapping each world to its temporal successor, $\lb$ is interpreted as `next' and $\lbm$ as `henceforth'. It makes sense under such an interpretation to assume $f$ to be a total function. Note that $\mathbf{IM_f}$ does not validate some of the standard tautologies of linear temporal logic, such as $\lb (p \vee q) \rightarrow (\lb p \vee \lb q)$.

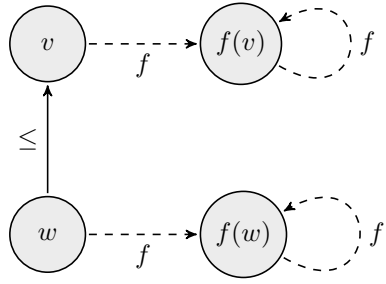
\begin{figure}[t!]
    \centering
    \begin{tikzpicture}[modal]
        \node[world](w){$w$} ;
        \node[world](v)[above=of w]{$v$};
        \node[world](fw) [right= of w]{$f(w)$};
        \node[world](fv) [right= of v]{$f(v)$};
        \path[->] (w) edge node[left] {$\leq$} (v);
        \path[->] (w) edge[dashed] node[below] {$f$} (fw);
        \path[->] (v) edge[dashed] node[below] {$f$} (fv);
        \path[->] (fw) edge[reflexive right, dashed] node[right] {$f$} (fw);
        \path[->] (fv) edge[reflexive right, dashed] node[right] {$f$} (fv);
    \end{tikzpicture}
    \caption{A (total) functional model falsifying $\lb(p \vee q) \rightarrow (\lb p \vee \lb q)$.}
    \label{f: functional model falsifying distribution of vee}
\end{figure}

\begin{lemma}
    $\lb (p \vee q) \rightarrow (\lb p \vee \lb q) \not \in \mathbf{IM_f}$.
\end{lemma}
\begin{proof}
    Consider the functional model $\fw =(W, \leq, f, V)$ depicted in Figure \ref{f: functional model falsifying distribution of vee}, where $V(w) = V(v) = \emptyset$, $V(f(w)) = \{p\}$ and $V(f(v)) = \{q\}$. Note that $\fw, w \models \lb (p \vee q)$ since $\fw, f(w) \models p$ and hence $\fw, f(w) \models p \vee q$ and $\fw, f(v) \models q$ and so $\fw, f(v) \models p \vee q$. However, $\fw, w \not \models \lb p$, since $w \leq v$ and $\fw, f(v) \not \models p$. Similarly, $\fw, w \not \models \lb q$ since $\fw, f(w) \not \models q$. Hence $\fw, w \not \models  \lb p \vee  \lb q$.
\end{proof}

We finish this section with the following lemma, stating that every formula valid over the class of dynamic models is also valid over the classes of functional and triangle models. Note that this trivially holds since both functional and triangle models are dynamic models and $\fw, w \models \varphi$ if and only if $\fw, w \models_t \varphi$ for any triangle model $\fw$ and any world $w$ by Corollary \ref{cor: triangle model classical truth conditions}.

\begin{lemma}\label{l: IM subset of IMf and IMt}
The following hold.
    \begin{enumerate}
        \item $\mathbf{IM} \subseteq \mathbf{IM_t}$
        \item $\mathbf{IM} \subseteq \mathbf{IM_f}$
    \end{enumerate}
\end{lemma}

\section{Expressivity}\label{c: IM, section model constructions IM}

 This section explores whether the language $\mathcal{L}_\IM$ can distinguish between the classes of dynamic, functional and triangle models. The main result is Theorem \ref{t: three classes one logic}, stating that all three classes have the same set of valid formulas, i.e.
\begin{equation*}
    \mathbf{IM} = \mathbf{IM_t} = \mathbf{IM_f}.
\end{equation*}

That $\mathbf{IM} = \mathbf{IM_t}$ is straightforward, and was already observed about the language of $\LIM$ without the master modality in~\cite{litak_2018}. The following lemma provides a brief proof sketch.

\begin{lemma}\label{l: birelational model induces triangle model}
    $\mathbf{IM} = \mathbf{IM_t}$.
\end{lemma}
\begin{proof}[Proof sketch]
    That $\mathbf{IM} \subseteq \mathbf{IM_t}$ is Lemma \ref{l: IM subset of IMf and IMt}. For the other inclusion let $\fw=(W, \leq, R, V)$ be a dynamic model and define $\fw' = (W, \leq, (R \circ {\leq}), V)$. Clearly,  $(R \circ {\leq})$ is triangle confluent and therefore $\fw'$ is a triangle model. We show by induction on the structure of $\varphi$ that for any $w \in W$ and any $\varphi \in \LIM$ hold that
    \begin{equation*}
        \fw, w \models \varphi \text{ iff } \fw', w \models_t \varphi
    \end{equation*}
     
     The base cases and the cases where the main connective of $\varphi$ belongs to $\{\wedge, \vee, \rightarrow \}$ are routine. The cases where $\varphi = \lb \psi$ or $\varphi = \lbm \psi$ then follow immediately from the observation that $\Tilde{R} = (R \circ {\leq})$. Therefore if a formula $\varphi$ is falsifiable over the class of dynamic models, then there exists a dynamic model $\fw$ and a world $w$ with $\fw, w \not \models \varphi$, implying that $\fw', w \not \models_t \varphi$. Hence $\varphi$ is falsifiable over the class of triangle models and so $\mathbf{IM_t} \subseteq \mathbf{IM}$.
\end{proof}

\subsection{Dynamic and Functional Models}

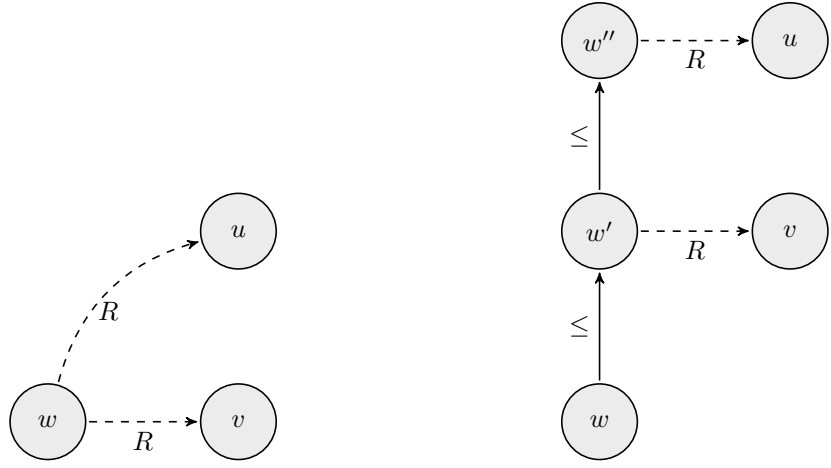
\begin{figure}[t!]
    \centering
    \begin{tikzpicture}[modal]
        \node[world](w){$w$} ;
        \node[world](v)[right=of w]{$v$};
        \node[world](u)[above=of v]{$u$};
        \path[->] (w) edge[dashed] node[below] {$R$} (v);
        \path[->] (w) edge[dashed, bend left] node[below] {$R$} (u);
    \end{tikzpicture}
    \hspace{100pt}
    \begin{tikzpicture}[modal]
        \node[world](w){$w$};
        \node[world](w1)[above=of w]{$w'$};
        \node[world](w2)[above=of w1]{$w''$};
        \node[world](v)[right=of w1]{$v$};
        \node[world](u)[right=of w2]{$u$};
        \path[->] (w) edge node[left] {$\leq$} (w1);
        \path[->] (w1) edge node[left] {$\leq$} (w2);
        \path[->] (w1) edge[dashed] node[below] {$R$} (v);
        \path[->] (w2) edge[dashed] node[below] {$R$} (u);
    \end{tikzpicture}
    \caption{Left: a dynamic model where $w$ has two modal successors. Right: the corresponding functional model where $w'$ and $w''$ are copies of $w$.}
    \label{f: constructing functional models}
\end{figure} 

The goal of this subsection is to establish $\mathbf{IM} = \mathbf{IM_f}$. By Lemma \ref{l: IM subset of IMf and IMt}, $\mathbf{IM} \subseteq \mathbf{IM_f}$. For the other inclusion, given a formula $\varphi$ and a dynamic model $\fw$ falsifying $\varphi$, we have to show how to construct a \emph{functional} model $\fw'$ falsifying $\varphi$. The functional model $\fw'$ will be constructed in stages: we start by adding a copy of the world $w$ of $\fw$ which falsifies $\varphi$. Then, in later steps, we add copies of worlds of $\fw$ in such a way that the copy of $w$ satisfies and falsifies the same formulas as $w$, and the resulting model is functional. The crucial insight is that if a world $w$ of $\fw$ has multiple modal successors, we can simply add, for each modal successor, a copy of $w$ as an intuitionistic successor of $w$ and for each copy one modal successor. Due to the intuitionistic truth conditions, both models then satisfy the same modal formulas. See Figure \ref{f: constructing functional models} for an illustration. In order to formalize the construction, we introduce \emph{induced structures}.

\begin{definition}\index{induced structure}
    Let $\fw=(W, \leq, R, V)$ be a dynamic model. An \emph{$\fw$-induced structure} is a tuple $\mathcal{I}=(I, \leq_I, R_I, V_I)$ together with a function $\pi: I \longrightarrow W$ such that the following hold.
    \begin{enumerate}
        \item $I$ is a finite set.
        \item $(I, \leq_I)$ is a poset and if $x \leq_I y$, then $\pi(x) \leq \pi(y)$.
        \item $R_I: I \longrightarrow I$ is a partial function such that if $R_I(x) = y$, then $\pi(x) \mathrel{R} \pi(y)$.
        \item $V_I = V \circ \pi$.
    \end{enumerate}
\end{definition}

Since we are only interested in falsifying the formula $\varphi$, it suffices to ensure that the constructed model evaluates formulas in the closure of $\varphi$ correctly. Therefore the following definition is given relative to a finite and closed set of formulas $\Sigma$.

\begin{definition}
   Let $\Sigma$ be a closed and finite set of formulas. Let $\fw=(W, \leq, R, V)$ be a dynamic model, $\langle \mathcal{I}=(I, \leq_I, R_I, V_I), \pi \rangle$ an $\fw$-induced structure and $x \in I$. 
    \begin{enumerate}
        \item An $\rightarrow$\emph{-defect}\index{defect!$\rightarrow$-defect} of $\mathcal{I}$ is a tuple $(x,\psi \rightarrow \chi)$ where $x \in I$ and $\psi \rightarrow \chi \in \Sigma$, such that $\fw, \pi(x) \not \models \psi \rightarrow \chi$, but there does not exist a world $y \in I$ such that $y \geq_I x$ and $\fw, \pi(y) \models \psi$ and $\fw, \pi(y) \not \models \chi$.
        \item A $\lb$\emph{-defect}\index{defect!$\lb$-defect} of $\mathcal{I}$ is a  tuple $(x,\lb \psi)$ where $x \in I$ and $\lb \psi \in \Sigma$, such that $\fw, \pi(x) \not \models \lb \psi$ but there does not exists a world $y \in I$ with $x \mathrel{( R_I \circ {\leq_I})}  y$ and $\fw, \pi(y) \not \models \psi$.
        \item A $\lbm$\emph{-defect}\index{defect!$\lbm$-defect} of $\mathcal{I}$ is a tuple $(x, \lbm \psi)$ where $x \in I$ and $\lbm \psi \in \Sigma$, such that $\fw, \pi(x) \not \models \lbm \psi$ but there does not exist a world $y \in I$ with $x \mathrel{( R_I \circ {\leq_I})^\ast} y$ and $\fw, \pi(y) \not \models \psi$.
    \end{enumerate}
\end{definition}

Observe that for any $\fw$-induced structure $\mathcal{I}$ and any closed and finite set of formulas $\Sigma$, there are only finitely many defects, due to $I$ and $\Sigma$ being finite. \smallskip

Suppose $\fw=(W, \leq, R, V)$ is a dynamic model and $w \in W$ a world such that $\fw,w \not \models \varphi$. We are going to construct a functional model $\mathcal{I}_\omega=(I_\omega, \leq_\omega, R_\omega, V_\omega)$ falsifying $\varphi$ in stages. To that end let $\Sigma = \Cl(\varphi)$. We define for each natural number $n$ an $\fw$-induced structure $\mathcal{I}_n$ as well as a first-in-first-out queue $D$ that stores the defects of $\mathcal{I}_n$ and we simultaneously prove that $\mathcal{I}_n$ is an $\fw$-induced structure. The functional model $\mathcal{I}_\omega$ is then defined as the limit of the sequence $(\mathcal{I}_n)_{n < \omega}$. \smallskip

\noindent \textbf{Stage 0:} Define the $\fw$-induced structure $\langle \mathcal{I}_0 =(I_0, \leq_0, R_0, V_0), \pi_0\rangle$ as follows.
\begin{itemize}
    \item $I_0 \coloneqq \{\Tilde{x}\}$ where $\Tilde{x}$ is a fresh world not contained in $W$.
    \item $\leq_0 \coloneqq \{(\Tilde{x},\Tilde{x})\}$.
    \item $R_0 \coloneqq \emptyset$.
    \item $V_0(\Tilde{x}) \coloneqq V(w)$.
    \item $\pi_0(\Tilde{x}) \coloneqq w$.
\end{itemize}

To complete stage 0, initialize the queue $D$ by adding all defects of $\mathcal{I}_0$ to $D$ in arbitrary order. It is immediate to check that $\mathcal{I}_0$ is an $\fw$-induced structure. \smallskip

\noindent \textbf{Stage n+1:} Suppose $\langle \mathcal{I}_n =(I_n, \leq_n, R_n, V_n), \pi_n \rangle$ has been defined and $D$ currently stores the defects of $\mathcal{I}_n$. We show first how to construct $\langle \mathcal{I}_{n+1}, \pi_{n+1}\rangle$ depending on the defect at the head of the queue $D$ and then how to update $D$. \smallskip

\noindent \textsc{($\rightarrow$-defects)} Suppose the defect at the head of the queue $D$ is an $\rightarrow$-defect $(y, \psi \rightarrow \chi)$. Then $y \in I_n$, $\psi \rightarrow \chi \in \Sigma$ and $\fw, \pi_n(y) \not \models \psi \rightarrow \chi$, however there does not exists a world $y' \in I_n$ with $y' \geq_n y$ such that $\fw, \pi_n(y') \models \psi$ and $\fw, \pi_n(y') \not \models \chi$. Since ${\fw, \pi_n(y) \not \models \psi \rightarrow \chi}$, there exists a world $u \geq \pi_n(y)$ such that $\fw,u \models \psi$ and $\fw,u \not \models \chi$. Let $y'$ be a fresh world not occurring in $I_n$ or $W$ and define
    \begin{itemize}
        \item $I_{n+1} \coloneqq I_n \cup \{y'\}$;
        \item let $\leq_{n+1}$ be the reflexive transitive closure of $\leq_n \cup \{(y, y')\}$;
        \item $R_{n+1} \coloneqq R_n$;
        \item $V_{n+1}$ is defined by \begin{equation*}
            V_{n+1}(z) := \begin{cases}
                V(u) & \text{if } z=y'\\
                V_n(z) & \text{otherwise;}\\
            \end{cases}
        \end{equation*}
        \item $\pi_{n+1}$ is defined by
        \begin{equation*}
                \pi_{n+1}(z) := \begin{cases}
                    u & \text{if } z= y'\\
                    \pi_n(z) & \text{otherwise.}\\
                \end{cases}
        \end{equation*}
    \end{itemize}
    To see that $\mathcal{I}_{n+1}$ is an $\fw$-induced structure, first note that by induction hypothesis $I_n$ is a finite set, and therefore $I_{n+1}$ is a finite set too. Suppose $x \leq_{n+1} z$. Then either $x \leq_n z$ which implies by induction hypothesis that $\pi_n(x) \leq \pi_n(z)$ and thus also $\pi_{n+1}(x) \leq \pi_{n+1}(z)$, or $x \leq_n y$ and $z= y'$, which, by construction and induction hypothesis, implies $\pi_{n+1}(x) \leq \pi_{n+1}(y) \leq \pi_{n+1}(z)$. That $(I_{n+1}, \leq_{n+1})$ is a poset follows immediately from the induction hypothesis and the construction. Next, $R_{n+1} = R_n$ by construction and thus, by induction hypothesis, $R_{n+1}$ is a partial function and if $R_{n+1}(x) = z$, then $\pi_{n+1}(x)\mathrel{R}\pi_{n+1}(z)$. If $x \in I_n$, then $V_n(x) = V(\pi_n(x))$ by induction hypothesis, so by construction $V_{n+1}(x) = V(\pi_{n+1}(x))$. Otherwise $x= y'$ and $V_{n+1}(x) = V(\pi_{n+1}(x))$ by construction. Thus $V_{n+1}= V \circ \pi_{n+1}$. Finally, by induction hypothesis $\pi_n: I_n \longrightarrow W$ is a function. Since $y'$ is a fresh world, $\pi_{n+1}:I_{n+1} \longrightarrow W$ is a function as well. Thus $\langle \mathcal{I}_{n+1}, \pi_{n+1}\rangle$ is an $\fw$-induced structure. \smallskip
    
    \noindent \textsc{($\lb$-defects)} Suppose the defect at the head of the queue $D$ is a $\lb$-defect $(y, \lb \psi)$. Then $y \in I_n$, $\lb \psi \in \Sigma$, $\fw, \pi_n(y) \not \models \lb \psi$ but there does not exist a world $y' \in I_n$ with $y \mathrel{(R_n \circ {\leq_n})} y'$ and $\fw, \pi_n(y') \not \models \psi$. Since $\fw, \pi_n(y) \not \models \lb \psi$, there exist worlds $u_0, u_1 \in W$ with $\pi_n(y) \leq u_0 \mathrel{R} u_1$ and $\fw, u_1 \not \models \psi$. Let $y_0, y_1$ be fresh worlds\footnote{Observe that $u_0$ might be identical to $\pi_n(y)$. Nevertheless, we add a fresh world $y_0$ to $I_n$ which in this case would be a copy of $y$.} not occurring in $I_n$ or $W$. Define
\begin{itemize}
    \item $I_{n+1} \coloneqq I_n \cup \{y_0, y_1\}$;
    \item let $\leq_{n+1}$ be the reflexive transitive closure of $\leq_n \cup \{(y, y_0)\}$;
    \item $R_{n+1} \coloneqq R_n \cup \{(y_0,y_1)\}$;
    \item $V_{n+1}$ is defined by
    \begin{equation*}
        V_{n+1}(z) := \begin{cases}
            V(u_0) & \text{if } z= y_0\\
        V(u_1) & \text{if } z= y_1 \\
        V_n(z) & \text{otherwise;}\\
        \end{cases}
    \end{equation*}
    \item $\pi_{n+1}$ is defined by
     \begin{equation*}
        \pi_{n+1}(z) := \begin{cases}
            u_0 & \text{if } z= y_0\\
        u_1 & \text{if } z= y_1 \\
        \pi_n(z) & \text{otherwise.}\\
        \end{cases}
    \end{equation*}
\end{itemize}
Let us check that $\mathcal{I}_{n+1}$ is an $\fw$-induced structure. That $I_{n+1}$ is finite, $x \leq_{n+1} y$ implies $\pi_{n+1}(x) \leq \pi_{n+1}(y)$ and that $(I_{n+1}, \leq_{n+1})$ is a poset follow by similar argument as above. Suppose $R_{n+1}(x) = z$. Then either $R_n(x) = z$, which by induction hypothesis implies that $\pi_n(x) \mathrel{R} \pi_n(y)$ and so $\pi_{n+1}(x) \mathrel{R} \pi_{n+1}(y)$ by construction, or $x=y_0$ and $z = y_1$. In that case we have by construction $\pi_{n+1}(x) \mathrel{R} \pi_{n+1}(y)$ as well. Moreover, by induction hypothesis and construction, $R_{n+1}$ is a partial function. That $V_{n+1} = V \circ \pi_{n+1}$ and that $\pi_{n+1}$ is a function follow by similar arguments as above.\smallskip

\noindent \textsc{($\lbm$-defects)} Suppose the defect at the head of the queue $D$ is a $\lbm$-defect $(y, \lbm \psi)$. Then ${y \in I_n}$, $\lbm \psi \in \Sigma$, $\fw, \pi_n(y) \not \models \lbm \psi$ but there does not exist a world $y' \in I_n$ with ${y \mathrel{(R_n \circ {\leq_n})^*} y'}$ such that $\fw, \pi_n(y') \not \models \psi$. Since $\fw, \pi_n(y) \not \models \lbm \psi$, there exists a world $u \in W$ with $\pi_n(y) \mathrel{(R \circ {\leq})^*} u$ and $\fw,u \not \models \psi$. Let $u_0, \ldots, u_{2k} \in W$ such that $u_0 = \pi_n(y)$, $u_{2k} = u$ and for all $0 \leq i < k$ holds that $u_{2i} \leq u_{2i+1}$ and $u_{2i+1} \Rel u_{2(i+1)}$. Observe that $k > 0$ as by assumption $\fw, \pi_n(y) \models \psi$ (again, some worlds $u_{2i+1}$ might be identical to $u_{2i}$). Let $y_i$ for $1 \leq i \leq 2k$ be fresh worlds not occuring in $I_n$ or $W$. Define
\begin{itemize}
    \item $I_{n+1} \coloneqq I_n \cup \{y_1, \ldots, y_{2k}\}$;
    \item let $\leq_{n+1}$ be the reflexive transitive closure of
    \begin{equation*}
       \leq_n \cup \{(y, y_1)\} \cup \{(y_{2i}, y_{2i+1}) \, \lvert \, 1 \leq i < k\};
    \end{equation*}
    \item $R_{n+1} \coloneqq R_n \cup \{(y_{2i+1}, y_{2(i+1)}) \, \lvert \, 0 \leq i < k\}$;
    \item $V_{n+1}$ is defined by
    \begin{equation*}
        V_{n+1}(z) \coloneqq \begin{cases}
            V(u_i) & \text{if } z= y_i \text{ for } 1 \leq i \leq 2k\\
            V_n(z) & \text{otherwise;}\\
        \end{cases}
    \end{equation*}
    \item $\pi_{n+1}$ is defined by
    \begin{equation*}
        \pi_{n+1}(z) \coloneqq \begin{cases}
            u_i & \text{if } z= y_i \text{ for } 1 \leq i \leq 2k\\
            \pi_n(z) & \text{otherwise.}\\
        \end{cases}
    \end{equation*}
\end{itemize}
Let us check that $\mathcal{I}_{n+1}$ is an $\fw$-induced structure. That $I_{n+1}$ is finite and $(I_{n+1}, \leq_{n+1})$ is a poset follows by similar arguments as above. Suppose $x \leq_{n+1} z$. Then either $x \leq_n z$, in which case by induction hypothesis and construction $\pi_{n+1}(x) \leq \pi_{n+1}(z)$, or $x\leq_n y$ and $z = y_1$, in which case by induction hypothesis and construction $\pi_{n+1}(x) \leq \pi_{n+1}(y) \leq \pi_{n+1}(z)$, or $x= y_{2i}$ and $z =y_{2i+1}$ for $1 \leq i \leq k$ (observe that by construction no world lies below $y_{2i}$ in the order $\leq_{n+1}$). In this case $\pi_{n+1}(x) \leq \pi_{n+1}(z)$ by construction. Thus $x \leq_{n+1} z$ implies $\pi_{n+1}(x) \leq \pi_{n+1}(z)$. By induction hypothesis $R_n$ is a partial function and so by construction $R_{n+1}$ is a partial function as well. Suppose $R_{n+1}(x) = z$. Then either $R_n(x) = z$, from which by induction hypothesis and construction follows that $\pi_{n+1}(x) \mathrel{R} \pi_{n+1}(z)$, or $x = y_{2i+1}$ and $z=y_{2(i+1)}$ for $1 \leq i < k$. In that case $\pi_{n+1}(x) \mathrel{R} \pi_{n+1}(z)$ by construction. Thus $R_{n+1}(x) = z$ implies $\pi_{n+1}(x) \mathrel{R} \pi_{n+1}(z)$. Finally, the cases for $V_{n+1}$ and $\pi_{n+1}$ follow by similar arguments as above. \smallskip

Having defined $\langle \mathcal{I}_{n+1}, \pi_{n+1}\rangle$, we update the queue $D$ as follows. First, delete every defect that has been resolved in the step from $n$ to $n+1$. Then add every new defect of $\mathcal{I}_{n+1}$ to the tail of the queue $D$. Observe that by induction hypothesis, the queue $D$ stores finitely many defects at stage $n$. After updating, we add at most finitely many new defects to $D$, since $I_{n+1}$ and $\Sigma$ are finite. We have therefore shown the following:

\begin{lemma}\label{l: I_n induced structure}
    For all $n < \omega$, $\langle \mathcal{I}_n, \pi_n\rangle$ is an $\fw$-induced structure.
\end{lemma}

The functional model $\mathcal{I}_\omega$ is now defined as the limit of the sequence $(\mathcal{I}_n)_{n < \omega}$.

\begin{definition}
    Let $\mathcal{I}_\omega = (I_\omega, \leq_\omega, R_\omega, V_\omega)$ be defined by
    \begin{equation*}
        \lambda_\omega \coloneqq \bigcup_{n < \omega} \lambda_n
    \end{equation*}
    where $\lambda \in \{I, \leq, R, V\}$. Additionally, let $\pi_\omega \coloneqq \bigcup_{n < \omega} \pi_n$.
\end{definition}

The structure $\mathcal{I}_\omega$ is called the $\fw$-\emph{induced model}.
Observe that $x \in I_\omega$ if and only if there exists $n < \omega$ such that $x \in I_n$. Moreover, $x \leq_\omega y$ if and only if there exists $n < \omega$ such that $x \leq_n y$ and similarly for $x \Rel_\omega y$ and $p \in V_\omega(x)$. 

\begin{lemma}\label{l: properties induced model}
    Let $x,y \in I_\omega$ and $n \leq m < \omega$. The following hold.
    \begin{enumerate}
        \item If $x \leq_n y$, then $x \leq_m y$.
        \item If $x \in I_n$, then $\pi_n(x) = \pi_m(x) = \pi_\omega(x)$.
        \item If $x \in I_n$, then $V_n(x) = V_m(x) = V_\omega(x)$.
    \end{enumerate}
\end{lemma}
\begin{proof}
    By construction.
\end{proof}

\begin{lemma}\label{l: I_omega is a functional model}
    $\mathcal{I}_\omega$ is a functional model and $\pi_\omega$ is a function.
\end{lemma}
\begin{proof}
By Lemma \ref{l: I_n induced structure} and the definition of an induced structure we have that $\leq_n$ is a partial order on $I_n$ for each $n < \omega$. This immediately implies that $\leq_\omega$ is reflexive and transitive. For antisymmetry, suppose $x,y \in I_\omega$ with $x \leq_\omega y$ and $y \leq_\omega x$. Then there are natural numbers $n,m$ such that $x \leq_n y$ and $y \leq_m x$. Suppose without loss of generality that $n \leq m$. By Lemma \ref{l: properties induced model}, $x \leq_m y$ and since $\leq_m$ is antisymmetric, $x = y$. Therefore $\leq_\omega$ is a partial order on $I_\omega$. Next, suppose that $x \mathrel{R_\omega} y$ and $x \mathrel{R_\omega} z$. Let $n$ be the least natural number such that $x,y,z \in I_n$. Recall that $R_n$ is a partial function. Moreover, for any $m > n$ the construction does not add modal successors to worlds in $I_n$.\footnote{It might be that at step $m > n$ a $\lb$-defect situated at a world $x$ in $I_n$ is resolved. But in that case the construction first adds a new world above $x$ and then a modal successor for that new world. The case for $\lbm$-defects is similar.} Therefore $x \mathrel{R_n} y$ and $x \mathrel{R_n} z$, implying that $y=z$. Thus $R_\omega$ is a partial function. Finally, suppose that $x \leq_\omega y$. Let $n$ be the least natural number such that $x,y \in I_n$. By construction $V_n(x) = V(\pi_n(x))$ and $V_n(y) = V(\pi_n(y))$. Since $x \leq_n y$, Property 2. of an $\fw$-induced structure guarantees that $\pi_n(x) \leq \pi_n(y)$. Since $V$ is monotone, $V(\pi_n(x)) \subseteq V(\pi_n(y))$. Thus $V_n(x) \subseteq V_n(y)$. Lemma \ref{l: properties induced model} then implies $V_\omega(x) \subseteq V_\omega(y)$, and so that $V_\omega$ is monotone in $\leq_\omega$.
\end{proof}
 
  It remains to show that the $\fw$-induced model preserves truth. Recall that $\varphi \in \LIM$ with $\fw, w \not \models \varphi$ and $\Cl(\varphi) = \Sigma$.

\begin{lemma}\label{l: I-omega falsifies phi}
  For any $\psi \in \Sigma$ and any $z \in I_\omega$, the following holds.
  \begin{equation*}
      \fw, \pi_\omega(z) \models \psi \text{ if and only if } \mathcal{I}_\omega, z \models \psi.
  \end{equation*}
\end{lemma}
\begin{proof}
    The proof proceeds by induction on $\psi$. The case for $\psi = \bot$ is trivial. Suppose $\psi = p$ for $p \in \Prop$. Using Lemma \ref{l: properties induced model} we obtain that $\fw, \pi_\omega(z) \models p$ if and only if $p \in V(\pi_\omega(z))$ if and only if $p \in V_\omega(z)$ if and only if $\mathcal{I}_\omega, z \models p$. For the induction step, the cases for $\psi = \chi \wedge \gamma$ and $\psi = \chi \vee \gamma$ follow immediately from the induction hypothesis and the fact that $\chi, \gamma \in \Sigma$ since $\Sigma$ is assumed to be closed. \smallskip

    \noindent \textsc{Case for $\rightarrow$.} Suppose $\psi = \chi \rightarrow \gamma$. Note that $\chi, \gamma \in \Sigma$. If $\mathcal{I}_\omega, z \not \models \chi \rightarrow \gamma$, then there exists a world $y$ such that $z \leq_\omega y$ and $\mathcal{I}_\omega, y \models \chi$ and $\mathcal{I}_\omega, y \not \models \gamma$. Let $n$ be the least natural number such that $z,y \in I_n$ and $z \leq_n y$. Thus $\pi_n(z) \leq \pi_n (y)$ and hence $\pi_\omega(z) \leq \pi_\omega (y)$. The induction hypothesis yields $\fw, \pi_\omega(y) \models \chi$ and $\fw, \pi_\omega(y) \not \models \gamma$. Hence $\fw, \pi_\omega(z) \not \models \chi \rightarrow \gamma$. For the other direction suppose  $\fw, \pi_\omega(z) \not \models \chi \rightarrow \gamma$. Let $n$ be the least natural number such that $z \in I_n$. Then either there exists $y \in I_n$ with $z \leq_n y$ and $\fw, \pi_n(y) \models \chi$ and $\fw, \pi_n(y) \not \models \psi$, or at the end of stage $n$ the queue $D$ is updated with the defect $(z, \chi \rightarrow \gamma)$. In that case there exists $m > n$ such that at stage $m$ the defect $(z, \chi \rightarrow \gamma)$ is resolved, i.e. there exists $y \in I_{m+1}$ with $z \leq_{m+1} y$ and $\fw, \pi_{m+1}(y) \models \chi$ and $\fw, \pi_{m+1}(y) \not \models \psi$. In either case $z \leq_\omega y$ and $\fw, \pi_\omega(y) \models \chi$ and $\fw, \pi_\omega(y) \not \models \psi$, so the induction hypothesis implies $\mathcal{I}_\omega, y  \models\chi$ and $\mathcal{I}_\omega, y  \not \models \gamma$ and so $\mathcal{I}_\omega, z \not \models\chi \rightarrow \gamma$. \smallskip
    
    \noindent \textsc{Case for $\lb$.} Suppose $\psi = \lb \gamma$ and note that $\gamma \in \Sigma$. If $\mathcal{I}_\omega, z \not \models \lb \gamma$, then there exists a world $y$ such that $z \mathrel{(R_\omega \circ {\leq_\omega})} y$ and $\mathcal{I}_\omega, y \not \models \gamma$. The induction hypothesis implies that $\fw, \pi_\omega(y) \not \models \gamma$. Let $n$ be the least natural number such that $z,y \in I_n$ and $z \mathrel{(R_n \circ {\leq_n})} y$. So $\pi_n(z) \mathrel{(R \circ {\leq})} \pi_n(y)$. Hence, $\pi_\omega(z) \mathrel{(R \circ {\leq})} \pi_\omega(y)$, implying that $\fw, \pi_\omega(z) \not \models \lb \gamma$. For the other direction suppose that $\fw, \pi_\omega(z) \not \models \lb \gamma$. Let $n$ be the least natural number such that $z \in I_n$. Either there exists $y \in I_n$ with $z \mathrel{(R_n \circ {\leq_n})} y$ and $\fw, \pi_n(y) \not \models \gamma$ or at the end of stage $n$, the $\lb$-defect $(z, \lb \gamma)$ is added to the tail of the queue $D$. In that case there exists $m > n$ such that at stage $m$ the defect $(z, \lb \gamma)$ is resolved. Thus there exists $y \in I_{m+1}$ with $z \mathrel{(R_{m+1} \circ {\leq_{m+1}})} y$ and $\fw, \pi_{m+1}(y) \not \models \gamma$. In either case $z \mathrel{(R_\omega \circ {\leq_\omega})} y$ and $\fw, \pi_\omega(y) \not \models \gamma$. By induction hypothesis $\mathcal{I}_\omega, y \not \models \gamma$ and so $\mathcal{I}_\omega, z \not \models \lb \gamma$. The case for $\varphi = \lbm \psi$ is similar and omitted.
\end{proof}

\begin{corollary}\label{c: falsifiable over b models implies falsifiable over f models}
$\mathbf{IM_f} \subseteq \mathbf{IM}$.
\end{corollary}
\begin{proof}
    Suppose $\varphi$ is falsifiable over the class of dynamic models. Then there exists a dynamic model $\fw=(W, \leq, R, V)$ and a world $w \in W$ such that $\fw,w \not \models \varphi$. Let $\Sigma = \Cl(\varphi)$ and consider the $\fw$-induced model $\mathcal{I}_\omega$ with $\pi_\omega(x) = w$ for some $x \in I_\omega$. By Lemma \ref{l: I_omega is a functional model}, $\mathcal{I}_\omega$ is a functional model. It then follows from Lemma \ref{l: I-omega falsifies phi} that $\mathcal{I}_\omega, x \not \models \varphi$. Hence $\varphi$ is falsifiable over the class of functional models.
\end{proof}

\begin{theorem}\label{t: three classes one logic}
Let $\varphi \in \LIM$. The following are equivalent.
\begin{enumerate}
    \item $\varphi$ is valid over the class of dynamic models.
    \item $\varphi$ is valid over the class of functional models.
    \item $\varphi$ is valid over the class of triangle models.
\end{enumerate}
\end{theorem}

Theorem \ref{t: three classes one logic} raises the question whether a similar result holds for \emph{serial} dynamic models (i.e. dynamic models with a serial modal accessibility relation), serial triangle models and total functional models? Clearly, transforming serial dynamic models into serial triangle models works by employing the same construction as given in Lemma \ref{l: birelational model induces triangle model}. For transforming serial dynamic models into total functional models, some small adaptions of the construction presented in the proof of Theorem \ref{t: three classes one logic} are required. We briefly sketch the basic idea. \smallskip

 In the construction of a functional model, $\lb$-defects $(x, \lb \varphi)$ were resolved by first adding an intuitionistic successor $y$ of $x$ and then a modal successor $z$ of $y$ to provide a witness where $\lb \varphi$ is falsified. This implies that many worlds in $\mathcal{I}_\omega$ do not have a modal successor. When constructing a total functional model we must guarantee that \emph{every} world has a unique modal successor. There are several ways how to achieve this. Perhaphs the easiest solution is to use the same construction as for functional models (i.e. $\lb$- and $\lbm$-defects are resolved by adding intuitionistic successors and then modal successors), but to consider an additional type of defect called a \emph{functional defect}. Such a defect is simply a world $x$ in an induced structure for which no modal successor exists. Functional defects are then resolved by considering the related world $\pi(x)$ in the serial dynamic model and chose an arbitrary modal successor $u$ of $\pi(x)$, which exists by seriality. Then add a fresh world $y$ to the induced structure as a modal successor of $x$ and define $\pi(y) = u$. The resulting structure is a total functional model which satisfies the corresponding version of Lemma \ref{l: I-omega falsifies phi}.

This explanation should suffice to convince us that $\LIM$ cannot distinguish between serial dynamic models, serial triangle models and total functional models either. 

\begin{remark}
    The previous observation discourages an intepretation of $\LIM$ over total functional models as an intuitionistic version of linear temporal logic, since the resulting logic is simply the logic obtained by evaluating $\LIM$ over serial dynamic models.
\end{remark}

\section{Axiomatization}\label{c: IM, section axiomatization}

This section introduces a Hilbert-style axiomatization capturing the $\LIM$-validities over the classes of dynamic / triangle / functional models. An axiomatization for a multi-modal version of $\IM$ was presented in~\cite{marti_2017}. Here, we present a different axiomatization, which is more in line with axiomatizations given later on in the thesis. The presented axiomatization is called $\mathrm{IM_H}$. We also establish basic properties of $\mathrm{IM_H}$, such as soundness and the Deduction Theorem.

\begin{definition}
    The axiomatization $\mathrm{IM_H}$\index{axiomatization!$\mathrm{IM_H}$} consists of the axiom schemes and rules depicted in Table \ref{t: axioms and rules of Hilbert system}. 
\end{definition}

  The axiom schemes are $\mathsf{Int}$, $\mathsf{K}$ and $\mathsf{Fix}$. Instances of $\mathsf{Int}$ are intuitionistic tautologies, i.e. $\LIM$-formulas obtained from a $\mathrm{IPL_H}$-derivable formula via uniform substitution; see Definition \ref{d: intuitionistic tautologies}. The axiom scheme $\mathsf{K}$ describes that $\lb$ distributes over implications while $\mathsf{Fix}$ describes that $\lbm \varphi$ is equivalent to its \emph{unfolding} $\varphi \wedge \lb \lbm \varphi$. The inference rules are $\mathsf{MP}$, $\mathsf{Nec}$, $\mathsf{Mon}$ and $\mathsf{Ind}$.  The rule $\mathsf{Ind}$ is the \emph{induction rule}. If we have a proof of $\varphi$ (the base case) and a proof of $\varphi \rightarrow \lb \varphi$ (the induction step), then we can apply $\mathsf{Ind}$ and $\mathsf{MP}$ to obtain a proof of $\lbm \varphi$.

\begin{table}
\centering
        \begin{tabular}{|l l|}
        \hline
         $\mathsf{Int}$:  Intuitionistic tautologies & \\
         $\mathsf{K}$: $\lb(\varphi \rightarrow \psi) \rightarrow (\lb \varphi \rightarrow \lb \psi)$ & \\
         $\mathsf{Fix}$: $\lbm \varphi \leftrightarrow (\varphi \wedge \lb \lbm \varphi)$ & \\
          & \\
         $\mathsf{MP}$: $\infer{\psi}{\varphi & \varphi \rightarrow \psi}$ & $\mathsf{Nec}$: $\infer{\lb \varphi}{\varphi}$ \\
         $\mathsf{Mon}$: $\infer{\lbm \varphi \rightarrow \lbm \psi}{\varphi \rightarrow \psi}$ &  $\mathsf{Ind}$: $\infer{\varphi \rightarrow \lbm \varphi}{\varphi \rightarrow \lb \varphi}$\\
         \hline
        \end{tabular}
     \caption{The Hilbert-style axiomatization $\mathrm{IM_H}$}  
     \label{t: axioms and rules of Hilbert system}
\end{table}

\begin{definition}
    Let $\Gamma \cup \{\varphi\} \subseteq \LIM$. A \emph{derivation of $\varphi$ with assumptions in $\Gamma$}\index{derivation!$\mathrm{IM_H}$-derivation} in $\mathrm{IM_H}$ is a finite tree $\pi$ labelled by formulas according to the rules of $\mathrm{IM_H}$ such that the following hold.
    \begin{enumerate}
        \item Every leaf is labelled by an axiom or by a formula $\psi \in \Gamma$.
        \item If a node $u \in \pi$ is labelled by the conclusion of a rule instance of $\mathsf{Nec}$, $\mathsf{Mon}$ or $\mathsf{Ind}$, then every leaf of the subtree of $\pi$ rooted at $u$ is labelled by an axiom.
    \end{enumerate}
    We write $\Gamma \vdash_\mathrm{IM_H} \varphi$ if there exists a derivation of $\varphi$ with assumptions in $\Gamma$ and $\vdash_\mathrm{IM_H} \varphi$ if $\Gamma = \emptyset$. If the proof system $\mathrm{IM_H}$ is clear from context, we also write $\Gamma \vdash \varphi$.
\end{definition}

If $\Gamma \vdash \varphi$, then $\varphi$ is called \emph{derivable from $\Gamma$}. If $\Gamma = \emptyset$, then $\varphi$ is simply called \emph{derivable}. Note that the rules $\mathsf{Nec}$, $\mathsf{Mon}$ and $\mathsf{Ind}$ cannot be applied to assumptions. We will use the standard proof theoretic conventions and denote sets of formulas with $\Gamma, \Delta$ etc. and we write $\Gamma, \varphi$ for $\Gamma \cup \{\varphi\}$. Moreover, derivations (with or without assumptions) are denoted by $\pi$ or $\tau$.
 
The following lemma serves as an example for derivations and, at the same time, establishes the derivability of some formulas that will become useful in the completeness proof.

\begin{lemma}\label{l: derivable formulas}
    The following hold:
    \begin{enumerate}
        \item\label{l: derivable formulas item 1} If $\vdash \varphi \rightarrow \psi$ and $\vdash \psi \rightarrow \chi$, then $\vdash \varphi \rightarrow \chi$.
        \item\label{l: derivable formulas item 2} $\vdash \lb (\varphi \wedge \psi) \rightarrow (\lb \varphi \wedge \lb \psi)$ and $\vdash (\lb \varphi \wedge \lb \psi) \rightarrow \lb (\varphi \wedge \psi)$.
        \item\label{l: derivable formulas item 3} $\vdash (\lb \varphi \vee \lb \psi) \rightarrow \lb (\varphi \vee \psi)$.\footnote{A semantic analysis immediately shows that the other direction, i.e. $\vdash \lb (\varphi \vee \psi) \rightarrow (\lb \varphi \vee \lb \psi)$, should \emph{not} hold. The soundness theorem (c.f. Theorem \ref{t: soundness}) will confirm this intuition.}
        \item\label{l: derivable formulas item 4} If $\vdash \varphi$, then $\vdash \psi \rightarrow \varphi$ for any $\psi \in \LIM$.
        \item\label{l: derivable formulas item 5} If $\vdash \varphi$, then $\vdash \lbm \varphi$ (necessitation of $\lbm$).
    \end{enumerate}
\end{lemma}
\begin{proof}
   1. It suffices to observe that $(\varphi \rightarrow \psi) \rightarrow ((\psi \rightarrow \chi) \rightarrow (\varphi \rightarrow \chi))$ is an intuitionistic tautology and hence derivable. The statement then follows by applying $\mathsf{MP}$ to the above formula and the formulas that are derivable by assumption. \smallskip
    
   \noindent 2. Observe that $(\varphi \wedge \psi) \rightarrow \varphi$ and $(\varphi \wedge \psi) \rightarrow \psi$ are intuitionistic tautologies. By $\mathsf{Nec}$, $\vdash \lb ((\varphi \wedge \psi) \rightarrow \varphi)$ and $\vdash \lb ((\varphi \wedge \psi) \rightarrow \psi)$. By the $\mathsf{K}$--axiom and $\mathsf{MP}$ we obtain $\vdash \lb(\varphi \wedge \psi) \rightarrow \lb \varphi$ and $\vdash \lb(\varphi \wedge \psi) \rightarrow \lb \psi$. Since 
    \begin{equation*}
     (\lb(\varphi \wedge \psi) \rightarrow \lb \varphi) \rightarrow ((\lb(\varphi \wedge \psi) \rightarrow \lb \psi) \rightarrow (\lb(\varphi \wedge \psi) \rightarrow (\lb \varphi \wedge \lb \psi)))
    \end{equation*}
    is an intuitionistic tautology, we obtain $\vdash \lb(\varphi \wedge \psi) \rightarrow (\lb \varphi \wedge \lb \psi)$ by applying $\mathsf{MP}$ twice. For the other direction observe that
    \begin{equation}\label{e: 1}
        \vdash \varphi \rightarrow (\psi \rightarrow (\varphi \wedge \psi))
    \end{equation}
    as this formula is an intuitionistic tautology. Applying $\mathsf{Nec}$ yields
    \begin{equation}\label{e: 2}
        \vdash \lb (\varphi \rightarrow (\psi \rightarrow (\varphi \wedge \psi))).
    \end{equation}
    Note that the following formula is an instance of $\mathsf{K}$ and is hence derivable:
    \begin{equation}\label{e: 3}
        \vdash \lb (\varphi \rightarrow (\psi \rightarrow (\varphi \wedge \psi))) \rightarrow (\lb \varphi \rightarrow \lb (\psi \rightarrow (\varphi \wedge \psi))).
    \end{equation}
    Applying $\mathsf{MP}$ to (\ref{e: 2}) and (\ref{e: 3}) yields
    \begin{equation}\label{e: 4}
        \vdash \lb \varphi \rightarrow \lb (\psi \rightarrow (\varphi \wedge \psi)).
    \end{equation}
    The following is an instance of $\mathsf{K}$:
    \begin{equation}\label{e: 5}
        \vdash \lb (\psi \rightarrow (\varphi \wedge \psi)) \rightarrow (\lb \psi \rightarrow \lb (\varphi \wedge \psi)).
    \end{equation}
    Item \ref{l: derivable formulas item 1} of this lemma applied to (\ref{e: 4}) and (\ref{e: 5}) yields
    \begin{equation}\label{e: 6}
        \vdash \lb \varphi \rightarrow (\lb \psi \rightarrow \lb (\varphi \wedge \psi)).
    \end{equation}
    Observe that the following is an intuitionistic tautology and thus derivable:
    \begin{equation}\label{e: 7}
        \vdash (\lb \varphi \rightarrow (\lb \psi \rightarrow \lb (\varphi \wedge \psi))) \rightarrow ((\lb \varphi \wedge \lb \psi) \rightarrow \lb (\varphi \wedge \psi))
    \end{equation}
    Applying $\mathsf{MP}$ to (\ref{e: 6}) and (\ref{e: 7}) thus yields
    \begin{equation}\label{e: 8}
       \vdash (\lb \varphi \wedge \lb \psi) \rightarrow \lb (\varphi \wedge \psi)
    \end{equation}
    which concludes the proof. \smallskip
    
    \noindent 3. Observe that $\varphi \rightarrow (\varphi \vee \psi)$ and $\psi \rightarrow (\varphi \vee \psi)$ are intuitionistic tautologies. By $\mathsf{Nec}$, $\vdash \lb(\varphi \rightarrow (\varphi \vee \psi))$ and $\vdash \lb(\psi \rightarrow (\varphi \vee \psi))$. By using the $\mathsf{K}$--axiom and $\mathsf{MP}$ we obtain $\vdash \lb \varphi \rightarrow \lb (\varphi \vee \psi)$ and $\vdash \lb \psi \rightarrow \lb (\varphi \vee \psi)$. Since
    \begin{equation*}
        (\lb \varphi \rightarrow \lb(\varphi \vee \psi)) \rightarrow ((\lb \psi \rightarrow \lb(\varphi \vee \psi)) \rightarrow ((\lb \varphi \vee \lb \psi) \rightarrow \lb(\varphi \vee \psi)))
    \end{equation*}
    is an intuitionistic tautology, we obtain $\vdash (\lb \varphi \vee \lb \psi) \rightarrow \lb (\varphi \vee \psi)$ by applying $\mathsf{MP}$ twice. \smallskip 
    
     \noindent 4. Suppose that $\vdash \varphi$ and let $\psi \in \LIM$. Since $\varphi \rightarrow (\psi \rightarrow \varphi)$ is an intuitionistic tautology, we obtain $\vdash \psi \rightarrow \varphi$ by $\mathsf{MP}$. \smallskip
    
    \noindent 5. Suppose that $\vdash \varphi$. By $\mathsf{Nec}$, $\vdash \lb \varphi$. By 4. we obtain $\vdash \varphi \rightarrow \lb \varphi$. Applying $\mathsf{Ind}$ yields $\vdash \varphi \rightarrow \lbm \varphi$. Finally, by applying $\mathsf{MP}$ we obtain $\vdash \lbm \varphi$.
\end{proof}

\begin{lemma}[Assumption weakening] \label{l: assumption weakening}
    If $\Gamma \vdash \varphi$ and $\Gamma \subseteq \Delta$, then $\Delta \vdash \varphi$.
\end{lemma}
\begin{proof}
    This follows immediately from the definition of a derivation with assumptions.
\end{proof}

We may now prove the Deduction Theorem, which will play a crucial role in the completeness proof.

\begin{theorem}[Deduction Theorem]\label{t: deduction theorem for IM}
    Let $\Gamma \cup \{\varphi, \psi\} \subseteq \LIM$. Then $\Gamma, \varphi \vdash \psi$ if and only if $\Gamma \vdash \varphi \rightarrow \psi$.
\end{theorem}
\begin{proof}
    For the direction from right to left suppose $\Gamma \vdash \varphi \rightarrow \psi$. Hence $\Gamma, \varphi \vdash \varphi \rightarrow \psi$ by Lemma \ref{l: assumption weakening}. Since $\Gamma, \varphi \vdash \varphi$, applying $\mathsf{MP}$ to $\varphi$ and $\varphi \rightarrow \psi$ yields $\Gamma, \varphi \vdash \psi$. \smallskip
    
    For the direction from left to right we proceed by induction on the height $h(\pi)$ of the derivation $\pi$ witnessing $\Gamma, \varphi \vdash \psi$. \smallskip 
    
    \noindent \textsc{$h(\pi) = 0$}: Then either $\psi$ is an instance of an axiom scheme or $\psi \in \Gamma \cup \{\varphi\}$. In the first case $\vdash \psi$ and thus by Lemma \ref{l: derivable formulas}, Item \ref{l: derivable formulas item 4}. $\vdash \varphi \rightarrow \psi$. By Lemma \ref{l: assumption weakening}, $\Gamma \vdash \varphi \rightarrow \psi$. In the second case first suppose $\psi = \varphi$. Then $ \varphi \rightarrow \psi$ is an intuitionistic tautology and therefore $\Gamma \vdash \varphi \rightarrow \psi$. Otherwise $\psi \in \Gamma$, implying that $\Gamma \vdash \psi$. Since $\psi \rightarrow (\varphi \rightarrow \psi)$ is an intuitionistic tautology, $\Gamma \vdash \psi \rightarrow (\varphi \rightarrow \psi)$. Applying $\mathsf{MP}$ yields $\Gamma \vdash \varphi \rightarrow \psi$. \smallskip 
    
   \noindent $h(\pi) > 0$: Consider the last rule applied in $\pi$ (i.e. the rule instance where the conclusion labels the root). \smallskip
    
     \noindent \textsc{1.} Suppose the last rule applied in $\pi$ is an instance of $\mathsf{MP}$:
        \begin{equation*}
            \infer{\psi}{\gamma & \gamma \rightarrow \psi}
        \end{equation*}
        By induction hypothesis there are derivations $\pi_0, \pi_1$ witnessing $\Gamma \vdash \varphi \rightarrow \gamma$ and $\Gamma \vdash \varphi \rightarrow (\gamma \rightarrow \psi)$. Consider the following derivation:
        \begin{prooftree}
            \AxiomC{$\pi_0$}
            \noLine
            \UnaryInfC{$\varphi \rightarrow \gamma$}
            \AxiomC{$\pi_1$}
            \noLine
            \UnaryInfC{$\varphi \rightarrow (\gamma \rightarrow \psi)$}
            \AxiomC{}
            \RightLabel{$\mathsf{Int}$}
            \UnaryInfC{$(\varphi \rightarrow (\gamma \rightarrow \psi)) \rightarrow ((\varphi \rightarrow \gamma) \rightarrow (\varphi \rightarrow \psi))$}
            \RightLabel{$\mathsf{MP}$}
            \BinaryInfC{$(\varphi \rightarrow \gamma) \rightarrow (\varphi \rightarrow \psi)$}
            \RightLabel{$\mathsf{MP}$}
            \BinaryInfC{$\varphi \rightarrow \psi$}
        \end{prooftree}
    Therefore, $\Gamma \vdash \varphi \rightarrow \psi$. \smallskip
    
    \noindent \textsc{2.} Suppose the last rule applied in $\pi$ is an instance of $\mathsf{R}$ for $\mathsf{R} \in \{\mathsf{Nec}, \mathsf{Mon}, \mathsf{Ind}\}$, with premise $\gamma$ and conclusion $\psi$. By definition of a derivation from assumptions, every leaf of $\pi$ is labelled by an axiom, implying that $\vdash \psi$. Lemma \ref{l: derivable formulas}, Item \ref{l: derivable formulas item 4} implies that $\vdash \varphi \rightarrow \psi$ and so, by Lemma \ref{l: assumption weakening}, $\Gamma \vdash \varphi \rightarrow \psi$. \qedhere
\end{proof}

We obtain the following useful corollary. Recall that $\bigwedge \emptyset = \top$.
\begin{corollary}\label{c: deduction theorem}
    For any finite set of formulas $\Gamma \cup \{\varphi\}$ the following holds:
    \begin{center}
        $\Gamma \vdash \varphi$ if and only if $\vdash \bigwedge \Gamma \rightarrow \varphi$.
    \end{center}
\end{corollary}
\begin{proof}
    By induction on $\lvert \Gamma \rvert$. For $\lvert \Gamma \rvert = 0$ we have that $\Gamma = \emptyset$ and so that $\vdash \varphi$ holds. Since $\varphi \rightarrow (\top \rightarrow \varphi)$ is an intuitionistic tautology, $\vdash \top \rightarrow \varphi$ is derived by an application of $\mathsf{MP}$. \smallskip \\
    Suppose $\lvert \Gamma \rvert = n+1$. Let $\Gamma =  \Gamma_0 \cup \Gamma_1$ where $\Gamma_0 = \{ \varphi_1, \ldots, \varphi_n \}$ and $\Gamma_1 = \{\varphi_{n+1}\}$. By the Deduction Theorem $\Gamma \vdash \varphi$ if and only if $\Gamma_0 \vdash \varphi_{n+1} \rightarrow \varphi$. By induction hypothesis
    \begin{equation*}
        \Gamma_0 \vdash \varphi_{n+1} \rightarrow \varphi \text{ if and only if } \vdash \bigwedge \Gamma_0 \rightarrow (\varphi_{n+1} \rightarrow \varphi).
    \end{equation*}
    It remains to show that
    \begin{equation*}\label{e: c: deduction theorem}
        \vdash \bigwedge \Gamma_0 \rightarrow (\varphi_{n+1} \rightarrow \varphi) \text{ if and only if } \vdash \bigwedge \Gamma \rightarrow \varphi.
    \end{equation*}
   Note that $(p \rightarrow (q \rightarrow r)) \rightarrow((p \wedge q) \rightarrow r)$ is derivable in $\mathrm{IPL_H}$. Thus
    \begin{equation*}
         (\bigwedge \Gamma_0 \rightarrow (\varphi_{n+1} \rightarrow \varphi)) \rightarrow (\bigwedge \Gamma \rightarrow \varphi)
    \end{equation*}
    is an intuitionistic tautology and hence derivable. Applying $\mathsf{MP}$ yields
    \begin{equation*}
        \vdash  \bigwedge \Gamma_0 \rightarrow (\varphi_{n+1} \rightarrow \varphi) \text{ implies }  \vdash \bigwedge \Gamma \rightarrow \varphi.
    \end{equation*}
    The direction from right-to-left is similar but uses $((p \wedge q) \rightarrow r) \rightarrow (p \rightarrow (q \rightarrow r))$ instead.  Hence,
    \begin{equation*}
        \Gamma \vdash \varphi \text{ if and only if } \vdash \bigwedge \Gamma \rightarrow \varphi.
    \end{equation*}
    which concludes the proof.
\end{proof}

We conclude this section by proving that $\mathrm{IM_H}$ is sound. First, let us check that all axioms and rules of $\mathrm{IM_H}$ are valid or validity preserving, respectively.

\begin{lemma}\label{l: axioms of IM are valid}
    Every instance of an axiom of $\mathrm{IM_H}$ is valid over the classes of dynamic models, functional models and triangle models.
\end{lemma}
\begin{proof}
    The case for intuitionistic tautologies follows from Theorem \ref{t: soundness and completeness of ipl}. Recall that by Theorem \ref{t: three classes one logic} all three classes of models have the same set of validities. Therefore we use triangle models to show the validity of the remaining axioms. \smallskip
    
    \noindent \textsc{Axiom $\mathsf{K}$.} Let $\fw=(W, \leq, R,V)$ be a triangle model, $w \in W$ a world and $\lb(\varphi \rightarrow \psi) \rightarrow (\lb \varphi \rightarrow \lb \psi)$ an instance of $\mathsf{K}$. Suppose $w \leq v$ and $\fw,v \models_t \lb (\varphi \rightarrow \psi)$. Thus for any world $u \in W$ with $v \mathrel{R} u$ holds $\fw,u \models_t \varphi \rightarrow \psi$. Now let $v' \in W$ be any world such that $v \leq v'$ and suppose that $\fw,v' \models_t \lb \varphi$. Then for any world $u' \in W$ with $v' \mathrel{R} u'$ holds $\fw, u' \models_t \varphi$. Since $v \leq v'$ and $v' \mathrel{R} u'$, triangle confluence implies that $v \mathrel{R} u'$. Hence $\fw, u' \models_t \varphi \rightarrow \psi$. As $\fw, u' \models_t \varphi$, also $\fw, u' \models_t \psi$. Thus $\fw,v' \models_t \lb \psi$. Therefore $\fw,v \models_t \lb \varphi \rightarrow \lb \psi$, implying that $\fw,w \models_t \lb (\varphi \rightarrow \psi) \rightarrow (\lb \varphi \rightarrow \lb \psi)$. Consequently, instances of $\mathsf{K}$ are valid. \smallskip

   \noindent \textsc{Axiom $\mathsf{Fix}$.} Let $\fw=(W, \leq, R,V)$ be a triangle model, $w \in W$ a world and $\lbm \varphi \leftrightarrow (\varphi \wedge \lb \lbm \varphi)$ an instance of $\mathsf{Fix}$. Suppose $w \leq v$ and $\fw,v \models_t \lbm \varphi$. So $\fw,u \models_t \varphi$ for all $u \in W$ with $v \mathrel{R^*} u$. In particular, $v \mathrel{R^*} v$ and so $\fw,v \models_t \varphi$. Let $u \in W$ be any world with $v \mathrel{R} u$ and let $u' \in W$ be any world with $u \mathrel{R^*} u'$. Then also $v \mathrel{R^*} u'$ and thus $\fw, u' \models_t \varphi$. Hence, $\fw,u \models_t \lbm \varphi$, implying that $\fw,v \models_t \lb \lbm \varphi$. Together, $\fw,v \models_t \varphi \wedge \lb \lbm \varphi$ and so $\fw,w \models_t \lbm \varphi \rightarrow (\varphi \wedge \lb \lbm \varphi)$. The other direction is similar. Thus instances of $\mathsf{Fix}$ are valid.
\end{proof}

\begin{lemma}\label{l: rules are valid}
    The rules $\mathsf{MP}$, $\mathsf{Nec}$, $\mathsf{Mon}$ and $\mathsf{Ind}$ preserve validity: if the premises are valid, then the conclusion is valid too.
\end{lemma}
\begin{proof}
 We show the contrapositive: if the conclusion is falsifiable, then so is one of the premises. Let $\fw=(W, \leq, R, V)$ be a triangle model and let $w \in W$ be a world. The cases for $\mathsf{MP}$ and $\mathsf{Nec}$ are standard and omitted. \smallskip
 
 
 \noindent \textsc{Rule $\mathsf{Mon}$.} Suppose that $\fw,w \not \models_t \lbm \varphi \rightarrow \lbm \psi$. Then there exists $v \in W$ with $v\geq w$ such that $\fw,v \models_t \lbm \varphi$ and $\fw,v \not \models_t \lbm \psi$. Therefore there exists $u \in W$ such that $v \mathrel{R^*} u$ and $\fw,u \not \models_t \psi$. However, $\fw,u \models_t \varphi$ by assumption and so $\fw,u \not \models_t \varphi \rightarrow \psi$. Therefore $\varphi \rightarrow \psi$ is falsifiable. \smallskip
 
 \noindent \textsc{Rule $\mathsf{Ind}$.} Suppose that $\fw,w \not \models_t \varphi \rightarrow \lbm \varphi$. So there exists $v \in W$ where $v \geq w$ such that $\fw, v \models_t \varphi$ and $\fw, v \not \models_t \lbm \varphi$. Let $n$ be the least natural number such that there exist $u_0, \ldots, u_n \in W$ with $u_0 = v$, for all $0 \leq i <n$ holds $u_i \mathrel{R} u_{i+1}$, and $\fw, u_n \not \models_t \varphi$. By assumption $n > 0$. Then $\fw, u_{n-1} \models_t \varphi$ but $\fw, u_{n-1} \not \models_t \lb \varphi$. Hence $\fw, u_{n-1} \not \models_t \varphi \rightarrow \lb \varphi$, implying that $\varphi \rightarrow \lb \varphi$ is falsifiable.
\end{proof}

Soundness of $\mathsf{IM_H}$ is then established by induction on the height of derivations. The proof is routine.

\begin{theorem}[Soundness of $\mathsf{IM_H}$]\label{t: soundness}
    The axiomatization $\mathrm{IM_H}$ is sound: If $ \vdash \varphi$, then $ \varphi$ is valid (over the classes of dynamic models, triangle models and functional models).
\end{theorem}

\section{Completeness of the Axiomatization}\label{c: IM, section completeness}

This section establishes completeness of $\mathrm{IM_H}$. The proof is by a canonical model construction, where we construct for each formula $\varphi$ a \emph{finite} canonical model. The presented proof closely follows the completeness proof for a multi-modal version of $\IM$ given in~\cite{marti_2017}. \smallskip

Fix an arbitrary finite and closed set of formulas $\Sigma$.

\begin{definition}\label{d: sigma prime theory}\index{prime theory!for $\IM$}
    A \emph{$\Sigma$-prime theory} is a set of formulas $\Gamma \subseteq \Sigma$ such that the following hold.
    \begin{enumerate}
        \item $\Gamma$ is \emph{consistent}: $\Gamma \not \vdash \bot$.
        \item $\Gamma$ is \emph{deductively closed} with respect to $\Sigma$: if $\Gamma \vdash \varphi$ and $\varphi \in \Sigma$, then $\varphi \in \Gamma$.
        \item $\Gamma$ satisfies the \emph{disjunction property}: if $\varphi \vee \psi \in \Gamma$, then $\varphi \in \Gamma$ or $\psi \in \Gamma$.
    \end{enumerate}
\end{definition}

$\Sigma$-prime theories will form the worlds of the canonical model. The following version of the Lindenbaum Lemma states that any consistent set of $\Sigma$-formulas can be extended into a $\Sigma$-prime theory.

\begin{lemma}[Lindenbaum]\label{l: Lindenbaum for IM}
    If $\Gamma \subseteq \Sigma$ and $\Gamma \not \vdash \varphi$, then there exists a $\Sigma$-prime theory $\Delta$ with $\Gamma \subseteq \Delta$ and $\Delta \not \vdash \varphi$.
\end{lemma}
\begin{proof}
Suppose $\Gamma \not \vdash \varphi$ and consider an enumeration $\psi_0, \ldots, \psi_k$ of the formulas in $\Sigma$. We first show how to construct a set of formulas $\Delta$ by induction on $n \leq k$.
\begin{itemize}
    \item Define $\Delta_0 \coloneqq \Gamma$.
    \item Define
    \begin{equation*}
        \Delta_{n+1} \coloneqq \begin{cases}
                               \Delta_n \cup \{\psi_n\}, \text{ if } \Delta_n, \psi_n \not \vdash \varphi \\\
                               \Delta_n, \text{ otherwise. }\\
                        \end{cases}
    \end{equation*}
\end{itemize}
Observe that $\Delta_n \subseteq \Delta_{n+1}$ for each $n \leq k$. Define $\Delta := \Delta_{k+1}$. \smallskip

A simple inductive proof shows that $\Delta_n \not \vdash \varphi$ for each $n \leq k+1$. In fact, the base case holds by assumption and the induction step follows immediately by construction and the induction hypothesis. Therefore $\Delta \not \vdash \varphi$, implying that $\Delta$ is consistent. Next, for showing that $\Delta$ is deductively closed, suppose that $\Delta \vdash \chi$ for some $\chi \in \Sigma$. Then there exists $n \leq k$ such that $\chi = \psi_n$. Consider $\Delta_{n+1}$. If $\psi_n \in \Delta_{n+1}$, then $\psi_n \in \Delta$ and we are done. Otherwise, $\Delta_n, \psi_n \vdash \varphi$, implying that $\Delta, \psi_n \vdash \varphi$. By the Deduction Theorem $\Delta \vdash \psi_n \rightarrow \varphi$. But since $\Delta \vdash \psi_n$ by assumption, $\Delta \vdash \varphi$ which is a contradiction. Hence $\psi_n \in \Delta_{n+1}$ and so $\psi_n \in \Delta$, implying that $\Delta$ is deductively closed with respect to $\Sigma$. For the disjunction property, suppose $\chi \vee \gamma \in \Delta$. Since $\Delta \subseteq \Sigma$, $\chi \vee \gamma \in \Sigma$ and since $\Sigma$ is closed also $\chi, \gamma \in \Sigma$. Therefore there are $n_1, n_2 \leq k$ with $\chi = \psi_{n_1}$ and $\gamma = \psi_{n_2}$. Suppose towards contradiction that $\psi_{n_1} \not \in \Delta$ and $\psi_{n_2} \not \in \Delta$. Thus $\Delta_{n_1}, \psi_{n_1} \vdash \varphi$ and $\Delta_{n_2}, \psi_{n_2} \vdash \varphi$. Therefore also $\Delta, \psi_{n_1} \vdash \varphi$ and $\Delta, \psi_{n_2} \vdash \varphi$. By the Deduction Theorem $\Delta \vdash \psi_{n_1} \rightarrow \varphi$ and $\Delta \vdash \psi_{n_2} \rightarrow \varphi$. Since $\psi_{n_1} \vee \psi_{n_2} \in \Delta$, also $\Delta \vdash \psi_{n_1} \vee \psi_{n_2}$. Observe that
\begin{equation*}
    (\psi_{n_1} \rightarrow \varphi) \rightarrow ((\psi_{n_2} \rightarrow \varphi) \rightarrow ((\psi_{n_1} \vee \psi_{n_2}) \rightarrow \varphi))
\end{equation*}
is an intuitionistic tautology. Hence, by applying $\mathsf{MP}$ three times we obtain $\Delta \vdash \varphi$, a contradiction. Therefore $\psi_{n_1} \in \Delta$ or $\psi_{n_2} \in \Delta$, implying that $\Delta$ satisfies the disjunction property. $\Delta$ is therefore a $\Sigma$-prime theory extending $\Gamma$, such that $\Delta \not \vdash \varphi$.
\end{proof}

\begin{lemma}\label{l: properties of prime theories IM}
    Let $\Gamma$ be a $\Sigma$-prime theory.
    \begin{enumerate}
        \item\label{l: properties of prime theories IM, Item 1} If $\varphi \wedge \psi \in \Sigma$, then $\varphi \wedge \psi \in \Gamma$ if and only if $\varphi \in \Gamma$ and $\psi \in \Gamma$.
        \item\label{l: properties of prime theories IM, Item 2} If $\varphi \vee \psi \in \Sigma$, then $\varphi \vee \psi \in \Gamma$ if and only if $\varphi \in \Gamma$ or $\psi \in \Gamma$.
        \item\label{l: properties of prime theories IM, Item 3}  $\lbm \varphi \in \Gamma$ if and only if $\varphi \in \Gamma$ and $\lb \lbm \varphi \in \Gamma$.
    \end{enumerate}
\end{lemma}
\begin{proof}
 Let $\Gamma$ be a $\Sigma$-prime theory and let $\varphi \ast \psi \in \Sigma$ for $\ast \in \{\wedge, \vee\}$. First of all, observe that since $\Sigma$ is closed, $\varphi, \psi \in \Sigma$. \smallskip \\
  1. For the direction from left-to-right suppose $\varphi \wedge \psi \in \Gamma$. Then $\Gamma \vdash \varphi \wedge \psi$. Since $(\varphi \wedge \psi) \rightarrow \varphi$ and $(\varphi \wedge \psi) \rightarrow \psi$ are intuitionistic tautologies, applying $\mathsf{MP}$ yields $\Gamma \vdash \varphi$ and $\Gamma \vdash \psi$. As $\Gamma$ is deductively closed with respect to $\Sigma$, $\varphi \in \Gamma$ and $\psi \in \Gamma$. For the other direction suppose $\varphi \in \Gamma$ and $\psi \in \Gamma$. Then $\Gamma \vdash \varphi$ and $\Gamma \vdash \psi$. Observe that $\varphi \rightarrow (\psi \rightarrow (\varphi \wedge \psi))$ is an intuitionistic tautology. Hence $\Gamma \vdash \varphi \wedge \psi$ by two applications of $\mathsf{MP}$. Thus, as $\Gamma$ is deductively closed with respect to $\Sigma$, $\varphi \wedge \psi \in \Gamma$. \smallskip \\
2. The direction from left-to-right follows from $\Gamma$ being a $\Sigma$-prime theory and therefore satisfying the disjunction property. For the other direction suppose without loss of generality that $\varphi \in \Gamma$. Observe that $\varphi \rightarrow (\varphi \vee \psi)$ is an intuitionistic tautology. Therefore $\Gamma \vdash \varphi \rightarrow (\varphi \vee \psi)$. Applying $\mathsf{MP}$ yields $\Gamma \vdash \varphi \vee \psi$ and thus, since $\Gamma$ is deductively closed with respect to $\Sigma$, $\varphi \vee \psi \in \Gamma$. \smallskip \\
3. Note that $\Sigma$ being closed implies that $\lbm \varphi \in \Sigma$ if and only if $\varphi,\lb \lbm \varphi \in \Sigma$. Both directions then follow immediately from the axiom $\mathsf{Fix}$. \qedhere
        
\end{proof}

For a set of formulas $\Gamma$, define the following notation.
\begin{equation*}
    \begin{split}
        \lb \Gamma & := \{ \lb \varphi \mid \varphi \in \Gamma \} \\
        \lb^{-1} \Gamma & := \{\varphi \mid \lb \varphi \in \Gamma\}\\
    \end{split}
\end{equation*}

We are ready to define the canonical model relative to a given finite and closed set of formulas $\Sigma$. From now on until the end of the section, the Greek letters $\Gamma, \Delta$ and $\Omega$ refer exclusively to $\Sigma$-prime theories for some fixed $\Sigma$.

\begin{definition}
    Let $\Sigma$ be a finite and closed set of formulas. The \emph{canonical model} relative to $\Sigma$ is given by $\fw_\Sigma = (W_\Sigma, \leq_\Sigma, R_\Sigma, V_\Sigma)$ where
    \begin{itemize}
        \item $W_\Sigma \coloneqq \{ \Gamma\mid \Gamma \text{ is a $\Sigma$-prime theory} \}$;
        \item $\Gamma \leq_\Sigma \Delta$ if and only if $\Gamma \subseteq \Delta$;
        \item $\Gamma \mathrel{R_\Sigma} \Delta$ if and only if $\lb^{-1} \Gamma \subseteq \Delta$;
        \item  $V_\Sigma(\Gamma) := \Gamma \cap \Prop$.
    \end{itemize}
\end{definition}

\begin{lemma}\label{l: canonical model is triangle model}\index{canonical model!for $\mathrm{IM_H}$}
    For a finite and closed set of formulas $\Sigma$, the canonical model $\fw_\Sigma$ is a finite triangle model.
\end{lemma}
\begin{proof}
    Let us first show that $\fw_\Sigma$ is a triangle model. By soundness, the empty set is consistent. The Lindenbaum Lemma therefore implies that there exists a $\Sigma$-prime theory and thus that $W_\Sigma \not = \emptyset$. That $(W_\Sigma, \leq_\Sigma)$ is a partial order follows immediately from the subset relation $\subseteq$ being reflexive, transitive and antisymmetric. For triangle confluence, suppose $\Gamma \leq_\Sigma \Delta$ and $\Delta \mathrel{R_\Sigma} \Omega$. By definition $\Gamma \subseteq \Delta$ and $\lb^{-1} \Delta \subseteq \Omega$. Therefore $\lb^{-1} \Gamma \subseteq \Omega$ and so $\Gamma \mathrel{R_\Sigma} \Omega$. It remains to show that $V_\Sigma$ is monotone in $\leq_\Sigma$, which follows immediately from definition. 
    
    Regarding the size of $\fw_\Sigma$, observe that there are at most $2^{\lvert \Sigma \rvert}$ many $\Sigma$-prime theories. Therefore, since $\Sigma$ is finite, so is $\fw_\Sigma$.
\end{proof}

The last step to obtain completeness is to establish the Truth Lemma, which states that for any formula $\varphi \in \Sigma$ and any $\Sigma$--prime theory $\Gamma$, $\Gamma \vdash \varphi$ if and only if $\fw_\Sigma, \Gamma \models_t \varphi$. For $\Gamma \in W_\Sigma$, let the \emph{reachable component} of $\Gamma$ be the set $\R_\Sigma^*(\Gamma) \coloneqq \{\Delta \in W_\Sigma \mid \Gamma \Rel_\Sigma^* \Delta\}$. Note that $\R_\Sigma^*(\Gamma)$ is finite.

\begin{definition}
    Let $\Gamma \in W_\Sigma$. The \emph{characteristic formula}\index{characteristic formula! for $\IM$} of $\Gamma$ is the formula
    \begin{equation*}
        \chi(\Gamma) \coloneqq \bigwedge \Gamma.
    \end{equation*}
    Furthermore, the \emph{reachable component formula}\index{reachable component formula} is the formula
    \begin{equation*}
        \gamma(\Gamma) \coloneqq \bigvee_{\Delta \in R_\Sigma^*(\Gamma)} \chi(\Delta).
    \end{equation*}
\end{definition}

The characteristic formula $\chi(\Gamma)$ characterizes the prime theory $\Gamma$. The reachable component formula $\gamma(\Gamma)$ characterizes the reachable component of $\Gamma$.

\begin{lemma}\label{l: properties characteristic formula IM}
    Let $\Gamma \in W_\Sigma$.  The following hold.
    \begin{enumerate}
        \item $\Gamma \vdash \chi(\Gamma)$.
        \item For any $\Delta \in R_\Sigma^*(\Gamma)$, $\Delta \vdash \gamma(\Gamma)$.
        \item If $\psi \in \LIM$ such that $\psi \in \Delta$ for every $\Delta \in R_\Sigma^*(\Gamma)$, then $\vdash \gamma(\Gamma) \rightarrow \psi$.
    \end{enumerate}
\end{lemma}
\begin{proof}
    1. is trivial. For 2. note that $\chi(\Delta)$ is one of the disjuncts of $\gamma(\Gamma)$, implying that $\chi(\Delta) \rightarrow \gamma(\Gamma)$ is an intuitionistic tautology. Hence, 1. implies that $\Delta \vdash \gamma(\Gamma)$. For 3. by assumption $\psi$ is a conjunct of each $\chi(\Delta)$ occurring as a disjunct of $\gamma(\Gamma)$, which immediately implies that $\gamma(\Gamma) \rightarrow \psi$ is an intuitionistic tautology and hence derivable.
\end{proof}

\begin{lemma}[Truth Lemma] \label{l: truth lemma}
Let $\Sigma$ be a finite and closed set of formulas. Let $\fw_\Sigma$ be the canonical model relative to $\Sigma$. Then for any $\Sigma$-prime theory $\Gamma$ and any formula $\varphi \in \Sigma$ the following holds.
\begin{center}
    $\varphi \in \Gamma$ if and only if $\fw_\Sigma, \Gamma \models_t \varphi$.
\end{center}
\end{lemma}
\begin{proof}
    We proceed by induction on the structure of $\varphi$. The case where $\varphi = \bot$ is trivial, and the case where $\varphi = p$ for $p \in \Prop$ follows immediately from the definition. For the induction step, the cases where $\varphi = \psi \wedge \chi$ or $\varphi = \psi \vee \chi$ follow immediately from Lemma \ref{l: properties of prime theories IM}, the induction hypothesis and $\Sigma$ being closed. \smallskip
    
    \noindent \textsc{($\rightarrow$)} Suppose $\varphi = \psi \rightarrow \chi$. For the left--to--right direction suppose $\psi \rightarrow \chi \in \Gamma$. Let $\Delta$ be any $\Sigma$-prime theory extending $\Gamma$, i.e. $\Gamma \subseteq \Delta$ and suppose $\fw_\Sigma, \Delta \models_t \psi$. By induction hypothesis $\psi \in \Delta$ implying that $\Delta \vdash \psi$. Since $\Delta$ extends $\Gamma$, we have $\psi \rightarrow \chi \in \Delta$ and thus $\Delta \vdash \psi \rightarrow \chi$. Applying $\mathsf{MP}$ yields $\Delta \vdash \chi$. As $\Delta$ is deductively closed with respect to $\Sigma$, $\chi \in \Delta$. The induction hypothesis yields $\fw_\Sigma, \Delta \models_t \chi$. $\Delta$ was an arbitrary $\Sigma$-prime theory extending $\Gamma$, therefore $\fw_\Sigma, \Gamma \models_t \psi \rightarrow \chi$. 
    For the right--to--left direction we proceed by contraposition. Suppose $\psi \rightarrow \chi \not \in \Gamma$. Then $\Gamma \not \vdash \psi \rightarrow \chi$. By the Deduction Theorem $\Gamma, \psi \not \vdash \chi$. By the Lindenbaum Lemma there exists a $\Sigma$-prime theory $\Delta$ with $\Gamma \cup \{\psi\} \subseteq \Delta$ and $\Delta \not \vdash \chi$. Hence, $\psi \in \Delta$ and $\chi \not \in \Delta$. The induction hypothesis yields $\fw_\Sigma, \Delta \models_t \psi$ and $\fw_\Sigma, \Delta \not \models_t \chi$. As $\Gamma \leq_\Sigma \Delta$, we have $\fw_\Sigma, \Gamma \not \models_t \psi \rightarrow \chi$. \smallskip
    
    \noindent \textsc{($\lb$)} Suppose $\varphi = \lb \psi$. For the left--to--right direction suppose $\lb \psi \in \Gamma$. Let $\Delta$ be any $\Sigma$-prime theory such that $\Gamma \mathrel{R_\Sigma} \Delta$ holds. By definition $\lb^{-1} \Gamma \subseteq \Delta$. Hence, $\psi \in \Delta$. By induction hypothesis $\fw_\Sigma, \Delta \models_t \psi$. As $\Delta$ was arbitrary, we conclude that $\fw_\Sigma, \Gamma \models_t \lb \psi$. For the right--to--left direction suppose $\fw_\Sigma, \Gamma \models_t \lb \psi$. Note that
    \begin{equation}\label{e: truth lemma1}
        \lb^{-1} \Gamma \vdash \psi
    \end{equation}
     as otherwise, by the Lindenbaum Lemma, there would exist a $\Sigma$-prime theory $\Delta$ with $\lb^{-1} \Gamma \subseteq \Delta$ and $\Delta \not \vdash \psi$. Hence $\psi \not \in \Delta$, and so by induction hypothesis $\fw_\Sigma, \Delta \not \models_t \psi$, contradicting that $\fw_\Sigma, \Gamma \models_t \lb \psi$. Therefore (\ref{e: truth lemma1}) holds. Since $\lb^{-1}\Gamma$ is a finite set, from (\ref{e: truth lemma1}) and Corollary \ref{c: deduction theorem} we obtain
    \begin{equation}\label{e: truthlemma2}
        \vdash \bigwedge \lb^{-1} \Gamma \rightarrow \psi.
    \end{equation}
    By $\mathsf{Nec}$ we obtain $\vdash \lb (\bigwedge \lb^{-1} \Gamma \rightarrow \psi) $. By the $\mathsf{K}$-axiom and $\mathsf{MP}$ we obtain $\vdash \lb (\bigwedge \lb^{-1} \Gamma) \rightarrow \lb \psi$. We claim that
    \begin{equation}\label{e: truthlemma3}
         \lb \lb^{-1} \Gamma \vdash \lb (\bigwedge \lb^{-1} \Gamma).
    \end{equation}
    If true, then applying $\mathsf{MP}$ to $\vdash \lb (\bigwedge \lb^{-1} \Gamma) \rightarrow \lb \psi$ and (\ref{e: truthlemma3}) gives $\lb \lb^{-1} \Gamma \vdash \lb \psi$. Lemma \ref{l: assumption weakening} then yields $\Gamma \vdash \lb \psi$ implying that $\lb \psi \in \Gamma$. We prove (\ref{e: truthlemma3}) by induction on the size of $\lb^{-1} \Gamma$. For $\lvert \lb^{-1} \Gamma \rvert = 0$, the statement reduces to $\emptyset \vdash \lb \top$ which is trivially true. For $\lvert \lb^{-1} \Gamma \rvert = k+1$ let $\lb^{-1}\Gamma_0 = \{ \varphi_1, \ldots, \varphi_k\}$ and let $\lb^{-1}\Gamma = \lb^{-1} \Gamma_0 \cup \{ \varphi_{k+1}\}$. By induction hypothesis
    \begin{equation*}
        \lb \lb^{-1} \Gamma_0 \vdash \lb (\bigwedge \lb^{-1} \Gamma_0).
    \end{equation*}
    Hence also $\lb \lb^{-1} \Gamma \vdash \lb (\bigwedge \lb^{-1} \Gamma_0)$. Since $\lb \lb^{-1} \Gamma \vdash \lb \varphi_{k+1}$ we also obtain
    \begin{equation*}
        \lb \lb^{-1} \Gamma \vdash \lb (\bigwedge \lb^{-1} \Gamma_0) \wedge \lb \varphi_{k+1}.
    \end{equation*}
     Lemma \ref{l: derivable formulas}, Item \ref{l: derivable formulas item 2}. yields
     \begin{equation*}
         \lb \lb^{-1} \Gamma \vdash \lb (\bigwedge \lb^{-1} \Gamma)
     \end{equation*}
     which finishes the proof of the claim. \smallskip
     
     \noindent \textsc{($\lbm$)} Suppose $\varphi = \lbm \psi$. For the left--to--right direction suppose $\lbm \psi \in \Gamma$. Let $\Delta \in R_\Sigma^*(\Gamma)$. Then there are prime theories $\Delta_0, \ldots, \Delta_n$ with $\Delta_0 = \Gamma$, $ \Delta_n = \Delta$ and for each $0 \leq i < n$ holds $\Delta_i \mathrel{R_\Sigma} \Delta_{i+1}$. We prove that $\psi \in \Delta_n$ and $\lbm \psi \in \Delta_n$ by induction on $n$. For $n=0$, $\lbm \psi \in \Delta_0 = \Gamma$ by assumption. Moreover, Lemma \ref{l: properties of prime theories IM}, Item \ref{l: properties of prime theories IM, Item 3}. implies that $\varphi \in \Gamma$. For $n = k+1$ we have that $\psi \in \Delta_k$ and $\lbm \psi \in \Delta_k$ by induction hypothesis. By Lemma \ref{l: properties of prime theories IM}, Item \ref{l: properties of prime theories IM, Item 3}. $\lb \lbm \psi \in \Delta_k$ and therefore $\lbm \psi \in \Delta_{k+1}$, implying that also $\psi \in \Delta_{k+1}$. Therefore $\psi \in \Delta$ for any $\Delta \in R_\Sigma^*(\Gamma)$. The induction hypothesis yields that $\fw_\Sigma, \Delta \models_t \psi$ for any such $\Delta$ and thus that $\fw_\Sigma, \Gamma \models_t \lbm \psi$. 
     
     For the right--to--left direction suppose that $\fw_\Sigma, \Gamma \models_t \lbm \psi$. For any $\Delta \in R_\Sigma^*(\Gamma)$ thus holds that $\fw_\Sigma, \Delta \models_t \psi$. By induction hypothesis $\psi \in \Delta$. In order to show that $\Gamma \vdash \lbm \psi$, we claim that $\vdash \gamma(\Gamma) \rightarrow \lbm \gamma(\Gamma)$. By the presence of the rule $\mathsf{Ind}$, it suffices to prove that $\vdash \gamma(\Gamma) \rightarrow \lb \gamma(\Gamma)$. To that end note that for any $\Delta \in R_\Sigma^*(\Gamma)$,
        \begin{equation*}
            \lb^{-1} \Delta \vdash \gamma(\Gamma).
        \end{equation*}
        In fact if $ \lb^{-1} \Delta \not \vdash \gamma(\Gamma)$ for some $\Delta \in R_\Sigma^*(\Gamma)$, then by the Lindenbaum Lemma there exists a prime theory $\Omega$ such that $\lb^{-1}\Delta \subseteq \Omega$ and $\Omega \not \vdash \gamma(\Gamma)$. By construction $\Delta \mathrel{R_\Sigma} \Omega$, implying that $\Omega \in R_\Sigma^*(\Gamma)$. By Lemma \ref{l: properties characteristic formula IM}, $\Omega \vdash \gamma(\Gamma)$ which is a contradiction. Thus $\lb^{-1} \Delta \vdash \gamma(\Gamma)$. By Corollary \ref{c: deduction theorem} and $\mathsf{Nec}$ we obtain
        \begin{equation*}
            \vdash \lb (\bigwedge \lb^{-1} \Delta \rightarrow \gamma(\Gamma)).
        \end{equation*}
    Using the $\mathsf{K}$-axiom and $\mathsf{MP}$ yields
    \begin{equation*}
        \vdash \lb \bigwedge \lb^{-1} \Delta \rightarrow \lb \gamma(\Gamma).
    \end{equation*}
    By (\ref{e: truthlemma3}),
    \begin{equation*}
        \lb \lb^{-1} \Delta \vdash \lb \gamma(\Gamma).
    \end{equation*}
    Hence, by Lemma \ref{l: assumption weakening} we obtain $\Delta \vdash \lb \gamma(\Gamma)$. By Corollary \ref{c: deduction theorem} $\vdash \chi(\Delta) \rightarrow \lb \gamma(\Gamma)$. As $\Delta$ was arbitrary, we obtain $\vdash \gamma(\Gamma) \rightarrow \lb \gamma(\Gamma)$ by repeatedly using the (correct substitution instance of the) intuitionistic tautology $(p \rightarrow r) \rightarrow ((q \rightarrow r) \rightarrow ((p \vee q) \rightarrow r)$. Thus applying \textsf{Ind} to $\gamma(\Gamma) \rightarrow \lb \gamma(\Gamma)$ yields $\vdash \gamma(\Gamma) \rightarrow \lbm \gamma(\Gamma)$. \smallskip
    
    \noindent By Lemma \ref{l: properties characteristic formula IM}, $\Gamma \vdash \gamma(\Gamma)$ and so we obtain
    \begin{equation*}
        \Gamma \vdash \lbm \gamma(\Gamma).
    \end{equation*}
   Since $\psi \in \Delta$ for every $\Delta \in R_\Sigma^*(\Gamma)$, Lemma \ref{l: properties characteristic formula IM} implies $\vdash \gamma(\Gamma) \rightarrow \psi$. Applying $\mathsf{Mon}$ gives
   \begin{equation*}
       \vdash \lbm \gamma(\Gamma) \rightarrow \lbm \psi.
   \end{equation*}
   Thus applying $\mathsf{MP}$ yields
   \begin{equation*}
       \Gamma \vdash \lbm \psi
   \end{equation*}
   and hence $\lbm \psi \in \Gamma$.
   \end{proof}

It remains to show completeness of $\mathrm{IM_H}$ with respect to the class of triangle models (and so also with respect to the classes of functional and of dynamic models).

\begin{theorem}[Completeness of $\mathsf{IM_H}$]\label{t: completeness of Hilbert style system}
    The axiomatization $\mathrm{IM_H}$ is complete: If $\varphi$ is valid (over the classes of triangle models, dynamic models and functional models), then $\vdash \varphi$.
\end{theorem}
\begin{proof}
    Let $\Sigma = \Cl(\varphi)$ and suppose $\not \vdash \varphi$. By the Lindenbaum Lemma there exists a $\Sigma$-prime theory $\Gamma$ such that $\Gamma \not \vdash \varphi$. By the Truth Lemma, $\varphi$ is falsified at world $\Gamma$ of the canonical model relative to $\Sigma$. As this model is a triangle model, $\varphi$ is not valid over the class of triangle models. Theorem \ref{t: three classes one logic} then implies that $\varphi$ is not valid over the classes of dynamic models and functional models as well.
\end{proof}

 The logic $\mathbf{IM_t}$ has the \emph{finite (triangle) model property} if every falsifiable formula over the class of triangle models is falsifiable over the class of \emph{finite} triangle models. Since the canonical model is finite, we obtain the finite (triangle) model property. As usual, the finite model property combined with a finite sound and complete axiomatization yields decidability.

\begin{theorem}[Finite model property and decidability]\index{finite model property!intuitionistic master modality}
    The logic $\mathbf{IM_t}$ has the finite model property and therefore the validity problems for $\mathbf{IM}, \mathbf{IM_t}$ and $\mathbf{IM_f}$ are decidable.
\end{theorem}

\section{Non-Wellfounded and Cyclic Proofs}\label{c: IM, section non-wellfounded and cyclic calculus for IM}

The axiomatization $\mathrm{IM_H}$ captures the validities of $\IM$ in a neat way. However, from a proof theoretic perspective, axiomatizations are unsatisfactory, since the presence of the modus ponens rule obstructs the proof theoretic analysis. As witnessed in the previous sections, to check whether an $\LIM$-formula $\varphi$ is derivable in $\mathrm{IM_H}$, we have to guess the right instances of intuitionistic tautologies in order to apply modus ponens. This drawback can be circumvented by passing to \emph{analytic} sequent calculi. Analytic means that every provable sequent $\sigma$ has a proof in which only formulas that are relevant to the formulas in $\sigma$ may occur. What `relevant' means depends on the specific logic at hand. For $\mathsf{IM}$ we count as relevant those formulas that belong to the closure of $\sigma$. The obstacle for constructing an analytic sequent calculus for $\mathsf{IM}$ lies in the presence of the master modality, which, as a fixed point operator, requires an \emph{induction rule} to treat its infinite behaviour (recall that $\mathrm{IM_H}$ featured the rule $\mathsf{Ind}$). A sequent calculus with an induction rule for a multi-modal version of $\IM$ was proposed by Jäger and Marti~\cite{jager-intuitionistic_2016}. Their induction rule translates to our setting as follows.
\begin{equation*}
    \infer[\mathsf{Ind_s}]{\Pi, \lbm \Gamma, \psi \Rightarrow \lbm \varphi, \Delta}{\lbm \Gamma, \psi \Rightarrow \varphi, \Delta & \lbm \Gamma, \psi \Rightarrow \lb \psi, \Delta}
\end{equation*}

This rule states that if we can prove $\varphi$ from $\psi$ and we can prove $\lb \psi$ from $\psi$, then we can also prove $\lbm \varphi$ from $\psi$. The drawback of induction rules is that their presence usually leads to a non-analytic system. For example, the calculus presented by Jäger and Marti is proven sound and complete, but the proof makes essential use of the cut rule and no cut elimination result is given. It is unclear whether completeness can be obtained without cut. For example, for a classical version of $\IM$ (namely common knowledge logic) evaluated over S5 models, a sequent calculus with induction rule is presented by Alberucci and Jäger~\cite{alberucci_2005}. Once again the cut rule is required, but the authors manage to obtain a partial cut elimination result. Nevertheless, a restriction to \emph{analytic cuts} (i.e. instances of the cut rule where the cut formula belongs to the set of formulas which is `relevant') was not achieved.
 
Let us pause here and briefly discuss why induction rules cause problems for obtaining analytic calculi. Proofs by induction are difficult to formalize due to the problem of finding the right induction hypothesis. A standard example is the following. Suppose we want to show that for all natural numbers $n$,
\begin{equation}\label{e: induction problem}
    s_n= \sum_{k \leq n} \frac{1}{k^2} < 2.
\end{equation}
 This statement does not admit a direct inductive proof: the induction hypothesis gives us that $s_n < 2$. But in order to prove that $s_{n+1} < 2$, we would require to know that $\sfrac{1}{(n+1)^2} < 2 - s_n$, which is not provided by the induction hypothesis. We may circumvent this problem by guessing a stronger statement which implies (\ref{e: induction problem}). Namely, the following statement suffices: 
 
 \begin{equation*}
    s_n= \sum_{k \leq n} \frac{1}{k^2} \leq 2 - \frac{1}{n}.
\end{equation*}
 
This is a problem frequently encountered in inductive proofs and translates into the realm of sequent calculi: the problem of guessing the right induction hypothesis results in an induction rule which is either non-analytic or requires the presence of the cut rule to obtain completeness.

One possible solution is to abandon induction rules for sequent calculi all together and replace them by a formal counterpart for \emph{proofs by infinite descent}. It is usually accepted that such proofs are equally strong as inductive proofs, see for example~\cite{brotherston}. Proofs by infinite descent are not formalized by adding a corresponding rule, but instead by the shape of the proof tree: we consider \emph{non-wellfounded proofs}\index{proof!non-wellfounded}, which are proofs in a sequent calculus as defined in Chapter \ref{c: preliminaries} but the proof tree is allowed to have infinitely long branches. The entire infinite branch is then the formal representation of an argument by infinite descent. To obtain infinite branches, we will use simple unfolding rules for $\lbm$ of the form
\begin{equation*}
    \infer{\lbm \varphi}{\varphi \wedge \lb \lbm \varphi}
\end{equation*}
which replace (when read bottom-up) a formula of the form $\lbm \varphi$ by its unfolding $\varphi \wedge \lb \lbm \varphi$. The presence of such a rule naturally leads to non-wellfounded proofs, as we can generate infinite branches by starting with a formula $\lbm \varphi$, unfold it into $\varphi \wedge \lb \lbm \varphi$ and then use the other rules of the calculus to decompose $\varphi \wedge \lb \lbm \varphi$ back into $\lbm \varphi$, creating a repetition. Note that between such a repetition the branch must pass through a modality rule to get from $\lb \lbm \varphi$ to $\lbm \varphi$. Intuitively, we may think of this step as taking a modal step in a dynamic model: if at $w$ the formula $\lb \lbm \varphi$ is falsified, then there exists a modal successor $v$ at which $\lbm \varphi$ is falsified. In order to seperate valid from invalid infinitary reasoning, infinite branches in a non-wellfounded proof must satisfy a \emph{global correctness criterion}. For $\IM$, this criterion roughly states that every infinite branch must progress infinitely often, where progress means passing through an unfolding of a $\lbm$-formula. Infinite branches therefore contain infinitely many instances of the rule for $\lbm$. Intuitively, this conditions implies that an infinite branch proves $\lbm \varphi$ by proving $\lb^n \varphi$ for any natural number $n$.

Non-wellfounded proofs therefore trade the finiteness of proofs for analyticity of the calculus. One may object that this seems hardly an improvement, but we would like to disagree. First, one goal of proof theory is to study mathematical proofs as mathematical objects. Proofs by infinite descent are infinite arguments that are commonly employed in mathematics; it seems natural to study such proofs in their own right. Second, some non-wellfounded calculi can be proven to be \emph{regularly complete}\index{completeness!regular}: every valid sequent has a regular non-wellfounded proof. Roughly, a non-wellfounded proof is \emph{regular}\index{proof!regular non-wellfounded} if it contains only finitely many subtrees up to isomorphism.  Regular non-wellfounded proofs can be depicted as a finite trees with back-edges: arrows that point from the leaves back into the tree. The intended meaning is that the infinite tree can be obtained by taking the finite tree with back-edges and co-recursively glue the trees rooted at the nodes to which the arrows point at on top of the corresponding leaf. Such finite trees with back-edges are called \emph{cyclic proofs}\index{proof!cyclic} and may be understood as finite representations of infinite proofs. Just as infinite branches in a non-wellfounded proof satisfy a global criterion, finite branches in a cyclic proof that end in a leaf with a back-edge must satisfy a \emph{local correctness criterion}. Cyclic calculi therefore enjoy the advantages of both sides: finite proofs and analytic rules. However, not every non-wellfounded calculus is regularly complete; we will see an example in Chapter \ref{c: iLTL}.

The goal of this section is to introduce a non-wellfounded sequent calculus called $\NWIM$ and a cyclic sequent calculus called $\CIM$ for $\IM$. We begin by introducing the basic definitions and the two calculi. The focus then lies on the cyclic calculus $\CIM$ which is proven to be sound and complete. The non-wellfounded calculus $\NWIM$ is only used as a tool to establish completeness for $\CIM$: we will use a proof search argument to show that $\NWIM$ is complete. The proof search argument provides another example of the adaptation of techniques which are successfully used for classical modal fixed point logics to the intuitionistic realm. However,  the presence of the intuitionistic order complicates the overall argument. Completeness of $\CIM$ is then derived by showing how to prune non-wellfounded proofs into cyclic proofs.

\subsection{Basic Definitions}

In order to formalize the aforementioned correctness criteria, the two calculi $\NWIM$ and $\CIM$ employ a simple \emph{formula annotation}.  Intuitively, this annotation keeps track of the development of a master modality formula throughout a branch. The annotation is used to show that the cyclic calculus is sound. Moreover, it simplifies the global correctness criterion for infinite branches in the non-wellfounded calculus, even though the use of annotations is not necessary to obtain a sound non-wellfounded calculus (see Chapter \ref{c: iLTL} for a non-wellfounded calculus without annotation).

\begin{definition}\label{d: annotated formula}
    An \emph{annotated formula}\index{annotated formula} is a pair $(\varphi, a)$ where $\varphi \in \LIM$ is a formula and $a \in \{\f,\u\}$ is an annotation. The annotation $\f$ designates that the formula is \emph{in focus} and $\u$ that the formula is \emph{unfocused}.
\end{definition}

 Annotated formulas are written as $\varphi^a$. We display annotated formulas without using brackets, e.g. the formula $\varphi \to \psi^\u$ should be read as $(\varphi \rightarrow \psi)^\u$. For the remainder of this chapter, the term \emph{formula} refers to annotated formula. Formulas without annotations are referred to as \emph{plain} formulas. Finite sets of annotated formulas are denoted by $\Gamma, \Delta, \Sigma$ and $\Pi$ with or without subscripts. For a set of annotated formulas $\Gamma$ define
\[
    \Gamma^- = \{\varphi \mid \varphi^a \in \Gamma\}
    \text{ and }
    \Gamma^\u = \{\varphi^\u \mid \varphi^a \in \Gamma\}.
\]
\begin{definition}\label{d: sequent for cim}
A \emph{sequent}\index{sequent!for $\IM$} is an ordered pair $\Gamma \Rightarrow \Delta$ where $\Gamma$ and $\Delta$ are finite sets of annotated formulas, such that the following conditions hold.
\begin{enumerate}
    \item Every formula in $\Gamma$ is unfocused.
    \item At most one formula in $\Delta$ is in focus.
    \item If a formula $\varphi$ is in focus, then $\varphi = \lbm \psi$ or $\varphi = \lb \lbm \psi$ for some formula $\psi$.
\end{enumerate}
\end{definition}
Sequents are denoted by $\sigma$, where $\Gamma_\sigma$  and $\Delta_\sigma$ denotes the left and right side of $\sigma$ respectively. 
The \emph{interpretation} of $\sigma$ is the formula 
$\sigma^I \coloneq \bigwedge \Gamma_\sigma^- \rightarrow \bigvee \Delta_\sigma^-$. Note that annotations convey no semantic meaning. Given a pointed model $(\fw,w)$ we write $\fw,w \models \sigma $ iff $\fw, w \models \sigma^I$, i.e. the notation for the interpretation is usually surpressed. Given a sequent $\sigma$, the \emph{closure} of $\sigma$ is the set $\Cl(\sigma) \coloneq \Cl(\Gamma^-_\sigma) \cup \Cl(\Delta^-_\sigma)$.

Note that sequents $\sigma$ are multi-conclusion, i.e. $\Delta_\sigma$ may contain more than one formula.
This streamlines the proof search argument for completeness as it allows writing the disjunction and left-implication rules in invertible form. 
But it is not an essential restriction; we showed in~\cite{afshari_intuitionistic_2024} how single-conclusion proofs are obtained from multi-conclusion proofs by a co-recursive proof translation.

In the next subsections we will define the non-wellfounded calculus $\NWIM$ and the cyclic calculus $\CIM$. Both calculi consist of the rules depicted in Table \ref{d: rules for IM}.

\begin{table}[h]
    \centering
    \small
        \begin{tabular}{|c c|}
        \hline
        & \\
         $\infer[\mathsf{id}]{\Gamma, \varphi^\u \Rightarrow \varphi^a, \Delta}{}$ 
         & $\infer[\bot]{\Gamma, \bot^\u \Rightarrow \Delta}{}$\\
         & \\
         $\infer[\wedge \mathsf{L}]{\Gamma, \varphi \wedge \psi^\u \Rightarrow \Delta}{\Gamma, \varphi^\u, \psi^\u \Rightarrow \Delta}$ 
         & $\infer[\wedge \mathsf{R}]{\Gamma\Rightarrow \varphi \wedge \psi^\u ,\Delta}{\Gamma \Rightarrow \varphi^\u, \Delta & \Gamma \Rightarrow \psi^\u, \Delta}$\\
         & \\
         $\infer[\vee \mathsf{L}]{\Gamma, \varphi \vee \psi^\u\Rightarrow \Delta}{\Gamma, \varphi^\u \Rightarrow \Delta & \Gamma, \psi^\u \Rightarrow \Delta}$ & $\infer[\vee \mathsf{R}]{\Gamma \Rightarrow \varphi \vee \psi^\u, \Delta}{\Gamma \Rightarrow \varphi^\u, \psi^\u, \Delta}$\\
         & \\
         $\infer[{\to} \mathsf{L}]{\Gamma, \varphi \rightarrow \psi^\u\Rightarrow \Delta}{\Gamma, \varphi \rightarrow \psi^\u \Rightarrow \varphi^\u, \Delta & \Gamma, \psi^\u \Rightarrow \Delta}$ 
         & $\infer[{\to} \mathsf{R}]{\Gamma \Rightarrow \varphi \rightarrow \psi^\u, \Delta}{\Gamma, \varphi^\u \Rightarrow \psi^\u}$\\
         & \\
         $\infer[\mathsf{u}]{\Gamma \Rightarrow \varphi^\f, \Delta}{\Gamma \Rightarrow \varphi^\u, \Delta}$ & $\infer[\mathsf{f}]{\Gamma \Rightarrow \varphi^\u, \Delta}{\Gamma \Rightarrow \varphi^\f, \Delta}$\\
         & \\
          $\infer[\lbm \mathsf{L}]{\Gamma, \lbm \varphi^\u \Rightarrow \Delta}{\Gamma, \varphi^\u, \lb \lbm \varphi^\u \Rightarrow \Delta}$ & $\infer[\lbm \mathsf{R}]{\Gamma \Rightarrow \lbm \varphi^a,\Delta}{\Gamma \Rightarrow \varphi^\u, \Delta & \Gamma \Rightarrow \lb \lbm \varphi^a, \Delta}$\\
         & \\
         $\infer[\lb]{\Pi, \lb \Gamma \Rightarrow \lb \varphi^a, \Sigma}{\Gamma \Rightarrow \varphi^a}$ &  \\
         & \\
         \hline
        \end{tabular}
    \caption{The rules of $\CIM$ and $\NWIM$. The symbols $\Gamma, \Delta, \Sigma$ and $\Pi$ range over finite sets of annotated formulas which may be empty.}
    \label{d: rules for IM}
\end{table}


The rules in Table~\ref{d: rules for IM} for $\wedge, \vee, \rightarrow$ as well as $\mathsf{id}$ and $\bot$ are annotated versions of the rules for $\mathrm{IPL_G}$. The annotations do not play a role for these rules, since only formulas of the form $\lbm \psi$ or $\lb \lbm \psi$ can be in focus. The rules $\mathsf{u}$ and $\mathsf{f}$ are called the \emph{focus rules} and are used to change the focus annotation of a formula. The rule $\lb$ is a standard modal rule, while $\lbm \mathsf{L}$ and $\lbm \mathsf{R}$ replace a master modality formula $\lbm \varphi$ - when read bottom-up - with its unfolding and thus reflect the equivalence $\lbm \varphi \leftrightarrow \varphi \wedge \lb \lbm \varphi$. Note that the annotations play a role in the rules $\lbm \mathsf{R}$ and $\lb$. All rules apart from $\mathsf{u}$ and $\mathsf{f}$ are called \emph{logical rules}.

\begin{definition}\label{d: invertible rule}
    A rule $\mathsf{R}$ is \emph{invertible} if whenever the conclusion is valid, then so are the premises.\footnote{This is a semantic notion of invertibility.}
\end{definition}

Note that the rules ${\rightarrow} \mathsf{R}$ and $\lb$ have single-conclusion premises. Therefore it is routine to check that these two rules are not invertible, while all other rules are.

\begin{lemma}
    All rules in Table \ref{d: rules for IM} apart from ${\rightarrow} \mathsf{R}$ and $\lb$ are invertible.
\end{lemma}

We will therefore refer to $\lb$ and ${\to}\mathsf{R}$ as the \emph{non-invertible} rules and to all other rules as \emph{invertible}.
It will also be useful to refer to formulas occurring in a rule instance according to their specific role. For each rule except $\lb$, the distinguished formula in the conclusion is called \emph{principal} and the distinguished formula(s) in the premises are called its \emph{residual(s)}.  For the rule $\lb$, all formulas in the conclusion are called principal and each formula in the premise is the residual of its corresponding boxed formula in the conclusion (formulas in $\Sigma$ and $\Pi$ have no residuals). 
In every rule instance, any formula that is neither principal nor residual is called a \emph{side formula}.

\begin{example}
    For an instance of ${\to}\mathsf{L}$ of the form
    \begin{equation*}
        \infer[{\to} \mathsf{L}]{\Gamma, \varphi \rightarrow \psi^\u\Rightarrow \Delta}{\Gamma, \varphi \rightarrow \psi^\u \Rightarrow \varphi^\u, \Delta & \Gamma, \psi^\u \Rightarrow \Delta} 
    \end{equation*}
    the principal formula is $\varphi\to \psi^\u$ in the conclusion, its residuals are $\varphi\to \psi^\u$ and $\varphi^\u$ in the left premise and $ \psi^\u$ in the right premise. The formulas in $\Gamma$ and $\Delta$ (both in the conclusion and the premises) are side formulas. For an instance of $\lb$ of the form
    \begin{equation*}
        \infer[\lb]{\Pi, \lb \psi_1^\u, \ldots, \lb \psi_k^\u \Rightarrow \lb \varphi^a, \Sigma}{\psi_1^\u, \ldots , \psi_k^\u \Rightarrow \varphi^a}
    \end{equation*}
    all formulas in the conclusion are principal. The formulas in $\Pi$ and $\Sigma$ have no residuals. The residual of $\lb \psi_i^\u$ for $1 \leq i \leq k$ is the formula $\psi_i^\u$ in the premise and the residual of $\lb \varphi^a$ is the formula $\varphi^a$ in the premise. There are no side formulas.
\end{example}

The condition that sequents have at most one formula in focus imposes restrictions on rule applications, as is illustrated by the following lemma. 
\begin{lemma}\label{l: proper application of rules}
    If in an instance of $\lbm \mathsf{R}$ the principal formula is in focus, then the left premise has no formula in focus.
\end{lemma}
\begin{proof}
    Suppose towards contradiction that there is an application of $\lbm \mathsf{R}$ in which the principal formula is in focus and the left premise has a formula in focus, too.
    \begin{prooftree}
        \AxiomC{$\Gamma \Rightarrow \varphi^\u, \Delta$}
        \AxiomC{$\Gamma \Rightarrow \lb \lbm \varphi^\f, \Delta$}
        \RightLabel{$\lbm \mathsf{R}$}
        \BinaryInfC{$\Gamma \Rightarrow \lbm \varphi^\f, \Delta$}
    \end{prooftree}
    Since $\varphi^\u$ is by definition not in focus, the formula in focus must occur in $\Delta$. Since the right premise cannot contain two formulas in focus, the formula in focus in $\Delta$ must be $\lb \lbm \varphi^\f$. But then the conclusion sequent contains two formulas in focus, a contradiction.
\end{proof}

We will usually construct proofs bottom-up. That is if $\sigma$ is a sequent and $\mathsf{r}$ a rule, then `$\mathsf{r}$ is applied to $\sigma$' means that we consider a rule instance of $\mathsf{r}$ with conclusion $\sigma$.

\subsection{Non-Wellfounded Proofs}

We introduce the non-wellfounded calculus $\NWIM$. The calculus is based on the rules from Table \ref{d: rules for IM}. We then define the notions of (non-wellfounded) \emph{pre-proof} and of (non-wellfounded) \emph{proof}, where a proof is defined as a pre-proof which satisfies certain correctness conditions on its finite and infinite branches. 

\begin{definition}
    The sequent calculus $\NWIM$\index{sequent calculus!$\NWIM$} consists of all rules depicted in Table \ref{d: rules for IM}.
\end{definition}

Non-wellfounded (pre-)proofs in $\NWIM$ are called $\NWIM$-(pre-)proofs. For the following it may be instructive to recall the basic definitions about trees, in particular the definitions of paths and branches (c.f. Chapter \ref{c: preliminaries}, Section \ref{c: preliminaries, section basic definitions}).

\begin{definition}
    An $\NWIM$-\emph{pre-proof} of a sequent $\sigma$ is a finite or countably infinite tree $\pi$ whose nodes are labelled by sequents according to the rules of $\NWIM$ and whose root is labelled by $\sigma$.
\end{definition}

Note that pre-proofs are finite-branching since every rule of $\NWIM$ has only finitely many premises and therefore, by König's Theorem, infinite pre-proofs contain infinite branches.

If there is no danger of confusion, we will tacitly identify a node in a pre-proof with the sequent labelling it and thereby paths and branches of a pre-proof with sequences of sequents. In order for a pre-proof to be a \emph{proof} it must satisfy a \emph{global correctness criterion} on its infinite branches: every infinite branch must contain a \emph{good suffix}\index{good suffix}.

\begin{definition}
    Let $\pi$ be an $\NWIM$-pre-proof and $\rho$ an infinite branch of $\pi$. A suffix $\rho'$ of $\rho$ is called \emph{good} if every sequent in $\rho'$ has a formula in focus and $\rho'$ passes through infinitely many applications of $\lbm \mathsf{R}$ where the principal formula is in focus. 
\end{definition}

\begin{definition}
    An $\NWIM$-\emph{proof}\index{proof!$\NWIM$-proof}  of a sequent $\sigma$ is an $\NWIM$-pre-proof of $\sigma$, such that every leaf is labelled by an axiom and every infinite branch $\rho$ has a \emph{good} suffix $\rho'$.
\end{definition}

An illustration of a non-wellfounded proof is provided in Example \ref{e: non-wellfounded and cyclic proof for IM} below. The following lemma establishes a straightforward yet crucial property of good suffixes.


\begin{lemma}\label{lemma good suffix contains inf many box appl}
    Every good suffix contains infinitely many applications of the rule $\lb$.
\end{lemma}
\begin{proof}
    Let $\rho$ be an infinite branch of an $\NWIM$-proof $\pi$ and let $\rho'$ be a good suffix of $\rho$. Suppose towards contradiction that $\rho'$ contains only finitely many applications of $\lb$. Let $\rho''$ be the suffix of $\rho'$ after the last application of $\lb$. Since $\rho'$ is good, every sequent in $\rho''$ has a formula in focus. This implies that the focus rules cannot be applied in $\rho''$. Furthermore, $\rho''$ contains infinitely many applications of $\lbm \mathsf{R}$ where the principal formula is in focus. Let $n$ be the least natural number such that $\rho''(n)$ is the conclusion of an application of $\lbm \mathsf{R}$ with the principal formula in focus. By Lemma \ref{l: proper application of rules}, $\rho''(n+1)$ must be the right premise. Thus the formula in focus is of the form $\lb \lbm \varphi^\f$ for some formula $\varphi$. By inspection of the rules of $\NWIM$, the only rule that can be applied to $\lb \lbm \varphi^\f$ is $\lb$, $\mathsf{u}$ or $\mathsf{f}$. Since there are no applications of $\lb$, $\mathsf{u}$ or $\mathsf{f}$ in $\rho''$, the formula in focus must be a side formula in any rule application after $\rho''(n+1)$, implying that $\rho''$ passes only through finitely many applications of $\lbm \mathsf{R}$ where the principal formula is in focus; a contradiction.
\end{proof}
%
\subsection{Cyclic Proofs}

Next, we introduce the cyclic calculus $\CIM$ which uses the same rules as $\NWIM$.  As before, we first introduce cyclic pre-proofs and then characterize the subset of pre-proofs that are proofs.

\begin{definition}
    The sequent calculus $\CIM$\index{sequent calculus!$\CIM$} consists of all rules depicted in Table \ref{d: rules for IM}.
\end{definition}

Cyclic (pre-)proofs in $\CIM$ are called $\CIM$-(pre-)proofs.

\begin{definition}
    A $\CIM$-\emph{pre-proof} of a sequent $\sigma$ is a finite $\NWIM$-pre-proof of $\sigma$. 
\end{definition}

In order to distinguish cyclic pre-proofs from cyclic proofs, we require to formulate a \emph{local correctness criterion} for those branches in a cyclic proof that do not end in an axiom. This criterion is essentially a finite version of the global correctness criterion for $\NWIM$-proofs.

\begin{definition}
    A path $\rho$ in a $\CIM$-pre-proof is \emph{successful} if the following hold.
    \begin{enumerate}
        \item Every sequent in $\rho$ has a formula in focus.
        \item The path $\rho$ passes through at least one instance of $\lbm \mathsf{R}$ where the principal formula is in focus.
    \end{enumerate}
\end{definition}

Given a $\CIM$-pre-proof $\pi$, a pair of nodes $(u,v)$ of $\pi$ is called a \emph{repetition} if $u \not= v$, there exists a path from $u$ to $v$ and both nodes are labelled by the same sequent. A repetition $(u,v)$ is \emph{successful}\index{successful repetition} if the path from $u$ to $v$ is successful.

\begin{definition}
    A $\CIM$-\emph{proof}\index{proof!$\CIM$-proof} of a sequent $\sigma$ is a $\CIM$-pre-proof  $\pi$ of $\sigma$ such that every leaf $l$ of $\pi$ is either labelled by an axiom or there exists a node $c(l)$ in $\pi$ such that $(c(l),l)$ is a successful repetition.
\end{definition}

Given a cyclic proof $\pi$, leafs labelled by axioms are called \emph{axiomatic leafs} and all other leafs are called \emph{non-axiomatic leafs}.





\begin{figure}[t]

\resizebox{\columnwidth}{!}{
\begin{tikzpicture}
\node[] (a) [] {
 \AxiomC{}
    \RightLabel{$\mathsf{id}$}
    \UnaryInfC{$\varphi^\u, \lbm \gamma^\u \Rightarrow \varphi^\u$}
    \AxiomC{}
    \RightLabel{$\mathsf{id}$}
    \UnaryInfC{$\varphi^\u, \gamma^\u, \lb \lbm \gamma^\u \Rightarrow \varphi^\u, \lb \lbm \varphi^\f$}
    \AxiomC{$\varphi^\u, \lbm \gamma^\u \Rightarrow \lbm \varphi^\f$}
    \RightLabel{$\lb$}
    \UnaryInfC{$\varphi^\u, \lb \varphi^\u, \lb \lbm \gamma^\u \Rightarrow \lb \lbm \varphi^\f$}
    \RightLabel{${\to} \mathsf{L}$}
    \BinaryInfC{$\varphi^\u,\gamma^\u, \lb \lbm \gamma^\u \Rightarrow \lb \lbm \varphi^\f$}
    \RightLabel{$\lbm \mathsf{L}$}
    \UnaryInfC{$\varphi^\u, \lbm \gamma^\u \Rightarrow \lb \lbm \varphi^\f$}
    \RightLabel{$\lbm \mathsf{R}$}
    \insertBetweenHyps{\hskip -24pt}
    \BinaryInfC{$\varphi^\u, \lbm\gamma^\u \Rightarrow \lbm \varphi^\f$}
    \RightLabel{$\mathsf{f}$}
    \UnaryInfC{$\varphi^\u, \lbm\gamma^\u \Rightarrow \lbm \varphi^\u$}
    \RightLabel{${\land}\mathsf{L}$}
    \UnaryInfC{$\varphi \wedge \lbm\gamma^\u \Rightarrow  \lbm \varphi^\u$}
    \RightLabel{${\to}\mathsf{R}$}
    \UnaryInfC{$\Rightarrow (\varphi \wedge \lbm\gamma) \rightarrow \lbm \varphi^\u$}
      \DisplayProof};

\draw[->,rounded corners=.25cm,dashed] (3,2.1) -- (3,2.6) -- (7,2.6) -- (7,-0.4) -- (-0.6,-0.4);
\end{tikzpicture}
}
\caption{A cyclic proof for the induction axiom.\label{fig:example-proof}}
\end{figure}
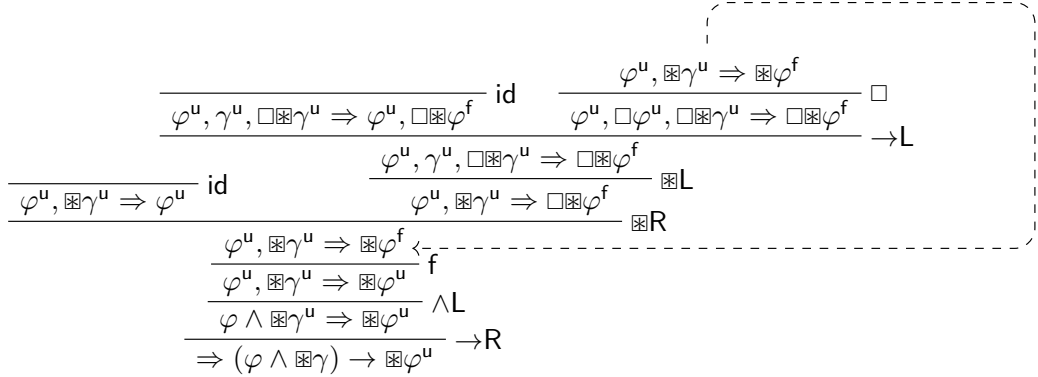

\begin{example}\label{e: non-wellfounded and cyclic proof for IM}
    The induction axiom $(\varphi \wedge \lbm (\varphi \rightarrow \lb \varphi)) \rightarrow \lbm \varphi$ for the master modality is provable in $\NWIM$ and $\CIM$. A cyclic proof is presented in Figure \ref{fig:example-proof}, where $\gamma$ stands for the formula $\varphi \rightarrow \lb \varphi$. The nodes distinguished by the arrow form a repetition. Note that the path from the lower to the higher node always has a formula in focus and passes through an instance of $\lbm \mathsf{R}$ where the principal formula is in focus. Therefore the repetition is successful. As every other leaf is labelled by an axiom, the depicted pre-proof is a cyclic proof. From this cyclic proof we obtain a non-wellfounded proof of the induction axiom by unfolding the tree infinitely often over the successful repetition.
\end{example}

\section{Soundness of the Cyclic Calculus}\label{c: IM, section soundness of cim}

The remainder of this chapter deals with establishing soundness and completeness of $\CIM$ for the classes of dynamic, functional and triangle models. Apart from simply providing soundness and completeness proofs, we also aim to give an alternative (proof-theoretic) proof of Theorem \ref{t: three classes one logic}. By Lemma \ref{l: IM subset of IMf and IMt}, we have $\mathbf{IM} \subseteq \mathbf{IM_t}$ and $\mathbf{IM} \subseteq \mathbf{IM_f}$. In the proof of Theorem \ref{t: three classes one logic}, we established the other inclusions by showing how to translate dynamic models into triangle models and into functional models. Here, we will establish the other inclusions by first showing that every $\CIM$-provable formula belongs to $\mathbf{IM}$, and then that every formula in $\mathbf{IM_t}$ and $\mathbf{IM_f}$ is $\CIM$-provable.

This section carries out the first step by proving soundness of $\CIM$ with respect to the class of dynamic models.\footnote{Although the calculus $\NWIM$ is only used to show completeness of $\CIM$, let us note here that $\NWIM$ is also sound: this follows from soundness of $\CIM$ and Lemma \ref{l: non-wellfounded proof implies cyclic proof}.} Therefore, the terms `valid', `invalid' and so on refer in this section exclusively to the class of dynamic models.

Since cyclic proofs may contain non-axiomatic leafs, a more complex argument than an induction on the height of a proof is required to establish soundness.
Intuitively, the cyclic calculus is sound because success of repetitions ensures that along every repetition, \emph{progress} is made in the form of a modal step: as a result, when proving $\lbm\phi$ we are essentially proving $\lb^n\phi$ for every $n<\omega$. To prove this formally, we argue by contraposition.  We first assign to each invalid sequent with a formula in focus a \emph{measure} in the form of a natural number. Then, assuming there is a cyclic proof $\pi$ of an invalid sequent, we show that there must be a successful repetition $(u,v)$ in $\pi$ of invalid sequents whose measure strictly decreases along the path $(u,v)$. As $u$ and $v$ are labelled by the same sequent, this contradicts the fact that measures are well-defined. The following proof is routine (c.f. Lemma \ref{l: axioms of IM are valid}).

\begin{lemma}\label{l: local soundness}
    Suppose
    \begin{equation*}
        \infer[\mathsf{r}]{\sigma}{\sigma_1 & \dotsm & \sigma_n}
    \end{equation*}
    is an instance of a rule $\mathsf{r}$ of $\CIM$. If $\sigma$ is invalid, then there is a natural number $i$ with $1 \leq i \leq n$ such that $\sigma_i$ is invalid. 
\end{lemma}

Let $\sigma$ be a sequent that has a formula in focus, i.e., $\Delta_\sigma$ contains a formula of the form $\lb^j \lbm \varphi^\f$ for $j \in \{0,1\}$. Denote by $\sigma(n)$ the sequent $\Gamma_\sigma \Rightarrow \Delta_\sigma, \lb^j \lb^n \varphi^\u$, i.e., the sequent expanding the right side of $\sigma$ by the formula $\lb^j \lb^n \varphi^\u$. 
\begin{lemma}\label{l: invalid sequent with formula in focus}
    If $\sigma$ has a formula in focus and is invalid, then there exists a natural number $n$, such that $\sigma(n)$ is invalid.
\end{lemma}
\begin{proof}
Let $\sigma$ be an invalid sequent with a formula in focus. Then there exists a formula $\lb^j \lbm \varphi^\f \in \Delta_\sigma$ for $j \in \{0,1\}$ and a pointed dynamic model $(\fw,w)$ with $\fw,w  \models \bigwedge \Gamma_\sigma^-$ and $\fw,w \not \models \bigvee \Delta_\sigma^-$. So in particular $\fw,w \not \models \lb^j \lbm \varphi$. If $j=0$, then there exists a world $v$ with $w \mathrel{(\Tilde{R})^*} v$ and $\fw,v \not \models \varphi$. Since $\leq$ is reflexive there are worlds $u_0, \ldots, u_{2n}$ such that $u_0 = w$, $u_{2n} = v$ and for all $0 \leq i < 2n$ it holds that if $i$ is even, then $u_i \leq u_{i+1}$ and if $i$ is odd, then $u_i \mathrel{R} u_{i+1}$. Therefore $w \mathrel{(\Tilde{R})^n} v$, implying that $\fw,w \not \models \lb^n \varphi$. If $j=1$, then there is a world $v$ with $w \mathrel{\Tilde{R}} v$ and $\fw,v \not \models \lbm \varphi$. By the previous case we have that $\fw,v \not \models \lb^n \varphi$ for some $n$. Hence $\fw,w \not \models \lb \lb^n \varphi$. Therefore $\fw,w \not \models \sigma(n)$ for some natural number $n$.
\end{proof}

As a consequence, every invalid sequent $\sigma$ with a formula in focus can be assigned a measure:
\begin{equation*}
    \mu(\sigma) \coloneqq \min \{n < \omega \mid \sigma(n) \text{ is invalid}\}.
\end{equation*}
We may now prove a strenghtening of Lemma \ref{l: local soundness}.

\begin{lemma}\label{l: global soundness}
    Suppose
    \begin{equation*}
        \infer[\mathsf{r}]{\sigma}{\sigma_1 & \dotsm & \sigma_n}
    \end{equation*}
    is an instance of a rule $\mathsf{r}$ of $\CIM$. If $\sigma$ is invalid, then there is a natural number $i$ with $1 \leq i \leq n$ such that $\sigma_i$ is invalid. 
    If both $\sigma$ and $\sigma_i$ have a formula in focus, then, moreover,
    \[
        \mu(\sigma_i) \leq \mu(\sigma),
    \]
    where the inequality is strict if $\mathsf{r} = \lbm \mathsf{R}$ and the principal formula is in focus.
\end{lemma}
\begin{proof}
    By Lemma \ref{l: local soundness} it suffices to only consider the case where both the conclusion and at least one premise have a formula in focus. We first treat the case that the formula in focus is not principal. Then $\mathsf{r}\notin \{{\to}\R, \lb\}$, as this would contradict the existence of a premise with a formula in focus. By inspection of the rules, note that then \emph{every} premise must have a formula in focus, and so the following is a correct rule instance of $\mathsf{r}$.
    \begin{equation*}
        \infer[\mathsf{r}]{\sigma(\mu(\sigma))}{\sigma_1(\mu(\sigma)) & \dotsm & \sigma_n(\mu(\sigma))}
    \end{equation*}
    By Lemma~\ref{l: local soundness}, since $\sigma(\mu(\sigma))$ is invalid, there exists a premise $\sigma_i(\mu(\sigma))$ that is invalid. Hence $\sigma_i$ is invalid and $\mu(\sigma_i) \leq \mu(\sigma)$. \smallskip
    
    Now suppose that the formula in focus is principal in $\mathsf{r}$. Then $\mathsf{r}=\lbm\mathsf{R}$ or $\mathsf{r}=\lb$. If $\mathsf{r} = \lbm \mathsf{R}$, then $\sigma$ is of the form $\Gamma \Rightarrow \lbm \varphi^\f, \Delta$ with premises $\sigma_1$ and $\sigma_2$ given by $\Gamma \Rightarrow \varphi^\u, \Delta$ and $\Gamma \Rightarrow \lb\lbm \varphi^\f, \Delta$, respectively.  As there exists a pointed dynamic model $(\fw,w)$ that falsifies $\sigma(\mu(\sigma))$, $w$ has an intuitionistic successor $v$ such that $\fw,v\models \bigwedge \Gamma^-$ and ${\fw,v\not\models \lbm \varphi \lor \lb^{\mu(\sigma)} \varphi\lor \bigvee\Delta^{-}}$. If $\mu(\sigma) = 0$, then $\fw,v \not \models \varphi $, so $(\fw,v)$ falsifies the left premise $\sigma_1$. By Lemma~\ref{l: proper application of rules}, $\sigma_1$ does not have a formula in focus, and so the statement of the lemma holds. If $\mu(\sigma) > 0$, then $\fw,v \not \models \lb \lb^{\mu(\sigma) - 1} \varphi$, implying that $\fw, v \not \models \lb \lbm \varphi$. Hence $(\fw,v)$ falsifies $\sigma_2(\mu(\sigma)-1)$. So $\sigma_2$ is invalid and we have  $\mu(\sigma_2) < \mu(\sigma)$. \smallskip

    If $\mathsf{r} = \lb$, then the conclusion $\sigma$ is of the form $\Pi, \lb \Gamma \Rightarrow \lb \lbm \varphi^\f, \Sigma$ and the single premise $\sigma_1$ is of the form $\Gamma \Rightarrow \lbm \varphi^\f$.  By assumption there exists a pointed dynamic model $(\fw, w)$ falsifying $\sigma(\mu(\sigma))$. Thus, there exists $w \leq v$ with $\fw, v  \models \bigwedge \lb \Gamma^-$ and $\fw, v \not \models \lb \lb^{\mu(\sigma)} \varphi$. This implies that there exists a world $u \in W$ with $v \mathrel{\Tilde{R}} u$ and $\fw, u \not \models \lb^{\mu(\sigma)} \varphi$. Note that $\fw, u \models \bigwedge \Gamma^-$. Therefore $(\fw, u)$ falsifies $\sigma_1(\mu(\sigma))$, implying that $\sigma_1$ is invalid and  $\mu(\sigma_1) \leq \mu(\sigma)$.
\end{proof}

\begin{theorem}[Soundness of $\CIM$ with respect to dynamic models]\label{t: CIM soundness}
   If there is a $\CIM$-proof of a sequent $\sigma$, then $\sigma$ is valid over the class of dynamic models.
\end{theorem}
\begin{proof}
    Let $\pi$ be a $\CIM$-proof of $\sigma$ and suppose for contradiction that $\sigma$ is invalid. By repeatedly applying Lemma~\ref{l: global soundness} we obtain a path of invalid sequents
    \[
        \rho = \sigma_1,\sigma_2, \ldots, \sigma_n
    \]
    through $\pi$ such that $\sigma=\sigma_1$ and  $\sigma_n$ is a leaf. As $\sigma_n$ cannot be an axiom and $\pi$ is a proof, there exists $i < n$ such that $(\sigma_i, \sigma_n)$ is a successful repetition. Then the path from $\sigma_i$ to $\sigma_n$ always has a formula in focus and passes through at least one instance of $\lbm \mathsf{R}$ in which the formula in focus is principal. Hence, by construction, we have $\mu(\sigma_n) < \mu(\sigma_i)$, contradicting that $\sigma_n=\sigma_i$. 
\end{proof}

Note that soundness of $\CIM$ with respect to the class of dynamic models implies soundness with respect to the classes of functional and of triangle models.

\begin{corollary}
   If there is a $\CIM$-proof of a sequent $\sigma$, then $\sigma$ is valid over the classes of functional models and of triangle models.
\end{corollary}

\section{Completeness of the Cyclic Calculus}\label{c: IM, section completeness of the cyclic calculus cIM}

This section establishes completeness of the cyclic calculus with respect to triangle and functional models. 
The argument proceeds in two steps.
First, we set up a general framework for proving completeness via proof search games, from which completeness of the non-wellfounded calculus $\NWIM$ is deduced. 
We then show how to transform an arbitrary $\NWIM$-proof into a $\CIM$-proof, whence obtaining completeness of $\CIM$.
\subsection{Proof Search Games}
Each sequent $\sigma$ will be associated with a \emph{proof search tree} which will form the arena of a two-player game between \emph{Prover}, whose winning strategies establish proofs of $\sigma$, and \emph{Refuter}, whose winning strategies describe countermodels for $\sigma$. 
Completeness then becomes a corollary of determinacy of the game.

A proof search tree for $\sigma$ is built by applying rules bottom-up to $\sigma$. The invertible rules are applied first until a \emph{saturated} sequent is obtained. 
\begin{definition}
    A sequent $\Gamma \Rightarrow \Delta$ is \emph{saturated}\index{sequent!saturated} if the following hold.
    \begin{enumerate}
        \item If $\varphi\wedge \psi^\u  \in \Gamma$, then $\varphi^\u  \in \Gamma$ and $\psi^\u  \in \Gamma$.
        \item If $\varphi \vee \psi^\u  \in \Gamma$, then $\varphi^\u  \in \Gamma$ or $\psi^\u  \in \Gamma$.
        \item If $\varphi \rightarrow \psi^\u  \in \Gamma$, then $\varphi^\u  \in \Delta$ or $\psi^\u  \in \Gamma$.
        \item If $\lbm \varphi^\u  \in \Gamma$, then $\varphi^\u  \in \Gamma$ and $\lb \lbm \varphi^\u  \in \Gamma$.
        \item If $\varphi \wedge \psi^\u \in \Delta$, then $\varphi^\u \in \Delta$ or $\psi^\u \in \Delta$.
        \item If $\varphi \vee \psi^\u \in \Delta$, then $\varphi^\u \in \Delta$ and  $\psi^\u \in \Delta$.
        \item If $\lbm \varphi^\u \in \Delta$, then $\varphi^\u \in \Delta$ or $\lb \lbm \varphi^\u \in \Delta$.
        \item $\lbm \varphi^\f \not \in \Delta$ for all formulas $\varphi$.
    \end{enumerate}
    Given a sequent $\sigma$, a formula occurring in $\sigma$ is said to be \emph{saturated} if $\sigma$ satisfies the corresponding clause above for that formula.
\end{definition}
As we are working with set sequents, formulas can simultaneously function as principal and as side formulas. We call an application of a rule \emph{preserving} if the principal formula(s) also occurs as a side formula. For example, in the following two applications of $\lbm \mathsf{L}$ to the sequent $\lbm \varphi^\u \Rightarrow \psi^\u$, the left application is preserving but the right application is not:
\begin{center}
    \begin{tabular}{l l}
       $\infer[\lbm \mathsf{L}]{\lbm \varphi^\u \Rightarrow \psi^\u}{\lbm \varphi^\u, \varphi^\u, \lb \lbm \varphi^\u \Rightarrow \psi^\u}$  &  $\infer[\lbm \mathsf{L}]{\lbm \varphi^\u \Rightarrow \psi^\u}{\varphi^\u, \lb \lbm \varphi^\u \Rightarrow \psi^\u}$\\
    \end{tabular}
\end{center}

When a rule is applied preservingly to a sequent $\sigma$, we ensure that the premise(s) preserve the information contained in $\sigma$ by preserving all formulas of $\sigma$. Note that applications of the non-invertible rules $\lb$ and ${\to}{\mathsf{R}}$ are never preserving. Moreover, applications of $\lbm \mathsf{R}$ where the principal formula is in focus are not preserving either (since sequents can only have one formula in focus).

We will now give a general definition of proof search trees, that will be employed in the following completeness arguments. The particular form of the proof search tree depends on the kind of countermodel one wants to obtain from a winning strategy of Refuter. In general, proof search trees are built by first applying invertible rules until a saturated sequent is encountered. Then a non-invertible rule must be applied. Since we do not know which non-invertible rule to apply to which formula, all possible non-invertible rule instances are merged into a new rule, called the \emph{choice rule}. Depending on the specific form of the choice rule, different kinds of countermodels can be read off from a failed proof search. Therefore, the following general definition of proof search tree is given relative to an arbitrary choice rule. For technical reasons proof search trees will be labelled by \emph{indexed sequents} $\Gamma \Rightarrow_k \Delta$, i.e. sequents decorated with a natural number $k$ which is called the \emph{index} of the sequent. Each rule is applied to indexed sequents just as it is applied to normal (non-indexed) sequents. Every proof search tree for a sequent $\sigma$ additionally contains an enumeration of the formulas in $\Cl(\sigma)$.

\begin{definition}\label{d: proof search tree nIM}
    Fix some inference rule $\mathsf{C}$ and a sequent $\sigma$. A \emph{proof search tree (with choice rule $\mathsf{C}$)}\index{proof search!tree} for $\sigma$ consists of an arbitrary enumeration of the formulas in $\Cl(\sigma)$ and a finite or countably infinite tree $\mathcal{T}$ whose nodes are labelled by indexed sequents according to $\mathsf{C}$  and the invertible logical rules of $\NWIM$ such that the following hold.
    \begin{enumerate}
        \item The root is labelled by $\Gamma_\sigma\Rightarrow_0 \Delta_\sigma$;
        \item Every invertible rule is applied preservingly, with the exception of $\lbm \mathsf{R}$ if the principal formula is in focus;
        \item No invertible rule is applied to a sequent in which the principal formula is already saturated;
        \item Invertible rules leave the index of the sequent unchanged;
         \item A node is a leaf if and only if it is labelled by an axiom or by a saturated sequent to which the $\mathsf{C}$-rule cannot be applied;
         \item The $\mathsf{C}$-rule is only applied to saturated sequents.
    \end{enumerate}
\end{definition}

 The $\C$-rule replaces both the non-invertible and the focus rules. Each completeness proof we present will be relative to a suitable `choice' rule $\C$. For completeness with respect to triangle models, the respective choice rule will leave the index of sequents unchanged, implying that every sequent in such a proof search tree will be indexed by $0$. Thus for that proof the index can be ignored, which we will do without further mention. The completeness proof for functional models on the other hand will make use of the index and the respective choice rule will change the index of sequents. Similarly, the enumeration of $\Cl(\sigma)$ will only be used in the completeness proof for functional models.
 
 Let $\sigma$ be a sequent and let $\Gamma_\sigma^{ns} \subseteq \Gamma_\sigma$ and $\Delta_\sigma^{ns} \subseteq \Delta_\sigma$ be the sets of formulas occurring on the left side and right side of $\sigma$, respectively, which are \emph{not} saturated and are not of the form $\lb \varphi$. Recall that $c(\Gamma)$ denotes the complexity of $\Gamma$ (c.f. Definition \ref{d: complexity of IM formula}).

\begin{lemma}\label{l: proof search trees exist}
   Every sequent $\sigma$ has a proof search tree. 
\end{lemma}
\begin{proof}[Proof sketch.]
    Let $\sigma$ be a sequent and suppose that $\sigma_1$ is a premise of some preserving rule application of an invertible logical rule to $\sigma$ where the principal formula is not saturated or a premise of a rule instance of $\lbm \mathsf{R}$ with the principal formula in focus. An inspection of the rules yields that 
    \begin{equation}\label{e: proof search tree exists}
        c(\Gamma_\sigma^{ns}) + c(\Delta_\sigma^{ns}) > c(\Gamma_{\sigma_1}^{ns}) + c(\Delta_{\sigma_1}^{ns}).
    \end{equation}
    In particular if $\sigma$ is $\Gamma, \lbm \varphi^\u \Rightarrow \Delta$ and the rule applied is $\lbm \mathsf{L}$ with $\lbm \varphi^\u$ principal, then the premise $\sigma_1$ is $\Gamma, \lbm \varphi^\u, \varphi^\u, \lb \lbm \varphi^\u \Rightarrow \Delta$. Note that $c(\varphi) < c(\lbm \varphi)$ and $\lbm \varphi, \lb \lbm \varphi \not \in \Gamma_{\sigma_1}^{ns}$, implying that (\ref{e: proof search tree exists}) holds. By induction on $c(\Gamma_\sigma^{ns}) + c(\Delta_\sigma^{ns})$ we can therefore prove that preservingly applying invertible logical rules bottom-up to non-saturated formulas starting at $\sigma$ results in a finite tree where each leaf is either an axiom or saturated.  \smallskip \\
    Leafs that are axioms or saturated sequents which are not conclusions of a $\C$-rule instance are closed. To any other leaf we apply the $\C$-rule bottom-up. We then co-recursively repeat this procedure for each open leaf, resulting in the construction of a proof search tree.
\end{proof}
 
  As a corollary of this construction we obtain the following result.
  
\begin{lemma}\label{lemma every inf branch inf many C}
    Every infinite branch of a proof search tree contains infinitely many applications of $\mathsf{C}$. 
\end{lemma}

Next, we define \emph{proof search games}\index{proof search!games}, which are played by two players called Prover and Refuter on a given proof search tree $\mathcal{T}$.

\begin{definition}\label{d: proof search tree for iltln}
    Let $\sigma$ be a sequent and $\mathcal{T}$ a proof search tree for $\sigma$ with choice rule $\C$. The \emph{proof search game} $\mathcal{G}(\mathcal{T}, \C)$ is played by two players called \emph{Prover} and \emph{Refuter}. The \emph{arena} is the proof search tree $\mathcal{T}$, where each \emph{position} is a node of $\mathcal{T}$. Prover \emph{owns} every position $t \in \mathcal{T}$ which is labelled by the conclusion of a $\mathrm{C}$-rule instance. Refuter owns every other position. The \emph{admissible moves} for each player are the children of every node player owns. A \emph{play} is a sequence of positions $(t_i)_i$ such that $t_0$ is the root of $\mathcal{T}$ and any two consecutive positions are related by an admissible move. A play is either finite and ends in a leaf of $\mathcal{T}$ or infinite. Note that every play is a branch of $\mathcal{T}$. The \emph{winning conditions are as follows}.
    \begin{enumerate}
        \item Prover wins a play $(t_i)_i$ if the play is finite and ends in an axiom or if it is infinite and $(t_i)_i$ has a good suffix.
        \item Refuter wins a play $(t_i)_i$ if the play is finite and ends in a non-axiomatic sequent or infinite and does not have a good suffix.
    \end{enumerate}
\end{definition}

 We will usually identify plays in $\mathcal{G}(\mathcal{T}, \mathsf{C})$ and branches of $\mathcal{T}$ without explicit mention. The following definitions are given for both players Prover and Refuter. We simply write Player instead.

\begin{definition}\label{d: strategy}
    Given a proof search game $\mathcal{G}(\mathcal{T}, \C)$, a \emph{strategy} for Player is a partial function $f : \mathcal{T} \longrightarrow \mathcal{T}$ such that for any node $u$ of $\mathcal{T}$, $\undefined{f(u)}$ if $u$ is not owned by Player or if $u$ is a leaf and $\exists f(u)$ where $f(u)$ is a child node of $u$ otherwise.
\end{definition}

A strategy thus tells Player how to move in every owned position. Note that each node $u \in \mathcal{T}$ is unique, implying that strategies as defined here are aware of the history of the game. Player \emph{uses strategy $f$} if whenever the current play is in position $u$ and $\exists f(u)$, then Player moves to $f(u)$.

\begin{definition}\label{d: winning strategy}
    A strategy $f$ for Player is \emph{winning} if Player wins every play in which $f$ is used.
\end{definition}

Given a game $\mathcal{G}(\mathcal{T}, \C)$ and a strategy $f$ for Player, the \emph{strategy tree} of $f$ is the subtree of $\mathcal{T}$ which consists of all plays that can occur when Player uses $f$. The formal definition is as follows.

\begin{definition}\label{d: strategy tree}\index{strategy tree}
    The \emph{strategy tree} $\mathcal{T}_f$ of a strategy $f$ for Player in $\mathcal{G}(\mathcal{T}, \C)$ is the subtree of $\mathcal{T}$ defined by
    \begin{enumerate}
        \item $\mathcal{T}_f$ contains the root of $\mathcal{T}$.
        \item If $u \in \mathcal{T}$ belongs to $\mathcal{T}_f$ and $\exists f(u)$, then for any child $v$ of $u$ in $\mathcal{T}$ holds that $v \in \mathcal{T}_f$ if and only if $v=f(u)$.
        \item If $u \in \mathcal{T}$ belongs to $\mathcal{T}_f$ and $\undefined{f(u)}$, then all children $v$ of $u$ in $\mathcal{T}$ belong to $\mathcal{T}_f$.
    \end{enumerate}
\end{definition}

Given a sequent $\sigma$ and a proof search tree $\mathcal{T}$ for $\sigma$, a \emph{refutation}\index{proof search!refutation} of $\sigma$ is defined as follows.

\begin{definition}\label{def refutation}
    A \emph{refutation} of a sequent $\sigma$ is a subtree $\mathcal{R}$ of a proof search tree $\mathcal{T}$ for $\sigma$ satisfying the following properties.
    \begin{enumerate}
        \item $\mathcal{R}$ contains the root of $\mathcal{T}$.
        \item No leaf of $\mathcal{R}$ is an axiom. 
        \item No infinite branch of $R$ has a good suffix.
        \item If $\mathcal{R}$ contains a node $u$ that is labelled by the conclusion of a $\C$-application, then $\mathcal{R}$ contains all children of $u$ in $\mathcal{T}$.
        \item If $\mathcal{R}$ contains a node $u$ that is labelled by the conclusion of any other rule than $\C$, $\mathsf{id}$ or $\bot$, then $\mathcal{R}$ contains exactly one child of $u$ in $\mathcal{T}$.
    \end{enumerate}
\end{definition}

It is immediate to see from the definition of the winning conditions of Refuter that the strategy trees of winning strategies for Refuter are refutations.

\begin{lemma}\label{l: winning strategy for refuter is refutation IM}
    Let $f$ be a winning strategy for Refuter in the game $\mathcal{G}(\mathcal{T}, \C)$. Then $\mathcal{T}_f$ is a refutation.
\end{lemma}

In the next sections we will show that when using specific choice rules $\C$, refutations of $\sigma$ correspond to countermodels for $\sigma$, while the winning strategies for Prover correspond to $\NWIM$-proofs of $\sigma$. To show that refutations correspond to countermodels, we will make use of the following `canonical' model construction. For this construction, we assume that the premises of the choice rule $\C$ have been partitioned into two groups: the \emph{intuitionistic premises} and the \emph{modal premises}.

\begin{definition}\label{d: canonical model for nIM}
    Let $\sigma$ be a sequent and let  $\mathcal{R}$ be a refutation of $\sigma$. The \emph{canonical model}\index{canonical model!for $\NWIM$} based on $\mathcal{R}$ is the model $\fw_\mathcal{R}=(W, \leq, R, V)$ defined as follows.
    \begin{enumerate}
        \item $W = \sfrac{\mathcal{R}}{\sim}$, where $s \sim t$ iff there exists a path between $s$ and $t$ in $\mathcal{R}$ in which no instance of the $\C$-rule occurs.

        \item $\leq$ is the reflexive, transitive closure of the relation $ {\leq_0} \subseteq W \times W$ given by
        \begin{equation*}
        \begin{split}
            w \leq_0 v \text{ iff } &\text{there exist } s \in w \text{ and } t \in v \text{ such that } s \text{ is the conclusion and }\\ &t \text{ an intuitionistic premise}
            \text{ of the same } \C\text{-rule instance.}
         \end{split}
        \end{equation*}

        \item $R\subseteq W\times W$ is such that
        \[
        \begin{split}
            w \mathrel{R} v \text{ iff } &\text{there exist } s \in w \text{ and } t \in v \text{ such that } s \text{ is the conclusion and }\\ &t \text{ is a modal} \text{ premise of the same } \C\text{-rule instance.}
         \end{split}
        \]

        \item $V \colon W \longrightarrow \mathcal{P}(\Prop)$ is defined by  $V(w) = \Gamma_w \cap \Prop$ where $\Gamma_w$ is the left side of the sequent labelling the unique node in \( w \) that is the conclusion of a $\mathsf{C}$-rule application or a leaf of $\mathcal{R}$.
    \end{enumerate}
\end{definition}
\noindent The construction of $\fw_\mathcal{R}$ reflects the idea that applying invertible rules to sequents in $\mathcal{R}$ provides more information about the current `world', while applying non-invertible rules (captured by the $\mathsf{C}$-rule) corresponds to taking either a modal or an intuitionistic step.
%
\subsection{Completeness for Triangle Models}
To show completeness of $\NWIM$ for triangle models, we consider the following choice rule $\mathsf{C_t}$. A \emph{$\lb$-formula} is a formula of the form $\lb \varphi$ for any formula $\varphi$. Similarly, a \emph{$\lbm$-formula} is a formula of the form $\lbm \varphi$ for any formula $\varphi$ and an \emph{$\rightarrow$-formula} is a formula of the form $\varphi \rightarrow \psi$ for any formulas $\varphi, \psi$.

\begin{definition}
    The choice rule $\Ct$\index{choice rule!$\Ct$} is given by
    \begin{equation*}
            \infer[\mathsf{C_t}]{\Pi, \lb \Gamma \Rightarrow \{(\varphi_i \rightarrow \psi_i)^{\u}\}_{i=0}^l,\{\lb\chi^{a_i}_i\}_{i=0}^m, \Sigma}{\Pi, \lb \Gamma, \varphi^\u_0 \Rightarrow \psi^{\u}_0 &\dotsm  & \Pi, \lb \Gamma, \varphi^\u_l \Rightarrow \psi^{\u}_l & \Gamma \Rightarrow \chi^{b_0}_0 &\dotsm& \Gamma \Rightarrow \chi^{b_m}_m}
        \end{equation*}
        where the annotations $b_i$ are equal to $\f$ whenever the underlying formula $\chi_i$ is a $\lbm$-formula, and equal to $\u$ otherwise. Moreover, $\Pi\cup\Sigma$ contains no $\lb$-formulas and $\Sigma$ contains no $\to$-formulas. The modal premises are those of the form $\Gamma\Rightarrow \chi^{b_i}_i$, and the others are the intuitionistic premises. In case the conclusion of a $\Ct$-instance has no $\rightarrow$- or $\lb$-formulas on the right-hand side, then we stipulate that $l = -1$ or $m=-1$, respectively.
\end{definition}

\begin{lemma}\label{lem WS for P with Ct correspond to proofs}
     If $\mathcal{T}$ is a proof search tree for $\sigma$ with choice rule $\Ct$ and Prover has a winning strategy in $\mathcal{G}(\mathcal{T}, \Ct)$, then $\sigma$ has a $\NWIM$-proof. 
 \end{lemma}
 
 \begin{proof}
     Let $f$ be a winning strategy for Prover in $\mathcal{G}(\mathcal{T}, \Ct)$ and $\mathcal{T}_f$ the strategy tree of $f$. By definition $\mathcal{T}_f$ contains the root of  $\mathcal{T}$ which is labelled by $\sigma$. By definition of $\mathcal{G}(\mathcal{T}, \Ct)$, the only positions owned by Prover are those nodes of $\mathcal{T}$ that are labelled by the conclusion of a $\Ct$-rule instance. Therefore whenever $\mathcal{T}_f$ contains a node $u$ labelled by the conclusion of an invertible rule instance, $\mathcal{T}_f$ contains all of $u$'s children. If $u$ is labelled by the conclusion of a $\Ct$-rule instance, then $\mathcal{T}_f$ contains exactly one child $v$ of $u$. Observe that if $v$ is an intuitionistic premise, then the sequents labelling $v$ and $u$ form an instance of the rule ${\rightarrow}\mathsf{R}$, and if $v$ is a modal premise, then the sequents labelling $v$ and $u$ form an instance of $\lb$, or an instance of $\lb$ followed by an instance of $\mathsf{f}$. Let $\pi$ be $\mathcal{T}_f$ where whenever $u$ is labelled by the conclusion of an instance of $\Ct$
     \begin{equation*}
         \Pi, \lb \Gamma \Rightarrow \{(\varphi_i \rightarrow \psi_i)^{\u}\}_{i=0}^l,\lb \chi_0^{a_0}, \ldots, \lb \chi_m^{a_m}, \Sigma
     \end{equation*}
     with $a_i = \u$ and its unique child $v$ is a modal premise labelled by
     \begin{equation*}
         \Gamma \Rightarrow \chi_i^\f
     \end{equation*}
     for $0 \leq i \leq m$, then $u$ has a unique child $u'$ in $\pi$ (where $u'$ does not occur in $\mathcal{T}_f$ or $\mathcal{T}$) labelled by
     \begin{equation*}
          \Gamma \Rightarrow \chi_i^\u
     \end{equation*}
     and $u'$ has a unique child which is $v$. Observe that $(u,u')$ form an instance of $\lb$ and $(u', v)$ form an instance of $\mathsf{f}$. Hence by construction $\pi$ is an $\NWIM$-pre-proof of $\sigma$. Since $\mathcal{T}_f$ is the strategy tree of a \emph{winning strategy} for Prover, every leaf of $\pi$ is labelled by an axiom and every infinite branch contains a good suffix. Thus $\pi$ is an $\NWIM$-proof of $\sigma$.
 \end{proof}

\begin{proposition}\label{p: refutation gives triangle model}
   If $\mathcal{T}$ is a proof search tree for $\sigma$ with choice rule $\mathsf{C_t}$ and Refuter has a winning strategy in $\mathcal{G}(\mathcal{T}, \mathsf{C_t)}$, then $\sigma$ is falsifiable over the class of triangle models. 
\end{proposition}

\begin{proof}
Let $f$ be a winning strategy for Refuter in $\mathcal{G}(\mathcal{T}, \mathsf{C_t})$ and $\mathcal{T}_f$ the strategy tree of $f$. By Lemma \ref{l: winning strategy for refuter is refutation IM},  $\mathcal{T}_f = \mathcal{R}$ is a subtree of $\mathcal{T}$ that is a refutation of $\sigma$. Let $\fw_\mathcal{R}=(W, \leq, R, V)$ be the canonical model based on $\mathcal{R}$. It is straightforward to check that $\fw_\mathcal{R}$ is a dynamic model. Let $\fw=(W,\leq, (R \circ {\leq}),V)$ be the induced triangle model.  For each $s \in \mathcal{T}$ let $\Gamma_s \Rightarrow \Delta_s$ be the sequent labelling $s$. For each $w\in W$, let $\Gamma_w \coloneqq \bigcup_{s \in w} \Gamma_s$ and let $\Delta_w \coloneqq \bigcup_{s \in w} \Delta_s$.

Let $\varphi$ be a formula. By induction on the structure of $\varphi$, we simultaneously prove that for any $w\in W$ holds
\begin{enumerate}
    \item[(a)] $\fw,w\models_t \varphi$ if $\varphi\in \Gamma^-_{w}$, and
    \item[(b)] $\fw,w\not\models_t \varphi$ if $\varphi\in \Delta^-_{w}$.
\end{enumerate}

 For any $w \in W$ let $t_w \in w$ be the unique node which is either a leaf of $\mathcal{R}$ or labelled by the conclusion of a $\Ct$-instance. The case for $p \in \Prop$ follows immediately from the defintion of the valuation $V$. For $\varphi = \bot$ note that if $\bot \in \Gamma_w^-$, then there exists $s \in w$ such that $\bot \in \Gamma_s^-$. Therefore $\Gamma_s \Rightarrow \Delta_s$ is an instance of the axiom $\bot$. By definition of a proof search tree, $s$ is a leaf of $\mathcal{R}$, implying that $\mathcal{R}$ contains an axiomatic leaf, which contradicts the assumption that $\mathcal{R}$ is a refutation. Hence $\bot \not \in \Gamma_w^-$. If $\bot \in \Delta_w^-$, then by definition $\fw, w \not \models_t \bot$. The cases for $\varphi = \psi \wedge \chi$ and $\varphi = \psi \vee \chi$ follow immediately from saturation and the induction hypothesis. For example if $\psi \wedge \chi \in \Gamma_w^-$, then, since invertible rules are applied preservingly, $\psi \wedge \chi \in \Gamma_{t_w}^-$. Since $\Gamma_{t_w} \Rightarrow \Delta_{t_w}$ is saturated, we have $\psi, \chi \in \Gamma_{t_w}^-$ and hence $\psi, \chi \in \Gamma_w^-$. By induction hypothesis $\fw, w \models_t \psi$ and $\fw, w \models_t \chi$, implying that $\fw, w \models_t \psi \wedge \chi$. The other cases involving $\psi \wedge \chi$ and $\psi \vee \chi$ are similar and omitted. \smallskip

\noindent \textsc{Case for $\rightarrow$.} For (a) if $\varphi\rightarrow \psi\in \Gamma^-_w$, then since rules are applied preservingly we have $\varphi \rightarrow \psi \in \Gamma_{t_w}^-$. If $v\geq w$, then since rules are applied preservingly and by definition of $\mathsf{C_t}$, it holds that $\varphi\to \psi\in \Gamma^-_{t_v}$. Since $\Gamma_{t_v} \Rightarrow \Delta_{t_v}$ is saturated, $\varphi \in \Delta^-_v$ or $\psi \in \Gamma^-_v$. By the induction hypothesis, $\fw,v\models_t \psi$ or $\fw,v\not\models_t \varphi$, so we obtain $\fw,w\models_t \varphi\to\psi$. 
   For (b) if $\varphi\to \psi\in \Delta_w^-$, then $\varphi \rightarrow \psi \in \Delta_{t_w}^-$, since $\varphi \rightarrow \psi$ may only be principal in instances of $\mathsf{C_t}$. By definition of $\Ct$ and construction of $\leq$ there exists a $v\geq_0 w$ such that $\varphi\in \Gamma^-_v$ and $\psi\in \Delta^-_v$. The induction hypothesis then implies that $\fw,v\models_t \varphi$ and $\fw,v\not\models_t\psi$, and so $\fw,w\not\models_t \varphi\to\psi$. \smallskip
   
\noindent \textsc{Case for $\lb$.} For (a) if $\lb\varphi\in \Gamma_w^-$, then $\lb \varphi \in \Gamma_{t_w}^-$. If  $w\mathrel{(R \circ {\leq}) }v$, then there are two cases. First if there exists a node $s \in v$ which is a modal premise of the $\mathsf{C_t}$-instance with $t_w$ as conclusion, then by definition of $\mathsf{C_t}$ we have $\varphi\in \Gamma_v^-$. Second if there exists $u \geq w$ and a node $s \in v$ which is the premise of a $\mathsf{C_t}$-instance with conclusion $t_u$, then by definition of $\mathsf{C_t}$ and the proof search tree $\lb \varphi \in \Gamma_{t_u}^-$ and hence, as before, $\varphi \in \Gamma_v^-$.\footnote{Since $\fw$ is a triangle model, we have $w \mathrel{(R \circ {\leq})} v$. However, in the refutation $v$ does not contain a node which is a modal premise of the same $\Ct$-rule instance of which $t_w$ is a conclusion of. Therefore this additional case has to be considered.} The induction hypothesis then implies  $\fw,v\models_t \varphi$, so we obtain $\fw,w\models_t \lb\varphi$. For (b) if $\lb\varphi\in \Delta_w^-$, then $\lb \varphi \in \Delta_{t_w}^-$. By definition of $\Ct$ and $R$ there exists a $v$ with $w\mathrel{R}v$ such that $\varphi\in \Delta_v^-$. By induction hypothesis $\fw,v\not\models_t \varphi$, so $\fw,w\not\models_t \lb\varphi$. \smallskip
    
\noindent \textsc{Case for $\lbm$.} For (a) if $\lbm\varphi\in \Gamma^-_w$, then $\lbm \varphi \in \Gamma_{t_w}^-$. Suppose $w\mathrel{(R\circ{\leq})^*}v$. Let 
\begin{equation*}
    w = u_0 \mathrel{(R \circ \leq)} u_1 \mathrel{(R \circ \leq)} \ldots \mathrel{(R \circ \leq)} u_n = v.
\end{equation*}
be the ${(R \circ \leq)}$-path from $w$ to $v$. We prove inductively that for each $i \leq n$ holds that $\varphi, \lb \lbm \varphi \in \Gamma_{u_i}^-$. For $u_0$, note that $u_0= w$ and so it follows from saturation that  $\varphi, \lb \lbm \varphi \in \Gamma_w^-$. For $i >0$ we have by (the inner) induction hypothesis that $\lb \lbm \varphi \in \Gamma_{u_{i-1}}^-$ and hence ${\lb \lbm \varphi \in \Gamma_{t_{u_{i-1}}}^-}$. If $u_i$ contains a node $t$ which is a modal premise of the same $\Ct$-instance of which $t_{u_{i-1}}$ is the conclusion of, then the definition of $\Ct$ yields that $\lbm \varphi \in \Gamma_t^-$ and thus $\lbm \varphi \in \Gamma_{u_i}^-$. By saturation, we have $\varphi, \lb \lbm \varphi \in \Gamma_{u_i}^-$. Otherwise $u_i$ is a modal successor of some world $u' \geq u_{i-1}$. Since $\lb \lbm \varphi \in \Gamma_{t_{u_{i-1}}}^-$, it follows from the definition of $\Ct$ and the proof search tree that $\lb \lbm \varphi \in \Gamma_{\Tilde{u}}^-$ for all $\Tilde{u} \geq u_{i-1}$. Hence $\lb \lbm \varphi \in \Gamma_{u'}^-$ and so $\varphi, \lb \lbm \varphi \in \Gamma_{u_i}^-$ by the same argument as in the previous case. We conclude that $\varphi \in \Gamma_v^-$ and so by induction hypotheses $\fw, v \models_t \varphi$. Hence $\fw, w \models_t \lbm \varphi$. \smallskip

For (b) if $\lbm\varphi\in \Delta^-_w$, then saturation implies $\varphi\in \Delta^-_{t_w}$ or $\lb\lbm\varphi\in \Delta^-_{t_w}$.\footnote{If $\lbm \varphi$ is unfocused, then the saturation clause 7. guarantees this. Otherwise the saturation clause 8. guarantees it. In the second case, there exists a node $s \in w$ with $\lbm \varphi^f \in \Delta_s$. Since $\lbm \varphi^f$ cannot occur in $\Delta_{t_w}$, there must be a (non-preserving) application of $\lbm \mathsf{R}$ with $\lbm \varphi^f$ as principal formula.} Suppose, for contradiction, that for all $w\mathrel{(R \circ {\leq})^*} v$ we have $\varphi \not \in \Delta^-_v$. This implies that $\lb \lbm \varphi \in \Delta_{t_w}$. Then we can define an infinite path $\rho$ in $\mathcal{R}$ starting from $t_w$ as follows: at each $\mathsf{C_t}$-application, we pick the modal premise that has $\lbm\varphi^\f$ as consequent. Note that since no $w \mathrel{(R \circ {\leq})^*} v$ satisfies $\varphi\in \Delta^-_v$, this is always possible. The path $\rho$ then forms a good suffix of the infinite branch of $\mathcal{R}$ in which it is contained, contradicting that $\mathcal{R}$ is a refutation. So there must be some $w \mathrel{(R \circ {\leq})^*} v$ with $\varphi\in \Delta^-_v$, and thus $\fw,v\not\models_t \varphi$ by the induction hypothesis. Therefore $\fw,w\not\models_t \lbm\varphi$.

Let $w \in W$ contain the root of $R$. By definition $\Gamma_\sigma \subseteq \Gamma_w$ and $\Delta_\sigma \subseteq \Delta_w$. Thus  $\fw,w \not \models \sigma$.
\end{proof}

We have established that winning strategies for Prover correspond to $\NWIM$-proofs and winning strategies for Refuter to (triangle) countermodels. In order to obtain completeness, we require that the game $\mathcal{G}(\mathcal{T}, \Ct)$ is determined\index{Gale--Steward games!determinacy}, which means that exactly one of the two players has a winning strategy. It is well-known that proof search games like $\mathcal{G}(\mathcal{T}, \Ct)$ can be reformulated as \emph{Gale--Steward games}\index{Gale--Steward games}. In fact, since Gale--steward games only permit infinite plays, it suffices to extend every finite branch in the proof search tree $\mathcal{T}$ by an infinite sequence of nodes labelled by the same sequent as the leaf and determine that such plays are then won by Prover if the suffix contains only axioms and by Refuter otherwise.  The resulting game is then a Gale--Steward game. It is a standard result in non-wellfounded proof theory that the set of winning plays for each player in such a game is Borel. In fact, the winning sets belongs to a low level of the Borel hierarchy, namely $\Delta_3$. Therefore we may apply the following theorem.

\begin{theorem}[Martin 1975~\cite{Martin_1975}]
    If $\mathcal{G}$ is a Gale--Steward game and the set of winning plays for each player is Borel, then $\mathcal{G}$ is determined.
\end{theorem}

To prove that the set of winning plays for each player is Borel is out of the scope of this thesis. We would require to build a Büchi automaton which recognizes the good branches in the adapted proof search tree. For a precise construction, see~\cite{menendez_2024}.

\begin{theorem}[Completeness of $\NWIM$ with respect to triangle models]\label{t: NIM triangle completeness}
    If $\sigma$ is valid over the class of triangle models, then $\sigma$ has a $\NWIM$-proof.
\end{theorem}
\begin{proof}
   Suppose a sequent $\sigma$ is valid over the class of triangle models. Let $\mathcal{T}$ be a proof search tree with choice rule $\mathsf{C_t}$ for $\sigma$ and consider the two-player game $\mathcal{G}(\mathcal{T}, \mathsf{C_t})$. By determinacy, exactly one of the two players Prover and Refuter has a winning strategy. By validity of $\sigma$ and Proposition \ref{p: refutation gives triangle model}, Refuter cannot have a winning strategy. Hence Prover has a winning strategy. Therefore, by Lemma \ref{lem WS for P with Ct correspond to proofs}, $\sigma$ is provable in $\NWIM$.
\end{proof}



\subsection{Completeness for Functional Models}\label{subsectin completeness functional models}
To obtain completeness with respect to functional models, it suffices to replace the choice rule $\Ct$ in the previous argument by a suitable choice rule $\Cf$. The design of $\Cf$ will guarantee that the canonical model (relative to $\Cf$) is functional. In the completeness proof for triangle models, when we reached a saturated sequent, the choice rule $\Ct$ created multiple premises corresponding to intuitionistic and modal successors of the current world. Here, when we reach a saturated sequent with multiple $\lb$-formulas on the right side, only one of these formulas can be used to create a modal successor, since we are building a functional model. This leaves the question how to deal with the remaining $\lb$-formulas on the right side, which also have to be falsified. In the proof of Theorem \ref{t: three classes one logic}, whenever a world in an $\fw$-induced structure had multiple $\lb$-defects, the issue of falsifying all of them was resolved by adding copies of that world as intuitionistic successors and resolving one $\lb$-defect in each of these copies. A similar strategy will work in the completeness proof here. Namely, the choice rule $\Cf$ will pick exactly one $\lb$-formula on the right (if one exists) to create a modal successor, while the other $\lb$-formulas generate intuitionistic successors, where they can be falsified as well. To keep track of which right $\lb$-formula has to be `taken care of' at a particular step, the proof search tree will make use of the indexes introduced earlier. \smallskip 

Recall that a proof search tree for $\sigma$ contains an enumeration $e$ of the formulas of $\Cl(\sigma)$. Suppose $\lvert \Cl(\sigma) \rvert = m$.
\begin{definition}
    The choice rule $\Cf$\index{choice rule!$\Cf$} is given by
    \begin{equation*}
            \infer[\mathsf{C_f}]{\Pi, \lb \Gamma \Rightarrow_k \{\varphi_i \rightarrow \psi^{\u}_i\}_{i=0}^l, \{\lb \chi^{a_i}_{i}\}_{i=0}^{r}, \Sigma}{\{\Pi, \lb \Gamma, \varphi^\u_i \Rightarrow_0 \psi^\u_i\}_{i=0}^{l}& \Gamma_{\tau}  \Rightarrow_{(k+1)_m}\Delta_{\tau} & \Gamma \Rightarrow_0 \chi^b_{i}}
        \end{equation*}
        where $\tau$ is the sequent labelling the conclusion (so $\Gamma_\tau$ and $\Delta_\tau$ denote the left and right side of the conclusion, respectively), and $b$ equals $\f$ if $\chi_{k}$ is a $\lbm$-formula and equals $\u$ otherwise. If there are no $\rightarrow$-formulas or no $\lb$-formulas on the right-hand side, then $l= -1$ or $r= -1$, respectively. Moreover, we require that $k < m+1$ and $e(\lb \chi_i) = k$. If there is no formula $\lb \chi_i$ on the right side of the conclusion with $e(\lb \chi_i) = k$, then the instance of $\Cf$ does not have the rightmost premise. The expression $(k+1)_m$ denotes `$k+1$ modulo $m$'. Furthermore, $\Pi\cup\Sigma$ contains no $\lb$-formulas and $\Sigma$ contains no $\to$-formulas. The rightmost premise is the modal premise and the others are intuitionistic premises, where we call the premise $\Gamma_\tau \Rightarrow_{(k+1)_m} \Delta_\tau$ the \emph{right intuitionistic premise} and all others the \emph{left intuitionistic premises}.
\end{definition}

The choice rule $\Cf$ thus chooses the $k$-th formula in the enumaration of $\sigma$ to create a modal successor if the $k$-th formula is a $\lb$-formula and present on the right side of the conclusion. Otherwise, the world $w$ containing the conclusion of $\Cf$ does not have a modal successor. Moreover, $\Cf$ adds an intuitionistic successor which is a copy of $w$ where the index is increased. Note that the premise $\Gamma_\tau  \Rightarrow_{(k+1)_m}\Delta_\tau$ differs from the conclusion only in the index and is therefore saturated, implying that it is the conclusion of another $\Cf$-instance, where the corresponding world either has no modal successor or a modal successor falsifying a different $\lb$-formula.

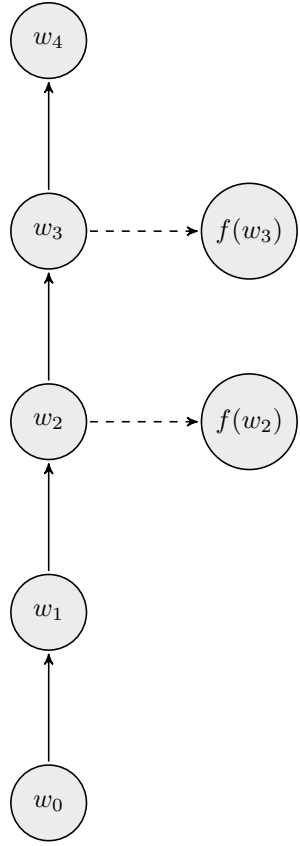
\begin{figure}[t!]
    \centering
    \begin{tikzpicture}[modal]
        \node[world](w0){$w_0$} ;
        \node[world](w1)[above=of w0]{$w_1$};
        \node[world](w2)[above=of w1]{$w_2$} ;
        \node[world](w3)[above=of w2]{$w_3$};
        \node[world](w4)[above=of w3]{$w_4$};
        \node[world](v2)[right=of w2]{$f(w_2)$};
        \node[world](v3)[right=of w3]{$f(w_3)$};
         \path[->] (w0) edge (w1);
        \path[->] (w1) edge (w2);
        \path[->] (w2) edge (w3);
        \path[->] (w3) edge (w4);
        \path[->] (w2) edge[dashed] (v2);
        \path[->] (w3) edge[ dashed] (v3);
    \end{tikzpicture}
    \caption{The induced functional model from Example \ref{ex: refutation}. Solid arrows indicate the $\leq$-relation and dashed arrows the partial function $f$. The worlds above $w_4$ are not depicted.}
    \label{f: example countermodel refutation}
\end{figure}

\begin{example}\label{ex: refutation}
    Consider the sequent $\sigma=  p^\u \Rightarrow \lb p^\u, \lb q^\u$ and let $e(p) = 0$, $e(q) = 1$, $e(\lb p) = 2$ and $e(\lb q) = 3$ be an enumeration of $\Cl(\sigma)$. The following is a proof search tree for $\sigma$.
    \begin{prooftree}
        \AxiomC{.}
        \noLine
        \UnaryInfC{.}
        \noLine
        \UnaryInfC{.}
        \noLine
        \UnaryInfC{$p^\u \Rightarrow_0 \lb p^\u, \lb q^\u$}
        \AxiomC{$\Rightarrow_0 q^\u$}
        \RightLabel{$\Cf$}
        \BinaryInfC{$p^\u \Rightarrow_3 \lb p^\u, \lb q^\u$}
        \AxiomC{$\Rightarrow_0 p^\u$}
        \RightLabel{$\Cf$}
        \BinaryInfC{$p^\u \Rightarrow_2 \lb p^\u, \lb q^\u$}
        \RightLabel{$\Cf$}
        \UnaryInfC{$p^\u \Rightarrow_1 \lb p^\u, \lb q^\u$}
        \RightLabel{$\Cf$}
        \UnaryInfC{$p^\u \Rightarrow_0 \lb p^\u, \lb q^\u$}
    \end{prooftree}
    Note that $\sigma$ is saturated. The first two $\Cf$-instances only produce a right-intuitionistic premise, where both times the index is increased by 1.  The third $\Cf$-instance has a conclusion with index $2$, which is the number of $\lb p$ in the enumeration of $\Cl(\sigma)$. Therefore this instance creates also a modal successor falsifying $p$. Note that the modal successor is saturated and thus a leaf. Once $\lb q$ is taken care of in the $\Cf$-instance, the index is reset to $0$. The resulting proof search tree is exactly a refutation for $\sigma$, yielding the countermodel depicted in Figure \ref{f: example countermodel refutation}. The valuation is given by $V(w_i)= \{p\}$ for $i < \omega$ and $V(f(w_2)) = V(f(w_3))= \emptyset$. Figure \ref{f: example countermodel refutation} only depicts the worlds up to $w_4$. It is straightforward to check that $\fw, w_0 \not \models \sigma^I$.
\end{example}

\begin{lemma}
     If $\mathcal{T}$ is a proof search tree for $\sigma$ with choice rule $\Cf$ and Prover has a winning strategy in $\mathcal{G}(\mathcal{T},\Cf)$, then $\sigma$ has a $\NWIM$-proof. 
 \end{lemma}
 \begin{proof}[Proof sketch]
     As in the proof of Lemma~\ref{lem WS for P with Ct correspond to proofs}, the strategy tree of a winning  strategy for Prover is a subtree $\mathcal{S}$ of $\mathcal{T}$ that contains exactly one child of each node $u$ labelled by a sequent which is the conclusion of a $\Cf$-rule. We obtain a derivation $\pi$ from $\mathcal{S}$ as follows. First of all delete all indices from sequents occurring in $\mathcal{S}$. Then in case the child of $u$ is not labelled by the premise of the form $\Gamma_{\tau}  \Rightarrow_{(k+1)_m}\Delta_{\tau}$, we can simply view this as an application of ${\rightarrow}\mathsf{R}$ or $\lb$ (possibly with a subsequent application of $\f$). In case the direct successor \emph{is} labelled by the premise of the form $\Gamma_{\tau}  \Rightarrow_{(k+1)_m}\Delta_{\tau}$, we simply view this as no rule application at all, as the left and right side of the sequent have not changed. It is clear that from these steps one can obtain a pre-proof $\pi$ of $\sigma$. Due to the winning conditions of Prover, every leaf of $\mathcal{S}$ must be an axiom and every infinite branch must have a good suffix. Thus $\pi$ is a $\NWIM$-proof of $\sigma$.
 \end{proof}

\begin{proposition}\label{p: indexed refutation yields countermodel}
If $\mathcal{T}$ is a proof search tree for $\sigma$ with choice rule $\Cf$ and Refuter has a winning strategy in $\mathcal{G}(\mathcal{T},\Cf)$, then $\sigma$ is falsifiable over the class of functional models. 
\end{proposition}

\begin{proof}
Let $\mathcal{T}$ be a proof search tree for $\sigma$ and $e$ be the enumeration of $\sigma$ where $\lvert \Cl(\sigma) \rvert = m$. Let $f$ be a winning strategy for Refuter in $\mathcal{G}(\mathcal{T},\Cf)$ and $\mathcal{T}_f$ the strategy tree of $f$. By Lemma~\ref{l: winning strategy for refuter is refutation IM}, $\mathcal{T}_f = \mathcal{R}$ is a refutation. Let $\fw_\mathcal{R}=(W, \leq, R, V)$ be the canonical model based on $\mathcal{R}$. It is straighforward to check that $\fw_\mathcal{R}$ is a dynamic model. Moreover, since every instance of $\Cf$ has at most one modal premise, each world in $\fw_\mathcal{R}$ has at most one modal successor, implying that $\fw_\mathcal{R}$ is functional.

We follow the proof of Proposition \ref{p: refutation gives triangle model}. For any formula $\varphi$, by induction on the structure on $\varphi$ we simultaneously prove that for any  $w\in W$ we have 
\begin{enumerate}
    \item[(a)] $\fw,w\models \varphi$ if $\varphi\in \Gamma^-_{w}$ and
    \item[(b)] $\fw,w\not\models \varphi$ if $\varphi\in \Delta^-_{w}$,
\end{enumerate}

where $\Gamma_w \coloneqq \bigcup_{s \in w} \Gamma_s$ and $\Delta_w \coloneqq \bigcup_{s \in w} \Delta_s$. For each $w \in W$ let $t_w$ be the unique node in $w$ which is labelled by the conclusion of a $\Cf$-instance. The base cases as well as the cases for $\varphi = \psi \ast \chi$ with $\ast \in \{\wedge, \vee, \rightarrow\}$ are identical to the corresponding cases in the proof of Proposition \ref{p: refutation gives triangle model}. \smallskip

\noindent \textsc{Case for $\lb$.} For (a) if $\lb\varphi\in \Gamma^-_w$, then $\lb \varphi \in \Gamma_{t_w}^-$. Suppose $w\leq v \mathrel{R} u$. Then, by definition of the $\mathsf{C_f}$-rule, $\varphi\in \Gamma^-_{u}$. The induction hypothesis then implies  $\fw,u\models \varphi$, so $\fw,w\models \lb\varphi$. For (b) if $\lb\varphi\in \Delta^-_w$, then $\lb \varphi \in \Delta_{t_w}^-$. As $\Gamma_{t_w}\Rightarrow \Delta_{t_w}$ is the conclusion of a $\mathsf{C_f}$-instance, it must be of the form $\Pi, \lb \Gamma \Rightarrow_k \{\varphi_i \rightarrow \psi_i^\u\}_{i=0}^l, \{\lb \chi_{j}^{a_j}\}_{j=0}^{r}, \lb \varphi^a, \Sigma$ with $e(\lb \varphi^a) = k'$ for some $k' \leq m$. Now, by construction of $\leq$ and the rule $\Cf$, it follows that there exists a $v\geq w$ such that  $\Gamma_v\Rightarrow \Delta_v$ is  equal to $\Pi, \lb \Gamma \Rightarrow_{k'} \{\varphi_i \rightarrow \psi_i^\u\}_{i=0}^l, \{\lb \chi_{j}^{a_j}\}_{j=0}^{r}, \lb \varphi^a, \Sigma$. So, by construction of $R$, there exists a $u$ with $v\mathrel{R} u$ and $\varphi \in \Delta^-_{u}$. The induction hypothesis then implies $\fw,u\not\models \varphi$, so $\fw,w\not\models \lb\varphi$. \smallskip

\noindent \textsc{Case for $\lbm$.} For (a) if $\lbm \varphi \in \Gamma_w^-$, then $\lbm \varphi \in \Gamma_{t_w}^-$. Suppose $w \mathrel{\Tilde{R}^*} v$. Let
\begin{equation*}
    w = u_0 \leq u_1 \mathrel R u_2 \leq ... \leq u_{n-1} \mathrel R u_n = v
\end{equation*}
be the $\Tilde{R}$-path from $w$ to $v$. We prove inductively that for each $i \leq n$ holds that $\varphi, \lb \lbm \varphi \in \Gamma^-_{u_i}$. For $u_0 = w$, saturation yields $\varphi, \lb \lbm \varphi \in \Gamma_w^-$. For $i >0$ we have by (the inner) induction hypothesis that $\varphi, \lb \lbm \varphi \in \Gamma_{u_{i-1}}^-$. If $u_{i-1} \Rel u_i$, then we have $\lbm \varphi \in \Gamma_{u_i}^-$, and so saturation yields $\varphi, \lb \lbm \varphi \in \Gamma_{u_i}^-$. Otherwise $u_{i-1} \leq u_i$. The definition of $\Cf$ and the fact that invertible rules are applied preservingly implies that $\varphi, \lb \lbm \varphi \in \Gamma_{u_i}^-$. Hence $\varphi \in \Gamma_v^-$, and so by induction hypothesis $\fw, v \models \varphi$. Hence $\fw, w \models \lbm \varphi$. \smallskip

For (b) if $\lbm \varphi \in \Delta_w^-$, then by saturation $\varphi \in \Delta_{t_w}^-$ or $\lb \lbm \varphi \in \Delta_{t_w}^-$. Suppose towards contradiction that for all $w \mathrel{\Tilde{R}^\ast} v$ holds that $\varphi \not \in \Delta_v^-$. This implies that $\varphi \not \in \Gamma_{t_w}^-$, hence $\lb \lbm \varphi \in \Gamma_{t_w}^-$. Then we can define an infinite path $\rho$ in $\mathcal{R}$ starting from $t_w$ as follows. At each $\Cf$-application, pick the modal premise in case $\lbm \varphi^\f$ is the formula in the consequent of its sequent and otherwise pick the right intuitionistic premise. Since $\lb \lbm \varphi \in \Delta_{t_w}^-$ and $e(\lb \lbm \varphi^a) =  k$ for some $k \leq m$, eventually, when following the path consisting of right intuitionistic premises starting in $t_w$, the index of the sequent $\Gamma_{t_w} \Rightarrow \Delta_{t_w}$ will become $k$, implying that eventually the modal premise of a $\Cf$-application will have the formula $\lbm \varphi^\f$ on the right side. On the path from the modal premise of one $\Cf$-instance to the conclusion of the next $\Cf$-instance, the formula $\lbm \varphi^\f$ must be principal in an instance of $\lbm \mathsf{R}$. Since $\varphi \not \in \Delta_v^-$ for any $w \mathrel{\Tilde{R}^*} v$, the right premise of this instance is chosen by Refuter, implying that $\lb \lbm \varphi^\f$ occurs on the right-hand side of the next saturated sequent and therefore the argument can be iterated. The generated branch $\rho$ has always a formula in focus (namely $\lbm \varphi$ or $\lb \lbm \varphi$) and passes infinitey often through $\lbm \mathsf{R}$, implying that it has a good suffix. This contradicts the assumption that $\mathcal{R}$ is a refutation. Thus there exists $w \mathrel{\Tilde{R}^*} v$ with $\varphi \in \Delta_v^-$. The induction hypothesis implies $\fw, v \not \models \varphi$, hence $\fw, w \not \models \lbm \varphi$.

Let $w \in W$ contain the root of $\mathcal{R}$. By definition $\Gamma_\sigma \subseteq \Gamma_w$ and $\Delta_\sigma \subseteq \Delta_w$. Thus  $\fw,w \not \models \sigma$.
\end{proof}

Combining these results, once again together with determinacy of the game, yields completeness of $\NWIM$ with respect to functional models. The proof of the following theorem is analogous to the proof of Theorem \ref{t: NIM triangle completeness}.

\begin{theorem}[Completeness of $\NWIM$ with respect to functional models]\label{t: NIM functional completeness}
   If a sequent $\sigma$ is valid over the class of functional models, then $\sigma$ has a $\NWIM$-proof.
\end{theorem}


\subsection{Translating Non-Wellfounded into Cyclic Proofs}\label{section completeness cyclic system}

With completeness of $\NWIM$ for triangle and functional models at hand, we now show that $\CIM$ is complete by translating non-wellfounded proofs into cyclic proofs. Due to the presence of the focus annotation, the translation is relatively straightforward.

\begin{lemma}\label{l: infinite branches contain good repetitions}
    Let $\rho$ be an infinite branch of an $\NWIM$-proof $\pi$. Then $\rho$ contains a successful repetition.
\end{lemma}
\begin{proof}
    Note that each rule of $\NWIM$ (and hence of $\CIM$) is \emph{analytic} in the following sense. If
    \begin{equation*}
        \infer[\mathsf{r}]{\sigma}{\sigma_1 & \ldots & \sigma_n}
    \end{equation*}
is a rule instance of $\mathsf{r} \in \NWIM$, then $\Cl(\sigma_i) \subseteq \Cl(\sigma)$ for all $1 \leq i \leq n$. This implies that every sequent occurring in $\rho$ consists of formulas of $\Cl(\sigma)$ where $\sigma$ is the sequent labelling the root of $\pi$. Since $\Cl(\sigma)$ is finite by Lemma \ref{l: closure finite}, there are only finitely many sequents occurring in $\rho$. Thus, there exists a sequent $\Gamma \Rightarrow \Delta$ which labels infinitely many nodes in $\rho$. Since $\pi$ is an $\NWIM$-proof, there exists a good suffix $\rho'$ of $\rho$, i.e. every sequent in $\rho'$ has a formula in focus and $\rho'$ passes infinitely often through $\lbm \mathsf{R}$ with the principal formula in focus. Let $i < \omega$ be the least natural number such that $\rho(i)$ belongs to $\rho'$ and is labelled by $\Gamma \Rightarrow \Delta$. Since there are infinitely many nodes labelled by $\Gamma \Rightarrow \Delta$ in $\rho'$ and $\rho'$ passes through infinitely many instances of $\lbm \mathsf{R}$ with the principal formula in focus, there exists a least natural number $i < j < \omega$ such that $\rho(j)$ is labelled by $\Gamma \Rightarrow \Delta$ as well and the path from $\rho(i)$ to $\rho(j)$ passes through an instance of $\lbm \mathsf{R}$ with the principal formula in focus. Therefore $(\rho(i), \rho(j))$ is a successful repetition occurring in the branch $\rho$.
\end{proof}

Note that the existence of a successful repetition in $\rho$ implies the existence of a \emph{lowermost} successful repetition in $\rho$.

\begin{definition}
    Let $\pi$ be an $\NWIM$-proof. The \emph{induced cyclic proof} $\pi_c$ is the subtree of $\pi$ obtained by pruning every infinite branch at the lowermost successful repetition.
\end{definition}

The following then follows immediately from Lemma \ref{l: infinite branches contain good repetitions}.

\begin{lemma}\label{l: non-wellfounded proof implies cyclic proof}
    If $\pi$ is an $\NWIM$-proof, then $\pi_c$ is a $\CIM$-proof.
\end{lemma}
\begin{proof}
    Since $\pi$ is an $\NWIM$-proof, every infinite branch contains a successful repetition by Lemma \ref{l: infinite branches contain good repetitions}. Therefore, by K\H{o}nig's Lemma, $\pi_c$ is finite and every branch $\rho$ of $\pi_c$ corresponds to an initial segment of a branch $\rho'$ of $\pi$. If $\rho'$ is a finite branch, then $\rho = \rho'$ and since $\pi$ is an $\NWIM$-proof, $\rho$ ends in an axiom. Otherwise $\rho'$ is an infinite branch and $\rho$ is the initial segment of $\rho'$ ending at the first successful repetition of $\rho'$. Thus every leaf of $\pi_c$ is an axiom or belongs to a successful repetition, implying that $\pi_c$ is a $\CIM$-proof.
\end{proof}

Completeness of $\CIM$ then follows as a corollary of Theorem \ref{t: NIM triangle completeness}, Theorem \ref{t: NIM functional completeness} and Lemma \ref{l: non-wellfounded proof implies cyclic proof}.

\begin{theorem}[Completeness of $\CIM$]\label{t: completeness CIM}
    If a sequent $\sigma$ is valid over the classes of dynamic models, of triangle models or of functional models, then $\sigma$ has a $\CIM$-proof.
\end{theorem}
\begin{proof}
    By Lemma \ref{l: IM subset of IMf and IMt} every formula valid over the class of dynamic models is also valid over the classes of functional and of triangle models. Thus suppose $\sigma$ is valid over the classes of triangle models or functional models. Theorem \ref{t: NIM triangle completeness} or Theorem \ref{t: NIM functional completeness} then imply that $\sigma$ has an $\NWIM$-proof $\pi$. Lemma \ref{l: non-wellfounded proof implies cyclic proof} implies that $\pi_c$ is a $\CIM$-proof of $\sigma$.
\end{proof}

Theorem \ref{t: CIM soundness} and Theorem \ref{t: completeness CIM} together provide an alternative proof of Theorem \ref{t: three classes one logic} that $\mathbf{IM}= \mathbf{IM_t} = \mathbf{IM_f}$.

\section{Conclusion}\label{c: IM, section conclusion IM}

We have introduced a simple intuitionistic dynamic logic called intuitionistic master modality. The following summarizes the results established in this chapter.

\begin{enumerate}
    \item We have presented three classes of models to evaluate formulas of $\IM$, namely dynamic models, triangle models and functional models.
    \item Theorem \ref{t: three classes one logic} shows that all three classes generate the same set of validities.
    \item We have presented a sound and complete axiomatization for $\IM$ and established several properties about derivations, such as the Deduction Theorem. The presented completeness proof closely follows the completeness proof for a multi--modal version of $\IM$ presented by Marti~\cite{marti_2017} (who uses a different axiomatization) and yields the finite model property for $\IM$ as well as decidability.
    \item We have developed  a sound and complete cyclic calculus for $\IM$. Completeness was established via a proof search argument, whose robustness was illustrated by giving an alternative proof of Theorem \ref{t: three classes one logic}.
\end{enumerate}

That the language of $\IM$ without the master modality cannot distinguish between dynamic and triangle models has already been observed by other authors, for example Litak and Visser~\cite{litak_2018}. Adding the master modality to the language does not impact this result, as in order to evaluate $\lb$ and $\lbm$, the truth conditions of dynamic models consider the same worlds as when one takes the closure under triangle confluence. The result that $\IM$ cannot distinguish between dynamic and functional models is new and perhaps more surprising. However, this is once again caused by the evaluation of the modalities, where in order to evaluate a $\lb$-formula at a world $w$, one has to consider all modal successors of all intuitionistic successors of $w$. This implies that it does not matter whether we consider models with one or multiple modal successors, as we can always add more intuitionistic successors to obtain additional modal successors. A consequence of Theorem \ref{t: three classes one logic} is that stronger confluence conditions are needed when we want to obtain an interesting intuitionistic linear temporal logic, namely conditions that allow us to evaluate modalities by only looking at the modal successor of the current world. This will be done in Chapter \ref{c: iLTL}, where we will study dynamic models where the modal accessibility relation is a total function $f$ which is order-preserving.

Our observations regarding the three classes of models naturally lead to the following question.

\begin{question}\label{q: IM}
    What logic is obtained when evaluating $\LIM$ over functional triangle models?
\end{question}

Over such models $(W, \leq, f, V)$ if $f$ is a total function, then note that for any world $w \in W$ and any $v \geq w$, $f(w) = f(v)$. In other words, the entire intuitionistic tree rooted at $w$ has the same modal successor. This implies that the modalities satisfy classical principles like the \emph{boxed law of excluded middle}: $\lb p \vee \neg \lb p$ is valid.\footnote{This requires $f$ to be a total function, as otherwise there could be maximal worlds in $\leq$ which have no successor.} A similar phenomenon will be encountered in Chapter \ref{c: ICK} when we combine triangle confluence with frame conditions.

The completeness proof for $\mathrm{IM_H}$ demonstrates a rather straightforward adaptation of the mathematical methods used to obtain completeness for classical modal logic with the master modality to the intuitionistic case. In fact, the presented canonical model construction combines the canonical model construction for intuitionistic logic (see~\cite{Bezhanishvili_?}) with the standard technique to handle the master modality from the classical realm (see e.g.~\cite[Chapter 7]{ditmarsch_2017}). Due to the lack of strong confluence properties, we can directly define finite canonical models for fragments of the language. As we will see in Chapter \ref{c: biLTL}, this is no longer possible when stronger confluence conditions are considered. 

The developed cyclic calculus $\CIM$ for $\IM$ is an extension of a multi-succedent calculus for $\mathsf{IPL}$ with rules for $\lb$ and $\lbm$. Many classical techniques, such as focus annotations and proof search games, are directly applicable to $\IM$ as demonstrated in this chapter. The main difference between the intuitionistic and the classical case is that the presence of the intuitionistic implication complicates the proof search argument, as we have to deal with falsifying $\rightarrow$-formulas in intuitionistic successors and $\lb$-formulas in modal successors, as opposed to only dealing with the latter in classical modal logic with the master modality (for an in depth study of cyclic proof systems for classical modal logic with the master modality, see the PhD thesis of Rooduijn~\cite{rooduijn_2024}; for proof search arguments for classical common knowledge logics, see~\cite{Marti_2018}). The increase in the mathematical difficulty is not substantial though, once again due to the lack of strong enough confluence conditions, which allows us to falsify $\rightarrow$- and $\lb$-formulas independently of one another. The system $\CIM$ is well-behaved: it is analytic and does not require a cut rule; furthermore it also does not require additional structure on sequents such as labels or nesting, and has a simple success condition on cyclic branches. Our work therefore lies the foundation for studying the proof theory of $\IM$ in more details. In particular, the following questions are left open.

\begin{question}
    Does $\IM$ have Craig/uniform/Lyndon interpolation?
\end{question}

Studer showed that classical common knowledge logics does not have Craig interpolation and hence also not uniform and Lyndon interpolation~\cite{studer_2009}. The provided counterexample however requires the language to feature at least two agents, which implies that there is hope for obtaining Craig interpolation for $\IM$. The cyclic system $\CIM$ is thereby a good starting point for studying this problem from a proof theoretic point of view. Another open proof theoretic question concerns the problem of cut elimination.

\begin{question}
    Do $\CIM$ and $\NWIM$ admit cut elimination?
\end{question}

In recent work, Sierra-Miranda and Studer showed cut elimination for a non-wellfounded proof system for classical modal logic with the master modality~\cite{miranda_2025}. The similarities between their system and $\NWIM$ suggest that their proof may be adaptable to the intuitionistic case.

\chapter{Intuitionistic Common Knowledge Logic}\label{c: ICK}

\section{Introduction}\label{c: ICK, Section Introduction}

Epistemic logic describes a family of logical systems used to reason about knowledge and belief of agents as well as studying various notions of group knowledge. Traditionally, epistemic logics are devised as (classical) modal logics. Given a finite set of agents $\A$, the language of epistemic logic extends propositional logic with modal operators $\K_i$ for $i \in \A$ which are interpreted as the knowledge operator of agent $i$. Thus a formula of the form $\K_i \varphi$ is read as `agent $i$ knows that $\varphi$'. Of particular interest is the study of nesting of knowledge among different agents, i.e. what agents know about the knowledge of other agents. For example, the formula $\K_i \K_j \neg \K_i \neg \varphi$ expresses that agent $i$ knows, that agents $j$ knows, that agent $i$ considers $\varphi$ to be possible. Such kind of reasoning naturally leads to the study of notions of group knowledge. Perhaps the most famous example of such a notion is \emph{common knowledge}. Informally, among a group of agents a formula $\varphi$ is common knowledge if every agent knows $\varphi$ and every agent knows, that every agent knows $\varphi$, and so on. Write $\Ek \varphi$ for `everybody (in the group) knows $\varphi$'. Formally, $\Ek \varphi$ is defined as
\begin{equation*}
    \Ek \varphi \coloneqq \bigwedge_{i \in \A} \K_i \varphi.
\end{equation*}
As usual, let $\Ek^0 \varphi \coloneqq \varphi$ and $\Ek^{n+1} \varphi \coloneqq \Ek \Ek^n \varphi$. Then `common knowledge of $\varphi$' is equivalent to the infinite conjunction
\begin{equation*}
    \bigwedge_{k < \omega} E^k \varphi.
\end{equation*}
In order to express common knowledge, we use a new operator $\C$ where $\C \varphi$ is given as the greatest fixed point of the propositional function $x \mapsto \varphi \wedge \Ek x$. Epistemic logic with common knowledge is called \emph{common knowledge logic} ($\mathsf{CK}$) and has found applications in computer science, such as in the study of distributed systems (see e.g.~\cite{Halpern_1990}).

Epistemic logic over an intuitionistic base logic, called \emph{intuitionistic epistemic logic} has been studied at least since the 1990's and the work by Williamson~\cite{Williamson_1992}. In the last decade, several systems of intuitionistic epistemic logic have been developed. These systems fall into two general categories. The first category aims to develop systems of intuitionistic epistemic logic to study knowledge from the point of view of intuitionism. Such systems usually attempt to provide a BHK interpretation of knowledge. In this category falls the work of Williamson~\cite{Williamson_1992}, the intuitionistic epistemic logic IEL developed by Artemov and Protopopescu~\cite{Artemov_2016}, as well as recent work about logics of knowing-wh (which are epistemic logics featuring operators to capture different types of knowledge such as knowing-how) and explicit interpretations of intuitionistic logic by Wang et al. (see e.g.~\cite{wang_2013, Wang_2017, Wang_2018, Wang_2021, wang_2022}). The second category is primarily motivated by mathematical reasons as well as applications to computer science and studies intuitionistic modal logics, where the $\lb$-operator is interpreted as a knowledge operator and which are extended by fixed point operators for group knowledge such as common knowledge or distributed knowledge. In particular, logics from this category do not attempt to give a BHK reading of knowledge and largely treat knowledge classically. To this category belongs most notably the intuitionistic common knowledge logic $\ICK$, developed by Jäger and Marti~\cite{jager-intuitionistic_2016}. Jäger and Marti focus on the mathematical theory of $\ICK$ by developing a sound and complete axiomatization and a sound and complete sequent calculus with induction rule. Moreover, they also study intuitionistic epistemic logic with distributed knowledge~\cite{jager-distributed_2016}, which is further studied in~\cite{murai_2022, murai_2024}. $\ICK$ is a multi-modal version of the logic $\IM$ studied in the previous chapter. Given a finite set of agent names $\mathsf{A}$, the language features a $\lb$-operator for each agent $i \in \mathsf{A}$, written as $\K_i$. In that setting the master modality becomes more expressive, as it is no longer simply a reflexive transitive closure operator of $\lb$, but instead of $\Ek$. For example, it is easily verified that over triangle models where the modal accessibility relation is reflexive and transitive, the master modality is definable in terms of $\lb$. This is not the case in the multi-modal setting of $\ICK$ with common knowledge. Formulas of $\ICK$ are evaluated on triangle models, satisfying standard frame conditions such as reflexive triangle models, S4 triangle models and S5 triangle models. Jäger and Marti provide a sequent calculus for $\ICK$ over the class of triangle and over the class of reflexive triangle models, but do not extend their work to the S4 and S5 case~\cite{jager-intuitionistic_2016}. Furthermore, Marti also provides an axiomatization~\cite{Marti_2018}. The presented sequent calculus characterizes common knowledge via an induction rule, and features a cut rule which is used in the completeness proof. Whether cut elimination is possible is not discussed, however in a similar calculus for \emph{classical} common knowledge logic presented by Alberucci and Jäger~\cite{alberucci_2005}, the cut rule cannot be eliminated (at least for the logic over S5 models) and no restriction to analytic cuts is possible, which sheds doubt on whether such an enterprise is possible for $\ICK$. 

In this chapter we aim to use the cyclic calculus developed in Chapter \ref{c: IM} to build a uniform proof theory for $\ICK$ evaluated over the classes of triangle models, reflexive triangle models, S4 triangle models and S5 triangle models. To that end we will first show how to translate $\CIM$ into a multi-modal version and then introduce rules for the frame conditions. As it turns out, the adaptation of the calculus to account for reflexive and S4 models is straightforward by adding the standard modal rules for reflexivity and transitivity to the calculus. Moreover, completeness can be established by following the proof search argument introduced in the previous chapter with only minor changes to the `choice rule'. The case for S5 is more difficult. First, the combination of triangle confluence with S5 frame conditions results in rather strange models where given a world $w$, the entire intuitionistic tree rooted at $w$ belongs to the equivalence class of $w$ under $R$. In such a setting knowledge essentially reduces to classical knowledge, which satisfies classical principles such as the `boxed law of excluded middle': for every proposition $p$, the formula $\K_i p \vee \neg \K_i p$ is valid. Second, the S5 conditions lead to the standard proof theoretical problems already encountered for classical S5 modal logic: we will show that the extension of $\CIM$ with rules for the frame conditions is not cut-free complete. This could perhaps be circumvented by passing to labelled or nested sequents, but we will follow the method presented in a joint publication with Rooduijn~\cite{rooduijn_analytic_2022} on classical common knowledge logic over S5 and instead extend the calculus with a cut rule. Analytic completeness is then established via a canonical model construction by restricting applications of the cut rule to \emph{analytic cuts}. Importantly, the extension with the cut rule allows us to prove completeness directly; a detour to a non-wellfounded system as for $\CIM$ is no longer required. It is in fact unclear whether the proof search argument can be adapted in the presence of the cut rule. Furthermore, we investigate the relationship between $\ICK$ over S5 and $\mathsf{CK}$ over S5 by developing a translation from $\mathsf{CK}$ into $\mathsf{ICK}$ and we show that proof search in the cyclic calculus for the S5 case can be automated by reducing the problem of finding proofs to the problem of solving a certain parity game. Finally, combining the translation and parity game, we show that the problem of finding proofs in the cyclic calculus for $\ICK$ over S5 triangle models is \textsc{ExpTime}-complete.

\section{Syntax and Semantics}\label{c: ICK, Section syntax}\index{logic!intuitionistic common knowledge}

Let $\A$ be a finite set of \emph{agent names}. The \emph{language} $\LICK$ extends $\LIPL$ by modal operators $\K_i$ for $i \in \A$ and the fixed point operator $\C$\index{fixed point operator!common knowledge}, where $\C \varphi$ is characterized as the greatest fixed point of the propositional function $x \mapsto \varphi \wedge \Ek x$. \emph{Formulas} of $\mathcal{L}_\mathrm{ICK}$ are defined by the following grammar in Backus-Naur form:
\begin{equation*}
    \varphi ::= \bot \, \lvert \, p \, \lvert \, \varphi \wedge \varphi \, \lvert \, \varphi \vee \varphi \, \lvert \, \varphi \rightarrow \varphi \, \lvert \, \K_i \varphi \, \lvert \, \C \varphi.
\end{equation*}
where $p \in \Prop$ and $i \in \A$. The modal operator $K_i$ is the \emph{knowledge operator} of agent $i$: $\K_i \varphi$ is read as `agent i knows $\varphi$'. The modal operator $\C$ is the \emph{common knowledge operator}: $\C \varphi$ is read as `$\varphi$ is common knowledge (among the agents in $\A$)'. The modality $\Ek$, read as `everybody knows', is defined by
\begin{equation*}
        \Ek \varphi  \coloneqq \bigwedge_{i \in \A} K_i \varphi.
\end{equation*}
Due to the presence of multiple knowledge operators, the definition of the closure of a formula (c.f. Definition \ref{d: closure IM formula}) is adjusted as follows.

\begin{definition}
         The \emph{closure}\index{closure!for $\LICK$} $\Cl(\varphi)$ of a formula $\varphi$ is defined by induction on $\varphi$ as follows.
     \begin{itemize}
        \item $\Cl(\bot) = \{\bot\}$
        \item $\Cl(p) = \{ p \}$ for $p \in \Prop$
        \item $\Cl(\varphi \ast \psi) = \Cl(\varphi) \cup \Cl(\psi) \cup \{ \varphi \ast \psi\}$ for $\ast \in \{\wedge, \vee, \rightarrow\}$
        \item $\Cl(\K_i \varphi) = \Cl(\varphi) \cup \{\K_i \varphi\}$
        \item $\Cl(\C \varphi) = \Cl(\varphi) \cup \{\C \varphi\} \cup \{ \K_i \C \varphi \mid i \in \A\}$
    \end{itemize}
    Given a set $\Gamma$ of formulas, the \emph{closure} $\Cl(\Gamma)$ of $\Gamma$ is defined by
    \begin{equation*}
        \Cl(\Gamma):= \bigcup_{\varphi\in\Gamma}\Cl(\varphi).
    \end{equation*}
    A set of formulas $\Gamma$ is \emph{closed} if $\Gamma = \Cl(\Gamma)$.
\end{definition}

  Additionally, we define the \emph{negation closure} of a formula, which will be used to treat $\ICK$ over S5 models.

\begin{definition}
    The \emph{negation closure}\index{closure!negation} $\Cl^\neg(\varphi)$ of a formula $\varphi$ is the least closed set of formulas containing $\varphi$, such that whenever $\K_i \psi \in \Cl^\neg(\varphi)$, then $\neg \K_i \psi \in \Cl^\neg(\varphi)$. The negation closure $\Cl^\neg(\Gamma)$ of a set of formulas $\Gamma$ is given by 
    \begin{equation*}
        \Cl^\neg(\Gamma) \coloneqq \bigcup_{\varphi \in \Gamma} \Cl^\neg (\varphi).
    \end{equation*} 
    A set $\Gamma$ is \emph{negation closed} if $\Gamma = \Cl^\neg(\Gamma)$.
\end{definition}

 The following lemma is easily verified by an induction on the structure of formulas.

\begin{lemma}\label{l: negation closure is finite}
    For any formula $\varphi$, $\Cl (\varphi)$ and $\Cl^\neg(\varphi)$ are finite. Moreover, if $\Gamma$ is a finite set of formulas, then $\Cl(\Gamma)$ and $\Cl^\neg(\Gamma)$ are finite.
\end{lemma}

The \emph{complexity} of a formula is defined as follows.

\begin{definition}\label{d: complexity of ICK formula}
    The \emph{complexity}\index{complexity!of $\LICK$-formula} $c(\varphi)$ of a formula $\varphi$ is defined inductively as follows.
    \begin{itemize}
        \item $c(\bot) = c(p) = 0$
        \item $c(\varphi \ast \psi) = c(\varphi) + c(\psi) +1$ for $\ast \in \{\wedge, \vee, \rightarrow\}$
        \item $c(\K_i \varphi) = c(\C \varphi) = c(\varphi) + 1$
    \end{itemize}
    Given a finite set of formulas $\Gamma$, the \emph{complexity} of $\Gamma$ is defined as 
    \begin{equation*}
       c(\Gamma) = \sum_{\varphi \in \Gamma}c(\varphi).
    \end{equation*}
\end{definition}

Formulas of $\mathcal{L}_\mathrm{ICK}$ are evaluated over triangle models equipped with a modal accessibility relations for each agent. We follow~\cite{jager-intuitionistic_2016} and call such models \emph{(intuitionistic) epistemic models}.

\begin{definition}
    An \emph{(intuitionistic) epistemic model}\index{model!epistemic} is a tuple $\fw=(W, \leq, \{R_i\}_{i \in \A}, V)$ such that $(W, \leq, R_i, V)$ is a triangle model for each $i \in \A$.
\end{definition}

 Given an epistemic model $\fw=(W, \leq, \{R_i\}_{i \in \A}, V)$, define the relation
\begin{equation*}
    R := \bigcup_{i \in A} R_i
\end{equation*}
and let $R^*$ be the reflexive and transitive closure of $R$.

\begin{definition}
    The \emph{truth relation} $\models_t$ between worlds $w$ of an epistemic model $\fw$ and formulas $\varphi$ extends Definition \ref{d: truth relation for intuitionistic Kripke models} by the following clauses.
\begin{center}
    \begin{tabular}{l l l}
    $M,w \models_t \K_i \varphi$ & iff & for all $v \in W$ if $w \mathrel{R_i} v$, then $M,v \models \varphi$ \\
    $M, w \models_t \C \varphi$ & iff & for all $v \in W$ if $w \mathrel{R^*} v$, then $M,v \models \varphi$
    \end{tabular}
\end{center}
\end{definition}

Since only triangle models are studied in this chapter, we will supress the index $t$ in $\models_t$ and simply write $\models$. The notions of \emph{satisfiability} and \emph{validity} are defined as before. Since $\mathsf{ICK}$ is simply a multi-modal version of $\mathsf{IM}$ it is obvious that the monotonicity lemma (c.f. Lemma \ref{l: monotonicity intuitionistic Kripke models}) still holds.

We will study $\LICK$ over the classes of epistemic models, reflexive epistemic models, S4 epistemic models and S5 epistemic models. The definition of these classes of models is a generalization of Definition \ref{d: frame conditions}.

\begin{definition}
    An epistemic model $\fw =(W, \leq, \{R_i\}_{i \in \A}, V)$ is called a
    \begin{enumerate}
        \item \emph{reflexive epistemic model} if for each $i \in \A$, $R_i$ is reflexive;
        \item \emph{S4 epistemic model} if for each $i \in \A$, $R_i$ is reflexive and transitive;
        \item \emph{S5 epistemic model} if for each $i \in \A$, $R_i$ is reflexive, transitive and symmetric.
    \end{enumerate}
\end{definition}

For simplicity we will usually refer to reflexive epistemic models simply as reflexive models throughout this chapter, and similarly for S4 and S5 epistemic models.

\begin{definition}
    Denote by $\mathbf{ICK}$ the set of valid formulas over the class of epistemic models. Furthermore, denote by $\mathbf{ICK_\ast}$ for $\mathbf{\ast} \in \{\mathbf{T}, \mathbf{S4}, \mathbf{S5}\}$ the set of valid formulas over the classes of reflexive, S4 and S5 models, respectively.
\end{definition}

Note that $\mathbf{ICK} \subseteq \mathbf{ICK_T} \subseteq \mathbf{ICK_{S4}} \subseteq \mathbf{ICK_{S5}}$. The following lemma shows that the logics $\mathbf{ICK}$, $\mathbf{ICK_T}$ and $\mathbf{ICK_{S4}}$ contain the standard validities of classical epistemic logic regarding the knowledge and common knowledge operators (see e.g. \cite{ditmarsch_2017} for axiomatizations of classical common knowledge logics over different frame conditions). The proof is by standard semantical arguments using the frame conditions that the models of the relevant logic satisfy and is omitted.

\begin{lemma}
    Let $\varphi, \psi$ be formulas and $i \in \A$. Then
    \begin{enumerate}
        \item $\K_i (\varphi \rightarrow \psi) \rightarrow (\K_i \varphi \rightarrow \K_i \psi) \in \mathbf{ICK}$;
        \item $\C (\varphi \rightarrow \psi) \rightarrow (\C \varphi \rightarrow \C \psi) \in \mathbf{ICK}$;
        \item $\K_i \varphi \rightarrow \varphi \in \mathbf{ICK_T}$;
        \item $\K_i \varphi \rightarrow \K_i \K_i \varphi \in \mathbf{ICK_{S4}}$.
    \end{enumerate}
\end{lemma}

The logic $\mathbf{ICK_{S5}}$, on the other hand, contains some validities that are rather unexpected in the context of intuitionistic logic. The combination of S5 frame conditions with triangle confluence has a strong impact on the models and thus on the resulting logic. Given worlds $w$ and $v$ if $w \leq v$, then by reflexivity $v \mathrel{R_i} v$ and so by triangle confluence $w \mathrel{R_i} v$ for all $i \in \A$. In other words, the intuitionistic order is a subset of each modal accessibility relation: ${\leq} \subseteq {R_i}$ for each $i \in \A$ (note that this already holds for S4 models). Moreover, by symmetry, we obtain that $v \mathrel{R_i} w$, implying that given $w$, the entire intuitionistic tree rooted at $w$ belongs to the equivalence class of $w$ under the equivalence relation induced by $R_i$.

This leads to an interesting observation. Since every intuitionistic successor $v$ of $w$ belongs to the equivalence class of $w$, every world accessible from $v$ is accessible from $w$ and vice versa, implying that for $w \leq v$, $w \models \K_i \varphi$ if and only $v \models \K_i \varphi$. Moreover, since truth is persistent in $\leq$ and falsity is persistent in $\geq$, implications with a `boxed' antecedent (i.e. formulas of the form $\K_i \varphi \rightarrow \psi$ or $\C \varphi \rightarrow \psi$) are evaluated classically, as shown in the following lemma.

\begin{lemma}\label{l: classical truth conditions for implication}
    Let $\fw$ be an S5 model, $w$ a world, $i \in \A$ and $\varphi, \psi \in \LICK$. Then the following holds.
    \begin{center}
        $\fw, w \models \K_i \varphi \rightarrow \psi$ if and only if $\fw, w \not \models \K_i \varphi$ or $\fw, w \models \psi$.
    \end{center}
\end{lemma}
\begin{proof}
    For the left--to--right direction suppose $\fw, w \models \K_i \varphi \rightarrow \psi$. Then for all $w \leq v$ if $\fw, v \models \K_i \varphi$, then $\fw, v \models \psi$. Thus, for $v=w$, $\fw, w \not \models \K_i \varphi$ or $\fw, w \models \psi$. For the right--to--left direction we proceed by contraposition. Suppose $\fw, w \not \models \K_i \varphi \rightarrow \psi$. Then there exists $w \leq v$ such that $\fw, v \models \K_i \varphi$ and $\fw, v \not \models \psi$. Since $w \leq v$, by the Monotonicity Lemma $\fw, w \not \models \psi$. Now suppose $w \Rel_i u$. Since $w \leq v $ and $v \Rel_i v$ by reflexivity of $R_i$, triangle confluence implies that $w \Rel_i v$. By symmetry of $R_i$ we $v \Rel_i w$ and so by transitivity of $R_i$, $v \Rel_i u$. Thus $\fw, u \models \varphi$, implying that $\fw, w \models \K_i \varphi$. 
\end{proof}

As a corollary we obtain that $\mathsf{ICK}$ over S5 models satisfies a boxed version of the law of excluded middle: for any formula $\varphi$ the formula $\K_i \varphi \vee \neg \K_i \varphi$ is valid. The proof is omitted.

\begin{corollary}\label{l: properties of K-implications}
    Let $\varphi$ be a formula and $i \in \A$. Then
    \begin{enumerate}
        \item $ \K_i \varphi \vee \neg \K_i \varphi \in \mathbf{ICK_{S5}}$;
        \item $\C \varphi \vee \neg \C \varphi \in \mathbf{ICK_{S5}}$;
        \item $ \neg \K_i \varphi \rightarrow \K_i \neg \K_i \varphi \in \mathbf{ICK_{S5}}$.
    \end{enumerate}
\end{corollary}


\section{Cyclic Calculi}\label{c: ICK, Section cyclic calculi ICK}

This section introduces four sequent calculi for $\ICK$ with the aim to capture the validities of $\ICK$ over the classes of all epistemic models, reflexive models, S4 models and S5 models. Each calculus shares the same set of basic rules as $\CIM$ and additional rules for the modalities, which differ depending on which frame condition ought to be captured. The calculus for $\ICK$ over the class of epistemic models is denoted $\CICK$ and is simply a multi-modal version of $\CIM$. The calculi $\CICK_{\mathrm{T}}$, $\CICK_{\mathrm{S4}}$ and $\CICK_{\mathrm{S5}}$ are modular extensions of $\CICK$. The focus lies mostly on $\CICK_{\mathrm{S5}}$, which is the extension of $\CICK$ by two rules for the modalities and, additionally, the cut rule and a special rule dealing with implications $\varphi \rightarrow \psi$ where $\varphi$ is a boxed formula. We begin by introducing the basic components of the calculi. Since $\CICK$ and its extensions are multi-modal versions of $\CIM$, most definitions are identical or similar to the case of $\IM$; for convenience we briefly repeat the basic definitions here and refer the reader to Chapter \ref{c: IM}, Section \ref{c: IM, section non-wellfounded and cyclic calculus for IM} for further details. 

An \emph{annotated formula} is a tuple $(\varphi, a)$, written $\varphi^a$, where $\varphi \in \LICK$ and $a \in \{\u, \f\}$. The annotation $\u$ designates that the formula is \emph{unfocused} and $\f$ that the formula is \emph{in focus}.

\begin{definition}
    A \emph{sequent}\index{sequent!for $\ICK$} is an ordered pair of finite sets of annotated formulas $\Gamma \Rightarrow \Delta$ such that
    \begin{enumerate}
    \item Every formula in $\Gamma$ is unfocused.
    \item At most one formula in $\Delta$ is in focus.
    \item If a formula $\varphi$ is in focus, then $\varphi = \C \psi$ or $\varphi = \K_i \C \psi$ for some formula $\psi$ and $i \in \A$.
\end{enumerate}
\end{definition}

We denote sequents by $\sigma$ and write $\Gamma_\sigma$ and $\Delta_\sigma$ for the left and right side of $\sigma$, respectively. The \emph{interpretaion} of a sequent $\sigma$ is the formula $\sigma^I \coloneqq \bigwedge \Gamma_\sigma^- \rightarrow \bigvee \Delta_\sigma^-$. The \emph{closure} $\Cl(\sigma)$ of a sequent $\sigma$ is defined as $\Cl(\sigma) \coloneqq \Cl(\Gamma_\sigma) \cup \Cl(\Delta_\sigma)$. Additionally, the \emph{negation closure} of a sequent $\sigma$ is defined as $\Cl^\neg (\sigma) \coloneqq \Cl^\neg(\Gamma_\sigma) \cup \Cl^\neg (\Delta_\sigma)$. Each calculus consists of all of the basic rules depicted in Table \ref{d: basic rules} and additionally some of the rules depicted in Table \ref{d: additional rules for ICK}.

\begin{definition}
    Consider the rules depicted in Table \ref{d: basic rules} and Table \ref{d: additional rules for ICK}. 
    \begin{enumerate}\index{sequent calculus!$\CICK$ and extensions}
        \item The calculus $\CICK$ consists of the basic rules from Table \ref{d: basic rules} as well as the rules $\mathsf{K_i}$ for $i \in \A$, $\C\mathsf{L}$ and $\C \mathsf{R}$.
        \item The calculus $\CICK_\mathrm{T}$ is the extension of $\CICK$ with the rules $\mathsf{T_i}$ for $i \in \A$.
        \item The calculus $\CICK_\mathrm{S4}$ is the calculus  $\CICK_\mathrm{T}$ with the rules $\mathsf{K_i}$ replaced by the rules $\mathsf{S4_i}$ for $i \in \A$.
        \item The calculus $\CICK_\mathrm{S5}$ is the calculus $\CICK_\mathrm{S4}$ with the rules $\mathsf{S4_i}$ replaced by the rules $\mathsf{S5_i}$ and extended by the rules $\mathsf{K_i} {\rightarrow}$ and $\mathsf{cut}$ for $i \in \A$.
    \end{enumerate}
\end{definition}

\begin{table}[t]
    \centering
        \begin{tabular}{|c c|}
        \hline
        & \\
         $\infer[\mathsf{id}]{\Gamma, \varphi^\u \Rightarrow \varphi^a, \Delta}{}$ 
         & $\infer[\bot]{\Gamma, \bot^\u \Rightarrow \Delta}{}$\\
         & \\
         $\infer[\wedge \mathsf{L}]{\Gamma, \varphi \wedge \psi^\u \Rightarrow \Delta}{\Gamma, \varphi^\u, \psi^\u \Rightarrow \Delta}$ 
         & $\infer[\wedge \mathsf{R}]{\Gamma\Rightarrow \varphi \wedge \psi^\u ,\Delta}{\Gamma \Rightarrow \varphi^\u, \Delta & \Gamma \Rightarrow \psi^\u, \Delta}$\\
         & \\
         $\infer[\vee \mathsf{L}]{\Gamma, \varphi \vee \psi^\u\Rightarrow \Delta}{\Gamma, \varphi^\u \Rightarrow \Delta & \Gamma, \psi^\u \Rightarrow \Delta}$ & $\infer[\vee \mathsf{R}]{\Gamma \Rightarrow \varphi \vee \psi^\u, \Delta}{\Gamma \Rightarrow \varphi^\u, \psi^\u, \Delta}$\\
         & \\
         $\infer[{\to} \mathsf{L}]{\Gamma, \varphi \rightarrow \psi^\u\Rightarrow \Delta}{\Gamma, \varphi \rightarrow \psi^\u \Rightarrow \varphi^\u, \Delta & \Gamma, \psi^\u \Rightarrow \Delta}$ 
         & $\infer[{\to} \mathsf{R}]{\Gamma \Rightarrow \varphi \rightarrow \psi^\u, \Delta}{\Gamma, \varphi^\u \Rightarrow \psi^\u}$\\
         & \\
         $\infer[\mathsf{u}]{\Gamma \Rightarrow \varphi^\f, \Delta}{\Gamma \Rightarrow \varphi^\u, \Delta}$ & $\infer[\mathsf{f}]{\Gamma \Rightarrow \varphi^\u, \Delta}{\Gamma \Rightarrow \varphi^\f, \Delta}$\\
         & \\
         \hline
        \end{tabular}
    \caption{The basic rules. The symbols $\Gamma$ and $\Delta$ range over finite sets of annotated formulas which may be empty.}
    \label{d: basic rules}
\end{table}

\begin{table}[t]
    \centering
        \begin{tabular}{|c c|}
        \hline
        & \\
        $\infer[\mathsf{K_i}]{\Pi, \K_i \Gamma \Rightarrow \K_i \varphi^a, \Sigma}{\Gamma \Rightarrow \varphi^a}$ & $\infer[\mathsf{cut}]{\Gamma \Rightarrow \Delta}{\Gamma, \varphi^\u \Rightarrow \Delta & \Gamma \Rightarrow \varphi^\u, \Delta}$\\
         & \\
        $\infer[\mathsf{T_i}]{\Gamma, \K_i \varphi^\u \Rightarrow \Delta}{\Gamma, \varphi^\u \Rightarrow \Delta}$ &  $\infer[\mathsf{S4_i}]{\Pi, \K_i \Gamma \Rightarrow \K_i \varphi^a, \Sigma}{ \K_i \Gamma \Rightarrow \varphi^a}$\\
        & \\
        $\infer[\mathsf{S5_i}]{\Pi, \K_i \Gamma \Rightarrow \K_i \varphi^a, \K_i \Delta, \Sigma}{ \K_i \Gamma \Rightarrow \varphi^a, \K_i \Delta}$ &  $\infer[\mathsf{K_i}{\rightarrow}]{\Gamma \Rightarrow \K_i \varphi \rightarrow \psi^\u, \Delta}{\Gamma, \K_i \varphi^\u \Rightarrow \psi^\u, \Delta}$\\
        & \\
        $\infer[\C \mathsf{L}]{\Gamma, \C \varphi^\u \Rightarrow \Delta}{\Gamma, \varphi^\u, \{\K_i \C \varphi^\u\}_{i \in \A} \Rightarrow \Delta}$ & $\infer[\C \mathsf{R}]{\Gamma \Rightarrow \C \varphi^a,\Delta}{\Gamma \Rightarrow \varphi^\u, \Delta & \{\Gamma \Rightarrow \K_i \C \varphi^a, \Delta\}_{i \in \A}}$\\
        & \\
        \hline
        \end{tabular}
    \caption{The additional rules. The symbols $\Gamma, \Delta, \Sigma$ and $\Pi$ range over finite sets of annotated formulas which may be empty.}
    \label{d: additional rules for ICK}
\end{table}

For every rule in Table \ref{d: basic rules} as well the rules $\mathsf{CL}$, $\mathsf{CR}$ and $\mathsf{K_i}{\rightarrow}$, the distinguished formula in the conclusion is called \emph{principal} and the distinguished formulas in the premises are its \emph{residuals}. For example, in the rule $\mathsf{\C L}$ the principal formula is $\C \varphi^\u$ and the residual formulas are $\varphi^\u$ and $\K_i \C \varphi^\u$ for $i \in \A$. The rule $\mathsf{cut}$ has no principal formulas, while both distinguished formulas in the premises are residual. These distinguished formulas are called the \emph{cut formulas}. For the knowledge rules for agent $i$ (i.e. $\mathsf{K_i}$, $\mathsf{T_i}$, $\mathsf{S4_i}$ and $\mathsf{S5_i}$) every formula in the conclusion is principal and every formula in the premise is the residual of the corresponding principal formula. Formulas in $\Sigma, \Pi$ have no residual formulas. All other formulas occurring in a rule are called \emph{side formulas}.

\subsection{Cyclic Proofs}

We define the notion of cyclic proof for each system introduced above. Once again, the definitions simply adapt the definitions for $\IM$ to the case for $\ICK$. 

Let $\mathsf{P} \in \{\CICK, \CICK_\mathrm{T}, \CICK_\mathrm{S4}, \CICK_\mathrm{S5}\}$. A $\mathsf{P}$\emph{-pre-proof} of a sequent $\sigma$ is a finite tree $\pi$ labeled by sequents according to the rules of $\mathsf{P}$ and whose root is labeled by $\sigma$.

\begin{definition}
    A path $\rho$ through a pre-proof $\pi$ is \emph{successful} if the following hold.
    \begin{enumerate}
        \item Every sequent in $\rho$ has a formula in focus.
        \item The path $\rho$ passes through an instance of the rule $\C \mathsf{R}$ where the principal formula is in focus.
    \end{enumerate}
\end{definition}

Given a pre-proof $\pi$, a pair of nodes $(u,v)$ of $\pi$ is a \emph{repetition} if $u \not = v$, there exists a path from $u$ to $v$ and both nodes are labeled by the same sequent. A repetition $(u,v)$ is \emph{successful} if the path from $u$ to $v$ is successful.

\begin{definition}
    A \emph{cyclic proof}\index{proof!$\CICK$-proof} of a sequent $\sigma$ in $\mathsf{P}$ is a $\mathsf{P}$-pre-proof $\pi$ of $\sigma$ such that every leaf $v$ of $\pi$ is labeled by an axiom or there exists a node $u$ in $\pi$ such that $(u,v)$ form a successful repetition. If a sequent $\sigma$ has a proof in $\mathsf{P}$, then we say that $\sigma$ is $\mathsf{P}$-\emph{provable}.
\end{definition}


    

An example of a cyclic proof showing that $\C \varphi^\u \Rightarrow \C (\varphi \vee \psi)^\u$ is $\CICK$-provable is displayed in Figure \ref{f: example ICKS4 proof}, where it is assumed that $\A =\{1,2\}$. By inspecting the proof it is obvious that the example can be generalized to $n$ agents for any natural number $n>0$. Note that the presented calculi combine standard rules for $\mathsf{IPL}$ and for common knowledge logic (see for example \cite{rooduijn_analytic_2022, Negri_2001}). The only exception is the calculus $\CICK_{\mathrm{S5}}$ which features the rule ${\rightarrow}\mathsf{K_i}$. This rule formalizes the observation from Lemma \ref{l: classical truth conditions for implication} that implications of the form $\K_i \varphi \rightarrow \psi$ behave classically. Observe that the rule is only applicable if the principal formula is of the form $\K_i \varphi \rightarrow \psi$. It is therefore a special case of the classical right implication rule. We now give two examples showing that  $\CICK_{\mathrm{S5}}$ proves the boxed law of excluded middle for $\K_i$ and $\C$, which illustrate how the rule ${\rightarrow}\mathsf{K_i}$ can be used to prove classical principles.

\begin{example}
    The sequent $\Rightarrow \K_i p \vee \neg \K_i p^\u$ is provable in $\CICK_\mathsf{S5}$. Recall that $\neg \varphi$ is a shortcut for $\varphi \rightarrow \bot$.
    \begin{prooftree}
        \AxiomC{}
        \RightLabel{$\mathsf{id}$}
        \UnaryInfC{$\K_i p^\u \Rightarrow \K_i p^\u, \bot^\u$}
        \RightLabel{${\rightarrow}\mathsf{K_i}$}
        \UnaryInfC{$\Rightarrow \K_i p^\u, \K_i p \rightarrow \bot^\u$}
        \RightLabel{$\vee \mathsf{R}$}
        \UnaryInfC{$\Rightarrow \K_ip \vee (\K_ip \rightarrow \bot)^\u$}
    \end{prooftree}
    Note that with the standard intuitionistic right implication rule, this proof is not possible: if ${\rightarrow}\mathsf{R}$ is applied (bottom-up) to the sequent $\Rightarrow \K_i p^\u, \K_i p \rightarrow \bot^\u$, the premise is $\K_i p^\u \Rightarrow \bot^\u$, which is not provable.
\end{example}


\begin{example}\label{e: law of excluded middle for C}
   The following is a proof of  $\Rightarrow \C p  \vee \neg \C p^\u$. For simplicity we assume that there is only one agent in the language, whose knowledge operator is $\K_i$. The proof is generalized in a straightforward way for multiple agents. Let $\pi$ be the following proof:
 
    \begin{prooftree}
    \small
        \AxiomC{}
        \RightLabel{$\mathsf{id}$}
        \UnaryInfC{$\neg \K_i \C p^\u, p^\u, \K_i \C p^\u \Rightarrow \K_i \C p^\u, \bot^\u$}
        \RightLabel{$\C \mathsf{L}$}
        \UnaryInfC{$\neg \K_i \C p^\u, \C p^\u \Rightarrow \K_i \C p^\u, \bot^\u$}
        \AxiomC{}
        \RightLabel{$\mathsf{id}$}
        \UnaryInfC{$\bot^\u, \C p^\u \Rightarrow \bot^\u$}
        \RightLabel{$\rightarrow \mathsf{L}$}
        \BinaryInfC{$\neg \K_i \C p^\u, \C p^\u \Rightarrow \bot^\u$}
        \RightLabel{$\rightarrow \mathsf{R}$}
        \UnaryInfC{$\neg \K_i \C p^\u \Rightarrow p^\u, \neg \C p^\u$}
        \AxiomC{}
        \RightLabel{$\mathsf{id}$}
        \UnaryInfC{$p^\u, \K_i \C p^\u \Rightarrow p^\u, \neg \C p^\u, \bot^\u$}
        \RightLabel{$\C \mathsf{L}$}
        \UnaryInfC{$\C p^\u \Rightarrow p^\u, \neg \C p^\u, \bot^\u$}
        \RightLabel{$\mathsf{T_i}$}
        \UnaryInfC{$\K_i \C p^\u \Rightarrow p^\u, \neg \C p^\u, \bot^\u$}
        \RightLabel{${\rightarrow}\K_i$}
        \UnaryInfC{$\Rightarrow p^\u, \neg \C p^\u, \neg \K_i \C p^\u $}
        \RightLabel{$\mathsf{cut}$}
        \BinaryInfC{$\Rightarrow p^\u, \neg \C p^\u$}
    \end{prooftree}

Let $\tau$ be the following proof:

    \begin{prooftree}
    \small
        \AxiomC{}
        \RightLabel{$\mathsf{id}$}
        \UnaryInfC{$\neg \K_i \C p^\u, p^\u, \K_i \C p^\u \Rightarrow \K_i \C p^\u, \bot^\u$}
        \RightLabel{$\C \mathsf{L}$}
        \UnaryInfC{$\neg \K_i \C p^\u, \C p^\u \Rightarrow \K_i \C p^\u, \bot^\u$}
        \AxiomC{}
        \RightLabel{$\mathsf{id}$}
        \UnaryInfC{$\bot^\u, \C p^\u \Rightarrow \bot^\u$}
        \RightLabel{${\rightarrow} \mathsf{L}$}
        \BinaryInfC{$\neg \K_i \C p^\u, \C p^\u \Rightarrow \bot^\u$}
        \RightLabel{${\rightarrow} \mathsf{R}$}
        \UnaryInfC{$\neg \K_i \C p^\u \Rightarrow \K_i \C p^\u, \neg \C p^\u$}
        \AxiomC{}
        \RightLabel{$\mathsf{id}$}
        \UnaryInfC{$\K_i \C p^\u \Rightarrow \K_i \C p^\u, \neg \K_i p^\u$}
        \RightLabel{${\rightarrow}\mathsf{K_i}$}
        \UnaryInfC{$\Rightarrow \neg \K_i \C p^\u, \K_i \C p^\u, \neg \C p^\u$}
        \RightLabel{$\mathsf{cut}$}
        \BinaryInfC{$\Rightarrow\K_i \C p^\u, \neg \C p^\u$}
    \end{prooftree}

    Finally, the following is then a cyclic proof of $\Rightarrow \C p \vee \neg \C p^\u$:
    \begin{prooftree}
        \AxiomC{$\pi$}
        \AxiomC{$\tau$}
        \RightLabel{$\C \mathsf{R}$}
        \BinaryInfC{$\Rightarrow \C p^\u, \neg \C p^\u$}
        \RightLabel{$\vee \mathsf{R}$}
        \UnaryInfC{$\Rightarrow \C p \vee \neg \C p^\u$}
    \end{prooftree}
\end{example}

\begin{figure}[t]
\centering
\resizebox{\columnwidth}{!}{
\begin{tikzpicture}
\node[] (a) [] {
\AxiomC{}
\RightLabel{$\mathsf{id}$}
\UnaryInfC{$\varphi^\u, \{\K_i \C \varphi^\u\}_{i \in \A} \Rightarrow \varphi^\u,\psi^\u$}
\RightLabel{$\mathsf{CL}$}
\UnaryInfC{$\C \varphi^\u \Rightarrow \varphi^\u,\psi^\u$}
\RightLabel{$\vee \mathsf{R}$}
\UnaryInfC{$\C \varphi^\u \Rightarrow \varphi \vee \psi^\u$}
\AxiomC{$\C \varphi^\u \Rightarrow \C(\varphi \vee \psi)^\f$}
\RightLabel{$\mathsf{K_1}$}
\UnaryInfC{$\varphi^\u, \{\K_i\C \varphi^\u\}_{i \in \A} \Rightarrow \K_1 \C(\varphi \vee \psi)^\f$}
\RightLabel{$\mathsf{CL}$}
\UnaryInfC{$\C \varphi^\u \Rightarrow \K_1 \C (\varphi \vee \psi)^\f$}
\AxiomC{$\C \varphi^\u \Rightarrow \C(\varphi \vee \psi)^\f $}
\RightLabel{$\mathsf{K_2}$}
\UnaryInfC{$\varphi^\u, \{\K_i\C \varphi^\u\}_{i \in \A} \Rightarrow \K_2 \C(\varphi \vee \psi)^\f$}
\RightLabel{$\mathsf{CL}$}
\UnaryInfC{$\C \varphi^\u \Rightarrow \K_2 \C (\varphi \vee \psi)^\f$}
\RightLabel{$\mathsf{CR}$}
\TrinaryInfC{$\C \varphi^\u \Rightarrow \C(\varphi \vee \psi)^\f$}
\RightLabel{$\mathsf{f}$}
\UnaryInfC{$\C \varphi^\u \Rightarrow \C(\varphi \vee \psi)^\u$}
\DisplayProof};

\draw[->,rounded corners=.25cm,dashed] (4.9,1.3) -- (4.9, 1.7) -- (9, 1.7) -- (9,-0.6) -- (1,-0.6);
\draw[->,rounded corners=.25cm,dashed] (-1.2,1.3) -- (-1.2, 1.7) -- (-9, 1.7) -- (-9,-0.6) -- (-1.8,-0.6);
\end{tikzpicture}
}
\caption{A $\CICK$-proof of $\C \varphi^\u \Rightarrow \C(\varphi \vee \psi)^\u$. The dashed arrows indicate the good repetitions.}
\label{f: example ICKS4 proof}
\end{figure}

Example \ref{e: law of excluded middle for C} hints towards the fact that $\CICK_\mathsf{S5}$ is not cut-free complete. This does not come as a surprise, given that it is well-known that classical S5 modal logic does not have a cut-free (plain) sequent calculus. A famous counterexample in the classical realm is the formula $p \rightarrow \Box \Diamond p$. In the intuitionistic setting, $\Box$ and $\Diamond$ are not interdefinable, therefore our language cannot express $\Diamond$. However, we may still use the formula as a counterexample, by identifying $\Box$ with $\K_i$ and $\Diamond$ with $\neg \K_i \neg$. An easy computation shows that $\neg \K_i \neg$ is not evaluated as a $\Diamond$, but instead its truth conditions are as follows: $\fw, w \models \neg \K_i \neg \varphi$  if and only if  there exist $u,u' \in W$ with $w \mathrel{R_i} u$, $u \leq u'$ and $\fw, u' \models \varphi$. Therefore the following holds.

\begin{lemma}
    The formula $p \rightarrow \K_i \neg \K_i \neg p$ is valid over the class of S5 models.
\end{lemma}
\begin{proof}
    Let $\fw=(W, \leq, \{R_i\}_{i \in \A}, V)$ be an arbitrary S5 model and $w \in W$ an arbitrary world. Let $v \geq w$ and suppose that $\fw,v \models p$. Let $u \in W$ be any world such that $v \mathrel{R_i} u$. We want to show that $\fw, u \models \neg \K_i \neg p$. By the previous observation it suffices to find worlds $s,t$ such that $u \mathrel{R_i} s$, $s \leq t$ and $\fw,t \models p$. By symmetry $u \mathrel{R_i} v$. Since $v \leq v$ and $\fw,v \models p$, choosing $s = t = v$ shows that $\fw, u \models \neg \K_i p$ and hence that $\fw,v \models \K_i \neg \K_i \neg p$. Hence $\fw,w \models p \rightarrow \K_i \neg \K_i \neg p$, implying that $p \rightarrow \K_i \neg \K_i \neg p$ is valid.
\end{proof}

\begin{proposition}\label{p: cicks5 not cutfree complete}
    $\CICK_\mathsf{S5}$ is not cut-free complete.
\end{proposition}
\begin{proof}[Proof sketch]
    Observe that the sequent  $\sigma = (p^\u \Rightarrow \K_i \neg \K_i \neg p^\u)$ does not contain a formula with a common knowledge operator. Therefore, the focus rules cannot be applied, implying that any proof of $\sigma$ must be a proof where every branch ends in an axiom. Applying rules bottom-up, by inspection of the rules, the only rule applicable to $\sigma$ (apart from $\mathsf{cut}$) is $\mathsf{S5_i}$. There are two possible such applications. The first application has the sequent $\Rightarrow \neg \K_i \neg p^\u$ as premise and the second has the sequent $\Rightarrow \neg \K_i \neg p^\u, \K_i \neg \K_i \neg p^\u$ as premise (this case happens if $\K_i \neg \K_i \neg p^\u$ in the conclusion is assumed to be contained in $\K_i \Delta$; see the definition of the rule $\mathsf{S5_i}$). It is routine to check that both sequents cannot be proven.\footnote{For the interested reader who is unwilling to try out the many possible and sensible candidates for a proof, let us remark here that both possible sequents in the premise are invalid. Thus the soundness result presented in the next section will provide an alternative and simpler proof.}\qedhere
\end{proof}

Note that the cut formulas used in the cyclic proof in Example \ref{e: law of excluded middle for C} belong to the negation closure of the root sequent. Similarly, the sequent $p^\u \Rightarrow K_i \neg K_i \neg p^\u$ is provable by using $\neg K_i \neg p^\u$ as cut formula, which also belongs to the negation closure of the root sequent. Following this observation we will show that $\CICK_\mathsf{S5}$ is complete when the cut-rule is restricted to \emph{analytic cuts}, i.e. instances of $\mathsf{cut}$ where the cut formula belongs to the negation closure of the root sequent. Therefore we will still obtain analytic completeness for $\CICK_\mathsf{S5}$.

\section{Soundness}\label{c: ICK, Section soundness}

This section establishes the soundness of all cyclic systems introduced before. We follow the same strategy as for the soundness proof of $\CIM$ in Section~\ref{c: IM, section soundness of cim}, but some small adaptions are needed. The following lemma is routine (c.f. Lemma~\ref{l: axioms of IM are valid}).

\begin{lemma}\label{l: S5 rules locally sound}
    All rules $\mathsf{r}$ depicted in Table \ref{d: basic rules} and Table \ref{d: additional rules for ICK} preserve validity (over the respective class of models): if the conclusion of $\mathsf{r}$ is invalid, the one of the premises is invalid.
\end{lemma}

 Let $\sigma$ be a sequent with a formula in focus, i.e. $\Delta_\sigma$ contains a formula of the form $\K_i^j \C \varphi^\f$ for $j \in \{0,1\}$. Denote by $\sigma(n)$ the sequent $\Gamma_\sigma \Rightarrow \Delta_\sigma, \K_i^j \Ek^n \varphi^\u$, i.e. the sequent expanding the right side of $\sigma$ by the formula  $\K_i^j \Ek^n \varphi^\u$. The following lemma is then proven as Lemma~\ref{l: invalid sequent with formula in focus}.

\begin{lemma}\label{l: invalid sequent existence number for ICK}
    If $\sigma$ has a formula in focus and is invalid (over one of the classes of epistemic, reflexive, S4 or S5 models), then there exists a natural number $n$ such that $\sigma(n)$ is invalid.
\end{lemma}

Therefore we may define the following measure,

\begin{equation*}
    \mu(\sigma) := \min \{n < \omega \, \mid \, \sigma(n) \text{ is invalid}\}
\end{equation*}

and then prove a strengthening of Lemma \ref{l: S5 rules locally sound}.

\begin{lemma}\label{l: global soundness for cick}
    Suppose
    \begin{equation*}
        \infer[\mathsf{r}]{\sigma}{\sigma_1 & \dotsm & \sigma_n}
    \end{equation*}
    is an instance of a rule in Table \ref{d: basic rules} or Table \ref{d: additional rules for ICK}. If $\sigma$ is invalid (over the respective class of models), then there is a natural number $1 \leq k \leq n$ such that $\sigma_k$ is invalid. 
    If both $\sigma$ and $\sigma_k$ have a formula in focus, then, moreover,
    \[
        \mu(\sigma_k) \leq \mu(\sigma),
    \]
    where the inequality is strict if $\mathsf{r} = \C \mathsf{R}$ and the principal formula is in focus.
\end{lemma}
\begin{proof}
    The proof is similar to the proof of Lemma \ref{l: global soundness}. We only treat the new cases in which the formula in focus is principal. This implies that $\mathsf{r} \in \{\mathsf{K_i}, \mathsf{S4_i}, \mathsf{S5_i}, \mathsf{CR}\}$. The case for $\mathsf{K_i}$ is similar to the case for $\lb$ in the proof of Lemma \ref{l: global soundness} and we omit it. \smallskip

    \noindent \textsc{Case for  $\mathsf{r} = \mathsf{S5_i}$.} In this case the conclusion $\sigma$ is of the form:
			\[
			\Pi, \K_i \Gamma \Rightarrow \K_i \C \psi^\f, \K_i \Delta, \Sigma.
			\]
			Since $\sigma$ is invalid, there is a pointed S5 model $(\fw, w)$  such that $\fw, w \not \models \sigma(n)$ where $n= \mu(\sigma)$. Thus there exists $w' \geq w$ such that $\fw, w' \models \bigwedge \Gamma_\sigma^-$ and $\fw, w' \not \models \bigvee \Delta_\sigma^- \vee \K_i \Ek^n \psi$. In particular it holds that
			\[
			\fw, w' \not \models \K_i \Ek^{n} \psi.
			\]
			It follows that there is a world $v \in W$ such that $w' \mathrel{R_i} v$ and $\fw, v \not \models \Ek^n \psi$. Clearly this also means that $\fw, v \not \models \C \psi$.  We claim that, in fact,
			\[
			\fw,v \not \models (\K_i \Gamma \Rightarrow \C \psi^\f, \Ek^n \psi^\u,  \K_i\Delta)^I,
			\]
			which implies that the premise $\sigma_1$ is invalid and that $\mu(\sigma_1) \leq \mu(\sigma)$.
			
			By the fact that $R_i$ is transitive, it holds for all $\varphi$ with $\fw, w' \models \K_i \varphi$ that $\fw, v \models \K_i \varphi$. Hence $M, v \models \K_i \varphi$ for each $\K_i \varphi^\u \in \K_i \Gamma$. Moreover, suppose that $\K_i \psi^a \in \K_i \Delta$. Then $\fw, w' \not \models \K_i \psi$. Thus there is a state $u \in W$ such that $w' \mathrel{R_i} u$ and $\fw, u \not \models \psi$. By symmetry and transitivity, we get $v \mathrel{R_i} u$, whence $\fw, v \not \models \K_i \psi$, as required. \smallskip

            \noindent \textsc{Case for $\mathsf{r} = \mathsf{S4_i}$.} For this case observe that $\mathsf{S4_i}$ is a special case of $\mathsf{S5_i}$, namely it consists of all instances of $\mathsf{S5_i}$ where $\K_i \Delta = \emptyset$. In the case above, the symmetry of $R_i$ is exclusively used to deal with formulas in $\K_i \Delta$. Hence, we obtain the result by following the previous case using a pointed S4 model instead of a pointed S5 model, and skipping the part dealing with $\K_i \Delta$. \smallskip

            \noindent \textsc{Case for $\mathsf{r} = \mathsf{CR}$.} In this case the conclusion $\sigma$ is of the form $\Gamma \Rightarrow \C \varphi^\f, \Delta$ with premises $\sigma_0$ given by $\Gamma \Rightarrow \varphi^\u, \Delta$, and $\sigma_i$ for $1 \leq i \leq n$ given by $\Gamma \Rightarrow \K_i\C \varphi^\f, \Delta$, respectively.  As there exists a pointed epistemic model $(\fw,w)$ that falsifies $\sigma(\mu(\sigma))$, $w$ has an intuitionistic successor $v$ such that $\fw,v\models \bigwedge \Gamma^-$ and $\fw,v\not\models \C \varphi \lor \Ek^{\mu(\sigma)} \varphi\lor \bigvee\Delta^{-}$. If $\mu(\sigma) = 0$, then $\fw,v \not \models \varphi $, so $(\fw,v)$ falsifies the left premise $\sigma_0$. By Lemma~\ref{l: proper application of rules}, $\sigma_0$ does not have a formula in focus, and so the statement of the lemma holds. If $\mu(\sigma) > 0$, then $\fw,v \not \models \K_i\Ek ^{\mu(\sigma) - 1} \varphi$ for some $i \in \A$. Hence $(\fw,v)$ falsifies $\sigma_i(\mu(\sigma)-1)$. So $\sigma_i$ is invalid and we have  $\mu(\sigma_i) < \mu(\sigma)$.
\end{proof}

Soundness for each cyclic calculus then follows by the same argument as used to prove Theorem \ref{t: CIM soundness}.

\begin{theorem}[Soundness of $\CICK_\ast$]\label{t: soundness cick}
   Let $\sigma$ be a sequent. The following hold.
   \begin{enumerate}
       \item If $\sigma$ is $\CICK$-provable, then $\sigma$ is valid over the class of epistemic models.
       \item If $\sigma$ is $\CICK_\mathrm{T}$-provable, then $\sigma$ is valid over the class of reflexive epistemic models.
       \item If $\sigma$ is $\CICK_\mathrm{S4}$-provable, then $\sigma$ is valid over the class of S4 epistemic models.
       \item If $\sigma$ is $\CICK_\mathrm{S5}$-provable, then $\sigma$ is valid over the class of S5 epistemic models.
   \end{enumerate}
\end{theorem}

\section{Completeness}\label{c: ICK, Section completeness}

This section establishes completeness of all cyclic calculi introduced above. Completeness of $\CICK, \CICK_\mathrm{T}$ and $ \CICK_\mathrm{S4}$ for their respective classes of models will be shown by using the proof search technique introduced in Section \ref{c: IM, section completeness}. Since the resulting proofs are very similar, only brief sketches are provided. The main part of this section is the completeness proof for $\CICK_\mathsf{S5}$. 

\subsection{Completeness for Epistemic, Reflexive and S4 Models}

In order to prove completeness of $\CICK$, $\CICK_\mathsf{T}$ and $ \CICK_\mathsf{S4}$, we define corresponding non-wellfounded calculi $\mathsf{nICK}$, $\mathsf{nICK_T}$ and $\mathsf{nICK_{S4}}$ following the definitions in Section \ref{c: IM, section non-wellfounded and cyclic calculus for IM}. We then employ the same proof search strategy, where the definition of a saturated sequent is adapted from the definition for $\NWIM$ in the obvious way. Proof search trees are defined in Definition \ref{d: proof search tree nIM} (by replacing $\lbm$ with $\C$). Neither the indexes nor the enumeration of the formulas in $\Cl(\sigma)$ are needed and so are ignored henceforth. Proof search games are defined as for $\NWIM$. The definition of the canonical model has to be adapted to the multi-modal case: given a choice rule $\C$, we assume that the modal premises are partitioned into $\lvert \A \rvert$ groups. Then replace clause 3. in Definition \ref{d: canonical model for nIM} by the following clause.
\begin{enumerate}
    \item[3'.] For each $i \in \A$, $R_i \subseteq W \times W$ is such that
    \begin{center}
        $w \mathrel{R_i} v$ iff there exists $s \in w$ and $t \in v$ such that $s$ is the conclusion and $t$ a modal premise from group $i$ of the same $\C$-rule instance.
    \end{center}
\end{enumerate}

The calculus $\mathsf{nICK}$ is simply a multi-modal version of $\NWIM$, so for the choice rule a slightly adapted version of $\Ct$ suffices to obtain completeness, namely the rule
 \begin{equation*}
            \infer[\mathsf{C_t'}]{\Pi, \{\K_i \Gamma_i\}_{i \in \A} \Rightarrow \{(\varphi_i \rightarrow \psi_i)^{\u}\}_{i=0}^l,\{\{\K_i\chi^{a_{i_j}}_{i_j}\}_{j=0}^{m_i}\}_{i \in \A}, \Sigma}{\Pi, \{\K_i \Gamma_i\}_{i \in \A}, \varphi^\u_0 \Rightarrow \psi^{\u}_0 &\dotsm  & \Pi, \{\K_i \Gamma_i\}_{i \in \A}, \varphi^\u_l \Rightarrow \psi^{\u}_l & \{ \{\Gamma_i \Rightarrow \chi^{b_{i_j}}_{i_j}\}_{j = 0}^{m_i}\}_{i \in \A}}
        \end{equation*}
        where the annotations $b_{i_j}$ are equal to $\f$ whenever the underlying formula $\chi_{i_j}$ is a  $\C$-formula, and equal to $\u$ otherwise. Moreover, we require that  $\Pi\cup\Sigma$ contains no $\K_i$-formulas for any $i \in \A$ and that $\Sigma$ contains no $\rightarrow$-formulas. The modal premises are those of the form $\Gamma_i \Rightarrow \chi^{b_{i_j}}_{i_j}$, where $\{\Gamma_i \Rightarrow \chi_{i_j}^{b_{i_j}}\}_{j = 0}^{m_i}$ is group $i$ and the others are the intuitionistic premises. In case the conclusion of a $\Ct'$-instance has no $\rightarrow$- or $\K_i$-formulas on the right-hand side for $i \in \A$, then we stipulate that $l = -1$ or $m_i=-1$, respectively.\smallskip

Observe that the conclusion and each modal premise form an instance of the rule $\mathsf{K_i}$. It is then routine to show that winning strategies for Prover correspond to $\CICK$-proofs and winning strategies for Refuter to refutations. Finally, when proving that refutions induce countermodels, as in the proof of Proposition \ref{p: refutation gives triangle model}, we take the canonical model induced by the refutation, close it under triangle confluence\footnote{We will close every canonical model under triangle confluence and not mention it in the next cases.} and then show that it falsifies the root sequent of the refutation. The proof is similar to the proof of Proposition \ref{p: refutation gives triangle model}.

For $\mathsf{nICK_T}$ the definition of saturated sequent has to be extended with the following clause:
\begin{enumerate}
    \item[8.] If $\K_i \varphi^\u \in \Gamma$, then $\varphi^\u \in \Gamma$.
\end{enumerate}
Note that the rule $\mathsf{T_i}$ is invertible. Hence the same choice rule $\Ct'$ can be used and when proving that refutations induce a countermodel, we only have to close the modal relations of the canonical model under reflexivity. The additional saturation clause 8. ensures that for any world $w$ of the canonical model $\fw$ and any formula $\K_i \varphi$ holds that if $\fw, w \models \K_i \varphi$, then $\fw, w \models \varphi$.

Finally, for $\mathsf{nICK_{S4}}$ we employ the following choice rule
\begin{equation*}
    \infer[\mathsf{C_{S4}}]{\Pi, \{\K_i \Gamma_i\}_{i \in \A} \Rightarrow \{(\varphi_i \rightarrow \psi_i)^{\u}\}_{i=0}^l,\{\{\K_i\chi^{a_{i_j}}_{i_j}\}_{j=0}^{m_i}\}_{i \in \A}, \Sigma}{\Pi, \{\K_i \Gamma_i\}_{i \in \A}, \varphi^\u_0 \Rightarrow \psi^{\u}_0 &\dotsm  & \Pi, \{\K_i \Gamma_i\}_{i \in \A}, \varphi^\u_l \Rightarrow \psi^{\u}_l & \{ \{\K_i \Gamma_i \Rightarrow \chi^{b_{i_j}}_{i_j}\}_{j = 0}^{m_i}\}_{i \in \A}}
\end{equation*}
with the same conditions as for $\Ct'$. Note that the only difference to $\Ct'$ is that the applications of $\mathsf{K_i}$ were replaced by applications of $\mathsf{S4_i}$. When proving that refutations induce countermodels, it then suffices to close the modal accessibility relations of the canonical model induced by the refutation under reflexivity and transitivity. Note that since formulas in $\K_i \Gamma$ are preserved when moving to a modal premise of group $i$, closing the modal accessibility relation $R_i$ under transitivity does not lead to the failure of any formula in $\K_i \Gamma$. Therefore we obtain completeness for the non-wellfounded calculi $\mathsf{nICK}$, $\mathsf{nICK_T}$ and $\mathsf{nICK_{S4}}$. By following the construction given in Lemma \ref{l: non-wellfounded proof implies cyclic proof} we obtain completeness for the cyclic calculi.

\begin{theorem}[Completeness of $\CICK$, $\CICK_{\mathsf{T}}$ and $\CICK_\mathsf{S4}$]
Let $\sigma$ be a sequent. The following hold.
\begin{enumerate}
    \item If $\sigma$ is valid over the class of epistemic models, then $\sigma$ is $\CICK$-provable.
     \item If $\sigma$ is valid over the class of reflexive epistemic models, then $\sigma$ is $\CICK_{\mathsf{T}}$-provable.
      \item If $\sigma$ is valid over the class of S4 epistemic models, then $\sigma$ is $\CICK_\mathsf{S4}$-provable.
\end{enumerate}
\end{theorem}

\subsection{Completeness for S5 Models}

 This section shows that $\CICK_\mathsf{S5}$ is complete with respect to the class of S5 epistemic models. Due to the presence of $\mathsf{cut}$ we can prove completeness directly via a canonical model construction, which arguably simplifies the completeness proof. To obtain \emph{analytic completeness}, applications of $\mathsf{cut}$ will be restricted to a finite set of formulas relevant to the root sequent. So far we counted as relevant the formulas in the closure of the root sequent. However, as we have seen, the logic $\mathbf{ICK_{S5}}$ satisfies many classical principles, implying that we need a stronger notion of closure to obtain completeness, namely the \emph{negation closure}. In the following we provide the basic definitions and lemmas to obtain the canonical model for $\CICK_\mathsf{S5}$.

\begin{definition}
Let $\Sigma$ be a non-empty, negation closed and finite set of formulas. A sequent $\sigma$ is called a
\begin{enumerate}
    \item \emph{$\Sigma$-sequent}\index{sequent!$\Sigma$-} if $\Gamma_\sigma \cup \Delta_\sigma \subseteq \Sigma$;
    \item \emph{$\Sigma$-provable}\index{sequent!$\Sigma$-provable} if there exists a $\CICK_\mathsf{S5}$-proof of $\sigma$ in which only $\Sigma$-sequents occur;
    \item \emph{$\Sigma$-saturated}\index{sequent!$\Sigma$-saturated} if $\sigma$ is $\Sigma$-unprovable and $\Gamma_\sigma \cup \Delta_\sigma = \Sigma$.
\end{enumerate} 
\end{definition}

$\Sigma$-saturated sequents will be the worlds of the canonical model. Let us first prove an analogue of the Lindenbaum Lemma for saturated sequents.

\begin{lemma}[Lindenbaum]\label{l: Lindenbaum for saturated sequents}
    If a $\Sigma$-sequent $\sigma$ is $\Sigma$-unprovable, then there exists a $\Sigma$-saturated sequent $\sigma'$ with $\Gamma_\sigma \subseteq \Gamma_\sigma'$ and $\Delta_\sigma \subseteq \Delta_\sigma'$.
\end{lemma}
\begin{proof}
    Suppose $\sigma$ is $\Sigma$-unprovable. Let $\pi$ be the pre-proof of $\sigma$ built as follows: starting with $\sigma$ at the root, repeatedly apply the rule $\mathsf{cut}$ bottom-up to introduce fresh $\Sigma$-formulas (i.e. formulas not yet occurring in the conclusion) until every leaf is labelled by a sequent $\Gamma \Rightarrow \Delta$ with $\Gamma \cup \Delta = \Sigma$. At least one of these sequents must be $\Sigma$-unprovable (and thus $\Sigma$-saturated), as otherwise $\pi$ could be extended into a $\Sigma$-proof of $\sigma$. Note that the saturated sequent extends $\sigma$.
\end{proof}
Next, we will establish some important properties of saturated sequents, that will become useful in the proof of the Truth Lemma (c.f. Lemma \ref{l: truth lemma for S5}).

\begin{lemma}\label{l: properties of saturated sequents}
    Let $\Gamma \Rightarrow \Delta$ be a $\Sigma$-saturated sequent. 
    \begin{enumerate}
        \item If $\varphi \wedge \psi \in \Sigma$, then $\varphi \wedge \psi \in \Gamma^-$ if and only if $\varphi \in \Gamma^-$ and $\psi \in \Gamma^-$. 
        \item If $\varphi \vee \psi \in \Sigma$, then $\varphi \vee \psi \in \Gamma^-$ if and only if $\varphi \in \Gamma^-$ or $\psi \in \Gamma^-$.
        \item If $\K_i \varphi \in \Gamma^-$, then $\varphi \in \Gamma^-$ for all $i \in \A$.
        \item $\K_i \varphi \in \Gamma^-$ if and only if $\neg \K_i \varphi \in \Delta^-$ for all $i \in \A$.
        \item $\C \varphi \in \Gamma^-$ if and only if $\varphi \in \Gamma^-$ and $\K_i \C \varphi \in \Gamma^-$ for all $i \in \A$.
    \end{enumerate}
\end{lemma}
\begin{proof}
    \textsc{1.} Let $\varphi \wedge \psi \in \Sigma$ and suppose towards contradiction that $\varphi \wedge \psi \in \Gamma^-$ but (without loss of generality) $\varphi \not \in \Gamma^-$. By saturation $\varphi \in \Delta^-$. Thus $\Gamma \Rightarrow \Delta$ is $\Sigma$-provable by applying $\wedge\mathsf{L}$ with principal formula $\varphi \wedge \psi$ and then the axiom $\mathsf{id}$.\\
    For the other direction suppose that $\varphi \wedge \psi \in \Sigma$ and both $\varphi \in \Gamma^-$ and $\psi \in \Gamma^-$. Assume towards contradiction that $\varphi \wedge \psi \not \in \Gamma^-$. By saturation $\varphi \wedge \psi \in \Delta^-$. Thus $\Gamma \Rightarrow \Delta$ is $\Sigma$-provable by applying $\wedge \mathsf{R}$ with principal formula $\varphi \wedge \psi$ and then the axiom $\mathsf{id}$ in both resulting branches. The proofs of 2. and 3. are analogous by using the rules for disjunction and the rule $\mathsf{T_i}$, respectively. \smallskip \\
    \textsc{4.} Suppose towards contradiction that $\K_i \varphi \in \Gamma^-$ and $\neg \K_i \varphi \not \in \Delta^-$. Note that $\K_i \varphi \in \Gamma$ implies $\K_i \varphi \in \Sigma$ and since $\Sigma$ is negation closed, $\neg \K_i \varphi \in \Sigma$ as well. By saturation $\neg \K_i \varphi \in \Gamma^-$. Then $\Gamma \Rightarrow \Delta$ is $\Sigma$-provable by applying ${\rightarrow}\mathsf{L}$ with principal formula $\K_i \varphi \rightarrow \bot^\u$ and then the axiom $\mathsf{id}$ in the left branch and the axiom $\bot$ in the right branch.
    For the other direction suppose towards contradiction that $\neg \K_i \varphi \in \Delta^-$ and $\K_i \varphi \not \in \Gamma^-$. By saturation $\K_i \varphi \in \Delta^-$. Then $\Gamma \Rightarrow \Delta$ is $\Sigma$-provable by applying ${\rightarrow}\mathsf{K_i}$ and then $\mathsf{id}$. The proof of 5. is standard and omitted. 
\end{proof}

Given a set of formulas $\Gamma$, let $\K_i^{-1} \Gamma \coloneqq \{\varphi \mid \K_i \varphi \in \Gamma\}$. The canonical model relative to $\Sigma$ is defined as follows.

\begin{definition}
    Let $\Sigma$ be a non-empty, negation closed and finite set of formulas. The \emph{canonical model}\index{canonical model!for $\CICK_{\mathrm{S5}}$} (relative to $\Sigma)$ is given by $\fw^\Sigma =(W^\Sigma, \leq^\Sigma, \{R_i^\Sigma\}_{i \in \A}, V^\Sigma)$ where
    \begin{itemize}
		\item $W^\Sigma := \{\Gamma^- \mid \Gamma \Rightarrow \Delta \text{ is a $\Sigma$-saturated sequent}\}$;
        \item $A \leq^\Sigma B$ if and only if $A \subseteq B$;
		\item $A \Rel_i^\Sigma B$ if and only if $\K_i \K_i^{-1} A = \K_i \K_i^{-1} B$;
		\item $V^\Sigma(A) := A \cap \Prop$; 
		\end{itemize}
        where $A, B \in W^\Sigma$.
\end{definition}

\begin{lemma}
    The canonical model is an S5 epistemic model.
\end{lemma}
\begin{proof}
    That $(W^\Sigma, \leq^\Sigma)$ is a partial order and each $R_i^\Sigma$ is reflexive, transitive and symmetric is immediate. Furthermore, by definition of $V^\Sigma$ and $\leq^\Sigma$ the valuation is monotone. It remains to show that each $R_i^\Sigma$ is triangle confluent. Suppose that $A \leq^\Sigma B$ and $B \mathrel{R_i^\Sigma} C$. By definition $\K_i \K_i^{-1} B = \K_i \K_i^{-1}C$. Moreover, $\K_i \K_i^{-1} A \subseteq \K_i \K_i^{-1} B$. Suppose towards contradiction that there exists a formula $\K_i \varphi \in  B$ such that $\K_i \varphi \not \in A$. Let $A = \Gamma^-$ where $\Gamma \Rightarrow \Delta$ is a saturated sequent. By saturation $\K_i \varphi \in \Delta^-$. By Lemma \ref{l: properties of saturated sequents}, $\neg \K_i \varphi \in \Gamma^-$. Hence $\neg \K_i \varphi \in B$, contradicting that $\K_i \varphi \in B$. Therefore $\K_i \K_i^{-1} A = \K_i \K_i^{-1} B = \K_i \K_i^{-1} C$, implying that $A \mathrel{R_i^\Sigma} C$.
\end{proof}

\begin{lemma}[Truth Lemma]\label{l: truth lemma for S5}
     Let $\Sigma$ be a non-empty, negation closed and finite set of formulas and $\fw^\Sigma =(W^\Sigma, \leq^\Sigma, \{R_i^\Sigma\}_{i \in \A}, V^\Sigma)$ the canonical model relative to $\Sigma$. Then for any $\varphi \in \Sigma$ and any $A \in W^\Sigma$ the following holds.
     \begin{center}
         $\varphi \in A$ if and only if $\fw^\Sigma, A \models \varphi$. 
     \end{center}
\end{lemma}

\begin{proof}
    By induction on the structure of $\varphi$. For $\varphi = \bot$ observe that if $\bot \in A$ and $A = \Gamma^-$ where $\Gamma \Rightarrow \Delta$ is $\Sigma$-saturated, then $\Gamma \Rightarrow \Delta$ is $\Sigma$-provable by applying $\bot$. Hence $\bot \not \in A$ and by definition $\fw^\Sigma, A \not \models \bot$. For $\varphi = p$ where $p \in \Prop$, observe that $p \in A$ if and only if $p \in V^\Sigma(A)$ if and only if $\fw^\Sigma, A \models p$. The cases for $\varphi = \psi \wedge \chi$ and $\varphi = \psi \vee \chi$ follow directly from Lemma \ref{l: properties of saturated sequents} and the induction hypothesis. \smallskip
    
    \noindent \textsc{Case for $\varphi = \psi \rightarrow \chi$.} First suppose that $\psi \rightarrow \chi \in A$. We want to show that ${\fw^\Sigma, A \models \psi \rightarrow \chi}$. Let $B \in W^\Sigma$ be any world such that $A \leq^\Sigma B$. Hence $A \subseteq B$ and so $\psi \rightarrow \chi \in B$. Suppose $\fw^\Sigma, B \models \psi$. By induction hypothesis $\psi \in B$. Let $\Gamma \Rightarrow \Delta$ be $\Sigma$-saturated such that $B = \Gamma^-$. Suppose towards contradiction that $\chi \in \Delta^-$. We may assume without loss of generality that $\chi$ occurs unfocused in $\Delta$. Write $\Gamma_0$ for $\Gamma \setminus \{\psi^\u, \psi \rightarrow \chi^\u\}$ and $\Delta_0$ for $\Delta \setminus \{\chi^\u\}$. The following is a $\Sigma$-proof of $\Gamma \Rightarrow \Delta$, which contradicts the assumption that $\Gamma \Rightarrow \Delta$ is $\Sigma$-saturated:
    \begin{prooftree}
        \AxiomC{}
        \RightLabel{$\mathsf{id}$}
        \UnaryInfC{$\Gamma_0, \psi^\u, \psi \rightarrow \chi^\u \Rightarrow \psi^\u, \chi^\u, \Delta_0 $}
        \AxiomC{}
        \RightLabel{$\mathsf{id}$}
        \UnaryInfC{$\Gamma_0, \psi^\u, \chi^\u \Rightarrow  \chi^\u, \Delta_0 $}
        \RightLabel{${\rightarrow} \mathsf{L}$}
        \BinaryInfC{$\Gamma_0, \psi^\u, \psi \rightarrow \chi^\u \Rightarrow \chi^\u, \Delta_0$} 
    \end{prooftree}
    Thus $\chi \in B$ and the induction hypothesis yields $\fw^\Sigma, B \models \psi$. Hence $\fw^\Sigma, A \models \psi \rightarrow \chi$. 
    
    For the other direction suppose $\psi \rightarrow \chi \not \in A$. Let $A = \Gamma^-$ and let $\Gamma \Rightarrow \Delta$ be $\Sigma$-saturated. By assumption $\psi \rightarrow \chi \in \Delta^-$. Apply the rule ${\rightarrow}\mathsf{R}$ to $\psi \rightarrow \chi$. The premise of this application is $\Gamma, \psi^\u \Rightarrow \chi^\u$ which must be $\Sigma$-unprovable. Lemma \ref{l: Lindenbaum for saturated sequents} implies that there exists a $\Sigma$-saturated sequent $\Gamma' \Rightarrow \Delta'$ with $\Gamma, \psi^\u \subseteq \Gamma'$. Let $B = \Gamma'^-$. By construction $A \leq^\Sigma B$ and $\psi \in B$ and $\chi \not \in B$. Hence by induction hypothesis $\fw^\Sigma, B \models \psi$ and $\fw^\Sigma, B \not \models \chi$. Thus $\fw^\Sigma, A \not \models \psi \rightarrow \chi$.  \smallskip
    
    \noindent \textsc{Case for $\varphi = \K_i \psi$.} First suppose that $\K_i \psi \in A$. Let $B \in W^\Sigma$ be any world such that $A \mathrel{R^\Sigma_i} B$. By definition $\K_i \psi \in B$ and so by Lemma \ref{l: properties of saturated sequents}, $\psi \in B$. Hence, by induction hypothesis, $\fw^\Sigma, B \models \psi$, implying that $\fw^\Sigma, A \models \K_i \psi$.
    
    For the other direction suppose $\K_i \psi \not \in A$. Let $\Gamma \Rightarrow \Delta$ be $\Sigma$-saturated such that $A = \Gamma^-$. By saturation $\K_i\psi^a \in \Delta$ for some $a \in \{\u, \f \}$. Let $\Delta_0 = \Delta \setminus \{\K_i \psi^a \}$. Observe that the sequent
    \begin{equation*}
        \K_i \K_i^{-1} \Gamma \Rightarrow \psi^a, \K_i \K_i^{-1} \Delta_0
    \end{equation*}
    is $\Sigma$-unprovable, as otherwise, by an application of $\mathsf{S5_i}$ we could derive $\Gamma \Rightarrow \Delta$. By Lemma~\ref{l: Lindenbaum for saturated sequents} there exists a saturated sequent $\Gamma' \Rightarrow \Delta'$ such that $\K_i\K_i^{-1}\Gamma \subseteq \Gamma'$ and, moreover, $ {\{\psi^a \} \cup \K_i \K^{-1} \Delta_0 \subseteq \Delta'}$. Let $B = \Gamma'^-$. By construction $\psi \not \in \Gamma'^-$ and so the induction hypothesis yields $\fw^\Sigma, B \not \models \psi$. It remains to show that $A \mathrel{R_i^\Sigma} B$. By construction $\K_i \K_i^{-1} A \subseteq \K_i \K_i^{-1} B$. For the other direction first observe that $\K_i \psi \not \in B$, as otherwise $\psi \in B$. Suppose $\K_i \gamma \in B$. Then $\K_i \gamma \not \in \Delta'^-$, and hence in particular $\K_i \gamma \not \in \Delta^-$. Thus, by saturation, $\K_i \gamma \in A$. Therefore $A \mathrel{R_i^\Sigma} B$ and so $\fw^\Sigma, A \not \models \K_i \psi$.\smallskip
    
    \noindent \textsc{Case for $\varphi = \C \psi$.} First suppose that $\C \psi \in A$. We want to show that $\fw^\Sigma, B \models \psi$ for any $B \in W^\Sigma$ such that $A \mathrel{(R^\Sigma)^*} B$. We prove by induction on the length $n$ of the path from $A$ to $B$ that $\psi \in B$ and $\C \psi \in B$. For the base case by assumption $\C \psi \in A$. Moreover, by Lemma \ref{l: properties of saturated sequents}, $\psi \in A$. For the induction step suppose that $A \mathrel{(R^\Sigma)^n} B_0$ and $B_0 \mathrel{R_i^\Sigma} B$ for some $i \in \A$. By (the inner) induction hypothesis $\psi \in B_0$ and $\C \psi \in B_0$. By Lemma \ref{l: properties of saturated sequents} also $\K_i \C \psi \in B_0$. Hence, by definition of $R_i^\Sigma$, we have that $\K_i \C \psi \in B$. Lemma \ref{l: properties of saturated sequents} implies that $\C \psi \in B$ and $\psi \in B$. Hence $\psi \in B$ for any $A \mathrel{(R^\Sigma)^*} B$. By (the outer) induction hypothesis $\fw^\Sigma, B \models \psi$ for any $A \mathrel{(R^\Sigma)^*} B$ and therefore $\fw^\Sigma, A \models \C \psi$. \smallskip
    
	In case $\C \psi \notin A$, consider a $\Sigma$-saturated sequent $\Gamma \Rightarrow \Delta$ such that $A = \Gamma^-$.  By the presence of the rules $\mathsf{u}$ and $\mathsf{f}$, we may assume without loss of generality that $\C \psi^\f \in \Delta$. Now suppose, towards a contradiction, that $\fw^\Sigma, A  \models \C \psi$. For every $A \mathrel{(R^\Sigma)^*} B$ then holds that $\fw^\Sigma, B \models \psi$. In particular, it follows that $\fw^\Sigma, A \models \psi$, whence, by the induction hypothesis, we have $\psi^\u \in \Gamma$.
			
	Let $\Delta_0 = \Delta \setminus \{\C \psi^\f\}$. Consider the following proof, where we assume without loss of generality that there are $n$ agents:
			\begin{prooftree}
			\AxiomC{}
			\RightLabel{$\mathsf{id}$}
			\UnaryInfC{$\Gamma \Rightarrow \psi^\u, \Delta_0$}
			\noLine
			\AxiomC{$\pi_1$}
			\UnaryInfC{$\Gamma \Rightarrow \K_1 \C \psi^\f, \Delta_0$}
			\AxiomC{}
			\noLine
			\UnaryInfC{$\cdots$}
			\AxiomC{$\pi_n$}
			\noLine
			\UnaryInfC{$\Gamma \Rightarrow \K_n \C \psi^\f, \Delta_0$}
			\RightLabel{$\C \mathsf{R}$}
			\QuaternaryInfC{$\Gamma \Rightarrow \C \psi^\f, \Delta_0$}
            \end{prooftree}
			 where each $\pi_i$ is constructed as follows:
			\begin{prooftree}
			\AxiomC{$\pi'$}
			\noLine
			\UnaryInfC{$\sigma'$}
			\AxiomC{$\pi'_1$}
			\noLine
			\UnaryInfC{$\sigma'_1$}
			\AxiomC{$\cdots$}
			\noLine
			\UnaryInfC{$\cdots$}
			\AxiomC{$\pi'_n$}
			\noLine
			\UnaryInfC{$\sigma'_n$}
			\RightLabel{$\C\mathsf{R}$}
			\QuaternaryInfC{$\K_i \K_i^{-1} \Gamma \Rightarrow \C \psi^\f, \K_i \K_i^{-1} \Delta_0$}
			\RightLabel{$\mathsf{S5_i}$}
			\UnaryInfC{$\Gamma \Rightarrow \K_i \C \psi^\f, \Delta_0$}
			\end{prooftree}
			In the above proof the sequent $\sigma'$ is given by
			\[
			\sigma' = \K_i \K_i^{-1} \Gamma \Rightarrow \psi^\u, \K_i \K_i^{-1} \Delta_0
			\]
			and the proof $\pi'$ is obtained from the $\Sigma$-provability of $\sigma'$. Indeed, if $\sigma'$ were not $\Sigma$-provable, then by applying Lemma \ref{l: Lindenbaum for saturated sequents} and the induction hypothesis, we would obtain a saturated sequent $\Gamma' \Rightarrow \Delta'$ extending $\sigma'$ and a world $B = \Gamma'^-$ such that $\fw^\Sigma, B \not \models \psi$. First note that $\K_i \C \psi \not \in B$, as otherwise $\psi \in B$, implying that the sequent $\Gamma' \Rightarrow \Delta'$ is an instance of $\mathsf{id}$ and hence $\Sigma$-provable. Therefore $\K_i \C \psi \in \Delta'^-$. Note that $A \mathrel{R_i^\Sigma} B$: by construction $\K_i \K_i^{-1} A \subseteq \K_i \K_i^{-1} B$. Suppose towards contradiction that there exists a formula $\K_i \chi \in B$ such that $\K_i \chi \not \in A$. Hence $\K_i \chi \not \in \Gamma^-$ and so by saturation $\K_i \chi \in \Delta^-$. Since $\chi \not = \C \psi$, $\K_i \chi \in \Delta_0$ and so $\K_i \chi \in \K_i \K_i^{-1} \Delta_0^-$.  Therefore $\K_i \chi \in \Delta'^-$, implying that $\K_i \chi \not \in B$, a contradiction. Therefore $A \mathrel{R_i^\Sigma} B$. Since $\fw^\Sigma, B \not \models \psi$, this contradicts the assumption that $\fw^\Sigma, A \models \C \psi$. Hence, $\sigma'$ must be $\Sigma$-provable. \smallskip
			
		    Furthermore, each sequent $\sigma'_l$ for $1 \leq l \leq n$ in the derivation $\pi_i$ is given by
			\[
			\sigma'_l = \K_i \K_i^{-1} \Gamma \Rightarrow \K_l \C \psi^\f, \K_i \K_i^{-1} \Delta_0
			\]
			and each derivation $\pi'_l$ is constructed by repeatedly applying $\mathsf{cut}$ to add formulas from $\Sigma$ until every leaf is either saturated or $\Sigma$-provable. To the leaves that are $\Sigma$-provable we append their respective proofs. Suppose $\Gamma' \Rightarrow \K_l \C \psi^\f, \Delta'$ is a saturated leaf. Note that by construction $\K_i \K_i^{-1} \Gamma = \K_i \K_i^{-1} \Gamma'$ and therefore $A \mathrel{R_i^\Sigma} C$ for $C = \Gamma'^-$. The assumption $\fw^\Sigma, A \models \C \psi$ therefore entails that $\fw^\Sigma, C \models \C \psi$ and thus $\fw^\Sigma, C \models \psi$ which, by induction hypothesis, implies that $\psi \in C$. Hence we can apply the same process to $\Gamma' \Rightarrow \K_l \C \psi^\f, \Delta'$ as we have just applied to $\Gamma \Rightarrow \K_i \C \psi^\f, \Delta_0$. \smallskip
			
			Since $\Sigma$ is finite, there are only finitely many distinct $\Sigma$-saturated sequents. This entails, by the pidgeonhole principle, that at some point one of the saturated leaves obtained from our construction must be identical to a saturated leaf reached earlier in the construction. Note that in this case the upward path from the earlier saturated leaf to the later one is successful. We then terminate the construction of this branch. Since every branch is terminated at some point, we end up with a $\Sigma$-proof of $\Gamma \Rightarrow \Delta$, a contradiction. 
\end{proof}

\begin{theorem}[Completeness of $\CICK_\mathsf{S5}$]\label{t: completeness of S5}
     If a $\Sigma$-sequent $\sigma$ is valid over the class of S5 epistemic models, then $\sigma$ is $\Sigma$-provable in $\CICK_\mathsf{S5}$.
\end{theorem}
\begin{proof}
    Suppose that $\sigma $ is not $\Sigma$-provable. By Lemma \ref{l: Lindenbaum for saturated sequents} there exists a saturated sequent $\Gamma \Rightarrow \Delta$ such that $\Gamma_\sigma \subseteq \Gamma$ and $\Delta_\sigma \subseteq \Delta$. Let $A \in W^\Sigma$ such that $A = \Gamma^-$. By the Truth Lemma we have that $\fw^\Sigma, A \models \varphi$ for each $\varphi \in \Gamma_\sigma$ and $\fw^\Sigma, A \not \models \varphi$ for each $\varphi \in \Delta_\sigma$. Hence, $\fw^\Sigma, A \not \models \sigma^I$, implying that $\sigma$ is not valid over the class of S5 epistemic models.
\end{proof}

\begin{corollary}[Finite model property]\label{c: small model property}\index{finite model property!intuitionistic common knowledge logic}
    If a formula $\varphi$ is falsifiable over the class of S5 epistemic models, then $\varphi$ is falsifiable in a finite S5 epistemic model.
\end{corollary}
\begin{proof}
    Suppose $\varphi$ is falsifiable over the class of S5 epistemic models. By Theorem~\ref{t: soundness cick} the sequent $\Rightarrow \varphi^\u$ is not provable in $\mathsf{cICK_{S5}}$. In particular, $\Rightarrow \varphi^\u$ is not $\Sigma$-provable for $\Sigma = \Cl^\neg(\varphi)$. Thus, by the same argument as in the proof of Theorem~\ref{t: completeness of S5}, $\varphi$ is falsified in some world of the canonical model $\fw^\Sigma$. Observe that the size of the canonical model (i.e. the number of worlds) is bounded by $2^{\lvert \Sigma \rvert}$, since every world is a subset of $\Sigma$, whereas $\Sigma$ is finite by Lemma~\ref{l: negation closure is finite}. Hence $\varphi$ is falsified on a finite S5 model.
\end{proof}

   \section{A Modal Variant of Kuroda's Translation}\label{c: ICK, Section negative tranlsation}

    Glivenko's Theorem~\cite{Glivenko_1929} shows that classical propositional logic ($\mathsf{CPC}$) can be embedded into $\mathsf{IPC}$ via a \emph{double-negation translation}, which maps every propositional formula $\varphi$ onto $\neg \neg \varphi$. The formula $\neg \neg \varphi$ is classically equivalent to $\varphi$ (but not intuitionistically) and furthermore satisfies the property that $\varphi$ is classically valid if and only if $\neg \neg \varphi$ is intuitionistically valid. A similar translation embedding classical first-order predicate logic into intuitionistic first-order predicate logic is Kuroda's translation~\cite{kuroda_1951}, which prefixes every formula with $\neg \neg$ and additional adds $\neg \neg$ after every universal quantifier. 
    
    This section shows that classical common knowledge logic ($\mathsf{CK}$) over S5 models can be embedded into $\mathsf{ICK}$ over S5. This will be achieved by constructing a modal variant of Kuroda's translation, which assigns to each formula $\varphi$ of $\LICK$ a formula $tr(\varphi)$ by adding $\neg \neg$ in front of $\varphi$ and after every knowledge and common knowledge operator. We will then show that for any formula $\varphi$,
    \begin{enumerate}
        \item $\varphi$ is classically equivalent to its translation $tr(\varphi)$ and
        \item  $\varphi$ is classically valid if and only $tr(\varphi)$ is intuitionistically valid,
    \end{enumerate}
    which entails that $\mathsf{CK}$ can be regarded as a fragment of $\mathsf{ICK}$. First, let us briefly recall some basic definitions for $\mathsf{CK}$.
    
    The \emph{language} of classical common knowledge logic is $\mathcal{L}_{\ICK}$. Formulas are evaluated on classical S5 epistemic models. In order to distinguish these models from the S5 epistemic models for $\mathsf{ICK}$, we will refer to the former as classical and to the latter as intuitionistic S5 epistemic models.
    
    \begin{definition}
        A \emph{classical S5 epistemic model} is a tuple $\fw = (W, \{R_i\}_{i \in \A}, V)$ where
        \begin{itemize}
            \item $W$ is a non-empty set of \emph{worlds};
            \item $R_i \subseteq W \times W$ is an equivalence relation for each $i \in \A$;
            \item $V: W \longrightarrow \mathcal{P}(\Prop)$ is a valuation function.
        \end{itemize}
    \end{definition}

In difference to intuitionistic epistemic models, classical epistemic models do not feature the intuitionistic order $\leq$. Formulas are evaluated on classical S5 epistemic models as follows. Let $\fw = (W, \{R_i\}_{i \in \A}, V)$ be a classical S5 epistemic model and $w \in W$ a world.
\begin{center}
    \begin{tabular}{l l l}
        $\fw, w \not \models \bot$ &&\\
        $\fw, w \models p$ & iff & $p \in V(w)$ \\
        $\fw, w \models \varphi \wedge \psi$ & iff & $\fw, w \models \varphi$ and $\fw, w \models \psi$\\
        $\fw, w \models \varphi \vee \psi$ & iff & $\fw, w \models \varphi$ or $\fw, w \models \psi$ \\
        $\fw, w \models \varphi \rightarrow \psi$ & iff & $\fw, w \not \models \varphi$ or $\fw, w \models \psi$ \\
        $\fw, w \models \K_i \varphi$ & iff & for all $v \in W$ if $w \mathrel{R_i}v$, then $\fw, v \models \varphi$ \\
        $\fw, w \models \C \varphi$ & iff & for all $v \in W$ if $w \mathrel{R^*} v$, then $\fw, v \models \varphi$.
    \end{tabular}
\end{center}
The notions of \emph{satisfiability} and \emph{validity} are defined as usual. Let $\mathbf{CK_{S5}}$ denote the set of valid $\mathcal{L}_{\ICK}$-formulas over the class of classical S5 models. It is easily checked that $\fw, w \models \neg \varphi$ where $\neg \varphi = \varphi \rightarrow \bot$ if and only if $\fw, w \not \models \varphi$ and therefore that $\fw, w \models \varphi$ if and only if $\fw, w \models \neg \neg \varphi$.

The following definition introduces a function $tr: \mathcal{L}_{\ICK} \longrightarrow \mathcal{L}_{\ICK}$ which serves as our modal variant of Kuroda's translation.

\begin{definition}
    Define the function $\tau: \mathcal{L}_{\ICK} \longrightarrow \mathcal{L}_{\ICK}$ by induction on $\varphi$ as follows.
    \begin{align*}
        \tau(\bot) & \coloneqq \bot  & \tau(p) & \coloneqq p \text{ for } p \in \Prop\\
        \tau(\varphi \wedge \psi) & \coloneqq \tau (\varphi) \wedge \tau(\psi) & \tau(\varphi \vee \psi) & \coloneqq \tau (\varphi) \vee \tau(\psi)\\
        \tau(\varphi \rightarrow \psi) & \coloneqq \tau (\varphi) \rightarrow \tau(\psi) & \tau(\C \varphi) & \coloneqq \C \neg \neg \tau(\varphi) \\
        \tau(\K_i \varphi) & \coloneqq \K_i \neg \neg \tau(\varphi) \text{ for } i \in \A 
    \end{align*}

Then define $tr: \mathcal{L}_{\ICK} \longrightarrow \mathcal{L}_{\ICK}$ by $tr(\varphi) \coloneqq  \neg \neg \tau(\varphi).$
\end{definition}

In the following we show that $tr$ satisfies Properties 1. and 2. stated above.

\begin{lemma}\label{l: tr is classicaly equivalent}
    For any $\mathcal{L}_\ICK$-formula $\varphi$, $\varphi  \leftrightarrow tr(\varphi) \in \mathbf{CK_{S5}}$. 
\end{lemma}
\begin{proof}
    We first prove by induction on the structure of $\varphi$ that for any classical S5 model $\fw=(W, \{R_i\}_{i \in \A}, V)$ and any world $w \in W$, $\fw,w \models \varphi$ if and only if $\fw,w \models \tau(\varphi)$. The base cases for $\varphi = \bot$ and $\varphi = p$ for $p \in \Prop$ are trivial. The cases for $\varphi = \psi \ast \chi$ where $\ast \in \{\wedge, \vee, \rightarrow\}$ follow immediately from the induction hypothesis. Suppose $\varphi = \K_i \psi$. By definition $\tau(\varphi) = \K_i \neg \neg \tau(\psi)$. Then $\fw,w \models K_i \psi$ if and only if $M,v \models \psi$ for all $v \in W$ with $w\mathrel{R_i} v$ if and only if (by induction hypothesis) $\fw,v \models \tau(\psi)$ for all $v \in W$ with $w \mathrel{R_i} v$ if and only if $\fw,v \models \neg \neg \tau(\psi)$ for all $v \in W$ with $w \mathrel{R_i} v$ if and only if $\fw,w \models \K_i \neg \neg \tau(\varphi)$. The case for $\varphi = \C \psi$ is similar. \smallskip

    By definition, $tr(\varphi) = \neg \neg \tau(\varphi)$. Then $\fw,w \models \varphi$ if and only if $\fw,w \models \tau(\varphi)$ if and only if $\fw,w \models \neg \neg \tau(\varphi)$. Therefore $\varphi \leftrightarrow tr(\varphi) \in \mathbf{CK_{S5}}$.
\end{proof}

Recall that $\mathbf{ICK_{S5}}$ has the finite model property. Therefore to check whether a formula is valid it suffices to consider the class of \emph{finite} intuitionistic S5 epistemic models. Let ${\fw=(W, \leq, \{R_i\}_{i \in \A}, V)}$ be a finite intuitionistic S5 epistemic model. Then $w \in W$ is called a \emph{maximal world} if for all $v \in W$, if $w \leq v$, then $w=v$. 

\begin{definition}
    Let $\fw=(W, \leq, \{R_i\}_{i \in \A}, V)$ be a finite intuitionistic S5 epistemic model. The \emph{$\fw$-induced model} $\fw^c =(W^c, \{R_i^c\}_{i \in \A}, V^c)$ is defined as follows.
    \begin{itemize}
        \item[] $W^c \coloneqq \{w \in W \, \lvert \, w \text{ is a maximal world}\}$
        \item[] $R_i^c \coloneqq R_i \cap (W^c \times W^c)$
        \item[] $V^c \coloneqq V \cap (W^c \times \mathcal{P}(\Prop))$
    \end{itemize}
\end{definition}

\begin{lemma}
    Let $\fw=(W, \leq, \{R_i\}_{i \in \A}, V)$ be a finite intuitionistic S5 epistemic model. The $\fw$-induced model is a classical S5 epistemic model.
\end{lemma}
\begin{proof}
    Since $\fw$ is a finite model, maximal worlds exist and therefore $W^c \not = \emptyset$. By definition $V^c: W^c \longrightarrow \mathcal{P}(\Prop)$. It remains to show that each $R_i^c$ is an equivalence relation on $W^c$. This follows immediately from the fact that $R_i^c$ is the restriction of $R_i$ to worlds in $W^c$, and $R_i$ is an equivalence relation on $W$.
\end{proof}

\begin{lemma}\label{l: tau_0 preserves truth}
     Let $\fw=(W, \leq, \{R_i\}_{i \in \A}, V)$ be a finite intuitionistic S5 epistemic model and let $\fw^c =(W^c, \{R_i^c\}_{i \in \A}, V^c)$ be the $\fw$-induced model. Let $w \in  W^c$. For any $\mathcal{L}_\ICK$-formula $\varphi$, $\fw,w \models \tau(\varphi)$ if and only if $\fw^c, w \models \varphi$.  
\end{lemma}
\begin{proof}
    We proceed by induction on the structure of $\varphi$. The base cases where $\varphi = \bot$ and $\varphi = p$ for $p \in \Prop$ are trivial. The cases where $\varphi = \psi \ast \gamma$ for $\ast \in \{\wedge, \vee\}$ follow immediately from the induction hypothesis. Suppose $\varphi = \psi \rightarrow \gamma$. Then $\tau(\varphi) = \tau(\psi) \rightarrow \tau(\gamma)$. Since $w$ is a maximal world, observe that $\fw,w \models \tau(\psi) \rightarrow \tau(\gamma)$ if and only if $\fw,w \not \models \tau(\psi)$ or $\fw,w \models \tau(\gamma)$. Therefore the claim follows immediately from the induction hypothesis.\smallskip 
    
    \noindent \textsc{Case for $\varphi = \K_i \psi$.} By definition, $\tau(\varphi) = \K_i \neg \neg \tau(\psi)$. Suppose first that $\fw^c, w \models \K_i \psi$. Then for all $v \in W^c$ if $w \mathrel{R_i^c} v$, then $\fw^c,v \models \psi$. Suppose $w \mathrel{R_i} u$. Let $v \geq u$ be a maximal world. Reflexivity and transitivity of $R_i$ as well as triangle confluence imply that $w \mathrel{R_i} v$. Therefore $w \mathrel{R_i^c} v$. By assumption $\fw^c, v \models \psi$ and hence, by induction hypothesis, $\fw,v \models \tau(\psi)$. Thus $\fw,u \models \neg \neg \tau(\psi)$ and so $\fw,w \models \K_i \neg \neg \tau(\psi)$. 
    
    For the other direction suppose that $\fw, w \models \K_i \neg \neg \tau(\psi)$. Then for all $v \in W$ if ${w \mathrel{R_i} v}$, then $\fw,v \models \neg \neg \tau(\psi)$. Let $v \in W^c$ such that $w \mathrel{R_i^c} v$. Then $w \mathrel{R_i} v$ and therefore ${\fw, v \models \neg \neg \tau(\psi)}$. Since $v$ is maximal it follows that $\fw,v \models \tau(\psi)$. By induction hypothesis $\fw^c, v \models \psi$. Hence $\fw^c, w \models \K_i \psi$.\smallskip

    \noindent \textsc{Case for $\varphi = \C \psi$.}  By definition, $\tau(\varphi) = \C \neg \neg \tau(\psi)$. Let $u_0, \ldots, u_n$ such that $u_0 = w$ and for each $0 \leq i < n$ there is $j \in \A$ with $u_i \mathrel{R_j} u_{i+1}$.
    
    We prove by induction on $n$ that there exists $v_0, \ldots, v_n \in W^c$ such that $v_0 = w$ and for all $0 \leq i \leq n$, $u_i \leq v_i$ and if $i <n$ and $u_i \mathrel{R_j} u_{i+1}$, then $v_i \mathrel{R_j^c} v_{i+1}$. For $n=0$ let $v_0 = w$. For $n>0$ the induction hypothesis yields that there are $v_0, \ldots, v_{n-1} \in W^c$ with $v_0 = w$, for all $0 \leq i \leq n-1$, $u_i \leq v_i$ and if $i < n-1$ and $u_i \mathrel{R_j} u_{i+1}$ for some $j \in \A$, then $v_i \mathrel{R_j^c} v_{i+1}$. Let $j \in \A$ such that $u_{n-1}\mathrel{R_j} u_n$. Let $v_n \in W$ be a maximal world with $u_n \leq v_n$. By reflexivity $v_n \mathrel{R_j} v_n$. By triangle confluence $u_n \mathrel{R_j} v_n$ and so by transitivity $u_{n-1} \mathrel{R_j} v_n$. Since $u_{n-1} \leq v_{n-1}$, by reflexivity and triangle confluence $u_{n-1} \mathrel{R_j} v_{n-1}$. By symmetry and transitivity $v_{n-1} \mathrel{R_j} v_n$. Hence, since $v_{n-1}, v_n \in W^c$, we have $v_{n-1} \mathrel{R_j^c} v_n$.

    Now suppose that $\fw^c, w \models \C \psi$. Then for all $v \in W^c$ if $w \mathrel{(R^c)^*} v$, then ${\fw,v \models \psi}$. Suppose that $w \mathrel{R^*} u$ for some $u \in W$. By the proof above there exists $v \in W^c$ with ${w\mathrel{(R^c)^*} v}$ and $u \leq v$. By assumption $\fw^c, v \models \psi$. Thus by induction hypothesis ${\fw,v \models \tau(\psi)}$. Since $u \leq v$ it follows that $\fw,u \models \neg \neg \tau(\psi)$. Therefore $\fw,w \models \C \neg \neg \tau(\psi)$. 

    For the other direction suppose that $\fw,w \models \C \neg \neg \tau(\varphi)$. Then for all $u \in W$ if $w \mathrel{R^*} u$, then $\fw,u \models \neg \neg \tau (\psi)$. Let $v \in W^c$ with $w \mathrel{(R^c)^*} v$. Then $w \mathrel{R^*} v$ and so $\fw,v \models \neg \neg \tau(\psi)$. Since $v$ is a maximal world, $\fw,v \models \tau(\psi)$. By induction hypothesis $\fw^c, v \models \psi$. Therefore $\fw^c, w \models \C \psi$.    
\end{proof}

Note that every classical S5 epistemic model $\fw =(W, \{R_i\}_{i \in \A}, V)$ induces an intuitionistic S5 epistemic model $\fw^i = (W, \leq, \{R_i\}_{i \in \A}, V)$ by setting $w \leq v$ if and only if $w = v$. Observe that $R_i$ is triangle confluent for each $i \in \A$.

\begin{lemma}\label{l: classical model induces intuitionistic model}
    If $\fw = (W, \{R_i\}_{i \in \A}, V) $ is a classical S5 epistemic model and $w \in W$, then for any formula $\varphi$, $\fw, w \models \varphi$ if and only if $\fw^i, w \models \varphi$.
\end{lemma}
\begin{proof}
    By definition of $\leq$, $\fw^i, w \models \varphi \rightarrow \psi$ if and only if $\fw^i, w \not \models \varphi$ or $\fw^i, w \models \psi$.
\end{proof}

\begin{theorem}\label{t: negative translation}
    For any $\LICK$-formula $\varphi$ the following hold.
    \begin{enumerate}
        \item $\varphi \leftrightarrow tr(\varphi) \in \mathbf{CK_{S5}}$.
        \item $\varphi \in \mathbf{CK_{S5}}$ if and only if $tr(\varphi) \in \mathbf{ICK_{S5}}$.
    \end{enumerate}
\end{theorem}
\begin{proof}
     1. is proven in Lemma \ref{l: tr is classicaly equivalent}. For 2. let $\varphi \in \mathcal{L}_\ICK$ be any formula. First suppose that $tr(\varphi) \in \mathbf{ICK_{S5}}$. Then $tr(\varphi)$ is valid over the class of intuitionistic S5 epistemic models. Thus, in particular, $tr(\varphi)$ is valid over the class of classical S5 epistemic models by Lemma~\ref{l: classical model induces intuitionistic model}. Item 1. of this theorem implies that $\varphi \in \mathbf{CK_{S5}}$. For the other direction suppose that $\varphi \in \mathbf{CK_{S5}}$. Let $\fw$ be an arbitrary finite intuitionistic S5 epistemic model and let $w$ be an arbitrary world. Let $\fw^c$ be the $\fw$-induced model. Let $u$ be a maximal world of $\fw$ with $w \leq u$. By assumption $\fw^c, u \models \varphi$. By Lemma \ref{l: tau_0 preserves truth}, $\fw, u \models \tau(\varphi)$. Therefore $\fw,w \models \neg \neg \tau(\varphi)$, i.e. $\fw,w \models tr(\varphi)$. Therefore $tr(\varphi)$ is valid over the class of finite intuitionistic S5 epistemic models. Corollary~\ref{c: small model property} yields that $tr(\varphi)$ is valid over the class of intuitionistic S5 epistemic models, i.e. $tr(\varphi) \in  \mathbf{ICK_{S5}}$.
\end{proof}

\section{Complexity}\label{c: ICK, Section complexity}

The last section of this chapter investigates the computational properties of the cyclic calculus $\CICK_{\mathsf{S5}}$. First, we show that proof search in $\CICK_{\mathsf{S5}}$ can be automated by reducing the problem of finding a $\CICK_\mathsf{S5}$-proof for a sequent to the problem of solving a certain \emph{parity game} (for which computationally well-behaved algorithms exist, see e.g.~\cite{calude_2017}). From this we obtain an exponential upper bound for solving the \emph{proof search problem} for $\CICK_\mathsf{S5}$ (see Table~\ref{tab: proof search problem}). By using the results from the previous section about the modal variant of Kuroda's translation as well as soundness and completeness, we then provide a polynomial reduction of the proof search problem for $\CICK_\mathsf{S5}$ to the \emph{validity problem} for $\mathbf{CK}$ (see Table~\ref{tab: validity problem CK}), which is known to be \textsc{ExpTime}-complete~\cite{halpern_1992}, thus establishing that the proof-search problem is \textsc{ExpTime}-complete as well. We assume familiarity with parity games; for an introduction the reader is referred to~\cite[Chapter 2]{graedel_2001}.
    
\begin{table}[t]
    \centering
    \begin{tabular}{|l l|}
    \hline
      \textbf{Input:}   &  A sequent $\sigma$.\\
       \textbf{Question:}  &  Is $\sigma$ $\CICK_{\mathsf{S5}}$-provable? \\
       \hline
    \end{tabular}
    \caption{The proof search problem for $\ICK_\mathsf{S5}$.}
    \label{tab: proof search problem}
\end{table}

\subsection{Proof Search Games}

 Each sequent $\sigma$ is associated with a parity game $\mathcal{G}_\sigma$ called the \emph{proof search game} associated to $\sigma$. This game is played two players, one called Prover and the other called Refuter. Intuitively, like before, Prover is trying to show that a given sequent is provable, while Refuter tries to show the opposite. In difference to the proof search games employed in Chapter \ref{c: IM}, the games here are parity games and not Gale--Steward games. The proof search games employed here are not used to prove completeness and hence we do not require to obtain a countermodel from a winning strategy for Refuter. Instead, it suffices to show that Prover has a winning strategy in $\mathcal{G}_\sigma$ if and only if $\sigma$ has a $\CICK_{\mathrm{S5}}$-proof. Crucially, parity games\index{parity game} are \emph{memoryless determined}, meaning that exactly one of the two players has a memoryless winning strategy (see e.g.~\cite[Chapter 6]{graedel_2001}). From this we will be able to obtain that $\CICK_{\mathrm{S5}}$ is \emph{uniformly complete}, c.f. Theorem \ref{t: strategy iff provable}. 
 
 Recall that a rule instance in a sequent calculus is a tuple $\langle \sigma, \langle \sigma_1, \ldots, \sigma_n \rangle \rangle$ and a rule is a set of rule instances (c.f. Definition \ref{d: rule instances and rules}).

 \begin{table}[t]
    \centering
    \begin{tabular}{|l l|}
    \hline
      \textbf{Input:}   &  An $\LICK$-formula $\varphi$.\\
       \textbf{Question:}  &  Is $\varphi \in \mathbf{CK_{S5}}$? \\
       \hline
    \end{tabular}
    \caption{The validity problem for $\mathbf{CK_{S5}}$.}
    \label{tab: validity problem CK}
\end{table}

\begin{definition}
		A \emph{rule position} in $\CICK_\mathsf{S5}$ is a triple $\langle \sigma, \mathsf{r}, \langle \sigma_1, \ldots, \sigma_n \rangle\rangle$ such that $\mathsf{r}$ is a rule of $\CICK_{\mathrm{S5}}$ and $\langle \sigma, \langle \sigma_1, \ldots, \sigma_n \rangle \rangle \in \mathsf{r}$.
	\end{definition}
    
	We use $\pi_n^k$ to denote the projection function which takes as input an $n$-tuple and outputs the $k$-th component. Thus for a rule position $i =\langle \sigma, \mathsf{r}, \langle \sigma_1, \ldots, \sigma_n \rangle\rangle$, $\pi_3^1(i) = \sigma$, $\pi_3^2(i) = \mathsf{r}$ and $\pi_3^3(i) = \langle \sigma_1, \ldots, \sigma_n \rangle$. Given a finite set of formulas $\Sigma$, a \emph{$\Sigma$-rule position} is a rule position involving only $\Sigma$-sequents. For the remainder of this section $\Sigma$ always denotes a finite and negation closed set of formulas. We associate to each $\Sigma$-sequent $\sigma$ a game as follows.
	\begin{definition}
		Let $\sigma$ be a $\Sigma$-sequent. The \emph{proof search game} $\mathcal{G}_\sigma$ associated to $\sigma$ takes positions in $S \cup I$, where $S$ is the set of $\Sigma$-sequents and $I$ is the set of $\Sigma$-rule positions in $\CICK_\mathsf{S5}$. The ownership function and admissible moves are as described in the following table:
		\begin{center}
			\begin{tabular}{|c|c|c|}
				\hline
				Position & Owner & Admissible moves \\
				\hline
				$\sigma$ & Prover & $\{i \in I \mid \pi_3^1(i) = \sigma\}$ \\
				$\langle \sigma, \mathsf{r}, \langle \sigma_1, \ldots, \sigma_n \rangle\rangle$ & Refuter & $\{\sigma_i \mid 1 \leq i \leq n\}$\\
				\hline 
			\end{tabular}
		\end{center}
		The positions are given the following priorities:
		\begin{enumerate}
			\item Every position of the form $\Gamma \Rightarrow \Delta^\u$ has priority $3$;
			\item Every position of the form $\langle\sigma, \mathsf{C R}, \langle \sigma_1, \ldots, \sigma_n \rangle\rangle$ where the principal formula is in focus has priority $2$;
			\item Every other position has priority $1$.
		\end{enumerate}
		A position is called a \emph{dead end} if its owner has no admissible moves in this position available. A \emph{play} in $\mathcal{G}_\sigma$ is a sequence of positions starting in $\sigma$, such that any two consecutive positions are related by an admissible move. A play is either finite and ends in a dead end or infinite. The \emph{winning conditions} for a play are as follows: Prover wins every finite play in which the dead end belongs to Refuter and every infinite play in which the highest priority encountered infinitely often is even. Refuter wins every finite play in which the dead end belongs to Prover and every infinite play in which the highest priority encountered infinitely often is odd.
	\end{definition}

Observe that the only dead ends are rule instances of axioms. Therefore Prover wins every finite play. It is straightforward to check that for every $\Sigma$-sequent $\sigma$ the game $\mathcal{G}_\sigma$ is a parity game. Given a set $X$, let $X^{< \omega} \coloneqq \bigcup_{n < \omega} X^n$. Therefore the set of all initial segments of all possible plays in a game is a \emph{subset} of $(S \cup I)^{< \omega}$. The following definitions apply to both Prover and Refuter. We write `Player' instead.

  \begin{definition}
      Let $\mathcal{G}_\sigma$ be a proof search game with positions $S \cup I$. 
      \begin{enumerate}
          \item A \emph{strategy}\index{strategy} for Player is partial function $\mathcal{S}: (S \times I)^{< \omega} \longrightarrow S \cup I$ which maps each initial segment of a play ending in a position owned by Player for which admissible moves exist onto an admissible move.
          \item A \emph{memoryless strategy}\index{strategy!memoryless} for Player is a partial function $\mathcal{S}: S \cup I \longrightarrow S \cup I$ which maps each position owned by Player onto an admissible move.
      \end{enumerate}
  \end{definition}

A strategy $\mathcal{S}$ for Player is \emph{winning} if Player wins every play in which $\mathcal{S}$ is \emph{used}. The \emph{strategy tree} of a memoryless strategy $\mathcal{S}$ is the tree of all possible plays that can occur when Player uses $\mathcal{S}$.

\begin{definition}
    Let $\sigma$ be a $\Sigma$-sequent and $\mathcal{G}_\sigma$ the proof search game associated to $\sigma$. Let $\mathcal{S}$ be a memoryless strategy for Player. The \emph{strategy tree}\index{strategy tree} $\mathcal{T}_\mathcal{S}$ is a $(S \cup I)$-labeled tree defined as follows.
    \begin{enumerate}
        \item The root of $\mathcal{T}_\mathcal{S}$ is labeled by $\sigma$.
        \item If a node $t$ of $\mathcal{T}_\mathcal{S}$ is labeled by $p \in S \cup I$ where $p$ is owned by Player and an admissible move exists, then $t$ has a unique child $u$ labeled by $\mathcal{S}(p)$.
        \item If a node $t$ of $\mathcal{T}_\mathcal{S}$ is labeled by $p \in S \cup I$ which is not owned by Player, then 
        \begin{enumerate}
            \item if $p \in S$ and $I_p = \{i \in I \mid \pi_3^1(i)=p\}$, then $t$ has $\cardinality{I_p}$ children, each labeled with a different $i \in I_p$.
            \item if $p \in I$ and $\pi_3^3(i) = \langle \sigma_1, \ldots, \sigma_n\rangle$, then $t$ has $n$ children, each labeled with a different $\sigma_i$.
        \end{enumerate}
    \end{enumerate}
\end{definition}

Note that due to $\Sigma$ being finite and rules having finitely many premises, strategy trees are finite branching (and possibly non-wellfounded). The goal is to show that a sequent $\sigma$ has a $\CICK_\mathrm{S5}$-proof if and only if Prover has a memoryless winning strategy in $\mathcal{G}_\sigma$. To that end we require to do some preliminary work. Recall that in a cyclic proof $\pi$ there exists for every non-axiomatic leaf $l$ a node $u$ such that $(u, l)$ is a successful repetition. As there might exist several candidates for the node $u$, we fix for each non-axiomatic leaf $l$ one candidate $c(l)$ which we call the \emph{companion} of $l$.
      
      \begin{definition}
          A \emph{finite tree with back edges} is a pair $(\mathcal{T},f)$ where $\mathcal{T}$ is a finite tree and $f: \mathcal{T} \longrightarrow \mathcal{T}$ a partial function, such that every $u \in dom(f)$ is a leaf of $\mathcal{T}$ and the node $f(u)$ is a proper ancestor of $u$.
      \end{definition}

     Observe that cyclic proofs can be considered to be finite trees with back edges which satisfy the property that if $u \in dom(f)$, then $u$ is a non-axiomatic leaf and $f(u)$ is the companion of $u$.
	
	  \begin{definition}
	      Let $(\mathcal{T}, f)$ be a finite tree with back edges. A \emph{looping path} $\rho$ through $(\mathcal{T},f)$ is a (possibly infinite) sequence $\rho = \rho(0), \rho(1),\ldots$ of nodes in $\mathcal{T}$ which satisfies the following properties:
	      \begin{enumerate}
	         \item $\rho(0)$ is the root of $\mathcal{T}$.
	         \item If $\rho(i)$ is a leaf $l$ of $\mathcal{T}$ and $l \not \in dom(f)$, then $\rho$ is finite and ends at $\rho(i)$.
	          \item If $\rho(i)$ is a leaf $l$ of $\mathcal{T}$ with $f(l) = u$, then $\rho(i+1)$ is a child node of $u$.
	          \item Otherwise, $\rho(i+1)$ is a child node of $\rho(i)$.
	      \end{enumerate}
	  \end{definition}
	  
	   If condition 3. applies, then we say that $\rho$ \emph{passes through the leaf} $l$. The following lemma states a first basic result about infinite looping paths through finite trees with back edges. The proof of the lemma follows immediately from the fact that such trees are finite.

	  \begin{lemma}\label{l: infinite path passes through leaf}
	      Suppose $(\mathcal{T},f)$ is a finite tree with back edges and $\rho$ is an infinite looping path through $(\mathcal{T},f)$. Then there exists a leaf $l \in dom(f)$ through which $\rho$ passes infinitely often.
	  \end{lemma}

      Next, we define a partial order on the range of the back edge function $f$.

      \begin{definition}
          Let $(\mathcal{T},f)$ be a finite tree with back edges. Define the \emph{one-step dependency order} $\preceq_1$ on $ran(f)$ as follows:
          \begin{center}
              $u \preceq_1 v$ $: \Leftrightarrow$ $u$ occurs on the path from $v$ to $v'$ for some $v' \in f^{-1}(v)$
          \end{center}
          Define the \textit{dependency order} $\preceq$ on $ran(f)$ as the transitive closure of $\preceq_1$.
      \end{definition}
      
      Note that $u \preceq v$ implies that $v$ is the companion of some leaf $v'$ and $u$ lies on the path from $v$ to $v'$. It is routine to check that the dependency order $\preceq$ is reflexive, transitive and antisymmetric, implying that $\preceq$ is a partial order on $ran(f)$. We will use the dependency order to show that for every infinite looping path there exists a lowermost companion through which the path passes infinitely often. Let $(\mathcal{T},f)$ be a finite tree with back edges and let $\rho$ be an infinite looping path through $(\mathcal{T},f)$. Denote by $Inf(\rho)$ the set of nodes of $\mathcal{T}$ that occur infinitely often in $\rho$.
      
      \begin{lemma}\label{l: existence of greatest element}
          Let $(\mathcal{T},f)$ be a finite tree with back edges and let $\rho$ be an infinite looping path through $(\mathcal{T},f)$. Then the set $Inf(\rho) \cap ran(f)$ has a $\preceq$-greatest element.
      \end{lemma}
      
      \begin{proof}
          Observe that the set $Inf(\rho) \cap ran(f)$ is finite since $\mathcal{T}$ is a finite tree. Furthermore, observe that $Inf(\rho) \cap ran(f)$ is non-empty, as $\rho$ must pass through some leaf $l \in dom(f)$ infinitely often (see Lemma \ref{l: infinite path passes through leaf}). It therefore suffices to prove that all $\preceq$-maximal elements in $Inf(\rho) \cap ran(f)$ are identical. To that end let $u \in Inf(\rho) \cap ran(f)$ be a $\preceq$-maximal element and let $\mathcal{T}_u$ be the subtree of $\mathcal{T}$ rooted at $u$. First of all, if $l \in Inf(\rho) \cap dom(f) \cap \mathcal{T}_u$, then $f(l)$ is a proper ancestor of $l$, implying that either $f(l)$ belongs to $\mathcal{T}_u$ or $f(l)$ is a proper ancestor of $u$. Note that $f(l) \in Inf(\rho)$, since $l \in Inf(\rho)$. Hence, since $u$ is $\preceq$-maximal, $f(l)$ cannot be a proper ancestor of $u$, as otherwise $u \prec f(l)$, implying that $f(l)$ must belong to $\mathcal{T}_u$. Note that there exists a natural number $n$ such that the suffix $\rho' = (\rho(i))_{i \geq n}$ only passes through leafs $l \in dom(f) \cap Inf(\rho)$. Let $k \geq n$ be the least natural number such that $\rho(k) = u$. Since for each $l \in dom(f) \cap Inf(\rho) \cap \mathcal{T}_u$ holds that $f(l) \in \mathcal{T}_u$, it follows that  $\rho(i)$ belongs to $T_u$ for all $i \geq k$. Therefore $Inf(\rho) \cap \mathcal{T}_u = Inf(\rho)$. Since $u$ is $\preceq$-maximal, it thus follows that $u$ is the $\preceq$-greatest element of $Inf(\rho) \cap ran(f)$.
      \end{proof}

     A slightly different version of the following lemma is proven in~\cite[Proposition 2]{Rooduijn_2021}.

     \begin{lemma}\label{l: infinite looping path has good suffix}
         Suppose $\pi$ is a $\CICK_\mathsf{S5}$-proof and $\rho$ is an infinite looping path through $\pi$. Then there exists a suffix $\rho'$ of $\rho$ in which every sequent has a formula in focus.
     \end{lemma}
     \begin{proof}
         Let $f$ be the back-edge function of $\pi$. By Lemma \ref{l: existence of greatest element} there exists a non-axiomatic leaf $l_0 \in \pi$ such that $c(l_0)$ is $\preceq$-greatest in $Inf(\rho) \cap ran(f)$. In particular, this implies that the following hold.
         \begin{enumerate}
             \item $\rho$ passes through $l_0$ infinitely often.
             \item If $Inf(\rho) \cap dom(f) = \{l_0, \ldots, l_n\}$, then there exists a path from $c(l_0)$ to $c(l_i)$ for all $1 \leq i \leq n$.
         \end{enumerate}
         Let $\pi_0$ be the subtree rooted at $c(l_0)$. We claim that every path from $c(l_0)$ to $l_i$ for $0 \leq i \leq n$ always has a formula in focus. The proof proceeds by induction on the cardinality of $Inf(\rho) \cap dom(f)$. The base case where $Inf(\rho) \cap dom(f) = \{l_0\}$ is trivial since the path from $c(l_0)$ to $l_0$ is successful by definition of a cyclic proof. For the induction step suppose that $Inf(\rho) \cap dom(f) = \{l_0, \ldots, l_n, l_{n+1}\}$. By induction hypothesis the path from $c(l_0)$ to $l_i$ for $0 \leq i \leq n$ always has a formula in focus. Consider the path $\rho'$ from $c(l_0)$ to $l_{n+1}$. If $c(l_{n+1})$ lies on the path from $c(l_0)$ to $l_0$, then every sequent occurring on the path from $c(l_0)$ to $c(l_{n+1})$ has a formula in focus and, by definition of a cyclic proof, also every sequent from $c(l_{n+1})$ to $l_{n+1}$. If the companion $c(l_{n+1})$ of $l_{n+1}$ does not lie on the path from $c(l_0)$ to $l_0$, then since $\rho$ passes infinitely often through $l_0$ and through $l_{n+1}$, there must be $1 \leq i \leq n$ such that $l_i$ belongs to the subtree $\pi_1$ of $\pi_0$ rooted at $c(l_{n+1})$ and $c(l_i)$ does not belong to $\pi_1$, as otherwise, once $\rho$ passes through $l_{n+1}$ it can never pass through $l_0$ again, contradicting the assumption that $l_0, l_{n+1} \in Inf(\rho)$. By induction hypothesis the path from $c(l_0)$ to $l_i$ has always a formula in focus. Observe that this path must contain the path from $c(l_0)$ to $c(l_{n+1})$ as an initial segment, since $l_i$ belongs to $\pi_1$. Hence the path from $c(l_0)$ to $l_{n+1}$ always has a formula in focus.

         Let $i$ be the least natural number such that the suffix $\rho'$ of $\rho$ starting at $\rho(i)$ only passes through non-axiomatic leafs in $Inf(\rho)$. Then, by the previous observation, $\rho'$ always has a formula in focus.
     \end{proof}

	 Recall that a play $m$ in a proof search game is a sequence  $m(0),m(1),m(2), \ldots$ of positions. Observe that all even positions $m(2i)$ are owned by Prover and all odd positions $m(2i+1)$ by Refuter. For an initial segment $m(0), \ldots m(i)$ of $m$ we say that its \emph{length} is $i$.
	
	\begin{definition}
	Let $\sigma$ be a $\Sigma$-sequent and let $\pi$ be a $\Sigma$-proof of $\sigma$ in $\CICK_\mathsf{S5}$. Let $\rho$ be an infinite looping path through $\pi$ and let $m$ be an infinite play in $\mathcal{G}_\sigma$.
	\begin{itemize}
	    \item An initial segment $m(0), \ldots, m(i)$ of $m$ \emph{corresponds} to an initial segment $\rho(0), \ldots, \rho(j)$ of $\rho$ if $i = 2j$ and for each $0 \leq k \leq i$ with $k = 2l$ it holds that $m(k)$ is the sequent that labels $\rho(l)$.
	    \item The play $m$ and the path $\rho$ are called \emph{corresponding} if every initial segment of $m$ with even length corresponds to an initial segment of $\rho$.
	\end{itemize}
	 
	\end{definition}

     \begin{proposition}\label{p: provable implies winning strategy}
		Let $\sigma$ be a $\Sigma$-sequent. If $\sigma$ is $\Sigma$-provable in $\CICK_\mathsf{S5}$, then Prover has a memoryless winning strategy in $\mathcal{G}_\sigma$.
	\end{proposition}
	
	\begin{proof}
	 Suppose that $\sigma$ is $\Sigma$-provable in $\CICK_\mathrm{S5}$ and let $\pi$ be a $\CICK_\mathrm{S5}$-proof of $\sigma$ in which every occurring sequent is a $\Sigma$-sequent. We denote the root of $\pi$ by $r_\pi$. We simultaneously define a strategy $\mathcal{S}$ for Prover in the game $\mathcal{G}_\sigma$ and show how to map each initial segment of a play of even length in which Prover uses $\mathcal{S}$ onto an initial segment of a looping path through $\pi$. The strategy $\mathcal{S}$ is a partial function which maps initial segments of plays $ m(0),\ldots, m(2i) $ of even length onto rule positions. Therefore strategy $\mathcal{S}$ uses memory.\smallskip
	 
	 For the base case observe that each play in $\mathcal{G}_\sigma$ begins in $\sigma$. Therefore $\langle m(0) \rangle$ for $m(0)= \sigma$ is an initial segment of every play in $\mathcal{G}_\sigma$. Similarly, every looping path through $\pi$ starts in $r_\pi$ which is labeled by $\sigma$. Thus $\langle \rho(0) \rangle$ for $\rho(0) = r_\pi$ is an initial segment of every looping path through $\pi$. Observe that $\langle m(0) \rangle$ corresponds to $\langle \rho(0) \rangle$.\smallskip
	 
	 For the inductive step suppose that we have already mapped the initial segment 
	 \begin{equation*}
	     m_i = \langle m(0), \ldots, m(2i) \rangle
	 \end{equation*}
	 of a play onto the initial segment
	 \begin{equation*}
	     \rho_i = \langle \rho(0),  \ldots, \rho(i) \rangle
	 \end{equation*}
	 of a looping path, such that $m_i$ corresponds to $\rho_i$, where $i \geq 0$. Let $j \in I$ be the $\Sigma$-rule position
	 \begin{equation*}
	     j = (m(2i), \mathsf{r}, \langle \sigma_1', \ldots, \sigma_k' \rangle),
	 \end{equation*}
	 which generates $\rho(i)$ in $\pi$ when read top down. Then define
	 \begin{equation*}
	     \mathcal{S}(m_i) = j.
	 \end{equation*}
	 Now suppose that Refuter extends the play by choosing premise $\sigma_l'$. Then let $m(2i+1) = j$ and let $m(2i+2) = \sigma_l'$ and extend the initial segment $m_i$ to
	 \begin{equation*}
	     m_{i+1} = \langle m(0),  \ldots, m(2i), m(2i+1), m(2i+2) \rangle
	 \end{equation*}
	 Furthermore, let $\rho(i+1)$ be the child of $\rho(i)$ which is labelled by $\sigma_l'$ and extend $\rho_i$ to
	 \begin{equation*}
	     \rho(i+1) = \langle \rho(0), \ldots, \rho(i), \rho(i+1) \rangle
	 \end{equation*}
	 Observe that $m(i+1)$ corresponds to $\rho(i+1)$.\smallskip
	 
	 Finally, in order to turn $\mathcal{S}$ into a function which maps \emph{every} initial segment of a play with even length (and thus every initial segment ending in a position owned by Prover) onto a rule instance (and not just those that correspond to initial segments of looping paths), we add the following clause. Fix a $\Sigma$-instance $j' \in I$. For any initial segment $m_i' = \langle m(0)', \ldots, m(2i)' \rangle$ of a play which is not covered in the above construction define $\mathcal{S}(m_i')= j'$. Observe that $\mathcal{S}$ is a well-defined strategy for Prover, which has the property that if $m$ is a play of $\mathcal{G}_\sigma$ in which Prover uses strategy $\mathcal{S}$, then there exists by construction a looping path $\rho$ through $\pi$ such that every initial segment of $m$ of even length corresponds to some initial segment of $\rho$. Therefore $m$ corresponds to $\rho$.\smallskip
	 
	 We show that $\mathcal{S}$ is a winning strategy for Prover. To that end let $m$ be a play in $\mathcal{G}_\sigma$ in which Prover uses strategy $\mathcal{S}$ and let $\rho$ be the looping path through $\pi$ which corresponds to $m$. In case $m$ is finite Prover wins by default. So suppose $m$ is infinite. Then $\rho$ is also infinite. By Lemma \ref{l: infinite looping path has good suffix}, $\rho$ has a suffix $\rho'$ in which every sequent has a formula in focus. Note that $\rho'$ passes through infinitely many rule instances of $\C \mathsf{R}$ where the principal formula is in focus: let $f(u)$ be the $\preceq$-greatest element of $Inf(\rho) \cap ran(f)$ where $f$ is the back-edge function of $\pi$. By assumption $\rho'$ passes infinitely often through the non-axiomatic leaf $u$. Between each two passings, $\rho'$ must pass through an instance of $\C \mathsf{R}$ with the principal formula in focus, since the path from $f(u)$ to $u$ is successful. Therefore, since $\rho$ corresponds to $m$, the play $m$ passes, after finitely many moves, only through positions with priority 1 or 2, and infinitely often through positions with priority 2. Hence, the highest priority encountered infinitely often is even and Prover wins the play. We conclude that $\mathcal{S}$ is a winning strategy for Prover. Finally, since in a given parity game exactly one of the two players has a \emph{memoryless} winning strategy~\cite{graedel_2001}, the existence of a winning strategy for Prover implies the existence of a memoryless winning strategy for Prover.
	 \end{proof}

Let us now consider the converse direction of Proposition \ref{p: provable implies winning strategy}.

	\begin{proposition}\label{proposition winning strategy implies provable}
		Let $\sigma$ be a $\Sigma$-sequent. If Prover has a memoryless winning strategy in $\mathcal{G}_\sigma$, then $\sigma$ is $\Sigma$-provable in $\CICK_\mathsf{S5}$. 
	\end{proposition}
	\begin{proof}
		Suppose that Prover has a memoryless winning strategy in $\mathcal{G}_\sigma$. Let $\mathcal{T}'$ be the corresponding strategy tree. Define the `condensed' labeled tree $\mathcal{T}=(T, \leq)$ of $\mathcal{T'}$ as follows.
        \begin{itemize}
            \item A node $t$ of $\mathcal{T}'$ is a node of $\mathcal{T}$ if and only if $t$ is labeled by $p \in S$ in $\mathcal{T}'$.
            \item A node $s$ in $\mathcal{T}$ is a child of a node $t$ in $\mathcal{T}$ if and only if there exists a node $s'$ in $\mathcal{T}'$ such that $s'$ is a child of $t$ and $s$ is a child of $s'$ in $\mathcal{T}'$.
            \item A node $t$ in $\mathcal{T}$ is labeled by $p \in S$ if and only if $t$ is labeled by $p$ in $\mathcal{T}'$. 
        \end{itemize}
        Note that $\mathcal{T}$ is a finite branching (and possibly non-wellfounded) tree labeled by $\Sigma$-sequents according to the rules of $\CICK_{\mathrm{S5}}$. Let $\pi$ be the finite subtree of $T$ obtained by pruning every infinite branch of $T$ at the first repetition. Observe that the root of $\pi$ is labelled by $\sigma$ and $\pi$ is generated by rules of $\CICK_\mathrm{S5}$. Furthermore, since $\mathcal{T}$ is finite branching, $\pi$ is finite by K\H{o}nig's Lemma. Therefore $\pi$ is a pre-proof. In order to show that $\pi$ is indeed a $\CICK_\mathsf{S5}$-proof, let $l$ be an arbitrary leaf of $\pi$. If $l$ is also a leaf of $\mathcal{T}$, then $l$ is a node in $\mathcal{T}'$ which has a unique child $l'$ labeled by a rule position of the form $\langle \sigma', \mathsf{id}, \langle \rangle \rangle$ or of the form $\langle \sigma', \bot, \langle \rangle \rangle$, implying that $\sigma'$ is an instance of $\mathsf{id}$ or of $\bot$. Hence, $l$ is an axiomatic leaf in $\pi$. Otherwise, $l$ was generated by pruning an infinite branch of $T$. In that case there exists a node $c(l)$, such that $\langle c(l), l \rangle$ is a repetition. It remains to show $\langle c(l), l \rangle$ is successful. Suppose towards a contradiction that $\langle c(l), l \rangle$ is not successful. Then on the path $\rho$ from $c(l)$ to $l$ either some sequent does not have a formula in focus or $\rho$ does not pass through an application of $\C \mathsf{R}$ where the principal formula is in focus. Let $\rho'$ be the finite sequence of positions in $S \cup I$ obtained from $\rho$ in the obvious way (i.e. after each position $\rho(i) \in S$ of $\rho$, add the rule position which has $\rho(i)$ as conclusion and $\rho(i+1)$ as premise). Then either there is a position occurring in $\rho'$ which has priority $3$, or every position in $\rho'$ has priority $1$. Since $\mathcal{T}'$ is the strategy tree of a \emph{memoryless} strategy, there exists an infinite branch in $\mathcal{T}'$ that has a suffix which is an infinite concatenation $\rho' \cdot \rho' \cdot \rho' \cdots$. On this path the highest priority encountered infinitely often is odd, contradicting the assumption that $\mathcal{T}'$ is the strategy tree of a winning strategy for Prover. Therefore each repetition is successful and so we conclude that $\pi$ is a $\Sigma$-proof of $\sigma$ in $\CICK_\mathrm{S5}$.
	\end{proof}
	
	Let $R_I$ be the set of all rule instances of $\CICK_{\mathrm{S5}}$ involving only $\Sigma$-sequents. Observe that the above constructed proof is \emph{uniform} in the following sense:
	
	\begin{definition}
		A $\Sigma$-proof $\pi$ is \emph{uniform} if there exists a function $f: S \longrightarrow R_I$ such that whenever a sequent $\sigma \in S$ occurs in $\pi$, it occurs as the conclusion of the rule instance $f(\sigma)$.
	\end{definition}
   Note that in a uniform proof the first repetition in each branch is successful. We conclude:
	\begin{theorem}\label{t: strategy iff provable}
		The following are equivalent for any sequent $\sigma$:
		\begin{enumerate}
			\item $\sigma$ is $\CICK_\mathsf{S5}$-provable.
            \item $\sigma$ has a $\Sigma$-proof in $\CICK_\mathsf{S5}$.
			\item Prover has a memoryless winning strategy in $\mathcal{G}_\sigma$.
			\item $\sigma$ has a uniform $\Sigma$-proof in $\CICK_\mathsf{S5}$.
		\end{enumerate}
	\end{theorem}

 It should be noted that the construction of the proof search games and the corresponding proofs are not depending on $\CICK_\mathsf{S5}$. In fact, the same proof holds for all proof systems for intuitionistic common knowledge logic (and of intuitionistic master modality) considered in this thesis.

\subsection{Exponential Completeness}

This subsection shows that the proof search problem for $\CICK_{\mathrm{S5}}$ is \textsc{ExpTime}-complete. Let us first establish an exponential upper bound. Recall the notion of \emph{complexity} of a formula and of a set of formulas (c.f. Definition \ref{d: complexity of ICK formula}). Given a sequent $\sigma$, define the \emph{complexity}\index{complexity!of $\CICK$-sequent} of $\sigma$ to be $c(\sigma) \coloneqq c(\Gamma_\sigma) + c(\Delta_\sigma)$. Note that if $\Sigma$ is the negation closure of $\Gamma^- \cup \Delta^-$, then $\lvert \Sigma \rvert$ is linear in $c(\Gamma \Rightarrow \Delta)$. 
    
	\begin{lemma}\label{l: number of positions in games}
		Given a $\Sigma$-sequent $\sigma$, the number of positions in $\mathcal{G}_\sigma$ is polynomially bounded in $\lvert \mathcal{P}(\Sigma) \rvert$.
	\end{lemma}
	
	\begin{proof}
		Observe that each unannotated $\Sigma$-sequent is an ordered pair of subsets of $\Sigma$. Therefore there are $\lvert \mathcal{P}(\Sigma) \lvert^2$ many unannotated $\Sigma$-sequents. By taking the focus annotations into account we obtain at most $\lvert \mathcal{P}(\Sigma) \lvert^3$ many $\Sigma$-sequents. Hence $\lvert S \rvert \leq \lvert \mathcal{P}(\Sigma) \lvert^3$. Next, note that for each $\Sigma$-sequent $\sigma'$ and each rule $\mathsf{r}$, there are at most $2 \cdot\lvert \Sigma \rvert$ many ways of applying $\mathsf{r}$ to $\sigma'$. We therefore obtain the upper bound
		\begin{equation*}
			\lvert I \rvert \leq 16 \cdot 2 \cdot \lvert \Sigma \rvert \cdot \lvert \mathcal{P}(\Sigma) \rvert^3  \in \mathcal{O}(\lvert \mathcal{P}(\Sigma) \rvert^4).
		\end{equation*}
		Together, the set of positions of $\mathcal{G}_\sigma$ is in  $\mathcal{O}(\lvert \mathcal{P}(\Sigma) \rvert^4)$, i.e. polynomial in $\lvert \mathcal{P}(\Sigma) \rvert$. \qedhere
	\end{proof}

	In order to get a polynomial bound for deciding the winner of a given proof search game we can now make use of one of the many existing algorithms for solving parity games. For instance the following result by Calude et al.~\cite{calude_2017}. 
	\begin{theorem}[{\cite[Theorem 2.9]{calude_2017}}]\label{Theorem 2.9 Calude}
		There is an algorithm which finds the winner of a parity game in time $\mathcal{O}(n^{log(m)+6})$ for a parity game with $n$ positions and priorities in $\{1,2,...,m\}$. Furthermore, the algorithm can compute a memoryless winning strategy for the winner in time $\mathcal{O}(n^{log(m)+7} \cdot log(n))$.
	\end{theorem}
	
	Let $\sigma$ be a sequent and let $\Sigma$ be its negation closure. By Lemma \ref{l: number of positions in games} the number $n$ of positions in $\mathcal{G}_\sigma$ is polynomial in the size of $\lvert \mathcal{P}(\Sigma) \lvert$. Since the number of different priorities in our games is constant, Theorem \ref{Theorem 2.9 Calude} implies that there is an algorithm deciding the winner of $\mathcal{G}_\sigma$ in time polynomial in $\lvert \mathcal{P}(\Sigma) \lvert$ and so exponential in $c(\sigma)$. By the same argument the above mentioned algorithm also computes a memoryless winning strategy in exponential time.
	
	\begin{corollary}\label{c: proof search problem is in exptime}
		For any sequent $\sigma$, there is an algorithm deciding whether $\sigma$ is $\CICK_\mathsf{S5}$-provable that runs in exponential time in $c(\sigma)$.
	\end{corollary}

   \begin{theorem}
       The proof search problem for $\mathsf{cICK_{S5}}$ is \textsc{ExpTime}-complete.
   \end{theorem}
   \begin{proof}
       That the proof search problem for $\mathsf{cICK_{S5}}$ belongs to \textsc{ExpTime} is Corollary \ref{c: proof search problem is in exptime}.  Deciding whether a formula of classical S5 common knowledge logic is valid is known to be \textsc{ExpTime}-complete~\cite{halpern_1992}. By Theorem \ref{t: negative translation}, Theorem \ref{t: soundness} and Theorem \ref{t: completeness of S5}, $\sigma^I \in \mathbf{CK_{S5}}$ if and only if $tr(\sigma^I) \in \mathbf{ICK_{S5}}$ if and only if $tr(\sigma) $ is provable in $\mathsf{cICK_{S5}}$, where $tr(\sigma) = tr(\Gamma_\sigma) \Rightarrow tr(\Delta_\sigma)$. Observe that the function mapping $\sigma$ to $tr(\sigma)$ is computable in polynomial time in the size of $\sigma$, and therefore is a polynomial-time reduction from the validity problem of $\mathbf{CK_{S5}}$ onto the proof search problem for $\mathsf{\CICK_{S5}}$, implying that the proof search problem is \textsc{ExpTime}-hard.
   \end{proof}

   \section{Conclusion}\label{c: ICK, Section conclusion}

   This chapter studied sound and complete analytic cyclic sequent calculi for intuitionistic common knowledge logic over epistemic models satisfying additional frame conditions. The introduced calculi are extensions of $\CIM$ with rules for multiple modalities and frame conditions. Our work extends the seminal studies by Jäger and Marti~\cite{jager-intuitionistic_2016, marti_2017} and uses techniques developed in~\cite{afshari_intuitionistic_2024, rooduijn_analytic_2022}. The following summarizes the main results.

   \begin{enumerate}
       \item We have presented the cyclic calculi $\CICK$, $\CICK_{\mathrm{T}}$, $\CICK_{\mathrm{S4}}$ and $\CICK_{\mathrm{S5}}$ and shown that $\CICK_{\mathrm{S5}}$ is not cut-free complete.
       \item We have proven soundness and completeness of all four calculi with respect to the classes of epistemic models, reflexive epistemic models, S4 epistemic models and S5 epistemic models, respectively. The completeness proof for the first three calculi was an adaptation of the proof search method for $\CIM$, thereby demonstrating the robustness of the method once more, while completeness for $\CICK_{\mathrm{S5}}$ was proven by employing the technique of analytic cuts and a canonical model construction.
       \item We have embedded classical common knowledge over S5 into intuitionistic common knowledge logic over S5 by adapting Kuroda's translation to the modal case.
       \item We have shown that proof search in $\CICK_\mathsf{S5}$ can be automated by translating the calculus into a parity game and we have established a precise complexity bound for the proof search problem of $\CICK_\mathsf{S5}$.
   \end{enumerate}

   The calculus $\CICK_{\mathrm{S5}}$ requires analytic cuts for completeness. An alternative solution would be to pass to calculi that manipulate sequents which have more structure, such as labelled or nested sequents. Labelling or nesting is a successful tool to handle frame conditions, thus the cut rule can be dropped. However, the combination of labelled/nested sequents with non-wellfounded proofs leads to the problem of how to detect successful repetitions in non-wellfounded branches of a proof and thus of how to obtain cyclic proof systems. The notion of repetition employed for $\CIM$ and $\CICK$ is too strong here, as in general it can not be expected to find a precise repetition in the presence of labels or nesting. Instead, a more likely solution is to weaken the notion of repetition. For example, since nested sequents are essentially finite trees of normal sequents, which might grow indefinitely along a non-wellfounded branch, a more promising notion of repetition would be a pair of sequents $(\sigma, \sigma')$ such that $\sigma$ can be embedded into $\sigma'$, implying that every rule applied between $\sigma$ and $\sigma'$ can be repeated at $\sigma'$. We have been unable to solve this problem so far, and to the best of my knowledge there is no satisfying solution proposed in the literature. We thus leave it as an open question for future research.

   \begin{question}\label{q: ICK proof theory}
       Is it possible to develop non-wellfounded labelled or nested proofs for $\ICK$ over S5 which are regularly complete? How should the notion of `repetition' be formulated to prove that every non-wellfounded branch has a successful repetition?
   \end{question}

   Another open question regards the S5 epistemic models studied in this chapter. As demonstrated, the interaction of S5 frame conditions with triangle confluence leads to models where the modalities behave essentially `classical'. While this is an interesting mathematical observation, it is unclear whether such a logic is interesting from an application driven point of view. A possible next step is to replace triangle confluence with forth-down confluence. It is not hard to see that forth-down confluent and triangle confluent models result in the same logic when the modal accessibility relation satisfies no frame conditions. However, for S5 frame conditions, when using forth-down confluent models, the observation that the entire intuitionistic tree of a world $w$ belongs to the same equivalence class no longer holds, implying that in such a setting triangle confluence is stronger than forth-down confluence. The following questions are thus left open by our work.

   \begin{question}
       What is the logic obtained from evaluating $\LICK$ over S5 forth-down confluent epistemic models? Are such models preferable over S5 epistemic (triangle confluent) models?
   \end{question}

\chapter{Intuitionistic Linear Temporal Logic}
\label{c: iLTL}

\section{Introduction} 

The intuitionistic dynamic logics studied so far are evaluated over dynamic models which satisfy weak confluence conditions. As seen in Theorem~\ref{t: three classes one logic} the language of $\IM$ (and thus also the language of $\ICK$) cannot distinguish between dynamic models, triangle models and in particular functional models. This observation discouraged an interpration of $\LIM$ as an intuitionistic linear temporal logic. In this chapter we will turn our attention to more expressive logics which are evaluated over classes of dynamic models satisfying stronger confluence properties. Specifically, we will study a version of intuitionistic linear temporal logic called $\iLTL$, whose formulas are evaluated over two classes of total functional models: those satisfying forward confluence, and those satisfying both forward and back-up confluence (c.f. Definition \ref{d: confluence conditions}). We follow~\cite{Balbiani_2019} and call such models \emph{expanding} and \emph{persistent}, respectively. The language of $\iLTL$ is capable of distinguishing between arbitrary (serial) dynamic models and expanding models, and also between expanding models and persistent models.

The origin of $\iLTL$ goes back to the \emph{dynamic topological logic project}. This project aims to develop computationally well-behaved logics to reason about topological dynamics\index{topological dynamics}: topological spaces equipped with a continuous function. Such dynamic systems are applied in diverse fields ranging from biology to physics and theoretical computer science. The dynamic topological logic project originated in the work of Artemov, Davoren and Nerode~\cite{Artemov_1999} who used modal logic - where $\lb$ is evaluated as the `interior' operator in the sense of Tarski~\cite{Tarski_1938} - extended with a temporal `next' operator to reason about the action of the continuous function. Their logic was extended by Kremer and Mints~\cite{Kremer_2005} who introduced the logic $\mathsf{DTL}$, which additionally features a fixed point operator from temporal logic to reason about the asymptotic behaviour of the continuous function. After $\mathsf{DTL}$ was proven undecidable~\cite{Konev_2006}, the focus of the project shifted to intuitionistic versions of $\mathsf{DTL}$ and thus to intuitionistic dynamic logics. 

The logic $\iLTL$ was studied in this context by Fern\'andez-Duque~\cite{fernandez-duque_2018}. For the following, familiarity with basic topology is assumed. A \emph{topological dynamic system} is a tuple $\mathcal{X}=(X, \tau, f)$ where $(X, \tau)$ is a topological space (e.g. the real line with the Euclidean topology) and $f: X \longrightarrow X$ is a continuous function acting on $X$. The function $f$ is thought of as a \emph{temporal function}, which maps each point of $X$ onto its temporal successor. A topological model then consists of a topological dynamic system $(X, \tau, f)$ equipped with a valuation function $V: \Prop \longrightarrow \tau$ which assigns to each proposition an open set, intuitively the set of points where the proposition holds. The language of $\iLTL$ extends $\LIPL$ by the temporal operators $\X$, $\E$ and $\G$, where $\X \varphi$ is read as `in the next time step $\varphi$ holds', $\E \varphi$ is read as `eventually in the future $\varphi$ holds' and $\G \varphi$ is read as `henceforth in the future $\varphi$ holds'. Temporal formulas of $\LILTL$ are evaluated using the temporal function $f$. More formally, we assign a \emph{truth set} to formulas, which is the set of points at which the formula is true, as follows, where $\circ$ denotes the \emph{interior operator}.
\begin{center}
    \begin{tabular}{l l l}
       $\llbracket \bot \rrbracket$ & $\coloneqq$ & $\emptyset$ \\
       $\llbracket p \rrbracket$ & $\coloneqq$ & $V(p)$ for $p \in \Prop$\\
       $\llbracket \varphi \wedge \psi \rrbracket$ & $\coloneqq$ & $\llbracket \varphi \rrbracket \cap \llbracket \psi \rrbracket$ \\
        $\llbracket \varphi \vee \psi \rrbracket$ & $\coloneqq$ & $\llbracket \varphi \rrbracket \cup \llbracket \psi \rrbracket$ \\
         $\llbracket \varphi \rightarrow \psi \rrbracket$ & $\coloneqq$ & $ ((X \setminus \llbracket \varphi \rrbracket) \cup \llbracket \psi \rrbracket)^\circ$ \\
          $\llbracket {\X \varphi} \rrbracket$ & $\coloneqq$ & $f^{-1}(\llbracket \varphi \rrbracket)$ \\
           $\llbracket {\E \varphi} \rrbracket$ & $\coloneqq$ & $\bigcup_{n < \omega} f^{-n}(\llbracket \varphi \rrbracket)$ \\
           $\llbracket {\G \varphi} \rrbracket$ & $\coloneqq$ & $(\bigcap_{n < \omega} f^{-n}(\llbracket \varphi \rrbracket))^\circ$ \\
    \end{tabular}
\end{center}

Note that the truth set of each formula is open. In particular for the temporal cases, this follows from $f$ being continuous, $\tau$ being closed under arbitrary unions and in the case for $\G$ from taking the interior of the infinite intersection. Write $\mathcal{X} \models \varphi$ if $\llbracket \varphi \rrbracket = X$ for any valuation $V$. The resulting logic (without $\G$) was shown to be decidable and capable of expressing interesting properties about the underlying topological dynamic system by Fern\'andez-Duque~\cite{fernandez-duque_2018}. For example, call a topological dynamic system $(X, \tau, f)$ \emph{Poincaré-recurrent} if any open set $U \subseteq X$ contains a \emph{recurrent point}, i.e. there exists $x \in U$ and a natural number $n > 0$ such that $f^n(x) \in U$.

\begin{theorem}[Fern\'andez-Duque~\cite{fernandez-duque_2018}]
    A topological dynamic system $\mathcal{X}$ is Poincaré-recurrent if and only if
    \begin{equation*}
        \mathcal{X} \models p \rightarrow \neg \neg {\X} \E p.
    \end{equation*}
\end{theorem}

 Furthermore, Boudou, Diéguez and Fern\'andez-Duque provided a sound and complete axiomatization $\mathrm{iLTL_H}$ for the $\G$-free fragment of $\iLTL$ over the class of topological dynamic models~\cite{Boudou_2022}, which is presented in Table \ref{tab: axiomatization iltl}.\footnote{We use different names for the system and some of the rules than~\cite{Boudou_2022}.}

\begin{table}
\centering
        \begin{tabular}{|l l|}
        \hline
         $\mathsf{Int}$:   & Intuitionistic tautologies  \\
         $\mathsf{K}$: & $\X(\varphi \rightarrow \psi) \rightarrow (\X \varphi \rightarrow \X \psi)$  \\
         $\mathsf{D}$: & $\neg \X \bot$ \\
         $\mathsf{Dist_\wedge}$: & $(\X \varphi \wedge \X \psi) \rightarrow \X (\varphi \wedge \psi)$ \\
         $\mathsf{Dist_\vee}$: & $\X (\varphi \vee \psi) \rightarrow (\X \varphi \vee \X \psi)$  \\
         $\mathsf{Fix}$: & $\varphi \vee \X \E \varphi \rightarrow \E \varphi $  \\
         \hline
        \end{tabular}
        \hspace{10pt}
         \begin{tabular}{|l l|}
        \hline
          $\mathsf{MP}$: $\infer{\psi}{\varphi & \varphi \rightarrow \psi}$ &  $\mathsf{Nec}$: $\infer{\X \varphi}{\varphi}$ \\
         $\mathsf{Mon}$: $\infer{\E \varphi \rightarrow \E \psi}{\varphi \rightarrow \psi}$ &  $\mathsf{Ind}$: $\infer{\E \varphi \rightarrow  \varphi}{\X \varphi \rightarrow  \varphi}$\\
         \hline
        \end{tabular}
     \caption{Axioms and rules of $\mathrm{iLTL_H}$}
     \label{tab: axiomatization iltl}\index{axiomatization!$\mathrm{iLTL_H}$}
\end{table}

 When identifying the modal box operator $\lb$ of $\IM$ with $\X$, the difference between the modal axioms of $\mathrm{IM_H}$ and $\mathrm{iLTL_H}$ is the presence of two additional axioms, namely $\mathsf{D}$ which expresses the seriality of the temporal function and $\mathsf{Dist_\vee}$ which expresses that $\X$ distributes over disjunctions, which is not the case for $\IM$. The axiom $\mathsf{Dist_\wedge}$ is also not present in $\mathrm{IM_H}$, but, as shown in Lemma \ref{l: derivable formulas}, is derivable in $\mathrm{IM_H}$ and thus also in $\mathrm{iLTL_H}$. 
 Apart from completeness with respect to the class of topological dynamic models, it was also shown in~\cite{Boudou_2022} that $\mathrm{iLTL_H}$ is sound and complete with respect to the class of expanding models. Furthermore, Balbiani, Boudou, Diéguez and Fern\'andez-Duque established that the language of $\iLTL$ including $\G$ is decidable over the class of expanding models~\cite{Balbiani_2019, Boudou_2017}.

 Several problems about $\iLTL$ remain open. The first is about finding a sound and complete axiomatization for the language of $\iLTL$ including $\G$ (only an infinitary axiomatization has been found in~\cite{Chopoghloo_2021}). This problem will be addressed in Chapter \ref{c: biLTL}, where we present a finitary axiomatization for the language which is additionally extended by the co-implication connective of bi-intuitionistic logic. Second, it is unknown whether $\iLTL$ over persistent models is decidable or even recursively enumerable and third, the proof theory of $\iLTL$ remains largely unexplored. This chapter addresses the third problem. We will study non-wellfounded proof systems for $\iLTL$ without $\G$. Instead of using $\E$ in the language, we consider a slightly stronger temporal operator $\U$ (`until') with the intended semantics that a formula $\varphi \U \psi$ holds at a world $w$ if $\psi$ holds eventually in the future and until then $\varphi$ holds. Thus $\E$ is definable in terms of $\U$.  We present well-behaved non-wellfounded proof systems for $\iLTL$ over the classes of expanding and persistent models, by adapting the techniques used for $\IM$ and $\ICK$. In order to capture the stronger confluence conditions of expanding and persistent models, our calculi employ a simple form of \emph{nested sequents}, such that formulas can be operated on at different time steps. Such nesting is inspired from the work of Kojima and Igarashi~\cite{Kojima_2011} on sequent calculi for intuitionistic linear temporal logics featuring $\X$ as the only temporal operator. The resulting non-wellfounded calculi are analytic and sound and complete. However, it turns out that both calculi are not complete when restricted to regular proofs, implying that we won't be able to prove completeness for a cyclic version of the calculus. In his recent PhD thesis, Men\'endez Turata~\cite{menendez_2024} improved upon our work and showed how to obtain complete cyclic proof systems for $\iLTL$ over expanding models for the full language, but did not address persistent models. 
 
 We begin by introducing syntax and semantics of $\iLTL$. Then we introduce two non-wellfounded sequent calculi $\niltle$ and $\niltlp$ which are shown to be sound and complete for the classes of expanding and persistent models, respectively. Soundness is shown by a proof by infinite descent, which nicely illustrates the connection between such proofs and non-wellfounded proofs. Completeness is shown by proof search. In difference to $\IM$, the proof search argument is significantly more complicated due to the confluence conditions of the models. Finally, we show that both calculi are not regularly complete by providing counterexamples.

\section{Syntax and Semantics}
 The language $\LILTL$ of intuitionistic linear-time temporal logic $\iLTL$ extends $\LIPL$ by the temporal operators $\X$ and $\U$\index{fixed point operator!until}. \emph{Formulas} are given by the following grammar in Backus--Naur form:
    \begin{equation*}
        \varphi \Coloneqq  \bot \mid p \mid \varphi \wedge \varphi \mid \varphi \vee \varphi \mid \varphi \rightarrow \varphi \mid \X \varphi \mid \varphi \U \varphi 
    \end{equation*}
    where $p \in \mathsf{Prop}$. The operator $\X$ is the `next'-operator: read $\X \varphi$ as `in the \emph{next} time step, $\varphi$ holds' and $\U$ is the `until'-operator: read $\varphi \U \psi$ as `$\varphi$ holds \emph{until} $\psi$ holds (and $\psi$ eventually holds)'. $\U$ is a least fixed point operator: $\varphi \U \psi$ is characterized as the least fixed point of the propositional function $x \mapsto  \psi \vee (\varphi \wedge \X x)$. The aforementioned temporal modality $\E$ (`eventually')\index{fixed point operator!eventually} can be defined in terms of $\U$ by
    \begin{equation*}
        \E \varphi \coloneqq \top \U \varphi.
    \end{equation*}
    The temporal modality $\G$ (`henceforth') is however not definable in this language~\cite{Balbiani_2019}. 
    \begin{definition}
        The \emph{complexity}\index{complexity!of $\LILTL$-formula} $c(\varphi)$ of a formula $\varphi$ is defined as in Definition \ref{d: complexity of ICK formula} with the modal cases replaced by 
        \begin{align*}
            c(\X \varphi) &= c(\varphi) + 1 \\
            c(\varphi \U \psi) &= c(\varphi) + c(\psi) + 1
        \end{align*}
    \end{definition}

    Formulas of $\iLTL$ are evaluated on two classes of dynamic models where the modal accessibility relation is a function which in one case is forward confluent and in the other case both forward and back-up confluent (c.f. Definition \ref{d: confluence conditions}). Recall that given a dynamic model $\fw =(W, \leq, f, V)$ where $f$ is a function, $f$ is forward-confluent if and only if $f$ is monotone in $\leq$, i.e. if $w \leq v$, then $f(w) \leq f(v)$ (c.f. Lemma \ref{l: functional models confluence conditions}).
    
\begin{definition}
Let $\fw =(W, \leq, f, V)$ be a total functional dynamic model.
\begin{enumerate}
    \item $\fw$ is \emph{expanding}\index{model!expanding} if $f$ is forward-confluent.
    \item $\fw$ is \emph{persistent}\index{model!persistent} if $\fw$ is expanding and, additionally, $f$ is back-up confluent.
\end{enumerate}
\end{definition}

The confluence conditions guarantee that classical truth conditions for the temporal modalities can be used without losing the monotonicity property.

\begin{definition}\label{d: truth relation expanding models}
Given an expanding (persistent) model $\fw=(W, \leq, V, f)$, the \emph{truth relation} $\models$ between worlds of $\fw$ and formulas extends Definition \ref{d: truth relation for intuitionistic Kripke models} by the following clauses for the temporal operators, where $w \in W$.
       \begin{center}
        \begin{tabular}{l @{\ }c @{\ } l}
      $\fw,w \models \X \varphi$ & iff & $\fw,f(w) \models \varphi$,\\
      $\fw,w \models  \varphi\U \psi$ & iff & there exists $n < \omega$ such that $\fw, f^n(w) \models \psi$ and for all\\
      & & $0 \leq m< n$, $\fw, f^m(w) \models \varphi$. 
        \end{tabular}
     \end{center}
\end{definition}

As usual, a formula $\varphi$ is \emph{satisfiable} over the class of expanding (persistent) models if there exists an expanding (persistent) model $\fw =(W, \leq, f, V)$ and a world $w \in W$ with $\fw, w \models \varphi$ and \emph{unsatisfiable} otherwise. A formula $\varphi$ is \emph{valid} over the class of expanding (persistent) models if for all expanding (persistent) models $\fw =(W, \leq, f, V)$ and all worlds $w \in W$ holds that $\fw, w \models \varphi$ and \emph{invalid} otherwise.
 
\begin{definition}
    Let
    \begin{enumerate}
        \item $\mathbf{iLTL_e}$ be the set of all valid $\LILTL$-formulas over the class of expanding models.
        \item $\mathbf{iLTL_p}$ be the set of valid $\LILTL$-formulas over the class of persistent models.
    \end{enumerate}
\end{definition}

Since every persistent model is an expanding model, clearly $\mathbf{iLTL_e} \subseteq \mathbf{iLTL_p}$ holds. The following lemma shows that the inclusion is strict.

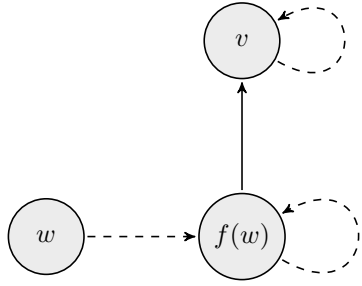
\begin{figure}[t!]
    \centering
    \begin{tikzpicture}[modal]
        \node[world](w){$w$} ;
        \node[world](w1)[right=of w]{$f(w)$};
        \node[world](v)[above=of w1]{$v$} ;
         \path[->] (w) edge[dashed] (w1);
        \path[->] (w1) edge (v);
        \path[->] (w1) edge[reflexive right, dashed] (w1);
         \path[->] (v) edge[reflexive right, dashed] (v);
    \end{tikzpicture}
    \caption{The countermodel from Lemma \ref{l: forward and backward confluence formulas}. Solid arrows indicate $\leq$ and dashed arrows indicate $f$.}
    \label{f: expanding model falsifies converse K}
\end{figure}

\begin{lemma}\label{l: forward and backward confluence formulas}
  For any $\varphi, \psi \in \LILTL$, the formula $(\X \varphi \rightarrow \X \psi) \rightarrow \X (\varphi \rightarrow \psi)$ is valid over the class of persistent models. However, $(\X p \rightarrow \X q) \rightarrow \X (p \rightarrow q)$ is not valid over the class of expanding models for $p,q \in \Prop$.
\end{lemma}
\begin{proof}
    Let $\fw=(W, \leq, f, V)$ be a persistent model and $w, v \in W$ such that $w \leq v$ and ${\fw, v \models \X \varphi \rightarrow \X \psi}$. Hence, for all $u \geq v$ if $\fw, u \models \X \varphi$, then $\fw, u \models \X \psi$. Let $u' \geq f(v)$ and suppose $\fw, u' \models \varphi$. By back-up confluence there exists $u \geq v$ with $f(u) = u'$. Hence $\fw, u \models \X \varphi$, implying $\fw, u \models \X \psi$. Thus $\fw, u' \models \psi$. Since $u' \geq f(v)$ was arbitrary, we have $\fw, f(v) \models \varphi \rightarrow \psi$ and so $\fw, v \models \X (\varphi \rightarrow \psi)$. Hence $\fw, w \models (\X \varphi \rightarrow \X \psi) \rightarrow \X (\varphi \rightarrow \psi)$. As $\fw$ and $w$ were arbitrary, $(\X \varphi \rightarrow \X \psi) \rightarrow \X (\varphi \rightarrow \psi)$ is valid over the class of persistent models. \smallskip

    To see that $(\X p \rightarrow \X q) \rightarrow \X (p \rightarrow q)$ is not valid over the class of expanding models, consider the expanding model $\fw=(W, \leq, f, V)$ depicted in Figure \ref{f: expanding model falsifies converse K}. Let $V(w) = V(f(w)) = \emptyset$ and let $V(v) = \{p\}$. Then it is straighforward to check that $\fw, w \models \X p \rightarrow \X q$ but $\fw, w \not \models \X (p \rightarrow q)$.
\end{proof}

\begin{lemma}\label{l: iteration of forward confluent function is forward confluent}
    Let $\fw =(W, \leq, f, V)$ be an expanding model and $w,v \in W$. If $w \leq v$, then $f^n(w) \leq f^n(v)$ for all $n < \omega$.
\end{lemma}
\begin{proof}
    By induction on $n$. For $n=0$ recall that $f^0(w) = w \leq v = f^0(v)$. For $n >0$ we have $f^{n-1}(w) \leq f^{n-1}(v)$ by induction hypothesis and therefore $f^n(w) \leq f^n(v)$ by forward confluence.
\end{proof}

\begin{lemma}\label{l: iteration of backwards confluent function is backward confluent}
Let $\fw =(W, \leq, f, V)$ be a persistent model, $w,v \in W$ and $n < \omega$. If $f^n(w) \leq v$, then there exists $u \in W$ with $w \leq u$ and $f^n(u) = v$.
\end{lemma}
\begin{proof}
    By induction on $n$. For $n= 0$ we have that $f^0(w) = w \leq v$, so let $u = v$. For $n > 0$ since $f(f^{n-1}(w)) = f^n(w)$ and $f^n(w) \leq v$, by back-up confluence there exists $u' \in W$ with $f^{n-1}(w) \leq u'$ and $f(u') = v$. By induction hypothesis there exists $u \in W$ with $w \leq u$ and $f^{n-1}(u) = u'$. Thus $f^n(u) = v$.
\end{proof}
 
 \begin{lemma}[Monotonicity]\label{l: monotonicity for expanding/persistent models}
    Let $\fw=(W, \leq, V, f)$ be an expanding (persistent) model, $w,v \in W$ and $\varphi$ a formula. If $\fw,w \models \varphi$ and $w \leq v$, then $\fw,v \models \varphi$.
\end{lemma}
\begin{proof}
    By induction on $\varphi$. The base cases as well as the cases where $\varphi = \psi \ast \chi$ for ${\ast \in \{\wedge, \vee, \rightarrow\}}$ follow from Lemma \ref{l: monotonicity for intuitionistic Kripke models}. The case for $\varphi = \X \psi$ follows from Lemma \ref{l: monotonicity for dynamic models}, Lemma \ref{l: confluence classical truth conditions} and Lemma \ref{l: functional models confluence conditions}. For $\varphi = \psi \U \chi$ if $\fw, w \models \psi \U \chi$, then there exists $n < \omega$ such that $f^n(w) \models \chi$ and for all $0 \leq m < n$, $\fw, f^m(w) \models \psi$. By Lemma \ref{l: iteration of forward confluent function is forward confluent}, $f^k(w) \leq f^k(v)$ for all $0 \leq k \leq n$. Thus by induction hypothesis $\fw, f^n(v) \models \chi$ and for all $0 \leq m < n$, $\fw, f^m(v) \models \psi$, implying that $\fw, v \models \psi \U \chi$.
\end{proof}
\section{Nested Non-Wellfounded Proofs}
\label{s-iLTLen}
This section presents non-wellfounded sequent calculi for $\iLTL$ over expanding models and over persistent models, respectively. In difference to the non-wellfounded calculus $\NWIM$ for $\IM$, the systems do not feature a focus annotation. Recall that the focus annotation in $\NWIM$ was used to prove soundness of the cyclic system. Since we only study non-wellfounded systems here, the focus annotation is no longer needed. However, due to the lack of a focus annotation, the global correctness criterion imposed on infinite branches is formulated differently using \emph{formula traces}. To ensure completeness, the calculi incorporate a simple form of \emph{nesting}, which is inspired by the work of Kojima and Igarashi~\cite{Kojima_2011} who developed a nested sequent calculus for a version of $\iLTL$ where the only temporal modality is $\X$.
 
\begin{definition}
A \emph{nested formula}\index{nested formula} is a tuple $(\varphi,n)$, denoted by $\varphi^n$, with $\varphi \in \LILTL$ and $n< \omega$. A \emph{sequent}\index{sequent!for $\iLTL$} is an ordered pair $\Gamma \Rightarrow \Delta$, where $\Gamma$ and $\Delta$ are finite sets of nested formulas.
\end{definition}

Note that in difference to the definition of sequents for $\NWIM$, every tuple $\Gamma \Rightarrow \Delta$ with $\Gamma, \Delta$ being finite sets of nested formulas is a sequent. As before, we denote sequents by $\sigma$ and $\Gamma_\sigma$, $\Delta_\sigma$ denote the left side and right side of $\sigma$, respectively. For the remainder of this chapter nested formulas are simply called formulas. Formulas that are not nested are called \emph{plain}. As for $\NWIM$ we consider multi-conclusion sequents. Once again, this is to simplify the proof search argument in the completeness proof. At first glance, nested formulas might look little different to annotated formulas, but this is misleading. The difference becomes clear in the interpretation of a sequent.

\begin{definition}
    The \emph{interpretation} of a nested formula $\varphi^n$ is the plain formula $\X^n \varphi$. The \emph{interpretation} of a sequent $\sigma$ is the plain formula
\[ 
	\sigma^I \coloneqq \bigwedge_{\varphi^n \in \Gamma_\sigma}\X^{n} \varphi \rightarrow \bigvee_{\psi^m \in \Delta_\sigma} \X^{m} \psi
\]
\end{definition} 

To simplify notation, we write $\fw,w \models \varphi^n$ if $M,w \models \X^n \varphi$ and $\fw,w\models \sigma$ if $M,w \models \sigma^I$. For any set $\Gamma$ of nested formulas, define 
\begin{equation*}
    \Gamma^{+1} \coloneqq \{\varphi^{n+1}: \varphi^n\in \Gamma\}.
\end{equation*} 

We will now introduce the sequent calculi $\niltle$ and $\niltlp$, which will be proven to be sound and complete with respect to the classes of expanding models and of persistent models, respectively. Both calculi are based on the basic rules depicted in Table \ref{tab: basic rules iltl} as well as some of the additional rules in Table \ref{tab: additional rules iltl}.

\begin{definition}\label{d: non-wellfounded calculus for iltl}
Consider the rules depicted in Table \ref{tab: basic rules iltl} and Table \ref{tab: additional rules iltl}.
\begin{enumerate}
    \item The sequent calculus $\niltle$\index{sequent calculus!$\niltle$} consists of the basic rules as well as the rules ${\rightarrow} \mathsf{R}$ and $\mathsf{S}$.
    \item The sequent calculus $\niltlp$\index{sequent calculus!$\niltlp$} consists of the basic rules as well as the rule ${\rightarrow} \mathsf{R_p}$.
\end{enumerate}
\end{definition}
\begin{table}[t]
\centering
\begin{tabular}{|c@{\quad}c|}
\hline
& \\
    $\infer[\mathsf{id}]{ \Gamma, \varphi^n \Rightarrow \varphi^n, \Delta}{}$
    &
    $\infer[\bot]{\Gamma, \bot^n \Rightarrow \Delta}{}$
    \\[1em]
    $\infer[\wedge \mathsf{L}]{\Gamma, \varphi \wedge \psi^n \Rightarrow \Delta}{\Gamma, \varphi^n,\psi^n \Rightarrow \Delta}$
    &
    $\infer[\wedge \mathsf{R}]{\Gamma \Rightarrow \varphi \wedge \psi^n, \Delta}{\Gamma \Rightarrow \varphi^n, \Delta & \Gamma \Rightarrow \psi^n, \Delta}$
    \\[1em]
    $\infer[\vee \mathsf{L}]{\Gamma, \varphi \vee \psi^n \Rightarrow \Delta}{\Gamma, \varphi^n \Rightarrow \Delta & \Gamma, \psi^n \Rightarrow \Delta}$
    &
    $\infer[\vee \mathsf{R}]{\Gamma \Rightarrow \varphi \vee \psi^n, \Delta}{\Gamma \Rightarrow \varphi^n,\psi^n, \Delta}$
    \\[1em]
    $\infer[\X \mathsf{L}]{\Gamma, \X \varphi^{n} \Rightarrow \Delta}{\Gamma,  \varphi^{n+1} \Rightarrow \Delta}$
    &
    $\infer[\X\mathsf{R}]{\Gamma\Rightarrow \X \varphi^{n} \!, \Delta}{\Gamma\Rightarrow  \varphi^{n+1} \!, \Delta}$
    \\[1em]
    $ \infer[\U \mathsf{L}]{\Gamma, \varphi \U \psi^n \Rightarrow \Delta}{\Gamma, \psi^n \Rightarrow \Delta & \Gamma, \varphi^n\! ,\X  (\varphi \U \psi)^n \Rightarrow \Delta}$
    &
    $\infer[\U \mathsf{R}]{\Gamma \Rightarrow \varphi \U \psi^n\!, \Delta}{\Gamma\Rightarrow \psi^n\!, \varphi^n\!, \Delta & \Gamma \Rightarrow \psi^n\! , \X (\varphi \U \psi)^n\!, \Delta}$
     \\[1em]
     $\infer[{\rightarrow} \mathsf{L}]{\Gamma, \varphi \rightarrow \psi^n \Rightarrow \Delta}{\Gamma, \varphi \rightarrow \psi^n \Rightarrow \varphi^n, \Delta & \Gamma, \psi^n \Rightarrow \Delta}$ & \\
     & \\
     \hline
\end{tabular}
\caption{The basic rules. The symbols $\Gamma$ and $\Delta$ range over arbitrary finite sets of nested formulas which may be empty.}
\label{tab: basic rules iltl}
\end{table}
\begin{table}[t]
\centering
\begin{tabular}{|c@{\quad}c|}
\hline
& \\
    $\infer[{\rightarrow} \mathsf{R}]{\Gamma \Rightarrow \varphi \rightarrow \psi^0, \Delta}{\Gamma, \varphi^0 \Rightarrow \psi^0}$
    &
    $\infer[{\rightarrow} \mathsf{R_p}]{\Gamma \Rightarrow \varphi \rightarrow \psi^n, \Delta}{\Gamma, \varphi^n \Rightarrow \psi^n}$
     \\[1em]
    $\infer[\mathsf{S}]{\Sigma, \Gamma^{+1} \Rightarrow  \Delta^{+1}, \Pi}{\Gamma \Rightarrow \Delta}$ & \\
    & \\
    \hline
\end{tabular}
\caption{The additional rules. The symbols $\Gamma, \Delta, \Sigma$ and $\Pi$ range over arbitrary finite sets of nested formulas which may be empty.}
\label{tab: additional rules iltl}
\end{table}
The rules $\mathsf{id}$ and $\bot$ are called \emph{axioms}. The basic rules for $\wedge, \vee, \rightarrow$ are nested versions of the corresponding rules in the sequent calculus for intuitionistic logic $\mathrm{IPL_s}$ (see Definition~\ref{d: sequent calculus for IPL}). The $\U$-rules capture the equivalence 
\begin{equation*}
    \varphi\U  \psi\equiv \psi\lor (\varphi\land \X(\varphi\U  \psi)).
\end{equation*} Note that the rules $\X \mathsf{L}$ and $\X \mathsf{R}$ are purely structural as $\X \varphi^n$ has the same interpretation as $\varphi^{n+1}$. Regarding the additional rules observe that the rule ${\rightarrow} \mathsf{R}$ can only be applied to implications with nesting {level $0$} while the rule ${\rightarrow} \mathsf{R_p}$ can be applied to implications of arbitrary nesting level. Interestingly, this is the crucial difference between the calculi $\niltle$ and $\niltlp$. To understand why, it is instructive to think semantically about what these rules express. Naturally, both rules should preserve validity when read top-down. The rule ${\rightarrow} \mathsf{R_p}$ then expresses that if $\X^n \varphi \rightarrow \X^n \psi$ is valid, so is $\X^n (\varphi \rightarrow \psi)$. By Lemma \ref{l: forward and backward confluence formulas} this rule preserves validity over persistent models, but not over expanding models. Thus, for $\niltle$, we must use the standard implication rule, where the nesting level is $0$. This restriction forces us to include an additional rule for $\niltle$, namely the `shift'-rule $\mathsf{S}$, that allows us to reduce the nesting level of formulas and captures necessitation. For $\niltlp$, the shift-rule is not required.

Recall the (semantic) definition of an invertible rule (c.f. Definition \ref{d: invertible rule}). The following lemma is routine.

\begin{lemma}
    All basic rules are invertible, while all additional rules are not invertible.
\end{lemma}

 We will therefore refer to the additional rules as the \emph{non-invertible rules} and to all other rules as the \emph{invertible rules}. 

 For each rule, the distinguished formula in the conclusion is called \emph{principal} and the distinguished formulas in the premises are called its \emph{residuals}. For $\mathrm{S}$, all formulas in the conclusion are principal and each formula in the premise is the residual of its corresponding principal formula; in particular, formulas in $\Sigma$ and $\Pi$ have no residual. In every rule application, any formula that is neither principal nor residual is called a \emph{side formula}. For the remainder of this section, let $\mathsf{P} \in \{\niltle, \niltlp\}$.

 \begin{definition}
     A \emph{$\mathsf{P}$-pre-proof} of a sequent $\sigma$ is a finite or infinite tree whose nodes are labeled according to the rules of $\mathsf{P}$ and whose root is labeled by $\sigma$.
 \end{definition}

   Recall the definition of a path through a tree (c.f. Section \ref{c: preliminaries, section basic definitions}). If there is no danger of confusion, paths will be identified with the sequence of sequents that label the nodes of the path. 

\begin{definition}
Let $\rho =(\rho(i))_i$ be a path through a pre-proof $\pi$. A \emph{(formula) trace}\index{trace} on $\rho$ is a finite or infinite sequence of nested formulas $\varphi_0^{n_0}, \varphi_1^{n_1},\dots$ such that for each index $i$ the following hold.
\begin{enumerate}
    \item $\varphi_i^{n_i}$ occurs on the left side of the sequent labelling $\rho(i)$;
    \item If $\varphi_{i}^{n_i}$ is a principal formula in the rule applied at $\rho(i)$, then $\varphi_{i+1}^{n_{i+1}}$ is a residual formula of $\varphi_i^{n_i}$ in $\rho(i+1)$;
    \item if $\varphi_{i}^{n_i}$ is a side formula in the rule applied at $\rho(i)$, then $\varphi_{i+1}^{n_{i+1}}=\varphi_{i}^{n_i}$. 
\end{enumerate}
\end{definition}
For any rule $\mathsf{r}$, a trace  $(\varphi_i^{n_i})_{i}$ \emph{passes actively through $\mathsf{r}$} if there is an index $j$ such that $\varphi_j^{n_j}$ is a principal formula in an instance of $\mathsf{r}$. To keep the notation simple, when reasoning about traces $(\varphi_i^{n_i})_i$, we will usually omit the index for the nesting and simply write $(\varphi_i)_i$. It will always be clear from context that we refer to nested formulas and not to plain formulas.

Note that the focus annotation employed in the non-wellfounded system $\NWIM$ in Chapter~\ref{c: IM} captures specific formulas traces. Namely a good suffix in a branch of an $\NWIM$-pre-proof always has a formula in focus which is of the form  $\lbm \psi^f$ or $\lb \lbm \psi^f$ for some formula $\psi$. Therefore we obtain a trace through a good suffix by considering the infinite sequence of focused formulas $(\varphi_i)_i$ occurring in the suffix. Note that this sequence satisfies the definition of a formula trace above (minus the first condition, which has to be replaced by the condition that the trace formula (i.e. the formula in focus) occurs on the right side of the sequent). The focus annotation thus keeps track of one trace through a suffix of an infinite branch, and in order for the suffix to be \emph{good}, it is required that the trace passes infinitely often through an instance of $\lbm \mathsf{R}$. Here, we define similarly the notion of a \emph{good trace}.

\begin{definition}
    A formula trace is \emph{good}\index{trace!good} if it actively passes through infinitely many applications of the rule $\U\mathrm{L}$.
\end{definition}

The following lemma describes a straightforward yet key property of good formula traces, where the complexity of a nested formula is simply the complexity of the underlying plain formula. We provide a proof sketch.

\begin{lemma}\label{lem fundamental property traces}
    If $(\varphi_i)_i$ is a good formula trace, then there is a plain formula of the form $ \varphi\U \psi$ and some $k < \omega$ such that for all $i\geq k$, $\varphi_i$ is of the form $\varphi\U \psi^{n_i}$ or $\X (\varphi\U \psi)^{n_i}$ for $n_i < \omega$. 
\end{lemma}
\begin{proof}[Proof Sketch.]
   Let $(\varphi_i)_i$ be a good formula trace. Then $(\varphi_i)_i$ actively passes through infinitely many applications of the rule $\U \mathsf{L}$. Let $(i_j)_j$ be the infinite sequence of natural numbers such that for all $i < \omega$ the trace formula $\varphi_{i}$ is principal in an instance of $\U \mathsf{L}$ if and only if $i = i_j$ for some $j < \omega$. We claim that for all $j < \omega$, $c(\varphi_{i_j}) \geq c(\varphi_{i_{j+1}})$. To show this let $j < \omega$ and consider $\varphi_{i_j}$ and $\varphi_{i_{j+1}}$. By assumption $\varphi_{i_j}$ is of the form $\varphi \U \psi^{n_{i_j}}$ for some plain formulas $\varphi$ and $\psi$ and $\varphi_{i_j+1}$ is either $\psi^{n_{i_j}}$ or $\varphi^{n_{i_j}}$ or $\X(\varphi \U \psi)^{n_{i_j}}$. 
   By inspection of the rules, note that the only rule that increases the complexity of a trace formula (when read bottom-up) is the rule $\U \mathsf{L}$ (recall that trace formulas cannot occur on the right side of a sequent). Since there are no applications of $\U \mathsf{L}$ between $\varphi_{i_j +1}$ and $\varphi_{i_{j+1}}$, we have $c(\varphi_{i_j +1}) \geq c(\varphi_{i_{j+1}})$. If $\varphi_{i_j + 1}$ is $\varphi^{n_{i_j}}$ or $\psi^{n_{i_j}}$, then $c(\varphi_{i_j}) > c(\varphi_{i_{j+1}})$. If $\varphi_{i_j + 1} = \X(\varphi \U \psi)^{n_{i_j}}$, then note that the only rules through which the trace can actively pass through between $i_j+1$ and $i_{j+1}$ are instances of $\mathsf{S}$ and of $\X \mathsf{L}$. Note that the trace must actively pass through exactly one instance of $\X \mathsf{L}$, implying that $c(\varphi_{i_j}) = c(\varphi_{i_{j+1}})$. Therefore $(c(\varphi_{i_j}))_{j}$ is an infinite sequence of decreasing natural numbers, which strictly decreases whenever $\varphi_{i_{j}}$ is the principal formula in an instance of $\U\mathsf{L}$ and $\varphi_{i_j+1} \not = \X \varphi_{i_j}$. Since natural numbers are well-founded, there exists a natural number $k$ such that $c(\varphi_{i_j}) = c(\varphi_{i_{k}})$ for all $j > k$. Thus the only rules through which the suffix $(\varphi_i)_{i \geq k}$ actively passes through are instances of $\mathsf{S},\X \mathsf{L}$ or instances of $\U \mathsf{L}$ where if $\varphi_i$ is principal, then $\varphi_{i+1} = \X \varphi_i$. Hence for all $i \geq k$, $\varphi_i = \varphi \U \psi^{n_i}$ or $\varphi_i = \X(\varphi \U \psi)^{n_i}$ for some plain formula $\varphi \U \psi$ and $n_i < \omega$.
\end{proof}

\begin{definition}\label{definition proof}
A \emph{$\mathsf{P}$-proof}\index{proof!$\niltle$-proof}\index{proof!$\niltlp$-proof} of a sequent $\sigma$ is a $\mathsf{P}$-pre-proof $\pi$ of  $\sigma$ such that
\begin{enumerate}
    \item every leaf in $\pi$ is labeled by an axiom;
    \item every infinite branch of $\pi$ contains a suffix that has a good formula trace.
\end{enumerate}
\end{definition}

\begin{example}
    The formula $(\X \varphi \rightarrow \X \psi) \rightarrow \X (\varphi \rightarrow \psi)^0$ is provable in $\niltlp$:
    \begin{prooftree}
        \AxiomC{}
        \RightLabel{$\mathsf{id}$}
        \UnaryInfC{$\X \varphi \rightarrow \X \psi^0, \varphi^1 \Rightarrow \varphi^1, \psi^1$}
        \RightLabel{$\X \mathsf{R}$}
        \UnaryInfC{$\X \varphi \rightarrow \X \psi^0, \varphi^1 \Rightarrow \X \varphi^0, \psi^1$}
        \AxiomC{}
        \RightLabel{$\mathsf{id}$}
        \UnaryInfC{$\psi^1, \varphi^1 \Rightarrow \psi^1$}
        \RightLabel{$\X \mathsf{R}$}
        \UnaryInfC{$\X \psi^0, \varphi^1 \Rightarrow \psi^1$}
        \RightLabel{${\rightarrow} \mathsf{L}$}
        \BinaryInfC{$\X \varphi \rightarrow \X \psi^0, \varphi^1 \Rightarrow \psi^1$}
        \RightLabel{${\rightarrow}\mathsf{R_p}$}
        \UnaryInfC{$\X \varphi \rightarrow \X \psi^0 \Rightarrow \varphi \rightarrow \psi^1$}
        \RightLabel{$\X \mathsf{R}$}
         \UnaryInfC{$\X \varphi \rightarrow \X \psi^0 \Rightarrow \X (\varphi \rightarrow \psi)^0$}
         \RightLabel{${\rightarrow}\mathsf{R_p}$}
         \UnaryInfC{$\Rightarrow (\X \varphi \rightarrow \X \psi) \rightarrow \X (\varphi \rightarrow \psi)^0$}
    \end{prooftree}
    Note that the same formula is not provable in $\niltle$. The lower-most application of ${\rightarrow}\mathsf{R_p}$ can be replaced by an application of ${\rightarrow} \mathsf{R}$, however, the next application of ${\rightarrow}\mathsf{R_p}$ cannot be replaced. If we apply $\mathsf{S}$, then the formula on the left-hand side is lost and the resulting sequent $\Rightarrow \varphi \rightarrow \psi^0$ is not provable. If we apply ${\rightarrow}\mathsf{L}$ first, then the resulting right premise $\X \psi^0 \Rightarrow \varphi \rightarrow \psi^1$ is provable by applying $\X \mathsf{L}$ to $\X \psi^0$, $\mathsf{S}$ and then ${\rightarrow} \mathsf{R}$, but the left premise $\X \varphi \rightarrow \X \psi^0 \Rightarrow \X \varphi^0, \varphi \rightarrow \psi^1$ is not provable, since ${\rightarrow} \mathsf{R}$ has a single-conclusion premise.
\end{example}

\section{Soundness}
\label{s-soundness}
This section establishes soundness of $\niltle$ with respect to the class of expanding models and of $\niltlp$ with respect to the class of persistent models. The proof is essentially an infinitary version of the soundness proof for $\CIM$. The measure $\mu(\sigma)$ assigned to sequents is generalised to \emph{signatures}: maps that associate a natural number to each `relevant' formula in a sequent $\sigma$.  We then assume towards a contradiction that there is a proof $\pi$ of an invalid sequent $\sigma$ and show, using a countermodel of $\sigma$, how to find an infinite path $\rho$ of invalid sequents in $\pi$ such that their signatures never increase but strictly decrease infinitely often, thus obtaining a contradiction to the well-foundedness of the natural numbers.
The aforementioned `relevant' formulas are called \emph{eventualities}.

\begin{definition}
    An \emph{eventuality}\index{eventuality} is a formula of the form $\X^j (\varphi\U \psi)^n$ with $n,j < \omega$. Given a sequent $\sigma$, a formula $E$ is an \emph{eventuality of $\sigma$} if  $E$ is an eventuality occurring in $\Gamma_\sigma$.
\end{definition}
Let $\U^k$ be the operator defined inductively by $\varphi\U^0 \psi=\psi$ and $\varphi\U^{k+1}\! \psi=\varphi\land \X(\varphi\U^{k}\! \psi)$. For an eventuality $E=\X^j (\varphi\U \psi)^n$ and $k < \omega$ define
\begin{align*}
    E[k]:=\X^j (\varphi\U^{k}\! \psi)^n. 
\end{align*}
Given a sequent $\sigma$, a \emph{signature for $\sigma$}\index{signature} is a map $\tau$ which assigns a natural number to each eventuality of $\sigma$. 
  Let $\Gamma_\sigma[\tau]$ be the set obtained from $\Gamma_\sigma$ by replacing each eventuality $E$ with $E[\tau(E)]$. Furthermore, let $\sigma[\tau]$ denote the sequent $\Gamma_\sigma[\tau]\Rightarrow \Delta_\sigma$.

  \begin{lemma}\label{l: invalid sequent existence signature}
      Let $\fw =(W, \leq, f, V)$ be an expanding (persistent) model and $E$ an eventuality. Then $\fw, w \models E$ if and only if there exists a natural number $k$ such that $\fw, w \models E[k]$.
  \end{lemma}
  \begin{proof}
     Let $E$ be an eventuality, $\fw=(W, \leq, f, V)$ an expanding (persistent) model and $w \in W$ a world with $\fw, w \models E$. There are formulas $\varphi, \psi$ and natural numbers $j,n$ such that $E = \X^j (\varphi \U \psi)^n$. If $\fw, w \models E$, then $\fw, f^{j+n}(w) \models \varphi \U \psi$. By definition there exists a natural number $k$ such that for all $j+n \leq i < k$ holds that $\fw, f^i(w) \models \varphi$ and $\fw, f^k(w) \models \psi$. Hence $\fw, f^{n+j}(w) \models \X^i \varphi$ for all $i < k -(j +n)$ and so $\fw, f^{j+n}(w) \models \varphi \U^k \psi$. Thus $\fw, w \models \X^j (\varphi \U^k \psi)^n$. For the other direction suppose there exists a natural number $k$ such that $\fw, w \models E[k]$. So $\fw, f^{j+n}(w) \models \varphi \U^k \psi$. An induction on $k$ shows that $\fw, f^{j+n}(w) \models \varphi \U \psi$. The proof is routine. Hence $\fw, w \models E$.
  \end{proof}

Note that Lemma \ref{l: invalid sequent existence signature} implies that given $\fw, w$ and $E$ with $\fw, w \models E$, there exists a least natural number $k$ with $\fw, w \models E[k]$. 

\begin{theorem}[Soundness for $\niltle$]\label{t: soundness for iltln}
    Every sequent provable in $\niltle$ is valid over the class of expanding models. 
\end{theorem}
\begin{proof}
    Let $\pi$ be an $\niltle$-proof of $\sigma$ and suppose, for contradiction, that $\sigma$ is not valid. Let $\fw=(W,\leq,f,V)$ be an expanding model and $w\in W$ such that $\fw,w\not\models\sigma$. For brevity, we will identify each node in $\pi$ with the sequent that labels it.

    We will inductively define an infinite path $(\sigma_i)_i$ of sequents through $\pi$, an infinite sequence of worlds $(w_i)_i$ in $\fw$ and an infinite sequence of signatures $(\tau_i)_i$ such that the following hold for every $i < \omega$:
    \begin{enumerate}
        \item $\tau_i$ is a signature for $\sigma_i$;
        \item  $w_i\models \bigwedge \Gamma_{\sigma_i}[\tau_i]$ and $w_i\not\models \bigvee \Delta_{\sigma_i}$ (thus $w_i\not\models \sigma_i[\tau_i]$ and so by Lemma \ref{l: invalid sequent existence signature}, $w_i\not\models \sigma_i$);
        \item for every eventuality $E$ of $\sigma_i$, the following hold:
        \begin{enumerate}
            \item $\tau_i(E)$ is the least natural number $k$ such that $w_i\models E[k]$;
            \item if $E$ is a side formula in the rule application with conclusion $\sigma_i$, then $E$ is an eventuality of $\sigma_{i+1}$ and $\tau_{i+1}(E)\leq\tau_i(E)$.
        \end{enumerate}
    \end{enumerate}
We define $(\sigma_i)_i$, $(w_i)_i$ and $(\tau_i)_i$ by induction on $i$ as follows. 

 Set $\sigma_0=\sigma$. Since $w\not\models \sigma$, there exists a $v\geq w$ such that $v\models \bigwedge \Gamma_\sigma$ and $v\not\models \bigvee \Delta_\sigma$. Set $w_0=v$ and for every eventuality $E$ in $\Gamma_\sigma$, define $\tau_0(E)$ to be the least $k$ such that $w_0 \models E[k]$ (which exists by Lemma \ref{l: invalid sequent existence signature}\footnote{This will not be mentioned again henceforth.}). Note that by construction $(\sigma_0,w_0, \tau_0)$ satisfies the properties 1. -- 3.
 
 Suppose $\sigma_i, w_i$ and $\tau_i$ are given and satisfy the properties 1. -- 3. To define $\sigma_{i+1}, w_{i+1}$ and $\tau_{i+1}$ we use a case distinction based on the rule applied at $\sigma_i$ in $\pi$ (i.e. the rule that has $\sigma_i$ as conclusion). Note that this rule cannot be an axiom, since $w_i\not\models \sigma_i$ and it is straightfoward to check that the conclusion of every instance of an axiom is valid. 

 \noindent \textsc{Case for $\wedge \mathsf{L}$.} Suppose $\sigma_i = (\Gamma, \varphi \wedge \psi^n \Rightarrow \Delta)$ with $\varphi \wedge \psi^n$ principal in the rule application. By assumption $w_i \models \varphi \wedge \psi^n$, implying that $f^n(w_i) \models \varphi \wedge \psi$. Hence $f^n(w_i) \models \varphi$ and $f^n(w_i) \models \psi$ and so $w_i \models \varphi^n$ and $w_i \models \psi^n$. Let $\sigma_{i+1} \coloneqq (\Gamma, \varphi^n, \psi^n \Rightarrow \Delta)$ and $w_{i+1} = w_i$. If $\varphi^n$ or $\psi^n$ is an eventuality, define $\tau_{i+1}(\varphi^n)$ to be the least natural number $k$ such that $w_{i+1} \models \varphi^n[k]$ and similarly for $\tau_{i+1}(\psi^n)$. On all other eventuality $E$, let $\tau_{i+1}(E) \coloneqq \tau_i(E)$. Note that the properties 1. -- 3. hold. \smallskip

 \noindent \textsc{Case for $\wedge \mathsf{R}$.} Suppose $\sigma_i = (\Gamma \Rightarrow \varphi \wedge \psi^n, \Delta)$ with $\varphi \wedge \psi^n$ principal in the rule application. By assumption $w_i \not \models (\varphi \wedge \psi^n)$, implying that $f^n(w_i) \not \models \varphi \wedge \psi$. Thus $f^n(w) \not \models \varphi$ or $f^n(w) \not \models \psi$, so $w_i \not \models \varphi^n$ or $w_i \not \models \psi^n$. If $w_i \not \models \varphi^n$, let $\sigma_{i+1} \coloneqq (\Gamma \Rightarrow \varphi^n, \Delta)$, $w_{i+1} \coloneqq w_i$ and $\tau_{i+1} \coloneqq \tau_i$. Note that the properties 1. -- 3. hold. The case where $w_i \not \models \psi^n$ is similar by choosing the sequent in the right premise of $\wedge \mathsf{R}$ as $\sigma_{i+1}$. \smallskip
 
 \noindent \textsc{Cases for $\vee \mathsf{L}$ and $\vee \mathsf{R}$.} The cases for $\vee \mathsf{L}$ and $\vee \mathsf{R}$ are symmetric to the cases for $\wedge \mathsf{R}$ and for $\wedge \mathsf{L}$, respectively. \smallskip
 
 \noindent \textsc{Case for ${\rightarrow} \mathsf{L}$.} Suppose $\sigma_i = (\Gamma, \varphi \rightarrow \psi^n \Rightarrow \Delta)$ and $\varphi \rightarrow \psi^n$ is principal in the rule application. By assumption $w_i \models \varphi \rightarrow \psi^n$, implying that $f^n(w_i) \models \varphi \rightarrow \psi$. If $f^n(w_i) \not \models \varphi$, then $w_i \not \models \varphi^n$ and so let $\sigma_{i+1} \coloneqq (\Gamma, \varphi \rightarrow \psi^n \Rightarrow \varphi^n, \Delta)$, $w_{i+1} \coloneqq w_i$ and $\tau_{i+1} \coloneqq \tau_i$. Note that the properties 1. -- 3. are satisfied. Otherwise $f^n(w_i) \models \psi$, implying that $w_i \models \psi^n$. Hence let $\sigma_{i+1} \coloneqq (\Gamma, \psi^n \Rightarrow \Delta)$, $w_{i+1} \coloneqq w_i$ and if $\psi^n$ is an eventuality, let $\tau_{i+1}(\psi^n)$ be the least natural number $k$ such that $w_{i+1} \models \psi^n[k]$. On all other eventualities $E$, $\tau_{i+1}(E) \coloneqq \tau_i(E)$. Note that the properties 1. -- 3. are satisfied. \smallskip

\noindent \textsc{Case for ${\to} \mathrm{R}$.} Suppose $\sigma_i= (\Gamma \Rightarrow \varphi\rightarrow \psi^0,\Delta)$ with $\varphi\rightarrow \psi^0$ principal in the rule application. Let $\sigma_{i+1}=(\Gamma, \varphi^0\Rightarrow \psi^0)$. Since $w_i \not \models \varphi \rightarrow \psi^0$, there exists a $v \geq w_i$ such that $v \models \varphi^0$ and $v\not\models \psi^0$. Let $w_{i+1} = v$. For any eventuality $E$ in $\Gamma\cup\{\varphi^0\}$, let $\tau_{i+1}$ map $E$ to the least $k$ such that $w_{i+1}\models E[k]$. Note that $\tau_{i+1}$ is a well-defined signature for $\sigma_{i+1}$. Moreover, by construction, the properties 2. and 3.(a) hold. Since $w_{i+1}\geq w_i$, by monotonicity (Lemma \ref{l: monotonicity for expanding/persistent models}) we have  $\tau_{i+1}(E)\leq \tau_{i}(E)$ for each eventuality $E$ in $\Gamma$, implying that 3.(b) holds as well. \smallskip

\noindent \textsc{Case for $\X \mathrm{L}$.} Suppose $\sigma_i= (\Gamma, \X \varphi^n\Rightarrow \Delta)$ with $\X \varphi^n$ principal in the rule application. Let $\sigma_{i+1}=(\Gamma,  \varphi^{n+1}\Rightarrow \Delta)$ and $w_{i+1}=w_i$. If $\varphi^{n+1}$ is an eventuality, then define ${\tau_{i+1}(\varphi^{n+1})\coloneqq  \tau_i(\X \varphi^n)}$. For any other eventuality $E$, let $\tau_{i+1}(E) \coloneqq \tau_i(E)$. Note once again that the properties 1.-3. hold. In particular, the formula $\varphi^{n+1}$ has the same interpretation as $\X \varphi^n$, implying that $\tau_{i+1}(\varphi^{n+1})$ is the least $k$ such that $w_{i+1} \models \varphi^{n+1}[k]$. \smallskip

\noindent \textsc{Case for $\mathrm{S}$.} Suppose $\sigma_i= (\Sigma,\Gamma^{+1} \Rightarrow \Delta^{+1},\Pi)$ such that $\Gamma \Rightarrow \Delta$ is the premise of the rule application. Let $\sigma_{i+1}=(\Gamma \Rightarrow \Delta)$, $w_{i+1}=f(w_i)$ and $\tau_{i+1}(E^n)=\tau_i(E^{n+1})$ for every eventuality $E^n$ in $\Gamma$. By construction $\tau_{i+1}$ is a signature. Note that since $w_i \models \Gamma^{+1}_{\sigma_i}[\tau_i]$, it holds that $f(w_i) \models \Gamma_{\sigma_i}[\tau_{i+1}]$. Therefore the property 2. holds as well. Property 3.(b) is given by construction, while 3.(a) holds since if there would exists a natural number $k < \tau_{i+1}[E]$ with $w_{i+1} \models E[k]$ for some eventuality $E$ in $\Gamma$, then $w_i \models E^{+1}[k]$, implying that $\tau_i$ does not satisfy property 3.(a); a contradiction. \smallskip

\noindent \textsc{Case for $\U \mathrm{L}$.} Suppose $\sigma_i= (\Gamma, \varphi\U \psi^n\Rightarrow \Delta)$ with $\varphi\U \psi^n$ principal in the rule application. We distinguish two cases.
\begin{enumerate} 
 \item If $\tau_i(\varphi\U \psi^n)=0$, let $\sigma_{i+1}=(\Gamma,  \psi^n\Rightarrow \Delta)$ and $w_{i+1}=w_i$. If $\psi^n$ is an eventuality and not in $\Gamma$, let $\tau_{i+1}$ map $\psi^n$ to the least $k$ such that $w_{i+1}\models \psi^n[k]$. On other eventualities, $\tau_{i+1}$ acts as $\tau_i$. Clearly, $\tau_{i+1}$ is a signature. Note that since $\tau_i(\varphi \U \psi^n) = 0$, we have $f^n(w_i) \models \psi$ since $\varphi \U^0 \psi = \psi$. Thus the property 2. holds. Properties 3. (a) and (b) hold by construction.
            
\item If $\tau_i(\varphi\U \psi^n)>0$, let $\sigma_{i+1}=(\Gamma, \varphi^n, \X(\varphi\U \psi)^n\Rightarrow \Delta)$ and $w_{i+1}=w_i$. If $\varphi^n$ is an eventuality and not in $\Gamma$, let $\tau_{i+1}$ map $\varphi^n$ to the least $k$ such that $w_{i+1}\models \varphi^n[k]$. Define $\tau_{i+1}(\X(\varphi\U \psi)^n) \coloneqq \tau_i(\varphi\U \psi^n)-1$. On other eventualities, $\tau_{i+1}$ acts as $\tau_i$. By construction the properties 1. and 3.(b) hold. Note that since $w_i \models \varphi \U \psi^n[k]$ where $k = \tau_i(\varphi\U \psi^n)$, we have that $w_{i+1} \models \X (\varphi \U \psi)^n[k-1]$, and $k-1$ is the least natural number with this property. Thus we obtain that the properties 2. and 3.(a) hold as well.
\end{enumerate}

Therefore we obtain $(\sigma_i)_i$, $(w_i)_i$ and $(\tau_i)_i$ satisfying the properties 1. -- 3. Since $\pi$ is a proof, the infinite branch  $(\sigma_i)_i$ must contain a good trace $(\varphi_i)_{i \geq j}$ starting in some sequent $\sigma_j$. By Lemma \ref{lem fundamental property traces}, we may assume that this trace only passes actively through the rules $\X \mathrm{L}$, $\mathrm{S}$ and $\U  \mathrm{L}$, and it cannot pass through the latter in a degenerative way.\footnote{Formally, a trace $(\varphi_i)_{i}$ \emph{passes degeneratively through $\U\mathrm{L}$} if there is a $\varphi_j$ of the form $\varphi\U \psi^n$ such that $\varphi_{j+1}\in\{\varphi^n,\psi^n\}$.} Now consider the infinite sequence $(\tau_i(\varphi_i))_{i \geq j}$ of natural numbers. Note that, by property 3(b), if  $\varphi_i$ is a side formula then $\tau_{i+1}(\varphi_{i+1})\leq \tau_{i}(\varphi_{i})$. Moreover, if $\varphi_i$ is principal in an application of $\X \mathrm{L}$ or $\mathrm{S}$ then $\tau_{i+1}(\varphi_{i+1})= \tau_{i}(\varphi_{i}),$ and if $\varphi_i$ is principal in a (non-degenerative) application of $\U   \mathrm{L}$ then $\tau_{i+1}(\varphi_{i+1})<\tau_{i}(\varphi_{i})$ by construction. As the trace is good, the latter case occurs infinitely often, and so we obtain an infinite, strictly decreasing sequence of natural numbers and thereby a contradiction. 
\end{proof}

We finish the section by proving that $\niltlp$ is sound with respect to the class of persistent models. The proof follows the same argument as for $\niltle$, with the only differences that the case for $\mathsf{S}$ can be skipped and the case for ${\rightarrow}\mathsf{R}$ must be replaced by the case for ${\rightarrow}\mathsf{R_p}$.

\begin{theorem}[Soundness for $\niltlp$]\label{t: soundness of iltlpn}
    Every sequent provable in $\niltlp$ is valid over the class of persistent models.
\end{theorem}

\begin{proof}
Let $\pi$ be an $\niltlp$-proof of $\sigma$ and suppose towards contradiction that $\sigma$ is not valid. Let $\fw =(W, \leq, f, V)$ be a persistent model and $w \in W$ such that $\fw, w \not \models \sigma$. As in the soundness proof for $\niltle$ we define inductively a path $(\sigma_i)_i$ of sequents through $\pi$, a sequence of worlds $(w_i)_i$ in $\fw$ and a sequence of signatures $(\tau_i)_i$ such that the properties 1. - 3. in the proof of Theorem \ref{t: soundness for iltln} hold. We consider the only new case. All other cases are identical to the cases shown above. \smallskip 

\noindent \textsc{Case for ${\rightarrow}\mathsf{R_p}$.} Suppose $\sigma_i = (\Gamma \Rightarrow \varphi \rightarrow \psi^n, \Delta)$ with $\varphi \rightarrow \psi^n$ principal in the rule application. Let $\sigma_{i+1} = (\Gamma, \varphi^n \Rightarrow \psi^n)$. Since $w_i \not \models \varphi \rightarrow \psi^n$, it holds that $f^n(w_i) \not \models \varphi \rightarrow \psi$. Hence there exists $v \geq f^n(w_i)$ with $v \models \varphi$ and $v \not \models \psi$. By Lemma \ref{l: iteration of backwards confluent function is backward confluent} we find a world $u \geq w_i$ such that $v = f^n(u)$. Then $u \models \varphi^n$ and $u \not \models \psi^n$. Let $w_{i+1}=u$. It is straightforward to check that $w_{i+1} \models \bigwedge\Gamma \wedge \varphi^n$ and $w_{i+1} \not \models \psi^n$. Let $\tau_{i+1}$ be the signature obtained by mapping each eventuality $E$ in $\Gamma \cup \{\varphi^n\}$ to the least natural number $k$ such that $w_{i+1} \models E[k]$. From monotonicity (Lemma \ref{l: monotonicity for expanding/persistent models}) then follows that $\tau_{i+1}(E) \leq \tau_i(E)$ for each eventuality $E$ in $\Gamma \cup \{\varphi^n\}$. \smallskip

A contradiction to the wellfoundnedness of the natural numbers is now obtained by the same argument as given in the proof of Theorem \ref{t: soundness for iltln}
\end{proof}
\section{Completeness}
\label{s-completeness}
This section establishes completeness of $\niltle$ with respect to the class of expanding models and of $\niltlp$ with respect to the class of persistent models. As for $\NWIM$ the proof proceeds via a proof search argument. For each sequent $\sigma$ we construct an infinite two-player game between \emph{Prover} and \emph{Refuter} such that a winning strategy for Prover corresponds to a proof of $\sigma$ and a winning strategy for Refuter to the existence of a countermodel for $\sigma$. The game will be played on a \emph{proof search tree}, which is a finitely branching, non-wellfounded tree that presents a systematic search for a proof of $\sigma$. The presented completeness proofs share many similarities with the completeness proof for $\NWIM$. However, there are some crucial differences, due to the presence of forward and back-up confluence, that the countermodels induced by a refutation must satisfy.\footnote{If the reader has skipped Chapter \ref{c: IM}, it may be informative to go back and briefly review the key ideas presented there.} For $\NWIM$, when a formula $\lbm \varphi$ was present on the right side of a sequent, it sufficed to unfold it once to saturate the formula, i.e. it was enough to guarantee that $\lbm \varphi$ was \emph{locally} correct in the world corresponding to the saturated sequent. In the presence of forward confluence, this does not suffice. Suppose for example that a formula of the form $\E (\varphi \vee \psi)^0$ (i.e. $\top \U (\varphi \vee \psi)^0$) occurs on the left side of a sequent $\sigma$ corresponding to the world $w$. The countermodel then must contain a temporal successor $f^n(w)$ which satisfies $\varphi \vee \psi$, which will be guaranteed by the fact that no infinite branch in a refutation can contain a good trace, i.e. after finitely many times unfolding $\E (\varphi \vee \psi)$ the Refuter must chose the left premise of $\U \mathsf{L}$, which will result in the corresponding world satisfying $\varphi \vee \psi^0$. Saturation then guarantees that $\varphi^0$ or $\psi^0$ is contained on the left side of the saturated sequent, implying that the corresonding world $f^n(w)$ satisfies $\varphi$ or $\psi$. However if there exists a world $v \geq w$, then by forward confluence $f^n(v) \geq f^n(w)$, so by monotonicity $f^n(v)$ must satisfy the same disjunct as $f^w(v)$. In the way proof search trees are defined, the finite segment in a branch corresponding to $f^n(v)$ is unrelated to the finite segment corresponding to $f^n(w)$ in the proof search tree. Thus it could happen that at the segment corresponding to $f^n(w)$, when we apply $\vee \mathsf{L}$ to $\varphi \vee \psi^0$, Refuter chooses $\varphi$, while at the segment corresponding $f^n(v)$, Refuter chooses $\psi$, implying that the resulting model is not monotone. To circumvent this problem, we will make use of the nesting and impose that in order to saturate a sequent, we must apply $\U \mathsf{L}$ finitely many times until the left premise is chosen, implying that some branches (namely those containing a good trace) will never encounter a saturated sequent. The nesting is crucial for this, as we are able to continue unfolding by eliminating $\X$-operators and instead increase the nesting level.

\subsection{Completeness for Expanding Models}

We start by giving a suitable notion of \emph{saturated sequent}.

\begin{definition}\label{def saturated}
A sequent $\Gamma\Rightarrow \Delta$ is \emph{left-saturated} if the following hold.
\begin{enumerate}
    \item if $\varphi\land \psi^n\in \Gamma$, then $\varphi^n,\psi^n\in\Gamma$;
    \item if $\varphi\lor \psi^n\in \Gamma$, then $\varphi^n\in\Gamma$ or $\psi^n\in\Gamma$;
    \item if $\varphi\to \psi^n\in \Gamma$, then $\varphi^n\in\Delta$ or $\psi^n\in\Gamma$;
    \item if $\X \varphi^n\in \Gamma$, then $\varphi^{n+1}\in\Gamma$;
    \item  if $\varphi\U \psi^n\in \Gamma$, then there exists an $m\geq n$ such that $\psi^m\in\Gamma$ and $\varphi^k\in \Gamma$ for all $n\leq k< m$.
\end{enumerate}
The sequent is \emph{saturated}\index{sequent!saturated} if, in addition,
\begin{enumerate}[resume]
    \item if $\varphi\land \psi^n\in \Delta$, then $\varphi^n\in\Delta$ or $\psi^n\in\Delta$;
    \item if $\varphi\lor \psi^n\in \Delta$, then $\varphi^n,\psi^n\in\Delta$;
    \item if $\X \varphi^n\in \Delta$, then $\varphi^{n+1}\in\Delta$;
    \item if $\varphi\U \psi^0\in \Delta$, then $\psi^0,\varphi^0\in\Delta$ or $\psi^0,\X(\varphi\U \psi)^0\in\Delta$.
\end{enumerate}
Given a sequent $\sigma$, we say that a formula $\varphi$ is \emph{saturated in $\sigma$} if $\sigma$ satisfies the relevant saturation clause for $\varphi$.
\end{definition}
Note that the saturation clause for right $\U$-formulas is restricted to the zeroth nesting level. As before, we call an application of a rule \emph{succinct} if the principal formula(s) is not also a side formula, and \emph{preserving} if the principal formula(s) is also a side formula. Note that rule applications of ${\to} \mathrm{R}$ and $\mathrm{S}$ are always succinct.

\begin{definition}\label{d: proof search tree for iltle}
A \emph{proof search tree}\index{proof search!tree} $\mathcal{T}$ for a sequent $\sigma$ is a finite or infinite tree whose nodes are labeled according to the rules of $\niltle$ and in which the following holds.
\begin{enumerate}
    \item The root of $\mathcal{T}$ is labeled by $\sigma$.
    \item A node of $\mathcal{T}$ is a leaf if and only if it is labeled by an axiom.
    \item Every left rule application is succinct.
    \item Every right rule application except ${\to} \mathrm{R}$ is preserving. 
    \item No invertible rule is applied to a sequent in which the principal formula is already saturated. 
    \item Instead of the rules ${\to} \mathrm{R}$ and $\mathrm{S}$, we have the rule\index{choice rule!$\mathsf{C}$}
       \begin{align*}
          \infer[\mathsf{C},]{\Sigma,\Gamma^{+1}\Rightarrow (\varphi_0\to \psi_0)^{0},\dots,(\varphi_k\to \psi_k)^{0},\Delta^{+1},  \Pi}{\Sigma,\Gamma^{+1}, \varphi_0^{0} \Rightarrow \psi_0^{0} &\dots&\Sigma,\Gamma^{+1}, \varphi_k^{0} \Rightarrow \psi_k^{0}& \Gamma \Rightarrow \Delta}
       \end{align*}
    where it is required that every formula in $\Sigma\cup \Pi$ is of nesting level $0$, $\Pi$ does not contain a formula of the form $\varphi\to \psi^0$ and the conclusion is a saturated sequent. We call the premises of the form $\Sigma,\Gamma^{+1}, \varphi_i^0 \Rightarrow \psi_i^0$ the \emph{intuitionistic premises}, and $\Gamma\Rightarrow \Delta$ the \emph{modal premise} of $\mathsf{C}$.
\end{enumerate}
\end{definition}
The choice rule $\mathsf{C}$ represents a choice between non-invertible rules that Prover has to make once the sequent is saturated. Note that the empty sequent is saturated; an empty sequent in a proof search tree  can only  be the conclusion of a $\mathrm{C}$-rule and has another empty sequent as its only direct successor. This is why, in difference to proof search trees for $\NWIM$, every leaf in a proof search tree is labeled by an axiom.

\begin{lemma}
    Every sequent $\sigma$ has a proof search tree.
\end{lemma}

 Intuitively, given a sequent $\sigma$, one can build a proof search tree as follows. First try to saturate all left formulas by succinctly applying invertible left rules. If a left-saturated sequent is obtained, saturate all right formulas by preservingly applying invertible right rules, then apply $\mathsf{C}$ and start over. Observe that it is possible that some branches in a proof search tree do not contain a saturated sequent due the fifth saturation clause. The proof is similar to the proof of Lemma~\ref{l: proof search trees exist} and omitted.

The following lemmas describe some key properties of proof search trees.
For a node $s$ in a proof search-tree we write $\Gamma_s\Rightarrow \Delta_s$ to denote the sequent labelling the node $s$. 

\begin{lemma}\label{lemma property (c) rule}
    If $\mathcal{T}$ is a proof search tree wherein $s\in \mathcal{T}$ is the conclusion of a $\mathsf{C}$-application with modal premise $t\in \mathcal{T}$, then the following hold.
    \begin{enumerate}
        \item $t$ is labeled by a left-saturated sequent;
        \item if $r\geq t$ and no $\mathsf{C}$-application occurs between $t$ and $r$, then $\Gamma_r=\Gamma_t$.    
    \end{enumerate}
\end{lemma}
\begin{proof} 
    Let $\Gamma_s \Rightarrow \Delta_s = \Sigma,\Gamma^{+1}\Rightarrow (\varphi_0\to \psi_0)^{0},\dots,(\varphi_k\to \psi_k)^{0},\Delta^{+1},  \Pi$ and let  $\Gamma_t \Rightarrow \Delta_t = \Gamma \Rightarrow \Delta$. That $\Gamma_t \Rightarrow \Delta_t$ is left-saturated follows directly from $\Gamma_s \Rightarrow \Delta_s$ being left-saturated: note that formulas in $\Sigma$ are of nesting level $0$ while each formula in $\Gamma^{+1}$ is of nesting level strictly larger than $0$. Hence for any formula in $\Gamma^{+1}$ its saturation clause is satisfied by formulas in $\Gamma^{+1}$ and so $\Gamma_t \Rightarrow \Delta_t$ is left-saturated. Suppose $r \geq t$ and no $\mathsf{C}$-application occurs between $t$ and $r$. By 1. $\Gamma_t$ is labeled by a left-saturated sequent. By Property 5. of a proof search tree no invertible rule is applied to a sequent in which the principal formula is already saturated. Furthermore, the only rule introducing new formulas on the left side of the sequent is ${\rightarrow}\mathsf{R}$, which is merged into the $\mathsf{C}$-rule. Therefore no rules can be applied to formulas in $\Gamma_s$ for any node $s$ between $t$ and the next node labeled by the conclusion of a $\mathrm{C}$-rule application. Therefore $\Gamma_t = \Gamma_r$.
\end{proof}

\begin{lemma}\label{lemma proof search tree inf UL or C}
    Every infinite branch of a proof search tree $\mathcal{T}$ contains infinitely many applications of $\U\mathsf{L}$ or $\mathsf{C}$.
\end{lemma}
\begin{proof}
    Let $(\rho_i)_{i}$ be an infinite branch of $\mathcal{T}$. Suppose there exists a suffix $(\rho_i)_{i \geq j}$ that contains no applications of $\U\mathsf{L}$ or $\mathsf{C}$. Due to Properties 3. to 5. of the proof search tree, there exists a $k\geq j$ such that all formulas except for left $\U$-formulas will be saturated in $\rho_k$. The only rules which may be applied at that point are $\U\mathsf{L}$ or $\mathsf{C}$, showing that $(\rho_i)_{i}$ must be finite.
\end{proof}

\begin{lemma}\label{lemma proof search tree inf (C) or good trace}
    Every infinite branch of a proof search tree $\mathcal{T}$ that contains only finitely many $\mathsf{C}$-applications has a suffix with a good formula trace.
\end{lemma}

\begin{proof}
    Let $\beta$ be an infinite branch of $\mathcal{T}$ with finitely many $\mathsf{C}$-applications. Let $\rho$ be a suffix of $\beta$ that starts after the last $\mathsf{C}$-application. By the previous lemma, $\rho$ must contain infinitely many applications of $\U\mathrm{L}$. We show that $\rho$ contains a good trace.

    Consider the tree $\mathcal{T}_{\rho}$ of formula traces on $\rho$ (add a fresh node as the root). Now let $\mathcal{T}'_{\rho}$ be the tree obtained from $\mathcal{T}_\rho$ by identifying consecutive nodes that are labeled by the same formula. Note that $\mathcal{T}'_\rho$ cannot be finite, since $\rho$ must contain infinitely many applications of $\U\mathsf{L}$ and this rule may not be applied to formulas that also function as a side formula. Moreover, by inspection of the rules note that $\mathcal{T}'_\rho$ is finitely branching. By K\"onig's Lemma, $\mathcal{T}'_\rho$ contains an infinite branch. Note that this branch corresponds to an infinite formula trace $(\varphi_i)_{i}$ on $\rho$ that does not stagnate on a side formula, that is,  $(\varphi_i)_{i}$ actively passes through a left rule infinitely often. Since left rules are applied succinctly and there are no $\mathsf{C}$-applications in $\rho$, the trace $(\varphi_i)_{i}$ must actively passes through $\U\mathsf{L}$ infinitely often, implying that $(\varphi_i)_i$ is a good trace. 
\end{proof}

Next we define \emph{proof search games}. As before, proof search games are played on proof search trees by two players called Prover and Refuter. Since only one frame condition is considered (and therefore only one choice rule is needed), the choice rule $\mathsf{C}$ is not displayed in the name of the proof search game.

\begin{definition}\label{d: proof search game for iltle}
    Let $\sigma$ be a sequent and $\mathcal{T}$ a proof search tree for $\sigma$. The \emph{proof search game}\index{proof search!games} $\mathcal{G}(\mathcal{T}, \sigma)$ is played by two players called \emph{Prover} and \emph{Refuter}. The \emph{arena} is the proof search tree $\mathcal{T}$, where each \emph{position} is a node of $\mathcal{T}$. Prover \emph{owns} every position $t \in \mathcal{T}$ which is labeled by the conclusion of a $\mathsf{C}$-rule instance. Refuter owns every other position. The \emph{admissible moves are as follows}: if Player owns position $t$, then Player plays by choosing any child node of $t$ (if one exists). A \emph{play} is a sequence of positions $(t_i)_i$ starting in the root of $\mathcal{T}$ such that any two consecutive positions are related by an admissible move. A play is either finite and ends in a leaf of $\mathcal{T}$ or infinite. The \emph{winning conditions are as follows}:
    \begin{enumerate}
        \item Prover wins a play $(t_i)_i$ if the play is finite or if it is infinite and $(t_i)_i$ contains a good trace.
        \item Refuter wins a play $(t_i)_i$ if the play is infinite and does not contain a good trace.
    \end{enumerate}
\end{definition}

Observe that every play in $\mathcal{G}(\mathcal{T}, \sigma)$ corresponds to a branch of $\mathcal{T}$. As for $\NWIM$ a \emph{strategy} for Player is a partial function $f$ which maps any position $t$ owned by Player onto a child node of $t$. Player \emph{uses} strategy $f$ if whenever a play is in position $t$ owned by Player, then Player chooses $f(t)$. A strategy $f$ is a \emph{winning strategy} if Player wins every play in which $f$ is used. The \emph{strategy tree}\index{strategy tree} of a strategy $f$ is the subtree of $\mathcal{T}$ consisting of all plays that can occur when Player uses $f$. For detailed definitions, see  Definition \ref{d: strategy}, Definition \ref{d: winning strategy} and Definition \ref{d: strategy tree}. The following lemma then follows directly from the winning conditions of Prover.

\begin{lemma}\label{l: winning strategy implies proof}
    Suppose Prover has a winning strategy in $\mathcal{G}(\mathcal{T},\sigma)$. Then $\sigma$ has a proof in $\niltle$.
\end{lemma}
\begin{proof}
    Let $\pi$ be the strategy tree of the winning strategy for Prover. By definition the root of $\pi$ is labeled by $\sigma$ and whenever $\pi$ contains a node $u$ labeled by a sequent which is the conclusion of an invertible rule, then all children of $u$ are included (since $u$ is owned by Refuter and thus Prover must have an answer to any admissible move Refuter makes at $u$). If $u$ is labeled by a sequent which is the conclusion of an instance of $\mathsf{C}$, then $u$ is owned by Prover and the strategy tells Prover which child of $u$ to choose. Suppose Prover chooses child node $u_i$. If $u_i$ is labeled by an intuitionistic premise of $\mathsf{C}$, note that the sequents labelling $u$ and $u_i$ form an instance of the rule ${\rightarrow}\mathsf{R}$. If $u_i$ is labeled by the modal premise of $\mathrm{C}$, then the sequents labelling $u$ and $u_i$ form an instance of the rule $\mathsf{S}$. Therefore $\pi$ is a pre-proof of $\sigma$ (modulo renaming of the edges). By definition of a proof search tree, every finite branch of $\pi$ ends in a leaf labeled by an axiom. Furthermore, since $\pi$ is the strategy tree of a \emph{winning strategy} for Prover, every infinite branch contains a path with a good formula trace. Therefore $\pi$ is a proof of $\sigma$.
\end{proof}

Winning strategies for Prover therefore correspond to proofs. Similarly, winning strategies for Refuter correspond to \emph{refutations}. 

\begin{definition}\label{definition refutation}
A \emph{refutation}\index{proof search!refutation} of a sequent $\sigma$ is a subtree $\mathcal{R}$ of a proof search tree $\mathcal{T}$ for $\sigma$ such that the following hold.
\begin{enumerate}
    \item $\mathcal{R}$ contains the root of $\mathcal{T}$. 
    \item Every branch of $\mathcal{R}$ is infinite.
    \item If a node $s$ in $\mathcal{R}$ is (labeled by) the conclusion of an application of $\mathsf{C}$ in $\mathcal{T}$, then $\mathcal{R}$ contains all children of $s$ in $\mathcal{T}$. 
    \item If a node $s$ in $\mathcal{R}$ is (labeled by) the conclusion of an application of any rule other than $\mathsf{C}$ in $\mathcal{T}$, then $\mathcal{R}$ contains exactly one child of $s$ in $\mathcal{T}$. 
    \item No infinite branch of $\mathcal{R}$ contains a path with a good formula trace. 
\end{enumerate}
\end{definition}

Note that the final condition above together with Lemma \ref{lemma proof search tree inf (C) or good trace} imply that every branch in a refutation must contain infinitely many applications of the $\mathsf{C}$-rule.

\begin{lemma}\label{l: winning strategy implies refutation}
    Suppose Refuter has a winning strategy in $\mathcal{G}(\mathcal{T}, \sigma)$. Then $\sigma$ has a refutation. 
\end{lemma}
\begin{proof}
    Let $\mathcal{R}$ be the strategy tree of the winning strategy for Refuter. By definition $\mathcal{R}$ contains the root of $\mathcal{T}$. For any node $u$ in $\mathcal{R}$ if $u$ is labeled by a sequent which is the conclusion of an instance of an invertible rule, then $u$ is owned by Refuter and therefore $\mathcal{R}$ contains exactly one child node of $u$ (namely the node obtained from the admissible move provided by the strategy). If $u$ is labeled by a sequent which is the conclusion of an instance of the $\mathsf{C}$-rule, then $u$ is owned by Prover and therefore every child node of $u$ is included in $\mathcal{R}$ (since the strategy must provide an answer to any admissible move Prover could make at $u$). Since $\mathcal{R}$ is the strategy tree of a \emph{winning strategy} for Refuter, $\mathcal{R}$ cannot contain a finite branch, as this would imply that Prover has a winning play when Refuter plays using the strategy. Furthermore, due to the same reason, no infinite branch can contain a path with a good formula trace. Hence $\mathcal{R}$ is a refutation.
\end{proof}

Fix a sequent $\sigma$ and a refutation $\mathcal{R}$ of $\sigma$. 

\begin{definition}
    The canonical model\index{canonical model!for $\niltle$} $\fw_c = (W, \leq, f, V)$ for $\sigma$ (relative to $\mathcal{R}$) is defined as follows. 
    \begin{enumerate}
    \item $W=\sfrac{\mathcal{R}}{\sim}$, where $s\sim t$ iff there exists a path between $s$ and $t$ in which no $\mathsf{C}$-application occurs. 
    \item Define the function $f$ by
    \begin{align*}
       f(w)=v\text{ iff }&\text{there exist $s\in w$ and $t\in v$ such that $s$ is the conclusion}
       \\&\text{and $t$ is the \emph{modal} premise of the same $\mathsf{C}$-application.}
    \end{align*}
    Note that $f$ is a total function, since every branch of $\mathcal{R}$ contains infinitely many $\mathsf{C}$-applications and every $\mathsf{C}$-application has a right premise.
    \item First define the relation $\leq_0$ on $W$ by 
    \begin{align*}
         w\leq_0 v \text{ iff }&\text{there exist $s\in w$ and $t\in v$ such that $s$ is the conclusion}
         \\&\text{and $t$ an \emph{intuitionistic} premise of the same $\mathrm{C}$-application.}
    \end{align*}
    Then let $\leq$ be the reflexive transitive closure of the relation
    \begin{align*}
        \leq_1\ \coloneqq \{(f^n(w),f^n(v)): w\leq_0 v \text{ and } n<\omega\}.
    \end{align*}
    \item Define the valuation by 
    \(
        V(w)=\{p\in \mathsf{Prop}: p^0\in \Gamma_{w}\}
    \)
    where 
    \(
        \Gamma_w = \bigcup_{s\in w} \Gamma_{s}.
    \)
\end{enumerate}
\end{definition}

Similar to $\Gamma_w$ we write $\Delta_w$ for $\bigcup_{s\in w} \Delta_{s}$. The following is straightforward to check.

\begin{lemma}
    For any world $w$ of the canonical model $\fw_c$, the sequent $\Gamma_w \Rightarrow \Delta_w$ is saturated.
\end{lemma}
\begin{proof}
    By definition of a proof search tree and $\sim$.
\end{proof}

\begin{lemma}\label{l: witnessing refutation model}
    Let $\fw_c$ be the canonical model for $\sigma$ relative to a refutation $\mathcal{R}$. Let $w \in W$ and let $s \in w$ be the unique node which is the conlcusion of a $\mathsf{C}$-instance. Then for any formula $\varphi$ any natural numbers $k \leq n < \omega$, $\varphi^n \in \Gamma_s$ if and only if $\varphi^{n-k} \in \Gamma_{f^k(w)}$.
\end{lemma}
\begin{proof}
     We proceed by induction on $k$. The base case for $k=0$ is trivial. For $k > 0$, by induction hypothesis $\varphi^n \in \Gamma_s$ if and only if $\varphi^{n-k+1} \in \Gamma_{f^{k-1}(w)}$ if and only if there exists $t_0 \in f^{k-1}(w)$ such that $\varphi^{n-k+1} \in \Gamma_{t_0}$. Let $t \in f^{k-1}(w)$ be the unique node which is the conclusion of a $\mathsf{C}$-rule instance. By Lemma \ref{lemma property (c) rule}, $\varphi^{n-k+1} \in \Gamma_{t_0}$ if and only if $\varphi^{n-k+1} \in \Gamma_t$. Let $t'$ be the modal premise of the $\mathsf{C}$-rule with $t$ as conclusion, then $\varphi^{n-k+1} \in \Gamma_t$ if and only if $\varphi^{n-k} \in \Gamma_{t'}$ if and only if $\varphi^{n-k} \in \Gamma_{f^k(w)}$. 
\end{proof}

\begin{lemma}\label{claim 1 completeness proof}
$\fw_c$ is an expanding model.
\end{lemma}
\begin{proof}
Forward confluence follows directly from the definition of $\leq_1$.  For monotonicity of the valuation, note that it suffices to show that the relation $\leq_1$ is monotone in $V$. 
In the following, we write $[t]$ for the equivalence class of $t$ with respect to $\sim$.

Let $w,v\in W$ with $w\leq_1 v$. Then there exist $n<\omega$ and $s,t\in \mathcal{R}$ such that $w=f^n([s])$, $v=f^n([t])$ and $t$ is an intuitionistic premise of a $\mathsf{C}$-application with conclusion $s$. 
By Lemma \ref{l: witnessing refutation model} for any proposition $p$,
\begin{alignat*}3\label{saturated sequent temporal successor property}
     p^0 &\in \Gamma_{f^n([s])}  &&\text{ implies }  p^n && \in \Gamma_{[s]},\tag{$1$}\\
     p^n &\in \Gamma_{[t]}  &&\text{ implies }  p^0 && \in \Gamma_{f^n([t])}.\tag{$2$}
\end{alignat*}
So we have the following chain of implications
\begin{align*}
    p^0\in \Gamma_{f^n([s])} \xRightarrow{(1)} p^n\in \Gamma_{[s]}\Longrightarrow p^n\in \Gamma_{[t]}\xRightarrow{(2)} p^0\in \Gamma_{f^n([t])},
\end{align*}
where the middle implication follows from the definition of $\mathsf{C}$. This shows $V(w)\subseteq V(v)$ as required.
\end{proof} 

\begin{proposition}\label{claim 2 completeness proof}
If a sequent $\sigma$ has a refutation, then $\sigma$ is falsified in an expanding model.
\end{proposition}
\begin{proof}
Let $\mathcal{R}$ be a refutation for $\sigma$, $\fw_c$ the canonical model for $\sigma$ relative to $\mathcal{R}$ and $\varphi$ a formula. By induction on the structure of $\varphi$, we simultaneously prove that for any $w\in W$ and $n<\omega$ holds that 
\begin{itemize}
    \item[(a)]  if $\varphi^n\in \Gamma_{w}$, then $\fw_c, w\models \varphi^n$ and
    \item[(b)] if $\varphi^n\in \Delta_{w}$, then $\fw_c, w\not\models \varphi^n$.
\end{itemize} 

 The proof relies on $\Gamma_w \Rightarrow \Delta_w$ being saturated for any $w \in W$. The case for $\varphi = \bot$ is straightforward, since $\bot^n \in \Gamma_{w}$ would imply that there exists $s \in w$ which is labeled by an axiom, contradicting the assumption that $\mathcal{R}$ is a refutation. Therefore $\bot^n \not \in \Gamma_{w}$. If $\bot^n \in \Delta_{w}$, then by definition $\fw_c, w \not \models \bot^n$. 
 \smallskip

\noindent \textsc{Case for $\varphi \in \Prop$.} Let $\varphi = p$ for $p \in \Prop$. For (a) if $p^n \in \Gamma_w$, then $p^0\in \Gamma_{f^n(w)}$ and thus, by definition, $\fw_c, f^n(w) \models p$, implying that $\fw_c, w \models p^n$. For (b) if $p^n \in \Delta_w$, then  $p^n\notin \Gamma_w$ since no sequent in $\mathcal{R}$ can be an axiom. By Lemma \ref{l: witnessing refutation model} we have $p^0\notin \Gamma_{f^n(w)}$ and thus $\fw_c, f^n(w) \not\models p$. Therefore $\fw_c,w \not \models p^n$. \smallskip

\noindent \textsc{Case for $\wedge$.} Let $\varphi = \psi \wedge \gamma$. For (a) if $\psi \wedge \gamma^n \in \Gamma_w$, then by saturation $\psi^n, \gamma^n \in \Gamma_w$ and so by induction hypothesis $\fw_c, w \models \psi^n$ and $\fw_c, w \models \gamma^n$. Hence $\fw_c, f^n(w) \models \psi$ and $\fw_c, f^n(w) \models \gamma$, implying that $\fw_c, f^n(w) \models \psi \wedge \gamma$. Thus $\fw_c, w \models \psi \wedge \gamma^n$. For (b) if $\psi \wedge \gamma^n \in \Delta_w$, then by saturation $\psi^n \in \Delta^w$ or $\gamma^n \in \Delta_w$. By induction hypothesis $\fw_c, w \not \models \psi^n$ or $\fw_c, w \not \models \gamma^n$. Hence $\fw_c, f^n(w) \not \models \psi$ or $\fw_c, f^n(w) \not \models \gamma$, implying that $\fw_c, f^n(w) \not \models \psi \wedge \gamma$ and thus $\fw_c, w \not \models \psi \wedge \gamma^n$. \smallskip

\noindent \textsc{Case for $\vee$.} This case is dual to the previous case. \smallskip

\noindent \textsc{Case for $\rightarrow$.} Let $\varphi = \psi \rightarrow \gamma$. For (a) suppose $\psi \rightarrow \gamma^n \in \Gamma_w$. By saturation either $\gamma^n \in \Gamma_w$ or $\psi^n \in \Delta_w$. In the first case we have $\fw_c, w \models \gamma^n$ by induction hypothesis. Therefore $\fw_c, f^n(w) \models \gamma$. By monotonicity, for any $v \geq f^n(w)$ holds $\fw_c, v \models \gamma$, implying that $\fw, f^n(w) \models \psi \rightarrow \gamma$ and so $\fw_c, w \models \psi \rightarrow \gamma^n$. 

In the second case note that by definition of ${\rightarrow} \mathsf{L}$, $\psi^n \in \Delta_w$ implies that $\psi \rightarrow \gamma^n \in \Gamma_s$, where $s \in w$ is the unique node which is the conclusion of a $\mathsf{C}$-instance. Hence we have $\psi \rightarrow \gamma^0\in \Gamma_{f^n(w)}$ by Lemma \ref{l: witnessing refutation model}. Define $u\coloneqq f^n(w)$ and let $v\geq u$. We restrict ourselves to the case where $v\geq_1 u$; the general case then follows from monotonicity. So there exists $r,t\in \mathcal{R}$ such that $t$ is an intuitionistic premise of a $\mathsf{C}$-instance with conclusion $r$ and $u = f^m([r])$ and $v = f^m([t])$ for some $m < \omega$.  Since $\psi \rightarrow \gamma^0\in \Gamma_u$, we have $ \psi \rightarrow \gamma^{m}\in \Gamma_{r}$ by Lemma \ref{l: witnessing refutation model}, which implies that $ \psi \rightarrow \gamma^{m}\in \Gamma_{t}$. 
As before, we then have $\fw_c, [t] \models \gamma^m$ or $ \psi \rightarrow \gamma^{m}\in \Gamma_{t'}$, where $t'\in [t]$ is the conclusion of a $\mathrm{C}$-instance. This implies $\fw_c, v\models \gamma$ or $\psi \rightarrow \gamma^{0}\in \Gamma_{v}$ by Lemma \ref{l: witnessing refutation model}. In the second case, saturation implies that $\gamma^0\in \Gamma_v$ or $\psi^0\in \Delta_v$. By induction hypothesis, in both cases we have $\fw_c, v\models \gamma^0$ or $\fw_c, v\not\models \psi^0$. Thus $\fw_c, u\models \psi \rightarrow \gamma^0$ and so $\fw_c, w\models \psi \rightarrow \gamma^n$.

For (b) suppose $\psi \rightarrow \gamma^n \in \Delta_w$. Note that in a proof search tree the only rule applicable to $\psi \rightarrow \gamma^n$ is the $\mathsf{C}$-rule, implying that $\psi \rightarrow \gamma^0 \in \Delta_{f^n(w)}$. Let $s \in f^n(w)$ be the unique node which is the conclusion of a $\mathsf{C}$-instance. Then $\psi \rightarrow \gamma^0 \in \Delta_s$, implying that there exists an intuitionistic premise $t$ with $\psi^0 \in \Gamma_t$ and $\chi^0 \in \Delta_t$. By letting $v = [t]$ we obtain $\psi^0 \in \Gamma_v$ and $\chi_0 \in \Delta_v$ and so the induction hypothesis yields $\fw_c, v \models \psi$ and $\fw_c, v \not \models \gamma$. By construction $f^n(w) \leq v$ and so $\fw_c, f^n(w) \not \models \psi \rightarrow \gamma$. Hence $\fw_c, w \not \models \psi \rightarrow \gamma^n$. \smallskip

\noindent \textsc{Case for $\X$.} Suppose $\varphi = \X \psi$. For (a) if $\X \psi^n \in \Gamma_w$, then by saturation $\psi^{n+1} \in \Gamma_w$ and so by induction hypothesis $\fw_c, w \models \psi^{n+1}$, implying that $\fw_c, w \models \X \psi^n$. For (b) if $\X \psi^n \in \Delta_w$, then by saturation $\psi^{n+1} \in \Delta_w$ and so by induction hypothesis $\fw_c, w \not \models \psi^{n+1}$, implying that $\fw_c, w \not \models \X \psi^n$. \smallskip 

\noindent \textsc{Case for $\U$.} Let $\varphi = \psi \U \gamma$. For (a) if $\psi \U \gamma^n \in \Gamma_w$, then by saturation there exists an $m\geq n$ such that $\gamma^m\in \Gamma_w$ and  $\psi^k\in \Gamma_w$ for all $n\leq k< m$. By induction hypothesis $\fw_c, w \models \psi^k$ for all $n \leq k < m$ and $\fw_c, w \models \gamma^m$, implying that $\fw_c, f^n(w) \models \psi \U \gamma$. Hence $\fw_c, w\models \psi \U \gamma^n$. For (b) if $\psi \U \gamma^n \in \Delta_w$, then $\psi \U \gamma^0 \in \Delta_{f^n(w)}$ because $\U  \mathsf{R}$-applications are preserving. Saturation and the induction hypothesis imply $\fw_c, f^n(w)\not\models \gamma^0$ and either $\fw_c, f^n(w)\not\models \psi^0$ or $\psi \U \gamma^1\in \Delta_{f^n(w)}$. Similarly, for every $m\geq n$, if $\psi \U \gamma^1\in \Delta_{f^{m}(w)}$ then $\fw_c, f^{m+1}(w)\not\models \gamma^0$ and either $\fw_c, f^{m+1}(w)\not\models \psi^0$ or $\psi \U \gamma^1\in \Delta_{f^{m+1}(w)}$. So either there exists an $m\geq n$ such that $\fw_c, f^m(w)\not\models \psi^0$ and $\fw_c, f^k(w)\not\models \gamma^0$ for all $n\leq k\leq m$, or $\fw_c, f^m(w)\not\models \gamma^0$ for all $m\geq n$. Either way, $\fw_c, f^n(w)\not\models \psi \U \gamma$, implying that $\fw_c, w \not \models \psi \U \gamma^n$. \smallskip

Let $w \in W$ be the world containing the root of $\mathcal{R}$ which is labeled by $\sigma$. For any $\varphi \in \Gamma_\sigma$, we have that $\varphi \in \Gamma_w$ and for any $\psi \in \Delta_\sigma$ we have that $\psi \in \Delta_w$. Hence, by (a) and (b), $\fw_c, w \models \bigwedge \Gamma_\sigma$ and $\fw_c, w \not \models \bigvee \Delta_\sigma$, implying that $\fw_c, w \not \models \sigma^I$. Since $\fw_c$ is an expanding model by Lemma \ref{claim 1 completeness proof}, it follows that $\sigma$ is falsified in an expanding model.
\end{proof} 

As for $\NWIM$, the set of winning plays for each player is Borel and therefore, by Martin's Determinacy Theorem~\cite{Martin_1975}, the game $\mathcal{G}(\mathcal{T}, \sigma)$ is determined. Thus we obtain completeness.

\begin{theorem}[Completeness of $\niltle$]
    Every sequent valid over the class of expanding models is provable in $\niltle$. 
\end{theorem}
\begin{proof}
    Suppose $\sigma$ is valid. Let $\mathcal{T}$ be a proof search tree for $\sigma$. By Proposition \ref{claim 2 completeness proof} Refuter cannot have a winning strategy in  $\mathcal{G}(\mathcal{T}, \sigma)$. By determinacy Prover must have a winning strategy in  $\mathcal{G}(\mathcal{T}, \sigma)$. By Lemma \ref{l: winning strategy implies proof}, $\sigma$ has an $\niltle$-proof.
\end{proof}

\subsection{Completeness for Persistent Models}

Completeness for $\niltlp$ is once again established using a proof search argument. Due to the more complex frame conditions on persistent models, an adaption of the argument given for $\niltle$ is needed. In particular, the canonical model relative to a refutation is defined differently. For $\niltle$, right premises of the $\mathsf{C}$-rule generated temporal successors of the current world. For $\niltlp$, right premises of (the appropriate version of the) $\mathsf{C}$-rule will instead serve as a further description of the current world. To formalize this, an improved notion of saturation is required.

\begin{definition}
    A \emph{pre-sequent} is a pair $\Gamma \Rightarrow \Delta$ such that $\Gamma, \Delta$ are sets of nested formulas.
\end{definition}

Note that a sequent is a pre-sequent where $\Gamma, \Delta$ are finite. 

\begin{definition}
Let $k<\omega$. A pre-sequent $\Gamma\Rightarrow \Delta$ is \emph{$k$-saturated}\index{sequent!$k$-saturated} if it satisfies the clauses 1. to 8. of Definition \ref{def saturated} and the following additional clause.

\begin{enumerate}[start=9]
    \item for all $n\leq k$, if $\varphi\U \psi^n\in \Delta$, then $\psi^n,\varphi^n\in\Delta$ or $\psi^n,\X(\varphi\U \psi)^n\in\Delta$.
\end{enumerate}
Given a pre-sequent $\sigma$, a formula $\varphi$ is \emph{$k$-saturated in $\sigma$} if $\sigma$ satisfies the relevant $k$-saturation clauses for $\varphi$. Furthermore $\sigma$ is \emph{strongly saturated}\index{sequent!strongly saturated} if $\sigma$ is $k$-saturated for all $k < \omega$.
\end{definition}

Note that $0$-saturation is equivalent to the notion of saturation provided in Definition~\ref{def saturated}. Furthermore note that if a pre-sequent $\sigma$ contains a formula $\varphi \U \psi^n$ on the right side, then if $\sigma$ is strongly saturated, $\Delta_\sigma$ must be infinite. The proof search tree defined below is labeled by \emph{indexed sequents} $\Gamma\Rightarrow_k \Delta$, that is, sequents decorated with a natural number $k<\omega$, as were used for obtaining completeness of $\NWIM$ with respect to the class of functional models. Formally, there is no difference between indexed sequents here and the ones used for $\NWIM$: both are simply sequents decorated with a natural number. However, here the indexed sequents will play a different role in the argument, as they are used to keep track of the saturation level of the current sequent, and not to keep track which $\lb$-formula has to be falsified in the $\mathsf{C}$-instance.

\begin{definition}
A \emph{(persistent) proof search tree}\index{proof search!tree} for a sequent $\Gamma\Rightarrow \Delta$ is a finite or infinite tree $\mathcal{T}$ whose nodes are labeled with indexed sequents following the rules of $\niltlp$ such that the following hold.
\begin{enumerate}
    \item The root of $\mathcal{T}$ is labeled by $\Gamma\Rightarrow_0 \Delta$.
    \item A node of $\mathcal{T}$ is a leaf if and only if it is labeled by an axiom.
    \item Invertible rule applications leave the index of a sequent unchanged.
    \item Every left rule application is succinct. 
    \item Every right rule application apart from ${\to}\mathrm{R}_\mathsf{p}$ is preserving. 
    \item No invertible rule is applied to a sequent of index $k$ in which the principal formula is already $k$-saturated. 
    \item In place of the rule ${\to}\mathrm{R}_\mathsf{p}$, the choice rule\index{choice rule!$\mathsf{C}_\mathsf{p}$}
       \[  
          \infer[\mathsf{C}_\mathsf{p}]{\Gamma\Rightarrow_k (A_0\to B_0)^{k}, \dots,(A_j\to B_j)^{k},\Delta}{\Gamma, A_0^{k} \Rightarrow_0 B_0^{k} & \dotsm & \Gamma, A_j^{k} \Rightarrow_0 B_j^{k}& \Gamma \Rightarrow_{k+1} (A_0\to B_0)^{k},\dots,(A_j\to B_j)^{k},\Delta}
       \]
       is utilised, where $\Delta$ may not contain a formula of the form $A\to B^k$ and the conclusion of the rule is a $k$-saturated sequent. The right-most premise is called the \emph{right premise} and all other premises are called \emph{left premises}.
\end{enumerate}
\end{definition}

Observe that the right premise of $\mathsf{C_p}$ is an instance of a structural rule which does not change any formulas but instead increases the index by $1$. Therefore, given that the conclusion is $k$-saturated, the right premise is still $k$-saturated, but not necessarily $(k+1)$-saturated.

The following lemmas are proven as before; proofs are omitted.

\begin{lemma}
    Every sequent has a proof search tree. 
\end{lemma}

\begin{lemma}\label{lemma property (c) rule persistent}
    Let $\mathcal{T}$ be a proof search tree and let $s\in \mathcal{T}$ be the conclusion of a $\mathsf{C_p}$-application with right premise $t\in \mathcal{T}$. Then the following hold:
    \begin{enumerate}
        \item $\Gamma_t=\Gamma_s$; 
        \item if $r\geq t$ and no $\mathsf{C_p}$-application occurs between $t$ and $r$, then $\Gamma_r=\Gamma_t$.    
    \end{enumerate}
\end{lemma}

\begin{lemma}
    Every infinite branch of a proof search tree $\mathcal{T}$ contains infinitely many applications of $\mathsf{UL}$ or $\mathsf{C_p}$.
\end{lemma}

\begin{lemma}\label{lemma proof search tree property}
    Every infinite branch of a proof search tree $\mathcal{T}$ that contains only finitely many $\mathsf{C_p}$-applications has a suffix with a good formula trace.
\end{lemma}

The game $\mathcal{G}(\mathcal{T}, \sigma)$ for $\mathcal{T}$ a proof search tree for $\sigma$ is defined as before (c.f. Definition~\ref{d: proof search game for iltle}) where $\mathsf{C}$ is replaced by $\mathsf{C_p}$.

\begin{lemma}
    If Prover has a winning strategy in $\mathcal{G}(\mathcal{T}, \sigma)$, then $\sigma$ has a $\niltlp$-proof.
\end{lemma}
\begin{proof}[Proof sketch.]
    Let $\pi$ be the strategy tree of a winning strategy for Prover in $\mathcal{G}(\mathcal{T}, \sigma)$, where the index of each sequent is removed. By definition the root of $\mathcal{T}$ labeled by $\sigma$ is the root of $\pi$. Whenever $\pi$ contains a node $u$ of $\mathcal{T}$ labeled by a sequent which is the conclusion of an invertible rule instance, then $\pi$ contains all children of $u$. If $u$ is labeled by a sequent which is the conclusion of a $\mathsf{C_p}$-instance, then $u$ is owned by Prover and therefore exactly one child $u_i$ of $u$ in $\mathcal{T}$ is included in $\pi$. If $u_i$ is a left premise, then note that the sequents labelling $u$ and $u_i$ form an instance of the rule ${\to}\mathsf{R_p}$. If $u_i$ is the right premise, then note that the sequents labelling $u$ and $u_i$ are identical. Therefore delete the node $u_i$ in $\pi$ and let the children of $u_i$ be the children of $u$. Denote the resulting labeled tree by $\pi'$. Clearly, $\pi'$ is a pre-proof of $\sigma$. Since $\pi$ is the strategy tree of a winning strategy for Prover, every leaf of $\pi'$ is labeled by an axiom and every infinite branch contains a good trace, implying that $\pi'$ is a proof.
\end{proof}

Winning strategies for Refuter, on the other hand, correspond to refutations which are defined as in Definition \ref{definition refutation}, with $\mathsf{C}$ replaced by $\mathsf{C_p}$. Note that the fifth condition together with Lemma \ref{lemma proof search tree property} imply that every branch in a refutation must contain infinitely many applications of the $\mathsf{C_p}$-rule. The proof of the following lemma is identical to the proof of Lemma \ref{l: winning strategy implies refutation}. 

\begin{lemma}
    If Refuter has a winning strategy in $\mathcal{G}(\mathcal{T}, \sigma)$, then $\sigma$ has a refutation.
\end{lemma}

It remains to show that the existence of a refutation for $\sigma$ implies the existence of a countermodel. In the following let $\sigma$ be a sequent and let $\mathcal{R}$ be a refutation for $\sigma$.

\begin{definition}
    The \emph{canonical model}\index{canonical model!for $\niltlp$} $\fw_c =(W, \leq, f, V)$ for $\sigma$ (relative to $\mathcal{R}$) is defined as follows.
    \begin{enumerate}
        \item Let $\sim_0 \subseteq \mathcal{R} \times \mathcal{R}$ be the equivalence relation given by
        \begin{center}
            $s \sim_0 t$ iff there exists a path between $s$ and $t$ in which no $\mathsf{C_p}$ application occurs.
        \end{center}
        Let $\sim_1 \subseteq \mathcal{R} \times \mathcal{R}$ be the equivalence relation given by
        \begin{center}
            $s \sim_1 t$ iff there exist $s' \in [s]$ and $t' \in [t]$ such that $s'$ is the conclusion and $t'$ is the right premise of a $\mathsf{C_p}$-application.
        \end{center}
        Let $\sim \subseteq \mathcal{R} \times \mathcal{R}$ be the smallest equivalence relation such that $(\sim_0 \cup \sim_1) \subseteq \sim$. Let
        \begin{equation*}
            W \coloneqq \sfrac{\mathcal{R}}{\sim} \cup \{w^n \, \lvert \, w \in \sfrac{\mathcal{R}}{\sim} \text{ and } 0 < n < \omega\}.
        \end{equation*}
        If a world $w \in \sfrac{\mathcal{R}}{\sim}$, then we write $w^0$ for $w$, i.e. each world in $W$ is decorated with a natural number.
        
        \item Let $\leq_0 \subseteq \sfrac{\mathcal{R}}{\sim} \times \sfrac{\mathcal{R}}{\sim}$ be given as follows.
        \begin{center}
            $w^0 \leq_0 v^0$ iff there exist $s \in w^0$ and $t \in v^0$ such that $s$ is the conclusion and $t$ a left premise of the same $\mathsf{C_p}$-application in $\mathcal{R}$.
        \end{center}
        Let $\leq \subseteq W \times W$ be the reflexive transitive closure of $\leq_1 \subseteq W \times W$ given as follows.
        \begin{equation*}
            \leq_1 \coloneqq \leq_0 \cup  \{ (w^n, v^n) \, \lvert \, w^0 \leq_0 v^0 \text{ and } 0 < n < \omega\}        
        \end{equation*}

        \item Let $f: W \longrightarrow W$ be given as follows where $n < \omega$:
        \begin{equation*}
            f(w^n) \coloneqq w^{n+1}.
        \end{equation*}

        \item Let $V: W \longrightarrow \mathcal{P}(\Prop)$ be given as follows.
        \begin{equation*}
            V(w^n) \coloneqq \{p \in \Prop \, \lvert \, p^n \in \Gamma_{w^0}\}.
        \end{equation*}

        where 
        \begin{equation*}
            \Gamma_{w^0} = \bigcup_{s \in w^0} \Gamma_s.\footnote{\text{ Observe that $\Gamma_{w^0}$ is in general infinite.}}
        \end{equation*}
    \end{enumerate}
\end{definition}

The construction of the canonical model formalizes the idea that the entire refutation corresponds to the intuitionistic tree rooted at time point $0$, while every temporal successor is added in `manually'.

   \begin{lemma}
       The canonical model $\fw_c$ is persistent.
   \end{lemma}
   \begin{proof}
       It is straightforward to check that $(W, \leq)$ is a partial order and that $f$ is well-defined. Note that it suffices to prove monotonicity as well as forward and backward confluence relative to $\leq_1$, since $\leq$ is the reflexive transitive closure of $\leq_1$. For forward confluence suppose that $w^i \leq_1 v^j$. From the definition of $\leq_1$ follows that  $i = j$ and $w^0 \leq_0 v^0$. Hence by definition of $f$ and $\leq_1$ it holds that $f(w^i) = w^{i+1} \leq_1 v^{i+1} = f(v^i)$.  For back-up confluence suppose that $v^{i+1} \geq_1 f(w^j)$. Again, by definition, $j =i$ and therefore $w^{i+1} \leq_1 v^{i+1}$. Hence $w^0 \leq_0 v^0$, and so $w^i \leq_1 v^i$ with $f(v^i)= v^{i+1}$. The valuation function is clearly well-defined. For monotonicity suppose that $w^i \leq_1 v^i$. Then $w^0 \leq_0 v^0$, implying that there exist $s \in w^0$ and $t \in v^0$ such that $s$ is the conclusion and $t$ a left premise of the same $\mathsf{C_p}$-application in $\mathcal{R}$. From the definition of $\mathsf{C_p}$ follows that $\Gamma_{w^0} \subseteq \Gamma_{v^0}$, implying that $V(w^i) \subseteq V(v^i)$.
   \end{proof}

For any $w^0 \in W$, denote by $\Delta_{w^0}$ the set of formulas
\begin{equation*}
    \Delta_{w^0} = \bigcup_{s \in w^0} \Delta_s.
\end{equation*}

As for $\Gamma_{w^0}$, note that $\Delta_{w^0}$ is in general infinite.

\begin{lemma}
    Given $w^0\in W$, the pre-sequent $\Gamma_{w^0} \Rightarrow \Delta_{w^0}$ is strongly saturated.
\end{lemma}
\begin{proof}
    By construction.
\end{proof}

\begin{proposition}\label{p: persistent refutation implies countermodel}
   If a sequent $\sigma$ has a refutation, then $\sigma$ is falsified in a persistent model.
\end{proposition}
\begin{proof}
    Let $\mathcal{R}$ be a refutation for $\sigma$, $\fw_c$ the canonical model for $\sigma$ relative to $\mathcal{R}$ and $\varphi$ a formula. By induction on the structure of $\varphi$, we simultaneously prove that for any $w^0 \in W$ and any $n < \omega$ holds that
    \begin{enumerate}
        \item[(a)]  if $\varphi^n \in \Gamma_{w^0}$, then $\fw_c, w^0 \models \varphi^n$, and
        \item[(b)] if $\varphi^n \in \Delta_{w^0}$, then $\fw_c, w^0 \not \models \varphi^n$. 
    \end{enumerate}

    The cases for $\varphi = \bot$, $\varphi = \X \psi$ and $\varphi = \psi \ast \gamma$ for $\ast \in \{\wedge, \vee, \rightarrow\}$ are similar to the corresponding cases in the proof of Proposition \ref{claim 2 completeness proof}. We check the remaining cases. \smallskip
    
    \noindent \textsc{Case for $\varphi \in \Prop$.} Let $\varphi = p$ for $p \in \Prop$. For (a) if $p^n \in \Gamma_{w^0}$, then $\fw_c, w^n \models p$ by definition of $V$, implying that $\fw_c, f^n(w^0) \models p$ and thus $\fw_c, w^0 \models p^n$. For (b) if $p^n \in \Delta_{w^0}$, then $p^n \not \in \Gamma_{w^0}$, as otherwise $\mathcal{R}$ would contain a branch ending in an axiom, and thus by definition of $V$, $\fw_c, w^n \not \models p$. Hence $\fw_c, f^n(w^0) \not \models p$, implying that $\fw_c, w^0 \not \models p^n$.\smallskip

    \noindent \textsc{Case for $\rightarrow$.} Let $\varphi = \psi \rightarrow \gamma$. Due to the presence of back-up confluence, this case is simpler than its corresponding case in Proposition \ref{claim 2 completeness proof}. For (a) suppose $\psi \rightarrow \gamma^n \in \Gamma_{w^0}$. Let $s \in w_0$ be the lowermost node in $\mathcal{R}$ which is the conclusion of a $\mathsf{C_p}$-instance. For any node $t \in w_0$ which is a descendant of $s$ in $\mathcal{R}$, note that by definition of $\mathsf{C_p}$ and Lemma \ref{lemma property (c) rule persistent}, $\Gamma_s = \Gamma_t$. Therefore there exists a node $s_0 \in w$ which is an ancestor of $s$ in $\mathcal{R}$ with $\psi \rightarrow \gamma^n \in \Gamma_{s_0}$. By definition of the rule ${\rightarrow}\mathsf{L}$ either $\psi \rightarrow \gamma^n \in \Gamma_s$ or $\gamma^n \in \Gamma_s$. In the second case $\gamma^n \in \Gamma_{w^0}$ and so by induction hypothesis $\fw_c, w^0 \models \gamma^n$. By monotonicity for all $v \geq w$ holds $\fw_c, v \models \gamma^n$. In the first case suppose $w \leq_0 v$. Then there exist nodels $t,t'$ in $\mathcal{R}$ such that $t \in w^0$, $t' \in v$ and $t$ is the conclusion and $t'$ a left premise of the same $\mathsf{C_p}$-instance. By the previous observation $\psi \rightarrow \gamma^n \in \Gamma_t$ and so by definition of $\mathsf{C_p}$ also $\psi \rightarrow \gamma^n \in \Gamma_{t'}$, implying that $\psi \rightarrow \gamma^n \in \Gamma_v$. By saturation of $\Gamma_v \Rightarrow \Delta_v$ holds that $\psi^n \in \Delta_v$ or $\gamma^n \in \Gamma_v$, implying that by induction hypothesis $\fw_c, v \not \models \psi^n$ or $\fw_c, v \models \gamma^n$. Note that by definition of $\leq_0$, $v = v^0$. Moreover, observe that $\leq$ restricted to $\sfrac{\mathcal{R}}{\sim}$ is the reflexive transitive closure of $\leq_0$. Thus the above argument generalizes to all worlds $v \geq w$ in the obvious way. Hence for all $v \geq w^0$ holds that $\fw_c, v \not \models \psi^n$ or $\fw_c, v \models \gamma^n$, implying that $\fw_c, w^0 \models \X^n \psi \rightarrow \X^n \gamma$. By Lemma \ref{l: forward and backward confluence formulas} the formula $(\X \psi \rightarrow \X \gamma) \rightarrow \X (\psi \rightarrow \gamma)$ is valid over the class of persistent models, hence $\fw_c, w^0 \models \psi \rightarrow \gamma^n$. 

    For (b) suppose $\psi \rightarrow \gamma^n \in \Delta_{w^0}$. So there exists a node $s_0 \in w^0$ with $\psi \rightarrow \gamma^n \in \Delta_{s_0}$. Let $s \in w^0$ be the unique conclusion of a $\mathsf{C_p}$-instance such that the sequent labelling $s$ is indexed by $n$. Note that the only rule applicable to $\psi \rightarrow \gamma^n$ is $\mathsf{C_p}$ if the index of the conclusion is $n$. Thus $\psi \rightarrow \gamma^n \in \Delta_t$ for any descendant $t \in w^0$ between $s_0$ and $s$, implying that $\psi \rightarrow \gamma^n \in \Delta_s$. By definition of $\mathsf{C_p}$ there exists a left premise $t$ of this rule instance with $\psi^n \in \Gamma_t$ and $\gamma^n \in \Delta_t$. For $v = [t]$ the induction hypothesis yields $\fw_c, v \models \psi^n$ and $\fw_c, v \not \models \gamma^n$. By construction $w^0 \leq v$ and so $\fw_c, w^0 \not \models \X^n \psi \rightarrow \X^n \gamma$. By Lemma \ref{l: forward and backward confluence formulas}, $\fw_c, w^0 \not \models \psi \rightarrow \gamma^n$. \smallskip
     
     \noindent \textsc{Case for $\U$.} Suppose $\varphi = \psi \U \gamma$. For (a) if $\psi \U \gamma^n \in \Gamma_{w^0}$, then by saturation there exists $m \geq n$ such that $\gamma^m \in \Gamma_{w^0}$ and $\psi^k \in \Gamma_{w^0}$ for all $n \leq k < m$. Thus by induction hypothesis $w^0 \models \psi \U \gamma^n$. For (b)  if $\psi\U \gamma^n\in\Delta_{w^0}$, then by strong saturation $\gamma^n,\psi^n\in\Delta_{w^0}$ or $\gamma^n,\X(\psi \U \gamma)^n\in \Delta_{w^0}$. By induction hypothesis we then obtain $\fw_c, w^0\not\models \gamma^n$ and either $\fw_c, w^0\not\models \psi^n$ or by saturation $\psi \U \gamma^{n+1}\in \Delta_{w^0}$. Similarly, by strong saturation, for every $m\geq n$ holds that $\psi\mathsf{U}\gamma^m\in \Delta_{w^0}$ implies that $\fw_c, w^0\not\models \gamma^m$ and either $\fw_c, w^0\not\models \psi^m$ or $\psi\U \gamma^{m+1}\in \Delta_{w^0}$. So either there exists an $m\geq n$ such that $\fw_c, w^0\not\models \psi^m$ and $\fw_c, w\not\models \gamma^k$  for all $n\leq k\leq m$, or $\fw_c, w^0\not\models \gamma^m$ for all $m\geq n$. From both cases follows $\fw_c, w^0\not\models \psi\U \gamma^n$.   
\end{proof}

Finally, as before, completeness follows from determinacy of $\mathcal{G}(\mathcal{T}, \sigma)$.

\begin{theorem}[Completeness of $\niltlp$]
Every sequent valid over the class of persistent models is provable in $\niltlp$.
\end{theorem}
\section{A Counterexample to Regular Completeness}
\label{c: iLTL, section counter example}

The previous section showed that the calculus $\niltle$ is complete with respect to the class of expanding models and $\niltlp$ is complete with respect to the class of persistent models. The proofs utilized a proof search argument. Unfortunately the presented proof does not yield regular completeness for either $\niltle$ or $\niltlp$, due to the presence of the nesting. In particular, the constructed proof search trees do not establish a bound on the nesting level occurring in the sequents in a proof. For $\niltle$ the $\mathsf{S}$-rule is only applied to saturated sequents (as part of the $\mathsf{C}$-rule), however successful branches may only pass through finitely many instances of $\mathsf{C}$, implying that the nesting level may grow indefinitely. The calculus $\niltlp$ does not even feature a rule which reduces the nesting level. This phenomenon causes issues when it comes to establishing cyclic completeness. In order to obtain a cyclic calculus, we would require a bound on the nesting depth to find successful repetitions. In this section we show that this is not possible by providing an example of a valid sequent which is not provable in $\niltle$ with a bounded nesting level. Consider the following sequent:
\begin{equation*}
    \Diamond (\varphi \vee \psi)^0 \Rightarrow \gamma \rightarrow \Diamond \varphi^0, \gamma \rightarrow \Diamond \psi^0.
\end{equation*}
 Recall that $\Diamond \varphi = \top \U \varphi$. First of all, let us show that $\sigma$ is $\niltle$-provable. In the proof depicted below, let $\Delta = \{\gamma \rightarrow \E \varphi^0, \gamma \rightarrow \E \psi^0\}$
\begin{prooftree}
    \AxiomC{$\pi_0$}
    \noLine
    \UnaryInfC{$ \varphi \vee \psi^0 \Rightarrow \Delta$}
    \AxiomC{$\pi_1$}
    \noLine
    \UnaryInfC{$\top, \varphi \vee \psi^1 \Rightarrow \Delta$}
    \AxiomC{$\vdots$}
    \noLine
    \UnaryInfC{$\top, \Diamond(\varphi \vee \psi)^2 \Rightarrow \Delta$}
    \RightLabel{$\X \mathsf{L}$}
    \UnaryInfC{$\top, \X \Diamond(\varphi \vee \psi)^1 \Rightarrow \Delta$}
    \RightLabel{$\U \mathsf{L}$}
    \BinaryInfC{$\top, \Diamond(\varphi \vee \psi)^1 \Rightarrow \Delta$}
    \RightLabel{$\X \mathsf{L}$}
    \UnaryInfC{$\top, \X \Diamond(\varphi \vee \psi)^0 \Rightarrow \Delta$}
    \RightLabel{$\U \mathsf{L}$}
    \BinaryInfC{$\Diamond(\varphi \vee \psi)^0 \Rightarrow \Delta$}  
\end{prooftree}
The subproof $\pi_0$ is given as follows.
\begin{scprooftree}{0.8}
\AxiomC{}
    \RightLabel{$\mathsf{id}$}
    \UnaryInfC{$\varphi^0, \gamma^0, \bot^0 \Rightarrow \bot^0$}
    \RightLabel{${\to}\mathsf{R}$}
    \UnaryInfC{$\varphi^0, \gamma^0 \Rightarrow \top^0, \psi^0$}
    \AxiomC{}
    \RightLabel{$\mathsf{id}$}
    \UnaryInfC{$\varphi^0, \gamma^0 \Rightarrow \varphi^0, \X \Diamond \varphi^0$}
    \RightLabel{$\U \mathsf{R}$}
    \BinaryInfC{$\varphi^0, \gamma^0 \Rightarrow \Diamond \varphi^0$}
    \RightLabel{${\to}\mathsf{R}$}
    \UnaryInfC{$\varphi^0 \Rightarrow \gamma \rightarrow \Diamond \varphi^0, \gamma \rightarrow \Diamond \psi^0$}
    \AxiomC{}
    \RightLabel{$\mathsf{id}$}
    \UnaryInfC{$\psi^0, \gamma^0, \bot^0 \Rightarrow \bot^0$}
    \RightLabel{${\to}\mathsf{R}$}
    \UnaryInfC{$\psi^0, \gamma^0 \Rightarrow \top^0, \psi^0$}
     \AxiomC{}
    \RightLabel{$\mathsf{id}$}
    \UnaryInfC{$\psi^0, \gamma^0 \Rightarrow \psi^0, \X \Diamond \psi^0$}
    \RightLabel{$\U \mathsf{R}$}
    \BinaryInfC{$\psi^0, \gamma^0 \Rightarrow \Diamond \psi^0$}
    \RightLabel{${\to} \mathsf{R}$}
    \UnaryInfC{$\psi^0 \Rightarrow \gamma \rightarrow \Diamond \varphi^0, \gamma \rightarrow \Diamond \psi^0$}
    \RightLabel{$\vee \mathsf{L}$}
    \BinaryInfC{$\varphi \vee \psi^0 \Rightarrow \gamma \rightarrow \Diamond \varphi^0, \gamma \rightarrow \Diamond \psi^0$}
\end{scprooftree}
The subproof $\pi_1$ is similar, the only difference being that the formulas $\Diamond \varphi^0$ and $\Diamond \psi^0$ have to be unfolded twice to reach an axiom instead of just once. In the same way, we obtain the subproofs $\pi_i$ for each $i<\omega$. 

Note that $\pi$ is indeed a proof, as it contains only one infinite branch and this branch contains a good trace, and that the nesting level in $\pi$ is unbounded. Furthermore, note that \emph{any} proof of this sequent will have an infinite branch on the right with unbounded nesting levels. Working bottom-up, applying any other rule than $\U \mathsf{L}$ to the root sequent results in an unprovable sequent, and applying any rule other than $\X \mathsf{L}$ to its right premise results in an unprovable sequent as well. The same argument applies to each sequent in the right-most branch of $\pi$. Therefore $\sigma$ cannot be proven with a bound on the nesting level, implying that $\niltle$ is not regularly complete.

\section{Conclusion}

This chapter studied non-wellfounded proof systems for intuitionistic linear temporal logic $\iLTL$ featuring the temporal operators $\X$ and $\U$. The following summarizes the main contributions.
\begin{enumerate}
    \item We have introduced the non-wellfounded nested calculus $\niltle$ and shown that it is sound and complete for the class of expanding models.
    \item We have introduced the non-wellfounded nested calculus $\niltlp$ and shown that it is sound and complete for the class of persistent models.
\end{enumerate}

The nesting is used to manipulate formulas within the scope of $\X$-operators, which is needed to deal with the confluence conditions imposed on expanding and persistent models. The main difference between the two calculi lies in the right implication rules: for $\niltle$, only implication formulas with nesting level $0$ can function as principal formulas; for $\niltlp$, any implication formula can be principal. The rule ${\rightarrow}\mathsf{R_p}$ directly corresponds to the dual $\mathsf{K}$-axiom $(\X \varphi \rightarrow \X \psi) \rightarrow \X (\varphi \rightarrow \psi)$ which is valid over the class of persistent models but not over the class of expanding models. Both systems are arguably elegant and easy to work with. The main drawback of our work is that the presented systems are not regularly complete, as shown in Section \ref{c: iLTL, section counter example}. The presented example shows that the problem of obtaining regular completeness lies in the non-invertibility of the right implication rule. In order to prove the sequent 
\begin{equation*}
    \Diamond (\varphi \vee \psi)^0 \Rightarrow \gamma \rightarrow \Diamond \varphi^0, \gamma \rightarrow \Diamond \psi^0
\end{equation*}
we cannot find a bound on the nesting level since whenever we apply ${\rightarrow}\mathsf{R}$ we lose one of the two formulas on the right side. Therefore, the only solution is to continue unfolding $\E(\varphi \vee \psi)$ on the left side, thereby increasing the nesting level indefinitely. Only once we follow a left branch after an instance of $\mathsf{UL}$ we may decompose the formula $\varphi \vee \psi^n$ and then in each branch chose the correct implication formula on the right side to obtain an axiom. 

Our observations imply that the presented calculi need to be adapted to obtain regular completeness. One possibility is to extend the system with the cut rule and attempt to prove regular completeness for an extended system using analytic cuts. As for $\ICK$, we would attempt to prove regular completeness by directly considering a cyclic version of $\niltle$ (or $\niltlp$). However, the situation is considerably harder than for $\ICK$. When defining a finite canonical model for a fragment of the language, the resulting structure does not preserve forward confluence. A similar issue will be encountered in Chapter \ref{c: biLTL} where a bi-intuitionistic version of temporal logic is studied. We were not able to solve this problem. In fact, it seems unlikely that regular completeness may be obtained by using analytic cuts, due to computational reasons. The precise complexity bound for checking validity for $\iLTL$ is unknown. The best known upper bound is non-elementary \cite{Boudou_2017}, meaning that the algorithm cannot be bounded by any tower of exponentials. If we would obtain completeness for a cyclic version of $\niltle$ with analytic cuts, we would obtain an exponential upper bound by the same argument as presented in Chapter \ref{c: ICK} for $\ICK$ over S5 models. It seems at least unlikely that such an improvement is possible; though certainty will only be obtained once a precise complexity bound for $\iLTL$ is established, which we leave as an open question. For $\iLTL$ over persistent models it is unknown whether the validity problem is even recursively enumerable. Once again, a cyclic version of $\niltlp$ with analytic cuts would lead to a decidability result and an exponential upper bound, which again seems rather unlikely. 

\begin{question}
    What is the precise complexity bound for the validity problem of $\iLTL$ over expanding models? Is $\iLTL$ over persistent models decidable?
\end{question}

A better solution was proposed by Men\'endez Turata in his recent PhD thesis \cite{menendez_2024}, who considered labeled instead of nested sequents. This change allows to write the right implication rule in invertible form, while structural rules manipulating labels guarantee that the monotonicity of the valuation of an induced countermodel is obtained in future time steps (a task that in our case the nesting fulfils). This change makes it possible to regularize non-wellfounded proofs: Men\'endez Turata presents a non-wellfounded and a cyclic proof system for the language of $\iLTL$ with the temporal operators $\X$, $\U$ and `release' $\mathsf{R}$. As for the cyclic proofs for $\IM$, a specific annotation of formulas is used to detect good cycles. For details, the reader is refered to \cite{menendez_2024}. The method used by Men\'endez Turata is perhaps adapable to our framework of using nested sequents. To that end we would need to add additional structure to sequents, namely a second type of nesting that allows for operating on formulas within the scope of implications. In that way a proof search algorithm could explore both the `future regions' and the `intuitionistic regions' in a proof search tree without getting stuck due to non-invertible rules. The details of such an adaptation are unknown to us and left for future work.

\chapter{Bi-Intuitionistic Modal Logic}\label{c: bi-int ml new}

\section{Introduction}

The remaining two chapters deal with intuitionistic dynamic logics which instead of being extensions of intuitionistic logic with modalities and fixed point operators are extensions of \emph{bi-intuitionistic logic}. As mentioned in Chapter \ref{c: Introduction}, the Kripke semantics for intuitionistic logic can be interpreted as an information ordering: each world is an information state and ascending the intuitionistic order $\leq$ corresponds to learning new information consistent with all information already available. Such `higher' worlds are accessed by the intuitionistic implication. A formula $\varphi \rightarrow \psi$ is true at a world $w$ if for all $v \geq w$ the classical implication holds: $v \not \models \varphi$ or $v \models \psi$. However, given a world $w$, the worlds below $w$ do not play any role in the evaluation of formulas at $w$. In other words, the language of intuitionistic logic is capable to express properties of worlds above, but not of worlds below. This situation is remedied in bi-intuitionistic logic which extends intuitionistic logic with an additional connective $\dimp$ called \emph{co-implication}. The evaluation of co-implication is obtained by dualizing the evaluation of the implication: a formula $\varphi \dimp \psi$ holds at $w$ if there exists a world $v \leq w$ such that the classical implication does not hold at $v$: $v \models \varphi$ and $v \not \models \psi$. The extended language is therefore capable of reasoning about worlds in an intuitionistic Kripke model that are below the current world. Bi-intuitionistic logic was introduced by Rauszer in 1974~\cite{Rauszer_1974a, Rauszer_1974b, Rauszer_1977}, who provided algebraic and Kripke semantics. Since then bi-intuitionistic logic has been studied extensively, with a primary focus on its proof theoretic properties (see e.g.~\cite{gore_2008, gore_2020, Tranchini_2017, pinto_2018, Lyon_2024}). More recently, extensions of bi-intuitionistic logic with modalities have gained increasing interest, such as bi-intuitionistic tense logic $\mathsf{BiKt}$ introduced by Goré, Postniece and Tiu~\cite{gore_2010}, which extends bi-intuitionistic logic by standard modalities $\lb$ and $\ld$ as well as `backwards looking' versions of $\lb$ and $\ld$. This language is evaluated over intuitionistic Kripke models extended with two modal accessibility relations $R_{\lb}$ and $R_{\ld}$, to evaluate $\lb$ and $\ld$, respectively. Crucially, while $R_{\lb}$ satisfies back-up confluence and $\R_{\ld}$ satisfies forth-up confluence, there are no interaction principles between $R_{\lb}$ and $R_{\ld}$. A different version of bi-intuitionistic tense logic was studied in~\cite{stell_2015}. 

This chapter introduces and studies bi-intuitionistic logic extended with two modalities $\lb$ and $\ld$. In difference to the tense logic $\mathsf{BiKt}$ from~\cite{gore_2010}, we evaluate formulas on dynamic models containing only one modal accessibility relation $R$, which is used to evaluate both $\lb$ and $\ld$. Moreover, $R$ satisfies forth-up and forth-down confluence. Modalities can therefore be evaluated by the classical truth conditions. We offer three reasons for why to study such a logic. First, we have proposed bi-intuitionistic modal logic as a framework for reasoning with incomplete information~\cite{fernandez-duque_family_2023, fernandez-duque_sound_2024}. When taking the view that intuitionistic Kripke models are information orderings, implications assert statements about \emph{acquiring} information while co-implications assert statements about \emph{relinquishing} information, since such formulas are evaluated in higher or lower worlds, respectively. If $\lb$ is interpreted as a knowledge operator, the resulting bi-intuitionistic modal logic formalizes an agents reasoning in a setting where information may be gained or lost. Similarly, when the modalities are interpreted as temporal operators, such a logic formalizes temporal reasoning with potentially faulty information. Such a framework might thus find applications in logic based artificial intelligence, more precisely in areas such as \emph{expert systems}, \emph{epistemic planning} or \emph{temporal logic programming}.

\begin{example}
In sensitive healthcare applications such as personalised cancer treatment, decisions must be made by considering various factors, including the type and stage of cancer and the patient's genetic profile. AI systems can aid in integrating such data from various sources, possibly including some that are more reliable than others or even mutually contradictory. An AI system might analyse a patient’s data and predict that chemotherapy should shrink the tumour significantly. However, some of the data is inconclusive and, should it be incorrect, there could be a substantial chance of severe side effects. In order to model this within our temporal logic framework, let us use the atom $c$ to mean that chemotherapy is applied, $r$ that the tumour is reduced and $h$ that the patient is healthy. Moreover $\lb$ is interpreted as `henceforth' and $\ld$ as `eventually'. The formula $(c\rightarrow \ld r)\wedge (c\dimp \lb h)$ then expresses the situation we have described: the intuitionistic implication tells us that, according to our current state of knowledge, chemotherapy will produce a reduction in the tumour eventually, but co-implication tells us that should we relinquish some of this information, there is a risk of the patient becoming unhealthy. By using bi-intuitionistic logic as a basis for temporal reasoning, such considerations regarding uncertainty are directly built into the language.
\end{example}

Second, extending intuitionistic modal logic with co-implication leads to an interesting mathematical system in its own right. We will provide sound and complete axiomatizations for the logics obtained by evaluating the language over six classes of dynamic models satisfying forth-up and forth-down confluence, where the modal accessibility relation satisfies any combination of reflexivity, seriality and transitivity. However, the focus of this chapter does not lie in providing completeness results; we will only introduce axiomatizations and then refer to our work in~\cite{fernandez-duque_family_2023} for completeness. Instead, the main contribution of this chapter is a uniform proof of the finite model property for all six logics. In difference to $\IM$ and $\ICK$, this is not a straightforward result following from the completeness proofs, due to the presence of the confluence conditions. Instead, we will employ an intricate combinatorial analysis of the dynamic models by passing to dynamic models whose worlds are labelled with sets of formulas, called labelled models. Such models are particularly amenable to the combinatorial analysis required for our proof. It will then be shown how to obtain finite models which `embed' into forth-up and forth-down confluent dynamic models and thus yield the finite model property. Our method also results in a computable bound on the size of the finite models, and so decidability is easily inferred. Therefore this work presents an interesting example of how stronger confluence conditions complicate the use of classical methods in an intuitionistic setting. Third, the construction of finite models will be readily applied in Chapter \ref{c: biLTL} to \emph{bi-intuitionistic linear temporal logic} and will be the key ingredient to prove completeness of our proposed axiomatization for the full language of intuitionistic linear temporal logic.

The next two sections introduce the six bi-intuitionistic modal logics, the corresponding axiomatizations and remark on their completeness. The remaining sections are devoted to prove the finite model property. First, we introduce the key concepts of our proof, namely labelled models, dynamic simulations and moments. Then we establish the finite model property by showing how to `trim' infinite models into finite models with bounded size.

\section{Syntax and Semantics}\label{SecBasic}\index{logic!bi-intuitionistic modal}

The language $\Lbi$ of \emph{bi-intuitionistic modal logic} extends $\LIPL$ by an additional connective $\dimp$ called \emph{co-implication} as well as the modal operators $\lb$ and $\ld$. \emph{Formulas} are defined by the following grammar in Backus--Naur form:

\[\varphi \ \coloneqq\ 
    \bot  \ |\ p   \ |\      \varphi\wedge\varphi  \ |\    \varphi\vee\varphi    \ |\    \varphi\rightarrow \varphi     \ |\    \varphi\dimp \varphi     \ |\    \ld \varphi    \ |\    \lb \varphi   \]
where $p\in \Prop$. The constant $\bot$ can be defined in terms of $\dimp$ by $\bot \coloneqq p \dimp p$ where $p \in \Prop$. Therefore we will treat $\bot$ as a defined constant throughout the rest of the chapter. Recall that $\neg$ is defined by $\neg \varphi \coloneqq \varphi \rightarrow \bot$. We call $\neg$ the \emph{strong negation} and additionally define \emph{weak negation} $\sim$ by ${\sim}\varphi \coloneqq \top \dimp \varphi$.

\begin{definition}
    Let $\varphi \in \Lbi$ be a formula and $\Sigma \subseteq \Lbi$ a set of formulas.
    \begin{enumerate}
        \item The \emph{closure} $\Cl(\varphi)$ of $\varphi$ is the least set of formulas containing $\varphi$, which is closed under taking subformulas.
        \item The \emph{closure} $\Cl(\Sigma)$ of $\Sigma$ is defined as $\Cl(\Sigma) \coloneqq \bigcup_{\varphi \in \Sigma}\Cl(\varphi)$.
        \item $\Sigma$ is \emph{closed} if $\Sigma = \Cl(\Sigma)$.
    \end{enumerate}
\end{definition}

Formulas of $\Lbi$ are evaluated on dynamic models $\fw=(W, \leq, R, V)$ where $R$ is \emph{forth-up} and \emph{forth-down} confluent (see Definition \ref{d: confluence conditions}). Recall that by Lemma \ref{l: functional models confluence conditions} if $R$ is a function, then $R$ is forth-up confluent if and only if $R$ is forth-down confluent if and only if $R$ is monotone in $\leq$. Thus we will call the models of bi-intuitionistic modal logic \emph{(non-deterministic) expanding models}.

\begin{definition}
A \emph{(non-deterministic) expanding model}\index{model!non-deterministic expanding} is a dynamic model ${\fw=(W, \leq, R, V)}$ where $R$ is forth-up and forth-down confluent.    
\end{definition}

We will usually call non-deterministic expanding models simply expanding models; it will always be clear from context whether the model under consideration is total functional or not. 

\begin{definition}
Let $\fw=(W, \leq, R, V)$ be an expanding model and $w \in W$. The \emph{truth relation} $\fw, w \models \varphi$ between worlds $w$ of $\fw$ and formulas extends Definition \ref{d: truth relation for intuitionistic Kripke models} by the following clauses.
\begin{center}
    \begin{tabular}{l l l}
    $\fw,w \models \varphi \dimp \psi$ & iff & there exists $v \leq w$ with $M,v \models \varphi$ and $M,v \not \models \psi$, \\
    $\fw,w \models \ld \varphi$ & iff & there exists $v \in W$ with $w \Rel v$ and $M,v \models \varphi$ \\
    $\fw,w \models \lb \varphi$ & iff & for all $v \in W$ if $w \Rel v$, then $M,v \models \varphi$.\\
    \end{tabular}
\end{center}
\end{definition}

As usual, a formula $\varphi$ is \emph{satisfiable} over the class of expanding models if there exists an expanding model $\fw=(W, \leq, R, V)$ and a world $w \in W$ such that $\fw, w \models \varphi$ and \emph{unsatisfiable} otherwise. A formula $\varphi$ is \emph{valid} over the class of expanding models if $\fw, w \models \varphi$ for any expanding model  $\fw=(W, \leq, R, V)$ and any world $w \in W$ and \emph{falsifiable} otherwise. Given a class $\mathcal{C}$ of expanding models and a set of formulas $\Gamma$, $\varphi$ is a \emph{local semantic consequence} of $\Gamma$ over $\mathcal{C}$, written $\Gamma \models_\mathcal{C} \varphi$ if for any expanding models $\fw \in \mathcal{C}$ and any world $w$ of $\fw$, $\fw,w \models \psi$ for all $\psi \in \Gamma$ implies $\fw, w \models \varphi$. Note that by a simple computation

\begin{center}
    $\fw, w \models {\sim}\varphi$ if and only if there exists $v \leq w$ with $\fw, v \not \models \varphi$,
\end{center}
showing that weak negation is order-dual to strong negation. Unsurprisingly, the standard monotonicity property holds.

\begin{lemma}[Monotonicity]\label{l: monotonicity for biml}
    Let $\fw=(W, \leq, R, V)$ be an expanding model and $w, v \in W$ worlds. For any formula $\varphi$ if $\fw,w \models \varphi$ and $w \leq v$, then $\fw,v \models \varphi$.
\end{lemma}
\begin{proof}
    By induction on $\varphi$. The cases for $\varphi = \bot$, $\varphi \in \Prop$ or $\varphi = \psi \ast \gamma$ for $\ast \in \{\wedge, \vee, \rightarrow\}$ follow from Lemma \ref{l: monotonicity for intuitionistic Kripke models}. The cases for $\varphi = \lb \psi$ and $\varphi = \ld \psi$ follow from Lemma \ref{l: confluence classical truth conditions}. Suppose $\varphi = \psi \dimp \gamma$ and $\fw, w \models \psi \dimp \gamma$. By definition there exists $u \leq w$ with $\fw, u \models \psi$ and $\fw, u \not \models \gamma$. Since $w \leq v$, clearly $u \leq v$, implying that $\fw, v \models \psi \dimp \gamma$ as well.
\end{proof}

\begin{lemma}\label{l: bi-int tautologies}
   The formula $p \rightarrow (q \vee (p\dimp q))$ is valid over the class of dynamic models (and hence, in particular, over the class of expanding models) for any $p,q \in \Prop$.
\end{lemma}
\begin{proof}
    Let $\fw=(W, \leq, R, V)$ be a dynamic model and $w,v \in W$ worlds, such that $w \leq v$. Suppose $\fw, v \models p$. Then either $\fw, v \models q$ or $\fw, v \not \models q$, where the latter implies that $\fw, v \models p \dimp q$. Hence $\fw,w \models p \rightarrow (q \vee (p \dimp q)$.
\end{proof}

\begin{definition}
We define the following logics:
\begin{enumerate}
    \item $\mathbf{biM_K}$ is the set of valid $\Lbi$-formulas over the class of expanding models.
    \item $\mathbf{biM_{KD}}$ is the set of valid $\Lbi$-formulas over the class of serial expanding models.
    \item $\mathbf{biM_{KT}}$ is the set of valid $\Lbi$-formulas over the class of reflexive expanding models.
     \item $\mathbf{biM_{K4}}$ is the set of valid $\Lbi$-formulas over the class of transitive expanding models.
      \item $\mathbf{biM_{KD4}}$ is the set of valid $\Lbi$-formulas over the class of serial and transitive expanding models.
       \item $\mathbf{biM_{S4}}$ is the set of valid $\Lbi$-formulas over the class of S4 expanding models.
\end{enumerate}
\end{definition}

We write $\Gamma \models_\ast \varphi$ for $\ast \{\mathbf{K}, \mathbf{KD}, \mathbf{KT}, \mathbf{K4}, \mathbf{KD4}, \mathbf{S4}\}$ if $\varphi$ is a local semantic consequence of $\Gamma$ over the class of all expanding models, of serial expanding models, and so on.

\section{Axiomatizations}

We characterize the logics defined semantically via Hilbert-style axiomatizations. The axioms and rules used in the axiom systems are depicted in Table \ref{tab:axioms for biml}. Recall that an intuitionistic tautology is any formula $\varphi$ obtained from a derivable formula in $\mathsf{IPL}$ via uniform substitutions. Here, we are going to use \emph{bi-intuitionistic tautologies}. Since we have not introduced an axiomatization for bi-intuitionistic logic (but see~\cite{Rauszer_1974b}), we are simply going to call a formula $\varphi$ a bi-intuitionistic tautology if $\varphi$ can be obtain from a valid bi-intuitionistic formula via a uniform substitution.

\begin{table}[t]
    \centering
    \begin{tabular}{|l l l l|}
    \hline
     $\mathsf{biPC}$ & all bi-intuitionistic tautologies & & \\
     $\mathsf{K_\Box}$ & $\Box(\varphi \rightarrow \psi) \rightarrow (\Box \varphi \rightarrow \Box \psi)$  & $\mathsf{K_\Diamond}$ & $\Box(\varphi \rightarrow \psi) \rightarrow (\Diamond \varphi \rightarrow \Diamond \psi)$\\
     $\mathsf{T_\Box}$ & $\Box \varphi \rightarrow \varphi$ & $\mathsf{T_\Diamond}$ & $\varphi \rightarrow \Diamond \varphi$ \\
     $\mathsf{4_\Box}$ & $\Box \varphi \rightarrow \Box \Box \varphi$ & $\mathsf{4_\Diamond}$ & $\Diamond \Diamond \varphi \rightarrow \Diamond \varphi$ \\
     $\mathsf{N}$ & $\neg \Diamond \bot$ & $\mathsf{D}$ & $\Diamond \top$ \\
     $\mathsf{DP}$ & $\Diamond (\varphi \vee \psi) \rightarrow (\Diamond \varphi \vee \Diamond \psi)$ & $\mathsf{RV}$ & $\Box (\varphi \vee \psi) \rightarrow (\Box \varphi \vee \Diamond \psi)$ \\
     &&&\\
     $\mathsf{DN}$ & $\infer{{\neg}{\sim} \varphi}{\varphi}$ & & \\
     $\mathsf{MP}$ & $\infer{\psi}{\varphi & \varphi \rightarrow \psi}$ & $\mathsf{Nec}$ & $\infer{\Box \varphi}{\varphi}$\\
     \hline
    \end{tabular}
    \caption{Axioms and rules for bi-intuitionistic modal logic}
    \label{tab:axioms for biml}
\end{table}

\begin{definition}
    Define the following axiom systems.\index{axiomatization!$\mathrm{biM_K}$ and extensions}
    \begin{enumerate}
        \item $\mathrm{biM_K}$ contains the axioms $\mathsf{biPC}$, $\mathsf{DP}$, $\mathsf{RV}$, $\mathsf{N}$, $\mathsf{K_{\Box}}$ and $\mathsf{K_{\Diamond}}$ and all rules from Table \ref{tab:axioms for biml}.
        \item $\mathrm{biM_{KD}}$ is the extension of $\mathrm{biM_K}$ with the axiom $\mathsf{D}$.
        \item $\mathrm{biM_{KT}}$ is the the extension of $\mathrm{biM_K}$ with the axioms $\mathsf{T_{\Box}}$ and $\mathsf{T_{\Diamond}}$.
        \item $\mathrm{biM_{K4}}$ is the extension of $\mathrm{biM_K}$ with the axioms $\mathsf{4_\Box}$ and $\mathsf{4_\Diamond}$.
        \item $\mathrm{biM_{KD4}}$ is the extension of $\mathrm{biM_{K4}}$ with the axiom $\mathsf{D}$.
        \item $\mathsf{biM_{S4}}$ is the extension of $\mathsf{biM_{K4}}$ with the axioms $\mathsf{T_\Box}$ and $\mathsf{T_\Diamond}$.
    \end{enumerate}
\end{definition}

Every axiom has two versions, one for $\lb$ and one for $\ld$, due to $\lb$ and $\ld$ not being interdefinable. Note that there are no modal axioms capturing the interaction between the co-implication and the modalities. The interaction arises implicitly and no explicit axioms are needed. This will be illustrated in Chapter \ref{c: biLTL}.

\begin{definition}\index{derivation!$\mathrm{biM_K}$-derivation}
Let $\Gamma \cup \{ \varphi \} \subseteq \Lbi$ and $\ast \in \{\mathsf{K}, \mathsf{KD}, \mathsf{KT}, \mathsf{K4}, \mathsf{KD4}, \mathsf{S4}\}$. A \emph{derivation of a formula $\varphi$ with assumptions in $\Gamma$} is a finite tree $\pi$ labelled by $\Lbi$-formulas according to the rules of $\mathsf{biM_\ast}$ such that
\begin{enumerate}
    \item every leaf is labelled by an axiom or by a formula in $\Gamma$, and
    \item if a node $u$ is labelled by the conclusion of a rule instance of $\mathsf{DN}$ or $\mathsf{Nec}$, then every leaf of the subtree of $\pi$ rooted at $u$ is labelled by an axiom.
\end{enumerate}
\end{definition}

We write $\Gamma \vdash_\mathrm{\ast} \varphi$ if there exists a $\mathrm{biM_\ast}$-derivation of $\varphi$ from $\Gamma$ and $\vdash_\ast \varphi$ if $\Gamma = \emptyset$. Note that, as for $\mathrm{IM_H}$, only the rule $\mathsf{MP}$ is applicable to assumptions, to ensure that the relation $\Gamma \vdash_\ast \varphi$ is sound (with respect to the local semantic consequence relation).

All six logics are strongly sound and complete. Soundness is established by a standard induction on the height of derivations, and completeness by a standard canonical model construction. For details, the reader is referred to~\cite{fernandez-duque_family_2023}.

\begin{theorem}[Strong soundness and completeness] \label{t: soundness and completeness for biml}
    Let $\Gamma \cup \{\varphi\} \subseteq \Lbi$. Then $\Gamma \vdash_\ast \varphi$ if and only if $\Gamma \models_\mathbf{\ast} \varphi$ for $\ast \in \{\mathsf{K}, \mathsf{KD}, \mathsf{KT}, \mathsf{K4}, \mathsf{KD4}, \mathsf{S4}\}$.
\end{theorem}

\section{Labelled Posets and Labelled Frames}

The remainder of this chapter consists of a proof of the finite model property for all six bi-intuitionistic modal logics introduced in the previous section. The proof shows, given an expanding model falsifying a formula $\varphi$, how to construct a finite model with a computable bound on its size which falsifies $\varphi$ as well. In the following sections we introduce all ingredients needed to construct such a finite model. Instead of directly working on expanding models, it will be useful to work with labelled structures instead, which are essentially partially evaluated models that are particularly amenable for a combinatorial analysis. This section introduces \emph{labelled models}, shows how to evaluate formulas on labelled models and finally establishes that a formula is valid over a class of expanding models if and only if it is valid over the corresponding class of labelled models.

\begin{definition}\label{d: type}
Let $\Sigma \subseteq \mathcal{L_\mathsf{bIM}}$ be closed under subformulas. A \emph{(two-sided) $\Sigma$-type}\index{type} is a pair  $\Phi = (\Phi^+, \Phi^-)$ of disjoint subsets of $\Sigma$ with the following properties:
\begin{enumerate}
    \item $\bot \not \in \Phi^+$.
    \item If $\varphi \wedge \psi \in \Phi^+$, then $\varphi, \psi \in \Phi^+$.
    \item If $\varphi \wedge \psi \in \Phi^-$, then $\varphi \in \Phi^-$ or $\psi \in \Phi^-$.
    \item If $\varphi \vee \psi \in \Phi^+$, then $\varphi \in \Phi^+$ or $\psi \in \Phi^+$.
    \item If $\varphi \vee \psi \in \Phi^-$, then $\varphi, \psi \in \Phi^-$.
    \item If $\varphi \rightarrow \psi \in \Phi^+$, then $\varphi \in \Phi^-$ or $\psi \in \Phi^+$. %
    \item If $\varphi \dimp \psi \in \Phi^-$, then $\varphi \in \Phi^-$ or $\psi \in \Phi^+$. 
\end{enumerate}
\end{definition}
The set of all $\Sigma$-types is denoted by $\mathrm{T}_\Sigma$. We emphasise that it is not necessary that $\Phi^+ \cup \Phi^- = \Sigma$; in this sense types are \emph{partial}\index{type!partial}. Define two partial orders on $\mathrm{T}_\Sigma$:
\begin{enumerate}
    \item $\Phi \leq_\type \Psi$ if and only if $\Phi^+ \subseteq \Psi^+$ and $\Psi^- \subseteq \Phi^-$,
    \item $\Phi \subseteq_\type \Psi$ if and only if $\Phi^+ \subseteq \Psi^+$ and $\Phi^- \subseteq \Psi^-$,
\end{enumerate}
and for a set of formulas $\Delta$ define $\Phi{\upharpoonright_\Delta} = (\Phi^+\cap\Delta,\Phi^-\cap \Delta)$.

\begin{definition}\label{d: defects}
    Given a type $\Phi=(\Phi^+, \Phi^-)$,
    \begin{enumerate}
        \item an \emph{$\rightarrow$-defect}\index{defect!$\rightarrow$-defect} of $\Phi$ is a formula $\varphi \rightarrow \psi \in \Phi^-$, such that $\varphi \not \in \Phi^+$.
        \item an \emph{$\dimp$-defect}\index{defect!$\dimp$-defect} of $\Phi$ is a formula $\varphi \dimp \psi \in \Phi^+$ such that $\psi \not \in \Phi^-$.
    \end{enumerate}
\end{definition}

The next definition introduces labelled structures which are called \emph{labelled posets}. Such posets are essentially intuitionistic Kripke models where the valuation is replaced by a labelling function.

\begin{definition} \label{d: labelled poset}\index{labelled poset}
Let $\Sigma \subseteq \mathcal{L}_\mathsf{bIM}$ be closed under subformulas. A $\Sigma$-\emph{labelled poset} is a tuple $\mathcal{X} = (X, \leq_\mathcal{X},  \ell_\mathcal{X})$ where:
\begin{itemize}
    \item $(X,\leq_\mathcal{X})$ is a poset.
    \item $\ell_\mathcal{X}: X \longrightarrow \type_\Sigma$ such that:
    \begin{itemize}
    \item For all $x,y \in  X$ if $x \leq_{\mathcal  X} y$, then $\ell_{\mathcal  X}(x) \leq_\type \ell_{\mathcal  X}(y)$. 
        \item If $\varphi \rightarrow \psi \in \ell_\mathcal{X}(x)^-$, then there exists $y \geq_\mathcal{X} x$ with $\varphi \in  \ell_\mathcal{X}(y)^+$ and $\psi \in  \ell_\mathcal{X}(y)^-$.
        \item If $\varphi \dimp \psi \in \ell_\mathcal{X}(x)^+$, then there exists $y \leq_\mathcal{X} x$ with $\varphi \in  \ell_\mathcal{X}(y)^+$ and $\psi \in  \ell_\mathcal{X}(y)^-$.
    \end{itemize}
\end{itemize}
\end{definition}

If the structure $\mathcal{X}$ is clear, then $\mathcal{X}$ may be dropped as subscript. Given a labelled poset $\mathcal{X}$ and $\varphi \rightarrow \psi \in \ell_\mathcal{X}(x)^-$ and $y \geq_\mathcal{X} x$ with $\varphi \in \ell_\mathcal{X}(y)^+$ and $\psi \in \ell_\mathcal{X}(y)^-$, we say that the defect $\varphi \rightarrow \psi$ at $x$ was \emph{resolved} at $y$, and similary for $\dimp$-defects.

Next we define conditions that will allow us to interpret the modalities on labelled posets.
For example, if $x \Rel y$ and $\lb\varphi\in \ell(x)^+$, we will want $\varphi\in \ell(y)^+$.
However, for transitive logics, it is \emph{also} convenient to have $\lb\varphi\in \ell(y)^+$.
In order to accommodate the possible variations that may be needed, we consider `sensibility conditions' that the pair $(\ell(x),\ell(y))$ must satisfy in order to relate them via $R$.

\begin{definition}\label{d: sensibility condition}\index{sensibility condition}
A binary relation $S  \subseteq \type_\Sigma \times \type_\Sigma$
is a \emph{sensibility condition} if whenever $\Phi\mathrel S  \Psi$ and $\Delta$ is any set of formulas closed under subformulas then ${\Phi{\upharpoonright_\Delta}} \mathrel S  {\Psi{\upharpoonright_\Delta}}$ and, moreover, if $\Phi\mathrel S  \Psi$ and $\Psi \subseteq_\type \Psi'$, then $\Phi\mathrel S  \Psi'$.
\begin{enumerate}
    \item The \emph{standard condition} is defined by setting $\Phi\mathrel S_\mathrm{st} \Psi$ if whenever $\ld \varphi\in \Phi^-$, it follows that $\varphi\in\Psi^-$, and whenever $\lb\varphi\in \Phi^+$, it follows that $\varphi\in\Psi^+$.

    \item The \emph{transitive condition} is defined by setting $\Phi\mathrel S_\mathrm{tr} \Psi$ if whenever $\ld \varphi\in \Phi^-$, it follows that $\varphi,\ld\varphi\in\Psi^-$, and whenever $\lb\varphi\in \Phi^+$, it follows that $\varphi,\lb\varphi\in\Psi^+$.
\end{enumerate}
\end{definition}

Note that both the standard condition and the transitive condition are sensibility conditions. 

\begin{definition}
Fix a sensibility condition $S$.
   Let $\Sigma  \subseteq \mathcal{L}_\mathsf{bIM}$ be subformula-closed, and let $\mathcal{X}=(X, \leq_\mathcal X, \ell_\mathcal X)$ and $\mathcal{Y}=(Y, \leq_\mathcal Y, \ell_\mathcal Y)$ be $\Sigma$-labelled posets. A relation  $R \subseteq X \times Y$ is \emph{sensible}\index{binary relation!sensible} (with respect to $S$) if the following hold.
   \begin{enumerate}
       \item  $R$ is forth-up and forth-down confluent with respect to $(\leq_X, \leq_Y)$.
       \item If  $w \mathrel R v$, then $\ell_\mathcal X(w) \mathrel S \ell_\mathcal Y(v)$.
   \end{enumerate}
\end{definition}

\begin{definition}\index{labelled frame}
Fix $\Sigma\subseteq \lanfull$ closed under subformulas and a sensibility condition $S $.
    A \emph{$\Sigma$-labelled frame with respect to $S$} is a $\Sigma$-labelled poset $\mathcal{X}=(X, \leq_\mathcal X, \ell_\mathcal X)$ equipped with a $S$-sensible relation $R_\mathcal X \subseteq X \times X$.
\end{definition}

When the sensibility condition is not relevant to the discussion we may omit mention of $S$ and write simply \emph{$\Sigma$-labelled frame}. 

Let $\mathcal{X}=(X, \leq_\mathcal X, \ell_\mathcal X, R_\mathcal{X})$ be a $\Sigma$-labelled frame and let $\varphi \in \Sigma$ be a formula. Then $\varphi$ is called \emph{satisfiable} on $\mathcal{X}$ if there exists a world $x \in X$ with $\varphi \in \ell_\mathcal{X}(x)^+$. In that case $\varphi$ is said to be \emph{true} at world $x$.  Similarly,  $\varphi$ is called \emph{falsifiable} on $\mathcal{X}$ if there exists a world $x \in X$ with $\varphi \in \ell_\mathcal{X}(x)^-$. In that case $\varphi$ is said to be \emph{false} at world $x$. Given a class of $\Sigma$-labelled frames $\mathcal{C}$, a formula $\varphi$ is \emph{satisfiable over $\mathcal{C}$} if it is satisfiable on some frame $\mathcal{X} \in \mathcal{C}$ and $\varphi$ is \emph{falsifiable over $\mathcal{C}$} if $\varphi$ is falsifiable on some frame $\mathcal{X} \in \mathcal{C}$. 

Observe that expanding models can be regarded as $\Sigma$-labelled frames by labelling worlds $w$ with the sets of formulas in $\Sigma$ which are true/false at $w$. The converse is not true in general! Specifically, if a $\Sigma$-labelled frame is regarded as a model, the Truth Lemma may fail: for example, $\ld p \in \ell(w)^+$ does not imply $w \models \ld p$, since no witness may be available. Therefore an additional condition on labelled frames is required.

\begin{definition}
Let $\mathcal{X}=(X, \leq_\mathcal X, \ell_\mathcal X, R_\mathcal X)$ be a $\Sigma$-labelled frame.
The relation $R_\mathcal X$ is \emph{witnessed}\index{binary relation!witnessed} if the following hold.
\begin{enumerate}

\item Whenever $\ld\varphi\in \ell_\mathcal X(w)^+$, there is $v$ such that $w\mathrel R_\mathcal X v$ and $ \varphi\in \ell_\mathcal X(v)^+$.

\item Whenever $\lb\varphi\in \ell_\mathcal X(w)^-$, there is $v$ such that $w\mathrel R_\mathcal X v$ and $ \varphi\in \ell_\mathcal X(v)^-$.
\end{enumerate}
If $R_\mathcal X$ is witnessed, then $\mathcal X$ is called a \emph{$\Sigma$-labelled model}.\index{labelled model}
\end{definition}

A $\Sigma$-labelled model $\mathcal X$ \emph{can} reasonably be regarded as an expanding model as shown in the following lemma.
\begin{lemma}\label{l: from labelled model to model}
Let $\mathcal{X}=(X, \leq, \ell, R)$ be a $\Sigma$-labelled model with respect to $S_\mathrm{st}$ or $S_\mathrm{tr}$ and let $\fw=(X, \leq, R, V)$ with $V(w) = \{p \in \Prop \mid p \in \ell_\mathcal X(w)^+\}$.\footnote{Other valuations compatible with the labelling are possible, the maximal such valuation being $V(w) = \{p \in \Prop \mid p \not\in \ell_\mathcal X(w)^-\}$.} Then $\fw$ is an expanding model and for any formula $\varphi \in \Sigma$,
\begin{equation}\label{e: labelled model expanding model}
\varphi \in \ell(w)^+ \implies \fw, w \models \varphi\mathrel{\text{ and }}\varphi \in \ell(w)^- \implies \fw, w \not\models \varphi.
\end{equation}
\end{lemma}
\begin{proof}
    That $\fw$ is an expanding model follows immediately from the definition of a $\Sigma$-labelled model and the fact that $V$ is monotone in $\leq$. For (\ref{e: labelled model expanding model}) we proceed by structural induction on $\varphi$. By definition $\bot \not \in \ell(w)^+$. If $\bot \in \ell(w)^-$, then $\fw, w \not \models \bot$ by definition. The case for $p \in \Prop$ follows immediately from the definition of $V$. The cases for $\varphi = \psi \ast \gamma$ for $\ast \in \{\wedge, \vee\}$ follow directly from the definition of a type and the induction hypothesis. Suppose $\varphi = \psi \rightarrow \gamma$. If $\psi \rightarrow \gamma \in \ell(w)^+$, then for any $w \leq v$, we have $\ell(w) \leq_\mathrm{T} \ell(v)$ and therefore $\psi \rightarrow \gamma \in \ell(v)^+$. By definition of a type $\psi \in \ell(v)^-$ or $\gamma \in \ell(v)^+$. Thus, by induction hypothesis, for all $v \geq w$, $\fw, v \not \models \psi$ or $\fw, v \models \gamma$, implying that $\fw, w \models \psi \rightarrow \gamma$. If $\psi \rightarrow \gamma \in \ell(w)^-$, then since $\mathcal{X}$ is a $\Sigma$-labelled poset there exists $v \geq w$ with $\psi \in \ell(v)^+$ and $\gamma \in \ell(v)^-$. By induction hypothesis $\fw, v \models \psi$ and $\fw, v \not \models \gamma$. Thus $\fw, w \not \models \psi \rightarrow \gamma$. The case for $\varphi = \psi \dimp \gamma$ is order dual. Suppose $\varphi = \lb \psi$. If $\lb \psi \in \ell(w)^+$, then let $v$ be any world with $w \Rel v$. Since $R$ is sensible with respect to $S \in \{S_\mathrm{st}, S_\mathrm{tr}\}$ we have that $\ell(w) \mathrel{S} \ell(v)$ and so $\psi \in \ell(v)^+$. By induction hypothesis $\fw, v \models \psi$, implying that $\fw, w \models \lb \psi$. If $\lb \psi \in \ell(w)^-$, then since $R$ is witnessed, there exists $v$ with $w \Rel v$ and $\psi \in \ell(v)^-$. Thus by induction hypothesis $\fw, v \not \models \psi$, implying that $\fw, w \not \models \lb \psi$. Finally, the case for $\varphi = \ld \psi$ is similar.
\end{proof}

The following lemma is now immediate by regarding expanding models as $\Sigma$-labelled models and vice-versa. An \emph{(expanding) frame} is an expanding model without the valuation function, i.e. a tuple $(M, \leq, R)$ where $(M, \leq)$ is a poset and $R$ is forth-up and forth-down confluent. A frame $(M, \leq, R)$ is called a $\mathbf{biM_\ast}$-frame for $\ast \in \{\mathbf{K},\mathbf{KD},\mathbf{KT}, \mathbf{K4},\mathbf{KD4},\mathbf{S4}\}$ if $R$ satisfies the corresponding frame condition.

\begin{lemma}\label{lemModiffLabel}
Let $\varphi$ be a formula and $\Sigma$ a subformula closed set of formulas with $\varphi \in \Sigma$.
\begin{enumerate}

\item If $\ast \in \{\mathbf{K},\mathbf{KD},\mathbf{KT}\}$, then $\varphi$ is valid over the class of $\mathbf{biM_\ast}$-models if and only if it is valid over the class of $\Sigma$-labelled models with respect to the standard condition $S_\mathrm{st}$ based on a $\mathbf{biM_\ast}$-frame.

\item If $\ast \in \{\mathbf{K4},\mathbf{KD4},\mathbf{S4}\}$, then $\varphi$ is valid over the class of $\mathbf{biM_\ast}$-models if and only if it is valid over the class of $\Sigma$-labelled models with respect to the transitive condition $S_\mathrm{tr}$ based on a $\mathbf{biM_\ast}$-frame.

\end{enumerate}
\end{lemma}

Thus our strategy for proving decidability will be to construct (for finite $\Sigma$) a finite $\Sigma$-labelled model from an arbitrary $\Sigma$-labelled model.

\section{Simulations}\label{s: simulations biml}

It is crucial for our proof to identify the correct notion of `embedding' in the setting of labelled models.
This is given by \emph{dynamic simulations}. We first define the component notion of \emph{simulation}. For this section, let $\Sigma \subseteq \Delta \subseteq \lanfull$ be closed under subformulas.

\begin{definition}\index{simulation}
Let $\mathcal{X} = (X, \leq_\mathcal{X}, \ell_\mathcal{X})$ and $\mathcal{Y} = (Y, \leq_\mathcal{Y}, \ell_\mathcal{Y})$ be $\Sigma$- and $\Delta$-labelled posets, respectively. A binary relation $E \subseteq X \times Y$ is a \emph{simulation} if the following hold.
\begin{enumerate}
    \item Whenever $x \mathrel E y$, we have $\ell_\mathcal{X}(x) \subseteq_\type \ell_\mathcal{Y}(y)$.
    \item $E$ is forth-up and forth-down confluent for $(\leq_\mathcal X,\leq_\mathcal Y)$.
    
\end{enumerate}
\end{definition}

A simulation $E \subseteq X \times Y$ is \emph{surjective} if $ran(E) = Y$. 

\begin{definition}\label{d: strongly surjective simulation}\index{simulation!strongly surjective}
    Let $\mathcal{X}=(X, \leq_\mathcal{X}, \ell_\mathcal{X})$ and $\mathcal{Y} = (Y, \leq_\mathcal{Y}, \ell_\mathcal{Y})$ be $\Sigma$- and $\Delta$-labelled posets, respectively. A simulation $E \subseteq X \times Y$ is \emph{strongly surjective} if for every $y \in Y$ there exists $x \in X$ with $x \mathrel 
E y$ and $\ell_\mathcal{X}(x) = \ell_\mathcal{Y} (y) \cap \Sigma$. 
\end{definition}

The following lemma establishes a simple but crucial property of simulations. The proof is straightforward and therefore omitted.

\begin{lemma}\label{l: unions and compositions of simulations}
    Unions and compositions of simulations are simulations.
\end{lemma}

Suppose $\mathcal X,\mathcal Y$ are labelled frames, $x\in X$ and $y\in Y$, and there is a simulation $E\subseteq X\times Y$ with $x\mathrel E y$.
Then in general $\mathcal X$ can be much smaller than $\mathcal Y$, and thus simulations help us to `compress' models.
However, it may be that $\mathcal Y$ is a labelled model, but $\mathcal X$ is only a labelled frame.
In order to avoid this situation, we work with dynamic simulations.

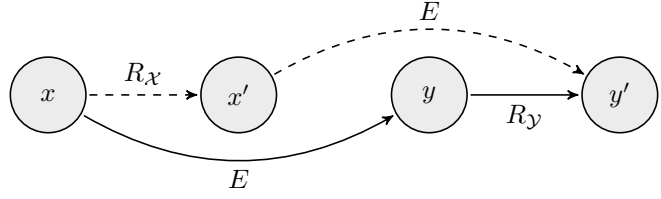
\begin{figure}[t!]
    \centering
    \begin{tikzpicture}[modal]
        \node[world](x){$x$} ;
        \node[world](x')[right=of x]{$x'$};
        \node[world](y)[right=of x']{$y$} ;
         \node[world](y')[right=of y]{$y'$} ;
         \path[->] (x) edge[dashed] node[above] {$R_\mathcal{X}$} (x');
        \path[->] (y) edge node[below] {$R_\mathcal{Y}$} (y');
        \path[->] (x) edge[bend right] node[below]{$E$} (y);
      \path[->] (x') edge[bend left, dashed] node[above]{$E$} (y');
    \end{tikzpicture}
    \caption{A dynamic simulation. The dashed arrows indicate the relations stipulated by the dynamic condition on $E$.}
    \label{f: dynamic simulation}
\end{figure}

\begin{definition}\index{simulation!dynamic}
Let $\mathcal{X} = (X, \leq_\mathcal{X},R_\mathcal X, \ell_\mathcal{X})$ and $\mathcal{Y} = (Y, \leq_\mathcal{Y}, R_\mathcal Y, \ell_\mathcal{Y})$ be $\Sigma$- and $\Delta$-labelled frames, respectively.
A simulation $E \subseteq X \times Y$ is a \emph{dynamic simulation} if whenever $x\mathrel E y \mathrel R_\mathcal Y y'$, then there is $x' \in X $ such that $x\mathrel R_\mathcal X x' \mathrel E y'$ (see Figure \ref{f: dynamic simulation}).
\end{definition}

The `dynamic part' of a dynamic simulation will help us prove that the substructure of $\mathcal{X}$ obtained by restricting $\mathcal{X}$ to the domain of the simulation is witnessed.

\section{Moments}\label{secMom}

The next step is to construct a $\Sigma$-labelled frame $\fw_\Sigma$ which satisfies the property that there exists a dynamic and strongly surjective simulation between $\moment_\Sigma$ and \emph{any} $\Sigma$-labelled model $\mathcal{X}$. This section constructs $\fw_\Sigma$, while the next section shows that such a simulation exists. In order to obtain the desired properties, the worlds of $\fw_\Sigma$ have an internal structure and are called \emph{moments}. We introduce the details below.

Let $(P,\leq)$ be a poset and $x, y \in P$. Then $y$ \emph{covers} $x$ if $x < y$ and there is no $z$ with $x < z < y$. The worlds $x$ and $y$ are called \emph{neighbours} if either $y$ covers $x$ or $x$ covers $y$. The poset $(P,\leq)$ is called \emph{discrete}\index{binary relation!partial order!discrete} if whenever $x< y$, then there are finitely many $x=x_0 < x_1< \ldots < x_n = y$ where $x_{i+1}$ covers $x_i$, for each $i$. 

\begin{remark}
    The completeness proof for all six bi-intuitionistic modal logics is based on a canonical model construction. Since the canonical model is discrete, we may restrict our attention here to discrete expanding models and thus to discrete labelled models. For the remainder of the chapter, every encountered labelled structure is assumed to be discrete, which will not always be mentioned explicitly.
\end{remark}

A \emph{zigzag path}\index{zigzag!path} through $(P,\leq)$ is a finite sequence $(x_i)_{0 \leq i \leq n}$ of elements of $P$ such that for all $0 \leq i < n$, $x_i$ and $x_{i+1}$ are neighbors. The \emph{length} of a zigzag path $(x_i)_{0 \leq i \leq n}$ is $n+1$. For technical reasons we allow the \emph{empty zigzag path} $\epsilon$ with length $0$.  Given two zigzag paths $\rho_1$ and $\rho_2$, denote by $\rho_1 \sqsubseteq \rho_2$ that $\rho_1$ is an initial segment of $\rho_2$. When $\rho_1 \sqsubseteq \rho_2$, denote by $\rho_2 - \rho_1$ the final segment of $\rho_2$ that strictly follows $\rho_1$. For example if $\rho_2 = (x_i)_{0 \leq i \leq n+k}$ and $\rho_1 =(x_i)_{0 \leq i \leq n}$, then $(\rho_2 - \rho_1) = (x_i)_{n < i \leq n+k}$. Furthermore, given a zigzag path $\rho = (x_i)_{i\leq n}$, write $\increasing(\rho)$ if $x_i < x_{i+1}$ for all $i$. Such a zigzag path is called \emph{increasing}. Similarly, write $\decreasing(\rho)$ if $x_i > x_{i+1}$ for each $i$ and call such a zigzag path \emph{decreasing}. The empty zigzag path is both increasing and decreasing and so is every zigzag path of length 1.

 \begin{definition}\index{zigzag!width hierarchy}
        Let $(P,\leq)$ be a poset. The \emph{zigzag width} hierarchy on zigzag paths is defined recursively as follows.
        \begin{itemize}
            \item The length $0$ zigzag path $\epsilon$ is both $\Pi_0$ and $\Sigma_0$.
            \item A zigzag path is $\Pi_1$ if it is decreasing and it is $\Sigma_1$ if it is increasing.
            \item  A zigzag path $\rho$ is $\Pi_{m+1}$ if there exists a decreasing zigzag path $\tau$ such that $\tau \sqsubseteq \rho$ and $\rho - \tau$ is $\Sigma_m$.
            \item  A zigzag path $\rho$ is $\Sigma_{m+1}$ if there exists an increasing zigzag path $\tau$ such that $\tau \sqsubseteq \rho$ and $\rho - \tau$ is $\Pi_m$.
        \end{itemize} 
    \end{definition}

\begin{remark}
    Note that zigzag paths of length 1 are both $\Sigma_1$ and $\Pi_1$. Moreover, the hierarchy determined by the $\Pi$ and $\Sigma$ classifications is \emph{cumulative}. That is, each $\Pi_{i+1}$ includes $\Sigma_{i}$, since every $\Sigma_i$ zigzag path can be viewed as the extension of itself by the empty zigzag path. For example, if $\rho$ is $\Sigma_i$, then the concatenation $\epsilon \circ \rho = \rho$ has a decreasing initial segment (namely $\epsilon$) while the final segment is $\Sigma_i$, implying that $\rho$ is $\Pi_{i+1}$. Similarly each $\Sigma_{i+1}$ includes $\Pi_i$. It is also the case (by induction) that each $\Pi_{i+1}$ includes $\Pi_i$ and each $\Sigma_{i+1}$ includes $\Sigma_i$.
\end{remark}

A zigzag path $(x_i)_{i \leq n}$ is called \emph{non-repeating}\index{zigzag!path!non-repeating} if all its elements are distinct. A discrete poset $(P, \leq)$ is called \emph{acyclic} if the \emph{undirected} graph induced by the covers relation is an acyclic graph (i.e. contains no cycles) and \emph{connected} if for any two points $x,y \in P$ there exists a zigzag path from $x$ to $y$. A discrete poset $(P, \leq)$ has \emph{zigzag width} $m$ if all non-repeating zigzag paths through $P$ are both $\Pi_m$ and $\Sigma_m$ and no lower classification in the hierarchy is possible.

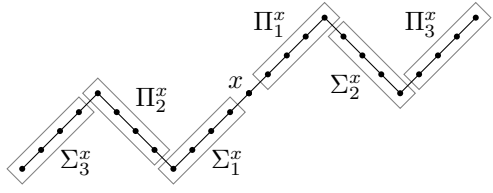
\begin{figure}\centering
\begin{tikzpicture}
\node at (.625,.625)[anchor=south east]{$\Pi^x_1$};
\node at (1.625,.375)[anchor=north east]{$\Sigma^x_2$};
\node at (2.625,.625)[anchor=south east]{$\Pi^x_3$};

\node at (-.625,-.625)[anchor=north west]{$\Sigma^x_1$};
\node at (-1.625,-.375)[anchor=south west]{$\Pi^x_2$};
\node at (-2.625,-.625)[anchor=north west]{$\Sigma^x_3$};
\draw[opacity=0.5] (0.05,.25)--(1,1.2)--(1.2,1)--(.25,0.05) -- cycle;
\draw[opacity=0.5] (1.25,.95)--(2.2,0)-- (2,-.2)-- (1.05,.75) -- cycle;
\draw[opacity=0.5] (2.05,.25)--(3,1.2)--(3.2,1)--(2.25,0.05) -- cycle;

\draw[opacity=0.5] (-0.05,-.25)--(-1,-1.2)--(-1.2,-1)--(-.25,-0.05) -- cycle;
\draw[opacity=0.5] (-1.25,-.95)--(-2.2,-0)-- (-2,.2)-- (-1.05,-.75) -- cycle;
\draw[opacity=0.5] (-2.05,-.25)--(-3,-1.2)--(-3.2,-1)--(-2.25,-0.05) -- cycle;

\draw  (-3,-1)-- (-2,0)-- (-1,-1) --(0,0) -- (1,1) -- (2,0) --(3,1) ;
\node at (.05,-.05)[anchor=south east]{$x$};
\foreach \j in {0}{
\foreach \i in {0,0.25,0.5, .75}
{
\node at (\i+\j,\i)[circle, fill, inner sep = .8pt]{};
\node at (\i+\j+1,1-\i)[circle, fill, inner sep = .8pt]{};
}}

\foreach \j in {2}{
\foreach \i in {0,0.25,0.5, .75}
{
\node at (\i+\j,\i)[circle, fill, inner sep = .8pt]{};
}}
\node at (3,1)[circle, fill, inner sep = .8pt]{};

\foreach \j in {0}{
\foreach \i in {0,0.25,0.5, .75}
{
\node at (-\i-\j,-\i)[circle, fill, inner sep = .8pt]{};
\node at (-\i-\j-1,-1+\i)[circle, fill, inner sep = .8pt]{};
}}
\foreach \j in {2}{
\foreach \i in {0,0.25,0.5, .75}
{
\node at (-\i-\j,-\i)[circle, fill, inner sep = .8pt]{};
}}
\node at (-3,-1)[circle, fill, inner sep = .8pt]{};
\end{tikzpicture}
\caption{Zigzag hierarchy relative to $x$}\label{fig:alternation}
\end{figure}

Given $x,y \in P$, $y$ is $\Pi^x_m$ if the non-repeating zigzag path from $y$ to $x$ is $\Pi_m$. Similarly, $y$ is $\Sigma^x_m$ if the non-repeating zigzag path from $y$ to $x$ is $\Sigma_m$. When we use this terminology, it should usually implicitly be understood that no lower classification in the hierarchy is possible (see Figure \ref{fig:alternation}). Write $zzw_x (y) = m$ if $m$ is the least natural number such that the non-repeating zigzag path from $y$ to $x$ is both $\Pi_m$ and $\Sigma_m$ (and no lower classification is possible) and let $zzw_x(y) = \omega$ if no such number exists. A \emph{pointed discrete poset} is a tuple $\mathcal{P}=(P, \leq, x)$ where $(P, \leq)$ is a discrete poset and $x \in P$. The \emph{zigzag radius}\index{zigzag!radius} $r(\mathcal{P})$ of a pointed discrete poset $\mathcal{P}=(P, \leq, x)$ is defined as
\begin{equation*}
    r(\mathcal{P}) = sup \{zzw_x(y) \mid y \in P\}.
\end{equation*}

The following definition introduces a measure to count the nesting level of implications and co-implications in $\Lbi$-formulas.
\begin{definition}
        The \emph{depth} $d(\varphi)$ of a formula $\varphi \in \lanfull$ is inductively defined as follows:
        \begin{center}
        \begin{tabular}{l l l l l}
            $d(\bot)$ & $\!=\!$ & $d(p)$ & $\!=\!$& $0$\\
            $d(\varphi \wedge \psi)$ & $\!=\!$ & $d(\varphi \vee \psi)$ & $\!=\!$ &$\max\{d(\varphi), d(\psi) \}$\\
            $d(\varphi \rightarrow \psi)$ & $\!=\!$ & $d(\varphi \dimp \psi)$ & $\!=\!$ &$\max\{d(\varphi), d(\psi) \} + 1$\\
            $d(\Diamond \varphi)$ & $\!=\!$ &   $d(\Box \varphi)$ &$\!=\!$ &$d(\varphi)$\\
        \end{tabular}
        \end{center}
        Given a \emph{finite} set of formulas $\Sigma$, the \emph{depth} $d(\Sigma)$ of $\Sigma$ is defined by
        \begin{equation*}
            d(\Sigma) = \max \{d(\varphi) \mid \varphi \in \Sigma \}.
        \end{equation*}
    \end{definition}

Let $\Sigma \subseteq \lanfull$ be a fixed finite and subformula closed set of formulas. Below we introduce the structure $\fw_\Sigma$ and show that it is a $\Sigma$-labelled frame. In the next section, we will then show that there exists a dynamic and surjective simulation between $\fw_\Sigma$ and any $\Sigma$-labelled frame. The worlds of $\fw_\Sigma$ have an internal structure and are called \emph{moments}.

\begin{definition}\label{d: moment}\index{moment}
    A \emph{$\Sigma$-moment} is a tuple $\mathcal{M} =(M, {\leq_\mathcal M},\allowbreak \ell_\mathcal M, m, n)$ where
    \begin{enumerate}
        \item\label{d: moment 1} $(M, {\leq_\mathcal M}, \ell_\mathcal M)$ is a $\Sigma$-labelled poset, such that the following hold for all $w \in M$:
        \begin{enumerate}
            \item\label{d: moment 1.1} if $\varphi \rightarrow \psi \in \ell_\fw(w)^-$, then there exists $v \geq w$ with $\varphi \in \ell_\fw(v)^+$, $\psi \in \ell_\fw(v)^-$ and $zzw_m(w) \leq zzw_m(v)$;
            \item\label{d: moment 1.2} if  $\varphi \dimp \psi \in \ell_\fw(w)^+$, then there exists $v \leq w$ with $\varphi \in \ell_\fw(v)^+$, $\psi \in \ell_\fw(v)^-$ and $zzw_m(w) \leq zzw_m(v)$;
        \end{enumerate}
        \item\label{d: moment 2} $(M, \leq_\mathcal M)$ is discrete, acyclic and connected;
         \item\label{d: moment 3} The zigzag radius of $(M, \leq, m)$ is bounded by $2d(\Sigma) +1$.
        \item\label{d: moment 4} $m \in M$ is called the \emph{generating world}.
        \item\label{d: moment 5} $n \in M$ is called the \emph{initial world}.
    \end{enumerate}
\end{definition}

 Note that moments are, in general, infinite. Let $\moment_\Sigma$ denote the class of all $\Sigma$-moments.\footnote{We treat $\moment_\Sigma$ as a set; it will not matter that it is a proper class.} Define a partial order ${\leq_\Sigma}$ on $\moment_\Sigma$ as follows:
\begin{center}
    $\mathcal{M} \leq_\Sigma \mathcal{N} \text{ iff } \mathcal{M} =(M, {\leq_\mathcal M}, \ell_\mathcal M, m, m')$, $\mathcal{N} =(M, {\leq_\mathcal M}, \ell_\mathcal M, m, n)$ and $m' \leq_\mathcal M n$.
\end{center}

Define a labelling function $\ell_\Sigma: \moment_\Sigma \longrightarrow \type_\Sigma$ as follows:
\begin{equation*}
    \ell_\Sigma(M, {\leq_\mathcal M}, \ell_\mathcal M, m, n) \coloneqq \ell_\mathcal M(n).
\end{equation*}
Fix a sensibility condition $S$.
Given two moments $\mathcal{M} =(M, {\leq_\mathcal M}, \ell_\mathcal M, m, m')$ and $\mathcal{N} =(N, {\leq_\mathcal N}, \ell_\mathcal N, n, n')$, the moment $\mathcal{N}$ is a \emph{modal successor} of $\mathcal{M}$ if there exists an $S$-sensible relation $R \subseteq M \times N$ such that $(m',n') \in R$.

Define the relation $R_\Sigma \subseteq \moment_\Sigma \times \moment_\Sigma$ as follows:
\begin{center}
    $(\mathcal{M}, \mathcal{N}) \in R_\Sigma \text{ iff } \mathcal{N}$ is a modal successor of $\mathcal{M}$.
\end{center}

\begin{definition}
    Define $\mathcal{M}_\Sigma \coloneqq (\moment_\Sigma, \leq_\Sigma, \ell_\Sigma, R_\Sigma)$.
\end{definition}

The following lemmas establish that $\mathcal{M}_\Sigma$ is a $\Sigma$-labelled frame.

\begin{lemma}
    $(\moment_\Sigma, \leq_\Sigma, \ell_\Sigma)$ is a $\Sigma$-labelled poset.
\end{lemma}

\begin{proof}
    First of all, observe that $\moment_\Sigma$ is non-empty by taking an arbitrary pointed model of small enough zigzag-width (say, a single reflexive world $w$ with all atoms false) and labelling $w$ according to the formulas of $\Sigma$ that are true/false on $w$.
    
    The relation $\leq_\Sigma$ is clearly reflexive. Transitivity and antisymmetry follow from the transitivity and antisymmetry of the partial orders from the given moments. Hence $\leq_\Sigma$ is a partial order on $\moment_\Sigma$.    
     
    Next, if $\mathcal{M} \leq_\Sigma \mathcal{N}$, then $\mathcal{M}=(M, {\leq_\mathcal M},\allowbreak \ell_\mathcal M,\allowbreak m, m')$ and $\mathcal{N}=(M, {\leq_\mathcal M}, \ell_\mathcal M, m, n)$ and $m' \leq_\mathcal M n$. As $\mathcal{M}$ is a $\Sigma$-labelled poset, $m' \leq_\mathcal M n$ implies that $\ell_\mathcal M(m') \leq_\mathrm{T} \ell_\mathcal M(n)$. Since $\ell_\mathcal M(m') = \ell_\Sigma(\mathcal{M})$ and $\ell_\mathcal M(n) = \ell_\Sigma(\mathcal{N})$ it follows that $\ell_\Sigma(\mathcal{M}) \leq_\mathrm{T} \ell_\Sigma(\mathcal{N})$.
    
    Let $\mathcal{M} =(M, {\leq_\mathcal M}, \ell_\mathcal M, m,n)$ be a moment and suppose that $\varphi \rightarrow \psi \in \ell_\Sigma(\mathcal{M})^-$. Recall that $\ell_\Sigma(\mathcal{M}) = \ell_\mathcal M(n)$. Therefore $\varphi \rightarrow \psi \in \ell_\mathcal M(n)^-$. As $\mathcal{M}$ is a labelled poset, there exists $n' \geq_\mathcal M n$ with $\varphi \in \ell_\mathcal M(n')^+$ and $\psi \in \ell_\mathcal M(n')^-$. Define $\mathcal{N}=(M, {\leq_\mathcal M}, \ell_\mathcal M, m, n')$. Observe that $\mathcal{N} \in \moment_\Sigma$. By definition $\mathcal{M} \leq_\Sigma \mathcal{N}$, $\varphi \in \ell_\Sigma(\mathcal{N})^+$ and $\psi \in \ell_\Sigma(\mathcal{N})^-$. The case where $\varphi \dimp \psi \in \ell_\Sigma(\mathcal{M})^+$ is similar. 
    \end{proof}

    \begin{lemma}
For any sensibility condition $S$, the relation $R_\Sigma \subseteq \moment_\Sigma \times \moment_\Sigma$ is sensible with respect to $S$.
\end{lemma}
\begin{proof}
        For forth-down confluence, suppose $\mathcal{M}, \mathcal{M}', \mathcal{N}' \in \moment_\Sigma$ are such that $\mathcal{M} \leq_\Sigma \mathcal{M}'$ and $\mathcal{M}' \Rel_\Sigma \mathcal{N}'$. By definition $\mathcal{M} =(M, {\leq_\mathcal M}, \ell_\mathcal M, m, m_0)$ and $\mathcal{M}' =(M, {\leq_\mathcal M}, \ell_\mathcal M, m, m_1)$ where $m_0 \leq_\mathcal M m_1$. Write $\mathcal{N}' =(N, {\leq_\mathcal N}, \ell_\mathcal N, n, n_1)$. By definition there exists a sensible relation $R \subseteq M \times N$ such that $m_1 \Rel n_1$. Since $m_0 \leq_\mathcal M m_1 \Rel n_1 $ and $R$ is forth-down confluent, there exists $n_0 \in N$ with $n_0 \leq_\mathcal N n_1$ and $m_0 \Rel n_0$. Define $\mathcal{N} =(N, {\leq_\mathcal N}, \ell_\mathcal N, n, n_0)$. Observe that $\mathcal{N} \in \moment_\Sigma$ and $\mathcal{N} \leq_\Sigma \mathcal{N}'$. Moreover, $\mathcal{M} \Rel_\Sigma \mathcal{N}$, witnessed by the sensible relation $R$. The case for forth-up confluence is similar.

       Now let $\mathcal{M}, \mathcal{N} \in \moment_\Sigma$ where $\mathcal{M} =(M, {\leq_\mathcal M},\allowbreak \ell_\mathcal M, m, m')$ and $\mathcal{N} =(N, {\leq_\mathcal N}, \ell_\mathcal N, n, n')$, and suppose $\mathcal{M} \mathrel{R_\Sigma} \mathcal{N}$. By definition there exists a sensible relation $R \subseteq M \times N$ such that $m' \Rel n'$.
It follows that $ \ell_\mathcal M(m')\mathrel S \ell_\mathcal N(n')$; hence $\ell_\Sigma (\mathcal M)\mathrel S\ell_\Sigma (\mathcal N)$. \qedhere
\end{proof}

Therefore, the following corollary is obtained.

\begin{corollary}
    $\mathcal{M}_\Sigma$ is a $\Sigma$-labelled frame.
\end{corollary}

\begin{lemma}
If the sensibility condition for $\mathcal{M}_\Sigma$ is $S_\mathrm{tr}$, then $\mathcal{M}_\Sigma$ is transitive.
\end{lemma}

\begin{proof}
Let $\mathcal{M}_1, \mathcal{M}_2$ and $\mathcal{M}_3$ be $\Sigma$-moments and let $R \subseteq M_1 \times M_2$ and $R' \subseteq M_2 \times M_3$ be sensible relations with respect to $S_\mathrm{tr}$. Then $R' \circ R \subseteq M_1 \times M_3$. It is immediate to check that $R' \circ R$ is forth-up and forth-down confluent, by using forth-up and forth-down confluence of $R$ and of $R'$. Suppose $w \in M_1$ and $v \in M_3$ such that $w \mathrel{(R' \circ R)}  v$. We need to show that $\ell_{\mathcal{M}_1}(w) \mathrel S_\mathrm{tr} \ell_{\mathcal{M}_3}(v)$. There exists $u \in M_2$ such that $w \mathrel R u$ and $u \mathrel R' v$. By assumption $R$ and $R'$ are sensible with respect to $S_\mathrm{tr}$. Suppose $\Box \varphi \in \ell_{\mathcal{M}_1}(w)^+$. Since $R$ is sensible, $\varphi, \Box \varphi \in \ell_{\mathcal{M}_2}(u)^+$. Since $R'$ is sensible, $\varphi, \Box \varphi \in \ell_{\mathcal{M}_3}(v)^+$. The case for $\Diamond \varphi \in \ell_{\mathcal{M}_1}(w)^-$ is similar. Therefore  $\ell_{\mathcal{M}_1}(w) \mathrel S_\mathrm{tr} \ell_{\mathcal{M}_3}(v)$, i.e. $(R' \circ R)$ is sensible with respect to $S_\mathrm{tr}$. It follows that $\mathcal{M}_3$ is a modal successor of $\mathcal{M}_1$ and so that $\mathcal{M}_1 \mathrel R_\Sigma \mathcal{M}_3$, i.e. $\mathcal{M}_\Sigma$ is transitive.
\end{proof}

\section{Constructing Surjective Dynamic Simulations}\label{s: constructing surjective dynamic simulations}

The goal of this section is to show that there exists a strongly surjective dynamic simulation between $\mathcal{M}_\Sigma$ and the path unravelling of \emph{any} $\Sigma$-labelled frame. We begin by describing the \emph{path unravelling} of a frame and prove that the frame and the unravelled frame satisfy/falsify the same formulas. As before, $\Sigma$ is a fixed finite and subformula closed set of formulas. Recall that if $\rho_1, \rho_2$ are zigzag paths with $\rho_1 \sqsubseteq \rho_2$, then $(\rho_2 - \rho_1)$ denotes the final segment of $\rho_2$ \emph{strictly} after $\rho_1$. For the following construction, we are going to slightly abuse the notation and let $(\rho_2 - \rho_1)$ denote the final segment of $\rho_2$ after $\rho_1$, \emph{including} the last point in $\rho_1$. For example if $\rho_1 =(x_i)_{i \leq n}$ and $\rho_2 = (x_i)_{i \leq n +k}$, then $(\rho_2 - \rho_1) = (x_i)_{n \leq i \leq n+k}$. 

    \begin{definition}\index{path unravelling}
        Let $\mathcal{Q}=(Q, \leq_{\mathcal Q}, \ell_\mathcal Q, R_{\mathcal Q})$ be a $\Sigma$-labelled frame with respect to $S$. Its \emph{path unravelling} is defined as $\mathcal{Q}^* =(Q^*, {\leq_{\mathcal Q^*}}, \allowbreak\ell_{\mathcal Q^*},\allowbreak R_{{\mathcal Q}^*})$ where
        \begin{enumerate}
            \item $Q^*$ is the set of all non-empty zigzag paths through $\mathcal{Q}$;
            \item ${\leq_{\mathcal Q^*}} \subseteq Q^* \times Q^*$ is defined by
            \begin{center}
                $\rho_1 \leq_{\mathcal Q^*} \rho_2 \text{ iff } (\rho_1 \sqsubseteq \rho_2$ and $\increasing(\rho_2 - \rho_1))$ or ($\rho_2 \sqsubseteq \rho_1$ and $\decreasing(\rho_1 - \rho_2))$;
            \end{center}
            \item For $\rho=(x_i)_{i \leq n} \in Q^*$, $\ell_{\mathcal Q^*}(\rho) \coloneqq \ell_\mathcal Q(x_n)$;
            \item\label{d: unravelling condition 4} For $\rho=(x_i)_{i \leq n}, \rho' = (y_i)_{i \leq m} \in Q^*$, $\rho \mathrel R_{{\mathcal Q}^*} \rho'$ if and only if the following hold:
            \begin{enumerate}
                \item\label{d: unravelling condition 4a} $x_0  \Rel_\mathcal{Q} y_0$ and $x_n  \Rel_\mathcal{Q} y_m$;
                \item\label{d: unravelling condition 4b} for all $0 < i < n$ there exists $0 \leq j \leq m$ such that $x_i  \Rel_\mathcal{Q} y_j$;
                \item\label{d: unravelling condition 4c} for all $0 \leq i \leq n$ and $0 \leq j,k \leq m$ if $x_i  \Rel_\mathcal{Q} y_j$ and $x_{i+1}  \Rel_\mathcal{Q} y_k$, then $j \leq k$ and
                \begin{enumerate}
                    \item if $x_i$ is covered by $x_{i+1}$, then for all $j \leq l < k$, $y_l$ is covered by $y_{l+1}$;
                    \item if $x_i$ covers $x_{i+1}$, then for all $j \leq l < k$, $y_l$ covers $y_{l+1}$.
                \end{enumerate}
            \end{enumerate}
        \end{enumerate}
    \end{definition}

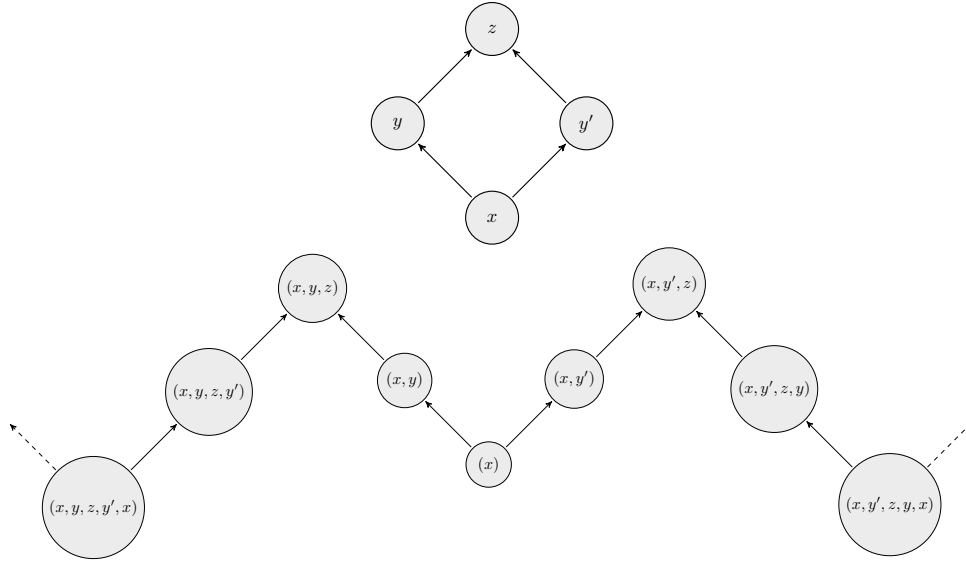
\begin{figure}[t!]
    \centering
    \scalebox{0.7}{
    \begin{tikzpicture}[modal]
        \node[world](x){$x$} ;
        \node[world](y)[above left=of x]{$y$};
        \node[world](y')[above right=of x]{$y'$} ;
         \node[world](z)[above right=of y]{$z$} ;
         \path[->] (x) edge (y);
        \path[->] (x) edge (y');
        \path[->] (y) edge (z);
      \path[->] (y') edge (z);
    \end{tikzpicture}}
    \scalebox{0.6}{
     \begin{tikzpicture}[modal]
        \node[world](x){$(x)$} ;
        \node[world](y)[above left=of x]{$(x,y)$};
        \node[world](z0)[above left=of y]{$(x,y,z)$};
        \node[world](y'0)[below left=of z0]{$(x,y,z,y')$};
        \node[world](x0)[below left=of y'0]{$(x,y,z,y',x)$};
        \node[world](y')[above right=of x]{$(x,y')$};
        \node[world](z1)[above right=of y']{$(x,y',z)$};
        \node[world](y1)[below right=of z1]{$(x,y',z,y)$};
        \node[world](x1)[below right=of y1]{$(x,y',z,y,x)$};
        \node (u)[above left=of x0]{};
        \node (v)[above right=of x1]{};
        \path[->] (x) edge (y);
        \path[->] (y) edge (z0);
        \path[->] (y'0) edge (z0);
      \path[->] (x0) edge (y'0);
       \path[->] (x) edge (y');
        \path[->] (y') edge (z1);
        \path[->] (y1) edge (z1);
      \path[->] (x1) edge (y1);
       \path[->] (x0) edge[dashed] (u);
      \path[->] (x1) edge[dashed] (v);
    \end{tikzpicture}}
    \caption{Above: A cycle in a labelled frame. Below: Its unravelling into an acyclic frame. Note that only some of the zigzag paths are depicted.}
    \label{f: path unravelling}
\end{figure}

Note that the zigzag paths in $\mathcal{Q}^*$ are allowed to be repeating. This is needed to take care of cycles in the underlying undirected graph of $(\mathcal{Q}, \leq_\mathcal{Q})$, where each such cycle generates an infinite acyclic graph with infinite zigzag width (see Figure \ref{f: path unravelling}). Regarding the abuse of notation for $(\rho_2 - \rho_1)$, note that if $(\rho_2 - \rho_1)$ denotes the final segment of $\rho_2$ \emph{strictly} after $\rho_1$, then the following problem occurs. If $\rho_1 =(x_i)_{i \leq n}$ and $\rho_2 =(x_i)_{i \leq n+1}$, then $(\rho_2 - \rho_1) = (x_{n+1})$. Recall that a zigzag path of length 1 is both increasing and decreasing. Hence $\rho_1 \sqsubseteq \rho_2$ and both $\increasing(\rho_2 - \rho_1)$ and $\decreasing (\rho_2 - \rho_1)$, implying that $\rho_1 \leq_{\mathcal Q^*} \rho_2$ and $\rho_2 \leq_{\mathcal Q^*} \rho_1$, even though $\rho_1 \not = \rho_2$. By taking the last point of $\rho_1$ into account, this case cannot happen. Note that if $\rho_1 = \rho_2$, then $(\rho_2 - \rho_1)$ contains the last point of $\rho_2$ and $\rho_1$, implying that $\rho_1 \leq_{\mathcal Q^*} \rho_2$ and $\rho_2 \leq_{\mathcal Q^*} \rho_1$, which ensures reflexivity. 

\begin{lemma}\label{l: unravelled frame is a frame}
        If $\mathcal{Q}$ is a  $\Sigma$-labelled frame, then $\mathcal{Q}^*$ is a $\Sigma$-labelled frame.
    \end{lemma}

    \begin{proof}
        Let $\mathcal{Q}=(Q, \leq_{\mathcal Q}, \ell_\mathcal Q, R_{\mathcal Q})$ be a $\Sigma$-labelled frame. We first show that $(Q^*, {\leq_{\mathcal Q^*}}, \ell_{\mathcal Q^*})$ is a $\Sigma$-labelled poset. To that end note that $\leq_{\mathcal Q^*}$ is reflexive. For transitivity, suppose $\rho_1,\rho_2,\rho_3 \in Q^*$ and $\rho_1 \leq_{\mathcal Q^*} \rho_2$ and $\rho_2 \leq_{\mathcal Q^*} \rho_3$. We distinguish four cases. \smallskip

        \noindent \textsc{Case 1.} Suppose $\rho_1 \sqsubseteq \rho_2$ and $\increasing(\rho_2 - \rho_1)$ as well as $\rho_2 \sqsubseteq \rho_3$ and $\increasing(\rho_3 - \rho_2)$. Then clearly $\rho_1 \sqsubseteq \rho_3$ and $\increasing(\rho_3 - \rho_1)$. Thus $\rho_1 \leq_{\mathcal Q^*} \rho_3$. \smallskip
        
        \noindent \textsc{Case 2.} Suppose $\rho_2 \sqsubseteq \rho_1$ and $\decreasing(\rho_1 - \rho_2)$ as well as $\rho_3 \sqsubseteq \rho_2$ and $\decreasing(\rho_2 - \rho_3)$. Then clearly $\rho_3 \sqsubseteq \rho_1$ and $\decreasing(\rho_1 - \rho_3)$. Thus $\rho_1 \leq_{\mathcal Q^*} \rho_3$. \smallskip
          
        \noindent \textsc{Case 3.} Suppose $\rho_1 \sqsubseteq \rho_2$ and $\increasing(\rho_2 - \rho_1)$ and $\rho_3 \sqsubseteq \rho_2$ and $\decreasing(\rho_2 - \rho_3)$. Then  $\increasing(\rho_2 - \rho_1)$ implies that there is a final segment of $\rho_2$ in which each step is upward and $\decreasing(\rho_2 - \rho_3)$ implies that there exists a final segment of $\rho_2$ in which each step is downward. This is only possible if either $\rho_1 = \rho_2$ or $\rho_2 = \rho_3$. In both cases it clearly holds that $\rho_1 \leq_{\mathcal Q^*} \rho_3$.\smallskip
        
       \noindent \textsc{Case 4.} The final case is similar to the previous one. \smallskip
    
       Therefore $\leq_{\mathcal Q^*}$ is transitive. That $\leq_{\mathcal Q^*}$ is antisymmetric follows immediately from the definition of $\leq_{\mathcal Q^*}$. Hence, $(Q^*, \leq_{\mathcal Q^*})$ is a poset. \smallskip
            
            Next, observe that $\ell_{\mathcal Q^*}$ assigns to each zigzag path $\rho \in Q^*$ a $\Sigma$-type, implying that $\ell_{\mathcal Q^*}: Q^* \longrightarrow \type_\Sigma$. Suppose $\rho_1 \leq_{\mathcal Q^*} \rho_2$. We distinguish two cases. \smallskip
          
            \noindent \textsc{Case a.} Suppose $\rho_1 \sqsubseteq \rho_2$ and $\increasing(\rho_2 - \rho_1)$ where $\rho_1 =(x_i)_{i \leq n}$ and $\rho_2 =(x_i)_{i \leq n+k}$. By definition $\ell_{\mathcal Q^*}(\rho_1) = \ell_\mathcal Q(x_n)$. As $\increasing(\rho_2 - \rho_1)$ it holds that $x_n \leq_{\mathcal Q} x_{n+k}$. As $\mathcal{Q}$ is a labelled frame, it follows that $\ell_\mathcal Q(x_n) \leq_\type \ell_\mathcal Q(x_{n+k})$. Since $\ell_{\mathcal Q^*}(\rho_2) = \ell_\mathcal Q(x_{n+k})$ it follows that $\ell_{\mathcal Q^*}(\rho_1) \leq_\type  \ell_{\mathcal Q^*}(\rho_2)$. \smallskip

             \noindent \textsc{Case b.} Suppose $\rho_2 \sqsubseteq \rho_1$ and $\decreasing(\rho_1 - \rho_2)$, where $\rho_1 = (x_i)_{i \leq n+k}$ and $\rho_2 = (x_i)_{i \leq n}$. By definition $\ell_{\mathcal Q^*}(\rho_1) = \ell_\mathcal Q(x_{n+k})$ and $\ell_{\mathcal Q^*}(\rho_2) = \ell_\mathcal Q(x_{n})$. Observe that since $\decreasing(\rho_1 - \rho_2)$, it follows that $x_{n+k} \leq_{\mathcal Q} x_{n}$. Therefore $\ell_\mathcal Q(x_{n+k}) \leq_\type \ell_\mathcal Q(x_n)$. Thus also $\ell_{\mathcal Q^*}(\rho_1) \leq_\type \ell_{\mathcal Q^*}(\rho_2)$. \smallskip
              
           Suppose $\rho_1 \in Q^*$ where $\rho_1 = (x_i)_{i \leq n}$ and $\varphi \rightarrow \psi \in \ell_{\mathcal Q^*}(\rho_1)^-$. By definition  $\varphi \rightarrow \psi \in \ell_\mathcal Q(x_n)^-$. As $\mathcal{Q}$ is a labelled frame, there exists $y \in Q$ with $x_n \leq_{\mathcal Q} y$ and $\varphi \in \ell_\mathcal Q(y)^+$ and $\psi \in \ell_\mathcal Q(y)^-$. Then define $\rho_2 = (x_i)_{i \leq n+k}$ such that $x_{n+k} = y$ and $(x_n, x_{n+1}, \ldots, x_{n+k})$ is the increasing zigzag path from $x_n$ to $y$. By definition $\rho_1 \sqsubseteq \rho_2$ and $\increasing(\rho_2 - \rho_1)$ and so $\rho_1 \leq_{\mathcal Q^*} \rho_2$. Furthermore, $\varphi \in \ell_{\mathcal Q^*}(\rho_2)^+$ and $\psi \in \ell_{\mathcal Q^*}(\rho_2)^-$. The case for $\varphi \dimp \psi$ is order dual. Therefore $(Q^*, {\leq_{\mathcal Q^*}}, \ell_{\mathcal Q^*})$ is a $\Sigma$-labelled poset. \smallskip

           It remains to show that $R_{\mathcal Q^*}$ is a sensible relation with respect to the sensibility condition $S$ of $\mathcal{Q}$. Suppose $\rho, \rho' \in Q^*$ such that $\rho \mathrel R_{\mathcal Q^*} \rho'$. Let $\rho = (x_i)_{i \leq n}$ and $\rho' = (y_i)_{i \leq m}$. By definition $x_n \Rel_\mathcal{Q} y_m$, implying that $\ell_\mathcal{Q}(x_n) \mathrel{S} \ell_\mathcal{Q}(y_m)$. Hence $ \ell_{\mathcal Q^*}(\rho) \mathrel S \ell_{\mathcal Q^*}(\rho')$. For forth-down confluence suppose that $\rho_0, \rho_1, \rho_1' \in Q^*$ with $\rho_0 \leq_{\mathcal{Q}*} \rho_1$ and $\rho_1 \Rel_{\mathcal{Q}^*} \rho_1'$. First, suppose $\rho_0 \sqsubseteq \rho_1$ and $\increasing(\rho_1 - \rho_0)$. Then $\rho_0 =(x_i)_{i \leq n}$, $\rho_1 = (x_i)_{i \leq n+k}$ and for all $n \leq i < n+k$, $x_i$ is covered by $x_{i+1}$. Let $\rho' = (y_i)_{i \leq m}$. Since $\rho_1 \Rel_{\mathcal{Q}^*} \rho_1'$, there exists $j \leq m$ with $x_n \Rel_\mathcal{Q} y_j$ and for all $j \leq l < m$ $y_l$ is covered by $y_{l+1}$. Let $\rho_0' = (y_i)_{i \leq j}$. It is straightforward to check that $\rho_0 \Rel_{\mathcal{Q}^*} \rho_0'$ and $\rho_0' \leq_{\mathcal{Q}^*} \rho_1'$. Second, suppose $\rho_1 \sqsubseteq \rho_0$ and $\decreasing(\rho_0 - \rho_1)$. Then $\rho_0 = (x_i)_{i \leq n+k}$, $\rho_1 = (x_i)_{i \leq n}$ and for each $n \leq i < n+k$, $x_i$ covers $x_{i+1}$. Let $\rho_1' = (y_i)_{i \leq m}$. By assumption $\rho_1 \Rel_{\mathcal{Q}}^* \rho_1'$. We define $\rho_0'$ by induction on $k$. If $k=0$, then let $\rho_0' = \rho_1'$. Clearly $\rho_0 \Rel_{\mathcal{Q}^*} \rho_0'$ and $\rho_0' \leq_{\mathcal{Q}^*} \rho_1'$. Now suppose $k > 0$. By induction hypothesis we have $(x_i)_{i \leq n+k-1} \Rel_{\mathcal{Q}^*} (y_i)_{i \leq m + l}$. Therefore $x_{n+k-1} \Rel_\mathcal{Q} y_{m+l}$. Since $x_{n+k-1}$ covers $x_{n+k}$, by forth-down confluence of $R_\mathcal{Q}$ find $y$ such $y_{m+l} \geq_{\mathcal{Q}} y$ and $x_{n+k} \Rel_\mathcal{Q} y$. Then define $\rho_0'= (y_i)_{i \leq m+l + j}$ such that $y_{m+l+j} = y$ and for each $m + l \leq i < m+l+j$ holds that $y_i$ covers $y_{i+1}$. It is routine to check that $\rho_0 \Rel_{\mathcal{Q}^*} \rho_0'$ and $\rho_0' \leq_{\mathcal{Q}^*} \rho_1'$.
           
           The argument for forth-up confluence is similar. We conclude that $R_{\mathcal Q^*}$ is sensible, and so that $\mathcal{Q}^*$ is a $\Sigma$-labelled frame.
    \end{proof}

    \begin{lemma}\label{Lemma unwindings are acyclic}
        The poset $({Q}^*,{\leq_{\mathcal Q^*}})$ is acyclic.
    \end{lemma}
    \begin{proof}
        By construction.
    \end{proof}

    \begin{lemma}\label{l: unravelled model is a model}
        If $\mathcal{Q}$ is a $\Sigma$-labelled model, then $\mathcal{Q}^*$ is a $\Sigma$-labelled model.
    \end{lemma}
    \begin{proof}
        That $\mathcal{Q}^*$ is a $\Sigma$-labelled frame is Lemma \ref{l: unravelled frame is a frame}. Suppose $\rho = (x_i)_{i \leq n} \in Q^*$ and $\lb \varphi \in \ell_{\mathcal{Q}^*}(\rho)^-$. Then $\lb \varphi \in \ell_Q(x_n)^-$. As $R_\mathcal{Q}$ is witnessed, there exists $y \in Q$ with $x_n \Rel_{\mathcal{Q}} y$ and $\varphi \in \ell_\mathcal{Q}(y)^-$. Let $\rho^{-1} = (x_{n-i})_{i \leq n}$. Using forth-up and forth down confluence of $R_\mathcal{Q}$ we find a zigzag path $\rho_0=(y_i)_{i \leq m}$ with $y_0 = y$ such that $\rho^{-1} \Rel_{\mathcal{Q}^*} \rho_0$. Then let $\rho' = \rho_0^{-1}$ and note that $\rho \Rel_{\mathcal{Q}^*} \rho'$. Moreover, $\ell_{\mathcal{Q}^*}(\rho') = \ell_{\mathcal{Q}}(y_0) = \ell_{\mathcal{Q}}(y)$ implying that $\varphi \in \ell_{\mathcal{Q}^*}(\rho')^-$. The case for $\ld \varphi \in \ell_{\mathcal{Q}^*}(\rho)^+$ is similar. Hence $R_{\mathcal{Q}^*}$ is witnessed and so $\mathcal{Q}^*$ is a $\Sigma$-labelled model.
    \end{proof}

     Clearly a formula $\varphi$ is falsified on $\mathcal Q$ if and only if it is falsified on $\mathcal Q^*$. 
     A labelled frame (model) $\mathcal{Q}^*$ which is the path unravelling of a labelled frame (model) $\mathcal{Q}$ is called an \emph{unravelled labelled frame (model)}. As a corollary of the previous lemma, we obtain the following proposition.

     \begin{proposition}\label{p: formula falsified on model iff on unravelled model}
         A formula $\varphi \in \Sigma$ is falsified over the class of $\Sigma$-labelled frames (models) if and only if $\varphi$ is falsified over the class of unravelled $\Sigma$-labelled frames (models).
     \end{proposition}

     For the remainder of this section fix a discrete unravelled $\Sigma$-labelled frame $\mathcal{Q}$ and recall that $\mathcal{Q}$ is by construction acyclic. We will now show that there exists a strongly surjective dynamic simulation $E \subseteq \moment_\Sigma \times Q$, i.e.~for each $\rho \in Q$ there exists a moment $\mathcal{M} \in \moment_\Sigma$ such that $\mathcal{M} \mathrel E \rho$ and $\ell_\Sigma(\fw) = \ell_\mathcal{Q}(\rho)$. To that end we are going to show that given $\rho \in Q$, we can inductively define a substructure of $Q$ that corresponds to a moment.

     \begin{definition}
     Let $\mathcal Q = (Q, {\leq_\mathcal Q}, \ell_\mathcal Q)$ be a discrete and acyclic $\Sigma$-labelled poset, and let $\tau \in Q$. The \emph{connected component} of $\tau$ is the substructure $(C(\tau), {{\leq_\mathcal Q}{\upharpoonright_{C(\tau)}}}, {{\ell_\mathcal Q}{\upharpoonright_{C(\tau)}}})$ where $C(\tau) \subseteq Q$ given by $\rho \in C(\tau)$ if and only if there exists a zigzag path from $\tau$ to $\rho$.
\end{definition}

Observe that $(C(\tau), {{\leq_\mathcal Q}{\upharpoonright_{C(\tau)}}}, \tau)$ does not in general have a finite zigzag radius. Given a finite and subformula closed set of formulas $\Sigma$, let
\begin{equation*}
    \Sigma(n) \coloneqq \{\varphi \in \Sigma \mid d(\varphi) \leq n\}
\end{equation*}

and let $\floor{\phantom x}$ denote the floor function. Note that $\Sigma = \Sigma(d(\Sigma))$. We define a pointed $\Sigma$-labelled substructure $\mathcal{M}^{C(\tau)}=(M, {\leq_\mathcal M}, \ell_\mathcal{M}, \tau)$ of  $(C(\tau), {{\leq_\mathcal Q}{\upharpoonright_{C(\tau)}}}, {{\ell_\mathcal Q}{\upharpoonright_{C(\tau)}}}, \tau)$ with bounded zigzag radius as follows:
\begin{itemize} 
 \item
 $M \coloneqq \{\rho \in C(\tau) \mid zzw_\tau(\rho) \leq 2d(\Sigma) +1\}$.
 \item $\leq_\mathcal M \coloneqq {\leq_\mathcal Q}{\upharpoonright_M}$.
 \item
 if $\rho \in M$ is $\Pi^\tau_m$, then $\ell_\mathcal{M}(\rho)^+ \coloneqq \ell_\mathcal Q(\rho)^+{\upharpoonright_{\Sigma(d(\Sigma)-\floor {m/2})}}$ and \\
 $\ell_\mathcal{M}(\rho)^- \coloneqq \ell_\mathcal Q(\rho)^-{\upharpoonright_{\Sigma(d(\Sigma)-\floor {(m+1)/2})}}$;
 
 if $\rho \in M$ is $\Sigma^\tau_m$, then $\ell_\mathcal{M}(\rho)^+ \coloneqq \ell_\mathcal Q(\rho)^+{\upharpoonright_{\Sigma(d(\Sigma)-\floor {(m+1)/2})}}$ and \\
 $\ell_\mathcal{M}(\rho)^- \coloneqq \ell_\mathcal Q(\rho)^-{\upharpoonright_{\Sigma(d(\Sigma)-\floor {m/2})}}$.
\end{itemize}

\begin{lemma}\label{lemma zigzag width bounded}
    The zigzag radius of $\mathcal{M}^{C(\tau)}$ is bounded by $2  d(\Sigma) + 1$.
\end{lemma}

\begin{proof}
    By construction.
\end{proof}

\begin{lemma}\label{lemma labelling function well defined}
    $\ell_\mathcal M: M \longrightarrow \type_\Sigma$ is well defined.
\end{lemma}

\begin{proof}
  We check that $\ell_\mathcal M$ assigns to each $\rho \in M$ a $\Sigma$-type. The conditions involving $\bot$, $\wedge$ and $\vee$ are immediate. Suppose $\psi \rightarrow \chi \in \ell_\mathcal M(\rho)^+$.   As $\ell_\mathcal Q(\rho)$ is a $\Sigma$-type, $\psi \in \ell_\mathcal Q(\rho)^-$ or $\chi \in \ell_\mathcal Q(\rho)^+$. As $d(\psi), d(\chi) < d(\psi \rightarrow \chi)$ it follows that $\psi \in \ell_\mathcal M(\rho)^-$ or $\chi \in \ell_\mathcal M(\rho)^+$ by construction.  The condition for $\psi \dimp \chi \in \ell_\mathcal M(\rho)^-$ is similar.
\end{proof}

In the following, given $\fw^{C(\tau)} = (M, \leq_\fw, \ell_\fw, \tau)$ and $\tau' \in M$, we write $\fw^{C(\tau)}_{\tau'}$ to denote the structure $(M, \leq_\fw, \ell_\fw, \tau, \tau')$. As it will turn out, $\fw^{C(\tau)}_{\tau'}$ is a $\Sigma$-moment for any $\tau' \in M$. Recall that $\mathcal{Q}$ is a discrete unravelled $\Sigma$-labelled frame.

\begin{lemma}\label{lemma:moment}
     For any $\tau' \in M$, $\mathcal{M}_{\tau'}^{C(\tau)}=(M, {\leq_\mathcal M}, \ell_\mathcal M, \tau, \tau')$ is a $\Sigma$-moment.
 \end{lemma}

\begin{proof}
    First we show that $(M, \leq_M, \ell_M)$ is a $\Sigma$-labelled poset satisfying Properties \ref{d: moment 1.1} and \ref{d: moment 1.2} of Definition \ref{d: moment}. Note that $\leq_\mathcal M$ is a partial order on $M$ as $\leq_\mathcal M = {\leq_\mathcal Q}{ \upharpoonright_M} $ and $\leq_\mathcal Q$ is a partial order on $Q$. Furthermore, $\ell_\mathcal M: M \longrightarrow \type_\Sigma$ is well defined by Lemma \ref{lemma labelling function well defined}.
    
    For monotonicity of $\ell_\mathcal M$, suppose $\rho \leq_\mathcal M \rho'$. If $\rho$ and $\rho'$ are both $\Pi^\tau_i$ or both $\Sigma^\tau_i$, for some $i$, then as $\ell_\mathcal Q(\rho) \leq_\type \ell_\mathcal Q(\rho')$, we have $\ell_\mathcal M(\rho) \leq_\type \ell_\mathcal M(\rho')$. Otherwise, either $\rho$ is $\Sigma^\tau_i$ and $\rho'$ is $\Pi^\tau_{i+1}$, or $\rho$ is $\Sigma^\tau_{i+1}$ and $\rho'$ is $\Pi^\tau_{i}$. In the first case $\ell_M(\rho)^+ = \ell_\mathcal Q(\rho)^+{\upharpoonright_{\Sigma(d(\Sigma)-\floor {(i+1)/2})}}$ and $\ell_M(\rho')^+ = \ell_\mathcal Q(\rho')^+{\upharpoonright_{\Sigma(d(\Sigma)-\floor {i+1/2})}}$.\footnote{ It may by instructive to refer again to Figure \ref{fig:alternation}.} Therefore if $\varphi \in \ell_M(\rho)^+$, then $\varphi \in \ell_\mathcal{Q}(\rho)^+$ and $d(\varphi) < d(\Sigma) - \floor {(i+1)/2}$. Similarly if $\varphi \in \ell_M(\rho')^+$, then $\varphi \in \ell_\mathcal{Q}(\rho)^+$ and $d(\varphi) < d(\Sigma) - \floor {(i+1)/2}$. Since $\ell_\mathcal{Q}(\rho)^+ \subseteq \ell_\mathcal{Q}(\rho')^+$ it follows that $\ell_M(\rho)^+ \subseteq \ell_M(\rho')^+$. If $\varphi \in \ell_M(\rho)^-$, then $d(\varphi) < d(\Sigma) - \floor{i/2}$ and if $\varphi \in \ell_M(\rho')^-$, then $d(\varphi) < d(\Sigma) - \floor{(i+2) / 2}$. Since $\ell_\mathcal{Q}(\rho')^- \subseteq \ell_M(\rho)^-$ it follows that $\ell_M(\rho')^- \subseteq \ell_M(\rho)^-$. Hence $\ell_M(\rho) \leq_\type \ell_M(\rho')$. The second case follows by a similar argument, once again following the definition of $\mathcal{M}^{C(\tau)}$ and, in particular, how formulas are deleted along the zigzag paths. 
            
    Suppose $\varphi \rightarrow \psi \in  \ell_\mathcal M(\rho)^-$. First note that $d(\varphi \rightarrow \psi) \geq 1$ and therefore $zzw_\tau(\rho) < 2d(\Sigma) +1$ (otherwise $\varphi \rightarrow \psi$ would have been deleted). Then  $\varphi \rightarrow \psi \in \ell_\mathcal Q(\rho)^-$. As $\mathcal{Q}$ is an unravelled $\Sigma$-labelled frame, there exists $\rho' \in Q$ with $\rho \leq_\mathcal Q \rho'$ and $zzw_\tau(\rho) \leq zzw_{\tau}(\rho') \leq zzw_\tau(\rho) +1$, such that $\varphi \in \ell_\mathcal Q(\rho')^+$ and $\psi \in \ell_\mathcal Q(\rho')^-$. Since $ d(\varphi), d(\psi) < d(\varphi \rightarrow \psi)$ and $zzw_\tau(\rho') \leq zzw_\tau(\rho) + 1 \leq 2d(\Sigma) +1$, we know $\rho' \in M$ and both $\varphi \in \ell_\mathcal M(\rho')^+$ and $\psi \in \ell_\mathcal M(\rho')^-$. The case for $\varphi \dimp \psi$ is treated similarly. Therefore $(M, \leq_M, \ell_M)$ is a $\Sigma$-labelled poset satisfying Property \ref{d: moment 1.1} and Property \ref{d: moment 1.2}. 

    Since $(Q, \leq_\mathcal Q)$ is discrete and acyclic and $(M, \leq_\fw)$ is a substructure of $(Q, \leq_\mathcal Q)$, it follows that $(M, \leq_\mathcal M)$ is discrete and acyclic as well. Moreover, since $M \subseteq C(\tau)$ and $\leq_\fw = \leq_Q \upharpoonright_M$, we have that $(M, \leq_\fw)$ is connected. The bound $2  d(\Sigma)  + 1$ on the zigzag radius follows from Lemma \ref{lemma zigzag width bounded}.\qedhere
    \end{proof}

     \begin{proposition}
        There exists a strongly surjective simulation $E \subseteq \moment_\Sigma \times Q$.
    \end{proposition}

    \begin{proof}
        Let $\rho \in Q$. By Lemma \ref{lemma:moment}, $\mathcal{M}_\rho^{C(\rho)} = (M, \allowbreak{\leq_\mathcal M},\allowbreak \ell_\mathcal M,\allowbreak \rho, \rho) \in \moment_\Sigma$. First define a simulation $E_\rho \subseteq \moment_\Sigma \times Q$ that includes the pair $(\mathcal{M}_\rho^{C(\rho)}, \rho)$ as follows:
        \begin{equation*}
            E_\rho \coloneqq \{(\fw_{\rho'}^{C(\rho)}, \rho') \mid \rho' \in M\}.
        \end{equation*}
    
        Clearly, $\mathcal{M}_\rho^{C(\rho)} \mathrel E_\rho \rho$ and $E_\rho \subseteq \moment_\Sigma \times Q$. In order to show that $E_\rho$ is a simulation we check the defining conditions. \smallskip
       
       \noindent \textsc{1.} Suppose $\mathcal{M}_{\rho'}^{C(\rho)} \mathrel{E_\rho} \rho'$. Then $\ell_\Sigma(\mathcal{M}_{\rho'}^{C(\rho)}) = \ell_\mathcal M(\rho') \subseteq_{\type} \ell_\mathcal Q(\rho')$. \smallskip

       \noindent \textsc{2.} For forth-up confluence, suppose $\mathcal{M}_{\rho'}^{C(\rho)} \mathrel{E_\rho} \rho'$ and $\mathcal{M}_{\rho'}^{C(\rho)} \leq_\Sigma \mathcal{M}$. By definition $\mathcal{M} = \fw_{\rho''}^{C(\rho)}$ for some $\rho'' \in M$ with $\rho' \leq_{\fw} \rho''$. Thus $\rho'' \in Q$ and $\rho' \leq_\mathcal Q \rho''$, and by definition $\mathcal{M}_{\rho''}^{C(\rho)} \mathrel E_\rho \rho''$. \smallskip

       \noindent \textsc{3.} The proof for forth-down confluence is similar. \smallskip
    
        Thus for each $\rho \in Q$ we have a simulation $E_{\rho}$ such that $\mathcal{M}_\rho^{C(\rho)} \mathrel E_{\rho} \rho$.
        Now define
        \begin{equation*}
            E \coloneqq \bigcup_{\rho \in Q} E_\rho.
        \end{equation*}
        By Lemma \ref{l: unions and compositions of simulations}, $E$ is a simulation, and by construction $E$ is strongly surjective.
    \end{proof}

 The next step is to show that $E$ is dynamic. Let $\rho, \tau \in Q$ such that $\rho \mathrel R_{\mathcal Q} \tau$, and write $\mathcal{M}_\rho^{C(\rho)} =(M, {\leq_\mathcal M}, \ell_\mathcal M, \rho, \rho)$ and $\mathcal{M}_\tau^{C(\tau)} = (N, {\leq_\mathcal N}, \ell_\mathcal N, \tau, \tau)$.

 \begin{lemma}
       There exists a sensible relation $R \subseteq M \times N$ such that $(\rho, \tau) \in R$. 
   \end{lemma}

    \begin{proof}
     We first show how to define partial functions $R_n: M \longrightarrow N$ for any $n \leq 2d(\Sigma) + 1$, such that whenever $\rho' \Rel_n \tau'$ holds, then $zzw_\tau(\tau') \leq zzw_\rho(\rho')$ and $\rho' \Rel_\mathcal{Q} \tau'$. \smallskip
        
    \noindent \textsc{Construction of $R_0$.}  Define $R_0 \coloneqq \{ (\rho, \tau) \}$. Observe that $zzw_\tau (\tau) \leq zzw_\rho (\rho)$ and by assumption $\rho \Rel_\mathcal{Q} \tau$.
    
    Now suppose that we have defined $R_n$, such that whenever $\rho' \mathrel R_n \tau'$, it holds that $\rho' \mathrel R_{\mathcal Q} \tau'$ and $zzw_\tau (\tau') \leq zzw_\rho (\rho')$. \smallskip
    
    \noindent \textsc{Construction of $R_{n+1}$. } Initialise $R_{n+1}$ to $R_n$. Then distinguish the following cases, for any $\rho', \rho'' \in M$ and $\tau' \in N$. \smallskip
       
    \noindent \textsc{1.}  If $\rho' \mathrel R_n \tau'$, $\rho' \leq_\mathcal M \rho''$ and $\rho'' \not \in dom(R_n)$, then by assumption $\rho' \Rel_\mathcal{Q} \tau'$ and so by forth-up confluence of $R_{\mathcal Q}$ there exists $\tau'' \in Q$ with $\tau' \leq_\mathcal Q \tau''$ and $\rho'' \mathrel R_{\mathcal Q} \tau''$. Observe that $zzw_\tau (\tau'') \leq zzw_\rho (\rho'') \leq 2  d(\Sigma) + 1$. Therefore $\tau'' \in N$. Add $(\rho'', \tau'')$ to $R_{n+1}$. \smallskip

    \noindent \textsc{2.} If $\rho' \mathrel R_n \tau'$, $\rho' \geq_M \rho''$ and $\rho'' \not \in dom(R_n)$, then by assumption $\rho' \Rel_\mathcal{Q} \tau'$ and so by forth-down confluence of $R_{\mathcal Q}$ there exists $\tau'' \in Q$ with $\tau' \geq_{\mathcal Q} \tau''$ and $\rho'' \mathrel R_{\mathcal Q} \tau''$. Observe that $zzw_\tau (\tau'') \leq zzw_\rho (\rho'') \leq 2  d(\Sigma) + 1$. Therefore $\tau'' \in N$. Add $(\rho'', \tau'')$ to $R_{n+1}$. \smallskip

    \noindent \textsc{3.} No other pairs are added to $R_{n+1}$. \smallskip
       
    Then define $R \coloneqq \bigcup_{n \leq 2d(\Sigma)} R_n$. By construction $(\rho, \tau) \in R$. It remains to show that $R$ is sensible.  Note that $R$ is forth-up and forth-down confluent. For example if $\rho' \leq_\fw \rho''$ and $\rho' \Rel \tau'$, then there exists a least natural number $n$ such that $\rho' \Rel_n \tau'$. If $zzw_\rho(\rho'') = zzw_\rho(\rho')$, then in the construction of $R_n$ the pair $(\rho'', \tau'')$ for some $\tau' \leq_{\fv} \tau''$ is added. Otherwise $\rho'' \not \in dom(R_n)$. The construction of $R_{n+1}$ then guarantees that $\rho''  \Rel_{n+1} \tau''$ for some $\tau' \leq_{\fv} \tau''$. Thus there exists $\tau' \leq_\fv \tau''$ with $\rho'' \Rel \tau''$. Let $S$ be the sensibility condition, and suppose $\rho' \Rel \tau'$. Then $\rho'  \Rel_{\mathcal Q} \tau'$,  so $\ell_{\mathcal Q}(\rho') \mathrel S \ell_{\mathcal Q}(\tau')$. Since $\ell_\mathcal M(\rho')$ and $\ell_\mathcal N(\tau')$ are given by restrictions of $\ell_{\mathcal Q}(\rho')$ and $\ell_{\mathcal Q}(\tau')$ respectively, and $\ell_{\mathcal Q}(\rho')$ is restricted more than $\ell_{\mathcal Q}(\tau')$ (since $zzw_\tau(\tau') \leq zzw_\rho(\rho')$), by the properties of sensibility conditions (namely, whenever $\Phi \mathrel{S} \Psi$ and $\Delta$ is closed under subformulas, then ${\Phi{\upharpoonright_\Delta}} \mathrel S  {\Psi{\upharpoonright_\Delta}}$ and, moreover, if $\Phi\mathrel S  \Psi$ and $\Psi \subseteq_\type \Psi'$, then $\Phi\mathrel S  \Psi'$), $\ell_\mathcal M(\rho') \mathrel S \ell_\mathcal N(\tau')$. Therefore $R$ is sensible.
    \end{proof}

 \begin{lemma}
        $dom(R) = M$.
    \end{lemma}

    \begin{proof}
        The construction of $R$ adds in step $0$ the world $\rho$ to $dom(R)$ and  in each step $n \geq 1$ every world $\rho'$ with $zzw_\rho(\rho') = n+1$. Therefore after $2  d(\Sigma)$ steps, $dom(R) = M$.
    \end{proof}

     Thus for every $\rho_0 \in M$, there exists $\tau_0 \in N$ with $(\rho_0, \tau_0) \in R$, yielding the following corollary.

        \begin{corollary}
        $\mathcal{N}_{\tau_0}^{C(\tau)}$ is a modal successor of $\mathcal{M}_{\rho_0}^{C(\rho)}$.
    \end{corollary}

    \begin{proof}
        Witnessed by the sensible relation $R \subseteq M \times N$ with $(\rho_0, \tau_0) \in R$.
    \end{proof}

    \begin{lemma}\label{l: how to construct modal successors}
        Suppose $\rho_0 \in \mathcal{M}_\rho^{C(\rho)}$ and $\rho_0 \mathrel R_{\mathcal Q} \tau_0$. Then there exists $\tau$ such that $\rho \mathrel R_{\mathcal Q} \tau$ and $\tau_0 \in \mathcal{N}_\tau^{C(\tau)}$. Moreover if $\alpha$ is the non-repeating zigzag path from $\rho_0$ to $\rho$ and $\beta$ the non-repeating zigzag path from $\tau_0$ to $\tau$, then $zzw(\beta) \leq zzw(\alpha)$.
    \end{lemma}

    \begin{proof}
        Since $\rho_0 \in \mathcal{M}_\rho^{C(\rho)}$, there exists a non-repeating zigzag path $\alpha = (x_i)_{i \leq n}$ such that $x_0 = \rho_0$ and $x_n = \rho$. Since $\rho_0 \mathrel R_{\mathcal Q} \tau_0$ we find, by using forth-up and forth-down confluence of $R_{\mathcal Q}$, a zigzag path $\beta =(y_i)_{i \leq m}$ such that $x_n \Rel_\mathcal{Q} y_m$. Denote $y_m$ by $\tau$ and we thus have $\rho \Rel_\mathcal{Q} \tau$. Note that since $\mathcal{Q}$ is an unravelled frame, we may assume that $\beta$ is non-repeating (if $x_i \Rel_\mathcal{Q} y_j$, $x_i$ is covered by $x_{i+1}$ and $x_{i+1} \Rel_{\mathcal{Q}} y_{j'}$ for $j' \geq j$, then $y_{j'} \geq_{\mathcal{Q}} y_j$ and if $y_{j'} = y_k$ for some $k < j$, then by $\mathcal{Q}$ being unravelled, there also exists a copy of $y_{j'}$ \emph{away} from $y_j$). Moreover, by construction the zigzag width of $\beta$ is at most the zigzag width of $\alpha$. Therefore, since $\rho_0 \in \mathcal{M}_\rho^{C(\rho)}$ and $\alpha$ is the path from $\rho_0$ to $\rho$, $\tau_0 \in \mathcal{N}_\tau^{C(\tau)}$.
    \end{proof}

    Putting everything together yields the desired result.

    \begin{proposition}\label{p: surjective dynamic simulation}
        The strongly surjective simulation $E \subseteq \moment_\Sigma \times Q$ is dynamic.
    \end{proposition}

    \begin{proof}
        Suppose $\mathcal{M} \mathrel E \rho_0$ and $\rho_0 \mathrel R_{\mathcal Q} \tau_0$. Then $\mathcal{M} = \mathcal{M}_{\rho_0}^{C(\rho)}$ for some $\rho \in Q$. Let $\tau \in Q$ such that $\rho \mathrel R_{\mathcal Q} \tau$ and $\tau_0 \in \mathcal{N}_\tau^{C(\tau)}$ (given by the previous lemma). As shown above, the moment $\mathcal{N}_{\tau_0}^{C(\tau)}$ is a modal successor of $\mathcal{M}_{\rho_0}^{C(\rho)}$. Moreover, $\mathcal{N}_{\tau_0}^{C(\tau)} \mathrel E \tau_0$.
    \end{proof}

We may now prove the main theorem of this section which states that the restriction of $\moment_\Sigma$ to the domain of the strongly surjective dynamic simulation $E$ constructed above is a labelled model. Denote by $\moment_\Sigma {\upharpoonright_{dom(E)}}$ the structure \begin{equation*}
(dom(E), \leq_\Sigma \cap (dom(E) \times dom(E)), \ell_\Sigma \cap (dom(E) \times \type_\Sigma), R_\Sigma \cap (dom(E) \times dom(E)).
\end{equation*}

Then we have the following.

\begin{theorem}\label{t: surjective simulation gives model}
    Let $\mathcal{Q}=(Q, \leq_\mathcal{Q}, \ell_\mathcal{Q}, R_\mathcal{Q})$ be a discrete unravelled $\Sigma$-labelled model with respect to a sensibility condition $S$ and $E \subseteq \mathbb{M}_\Sigma \times Q$ the strongly surjective dynamic simulation between $\moment_\Sigma$ and $\mathcal{Q}$ as defined above. Then $\moment_\Sigma{\upharpoonright_{dom(E)}}$ is a $\Sigma$-labelled model.
\end{theorem}
\begin{proof}
Denote $\moment_\Sigma{\upharpoonright_{dom(E)}}$ by $\mathcal{Z} = (Z, \leq_\mathcal{Z}, \ell_\mathcal{Z}, R_\mathcal{Z})$. First, we check that $(Z, \leq_\mathcal{Z}, \ell_\mathcal{Z})$ is a $\Sigma$-labelled poset. To that end note that $(Z, \leq_\mathcal{Z})$ is a poset as it is the restriction of $(\mathbb{M}_\Sigma, \leq_\Sigma)$ to $Z$. Suppose $\fw, \fv \in Z$ and $\fw \leq_\mathcal{Z} \fv$. Then $\fw \leq_\Sigma \fv$ and therefore $\ell_\Sigma(\fw) \leq_\type \ell_\Sigma(\fv)$. Thus $\ell_\mathcal{Z}(\fw) \leq_\type \ell_\mathcal{Z}(\fv)$. Let $\varphi \rightarrow \psi \in \ell_\mathcal{Z}(\fw)^-$. Then $\varphi \rightarrow \psi \in \ell_\Sigma (\fw)^-$. Therefore there exists $\fw' \in \mathbb{M}_\Sigma$ with $\fw \leq_\Sigma \fw'$ and $\varphi \in \ell_\Sigma (\fw')^+$ and $\psi \in \ell_\Sigma (\fw')^-$. Since $E$ is a simulation and $\fw \in dom(E)$ there exists $\rho \in Q$ with $\fw \mathrel E \rho$. By forth-up confluence there exists $\rho' \in Q$ with $\rho \leq_\mathcal{Q} \rho'$ and $\fw' \mathrel E \rho'$. Thus $\fw' \in dom(E)$ and $\fw \leq_\mathcal{Z} \fw'$. Finally, since $\ell_\Sigma(\fw') = \ell_\mathcal{Z}(\fw')$ it follows that $\varphi \in \ell_\mathcal{Z}(\fw')^+$ and $\psi \in \ell_\mathcal{Z}(\fw')^-$. The case for $\varphi \dimp \psi \in \ell_\mathcal{Z}(\fw)^+$ is similar, but using forth-down confluence of $E$ instead of forth-up confluence. Thus we conclude that $(Z, \leq_\mathcal{Z}, \ell_\mathcal{Z})$ is a $\Sigma$-labelled poset. 

Next, we show that $R_\mathcal{Z}$ is sensible. Recall that $S$ is the sensibility condition for $\moment_\Sigma$ and $\mathcal{Q}$. Suppose $\fw \leq_\mathcal{Z} \fw'$ and $\fw \mathrel R_\mathcal{Z} \fv$. Then $\fw \leq_\Sigma \fw'$ and $\fw \mathrel R_\Sigma \fv$. By forth-up confluence of $R_\Sigma$ there exists $\fv' \in \mathbb{M}_\Sigma$ with $\fv \leq_\Sigma \fv'$ and $\fw' \mathrel R_\Sigma \fv'$. By assumption $\fv \in Z$. Therefore there exists $\rho \in Q$ with $\fv \mathrel E \rho$. By forth-up confluence of $E$ it follows that there exists $\rho' \in Q$ with $\rho \leq_\mathcal{Q} \rho'$ and $\fv' \mathrel E \rho'$. Thus $\fv' \in Z$ and therefore there exists $\fv' \in Z$ with $\fv \leq_\mathcal{Z} \fv'$ and $\fw' \mathrel R_\mathcal{Z} \fv'$, i.e. $R_\mathcal{Z}$ is forth-up confluent. The case for forth-down confluence is similar. Finally, suppose $\fw \mathrel R_\mathcal{Z} \fv$. Then $\fw \mathrel R_\Sigma \fv$ and therefore $\ell_\Sigma(\fw) \mathrel S \ell_\Sigma(\fv)$. Hence, $\ell_\mathcal{Z}(\fw) \mathrel S \ell_\mathcal{Z}(\fv)$. It follows that $R_\mathcal{Z}$ is sensible and so that $\mathcal{Z}$ is a $\Sigma$-labelled frame. 

It remains to show that $R_\mathcal{Z}$ is witnessed. Suppose $\ld \varphi \in \ell_\mathcal{Z}(\fw)^+$. Let $\rho_0 \in Q$ such that $\fw \mathrel{E} \rho_0$. Then $\fw = \fw_{\rho_0}^{C(\rho)}$ for some $\rho \in Q$. By definition of a simulation $\ld \varphi \in \ell_{\mathcal{Q}}(\rho_0)^+$. Since $R_\mathcal{Q}$ is witnessed, there exists $\tau_0 \in Q$ with $\rho_0 \Rel_\mathcal{Q} \tau_0$ and $\varphi \in \ell_\mathcal{Q}(\tau_0)^+$. Since $E$ is dynamic, there exists $\fv \in Z$ with $\fv = \fv_{\tau_0}^{C(\tau)}$ for some $\tau \in Q$ such that $ \fv_{\tau_0}^{C(\tau)}$ is a modal successor of $\fw_{\rho_0}^{C(\rho)}$, i.e. $\fw_{\rho_0}^{C(\rho)} \Rel_{\mathcal{Z}} \fv_{\tau_0}^{C(\tau)}$ and $\fv_{\tau_0}^{C(\tau)} \mathrel{E} \tau_0$. By Lemma \ref{l: how to construct modal successors}, the zigzag width of the non-repeating zigzag path from $\tau_0$ to $\tau$ is at most the zigzag width of the non-repeating zigzag path from $\rho_0$ to $\rho$. Suppose $\tau_0$ is $\Sigma^\tau_{n+1}$ and $\Pi^\tau_{n+1}$ and $\rho_0$ is $\Sigma^\rho_{m+1}$ and $\Pi^\rho_{m+1}$ for $n \leq m$.  Then $n$ is the least natural number such that $\tau_0$ is $\Sigma_n^\tau$ or $\Pi_n^\tau$ and similarly for $m$. Suppose first that $\tau_0$ is $\Sigma_n^\tau$. Recall that
\begin{equation*}
\begin{split}
    \ell_\mathcal{Z}(\fv_{\tau_0}^{C(\tau)}) & = \ell_\Sigma(\fv_{\tau_0}^{C(\tau)})\\
    & = \ell_{\fv_{\tau_0}^{C(\tau)}}(\tau_0) \\
    &= (\ell_\mathcal{Q}(\tau_0)^+{\upharpoonright_{\Sigma(d(\Sigma) - \floor{((n+1)/2})}}, \ell_\mathcal{Q}(\tau_0)^-{\upharpoonright_{\Sigma(d(\Sigma) - \floor{n/2}})}.
\end{split}
\end{equation*}
Similarly if $\rho_0$ is $\Sigma_m^\rho$, then
\begin{equation*}
    \ell_\mathcal{Z}(\fw_{\rho_0}^{C(\rho)}) = (\ell_\mathcal{Q}(\rho_0)^+{\upharpoonright_{\Sigma(d(\Sigma) - \floor{((m+1)/2})}}, \ell_\mathcal{Q}(\rho_0)^-{\upharpoonright_{\Sigma(d(\Sigma) - \floor{m/2}})}.
\end{equation*}

Notice that $d(\ld \varphi) = d(\varphi)$. Therefore, since $n \leq m$ and $\ld \varphi \in \ell_\mathcal{Q}(\rho_0)^+{\upharpoonright_{\Sigma(d(\Sigma) - \floor{((m+1)/2})}}$, we have
\begin{equation*}
    d(\ld \varphi) \leq  d(\Sigma) - \floor{((m+1)/2}
\end{equation*}
implying that
\begin{equation*}
    d(\varphi) \leq d(\Sigma) - \floor{((n+1)/2},
\end{equation*}
and so that $\varphi \in \ell_\mathcal{Q}(\tau_0)^+{\upharpoonright_{\Sigma(d(\Sigma) - \floor{((n+1)/2})}}$. Hence $\varphi \in \ell_\mathcal{Z}(\fv_{\tau_0}^{C(\tau)})^+$. Next suppose that $\tau_0$ is $\Sigma_n^\tau$ and $\rho_0$ is $\Pi_m^\rho$. Then note that $n < m$ (see the proof of Lemma \ref{l: how to construct modal successors}). By definition 
\begin{equation*}
    \ell_\mathcal{Z}(\fv_{\tau_0}^{C(\tau)})^+ =  \ell_\mathcal{Q}(\tau_0)^+{\upharpoonright_{\Sigma(d(\Sigma) - \floor{((n+1)/2})}}
\end{equation*}
and
\begin{equation*}
    \ell_\mathcal{Z}(\fw_{\rho_0}^{C(\rho)})^+ = \ell_\mathcal{Q}(\rho_0)^+{\upharpoonright_{\Sigma(d(\Sigma) - \floor{ m/2})}}.
\end{equation*}
Since $n < m$, we have $n+1 \leq m$ and thus since $\ld \varphi \in \ell_\mathcal{Z}(\fw_{\rho_0}^{C(\rho)})^+$ and $d(\ld \varphi) = d(\varphi)$ we have $\varphi \in \ell_\mathcal{Z}(\fv_{\tau_0}^{C(\tau)})^+$. The remaining cases where $\tau_0$ is $\Pi_m^\tau$ and $\rho_0$ is either $\Sigma_m^\rho$ or $\Pi_m^\rho$ are similar. Moreover the case for $\lb \varphi \in \ell_{\mathcal{Z}}(\fw)^-$ is similar as well. Therefore $R_\mathcal{Z}$ is witnessed and so $\mathcal{Z}$ is a $\Sigma$-labelled model.
\end{proof}

Note that the zigzag width of $\moment_\Sigma{\upharpoonright_{dom(E)}}$ is bounded by $4 d(\Sigma) + 1$. Thus we have obtained an intermediate result on the way to establish the finite model property: if $\varphi$ is falsifiable over the class of labelled models, then $\varphi$ is falsifiable on a labelled model with bounded zigzag width. The next section will show that we can obtain a substructure of $\moment_\Sigma{\upharpoonright_{dom(E)}}$ which is a finite labelled model and still falsifies $\varphi$. The bound on the zigzag width will be of crucial importance for the proof.

\section{Succinct Moments}\label{SecIrred}

In order to obtain \emph{finite} $\Sigma$-labelled frames, we will restrict the class $\moment_\Sigma$ of all $\Sigma$-moments to those moments that are, in a sense, no bigger than necessary. Specifically, they should not be `bimersive' to a moment of strictly smaller cardinality. Below, we make this precise.

To prove the main results in this section, we will need to consider labels that are not necessarily types. Let $\mathcal C = (C,\leq)$ be a finite poset, referred to as \emph{poset of colors}. Each element $c \in C$ is called a \emph{color}.

\begin{definition}\index{moment!$\mathcal{C}$-moment}
    A \emph{$\mathcal C$-moment} is a tuple $\mathcal{M} = (M,\leq_\fw,\ell_\fw,r)$, where $(M,\leq_\fw)$ is an acyclic, discrete, connected and non-empty poset, $r \in M$ is called the \emph{root} and $\ell_\fw: M \longrightarrow C  $ is order preserving: $w\leq_\fw v \implies \ell_\fw(w) \leq \ell_\fw( v)$.
\end{definition}  

The \emph{zigzag radius} of a $\mathcal{C}$-moment $\fw=(M, \leq_\fw, \ell_\fw, r)$ is the zigzag radius of $(M, \leq_\fw, r)$. 
Observe that $\Sigma$-moments can be regarded as $\mathcal{C}$-moments (by deleting the initial world) where $\mathcal{C}=(\mathrm{T}_\Sigma, \leq_{\mathrm{T}_\Sigma})$. 

\begin{definition}\index{moment!tree-like}
A $\mathcal{C}$-moment $\fw = (M, \leq_\fw, \ell_\fw, r)$ is \emph{tree-like with root $r$} if either for all $ w\in M$, $w\leq r$ or else for all $ w\in M$,  $r\leq w$.
\end{definition}

We introduce the following notion of `structure preserving' map between moments.

\begin{definition}\label{DefSim}\index{bimersion}
Let $\mathcal{M}=(M, \leq_\fw, \ell_\fw, r)$ and $\mathcal{N}=(N, \leq_\fv, \ell_\fv, r')$ be $\mathcal C$-moments where $\mathcal C$ is a poset of colors. An \emph{immersion}\index{immersion} is a function $\sigma: M \longrightarrow N$ such that the following hold.
\begin{enumerate}
    \item $\sigma$ is order-preserving: for all $w,v \in M$ if $w \leq_\fw v$, then $\sigma(w) \leq_\fv \sigma(v)$.
    \item For all $w \in M$, $\ell_\fw(w) = \ell_\fv(\sigma(w))$.
    \item $\sigma(r) = r'$.
\end{enumerate}
If an immersion $\sigma: \fw\longrightarrow \fv $ exists, we write $\fw\redu \fv $.
If, in addition, there is an immersion $\tau: \fv \longrightarrow \fw$,
then $\fw$ and $\fv$ are \emph{bimersive},
write $\fw\simm \fv $, and the pair $(\sigma,\tau)$ is called a \emph{bimersion}.
\end{definition}

\begin{definition}\index{moment!succinct}
    A $\mathcal{C}$-moment $\mathcal{M}=(M, \leq_\fw, \ell_\fw, r)$ is \emph{succinct} if for every $\mathcal{C}$-moment $\mathcal{N}=(N, \leq_\fv, \ell_\fv, r')$, $\fw \simm \fv$ implies $\lvert M \rvert \leq \lvert N \rvert$.
\end{definition}

Observe that every $\mathcal{C}$-moment is bimersive to itself. Furthermore it can easily be checked that the bimersion relation is transitive and symmetric, i.e. an equivalence relation on the class of $\mathcal{C}$-moments. The equivalence classes of $\simm$ are called \emph{bimersion classes}. Observe that any bimersion class must contain a smallest moment, since cardinal numbers are wellfounded. Therefore the following lemma holds.

\begin{lemma}
    Let $\fw $ be a $\mathcal{C}$-moment. There exists a succinct $\mathcal{C}$-moment $\fv$ such that $\fw \simm \fv$.
\end{lemma}

We wish to show that the number of bimersion classes of $\Sigma$-moments is computably bounded, and that there is a computable bound on the cardinality of the succinct $\Sigma$-moments. 
We will prove this via an inductive argument, in which worlds of a $\mathcal{C}$-moment for an appropriate choice of $\mathcal{C}$ of maximal zigzag radius are labelled by $\Sigma$-moments of smaller zigzag radius, in order to apply the induction hypothesis and reduce these simpler moments. Thus the following lemma must be stated for an arbitrary finite poset of colors.

\begin{lemma}\label{LemIrrBound}
There exist computable functions $F$ and $G$ such that for any finite poset of colors $\mathcal C = (C, \leq)$ the following hold.
\begin{enumerate}
\item Any tree-like $\mathcal C$-moment $\fw$ is bimersive to a tree-like $\mathcal C$-moment $\fw'$ of cardinality bounded by $F(\lvert C\rvert)$;

\item The immersion from $\fw$ to $\fw'$ is surjective.

\item \label{ItKruskal} There are at most $G(\lvert C \rvert)$ bimersion classes of tree-like $\mathcal C$-moments.
\end{enumerate}
\end{lemma}

\begin{proof}
We prove the result for tree-like moments where the root is the least element. For the general result (where the root may alternatively be the greatest element), by order duality, the bound for $F$ remains the same and the bound for $G$ only needs to be doubled. 

Let the \emph{height} of a finite poset be the maximum length $h$ of a chain $w_1 < w_2 < \dots < w_h$ of elements of the poset. The proof is by induction on the height $h$ of $\mathcal C$. Write $\exp_2$ for the exponential function $m \mapsto 2^m$. Define the \emph{iterated exponential function} $exp_2^n$ inductively by $exp_2^0(m) = 1$ and $exp_2^{n+1}(m) = exp_2(exp_2^n(m))$. We claim that $exp_2^h(\cardinality C)$ is a valid choice for both $F$ and $G$.

Let $h = 1$, and let $\mathcal M$ be a $\mathcal C$-labelled tree-like moment. Then the labelling function must be constant, with value $c$ for some $c \in C$. Let $\mathcal M_c$ be the $\mathcal C$-labelled tree-like moment consisting of a single world labelled by $c$. Clearly the unique function $ \sigma:\mathcal M \longrightarrow \mathcal M_c$ is a surjective immersion. In the other direction, the function $\tau:\mathcal M_c \longrightarrow \mathcal M$ mapping the only world of $M_c$ onto the root of $M$ is also clearly an immersion. Thus $\mathcal M$ and $\mathcal M_c$ are bimersive. Hence
\begin{enumerate}
\item 
any tree-like $\mathcal C$-moment $\fw$ is bimersive to a (tree-like) $\mathcal C$-moment $\fw'$ of cardinality~1;
\item the immersion from $\fw$ to $\fw'$ is surjective;
\item \label{moment2} there are at most $\cardinality C$ bimersion classes of tree-like $\mathcal C$-moments.
\end{enumerate}
Since $1 \le \cardinality C \le 2^{\cardinality C} = \exp_2^1(\cardinality C)$, the bounds for both $F$ and $G$ are satisfied.

Now let $\mathcal C$ be of height $h+1$, and assume the result for posets of colors of height at most $h$. For each choice of $c \in C$, the poset of colors $\mathcal C_c \coloneqq (c^\uparrow \setminus \{c\}, \le)$ is of height at most $h$ and cardinality at most $\cardinality C - 1$, where $c^\uparrow = \{ c' \in C \mid c \leq c'\}$. So by induction hypothesis we can find a set $\mathsf S_c$ of at most $\exp_2^h(\cardinality C - 1)$ tree-like $\mathcal C$-moments each of cardinality  at most $\exp_2^h(\cardinality C)$, such that any tree-like $\mathcal C_c$-moment is bimersive to an element of $\mathsf S_c$.

Let $\mathcal M$ be a $\mathcal C$-labelled tree-like moment with root $r$.  Let $c$ be the colour of $r$, and let $X$ be the set of minimal elements of $\mathcal M$ that do  not have colour $c$. Then as $\mathcal M$ is tree-like, the sets of the form $x^\uparrow$ for $x\in X$ are disjoint. Further, for each $x \in X$, note that each world in $x^\uparrow$ is colored using colors in $c^\uparrow \setminus \{c\}$, implying that the restriction of $\fw$ to $x^\uparrow$ is a tree-like $\mathcal{C}_c$-moment, and hence bimersive to a $\mathcal C_c$-moment in $\mathsf S_c$.

Let $\mathcal M'$ be the following $\mathcal C$-labelled tree-like moment. The root of $\mathcal M'$ is coloured $c$ and for each bimersion class in 
\begin{equation*}
\{ \mathsf B \mid \text{ there exists } x \in X : \mathsf B\text{ is the bimersion class of $x^\uparrow$}\}
\end{equation*}
the corresponding representative in $\mathsf S_c$ is added above the root. 
Construct a bimersion $(\sigma, \tau)$ from $\mathcal M$ to $\mathcal M'$ as follows. The function $\sigma$ maps all $c$-colored elements of $\mathcal M$ to the root of $\mathcal M'$, and $\tau$ maps the root of $\mathcal M'$ to the root of $\mathcal M$. The values of $\sigma$ on the remaining elements of $\mathcal M$ are defined using the disjoint union of immersions from the $x^\uparrow$'s to elements of $\mathsf S_c$ in the obvious way. Note that by induction hypothesis the immersions from the $x^\uparrow$'s to elements of $\mathsf{S_c}$ are surjective, implying that $\sigma$ is surjective as well. The values of $\tau$ on the remaining elements of $\mathcal M'$ are defined similarly (for each $\mathcal S \in \mathsf S_c$ an arbitrary choice of $x \in X$ such that $x^\uparrow$ is bimersive to $\mathcal S$ is made, and an immersion into $x^\uparrow$ is used). It is easy to check that $(\sigma, \tau)$ is a bimersion. 

The cardinality of $\mathcal M'$ is at most $1 + \exp_2^h(\cardinality C -1) \cdot \exp_2^h(\cardinality C) \leq \exp_2^{h+1}(\cardinality C)$, proving the claimed bound for $F$. For each possible value of $c$, the number of distinct $\mathcal M'$s we may encounter is bounded by $2^{\exp_2^h(\cardinality C -1)} = \exp_2^{h+1}(\cardinality C - 1)$. So the total number of moments we need to represent all bimersion classes of tree-like $\mathcal C$-moments is bounded by $\cardinality C \cdot \exp_2^{h+1}(\cardinality C - 1) \le \exp_2^{h+1}(\cardinality C)$, proving the claimed bound for $G$.
\end{proof}

\begin{remark}
A more general version of Lemma \ref{LemIrrBound} is given in ~\cite[Theorem~23]{Boudou_2017} (without explicitly calculating bounds or explicitly mentioning the surjectivity of one of the immersions). There, the authors do not assume that $C$ is partially ordered, but instead consider labelling functions of a fixed \emph{level} $k$; the level of a labelling function $\ell$ is the maximal length of a chain $w_1,\ldots,w_k$ of worlds such that $\ell(w_i)\neq \ell(w_{i+1})$.
The $\mathcal C$-moments considered here automatically have level at most $\cardinality C$, since the labelling functions are monotone. 
\end{remark}

With Lemma \ref{LemIrrBound} proven, we shall show that a bound on the cardinality and the number of bimersion classes of $\Sigma$-moments can also be obtained. Recall that the zigzag radius of $\Sigma$-moments is bounded by $2d(\Sigma) +1$. As mentioned above, we can regard $\Sigma$-moments as $\mathcal{C}$-moments for $\mathcal{C}=(\type_\Sigma, \leq_\Sigma)$ by deleting the initial world. In the following proof, we will generally not mention the initial world, as it does not play a role in any of the arguments. Note that on the converse \emph{not} every $\mathcal{C}$-moment is a $\Sigma$-moment, since the definition of $\mathcal{C}$-moment does not deal with defects. The proof proceeds by induction on the zigzag radius $n$. Note that $\Sigma$-moments with zigzag radius $1$ contain only a single point $r$, due to connectedness. If there would exist a second point $w \geq r$, then $w$ is $\Pi_r^1$ and hence $\Sigma_r^2$, implying that the zigzag radius is at least $2$ and similarly for $w \leq r$. Let $\moment_\Sigma^n$ be the class of $\Sigma$-moments with zigzag radius bounded by $n$.

\begin{proposition}\label{propIrrFin}
There are computable functions $\kappa$ and $\lambda$ such that 
\begin{enumerate}
    \item if $\fw \in \moment_\Sigma^n$, then there exists $\fv \in \moment_\Sigma^n$ with $\fw \simm \fv$ and the size of $\fv$ is bounded by $\kappa(n)$;
    \item there are at most $\lambda(n)$ bimersion classes of $\moment_\Sigma^n$.
\end{enumerate}
\end{proposition}

\begin{proof}
Let $\fw = (M, \leq_\fw, \ell_\fw, r) \in \moment_\Sigma^n$. The proof proceeds by induction on $n$. For the base case where $n = 1$, $\fw$ consists of a single point, i.e. $M = \{ r \}$ by the previous observation. Clearly, $\fw$ is bimersive to itself and succinct. Hence $\kappa(1) = 1$ and there are at most $\lambda(1) = \cardinality{\type_\Sigma}$ bimersion classes of $\moment_\Sigma^1$. \smallskip

For the induction step suppose $n > 1$ and Items 1. and 2. hold for any $\Sigma$-moment with zigzag radius bounded by $n-1$. Let $\mathcal{C}= (C, \leq_\mathcal{C})$ where $C = \type_\Sigma$ and $\leq_\mathcal{C} = \leq_\type$. We define two $\mathcal{C}$-moments $\hat{\fw}$ and $\check{\fw}$ as follows. Given a non-repeating zigzag path $\rho$, write $U(\rho)$ if $\rho$ contains a non-empty initial segment $\rho'$ with ${\uparrow} \rho'$ and $D(\rho)$ if $\rho$ contains a non-empty initial segment $\rho'$ with ${\downarrow} \rho'$. Furthermore if $\rho = (x_i)_{0 \leq i \leq n}$, write $x_0 \mathrel{\rho} x_n$ to denote that $\rho$ is the zigzag path from $x_0$ to $x_n$. Define
\begin{equation*}
    \begin{split}
        \hat{M} & \coloneqq \{w \in M \mid \text{ there exists } \rho \text{ such that } r \mathrel{\rho} w \text{ and } U(\rho)\}; \\
        \check{M} & \coloneqq \{w \in M \mid \text{ there exists } \rho \text{ such that } r \mathrel{\rho} w \text{ and } D(\rho)\}. \\
    \end{split}
\end{equation*}

Therefore, $\hat{M}$ contains all points of $\fw$ reachable from the root $r$ over a zigzag path that first goes up and $\check{M}$ contains all points of $\fw$ reachable from the root $r$ over a zigzag path that first goes down. Note that $r$ is contained in both $\hat{M}$ and $\check{M}$. Then let
\begin{equation*}
    \begin{split}
        \hat{\fw} & := \fw_{\upharpoonright \hat{M}} = (\hat{M}, \leq_\fw \cap (\hat{M} \times \hat{M}), \ell_\fw \cap (\hat{M} \times C), r) \\
        \check{\fw} & := \fw_{\upharpoonright \check{M}} = (\check{M}, \leq_\fw \cap (\check{M} \times \check{M}), \ell_\fw \cap (\check{M} \times C), r).\\
    \end{split}
\end{equation*}

Denote $\hat{\fw} = (\hat{M}, \leq_{\hat{\fw}}, \ell_{\hat{\fw}}, r)$ and $\check{\fw} = (\check{M}, \leq_{\check{\fw}}, \ell_{\check{\fw}}, r)$. Observe that $\hat{\fw}$ and $\check{\fw}$ are $\mathcal{C}$-moments with zigzag radius at most $n$.\footnote{Note that $\hat{\fw}$ and $\check{\fw}$ are not necessarily $\Sigma$-moments, since some defects in $\check{M}$ might be resolved in $\hat{M}$ and vice versa.} Furthermore, let 
\begin{equation*}
    \hat{\fw} \cup \check{\fw} := (\hat{M} \cup \check{M}, (\leq_{\hat{\fw}} \cup \leq_{\check{\fw}})^*, \ell_{\hat{\fw}} \cup \ell_{\check{\fw}}, r)
\end{equation*}

where $(\leq_{\hat{\fw}} \cup \leq_{\check{\fw}})^*$ is the transitive closure of $\leq_{\hat{\fw}} \cup \leq_{\check{\fw}}$. Observe that  $\hat{\fw} \cup \check{\fw} = \fw$.

We will first show $\hat{\fw}$ is bimersive to a $\mathcal{C}$-moment $\hat{\fv}$ whose size is bounded. Recall that for any $u \in \hat{M}$, 
\begin{equation*}
    \begin{split}
        u^\uparrow &:= \{w \in \hat{M} \, \lvert \, u \leq_{\hat{\fw}} w\}\\
         u^\downarrow & := \{w \in \hat{M} \, \lvert \, u \geq_{\hat{\fw}} w\}.\\
    \end{split}
\end{equation*}

 Let $w \in r^\uparrow$ and let $\rho = (w_0, \ldots, w_k)$ be the unique non-repeating zigzag path from $r$ to $w$, i.e. $w_0 = r$ and $w_k = w$. For any $v \in \hat{M}$ which is covered by $w_i$ for some $0 < i \leq k$ (note that $v$ cannot be covered by $r$, as otherwise $v \in \check{M}$), let $\fw_v = (M_v, \leq_v, \ell_v, v)$ be the connected component of $v$ in $\hat{M} \setminus \{w_0, \ldots, w_k\}$, i.e.
\begin{itemize}
    \item $u \in M_v$ iff $u \in \hat{M}$ and there exists a zigzag path $\rho=(u_0, \ldots, u_l)$ with $u_0 = v$, $u_l = u$ and for all $0 \leq i \leq l$, $u_i \not \in \{w_0, \ldots, w_k\}$;
    \item $\leq_v = \leq_{\hat{\fw}} \cap (M_v \times M_v)$;
    \item $\ell_v = \ell_{\hat{\fw}} \cap (M_v \times C)$.
\end{itemize}

Observe that $\fw_v$ is a $\Sigma$-moment: clearly, $\fw_v$ is a $\mathcal{C}$-moment, since it is the restriction of $\hat{\fw}$ to $M_v$; furthermore, due to $\fw$ being a $\Sigma$-moment and Properties \ref{d: moment 1.1} and \ref{d: moment 1.2} of a $\Sigma$-moment, every defect of a world in $\fw_v$ is resolved in $\fw_v$. We claim that the zigzag radius of $M_v$ is at most $n-1$. Suppose towards contradiction that the zigzag radius of $\fw_v$ is $m \geq n$. Recall that $v$ is covered by one of the worlds in  $\{w_1, \ldots, w_k\}$, say by $w_i$ for $i > 0$. Let $u \in M_v$ such that $u$ is $\Pi_m^v$ and $\Sigma_m^v$. So there exists a non-repeating zigzag path $\rho$ with $u \mathrel{\rho} v$ and $\rho$ is $\Pi_m$ and $\Sigma_m$. Let $\rho' = \rho \circ (w_i, w_{i-1}, \ldots, w_0)$ where $\circ$ denotes concatenation. Then $\rho'$ is a zigzag path from $u$ to $r$ and, since $w_i > v$, $w_0 < w_i$ and no element of $\rho$ occurs in $\{w_0, \ldots, w_i\}$, $\rho'$ is non-repeating and both $\Sigma_l$ and $\Pi_l$ for $l > m \geq n$. Therefore the zigzag radius of $\hat{\fw}$ is strictly larger than $n$, which contradicts our initial assumption. Hence, the zigzag radius of $\fw_v$ is $k \leq n-1$ and so by induction hypothesis, each such $\fw_v$ is bimersive to a $\Sigma$-moment of size bounded by $\kappa(n-1)$.

Define a poset of colors $\mathcal{C'} = (C', \leq_\mathcal{C'})$, where
\begin{itemize}
    \item $C' = \{ (c, B) \, \lvert \, c \in C \text{ and } B \text{ is a set of bimersion classes of } \moment_\Sigma^{n-1}\}$;
    \item $(c, B) \leq_{\mathcal{C'}} (c', B')$ if and only if $c \leq_\mathcal{\mathcal{C}} c'$ and $B \subseteq B'$.
\end{itemize}

Observe that $\mathcal{C'}$ is finite since $C = \type_\Sigma$ is finite and there are only finitely many bimersion classes of $\moment_\Sigma^{n-1}$ by induction hypothesis. Let $\fw' := (M', \leq', \ell', r)$ where
\begin{itemize}
    \item $M' = \hat{M} \cap r^\uparrow$;
    \item $\leq' = \leq_{\hat{\fw}} \cap (r^\uparrow \times r^\uparrow)$;
    \item $l': M' \longrightarrow C'$ given by $l'(w) = (l_{\hat{\fw}}(w), B(w))$
    where $B(w)$ is the set of bimersion classes of $\Sigma$-moments $\fw_v$ for $v$ covered by one of $\{w_1, \ldots, w_k\}$ as defined above.
\end{itemize}

We claim that $\fw'$ is a tree-like $\mathcal{C'}$-moment. That $\fw'$ is tree-like is easy to see, since $M'$ is simply $r^\uparrow$ within $\hat{M}$. This implies that $(M', \leq')$ is acyclic, discrete and connected. The labelling function $\ell'$ assigns to each $w \in M'$ a color in $C'$. It remains to show that $\ell'$ is order-preserving for $\leq'$. Suppose $w \leq' u$. Then $w \leq_{\hat{\fw}} u$ and hence $\ell_{\hat{\fw}}(w) \leq_\mathcal{C} \ell_{\hat{\fw}}(u)$. Suppose $\alpha \in B(w)$. Then $\alpha$ is the bimersion class generated by some $\Sigma$-moment $M_v$ for $v$ covered by one of $\{w_1, \ldots, w_k\}$ where $(w_0, \ldots, w_k)$ is the zigzag path from $r$ to $w$. Since $w \leq_{\hat{\fw}} u$, we have $\alpha \in B(u)$ as well. Hence $B(w) \subseteq B(u)$ and so $\ell'(w) \leq_\mathcal{C'} \ell'(u)$. \smallskip

By Lemma \ref{LemIrrBound}, $\fw'$ is bimersive to a tree-like $\mathcal{C'}$-moment $\mathcal{N}'=(N', \leq_{\fv'}, \ell_{\fv'}, r')$ of size bounded by $F(\lvert C'\rvert)$. Let $(\sigma, \tau)$ be the bimersion between $\fw'$ and $\fv'$. For each bimersion class $\alpha$ of $\Sigma$-moments with zigzag radius at most $n-1$, 
choose one succinct moment in $\alpha$ and denote it by $\fv_\alpha=(N_\alpha, \leq_\alpha, \ell_\alpha, r_\alpha)$. 
 \smallskip

Define a $\mathcal{C}$-moment $\hat{\fv}=(N, \leq_{\hat{\fv}}, \ell_{\hat{\fv}}, r')$ as follows. Let 
\begin{equation*}
    N := N' \uplus \biguplus_{w \in N'} \biguplus_{\alpha \in B(w)} N_\alpha.
\end{equation*}

Since we add for every $w$ for which $\alpha \in B(w)$ a copy of $\fv_\alpha$, we will denote by $\fv_\alpha^w =(N_\alpha^w, \leq_\alpha^w, \ell_\alpha^w, r_\alpha^w)$ the copy of $\fv_\alpha$ added for $w$. Next, define

\begin{equation*}
    \leq_0 := \leq_{\fv'} \uplus \biguplus_{w \in N'} \biguplus_{\alpha \in B(w)} \leq_\alpha^w.
\end{equation*}

 Then let $\leq_{\hat{\fv}}$ be the transitive closure of
\begin{equation*}
     \leq_0 \cup \{(r_\alpha^w,w) \, \lvert \,  w\in N' \text{ and } \alpha \in B(w)\}
\end{equation*}

Finally, define $\ell_{\hat{\fv}}: N \longrightarrow C$ by
\begin{equation*}
    \ell_{\hat{\fv}} (u) = \begin{cases}
        \pi_2^1(\ell_{\fv'}(u)) & \text{if } u \in N' \\
        \ell_\alpha(u) & \text{if } u \in N_\alpha \text{ for } \alpha \in B(w) \text{ for some } w \in N'.
    \end{cases}
\end{equation*}

where $\pi_2^1 (x,y) = x$ is the projection to the first component of the tuple. \smallskip

We claim that $\hat{\fv}$ is a $\mathcal{C}$-moment bimersive to $\hat{\fw}$. Observe that $(N, \leq_{\hat{\fv}})$ is a discrete and acyclic poset, since each component of $\leq_{\hat{\fv}}$ is discrete and acyclic on the respective part of $N$. Moreover, since each component of $\hat{\fv}$ is a $\mathcal{C}$-moment, it follows from the definition of $\leq_{\hat{\fv}}$ that $\hat{\fv}$ is connected. By induction hypothesis the zigzag radius of each $\fv_\alpha$ is at most $n-1$. Thus, by construction, the zigzag radius of $(N, \leq_{\hat{\fv}}, r')$ is at most $n$. By construction the labelling function $\ell_{\hat{\fv}}$ maps each $w \in N$ onto a color $c \in C$. If $w \leq_{\hat{\fv}} v$, then there are three cases to consider. First if $w,v \in N'$ and $w \leq_{\fv'} v$, then $\ell_{\fv'}(w) \leq_{\mathcal{C'}} \ell_{\fv'}(v)$ and therefore, by definition of $\mathcal{C'}$ and $\ell_{\hat{\fv}}$, also $\ell_{\hat{\fv}}(w) \leq_\mathcal{C} \ell_{\hat{\fv}}(v)$. Second if $w,v \in N_\alpha$ for $\alpha \in B(u)$ for some $u \in N'$, then $w \leq_\alpha v$ and since $\fv_\alpha$ is a $\mathcal{C}$-moment, it follows that $\ell_{\hat{\fv}}(w) \leq_\mathcal{C} \ell_{\hat{\fv}}(v)$. Third if $w \in N_\alpha^u$ where $\alpha \in B(u)$ for some $u \in N'$ and $v \in N'$, then $v \geq_{\fv'} u$ and $w \leq_\alpha r_\alpha^u$. Consider $\tau(u) \in M'$. By definition of an immersion we have that $\alpha \in B(\tau(u))$, therefore there exists a $\mathcal{C}$-moment $\fw_\alpha$ occurring as a substructure of $\hat{M}$ below $\tau(u)$, such that $\fw_\alpha \simm \fv_\alpha$ via the bimersion $(\sigma_\alpha, \tau_\alpha)$. Hence $\tau_\alpha(r_\alpha^u)$ is the root of $\fw_\alpha$, which, by construction, is covered by $\tau(u)$ or by a world $u_i$ on the zigzag path from $r$ to $\tau(u)$. In either case we have that $\ell_{\hat{\fw}}(\tau_\alpha(r_\alpha^u)) \leq_\mathcal{C} \ell_{\hat{\fw}}(\tau(u))$. Thus also $\ell_{\hat{\fv}}(r_\alpha^u) \leq_\mathcal{C} \ell_{\hat{\fv}}(u)$, implying that $\ell_{\hat{\fv}}(w) \leq_\mathcal{C} \ell_{\hat{\fv}}(v)$ by applying the previous two cases. Thus $\hat{\fv}$ is a $\mathcal{C}$-moment. 

Recall that $\fv'$ is bimersive to $\fw'$, witnessed by $(\sigma, \tau)$. Moreover, each moment $\fw_v$ for $v$ covered by one of  $\{w_1, \ldots, w_k\}$ where $w = w_k \in r^\uparrow$ is bimersive to $\fv_\alpha$ for some bimersion class $\alpha$. Define an immersion $\hat{\sigma}$ from $\hat{\fw}$ to $\hat{\fv}$ as follows: if $u \in r^\uparrow$, then $\hat{\sigma}(u) = \sigma(u)$; if $u \in M_v$ for some $v$ covered by $w \in r^\uparrow$, then let $\alpha \in B(w)$ such that $\fw_v \in \alpha$ and let $\sigma_\alpha$ be the immersion from $\fw_v$ to the copy of $\fv_\alpha$ whose root is covered by $\sigma(w)$ in $\hat{\fv}$ and define $\hat{\sigma}(u) = \sigma_\alpha(u)$. 

Observe that $\ell_{\hat{\fw}}(u) = \ell_{\hat{\fv}}(\hat{\sigma}(u))$ since $\sigma$ and $\sigma_\alpha$ are immersions. Moreover, $dom(\hat{\sigma}) = \hat{M}$ and it is easy to check that $\hat{\sigma}$ is forth-up and forth-down confluent for $\leq_{\hat{\fw}}$ and $\leq_{\hat{\fv}}$. For example, for forth-up confluence, suppose $w \leq_{\hat{\fw}} w'$. If both $w,w' \in r^\uparrow$, then $\hat{\sigma}(w) \leq_{\hat{\fv}} \hat{\sigma}(w')$ since $\sigma$ is an immersion. Similarly if both $w,w' \in M_v$, then forth-up confluence follows from $\sigma_\alpha$ being an immersion. If $w \in M_v$ and $w' \in r^\uparrow$, then $w \leq_{\hat{\fw}} v$, $v$ is covered by some $w_i$ and $w' \geq_{\hat{\fw}} w_i$. By definition of an immersion, $\sigma_\alpha(w) \leq_\alpha \sigma_\alpha(v)$, where $\sigma_\alpha(v)$ is the root of $\fv_\alpha$, implying that $\hat{\sigma}(w) \leq_{\hat{\fv}} \hat{\sigma}(v)$. Similarly, $\sigma(w_i) \leq_{\hat{\fv'}} \sigma(w')$, implying that $\hat{\sigma}(w_i) \leq_{\hat{\fv}} \hat{\sigma}(w')$. Moreover, by construction, $\hat{\sigma}(v)$ is covered by $\hat{\sigma}(w_i)$, implying that $\hat{\sigma}(w) \leq_{\hat{\fv}} \hat{\sigma}(w')$.

Similarly, define an immersion $\hat{\tau}$ from $\hat{\fv}$ to $\hat{\fw}$ as follows. If $u \in r'^\uparrow$, let $\hat{\tau}(u) = \tau(u)$. If $u \in N_\alpha$ for some succinct moment $\fv_\alpha$ whose root is covered by some $w \in N'$, chose $\fw_v$ such that $v$ is covered by some $w_i$ on the zigzag path from $r$ to $\tau(w)$. Let $\tau_\alpha$ be the immersion from $\fv_\alpha$ to $\fw_v$ and define $\hat{\tau}(u) = \tau_\alpha(u)$.


The proof that $\hat{\tau}$ is an immersion is similar as for $\hat{\sigma}$ and omitted. We conclude that $\hat{M} \simm \hat{\fv}$. \smallskip

We claim that the size of $\hat{\fv}$ is bounded. Recall that $\lvert N' \rvert \leq F(\lvert C' \rvert)$. By induction hypothesis $\lvert N_\alpha \rvert \leq \kappa(n-1)$ for each bimersion class $\alpha$. Since there are at most $\lambda(n-1)$ bimersion classes of $\Sigma$-moments with zigzag radius bounded by $n-1$ and each $\fv_\alpha$ occurs at most $F(\lvert C' \rvert)$ many times, we obtain
\begin{equation*}
    \lvert N \rvert \leq F(\lvert C' \rvert) \cdot \lambda(n-1) \cdot \kappa(n-1).
\end{equation*}

Note that $\lvert C' \rvert = \lvert C \rvert \cdot 2^{\lambda_\mathcal{C}(n-1)}$, implying that
\begin{equation*}
    \lvert N \rvert \leq F(\lvert C \rvert \cdot 2^{\lambda(n-1)}) \cdot \lambda(n-1) \cdot \kappa(n-1)
\end{equation*}
which establishes the bound for $\hat{N}$. \smallskip

Next, the case for $\check{M}$ is order-dual and we obtain a $\mathcal{C}$-moment $\check{\fv}$ bimersive to $\check{M}$, witnessed by $(\check{\sigma}, \check{\tau})$, and bounded by the same bound as $\hat{\fv}$. \smallskip

Finally, let $\fv = \hat{\fv} \cup \check{\fv}$. Note that $\fv$ is a $\mathcal{C}$-moment with zigzag radius bounded by $n$ and whose size is bounded by
\begin{equation*}
   \kappa(n) = 2 \cdot F(\lvert C \rvert \cdot 2^{\lambda(n-1)}) \cdot \lambda(n-1) \cdot \kappa(n-1).
\end{equation*}

Moreover, $\fw \simm \fv$ by taking the bimersion $(\Tilde{\sigma}, \Tilde{\tau}) =(\hat{\sigma} \cup \check{\sigma}, \hat{\tau} \cup \check{\tau})$. It remains to show that $\fv$ is a $\Sigma$-moment. Recall that every $\fv_\alpha$ is a $\Sigma$-moment and hence resolves all its defects. Suppose $x \in r'^\uparrow$ and $\varphi \rightarrow \psi \in \ell_\fv(x)^-$. Recall that $\sigma$ is the immersion from $\fw'$ to $\fv'$, which are the tree-like $\mathcal{C'}$-moments. By Lemma \ref{LemIrrBound}, $\sigma$ is surjective. Furthermore, by construction, for each $w \in r^\uparrow$, $\sigma(w) = \Tilde{\sigma}(w)$. Thus there exists $w \in r^\uparrow$ with $\Tilde{\sigma}(w) = x$. By the properties of an immersion, $\varphi \rightarrow \psi \in \ell_\fw(w)^-$. Since $\fw$ is a $\Sigma$-moment, there exists $u \geq_\fw w$ with $\varphi \in \ell_\fw(u)^+$ and $\psi \in \ell_\fw(u)^-$. Thus $\Tilde{\sigma}(u) \geq_\fv x$ and $\varphi \in \ell_\fv(\Tilde{\sigma}(u))^+$ and $\psi \in \ell_\fv(\Tilde{\sigma}(u))^-$. If $\varphi \dimp \psi \in \ell_\fv(x)^+$ and $\Tilde{\sigma}(w) = x$, we have $\varphi \dimp \psi \in \ell_\fw(w)^+$. Since $\fw$ is a $\Sigma$-moment, there exists $u \leq_\fw w$ resolving the defect. There are three cases. First if $u \in r^\uparrow$, then the argument is as for implication defects. Second if $u$ belongs to a $\Sigma$-moment of the form $\fw_{v}$ for $v$ covered by one of the worlds on the zigzag path from $r$ to $w$, then there exists a $\Sigma$-moment $\fv_\alpha$ bimersive to $\fw_v$ below $x$. Note that $\Tilde{\sigma}\upharpoonright_{\fw_v} = \sigma_\alpha$. Since $\sigma_\alpha(v)$ is the root of $\fv_\alpha$ which is covered by $x$, $u \leq_\fw v$ and $\sigma_\alpha$ is order preserving, we have that $\sigma_\alpha(u) \leq_\fv \sigma_\alpha(v) \leq_\fv x$ and $\varphi \in \ell_\fv(\sigma_\alpha(u))^+$ and $\psi \in \ell_\fv(\sigma_\alpha(u))^-$. Third if $u \in r^\downarrow$, by order preservation of $\Tilde{\sigma}$ we have that $\Tilde{\sigma}(u) \leq_\fv x$ and $\varphi \in \ell_\fv(\Tilde{\sigma}(u))^+$ and $\psi \in \ell_\fv(\Tilde{\sigma}(u))^-$. The cases where $x \in r'^\downarrow$ are symmetric and so we conclude that defects in $\fv$ are resolved. Finally, note that in all of the above cases Properties \ref{d: moment 1.1} and \ref{d: moment 1.2} of a $\Sigma$-moment are satisfied. We conclude that $\fv$ is a $\Sigma$-moment.

Finally, each bimersion class of a $\Sigma$-moment $\fw$ with zigzag radius $n$ is determined by the bimersion classes of $\hat{\fw}$ and $\check{\fw}$. Recall that there are at most $G(\lvert C'\rvert)$ bimersion classes of tree-like $\mathcal{C}'$-moments. Since the coloring $\mathcal{C}'$ encodes all information about the corresponding $\mathcal{C}$-moment, there are at most 
\begin{equation*}
    \lambda(n)=G(\lvert C' \rvert)^2 =G(\lvert C \rvert \cdot 2^{\lambda_(n-1)})^2
\end{equation*}
bimersion classes of $\moment_\Sigma^n$. Note that $\kappa$ and $\lambda$ are definable by simultaneous recursion, using  the computable functions $F$ and $G$, implying that they are $\mu$-recursive functions and hence computable.
\end{proof}

\begin{remark}\label{r: bimersions initial worlds}
    Let $\fw=(M, \leq_\fw, \ell_\fw, m,m)$ and $\fv=(N, \leq_\fv, \ell_\fv, n,n)$ be $\Sigma$-moment which are bimersive, witnessed by the bimersion $(\sigma, \tau)$. Note that bimersions are independent of the choice of initial worlds for $\fw$ and $\fv$. Therefore if we denote by $\fw_{m'}$ and $\fv_{n'}$ the moments obtained by changing the initial worlds from $m$ to $m'$ and from $n$ to $n'$, respectively, we have that $\fw \simm \fv$ implies $\fw_{m'} \simm \fv_{n'}$ for all $m' \in M$ and $n' \in N$. In the proof of the next theorem (Theorem \ref{t: restriction to E0 is a model}) we will encounter the case that a succinct moment $\fw$ is bimersive to a moment $\fv$. We will therefore be allowed to assume that the immersion from $\fw$ to $\fv$ maps the initial world of $\fw$ onto the initial world of $\fv$. If this was not the case, we may simply take the moment $\fv$ with initial world changed to whatever world the immersion from $\fw$ to $\fv$ maps the initial world of $\fw$ to. Note that this is unproblematic as long as we only consider one direction of the bimersion. 
\end{remark}
Let $\irr\Sigma$ be the substructure formed by restricting $\moment_\Sigma $ to succinct moments (and with isomorphic moments identified). 
It follows from Proposition \ref{propIrrFin} that $\irr\Sigma$ is finite and only contains finite moments.
Next, it would be convenient if, whenever $E$ is a simulation from $\moment_\Sigma$ to some other labelled poset, and $\fw$ is any $\Sigma$-moment such that $\fw\mathrel E x$, we could replace $\fw$ by some succinct $\fw'\simm \fw$ and still have $\fw'\mathrel E x$. The following operations on simulations will help us achieve this.

\begin{definition}
Let $\Sigma \subseteq \lanfull$ be finite and closed under subformulas, $\mathcal{X} =(X, \leq_\mathcal{X}, \ell_\mathcal{X})$ be a $\Sigma$-labelled poset, and $E\subseteq \moment_\Sigma\times X$.
\begin{enumerate}

\item The \emph{bimersion closure} of $E$ is defined by $\check E\coloneqq E \circ { \simm }$. If $\check E=E$, then $E$ is called \emph{bimersion closed}.

\item The \emph{succinct part} $E_0$ of $E$ is defined by $E_0 \coloneqq E{\upharpoonright_{\irr\Sigma}}$.
\end{enumerate}
\end{definition}

In other words, $\fw \mathrel{\check E} x$ means that there exists $\fv$ such that $\fw \simm \fv \mathrel E x$. The following lemma then establishes the operations described above.

\begin{lemma}\label{l: bimersion closure of simulations}
Let $\Sigma\subseteq\Lbi$ be finite and closed under subformulas, $\mathcal{X}$ be a $\Sigma$-labelled frame, and $E\subseteq \moment_\Sigma\times X$ be a simulation. Then
\begin{enumerate}
\item 
\begin{enumerate}
    \item\label{l: bimersion closure of simulations 1a} 
    $\check E$ is also a simulation;
    \item \label{l: bimersion closure of simulations 1b}
    if $E$ is dynamic, then $\check E$ is dynamic.
\end{enumerate}

\item\label{l: bimersion closure of simulations 2}If $E$ is bimersion closed, then
\begin{enumerate}
\item \label{l: bimersion closure of simulations 2a}
$E_0$ is also a simulation;

\item\label{l: bimersion closure of simulations 2b} 
if $E$ is dynamic, then $ E_0$ is dynamic;

\item\label{l: bimersion closure of simulations 2c}
if $E$ is strongly surjective (see Definition \ref{d: strongly surjective simulation}) then $E_0$ is strongly surjective.
\end{enumerate}
\end{enumerate}
\end{lemma}

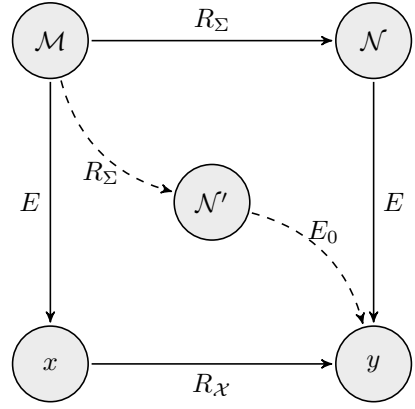
\begin{figure}[t!]
    \centering
    \begin{tikzpicture}[modal]
        \node[world](w0){$x$};
        \node (w1) [above=of w0]{};
        \node[world](w2)[above=of w1]{$\fw$};
        \node (v0) [right=of w0]{};
        \node[world] (v1) [above=of v0]{$\fv'$};
        \node (v2) [above=of v1]{};
        \node[world] (u0) [right=of v0]{$y$};
        \node (u1) [above=of u0]{};
        \node[world] (u2) [above=of u1]{$\fv$};
         \path[->] (w2) edge node[left] {$E$} (w0);
         \path[->] (w0) edge node[below] {$R_\mathcal{X}$} (u0);
         \path[->] (w2)  edge node[above] {$R_\Sigma$} (u2);
         \path[->] (u2) edge node[right] {$E$} (u0);
         \path[->] (w2) edge[dashed, bend right] node[below]{$R_\Sigma$} (v1);
         \path[->] (v1) edge[dashed, bend left] node[above] {$E_0$} (u0);
    \end{tikzpicture}
    \caption{A diagram illustrating Lemma \ref{l: bimersion closure of simulations}.\ref{l: bimersion closure of simulations 2b}.}
    \label{figure:commute}
\end{figure}

\begin{proof}
\ref{l: bimersion closure of simulations 1a}: To check that $\check E$ is a simulation, we check the two defining conditions of a simulation. First suppose there are $(N, \allowbreak {\leq_\mathcal N},\allowbreak \ell_\mathcal N, n)$ and $(M, \leq_\fw, \ell_\fw, m)$ and a bimersion  $(\sigma, \tau)$ between $(M, \allowbreak {\leq_\mathcal M},\allowbreak \ell_\mathcal M,m)$ and $(N, {\leq_\mathcal N}, \ell_\mathcal N,n)$ such that $(N, {\leq_\mathcal N}, \ell_\mathcal N, n, \sigma(m)) \mathrel E x$. We let $\mathcal{M} =(M, {\leq_\mathcal M},\allowbreak \ell_\mathcal M, m, m)$ and $\mathcal{N} =(N, {\leq_\mathcal N}, \ell_\mathcal N, n,\sigma(m))$ be the corresponding $\Sigma$-moments.
Then by the definition of the labelling $\ell_\Sigma$ on $\moment_\Sigma$ and the fact that $\sigma$ is an immersion,
we have $\ell_\Sigma(\mathcal M) = \ell_\mathcal M (m) = \ell_\mathcal N(\sigma(m)) = \ell_\Sigma(\mathcal N)$. Since $E$ is a simulation, we have $\ell_\Sigma(\mathcal N) \subseteq_\type \ell_\mathcal X(x)$. Thus $\ell_\Sigma(\mathcal M) \subseteq_\type \ell_\mathcal X(x)$. 

Now for forth-up confluence let $\mathcal{M}' =(M, {\leq_\mathcal M}, \ell_\mathcal M, m, m')$ be an arbitrary moment with $\mathcal M \leq_\Sigma \mathcal M'$; so $m \leq_\mathcal M m'$. Thus $\sigma(m) \leq_\mathcal N \sigma(m')$, and hence as $E$ is a simulation there exists $x' \ge x$ with $(N, {\leq_\mathcal N}, \ell_\mathcal N, n, \sigma(m')) \mathrel E x'$. Then by the definition of $\check E$, we have $\mathcal M' \mathrel{\check E} x'$. The case for forth-down confluence is similar. \smallskip

\noindent \ref{l: bimersion closure of simulations 1b}:
Let $\mathcal{M} =(M, {\leq_\mathcal M},\allowbreak \ell_\mathcal M, m, m_0)$,  $x$, and $y$ be such that $\mathcal M\mathrel {\check E} x \Rel_\mathcal{X} y$.
Then by the definition of $\check E$, there exists $(M', {\leq_{\mathcal M'}}, \ell_{\mathcal M'}, m')$ and a bimersion  $(\sigma, \tau)$ between $(M, {\leq_{\mathcal M}}, \ell_{\mathcal M},m)$ and $(M', {\leq_{\mathcal M'}}, \ell_{\mathcal M'}, m')$ such that $(M', {\leq_{\mathcal M'}}, \ell_{\mathcal M'}, m', \sigma(m_0)) \mathrel E x$. Let $\mathcal M' = (M', {\leq_{\mathcal M'}}, \ell_{\mathcal M'}, \allowbreak m', \sigma(m_0)) $.
Then as $E$ is dynamic, there exists $\mathcal{N} =(N, {\leq_\mathcal N}, \ell_\mathcal N, n, n')$ such that $\mathcal M'\mathrel R_\Sigma \mathcal N\mathrel {E} y$. Then as $\check E$ includes $E$, we also have $\mathcal N \mathrel {\check E} y$.
Let $(\sigma, \tau)$ be a bimersion from $(N, {\leq_\mathcal N}, \ell_\mathcal N)$ to a succinct $ (N', {\leq_\mathcal N}', \ell'_\mathcal N)$.
Then as $E$ is bimersion closed, we have that ${(N', {\leq'_\mathcal N}, \ell'_\mathcal N, \sigma(n)) \mathrel E y}$, and we define $\mathcal{N}' =(N', {\leq'_\mathcal N}, \ell'_\mathcal N, \sigma(n))$ (see Figure \ref{figure:commute}). By succinctness of $ (N', {\leq_\mathcal N}', \ell'_\mathcal N)$ we have $\mathcal N' \mathrel E_0 y$.
It remains to show that $\mathcal M \mathrel R_\Sigma \mathcal N$.  As $\mathcal M' \mathrel R_\Sigma \mathcal N$, there exists a sensible relation $R \subseteq M' \times N$ such that $(\sigma(m_0), n') \in R$. Then it is easy to check that the composition $ 
R \circ \sigma \subseteq M \times N'$ is a sensible relation. Further, $(m, n') \in R \circ \sigma$, so $R \circ \sigma$ witnesses that $\mathcal M \mathrel R_\Sigma \mathcal N$. \smallskip

\noindent \ref{l: bimersion closure of simulations 2a}: To check that $E_0$ is a simulation, first note that clearly $E_0$ preserves labels, since it is a restriction of $E$. For forth-up confluence suppose $\mathcal M \mathrel E_0 x$ and $\mathcal M \leq_\Sigma \mathcal M'$. Thus $\mathcal M \mathrel E x$, and since $E$ is a simulation, we obtain $x' \ge x$ with $\mathcal M' \mathrel E x'$. From $\mathcal M \mathrel E_0 x$ we conclude that $\mathcal M$ is succinct; thus $\mathcal M'$ is also succinct. Hence $\mathcal M' \mathrel E_0 x'$. The case for forth-down confluence is similar, thus proving that $E_0$ is a simulation.

\noindent \ref{l: bimersion closure of simulations 2b}: Let $\mathcal{M} =(M, {\leq_\mathcal M},\allowbreak \ell_\mathcal M, m, m')$,  $x$, and $y$ be such that $\mathcal M\mathrel E_0 x \Rel_{\mathcal{X}} y$. Then $\mathcal M\mathrel E x \Rel_\mathcal{X} y$, so as $E$ is dynamic, there exists $\mathcal{N} =(N, {\leq_\mathcal N}, \ell_\mathcal N, n, n')$ such that $\mathcal M\mathrel R_\Sigma \mathcal N\mathrel {E} y$. Let $(\sigma, \tau)$ be a bimersion from $(N, {\leq_\mathcal N}, \ell_\mathcal N, n)$ to a succinct $ (N', {\leq_\mathcal N}', \ell'_\mathcal N, \sigma{n})$. Then as $E$ is bimersion closed, we have $(N', {\leq'_\mathcal N}, \ell'_\mathcal N, \sigma(n), \sigma(n')) \mathrel E y$, and we define $\mathcal{N}' =(N', {\leq'_\mathcal N}, \ell'_\mathcal N, \sigma(n), \sigma(n'))$ (see Figure \ref{figure:commute}). By succinctness of $ (N', {\leq_\mathcal N}', \ell'_\mathcal N, \sigma(n))$ we have $\mathcal N' \mathrel E_0 y$. It remains to show that $\mathcal M \mathrel R_\Sigma \mathcal N'$. As $\mathcal M \mathrel R_\Sigma \mathcal N$, there exists a sensible relation $R \subseteq M \times N$ such that $(m', n') \in R$. Then it is easy to check that the composition $\sigma \circ R \subseteq M \times N'$ is a sensible relation. Further, $(m', \sigma(n')) \in \sigma \circ R$, so $ \sigma \circ R$ witnesses that $\mathcal M \mathrel R_\Sigma \mathcal N'$. \smallskip

\noindent \ref{l: bimersion closure of simulations 2c}: To check that $E_0$ is strongly surjective, let $x \in X$. Then as $E$ is strongly surjective, there exists $\mathcal M = (M, {\leq_\mathcal M},\allowbreak \ell_\mathcal M, m, m')$  such that $\mathcal M \mathrel E x$ and $\ell_\Sigma(\mathcal M) = \ell_\mathcal X (x)$.  We can find a succinct $(N, {\leq_\mathcal N}, \ell_\mathcal N, n)$  and a bimersion $(\sigma, \tau)$ between $(M, {\leq_\mathcal M}, \ell_\mathcal M, m)$ and $(N, {\leq_\mathcal N}, \ell_\mathcal N, n)$. Then $(N, {\leq_\mathcal N}, \ell_\mathcal N, n, \sigma(m')) \mathrel {\check E} x$, so as $E$ is bimersion closed $(N, {\leq_\mathcal N}, \ell_\mathcal N, n, \sigma(m')) \mathrel { E_0} x$, and as $\sigma$ is an immersion, $\ell_\Sigma((N, {\leq_\mathcal N}, \ell_\mathcal N, n, \sigma(m')) ) = \ell_\mathcal M(m') = \ell_\mathcal X(x)$. 
Thus $x \in E_0(\irr\Sigma)$, and we conclude that $E_0(\irr\Sigma)\supseteq E(\moment_\Sigma)$. The reverse inclusion is obvious.
\end{proof}

\begin{tikzpicture}
\end{tikzpicture}

Fix a discrete unravelled $\Sigma$-labelled model $\mathcal{Q}=(Q, \leq_\mathcal{Q}, \ell_\mathcal{Q}, R_\mathcal{Q})$. Let $E$ be the strongly surjective dynamic simulation between $\moment_\Sigma$ and $\mathcal{Q}$ constructed in Section \ref{s: constructing surjective dynamic simulations}. Let $\check{E}$ be the bimersion closure of $E$ and let $\check{E}_0$ be the succinct part of $\check{E}$. From the previous lemma we obtain the following corollary

\begin{corollary}\label{c: E0 is dynamic simulation}
    $\check{E}_0 \subseteq \irr\Sigma \times Q$ is a strongly surjective dynamic simulation.
\end{corollary}

\begin{theorem}\label{t: restriction to E0 is a model} $\irr\Sigma{\upharpoonright_{dom(\check{E}_0)}}$ is a $\Sigma$-labelled model.
\end{theorem}
\begin{proof}
 Denote $\irr\Sigma{\upharpoonright_{dom(\check{E}_0)}}$ by $\mathcal{X}=(X, \leq_\mathcal{X}, \ell_\mathcal{X}, R_\mathcal{X})$ and $\moment_\Sigma{\upharpoonright_{dom(E)}}$ by $\mathcal{Z}=(Z, \leq_\mathcal{Z}, \ell_\mathcal{Z}, R_\mathcal{Z})$. That $\mathcal{X}$ is a $\Sigma$-labelled frame is proven as for $\mathcal{Z}$, see the proof of Theorem \ref{t: surjective simulation gives model}. In particular note that the proof only makes use of the fact that $E$ is a simulation. Since $\check{E}_0$ is a simulation as well, the same proof suffices. It remains to show that $R_\mathcal{X}$ is witnessed. Suppose $\fw_s \in X$ and $\ld \varphi \in \ell_\mathcal{X}(\fw_s)^+$. By definition $\fw_s \in dom(\check{E}_0)$ implying that there exists $\rho \in Q$ with $\fw_s \mathrel{\check{E}_0} \rho$. This implies that there exists $\fw \in Z$ with $\fw_s \simm \fw$ and $\fw \mathrel{E} \rho$. Let $(\sigma, \tau)$ be the bimersion between $\fw_s$ and $\fw$. By Remark \ref{r: bimersions initial worlds} we assume that the immersion $\sigma$ maps the initial world of $\fw_s$ onto the initial world of $\fw$ and so that $\ld \varphi \in \ell_\mathcal{Z}(\fw)^+$. By Theorem \ref{t: surjective simulation gives model}, $\mathcal{Z}$ is a $\Sigma$-labelled model, implying that $R_\mathcal{Z}$ is witnessed. Therefore there exists $\fv \in Z$ with $\fw \Rel_\mathcal{Z} \fv$ and $\varphi \in \ell_\mathcal{Z}(\fv)^+$. Since every moment is bimersive to a succinct moment, there exists $\fv_s \in X$ with $\fv_s \simm \fv$. Let $(\sigma', \tau')$ be the bimersion between $\fv_s$ and $\fv$. Once again, we assume that the immersion $\tau'$ maps the initial world of $\fv$ onto the initial world of $\fv_s$. We claim that $\fw_s \mathrel{R_\mathcal{X}} \fv_s$. Recall that $R_\mathcal{X} = R_\Sigma{\upharpoonright_{dom(\check{E}_0)}}$. Therefore it suffices to show that there exists a sensible relation (relative to some fixed sensibility condition $S$) between $\fw_s$ and $\fv_s$. Let $R$ be the sensible relation between $\fw$ and $\fv$ witnessing that $\fw \Rel_\mathcal{Z} \fv$. Then define $R_s$ as follows, where we write $\fw_s =(M_s, \leq_{\fw_s}, \ell_{\fw_s}, w^g_s, w^i_s)$, $\fw =(M, \leq_{\fw}, \ell_{\fw}, w^g, w^i)$ and similarly, $\fv_s =(N_s, \leq_{\fv_s}, \ell_{\fv_s}, v^g_s, v^i_s)$ and $\fv =(N, \leq_{\fv}, \ell_{\fv}, v^g, v^i)$. For any $w \in M_s$ and any $v \in N_s$,
 \begin{center}
     $w \Rel_s v$ if and only if there exists $v' \in N$ such that $\sigma(w) \Rel v'$ and $\tau'(v') = v$.
 \end{center}
We claim that $R_s$ is a sensible relation between $\fw_s$ and $\fv_s$. First of all, by the previous observation we have that $\sigma(w_s^i) = w^i$, $w^i \Rel v^i$ by definition of a sensible relation and $\tau'(v^i) = v_s^i$. Hence $w_s^i \Rel_s v_s^i$. Next, suppose $w \in M_s$ and $v \in N_s$ and $w \Rel_s v$. Let $S$ be the sensibility condition and let $v' \in N$ such that $\sigma(w) \Rel v'$ and $\tau'(v') = v$. By definition of a bimersion,
\begin{equation*}
    \ell_{\fw_s}(w) = \ell_\fw(\sigma(w)) \mathrel{S} \ell_\fv(v') = \ell_{\fv_s}(\tau'(v')) = \ell_{\fv_s}(v).
\end{equation*}
Therefore $\ell_{\fw_s}(w) \mathrel{S} \ell_{\fv_s}(v))$ and it remains to show that $R_s$ is forth-up and forth-down confluent. For forth-up confluence suppose $w,w' \in M_s$, $w \leq_{\fw_s} w'$ and $v \in N_s$ with $w \Rel_s v$. Let $u \in N$ with $\sigma(w) \Rel u$ and $\tau'(u) = v$. By order-preservation of $\sigma$, $\sigma(w) \leq_\fw \sigma(w')$. By forth-up confluence of $R$ there exists $u' \in N$ with $u \leq_\fv u'$ and $w' \Rel u'$. By order-preservation of $\tau'$ we have $v = \tau'(u) \leq_{\fv_s} \tau'(u')$. By definition of $R_s$ we have $w' \Rel_s \tau'(u')$. For forth-down confluence suppose $w \leq_{\fw_s} w'$ and $w' \Rel_s v'$. Then there exists $u' \in N$ with $\sigma(w') \Rel u'$ and $\tau'(u')=v'$. By order-preservation of $\sigma$, $\sigma(w) \leq_\fw \sigma(w')$. By forth-down confluence of $R$ there exists $u \leq_\fv u'$ with $\sigma(w) \Rel u$. Hence $\tau'(u) \leq v'$ and $w \Rel_s \tau'(u)$. We conclude that $\fv_s$ is a modal successor of $\fw_s$ and hence $\fw_s \Rel_\mathcal{X} \fv_s$. Finally, since $\varphi \in \ell_\mathcal{Z}(\fv)^+$ and $\tau'$ maps the initial world of $\fv$ onto the initial world of $\fv_s$, we have $\varphi \in \ell_\mathcal{X}(\fv_s)^+$. The case for $\lb \varphi \in \ell_\mathcal{X}(\fw_s)^-$ is similar. We conclude that $\mathcal{X}$ is witnessed and hence a $\Sigma$-labelled model. \qedhere
\end{proof}

Finally, we are able to prove the main result.

\begin{theorem}\index{finite model property!bi-intuitionistic modal logic}
Each  $\mathbf{Q}\in \{\mathbf{biK},\mathbf{biKD},\allowbreak\mathbf{biKT},\mathbf{biK4},\mathbf{biKD4},\mathbf{biS4}\}$ has the finite model property, with a computable bound on the cardinality of the model, and hence is decidable.
\end{theorem}

\begin{proof}
Let $\mathbf{Q} \in\{\mathbf{biK},\mathbf{biKD},\allowbreak\mathbf{biKT},\mathbf{biK4},\mathbf{biKD4},\mathbf{biS4}\}$. If $\varphi$ is valid over the class of $\mathbf{Q}$-models, then $\varphi$ is clearly also valid over the class of finite $\mathbf{Q}$-models. For the other direction we show the contrapositive. Suppose $\varphi$ is not valid over the class of $\mathbf{Q}$-models. Then there exists a discrete $\mathbf{Q}$-model $\fw=(W, \leq, R, V)$ and a world $w \in W$ with $\fw, w \not \models \varphi$. Let $\Sigma$ be the least set of formulas containing $\varphi$ and closed under subformulas and consider the $\Sigma$-labelled model $\fw_i = (W, \leq, R, \ell)$ obtained from $\fw$ by labelling every world with the formulas in $\Sigma$ that are true/false. Hence $\varphi \in \ell(w)^-$. Let $\fw_i^* =(W^*, \leq_*, R_*, \ell_*)$ be the unravelling of $\fw_i$. By Lemma \ref{l: unravelled model is a model}, $\fw_i^*$ is a $\Sigma$-labelled model. Furthermore, by Proposition \ref{p: formula falsified on model iff on unravelled model}, $\varphi$ is falsified on $\fw_i^*$. Let $\rho$ be the world of $\fw_i^*$ with $\varphi \in \ell_*(\rho)^-$. Let $E$ be the strongly surjective dynamic simulation between $\moment_\Sigma$ and $\fw_i^*$ constructed in Section \ref{s: constructing surjective dynamic simulations}. By Corollary \ref{c: E0 is dynamic simulation} the relation $\check{E}_0$ is a simulation between $\irr\Sigma$ and $\fw_i^*$. Let $\moment_\Sigma{\upharpoonright_{dom(E)}} = \mathcal{Z}=(Z, \leq_\mathcal{Z}, \ell_\mathcal{Z}, R_\mathcal{Z})$ and $\irr\Sigma{\upharpoonright_{dom(\check{E}_0)}}=\mathcal{X}=(X, \leq_\mathcal{X}, \ell_\mathcal{X}, R_\mathcal{X})$. By Theorem \ref{t: surjective simulation gives model} and Theorem \ref{t: restriction to E0 is a model}, $\mathcal{X}$ and $\mathcal{Z}$ are $\Sigma$-labelled models. Since $E$ is strongly surjective, there exists $z \in Z$ with $z \mathrel{E} \rho$ and $\ell_\mathcal{Z}(z) = \ell_*(\rho)$.  Hence $\varphi \in \ell_{\mathcal{Z}}(z)^-$. By definition of $E_0$, there exists $x \in X$ with $x \simm z$ such that $\varphi \in \ell_\mathcal{X}(x)^-$, implying that $\varphi$ is falsified on $\irr\Sigma{\upharpoonright_{dom(\check{E}_0)}}$. Finally, since $\irr\Sigma$ is finite and its size bounded by a computable function, so is the size of $\irr\Sigma {\upharpoonright}_{dom(\check{E}_0)}$. By taking the corresponding expanding model (see Lemma \ref{l: from labelled model to model}), we conclude that $\varphi$ is falsified over a finite model with a computable bound on its size. 
\end{proof}

\section{Conclusion}

We have introduced six bi-intuitionistic modal logics which are evaluated over expanding models satisfying various frame conditions. The main contributions of this chapter are listed below.

\begin{enumerate}
    \item We have presented sound and complete axiomatizations for all six logics.

    \item We have given a uniform proof of the finite model property for all six logics, including a computable bound on the size of the finite models, thus obtaining decidability.
\end{enumerate}

We did not provide the soundness and completeness proofs for the axiomatizations, since each of them follows by a standard canonical model construction and combines ideas from canonical model constructions for bi-intuitionistic logic (see e.g.~\cite{gore_2020}) and of modal logic. However, the interested reader may consult the joint publication with Fern\'andez-Duque and McLean~\cite{fernandez-duque_family_2023} to see details about the completeness proofs. The main contribution consists of the uniform proof of the finite model property for all six logics. The basic idea to pass to labelled structures is not new and has been successfully applied to intuitionistic linear temporal logics without the co-implication~\cite{fernandez-duque_2018, Boudou_2017, Balbiani_2019} to obtain the finite model property. Nevertheless, the addition of the co-implication has increased the complexity of the overall argument, as we have to consider zigzag paths to evaluate formulas correctly, leading to the definition of a moment. 

Our work leaves several questions open, which we hope to solve in the future. First, our decidability proof provides an upper bound on the size of the finite models and thus also on the complexity of checking validity. However, the bound is rather large: The functions $\kappa$ and $\lambda$ depend on $F$ and $G$ which are \emph{towers of exponentials} whose length is bounded by the height of the poset of colors (and thus by the cardinality of the poset of colors). Since for $\Sigma$-moments the coloring is $(\type_\Sigma, \leq_\Sigma)$ and the cardinality of $\type_\Sigma$ depends on $\Sigma$, it follows that no tower of exponentials can bind the size of the finite models obtained from our proof, implying that the complexity bound is \emph{non-elementary}. Whether a better complexity bound is possible remains an open question.

\begin{question}
    What is the optimal complexity bound for the validity problem for the considered bi-intuitionistic modal logics?
\end{question}

Second, the methods employed in this chapter are model-theoretic. While we do provide an axiomatization, we have not yet addressed the proof theory (in terms of sequent calculi) for bi-intuitionistic modal logic. Due to the lack of fixed point operators, we expect that sequent calculi admitting standard finite proofs suffice, however in order to deal with the co-implication and the confluence conditions, it is likely that nested or labelled sequents will be needed. A possible avenue to tackle this problem is by extending known nested calculi for bi-intuitionistic logic (e.g. the calculus presented in~\cite{gore_2008}) by rules for modalities. A similar approach was taken in~\cite{gore_2010} which presents a nested calculus for bi-intuitionistic tense logic. Due to the presence of the confluence conditions and the resulting interaction between $\lb$ and $\ld$, we expect that constructing sequent calculi could be further complicated.

\begin{question}\label{q: proof theory for biml}
    Do bi-intuitionistic modal logics admit a `nice' proof theory?
\end{question}

Third, we are interested in studying other versions of bi-intuitionistic modal logics. For example, extending intuitionistic common knowledge logic $\ICK$ with the co-implication connective results in an epistemic logic with two notions of knowledge: explicit knowledge formalized by the knowledge operator (which corresponds to $\lb$) and implicit knowledge formalized by the intuitionistic order and thus the current information state in which the agent is in. The implication and co-implication then assert statements about gaining or losing information by changing the information state, while the knowledge operator asserts statements about the worlds accessible over the modal accessibility relation. Thus such a logic could reason about an agents knowledge in situations where information can be gained or lost. This leads to the natural question whether the implicit notion of knowledge can be made explicit. 

\begin{question}
    What is the logic obtained by extending $\ICK$ by co-implication? Can the implicit notion of knowledge formalized by the intuitionistic order be made explicit?
\end{question}

\chapter{Bi-Intuitionistic Linear Temporal Logic}\label{c: biLTL}

\section{Introduction}

The final contribution of this thesis is to present a sound and complete axiomatization for the language of intuitionistic linear temporal logic with the temporal operators `next', `eventually' and `henceforth' evaluated over (total functional) expanding models. Recall the axiomatization $\mathrm{iLTL_H}$ for $\iLTL$ without `henceforth' depicted in the introduction of Chapter \ref{c: iLTL}, which was originally presented and proven sound and complete in~\cite{Boudou_2022}. It was shown in~\cite{Boudou_2022} that $\mathrm{iLTL_H}$ is sound and complete with respect to topological semantics and with respect to expanding models, showing that the language without `henceforth' cannot distinguish between these two classes. The problem of axiomatizing the full language including `henceforth' has remained open for several years. A natural guess for obtaining a sound and complete system for $\iLTL$ with henceforth would be to extend $\iLTL_\mathrm{H}$ with a standard fixed point axiom for henceforth as well as rules for monotonicity and induction, i.e. extend $\iLTL_\mathrm{H}$ with the following axioms and rules.
\begin{center}
    \begin{tabular}{l l }
    $\mathsf{Fix}_{\G}$ & $\G \varphi \rightarrow (\varphi \wedge {\X} {\G} \varphi)$ \\
    $\mathsf{Mon}_{\G}$ & $\infer{\G\varphi\to \G\psi }{ \varphi\to\psi}$ \\
    $\mathsf{Ind}_{\G}$ & $\infer{ \varphi\to 
                  \G \varphi }{ \varphi\to\X \varphi}$  \\
    \end{tabular}
\end{center}

However, it was shown in~\cite{Boudou_2021} that the axioms and rules above are valid and validity preserving both over topological dynamic and expanding models. In the presence of `henceforth', the sets of validities over these classes of models are not identical, as opposed to the language without `henceforth'. This can be seen, for example, from the \emph{Rodr\'iguez--Vidal formula} $\mathbf{RV} \coloneqq \G (p \vee q) \rightarrow (\G p \vee \E q)$, expressing that if $p \vee q$ is true henceforth, then $q$ is true eventually or henceforth $p$ is true. Clearly, this formula is valid over expanding models: if $\fw, f^n(w) \models p \vee q$ for all $n < \omega$, then either there is $n < \omega$ such that $\fw, f^n(w) \models q$ or otherwise $\fw, f^n(w) \models p$ for all $n < \omega$. Over topological models, the formula $\mathbf{RV}$ is \emph{not} valid, as shown in the following example.

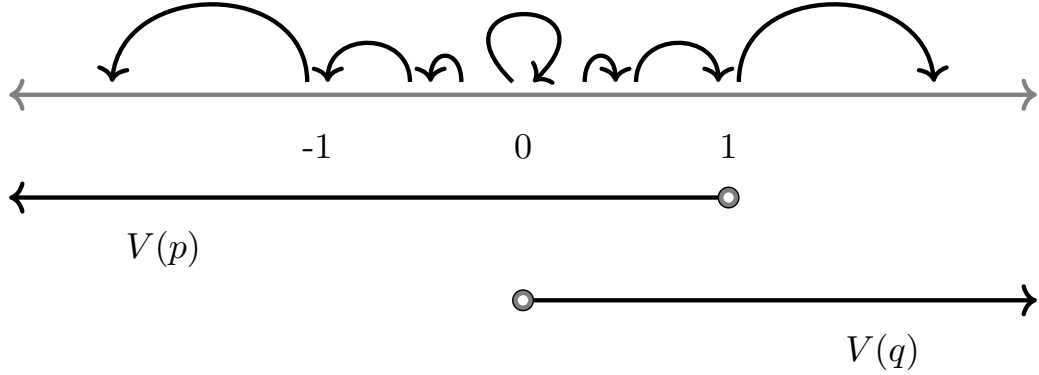
\begin{figure}

\def\arhgt{1/8}
\def\dx{.1}

\resizebox{\columnwidth}{!}{
\begin{tikzpicture}

\draw[<->,very thick, gray] (-5,0) -- (5,0);

\draw[->,very thick] (0-\dx, \arhgt) .. controls (-1,1) and (1,1) ..  (0+\dx,\arhgt);

\draw[->,very thick] (1/2+\dx,\arhgt) .. controls (1/2+\dx,\arhgt+1/3) and (1-\dx,\arhgt+1/3) ..  (1-\dx,\arhgt);

\draw[->,very thick] (1+\dx,\arhgt) .. controls (1+\dx,\arhgt+1/2) and (2-\dx,\arhgt+1/2) .. (2-\dx,\arhgt);

\draw[->,very thick] (2+\dx,\arhgt) .. controls (2+\dx,\arhgt+1) and (4-\dx,\arhgt+1) .. (4,\arhgt);

\draw[->,very thick] (-1/2-\dx,\arhgt) .. controls (-1/2-\dx,\arhgt+1/3) and (-1+\dx,\arhgt+1/3) ..  (-1+\dx,\arhgt);

\draw[->,very thick] (-1-\dx,\arhgt) .. controls (-1-\dx,\arhgt+1/2) and (-2+\dx,\arhgt+1/2) .. (-2+\dx,\arhgt);

\draw[->,very thick] (-2-\dx,\arhgt) .. controls (-2-\dx,\arhgt+1) and (-4+\dx,\arhgt+1) .. (-4,\arhgt);

\draw (0,-1/2) node {0};

\draw (2,-1/2) node {1};

\draw (-2,-1/2) node {-1};

\def\ydist{1}

\draw[<-,very thick] (-5,-\ydist) -- (2,-\ydist);

\draw (-3.5,-\ydist-1/2) node {$V(p)$};

\draw[fill=gray] (2,-\ydist) circle (.1);

\fill[white] (2,-\ydist) circle (.05);

\draw[->,very thick] (0,-2*\ydist) -- (5,-2*\ydist);

\draw (3.5,-1/2-2*\ydist) node {$V(q)$};

\draw[fill=gray] (0,-2*\ydist) circle (.1);

\fill[white] (0,-2*\ydist) circle (.05);

\end{tikzpicture}
}
\caption{A model based on the real line.}
\label{figLine}
\end{figure}

\begin{example}\label{ex: RV not valid over topological spaces}
Recall that a topological dynamic system is a tuple $(X, \tau, f)$ where $(X, \tau)$ is a topological space and $f: X \longrightarrow X$ a continuous function. A topological model is a topological dynamic system equipped with a valuation assigning to each proposition an open set. For details about the evaluation of formulas on a topological model, see the introduction to Chapter \ref{c: iLTL}. For the example here, it suffices to remember that
\begin{center}
\begin{tabular}{l l l}
 $\llbracket {\E \varphi} \rrbracket$ & $\coloneqq$ & $\bigcup_{n < \omega} f^{-n}(\llbracket \varphi \rrbracket)$ \\
           $\llbracket {\G \varphi} \rrbracket$ & $\coloneqq$ & $(\bigcap_{n < \omega} f^{-n}(\llbracket \varphi \rrbracket))^\circ$ \\
\end{tabular}
\end{center}

where $A^\circ$ denoted the interior of $A \subseteq X$. A topological model falsifying $\mathbf{RV}$ is obtained by considering the reals $\mathbb{R}$ with the standard Euclidean topology. Thus open sets are (unions) of open intervals $(x- \epsilon, x + \epsilon)$ for some $\epsilon > 0$. We obtain a topological model by adding the continuous function $f$ on $\mathbb{R}$ with $f(x) = 2x$ and defining the valuation $V$ by letting $V(p) = (- \infty, 1)$ and $V(q) = (0, \infty)$ (see Figure \ref{figLine}).

By construction $ \llbracket {p\vee q} \rrbracket = \mathbb{R}$, so $\llbracket {\G (p \vee q)} \rrbracket = \mathbb{R}$ as well. We claim that $0 \not \in \llbracket \mathbf{RV} \rrbracket$ and hence that $\llbracket \mathbf{RV} \rrbracket \not = \mathbb{R}$. It suffices to show that $0 \not \in \llbracket {\G p \vee \E q} \rrbracket$. It is clear that $ 0 \not \in \llbracket{\E q} \rrbracket$ simply because $f^n(0) = 0 \not \in V(q)$ for all $n$. Now since $\llbracket {\G p} \rrbracket$ is open,  if $0 \in \llbracket {\G p}\rrbracket$, then there exists an $\epsilon > 0$ such that $(- \epsilon, \epsilon) \subseteq \llbracket {\G p} \rrbracket$. But for every $x > 0$ there exists $n < \omega$ with $f^n(x) > 1$, which implies that $f^n(x) \not \in\llbracket {p} \rrbracket$ and so that $x \not \in \llbracket {\G p }\rrbracket$. Thus such an interval can not exist, implying that $0 \not \in \llbracket {\G p} \rrbracket$ and so that $\mathbf{RV}$ is not valid over the class of topological models.
\end{example}

This implies that the proposed axiomatization above cannot be complete for expanding models, as otherwise there would exist a derivation of $\mathbf{RV}$ and since the axiomatization is sound with respect to topological models, $\mathbf{RV}$ would be valid over topological models. Instead of extending $\mathrm{IM_H}$ with different axioms, we present an alternative and perhaps surprising solution. When extending the language of $\iLTL$ with the co-implication connective $\dimp$, the natural axiomatization becomes `magically' complete. This is because the presence of $\dimp$ and temporal modalities creates implicit interaction, which can be used to derive formulas such as $\mathbf{RV}$. Thus the derivation of $\mathbf{RV}$ (see Example \ref{ex: RV is derivable}) features formulas with co-implications, even though $\mathbf{RV}$ itself belongs to the $\dimp$-free fragment. Thus, our axiomatization is not conservative over the $\dimp$-free fragment. The resulting logic is called \emph{bi-intuitionistic linear temporal logic} $\biltl$, which contains $\iLTL$ as a fragment.

The next two sections introduce the syntax and semantics of $\biltl$, as well as the axiomatization $\mathrm{biLTL_H}$. We will then establish soundness as well as completeness for the axiomatization obtained from $\mathrm{biLTL_H}$ restricted to the language without $\G$ and $\E$. The final sections then establish completeness for $\mathrm{biLTL_H}$.

\section{Syntax and Semantics}

The \emph{language} of bi-intuitionistic temporal logic $\Lbiltl$ extends $\LIPL$ by the temporal operators $\X$, $\E$ and $\G$, as well as by the binary connective $\dimp$. The operators $\E$ and $\G$ are fixed point operators. $\E \varphi$ is characterized as the least fixed point of the propositional function $x \mapsto \varphi \vee \X x$.\index{fixed point operator!eventually} $\G \varphi$ is characterized as the greatest fixed point of the propositional function $x \mapsto \varphi \wedge \X x$.\index{fixed point operator!henceforth} \emph{Formulas} are given by the following grammar in Backus--Naur form (where $p \in \Prop$):
\begin{equation*}
        \varphi \coloneqq  \bot \mid p \mid \varphi \wedge \varphi \mid \varphi \vee \varphi \mid \varphi \rightarrow \varphi \mid \varphi \dimp \varphi \mid \X \varphi \mid \E \varphi \mid \G \varphi
\end{equation*}
As before, the connective $\dimp$ is called \emph{co-implication} and the temporal operators $\X$, $\E$ and $\G$ are read as `next', `eventually' and `henceforth', respectively.  The $\E$- and $\G$-free fragment of $\Lbiltl$ is denoted by $\Lbiltlnext$. As for bi-intuitionistic modal logic in the previous chapter, the constant $\bot$ can be defined in terms of $\dimp$ by setting $\bot \coloneqq p \dimp p$ for $p \in \Prop$ and we will treat $\bot$ as a defined constant henceforth. Moreover, we also once again define weak negation $\sim$ by ${\sim} \varphi \coloneqq \top \dimp \varphi$. Formulas of $\Lbiltl$ are evaluated over (total functional) expanding models as used for $\iLTL$ in Chapter \ref{c: iLTL}. Let us briefly recall the definition.

\begin{definition}
    An \emph{expanding model} is a total functional dynamic model ${\mathcal M=(M,\allowbreak {\leq},\allowbreak f,\allowbreak V)}$, where $f$ is forward-confluent.
\end{definition}

Recall that $f$ being forward confluent is equivalent to $f$ being order-preserving. 

\begin{definition}
Let ${\fw=(W, \leq, f, V)}$ be an expanding model. The \emph{truth relation} $\models$ between worlds of $\fw$ and formulas is defined by extending Definition \ref{d: truth relation for intuitionistic Kripke models} with the following clauses, where $w \in W$.
\begin{center}
    \begin{tabular}{l l l}
   $\mathcal M,w \models \varphi \dimp \psi$ & iff & $\exists v \leq w$ : $\mathcal M,v \models \varphi$ and $\mathcal M,v \not \models \psi$\\
   $\mathcal M,w \models \E \varphi$ & iff &  there exists $ n < \omega$ : $\mathcal M,f^n(w) \models \varphi$\\
   $\mathcal M,w \models \G \varphi$ & iff & for all $ n < \omega$ : $\mathcal M,f^n(w) \models \varphi$\\
\end{tabular}
\end{center}
\end{definition}
A formula $\varphi$ is \emph{satisfiable} over the class of expanding models if there exists an expanding model $\fw$ and a world $w$ with $\fw, w \models \varphi$ and \emph{unsatisfiable} otherwise. Moreover, $\varphi$ is \emph{valid} if $\fw, w \models \varphi$ for all expanding models $\fw$ and all worlds $w$ and \emph{falsifiable} otherwise.  

\begin{lemma}[Monotonicity]
    Let $\mathcal M=(W, {\leq}, f,V)$ be an expanding model, $w \in W$, and $\varphi$ be a formula. If $\mathcal M,w \models \varphi$ and $w \leq v$, then $\mathcal M,v \models \varphi$.
\end{lemma}
\begin{proof}
    By induction on the structure of $\varphi$. The base cases as well as the cases for $\varphi = \psi \ast \gamma$ for $\ast \in \{\wedge, \vee, \rightarrow\}$ are covered in Lemma \ref{l: monotonicity intuitionistic Kripke models}. The case for $\varphi = \psi \dimp \gamma$ is covered in Lemma \ref{l: monotonicity for biml} and the cases for $\varphi = \X \psi$ and $\varphi = \E \psi$ are covered in Lemma \ref{l: monotonicity for expanding/persistent models} (recall that $\E$ is definable in terms of $\U$). Suppose $\varphi = \G \psi$ and $\fw, w \models \G \psi$. Then for all $n < \omega$, $\fw, f^n(w) \models \psi$. Since $w \leq v$, Lemma \ref{l: iteration of forward confluent function is forward confluent} implies that for all $n < \omega$, $f^n(w) \leq f^n(v)$. By induction hypothesis $\fw, f^n(v) \models \psi$ for all $n < \omega$, hence $\fw, v \models \G \psi$.
\end{proof}

\begin{definition}
    The set of valid $\Lbiltl$-formulas over the class of expanding models is denoted by $\mathbf{biLTL}$.
\end{definition}

\section{Axiomatization}\label{sec:axioms}

This section introduces the axiomatization $\biltlH$ that captures $\Lbiltl$-validities over expanding models. As mentioned in the introduction, to obtain a sound and complete system in the presence of co-implication it suffices to expand $\mathrm{iLTL_H}$ by standard rules for $\G$. This implies that the system $\biltlH$ does not contain any axioms or rules for the interaction between co-implication and the temporal modalities. 

\begin{definition}\index{axiomatization!$\biltlH$}
    The Hilbert-style axiomatization $\biltlH$ consists of the axiom schemes and rules depicted in Table \ref{tab:axioms and rules of biltl}.
\end{definition}

\begin{table}[t]
    \centering
    \begin{tabular}{|l l|}
    \hline
       $\mathsf{biPC}$ & all bi-intuitionistic tautologies\\
       $\mathsf{D}$ &  ${\neg} {\X} \bot$\\
       $\mathsf{Dist}$ & $\X (\varphi \vee \psi) \to (\X \varphi \vee \X \psi)$\\
       $\mathsf{K}$ & $\X (\varphi \rightarrow \psi) \rightarrow (\X \varphi \rightarrow \X \psi)$\\
       $\mathsf{Fix}_{\E}$ & $(\varphi\vee\X\E \varphi)\to\E \varphi$ \\
              $\mathsf{Fix}_{\G}$ & $\G\varphi\to (\varphi\wedge{\X}{\G}\varphi)$ \\
              \hline
    \end{tabular}
     \hspace{0.5cm}
\begin{tabular}{|l l l l|}
\hline
        $\mathsf{MP}$ & $\infer{\psi}{\varphi & \varphi \rightarrow \psi}$ &  &\\
        $\mathsf{Nec}$ & $\infer{\X \varphi}{\varphi}$ &  $\mathsf{DN}$ & $\infer{\neg {\sim} \varphi}{\varphi }$ \\
        $\mathsf{Mon}_{\E}$ & $\infer{\E\varphi\to \E\psi }{ \varphi\to\psi}$ & $\mathsf{Mon}_{\G}$ & $\infer{\G\varphi\to \G\psi }{ \varphi\to\psi}$  \\
         $\mathsf{Ind}_{\E}$ & $\infer{\E\varphi\to \varphi }{\X\varphi\to\varphi}$ & 
                  $\mathsf{Ind}_{\G}$ & $\infer{ \varphi\to 
                  \G \varphi }{ \varphi\to\X \varphi}$  \\
                  \hline
    \end{tabular}
\caption{Left: the axioms and right: the rules of $\biltlH$}
    \label{tab:axioms and rules of biltl}
\end{table}

 We also define the Hilbert-style axiomatization $\biltlHX$ which is obtained from $\biltlH$ by restricting the language (and therefore the axioms and rules) to $\Lbiltlnext$. For example, a formula $\varphi$ is an instance of an axiom scheme of $\biltlHX$ if $\varphi \in \Lbiltlnext$ and is an instance of an axiom scheme of $\biltlH$.

 \begin{definition}\index{axiomatization!$\biltlHX$}
     The Hilbert-style axiomatization $\biltlHX$ consists of the axioms $\mathsf{biPC}$, $\mathsf{D}$, $\mathsf{Dist}$, $\mathsf{K}$ and the rules $\mathsf{MP}$, $\mathsf{Nec}$ and $\mathsf{DN}$.
 \end{definition}
 
\begin{definition}\index{derivation!$\biltlH$-derivation}
   Let $\Gamma \cup \{\varphi\}$ be a set of $\Lbiltl$-formulas or of $\Lbiltlnext$-formulas, respectively. A \emph{derivation with assumptions in $\Gamma$} of $\varphi$ in $\biltlH$ or $\biltlHX$ is a finite tree $\pi$ labelled by $\Lbiltl$-formulas or $\Lbiltlnext$-formulas, respectively, and according to the rules of $\biltlH$, such that 
   \begin{enumerate}
       \item every leaf of $\pi$ is labelled by an axiom or by a formula in $\Gamma$, and
       \item if a node $u$ is labelled by the conclusion of a rule instance of a rule different than $\mathsf{MP}$, then every leaf of the subtree of $\pi$ rooted at $u$ is labelled by an axiom.
   \end{enumerate}
\end{definition}

We write $\Gamma \vdash \varphi$ if $\varphi$ has a derivation with assumptions in $\Gamma$ and $\vdash \varphi$ if $\Gamma = \emptyset$. Furthermore, we write $\Gamma \vdash \Delta$ where $\Delta$ is any set of formulas if there exists a finite subset $\Delta_0 \subseteq \Delta$ with $\Gamma \vdash \bigvee \Delta_0$. Note that every $\biltlHX$-derivation is a $\biltlH$-derivation. Similarly, every $\biltlH$-derivation in which only formulas of $\Lbiltlnext$ occur is a $\biltlHX$-derivation. Thus we will not display in the notation $\Gamma \vdash \varphi$ whether derivability is in $\biltlH$ or $\biltlHX$, however, it will always be clear from context. The Deduction Theorem for $\biltlH$ and $\biltlHX$ is proven as Theorem \ref{t: deduction theorem for IM}; the proof is omitted.

\begin{theorem}[Deduction Theorem]
   Let $\Gamma\cup \{\varphi, \psi\}$ be a set of $\Lbiltl$- or $\Lbiltlnext$-formulas. Then $\Gamma, \varphi \vdash \psi$ if and only if $\Gamma \vdash \varphi \rightarrow \psi$.
\end{theorem}

The following lemma is proven in~\cite[Proposition 7.2]{gore_2020}.

\begin{lemma} \label{Lemma exclusion property}
For arbitrary formulas in $\Lbiltl$ the following hold.
\begin{enumerate}

\item $
 \vdash \varphi \rightarrow ( \psi \vee \chi) \text{ if and only if }    {\vdash  (\varphi \dimp \psi) \rightarrow \chi }
$.

\item\label{itDimpMon} If $\vdash \varphi\to \varphi'$ and $\vdash \psi'\to\psi$, then $\vdash (\varphi\dimp\psi) \to(\varphi'\dimp \psi')$.
\end{enumerate}
\end{lemma}

We will now illustrate how (implicit) interaction between co-implication and temporal modalities arise in proofs of our calculus.

\begin{example}\label{ex: dual version of K is derivable}

One might expect that in the presence of co-implication a dual version of the standard $\mathsf{K}$-axiom, namely $(\X \varphi \dimp \X \psi) \rightarrow \X (\varphi \dimp \psi)$ is required in the axiomatization. However, this formula is in fact derivable in $\biltlHX$. To see this, observe that the formula $\varphi\rightarrow (\psi\vee (\varphi\dimp\psi)) $ is a substitution instance of a bi-intuitionistic tautology and hence derivable. Applying $\mathsf{Nec}$ yields 
\begin{equation*}
    \X (\varphi\rightarrow (\psi\vee (\varphi\dimp\psi))).
\end{equation*}
Using axiom $\mathsf{K}$ and $\mathsf{MP}$ we obtain
\begin{equation*}
    \X \varphi \rightarrow \X (\psi \vee (\varphi \dimp \psi))
\end{equation*}
Using axiom $\mathsf{Dist}$ and $\mathsf{MP}$ yields that
\begin{equation*}
    \X \varphi\to (\X \psi\vee \X (\varphi\dimp\psi)) 
\end{equation*}
is derivable. By Lemma \ref{Lemma exclusion property}, this shows that $(\X \varphi\dimp \X \psi )\rightarrow \X (\varphi\dimp\psi) $ is derivable as well.
\end{example}

\begin{example}\label{ex: RV is derivable}
The formula $\mathbf{RV}$, i.e.~$\G(\varphi\vee\psi)\rightarrow (\E\varphi\vee\G\psi)$, is derivable in $\biltlH$. To see this, note that by Lemma \ref{l: bi-int tautologies} the formula
\begin{equation*}
\G(\varphi\vee\psi)\rightarrow(\E\varphi\vee(\G(\varphi\vee\psi)\dimp\E\varphi))
\end{equation*} 
is a substitution instance of a bi-intuitionistic tautology, so it suffices to check that
\begin{equation}\label{e: example RV 1}
 (\G(\varphi\vee\psi)\dimp\E\varphi)\rightarrow \G\psi
\end{equation}
is derivable. By $\mathsf{Fix_{\E/ \G}}$, the formulas $\G  (\varphi\vee\psi) \rightarrow {\X \G}  (\varphi\vee\psi)$ and $\X \E\varphi \rightarrow \E\varphi $ are derivable. So Lemma \ref{Lemma exclusion property} implies that 
\begin{equation*}
    (\G(\varphi\vee\psi)\dimp\E\varphi)\rightarrow ({\X\G}(\varphi\vee\psi)\dimp\X\E\varphi))
\end{equation*} 
is derivable as well. Hence by Example \ref{ex: dual version of K is derivable},
\begin{equation*}
    (\G(\varphi\vee\psi)\dimp\E\varphi)\to \X ( \G(\varphi\vee\psi)\dimp \E\varphi)
\end{equation*} 
is derivable. Using $\mathsf{Ind}_{\G}$ gives that
\begin{equation*}
    (\G(\varphi\vee\psi)\dimp\E\varphi)\rightarrow \G ( \G(\varphi\vee\psi)\dimp \E\varphi)
\end{equation*}
is derivable, implying that in order to prove (\ref{e: example RV 1}) it suffices to show that 

\begin{equation}\label{e: example RV 2}
    (\G(\varphi\vee\psi)\dimp\E\varphi)\rightarrow \psi
\end{equation}

is derivable. Note that $\G(\varphi\vee\psi)\rightarrow \varphi\vee\psi$ and $\varphi\to\E\varphi$  are derivable by $\mathsf{Fix_{\G}}$ and $\mathsf{Fix_{\E}}$. Thus we obtain that
\begin{equation*}
    ( \G(\varphi\vee\psi)\dimp \E\varphi) \rightarrow ((\varphi\vee \psi )\dimp\varphi)
\end{equation*}
is derivable by Lemma \ref{Lemma exclusion property}. But $((\varphi\vee \psi )\dimp\varphi) \rightarrow \psi$ is a substitution instance of a bi-intuitionistic tautology; hence $( \G(\varphi\vee\psi)\dimp \E\varphi) \rightarrow  \psi  $ is derivable, as desired. 
\end{example}

The rest of this section establishes that $\biltlH$ (and hence also $\biltlHX$) is sound with respect to expanding models. 

\begin{lemma}\label{l: axioms and rules are sound}
    The axioms of $\biltlH$ are valid over the class of expanding models and the rules preserve validity.
\end{lemma}
\begin{proof}
    For the axioms we only check the cases for the modalities. Let $\fw=(W, {\leq}, f,V)$ be an expanding model and $w \in W$ a world. \smallskip
    
   \noindent \textsc{Axiom $\mathsf{D}$.} Observe that for any $v \geq w$ the world $f(v)$ exists and therefore $\fw,v \not \models \X \bot$. Hence $\fw,w \models {\neg} {\X} {\bot}$. \smallskip

    \noindent \textsc{Axiom $\mathsf{Dist}$.} Let $v \geq w$ be any world such that $\fw,v \models \X (\varphi \vee \psi)$. Thus $\fw, f(v) \models \varphi \vee \psi$ and so $\fw,f(v) \models \varphi$ or $\fw,f(v) \models \psi$. Therefore $\fw, v \models \X \varphi$ or $\fw,v \models \X \psi$ and so $\fw,v \models \X \varphi \vee \X \psi$. We conclude that $\fw,w \models \X(\varphi \vee \psi) \rightarrow (\X \varphi \vee \X \psi)$. \smallskip

     \noindent \textsc{Axiom $\mathsf{K}$.} Let $v \geq w$ be any world such that $\fw,v \models \X (\varphi \rightarrow \psi)$. Then $\fw, f(v) \models \varphi \rightarrow \psi$. Suppose $u \geq v$ and $\fw,u \models \X \varphi$. Hence $\fw, f(u) \models \varphi$. By forward confluence, $f(v) \leq f(u)$ and since $\fw,f(v) \models \varphi \rightarrow \psi$, $\fw, f(u) \models \psi$. Thus $\fw,u \models \X \psi$ and so $\fw,v \models \X \varphi \rightarrow \X \psi$. From this we conclude that $\fw,w \models \X (\varphi \rightarrow \psi) \rightarrow (\X \varphi \rightarrow \X \psi)$.\smallskip

     \noindent \textsc{Axiom $\mathsf{Fix_\E}$.} Let $v \geq w$ be any world such that $\fw,v \models \varphi \vee \X \E \varphi$. Then $\fw,v \models \varphi$ or $\fw,v \models \X \E \varphi$. In both cases, $\fw,v \models \E \varphi$. Hence $\fw,w \models (\varphi \vee \X \E \varphi) \rightarrow \E \varphi$. \smallskip

      \noindent \textsc{Axiom $\mathsf{Fix_{\G}}$.} Let $v \geq w$ be any world such that $\fw,v \models \G \varphi$. Then $\fw,f^n(v) \models \varphi$ for all $n < \omega$. In particular, $\fw,v \models \varphi$ and $\fw,f^n(f(v)) \models \varphi$ for all $n < \omega$. Thus $M,f(v) \models \G \varphi$ and so $\fw,v \models {\X} {\G} \varphi$. Together, $\fw,v \models \varphi \wedge {\X} {\G} \varphi$, and so $\fw,w \models {\G} \varphi \rightarrow \varphi \wedge {\X} {\G} \varphi$. \smallskip

    For the rules we check all cases with the exception of $\mathsf{MP}$ and $\mathsf{Nec}$.\smallskip
    
    \noindent \textsc{Rule $\mathsf{DN}$.} Suppose that $\varphi$ is valid. Let $v$ be any world such that $v \geq w$. Let $u$ be any world such that $u \leq v$. Since $\varphi$ is valid, $\fw,u \models \varphi$. As $u$ is an arbitrary world, we conclude that $\fw,v \not \models {\sim} \varphi$. As $v$ is an arbitrary world as well, it follows that $\fw,w \models {\neg} {\sim} \varphi$. Hence ${\neg} {\sim} \varphi$ is valid. \smallskip

    \noindent \textsc{Rule $\mathsf{Mon}_\E$.} Suppose $\varphi \rightarrow \psi$ is valid. Let $v \geq w$ be any world and suppose $\fw,v \models \E \varphi$. So there exists $n < \omega$ such that $\fw,f^n(v) \models \varphi$. Since $\fw,f^n(v) \models \varphi \rightarrow \psi$, also $\fw,f^n(v) \models \psi$. Hence, $\fw,v \models \E \psi$, and so $\fw,w \models \E \varphi \rightarrow \E \psi$. We conclude that $\E \varphi \rightarrow \E \psi$ is valid. \smallskip

    \noindent \textsc{Rule $\mathsf{Mon}_{\G}$.} Suppose $\varphi \rightarrow \psi$ is valid. Let $v \geq w$ be any world and suppose that $\fw,v \models \G \varphi$. So for all $n < \omega$ it holds that $\fw,f^n(v) \models \varphi$. Since for each $n < \omega$, $\fw, f^n(v) \models \varphi \rightarrow \psi$, it also holds that $\fw,f^n(v) \models \psi$. Thus $\fw,v \models \G \psi$, and so $\fw,w \models \G \varphi \rightarrow \G \psi$. We conclude that $\G \varphi \rightarrow \G \psi$ is valid. \smallskip

    \noindent \textsc{Rule $\mathsf{Ind}_\E$.} Suppose $\X \varphi \rightarrow \varphi$ is valid. We first show that for any $n < \omega$, $\fw,f^n(w) \models \varphi$ implies $\fw,w \models \varphi$ by induction on $n$. The base case where $n=0$ is trivial. For the induction step suppose that the claim holds for $n=k$ and consider the case where $n=k+1$. If $\fw,f^{k+1}(w) \models \varphi$, then $\fw,f^k(w) \models \X \varphi$. Since $\X \varphi \rightarrow \varphi$ is valid, we have $\fw,f^k(w) \models \varphi$. By the induction hypothesis, $\fw,w \models \varphi$, which concludes the proof of the claim. Next, let $v \geq w$ be any world and suppose $\fw,v \models \E \varphi$. Then there exists $n  < \omega$ with $\fw,f^n(v) \models \varphi$. By the previous claim it follows that $\fw,v \models \varphi$. Hence $\fw,w \models \E \varphi \rightarrow \varphi$, and so $\E \varphi \rightarrow \varphi$ is valid. \smallskip

    \noindent \textsc{Rule $\mathsf{Ind}_{\G}$.} Suppose $ \varphi \rightarrow \X \varphi$ is valid. We first show that $\fw,w \models \varphi$ implies $\fw,f^n(w) \models \varphi$ for all $n < \omega$ by induction on $n$. The base case where $n=0$ is trivial. For the induction step suppose the claim holds for $n=k$ and consider the case for $n=k+1$. By the induction hypothesis $\fw,f^k(w) \models \varphi$. Since $\varphi \rightarrow \X \varphi$ is valid, $\fw,f^k(w) \models \X \varphi$, and therefore $\fw,f^{k+1}(w) \models \varphi$, which concludes the proof of the claim. Next, let $v \geq w$ be any world such that $\fw,v \models  \varphi$. By the previous claim it follows that $\fw,f^n(v) \models \varphi$ for all $n <  \omega$, and thereby that $\fw,v \models \G \varphi$. Thus $\fw,w \models \varphi \rightarrow \G \varphi$, implying that $\varphi \rightarrow \G \varphi$ is valid. \qedhere
\end{proof}

From the previous lemma, soundness of $\biltlH$ is inferred by a standard induction on the height of proofs. 

\begin{theorem}[Soundness of $\biltlH$]\label{t: soundness of biltl}
    For any $\Lbiltl$-formula $\varphi$ if $ \varphi$ is provable in $\biltlH$, then $\varphi$ is valid over the class of expanding models.
\end{theorem}

\section{Completeness for the Next-Fragment}

This section establishes completeness of $\biltlHX$ by emplyoing a standard canonical model construction. As the canonical model is also used later on in the completeness proof of $\biltlH$, the following definitions and lemmas apply to both $\Lbiltl$ and $\Lbiltlnext$. Write $\mathcal{L}$ for ${\mathcal{L} \in \{ \Lbiltl,\Lbiltlnext \}}$.

\begin{definition}\index{prime theory!for $\biltl$}
    An \emph{$\mathcal{L}$-prime theory} is a set of $\mathcal{L}$-formulas $\Gamma$, such that the following hold.
    \begin{enumerate}
        \item $\Gamma$ is deductively closed: if $\Gamma \vdash \varphi$, then $\varphi \in \Gamma$.
        \item $\Gamma$ satisfies the disjunction property: if $\varphi \vee \psi \in \Gamma$, then $\varphi \in \Gamma$ or $\psi \in \Gamma$.
        \item $\Gamma$ is consistent: $\Gamma \not \vdash \bot$.
    \end{enumerate}
\end{definition}

The following properties are readily checked for any prime theory $\Gamma$ (see Lemma \ref{l: properties of prime theories IM}).

\begin{lemma}\label{l: properties prime theory biLTL}
    Let $\Gamma$ be a $\mathcal{L}$-prime theory and $\varphi, \psi \in \mathcal{L}$.
    \begin{enumerate}
        \item $\varphi \wedge \psi \in \Gamma$ if and only if $\varphi \in \Gamma$ and $\psi \in \Gamma$.
        \item $\varphi \vee \psi \in \Gamma$ if and only if $\varphi \in \Gamma$ or $\psi \in \Gamma$.
        \item If $\varphi \rightarrow \psi \in \Gamma$, then $\varphi \not \in \Gamma$ or $\psi \in \Gamma$.
        \item If $\varphi \dimp \psi \in \Gamma$, then $\varphi \in \Gamma$.
        \item If $\E \varphi \in \Gamma$, then $\varphi \in \Gamma$ or ${\X}{\E}\varphi \in \Gamma$.
        \item If $\G \varphi \in \Gamma$, then $\varphi \in \Gamma$ and ${\X}{\G}\varphi \in \Gamma$.
    \end{enumerate}
\end{lemma}

Given a set of formulas $\Gamma$, define $\X^{-1} \Gamma \coloneqq \{\varphi \mid \X \varphi \in \Gamma\}$.

\begin{lemma}\label{prime theory lemma}
    If $\Gamma$ is a prime theory, then $\X^{-1}\Gamma$ is a prime theory as well.
\end{lemma}
\begin{proof}
    Suppose $\Gamma$ is a prime theory. We check all three properties of a prime theory. \smallskip
    
    \noindent \textsc{1.} Suppose $\X^{-1}\Gamma \vdash \varphi$ by derivation $\pi$, where $\pi$ is a $\biltlH$- or a $\biltlHX$-derivation. We proceed by induction on the height of $\pi$. In the base case $\varphi$ is an axiom or belongs to $\X^{-1} \Gamma$. If $\varphi$ is an axiom, then by necessitation, $\Gamma \vdash \X \varphi$ and therefore by the fact that $\Gamma$ is deductively closed $\X \varphi \in \Gamma$. Hence $\varphi \in \X^{-1} \Gamma$.
    If instead $\varphi$ belongs to $\X^{-1} \Gamma$, then there is nothing left to show. For the induction step first suppose the last rule applied in $\pi$ is $\mathsf{MP}$ with conclusion $\varphi$ and premises $\psi$ and $\psi \rightarrow \varphi$. By induction hypothesis $\psi, \psi \rightarrow \varphi \in \X^{-1} \Gamma$. Hence $\X \psi, \X (\psi \rightarrow \varphi) \in \Gamma$. By the presence of the $\mathsf{K}$-axiom and $\mathsf{MP}$ we conclude that $\X \varphi \in \Gamma$, and thus $\varphi \in \X^{-1} \Gamma$. Next, suppose the last rule applied in $\pi$ is $\mathsf{R}$ where $\mathsf{R} \not = \mathsf{MP}$, with premise $\psi$ and conclusion $\gamma$. By definition of a derivation from assumptions it must hold that $\psi$ is provable from the empty set, i.e.~$\vdash \psi$, and hence $\vdash \gamma$ as well. Thus applying $\mathsf{Nec}$ yields $\vdash \X \gamma$, and so $\Gamma \vdash \X \gamma$. Since $\Gamma$ is deductively closed, $\X \gamma \in \Gamma$, implying that $\gamma \in \X^{-1} \Gamma$. \smallskip

    \noindent \textsc{2.} Suppose $\varphi \vee \psi \in \X^{-1} \Gamma$. Then $\X (\varphi \vee \psi) \in \Gamma$. By $\mathsf{Dist}$ and $\mathsf{MP}$, $\X \varphi \vee \X \psi \in \Gamma$. As $\Gamma$ satisfies the disjunction property, $\X \varphi \in \Gamma$ or $\X \psi \in \Gamma$. Hence $\varphi \in \X^{-1} \Gamma$ or $\psi \in \X^{-1} \Gamma$. \smallskip

    \noindent \textsc{3.} Suppose towards contradiction that $\X^{-1} \Gamma \vdash \bot$. Then $\bot \in \X^{-1} \Gamma$ as $\X^{-1} \Gamma$ is deductively closed. Thus $\X \bot \in \Gamma$. But $\Gamma \vdash {\neg} {\X} \bot$, and thus $\Gamma \vdash \bot$; a contradiction. \qedhere
\end{proof}

Next, the standard Lindenbaum Lemma holds for $\biltlH$ and $\biltlHX$. The proof is from~\cite[Lemma 8.6]{gore_2020}.

\begin{lemma}[Lindenbaum]\label{l: Lindenbaum for biltl}
    Let $\Gamma \cup \{\gamma\} \subseteq \mathcal{L}$. If $\Gamma \not \vdash \gamma$, then there exists a $\mathcal{L}$-prime theory $\Delta$ with $\Gamma \subseteq \Delta$ and $\Delta \not \vdash \gamma$. 
\end{lemma}
\begin{proof}
Suppose $\Gamma \not \vdash \gamma$ and consider an arbitrary enumeration of the formulas of $\mathcal{L}$. We first show how to construct a set of formulas $\Delta_n$ for each $n < \omega$ by induction on $n$.
\begin{itemize}
    \item Define $\Delta_0 \coloneqq \Gamma$
    \item Suppose we have defined $\Delta_n$. Let $\varphi \vee \psi$ be the first formula in the enumeration that has a disjunction as its principal connective, has not yet been treated in a previous step and $\Delta_n \vdash \varphi \vee \psi$. Then define $\Delta_{n+1}$ as follows:
    
    \begin{equation*}
        \Delta_{n+1} \coloneqq \begin{cases}
                               \Delta_n \cup \{\varphi\} \text{ if } \Delta_n, \varphi \not \vdash \gamma\\
                               \Delta_n \cup \{\psi\} \text{ otherwise. }\\
                        \end{cases}
    \end{equation*}
\end{itemize}
Observe that for each $n < \omega$ holds that $\Delta_n \subseteq \Delta_{n+1}$. Define $\Delta$ to be
\begin{equation*}
    \Delta \coloneqq \bigcup_{n < \omega} \Delta_n
\end{equation*}
By construction $\Gamma \subseteq \Delta$. In order to show that $\Delta$ is a prime theory, let us first check whether it satisfies the disjunction property. If $\varphi \vee \psi \in \Delta$, then $\Delta \vdash \varphi \vee \psi$. Let $n$ be the least natural number such that $\Delta_n \vdash \varphi \vee \psi$. By construction there exists $m \geq n$ such that either $\varphi$ or $\psi$ is added to $\Delta_{m+1}$. Thus, since $\Delta_{m+1} \subseteq \Delta$ it holds that $\varphi \in \Delta$ or $\psi \in \Delta$. In order to show that $\Delta$ is deductively closed, suppose that $\Delta \vdash \varphi$. Then $\Delta \vdash \varphi \vee \varphi$, and thus $\varphi \in \Delta$ by the same argument as above. In order to show that $\Delta \not \vdash \gamma$, we first prove by induction on $n$ that each $\Delta_n \not \vdash \gamma$. The base case holds by assumption. So suppose that $\Delta_n \not \vdash \gamma$ and the formula treated at that step is $\varphi \vee \psi$. If $\Delta_{n+1} = \Delta_n \cup \{\varphi\}$, then $\Delta_{n+1} \not \vdash \gamma$ by construction. Otherwise  $\Delta_n \cup \{\varphi\} \vdash \gamma$ and $\Delta_{n+1} = \Delta_n \cup \{\psi\}$. Suppose towards contradiction that $\Delta_{n+1} \vdash \gamma$. Therefore
\begin{equation*}
    \begin{split}
        \Delta_n, \varphi \vdash \gamma\\
        \Delta_n, \psi \vdash \gamma\\
    \end{split}
\end{equation*}
By the Deduction Theorem $\Delta_n \vdash \varphi \rightarrow \gamma$ and $\Delta_n \vdash \psi \rightarrow \gamma$. Therefore $\Delta_n \vdash (\varphi \vee \psi) \rightarrow \gamma$. Since $\Delta_n \vdash \varphi \vee \psi$ by assumption it therefore follows that $\Delta_n \vdash \gamma$, a contradiction. Hence $\Delta_{n+1} \not \vdash \gamma$. Finally if $\Delta \vdash \gamma$, then $\Delta_n \vdash \gamma$ for some $n$. As this cannot be the case, we conclude that $\Delta \not \vdash \gamma$. Observe that this also implies that $\Delta$ is consistent and hence a prime theory.
\end{proof}

We are now ready to define the canonical models for $\Lbiltl$ and $\Lbiltlnext$.

\begin{definition}\index{canonical model!for $\biltl$}
Let $\mathcal L$ be either $\Lbiltlnext$ or $\Lbiltl$.
    The \emph{canonical model} for $\mathcal L$ is defined to be $\mathcal M_\mathrm{c}=(W_\mathrm{c}, {\leq_\mathrm{c}}, f_\mathrm{c}, V_\mathrm{c})$ where
    \begin{itemize}
        \item $W_\mathrm{c} = \{ \Gamma \subseteq \mathcal L \mid \Gamma$ is a $\mathcal{L}$-prime theory$\}$,
        \item $\Gamma \leq_\mathrm{c} \Gamma'$ $\iff$ $\Gamma \subseteq \Gamma'$,
        \item $f_\mathrm{c}(\Gamma) = \X^{-1} \Gamma$, 
        \item $V_\mathrm{c}(\Gamma) = \{ p \in \Prop \mid p \in \Gamma\}$.
    \end{itemize}
\end{definition}

\begin{lemma}
    The canonical model for either $\Lbiltlnext$ or $\Lbiltl$ is an expanding model.
\end{lemma}
\begin{proof}
    By definition of $\leq_c$ holds that $(W_c, \leq_c)$ is a poset and $V_c$ is monotone. Note that for each $\Gamma, \Gamma' \in W_c$ if $\X^{-1} \Gamma = \X^{-1} \Gamma'$, then $\Gamma = \Gamma'$. Moreover by Lemma \ref{prime theory lemma}, $\X^{-1} \Gamma \in W_c$, implying that $f_c: W_c \longrightarrow W_c$ is well-defined. If $\Gamma \leq_c \Gamma'$, then $\Gamma \subseteq \Gamma'$ and so $\X^{-1} \Gamma \subseteq \X^{-1} \Gamma'$, implying that $f_c (\Gamma) \leq_c f_c (\Gamma')$. Therefore $\fw_c$ is an expanding model.
\end{proof}

The proof of the first part of the following lemma is standard, while the second part was proven in~\cite[Lemma 8.7]{gore_2020}.

\begin{lemma}[Witnessing Lemma]\label{l: witnessing lemma for biltl}
    Let $\mathcal L$ be either $\Lbiltlnext$ or $\Lbiltl$. Let $\Gamma$ be a $\mathcal{L}$-prime theory and $\psi, \gamma$ any $\mathcal{L}$-formulas. The following hold.
    \begin{enumerate}
        \item  $\psi \rightarrow \gamma \not \in \Gamma$ if and only if there exists a prime theory $\Delta$ with $\Gamma \leq_c \Delta$ and $\psi \in \Delta$ and $\gamma \not \in \Delta$.
        \item $\psi \dimp \gamma \in \Gamma$ if and only if there exists a prime theory $\Delta$ with $\Delta \leq_c \Gamma$ and $\psi \in \Delta$ and $\gamma \not \in \Delta$.
    \end{enumerate}
\end{lemma}
\begin{proof}
    For 1. suppose that $\psi \rightarrow \gamma \not \in \Gamma$. So $\Gamma \not \vdash \psi \rightarrow \gamma$. The Deduction Theorem implies that $\Gamma, \psi \not \vdash \gamma$. By the Lindenbaum Lemma there exists a prime theory $\Delta$ such that $\Gamma \cup \{\psi\} \subseteq \Delta$ and $\gamma \not \in \Delta$. Note that $\Gamma \leq_c \Delta$. For the other direction suppose $\psi \rightarrow \gamma \in \Gamma$. Let $\Delta$ be any prime theory with $\Gamma \leq_c \Delta$. Then $\psi \rightarrow \gamma \in \Delta$ and so by Lemma \ref{l: properties prime theory biLTL}, $\psi \not \in \Delta$ or $\gamma \in \Delta$. \smallskip

    For 2. suppose that $\psi \dimp \gamma \in \Gamma$. Suppose towards contradiction that $\psi \vdash \gamma, \Gamma^c$ where $\Gamma^c = \mathcal{L} \setminus \Gamma$. Hence there exists a finite $\Omega \subseteq \Gamma^c$ such that $\psi \vdash \gamma \vee \bigvee \Omega$ and so by the Deduction Theorem
        \begin{equation*}
            \vdash \psi \rightarrow (\gamma \vee \bigvee \Omega)
        \end{equation*}
        Lemma \ref{Lemma exclusion property} then implies that
        \begin{equation*}
            \vdash (\psi \dimp \gamma) \rightarrow \bigvee \Omega
        \end{equation*}
        So also $\Gamma \vdash (\psi \dimp \gamma) \rightarrow \bigvee \Omega$. But as $\psi \dimp \gamma \in \Gamma$ we conclude that
        \begin{equation*}
            \Gamma \vdash \bigvee \Omega,
        \end{equation*}
        which implies that there exists $\delta \in \Omega$ such that $\delta \in \Gamma$ as $\Gamma$ is prime, contradicting $\Gamma \cap \Gamma^c = \emptyset$. Therefore $\psi \not \vdash \gamma,\Gamma^c$. By the Lindenbaum Lemma there exists a prime theory $\Delta$ such that $\psi \in \Delta$ and $(\Gamma^c \cup \{\gamma\}) \cap \Delta = \emptyset$. Therefore $\Delta \subseteq \Gamma$, implying $\Delta \leq_\mathrm{c} \Gamma$. Finally, observe that by construction $\psi \in \Delta$ and $\gamma \not \in \Delta$.
        
        For the other direction suppose that $\psi \dimp \gamma \not \in \Gamma$. So $\Gamma \not \vdash \psi \dimp \gamma$. Moreover, for each $\Delta \subseteq \Gamma$ it holds that $\Delta \not \vdash \psi \dimp \gamma$. Observe that
        \begin{equation*}
            \vdash \psi \rightarrow (\gamma \vee (\psi \dimp \gamma)).
        \end{equation*}
        Therefore for every $\Delta \subseteq \Gamma$ if $\Delta \vdash \psi$, then $\Delta \vdash \gamma$ i.e.~if $\psi \in \Delta$, then $\gamma \in \Delta$.
\end{proof}

For the remainder of this section we work exclusively in the language $\Lbiltlnext$ and show that $\biltlHX$ is complete.

\begin{lemma}[Truth Lemma]\label{l:truth}
Let $\mathcal M_\mathrm{c}$ be the canonical model for $\Lbiltlnext$. For every $\Gamma \in W_\mathrm{c}$ and any $\Lbiltlnext$-formula $\varphi$, it holds that
    \begin{center}
        $\varphi \in \Gamma \text{ if and only if } \mathcal M_\mathrm{c}, \Gamma \models \varphi$.
    \end{center}
\end{lemma}

\begin{proof}
We proceed by induction on the structure of $\varphi$.
The base case follows immediately from the definition of the valuation $V_\mathrm{c}$ and the fact that prime theories are consistent.
For the induction step the cases where $\varphi = \psi \wedge \chi$ and $\varphi = \psi \vee \chi$ are standard and omitted. The cases for $\varphi= \psi \rightarrow \chi$ and $\varphi = \psi \dimp \chi$ follow immediately from Lemma \ref{l: witnessing lemma for biltl} and the induction hypothesis. \smallskip    
\noindent \textsc{$\varphi = \X \psi$:} For the direction from left to right suppose that $\X \psi \in \Gamma$. Then, by definition, $\psi \in \X^{-1} \Gamma$. By the induction hypothesis $\mathcal{M}_\mathrm{c}, \X^{-1} \Gamma \models \psi$. As $f_\mathrm{c}(\Gamma) = \X^{-1} \Gamma$ we conclude that $\mathcal{M}_\mathrm{c}, \Gamma \models \X \psi$.
For the direction from right to left suppose that $\mathcal{M}_\mathrm{c}, \Gamma \models \X \psi$. This implies that $\mathcal{M}_\mathrm{c},f_\mathrm{c}(\Gamma) \models \psi$. By construction $f_\mathrm{c}(\Gamma) = \X^{-1} \Gamma$. By the induction hypothesis $\psi \in \X^{-1} \Gamma$. Hence by definition $\X \psi \in \Gamma$. \qedhere
\end{proof}

\begin{theorem}[Completeness of $\biltlHX$]
    If a $\Lbiltlnext$-formula $\varphi$ is valid over the class of expanding models, then $\varphi$ is $\biltlHX$-derivable.
\end{theorem}

\begin{proof}
    Suppose $\varphi$ is not $\biltlHX$-derivable, i.e.~$\emptyset \not \vdash \varphi$. By the Lindenbaum Lemma there exists a prime theory $\Gamma$ with $\Gamma \not \vdash \varphi$. Hence $\varphi \not \in \Gamma$, and so by the Truth Lemma, $\mathcal M_\mathrm{c}, \Gamma \not \models \varphi$. Therefore $\varphi$ is not valid. 
\end{proof}

\section{Proof Strategy for the Full Language}

The remainder of this chapter is devoted to prove that $\biltlH$ is  complete for the class of expanding models.
Unlike for $\biltlHX$, the Truth Lemma for the canonical model does not hold, since it may be for example that $\E \varphi\in \Gamma $, but there is no $n$ such that $\varphi\in f^n_\mathrm{c}(\Gamma)$. This is because $\E\varphi \vdash \bigvee _{i<n} \X^i \varphi$ is not derivable for any specific $n$, and derivations are finite.
Hence it is possible for $\E\varphi$ to hold but each individual $\X^n \varphi$ to fail in a prime theory.

A similar situation occurs for classical linear temporal logic ($\mathsf{LTL}$), but one can then pass to a filtration $\sfrac{\mathcal M_\mathrm{c}}\Sigma$ of $\mathcal M_\mathrm{c}$, i.e.~the quotient of $\mathcal M_\mathrm{c}$ modulo the equivalence relation given by 
\begin{equation*}
    \Gamma\sim\Gamma' \text{ if and only if } \Gamma\cap \Sigma=\Gamma'\cap \Sigma.
\end{equation*}
Assuming $\Sigma$ is finite, the equivalence class of each prime theory $\Gamma$ is determined by its \emph{characteristic formula} $\chi(\Gamma)\coloneqq \bigwedge ( \Gamma\cap \Sigma)$.\footnote{Note that this is essentially what was done in the completeness proof for $\mathrm{IM_H}$, but we defined $\sfrac{\fw_c}{\sim}$ directly instead of via a filtration.}
The filtrated model \emph{does} respect the semantics of $\E$. More precisely, $\sfrac{\mathcal M_\mathrm{c}}\Sigma$ satisfies a version of the Truth Lemma  restricted to formulas of $\Sigma$, which is sufficient as long as $\Sigma$ contains enough formulas.
The tradeoff is that $\sfrac{\mathcal M_\mathrm{c}}\Sigma$ is no longer equipped with a \emph{function}, as the quotient may assign more than one temporal successor to each equivalence class of prime theories. To see why, recall that if $\Gamma \sim \Gamma'$, then $\Gamma \cap \Sigma = \Gamma' \cap \Sigma$. Since $\Sigma$ is a finite set, it is possible that there exists a formula $\varphi \in \Sigma$ such that $\X \varphi \not \in \Sigma$, and therefore that $\X \varphi \in \Gamma$ but $\X \varphi \not \in \Gamma'$. This implies that $f_c(\Gamma) \not \sim f_c(\Gamma')$. Accordingly, the quotient model must relate the equivalence class of $\Gamma$ with both equivalences of $f_c(\Gamma)$ and of $f_c(\Gamma')$.

However, this is not a problem, since in a later phase one can choose a path $\Gamma_0,\Gamma_1,\Gamma_2,\dots$ that constitutes a genuine $\mathsf{LTL}$-model.
In particular, if $\varphi$ is not derivable, we can choose $\Sigma$ to be the set of subformulas of $\varphi$ and their negations and $\Gamma_0$ so that $ \varphi\not\in \Gamma_0$, thereby obtaining a model falsifying $\varphi$.

We wish to adapt this strategy, but there is an issue: filtration in general does not conserve order-preservation of the temporal dynamics (i.e.~$w \leq v$ implies $f(w) \leq f(v)$), so we must define $\sfrac{\mathcal M_\mathrm{c}}\Sigma$ differently. This structure should be a \emph{quasimodel}, which is similar to a model except that the temporal transition \emph{function} is replaced by a non-deterministic \emph{relation}.
Each point in a quasimodel is assigned a \emph{type}, which is similar to a prime theory except that a type only decides a finite set of formulas; i.e., a type is a pair $\Phi=(\Phi^+,\Phi^-)$ of (usually) finite sets of formulas for which a `Truth Lemma' should hold. In other words, a quasimodel is essentially what was called a \emph{labelled model} in Chapter \ref{c: bi-int ml new}. Quasimodels are designed so that they can be `unwound' into a genuine model, much like for the filtrated model of classical $\mathsf{LTL}$. 

To construct $\sfrac{\mathcal M_\mathrm{c}}\Sigma$ in the bi-intuitionistic setting, we use the tools and results from Chapter \ref{c: bi-int ml new}, where a finite labelled model  $\irr\Sigma$ was constructed and related to any given model via a dynamic and strongly surjective simulation $\check{E}_0$. The structure $\irr\Sigma\upharpoonright_{dom(\check{E}_0)}$ was shown to be a labelled model, and will play the role of  $\sfrac{\mathcal M_\mathrm{c}}\Sigma$ here. These results were given in a general modal setting, however we will show how they can be readily applied to the specific case of $\biltl$. 

Much as the characteristic formula $\chi(\Gamma)$ determines the equivalence class of $\Gamma$ in the classical setting, we can characterise which points of $\irr\Sigma$ are $\check{E}_0$-related to a given prime theory $\Gamma$ using \emph{simulation formulas}. Unlike the classical setting, we need two distinct formulas, $\chi^+$ and $\chi^-$, to capture respectively the `positive' and `negative' information determining a simulation. These simulation formulas enable us to prove that $\sfrac{\mathcal M_\mathrm{c}}\Sigma$ is indeed a quasimodel.

At the end of the chapter we put all the ingredients together to show that $\biltlH$ is complete for the class of expanding models: the argument is that if $\varphi$ is not derivable then we can find a prime theory $\Gamma$ with $\varphi\in \Gamma ^-$.
By choosing a point $w$ of $\sfrac{\mathcal M_\mathrm{c}}\Sigma$ with $w\mathrel {\check{E}_0} \Gamma$, we see that $\varphi$ is falsified on $\sfrac{\mathcal M_\mathrm{c}}\Sigma$.
By applying the unwinding procedure to $\sfrac{\mathcal M_\mathrm{c}}\Sigma$, we obtain a genuine model falsifying $\varphi$. Thus every formula that is not derivable can be falsified in some expanding model, implying that~$\biltlH$ is complete.

\section{Types, Labelled Structures and Quasimodels}

This section introduces the aforementioned relational structures called \emph{quasimodels}. Many definitions are akin to the definitions of labelled models in Chapter \ref{c: bi-int ml new}, but adjusted to the case of $\biltl$. From now on, $\Sigma$ denotes a set of $\Lbiltl$-formulas closed under subformulas.

\begin{definition}\label{d: Sigma type}\index{type}
Let $\Phi^+, \Phi^- \subseteq \Sigma$. A \emph{$\Sigma$-type} is a pair $\Phi = (\Phi^+, \Phi^-)$ of disjoint subsets of $\Sigma$  satisfying the properties 2. -- 7. of Definition \ref{d: type} and the following properties.
\begin{description}[font=\normalfont,
  leftmargin=!,     
  labelwidth=\widthof{$\dimp^+$.}]
 
    \item[8.] If $\G \varphi \in \Phi^+$, then $\varphi \in \Phi^+$.
    \item[9.] If $\E \varphi \in \Phi^-$, then $\varphi \in \Phi^-$.
\end{description}
\end{definition}

As before, it is not necessary that $\Phi^+ \cup \Phi^- = \Sigma$. Thus types are once again `partial'. The set of all $\Sigma$-types is denoted by $\type_\Sigma$. The partial orders $\leq_{\type}$ and $\subseteq_\type$ are defined as before. Moreover, the notion of a \emph{defect} of a type is also defined as before, c.f. Definition \ref{d: defects}.

Recall the definition of a \emph{$\Sigma$-labelled poset} $\mathcal{X}=(X, \leq_{\mathcal{X}}, \ell_{\mathcal{X}})$ (c.f. Definition \ref{d: labelled poset}). In Chapter \ref{c: bi-int ml new} sensible relations were defined with respect to a \emph{sensibility condition} to have a general argument for the multiple frame conditions considered. Here, we only consider one class of frames and therefore only require one specific sensibility condition.

\begin{definition}\index{binary relation!sensible}
    Let $\Phi, \Psi$ be $\Sigma$-types. The pair $(\Phi, \Psi)$ is called \emph{sensible} if the following conditions hold.
    \begin{enumerate}
        \item If $\X \varphi \in \Phi^+$, then $\varphi \in \Psi^+$.
        \item If $\X \varphi \in \Phi^-$, then $\varphi \in \Psi^-$.
        \item If $\E \varphi \in \Phi^+$, then $\varphi \in \Phi^+$ or $\E \varphi \in \Psi^+$.
        \item If $\E \varphi \in \Phi^-$, then $\varphi \in \Phi^-$ and $\E \varphi \in \Psi^-$.
        \item If $\G \varphi \in \Phi^+$, then $\varphi \in \Phi^+$ and $\G \varphi \in \Psi^+$.
        \item If $\G \varphi \in \Phi^-$, then $\varphi \in \Phi^-$ or $\G \varphi \in \Psi^-$.
    \end{enumerate}

    Given a $\Sigma$-labelled poset $\mathcal{X}=(X, {\leq}, \ell)$, a pair $(x,y) \in X \times X$ is called \emph{sensible} if $(\ell(x), \ell(y))$ is sensible. A relation $R \subseteq X \times X$ is called sensible if $R$ is forth-up and forth-down confluence (with respect to $(\leq_{\mathcal{X}}, \leq_{\mathcal{X}}) $) and every pair $(x,y) \in R$ is sensible.
    \end{definition}

    One can easily check that the relation $S \subseteq \type_\Sigma \times \type_\Sigma$ given by 
    \begin{center}
        $(\Phi, \Psi) \in S$ if and only if $(\Phi, \Psi)$ is sensible
    \end{center} 
    is a sensibility condition. Therefore every sensible structure considered henceforth is $S$-sensible.

     \begin{definition}\index{binary relation!$\omega$-sensible}
    Given a $\Sigma$-labelled poset $\mathcal{X}=(X, \allowbreak {\leq},\allowbreak \ell)$, a sensible relation $R \subseteq X \times X$ is called \emph{$\omega$-sensible} if the following hold.
    \begin{enumerate}
        \item If $\E \varphi \in \ell(x)^+ $, then there are $n < \omega$ and $y \in X$ such that $x \mathrel R^n y$ and $\varphi \in \ell(y)^+$.
        \item If $\G \varphi \in \ell(x)^-$, then there are $n < \omega$ and $y \in X$ such that $x \mathrel R^n y$ and $\varphi \in \ell(y)^-$.
    \end{enumerate}
\end{definition}

\begin{definition}\label{def:quasimodel}\index{labelled system}
    A \emph{$\Sigma$-labelled system} is a tuple $\mathcal{X}=(X,\allowbreak  {\leq},\allowbreak \ell,\allowbreak  R)$ consisting of a labelled poset equipped with a sensible relation $R \subseteq X \times X$. If moreover $R$ is serial and $\omega$-sensible, then $\mathcal X$ is a \emph{$\Sigma$-quasimodel}.\index{quasimodel}
\end{definition}

We may write simply \emph{labelled system} or \emph{quasimodel} when $\Sigma$ is clear from context.
A formula $\varphi$ is \emph{falsified} at world $x$ of a $\Sigma$-labelled system $\mathcal{X}=(X, {\leq}, \ell, R)$ if $\varphi \in \ell(x)^-$, and \emph{satisfied} if $\varphi \in \ell(x)^+$. A formula $\varphi$ is falsifiable over the class of $\Sigma$-quasimodels if there exists a $\Sigma$-quasimodel $\mathcal{X}=(X, {\leq}, \ell, R)$ and a world $x \in X$ such that $\varphi$ is falsified at $x$. \smallskip


 Observe that every expanding model can be regarded as a $\Sigma$-quasimodel by simply labelling each world with those formulas in $\Sigma$ that are true or false respectively. Thus we obtain the following result.

\begin{lemma}\label{l: falsifiable over expanding models}
    If $\varphi \in \Sigma$ is falsifiable over the class of expanding models, then $\varphi$ is falsifiable over the class of $\Sigma$-quasimodels.
\end{lemma}

The converse of Lemma \ref{l: falsifiable over expanding models} is also true, but establishing it requires some work. This will be done in the next section. However, we can already state the result for a particular subclass of quasimodels.

\begin{definition}
    A $\Sigma$-quasimodel $\mathcal{X}=(X, {\leq}, \ell, R)$ is \emph{total functional} if $R \subseteq X \times X$ is a function.
\end{definition}

\begin{lemma}\label{l: falsifiability of functional quasimodels}
    If a formula is falsifiable over the class of total functional $\Sigma$-quasimodels, then it is falsifiable over the class of expanding models.
\end{lemma}

\begin{proof}
    Given a total functional $\Sigma$-quasimodel $\mathcal{X}=(X, {\leq}, \ell, f)$ define an expanding model $\mathcal{M}=(X, {\leq}, V, f)$ where $V(y) \coloneqq \ell(y)^+ \cap \Prop$. Observe that $\mathcal{M}$ is well-defined. We prove simultaneously that for $\varphi \in \Sigma$ if $\varphi \in \ell(x)^+$, then $\fw, x \models \varphi$ and if $\varphi \in \ell(x)^-$, then $\mathcal{
    M}, x \not \models \varphi$ by induction on $\varphi$. The base case where $\varphi \in \Prop$ follows directly from the definition of the valuation. The cases for $\varphi = \psi \ast \gamma$ for $\ast \in \{\wedge, \vee, \rightarrow, \dimp\}$ follow from the definition of a type and a labelled poset. For example if $\psi \wedge \gamma \in \ell(x)^-$, then $\psi \in \ell(x)^-$ or $\gamma \in \ell(x)^-$. By induction hypothesis $\mathcal{M}, x \not \models \psi$ or $\mathcal{M}, x \not \models \gamma$. Thus $\mathcal{M}, x \not \models \psi \wedge \gamma$. Suppose $\varphi = \X \psi \in \ell(x)^-$. By sensibility of $f$, $\psi \in \ell(f(x))^-$ and so by induction hypothesis $\mathcal{M}, f(x) \not \models \psi$. Therefore $\mathcal{M}, x \not \models \X \psi$. The case for $\X \psi \in \ell(x)^+$ is similar. Suppose $\varphi = \E \psi \in \ell(x)^-$. By definition of a type $\psi \in \ell(x)^-$ and $\E \psi \in \ell(f(x))^-$. By a simple induction on $n$ it follows that $\psi, \E \psi \in \ell(f^n(x))^-$ for all natural numbers $n$. Thus by induction hypothesis $\mathcal{M}, f^n(x) \not \models \psi$ for all $n$, implying that $\mathcal{M},x \not \models \E \psi$. Suppose $\E \psi \in \ell(x)^+$. Then by $\omega$-sensibility there exists a natural number $n$ such that $\psi \in \ell(f^n(x))$. By induction hypothesis $\fw, f^n(x) \models \psi$ and therefore $\fw, x \models \E \psi$. The cases for $\varphi = \G \psi$ are similar.
\end{proof}

\section{From Quasimodels to Expanding Models}\label{secUnwinding}

This section establishes that if a formula in $\Lbiltl$ is falsifiable over the class of quasimodels, then it is falsifiable over the class of expanding models. Given a quasimodel falsifying a formula $\varphi$, we will show how to construct a total functional quasimodel which falsifies $\varphi$ as well. The construction is similar to the construction of a functional model from a dynamic model presented in Chapter \ref{c: IM}, Section \ref{c: IM, section model constructions IM}, however the confluence conditions complicate the argument. Applying Lemma \ref{l: falsifiability of functional quasimodels} then yields an expanding model which falsifies $\varphi$. For the construction it is useful to observe that forward confluence can be iterated, thereby yielding the following lemma for finite $R$-paths, which is a generalization of Lemma \ref{l: iteration of forward confluent function is forward confluent}.

\begin{lemma}\label{lemPathExtend}
Let $\mathcal X=(X, {\leq}, \ell, R)$ be a quasimodel and $w_0, \ldots w_n \in X$ such that $w_0\mathrel R w_1 \mathrel R \dots \mathrel R w_n$.
\begin{enumerate}
    
    \item If $w_0\leq u_0$, then there exist $u_0\mathrel R u_1 \mathrel R \dots \mathrel R u_n$ such that $w_i\leq u_i$ for all $i\leq n$.

\item If $u_0\leq w_0$ then there exist $u_0\mathrel R u_1 \mathrel R \ldots \mathrel R u_n$ such that $u_i\leq w_i$ for all $i\leq n$.   
\end{enumerate}
\end{lemma}
\begin{proof}
    For 1. proceed by induction on the length $n$ of the path $w_0, \ldots, w_n$. For $n=0$ the statement is vacuously true. For $n> 0$ there exist $u_0 \mathrel R u_1 \mathrel R \ldots \mathrel R u_{n-1}$ with $w_i \leq u_i$ for all $i < n$ by induction hypothesis. In particular, $w_{n-1} \leq u_{n-1}$. Since $w_{n-1} \mathrel R w_n$, forth-up confluence yields $u_n \in X$ with $u_{n-1} \mathrel R u_n$ and $w_n \leq u_n$. The same argument using forth-down confluence instead of forth-up confluence suffices to also show 2.
\end{proof}

For the remainder of this section let $\mathcal{X}=(X, {\leq}, \ell, R)$ be a fixed $\Sigma$-quasimodel. Suppose that $\mathcal{X}$ falsifies some formula $\varphi \in \Sigma$. We are now going to show how to construct from $\mathcal X$ a \emph{total functional} $\Sigma$-quasimodel falsifying $\varphi$. Recall that $\defined{f(x)}$ denotes that $x \in dom(f)$ and $\undefined{f(x)}$ that $x \not \in dom(f)$ for $f$ a partial function.

\begin{definition}\index{induced structure}
    An \emph{$\mathcal{X}$-induced structure} is a tuple $\mathcal{I} =(I, {\leq_I}, \ell_I, f_I)$ together with a map $\pi\colon I\to X$ where:
    \begin{enumerate}
        \item $I$ is finite,
        \item $\leq_I$ is acyclic and if $w \leq_I v $ then $ \pi(w)\leq \pi(v)$,
        \item $\ell_I = \ell_X \circ \pi$, and
        \item $f_I: I \to I$ is a partial function such that:
        \begin{enumerate}
            \item If $\defined{f_I(x)}$, then $\pi(x)
        \mathrel R \pi ( f_I(x))$.

         \item If $x \leq_I y$ then $\defined{f(x)} \iff \defined{f(y)}$.
         
            \item If $x \leq_I y$ and $\defined{f(x)}$, then $f_I(x) \leq_I f_I(y)$.

            \item\label{maximal} For each $x\in I$ there is a maximal $k$ such that $\defined{f^k(x)}$.
        \end{enumerate}
    \end{enumerate}
\end{definition}

In view of \eqref{maximal} and the assumption that $I$ is finite, there is a maximal $k$ such that $f^k(x)$ is defined for \emph{any} $x\in I$, and hence we may define $W_{i} $ to be the set of $x\in I$ such that $f^{k-i}(x)$ is defined but $f^{k-i+1}(x)$ is not.
This partitions $I$ into sets $W_0,\ldots,W_k$, and it is easy to see that $x \leq_I y$ implies that $x,y\in W_i$ for some $i$, and moreover $f[W_i]\subseteq W_{i+1}$ for all $i$.

A \emph{defect} of an $\mathcal{X}$-induced structure records that a claim made by its labelling $\ell_I$ lacks a witness. 

\begin{definition}
    Let $\mathcal{I}$ be an $\mathcal{X}$-induced structure.
    \begin{enumerate}
        \item A $\rightarrow$-\emph{defect}\index{defect!$\rightarrow$-defect} is a pair $(x, \varphi \rightarrow \psi)$ where $x \in I$ and $\varphi \rightarrow \psi \in \ell_I(x)^-$, but there is no $y \geq_I x$ with $\varphi \in \ell_I(y)^+$ and $\psi \in \ell_I(y)^-$.
        \item A $\dimp$-\emph{defect}\index{defect!$\dimp$-defect} is a pair $(x, \varphi \dimp \psi)$ where $x \in I$ and $\varphi \dimp \psi \in \ell_I(x)^+$, but there is no $y \leq_I x$ with $\varphi \in \ell_I(y)^+$ and $\psi \in \ell_I(y)^-$.
                \item A $\X$-\emph{defect}\index{defect!$\X$-defect} is a world $x \in I$ with $\undefined{f_I(x)}$.
        \item A $\E$-\emph{defect}\index{defect!$\E$-defect} is a pair $(x, \E \varphi)$ where $x \in I$, $\undefined{f_I(x)}$, and $\E \varphi \in \ell_I(x)^+$, but $\varphi \not \in \ell_I(x)^+$.
        \item A $\G$-\emph{defect}\index{defect!$\G$-defect} is a pair $(x, \G \varphi)$ where $x \in I$, $\undefined{f_I(x)}$, and $\G \varphi \in \ell_I(x)^-$, but $\varphi \not \in \ell_I(x)^-$.
    \end{enumerate}
\end{definition}

Let $x \in X$ be such that $\varphi \in \ell(x)^-$. We build a total functional $\Sigma$-quasimodel falsifying $\varphi$ in stages. We start with an $\mathcal{X}$-induced structure $\mathcal{I}_0$ consisting of a single world and then construct in the  step $n+1$ an $\mathcal{X}$-induced structure $\mathcal{I}_{n+1}$ extending $\mathcal{I}_n$. We make use of a first-in-first-out queue $D$ that stores the defects of the current $\mathcal{X}$-induced structure. Observe that for any $\mathcal{X}$-induced structure, the set of defects of said structure is always finite (since the structure and $\Sigma$ are finite) and non-empty (due to $\X$-defects). The $\mathcal{X}$-induced structure $\mathcal{I}_n$ is defined by induction on $n$ as follows.\smallskip

\noindent \textsc{Base case:} Define $\mathcal{I}_0=(I_0, {\leq_0}, \ell_0, f_0)$, where $I_0 = \{x'\}$ (where $x'$ is a fresh world not occurring in $X$), ${\leq_0} = \{(x',x')\}$, $\ell_0(x') = \ell(x)$, $f_0 = \emptyset$, and $\pi_0(x') = x$. It is straightforward to check that $(\mathcal{I}_0, \pi_0)$ is an $\mathcal{X}$-induced structure. Initialise $D$ with all defects of $\mathcal{I}_0$ in arbitrary order.\smallskip

\noindent \textsc{Induction step:} Suppose we have defined $ \mathcal{I}_{n}=(I_n, {\leq_n}, \ell_n, f_n)$ and $\pi_n$, and shown that $(\mathcal{I}_{n},\pi_{n})$ is an $\mathcal{X}$-induced structure. By induction hypothesis, $D$ currently stores all defects of $\mathcal{I}_n$. We first show how to define $(\mathcal{I}_{n+1},\pi_{n+1})$ and then how to update the queue $D$. We start by setting $I_{n+1} = I_n$ and proceed by a case distinction on the defect at the head of the queue $D$.
\smallskip

\noindent \textsc{($\X$-defects)}  Suppose the defect at the head of  $D$ is a $\X$-defect $y \in I_n$. Choose any $u \in X$ with $\pi_n(y)\mathrel R u$. Add a new point $u'$ to $I_{n+1}$ and define $f_{n+1}(y)= u'$, $\ell_{n+1}(u') = \ell(u)$, and $\pi_{n+1}(u')=u$. We extend $f_{n+1}$ to the connected component of $y$ by adding new worlds, working first `bottom up' starting with worlds covering $y$.
           If $y'$ covers $y$, use forth-up confluence to find $z \in X$ with $\pi_n(y')\mathrel R z$. Add a fresh node $z'$ to $\mathcal{I}_{n+1}$ and define $f_{n+1}(y') = z'$, $u' \leq_{n+1} z'$  and close $\leq_{n+1}$ under transitivity and reflexivity,\footnote{We will always close $\leq_{n+1}$ under transitivity and reflexivity and will not mention it in the following items.}, $\pi_{n+1}(z')= z$, and $\ell_{n+1}(z')=\ell(z)$.
           Then for $y''$ covering such $y'$, use forth-up confluence again (relative to $y'$) to define $f_{n+1}(y'')$, and so on. Next repeat the process for those worlds below $\{y' \mid y' \geq y\}$ where $f_{n+1}$ is not yet defined, this time working `top down'  and using forth-down confluence. 
           Continue alternating between `bottom up' and `top down' until $f_{n+1}$ is defined on the connected component of $y$. 

This process terminates because the newly added points are not in the connected component of $y$, which is finite. Thus if the connected component of $y$ in $\mathcal I_n$ is of size $m$, then $m$ new points are added.

\smallskip  

\noindent \textsc{($\rightarrow$-defects)} Suppose the defect at the head of $D$ is a $\rightarrow$-defect $(y, \psi \rightarrow \chi)$. Then $\psi \rightarrow \chi \in \ell_n(y)^-$, but there is no $z' \in I_n$ with $y \leq_n z'$, and $\varphi \in \ell_n(z')^+$ and $\psi \in \ell_n(z')^-$. As $\mathcal{X}$ is a quasimodel, there exists $\pi_n(y) \leq z \in X$ with $\psi \in \ell(z)^+$ and $\chi \in \ell(z)^-$. Let $k $ be largest natural number such that $f^k_n(y)$ is defined. Using Lemma \ref{lemPathExtend}, find $z=z_0\mathrel R z_1\mathrel R\dots \mathrel R z_k$ with $\pi_n(f^i_n(y)) \leq z_i$. Then add fresh points $z'_0,\ldots,z'_k$ to $I_{n+1}$ and extend $\pi_n$, $\leq_n$, $f_n$ and $\ell_n$ by setting $\pi_{n+1}(z'_i)=z_i$, $f^i_n(y) \leq_{n+1} z'_i$, $f_{n+1}(z'_i)=z'_{i+1}$ and $\ell_{n+1}(z'_i)=\ell(z_i)$.

This process terminates because there is a maximal number $l$ such that every point in $I_n$ has at most $l$ temporal successors. Therefore $k \leq l$ new points are added.

\smallskip  

\noindent \textsc{($\dimp$-defects)}
        Suppose the defect at the head of $D$ is a $\dimp$-defect $(y, \psi \dimp \chi)$. The construction of $(\mathcal{I}_{n+1}, \pi_{n+1})$ is then just as for $\rightarrow$-defects.

\smallskip

\noindent \textsc{($\E$-defects)} Suppose the defect at the head of  $D$ is a $\E$-defect $(y, \E \psi)$. Then $\undefined{f_n(y)}$ and $\E \psi \in \ell_n(y)^+$, but $\psi \not \in \ell_n(y)^+$. As $\mathcal{X}$ is a quasimodel we find  $u_1, \ldots, u_n \in X$ with $\pi_n(y)\mathrel R u_1\mathrel R u_2\mathrel R \ldots\mathrel R u_n$ and $\psi \in \ell(u_n)^+$. We add worlds $u'_1, \ldots, u'_n$ to $I_{n+1}$ with $\pi_{n+1}(u'_i) = u_i$, $f_{n+1}(y) = u'_{1}$ and $f_{n+1}(u'_i) = u'_{i+1}$, and $\ell_{n+1}(u'_i) = 
        \ell (u_i)$.
        Then we proceed as in the case of a $\X$-defect to define $f_{n+1}$ on the connected component of $y$, and proceed inductively to define $f_{n+1}$ on the connected component of each $u'_i$.
        In this case, we must add $n$-many components for some natural number $n$.
        Hence, the construction for ‘next’-defects must be repeated $n$-many times.
        Thus the termination of this process is proven by induction on $n$, with a secondary induction on the number of worlds in a component as in the $\X$-defect case.

        \smallskip  

\noindent \textsc {($\G$-defects)}
        Suppose the defect at the head of $D$ is a $\G$-defect $(y, \G \psi)$. The construction is then as for $\E$-defects. \smallskip

\noindent \textsc{Updating D} We have shown how to construct $(\mathcal{I}_{n+1}, \pi_{n+1})$ from $(\mathcal{I}_n, \pi_n)$. Next we show how to update the queue $D$. First, delete every defect from $D$ that has been resolved in the construction of $(\mathcal{I}_{n+1}, \pi_{n+1})$ (observe that in each of the above cases it is possible that multiple defects have been resolved at once). Then each remaining defect is rewritten as follows. Remaining $\rightarrow$- or $\dimp$-defects are left unchanged. For an $\E$-defect $(y, \E \psi)$ if $\undefined{f_{n+1}(y)}$ holds, then the defect is left unchanged. Otherwise there are $u_1, \ldots, u_k$ with $f_{n+1}(y)=u_1, f_{n+1}(u_1)=u_2, \ldots, f_{n+1}(u_{k-1}) = u_k$ and $\undefined{f_{n+1}(u_k)}$. By assumption $\E \psi \in \ell_{n+1}(u_k)^+$ and $\psi \not \in \ell_{n+1}(u_k)^+$. Thus overwrite $(y, \E \psi)$ with $(u_k, \E \psi)$. The $\G$-defects and $\X$-defects are overwritten in the same way. Finally, add all \emph{new} defects of $\mathcal{I}_{n+1}$ to the tail of the queue.

\begin{lemma}
    For each $n < \omega$, $\mathcal{I}_n$ is an $\mathcal{X}$-induced structure.
\end{lemma}
\begin{proof}
    Proceed by induction on $n$. For the base case it is easily checked that $\mathcal{I}_0$ is an $\mathcal{X}$-induced structure. Suppose $\mathcal{I}_n$ is an $\mathcal{X}$-induced structure and consider $\mathcal{I}_{n+1}$. As shown in the construction above, $\mathcal{I}_{n+1}$ is obtained from $\mathcal{I}_n$ by resolving a defect of $\mathcal{I}_n$. Furthermore it was shown that for each defect only finitely many new points were added. Therefore $I_{n+1}$ is finite. That $\leq_{n+1}$ is acyclic follows from $\leq_n$ being acyclic and the fact that whenever a new point $x'$ is added above or below an existing point $x$, then $x'$ is only related (via $\leq_{n+1}$) to $x$ (and every point reachable via transitivity and reflexivity) but unrelated to any other point in $I_n$. For example, if $\mathcal{I}_{n+1}$ is obtained from resolving an implication defect $(x, \varphi \rightarrow \psi)$, it is possible that the construction finds $z \geq \pi_n(x)$ and there exists $y \geq_n x$ with $\pi(y) \geq z$. In that case, the construction does not add a new point $x'$ to $I_n$ with $x \leq_{n+1} x' \leq_{n+1} y$, but instead adds a fresh copy $x'$ to $I_n$ with $x' \geq_{n+1} x$ and $x'$ unrelated to $y$. This ensures that no cycles are created in the construction for each defect. If $x \leq_{n+1} y$, then either $x \leq_n y$ or (without loss of generality) $y$ was added in the construction to generate $\mathcal{I}_{n+1}$. In the first case $\pi_{n+1}(x) \leq \pi_{n+1}(y)$. In the second case, the construction for each defect guarantees that $\pi_{n+1}(x) \leq \pi_{n+1}(y)$. That $\ell_{n+1}(x) = \ell(\pi(x))$ follows directly from the induction hypothesis (in case $x \in I_n$) and from the construction otherwise. For 4. each item follows directly from the construction.
\end{proof}

Observe that by construction,  each  $\mathcal{I}_{n+1}$ contains $\mathcal{I}_n$ as a substructure. Define $(\mathcal{I}_\omega \coloneqq (I_\omega, {\leq_\omega}, \ell_\omega, f_\omega), \pi_\omega)$ where
\begin{equation*}
    \lambda_\omega = \bigcup_{n < \omega} \lambda_n
\end{equation*}
for $\lambda \in \{I, {\leq}, \ell, f, \pi\}$. Observe that $x' \in I_\omega$ with $\pi_\omega(x') = x$ and therefore $\varphi \in \ell_\omega(x')^-$. Thus $\mathcal{I}_\omega$ falsifies $\varphi$.

\begin{lemma}\label{l: I-omega is a labelled poset}
    The structure $(I_\omega, \leq_\omega, \ell_\omega)$ is a $\Sigma$-labelled poset.
\end{lemma}
\begin{proof}
    That $(I_\omega, \leq_\omega)$ is a partial order follows from the fact that $\leq_\omega  $ is an increasing union of partial orders.
    The labelling function $\ell_\omega$ satisfies the property that for each $y \in I_\omega$, $\ell_\omega(y) = \ell(\pi_\omega(y))$.
    Hence $\ell_\omega$ assigns $\Sigma$-types to worlds in $I_\omega$ and so $(I_\omega, \leq_\omega, \ell_\omega)$ is $\Sigma$-labelled. Suppose $y \leq_\omega z$. Then $\pi_\omega(y) \leq \pi_\omega(z)$, and thus $\ell(\pi_\omega(y)) \leq_{\mathrm{T}} \ell(\pi_\omega(z))$. From the previous observation, it follows that  $\ell_\omega(y) \leq_{\mathrm{T}} \ell_\omega(z)$. Suppose $y \in I_\omega$ and $\psi \rightarrow \chi \in \delta \ell_\omega(y)$. So $\psi \rightarrow \chi \in \ell_\omega(y)^-$ but $\psi \not \in \ell_\omega(y)^+$. Let $n$ be the least natural number such that $y \in I_n$. First suppose that there exists a world $z \in I_n$ with $y \leq_n z$ and $\psi \in \ell_n(z)^+$ and $\chi \in \ell_n(z)^-$. Then $z \in I_\omega$ and since $\ell_n (z) = \ell_\omega(z)$ we are done. Otherwise at the end of step $n$ the queue $D$ is updated and the defect $(y, \psi \rightarrow \chi)$ is added to $D$. Then there exists $k \geq n$ such that at step $k$ the defect $(y, \psi \rightarrow \chi)$ is resolved (either the defect is at the head of $D$ and is resolved \emph{actively} or another defect is resolved that also resolves $(y, \psi \rightarrow \chi)$). Thus in $\mathcal{I}_{k+1}$ there exists a world $z \geq_{k+1} y$ with $\psi \in \ell_n(z)^+$ and $\chi \in \ell_n(z)^-$. By construction $z \in I_\omega$. The case for co-implication defects is analogous.
\end{proof}

\begin{lemma}\label{l: f-omega is a total function}
    $f_\omega$ is a total function. 
\end{lemma}
\begin{proof}
    By construction $f_\omega \subseteq I_\omega \times I_\omega$. We check that $f_\omega$ is total and functional. For totality, let $y \in I_\omega$, and let $n$ be the least natural number for which $y \in I_n$. Then either $\defined{f_n(y)}$ or at step $n$ $y$ is added to $D$ as a $\X$-defect. In that case there exists $m \geq n$ at which $y$ is resolved, implying that $\defined{f_m(y)}$. Both cases imply that $\defined{f_\omega(y)}$. For functionality if $f_n(y) = z$, let us call $z$ a 'temporal successor of $y$'. It then suffices to observe---by inspection of the construction and property 4(b) of an $\mathcal{X}$-induced structure---that the construction adds for each world at most one temporal successor. 
\end{proof}

Combining these results we obtain the following lemma.

\begin{lemma}\label{l: I-omega is a functional quasimodel}
    $\mathcal{I}_\omega$ is a functional $\Sigma$-quasimodel falsifying $\varphi$.
\end{lemma}

\begin{proof}
That $(I_\omega, \leq_\omega, \ell_\omega)$ is a $\Sigma$-labelled poset is Lemma \ref{l: I-omega is a labelled poset}. That $f_\omega : I_\omega \to I_\omega$ is a total function is Lemma \ref{l: f-omega is a total function}. For forth--up and forth--{\allowbreak}down confluence observe that $f_\omega$ satisfies both of these confluence properties if and only if $f_\omega$ is order-preserving, due to the fact that $f_\omega$ is a function (c.f. Lemma \ref{l: functional models confluence conditions}). Thus suppose that $y \leq_\omega z$. Let $n$ be the least natural number for which $y,z \in I_n$ and $\defined{f_n(y)}$ and $\defined{f_n(z)}$. Observe that $y \leq_n z$. Since $\mathcal{I}_n$ is an $\mathcal{X}$-induced structure, Property 4(c) of such a structure guarantees that $f_n(y) \leq_n f_n(z)$. Hence $f_\omega(y) \leq_\omega f_\omega(z)$, i.e.~$f_\omega$ is order-preserving. Next, for any $y \in I_\omega$ by construction $\pi_\omega(y)\mathrel R \pi_\omega(f_\omega(y))$, and so $f_\omega$ is sensible. Next, suppose that for $y \in I_\omega$ and $\E \psi \in \Sigma$ it holds that $\E \psi \in \ell_\omega(y)^+$. If $\psi \in \ell_\omega(y)^+$, then we are done. Otherwise let $n$ be the least natural number for which $y \in I_n$. Since $\ell_n(y) = \ell_\omega(y)$ we have that $\E \psi \in \ell_n(y)^+$ and $\psi \not \in \ell_n(y)^+$. Suppose $f_n^k(y) = y_k$ and $\undefined{f_n(y_k)}$. Then either there exists $1 \leq i \leq k$ with $\psi \in \ell_n(f_n^i(y))^+$ or $(y_k, \E \psi)$ is added to the queue $D$ as a $\E$-defect. In the first case $\psi \in \ell_\omega(f_\omega^i(y))^+$. In the second case let $l$ be the natural number bigger $n$ such that in the $l$-th step the defect $(y_k, \E \psi)$ (or its corresponding defect obtained from rewriting $(y_k, \E \psi)$ finitely many times) is deleted from $D$. Thus $\mathcal{I}_l$ contains a world $u$ such that $f_l^j(y) = u$ for some $j \geq 1$ and $\psi \in \ell_l(u)^+$. Hence there exists $j < \omega$ with $\psi \in \ell_\omega(f_\omega^j(y))^+$. The case for $\G \psi$ is similar and therefore $f_\omega$ is $\omega$-sensible. Finally, by construction, $x' \in I_\omega$ and $\varphi \in \ell_\omega(x')^-$. Putting everything together, $\mathcal{I}_\omega$ is a functional $\Sigma$-quasimodel falsifying $\varphi$.
\end{proof}

Finally, we obtain the main result of this section:

\begin{theorem}\label{theoQuasiToMod}
    A formula $\varphi$ is falsifiable over the class of expanding models if and only if $\varphi$ is falsifiable over the class of $\Sigma$-quasimodels.
\end{theorem}

\begin{proof}
    The left-to-right direction is Lemma \ref{l: falsifiable over expanding models}. For the right-to-left direction, suppose  $\varphi$ is falsifiable over the class of $\Sigma$-quasimodels. Hence there exists a $\Sigma$-quasimodel $\mathcal{X}=(X, {\leq}, \ell, R)$ and $x \in X$ with $\varphi \in \ell(x)^-$. 
    By Lemma \ref{l: I-omega is a functional quasimodel} there exists a functional $\Sigma$-quasimodel falsifying $\varphi$. So by Lemma \ref{l: falsifiability of functional quasimodels} there exists an expanding model falsifying $\varphi$. Thus $\varphi$ is falsifiable over the class of expanding models.
\end{proof}

\section{Simulations}\label{SecSim}

A key ingredient in our completeness proof will be to relate worlds in a finite quasimodel to prime theories in the canonical model. As in chapter \ref{c: bi-int ml new} \emph{dynamic simulations} will be used to achieve this. Let us briefly recall the definitions.

\begin{definition}\index{simulation}
Let $\Sigma \subseteq \Delta \subseteq \Lbiltl$ be subformula closed, and let $\mathcal{X} = (X, {\leq_\mathcal{X}}, \ell_\mathcal{X})$ and $\mathcal{Y} = (Y, {\leq_\mathcal{Y}}, \ell_\mathcal{Y})$ be $\Sigma$-labelled and $\Delta$-labelled posets respectively. A binary relation $E \subseteq X \times Y$ is a \emph{simulation} if the following hold:
\begin{enumerate}
    \item If $x \mathrel E y$, then $\ell_\mathcal{X}(x) \subseteq_\mathrm{T} \ell_\mathcal{Y}(y)$.
    
    \item $E$ is forth-up and forth-down confluence with respect to $(\leq_\mathcal{X}, \leq_\mathcal{Y})$.
\end{enumerate}
If there exists a simulation $E$ such that $x \mathrel E y$, then we write $(\mathcal{X},x) \rightharpoonup (\mathcal{Y},y)$.
\end{definition}

The following result is crucial for the completeness argument, yet straighforward to prove. Given labelled systems $\mathcal{X}=(X, \leq_\mathcal{X}, \ell_\mathcal{X}, R_\mathcal{X})$ and $\mathcal{Y}=(Y, \leq_\mathcal{Y}, \ell_\mathcal{Y}, R_\mathcal{Y})$ and a simulation $E \subseteq X \times Y$, let $X{\upharpoonright_{dom(E)}}$ be the structure obtained from restricting $\mathcal{X}$ to the domain of $E$, as before.

\begin{lemma}\label{isquasi}
Let $\mathcal X$, $\mathcal Y$ be labelled systems and $E \subseteq X\times Y$  a simulation.
Then $\mathcal X{\upharpoonright_{dom(E)}}$ is a labelled system.
\end{lemma}
\begin{proof}
    Denote $\mathcal X{\upharpoonright_{dom(E)}}$ by $\mathcal{Z}=(Z, \leq_\mathcal{Z}, \ell_\mathcal{Z}, R_\mathcal{Z})$ where $Z = dom (E)$. That $(Z, \leq_\mathcal{Z})$ is a poset follows directly from $(X, \leq_\mathcal{X})$ being a poset and $(Z, \leq_\mathcal{Z})$ being the restriction of $(X, \leq_\mathcal{X})$ to the domain of $E$. If $x \leq_\mathcal{Z} z$, then $x \leq_\mathcal{X} z$ and therefore $\ell_\mathcal{X}(x) \leq_{\mathrm{T}} \ell_\mathcal{X}(z)$. Since $\ell_\mathcal{X}(x') = \ell_\mathcal{Z}(x')$ for all $x' \in Z$ it follows that $\ell_\mathcal{Z}(x) \leq_{\mathrm{T}} \ell_\mathcal{Z}(z)$. For the defects suppose that $\varphi \rightarrow \psi \in \ell_\mathcal{Z}(x)^-$. Then $\varphi \rightarrow \psi \in \ell_\mathcal{X}(x)^-$. Since $\mathcal{X}$ is a labelled system, there exists $z \in X$ with $x \leq_\mathcal{X} z$ and $\varphi \in \ell_\mathcal{X}(z)^+$ and $\psi \in \ell_\mathcal{X}(z)^-$. As $x \in Z$ there exists $y \in Y$ with $x \mathrel{E} y$. By forth-up confluence there exists $y' \in Y$ with $y \leq_\mathcal{Y} y'$ and $z \mathrel{E} y'$. Hence $z \in Z$, $x \leq_\mathcal{Z} z$ and $\varphi \in \ell_\mathcal{Z}(z)^+$ and $\psi \in \ell_\mathcal{Z}(z)^-$. The case for $\dimp$-defects is similar. Hence $(Z, \leq_\mathcal{Z}, \ell_\mathcal{Z})$ is a labelled poset and it remains to show that $R_\mathcal{Z}$ is sensible. This follows immediately from the fact that $R_\mathcal{X}$ is sensible and $R_\mathcal{Z} = R_\mathcal{X} \cap (Z \times Z)$.
\end{proof}

 Finally, let us briefly recall the definition of a \emph{dynamic simulation}.

\begin{definition}\label{d-simulation}\index{simulation!dynamic}
Let $\mathcal X =(X,{\leq_\mathcal X},\ell_\mathcal X,R_\mathcal X)$ and $\mathcal Y=(Y,{\leq_\mathcal Y},\ell_\mathcal Y,R_\mathcal Y)$ be labelled systems. A \emph{dynamic simulation} is a simulation $E \subseteq X\times Y$ such that whenever $x\mathrel E y \mathrel R_\mathcal Y y'$ then there exists $x'$ such that $x\mathrel R_\mathcal X x'\mathrel E y'$.
\end{definition}

\section{Finite Quasimodel Property}\label{secUSigma}

In this section, we show that every falsifiable formula is falsified on a finite quasimodel. We wish to apply the techniques and results from Chapter \ref{c: bi-int ml new} to $\biltl$, where we proved that every $\Lbi$-formula falsified on an expanding model is falsified in a finite labelled model. The caveat is that the finite model property was proven for a language with two modalities, which here we denote $\blacklozenge$ and $\blacksquare$. These do not correspond to our own $\E$ and $\G$, but rather \emph{both} correspond to $\X$, which, being a functional modality, is self dual.

In order to apply the results from Chapter \ref{c: bi-int ml new}, let us start by considering only the language $\Lbiltlnext$. Recall the definition of a $\Sigma$-type for the language $\Lbi$ (c.f. Definition \ref{d: type}) and identify both $\blacklozenge$ and $\blacksquare$ with $\X$. To avoid confusion, we denote these types by $\type_{\Sigma}^{\X}$. Recall the definition of a sensibility condition (c.f. Definition \ref{d: sensibility condition}). The following theorem combines the results proven in Chapter \ref{c: bi-int ml new} and applied to $\Lbiltlnext$.


\begin{theorem}\label{thmISig}
Let $\Sigma\subseteq \Lbiltlnext$ be finite and closed under subformulas and $S \subseteq \type_\Sigma^{\X} \times \type_\Sigma^{\X}$ a  sensibility condition.
Then there exists a finite acyclic $\Sigma$-labelled system $\mathbb U= \mathbb U(\Sigma,S)$ that is sensible with respect to $S$ and such that for every $\Sigma$-labelled system $\mathcal X=(X,{\leq_\mathcal X},\ell_\mathcal X, R_\mathcal X)$ there exists a strongly surjective dynamic simulation $E_* \subseteq U \times X$, where $U$ is the set of worlds of $\mathbb{U}$.
\end{theorem}

Note that the structure $\mathbb{U}$ yields finite quasimodels for $\Lbiltlnext$. In order to apply Theorem \ref{thmISig} to the full language $\Lbiltl$, the idea is to regard each formula of the form $\E \varphi$ or $\G \varphi$ as a proposition, but replace e.g.~$\G \varphi$ by $\varphi\wedge\G\varphi$ in order to ensure that our types still satisfy the correct conditions for $\G$ and $\E$. To that end we introduce the concept of a \emph{fleeting type}. The intuition is that a fleeting type is a $\Sigma$-type which does not satisfy the conditions involving $\E$ and $\G$. Therefore, when uniformly replacing instances of $\E$ and $\G$ by fresh propositions, a fleeting type becomes simply a $\type_\Sigma^{\X}$-type.

\begin{definition}\index{type!fleeting}
    Let $\Sigma$ be a finite and subformula closed set of $\Lbiltl$-formulas and $\Phi^+, \Phi^- \subseteq \Sigma$. The pair $(\Phi^+, \Phi^-)$ is a \emph{fleeting type} if it satisfies the conditions 1. - 7. of Definition \ref{d: Sigma type}. Denote the set of all fleeting $\Sigma$-types by $\type_\Sigma^{\mathsf{f}}$.
\end{definition}

 A \emph{fleetingly labelled poset} is then a labelled poset, albeit labelled by fleeting types. The formal definition is as follows.

\begin{definition}\index{labelled poset!fleetingly}
    A \emph{fleetingly labelled poset} is  a tuple $\mathcal{X}=(X, \leq_{\mathcal{X}}, \ell_{\mathcal{X}})$ such that
    \begin{enumerate}
        \item $(X, \leq_\mathcal{X})$ is a poset.
        \item $\ell_{\mathcal{X}}: X \longrightarrow \type_\Sigma^{\mathsf{f}} \times \type_\Sigma^{\mathsf{f}}$ such that
        \begin{enumerate}
            \item if $x \leq_{\mathcal{X}} y$, then $\ell_{\mathcal{X}}(x) \leq_\type \ell_{\mathcal{X}}(y)$.
             \item if $\varphi \rightarrow \psi \in \ell_\mathcal{X}(x)^-$, then there exists $y \geq_\mathcal{X} x$ with $\varphi \in  \ell_\mathcal{X}(y)^+$ and $\psi \in  \ell_\mathcal{X}(y)^-$.
        \item if $\varphi \dimp \psi \in \ell_\mathcal{X}(x)^+$, then there exists $y \leq_\mathcal{X} x$ with $\varphi \in  \ell_\mathcal{X}(y)^+$ and $\psi \in  \ell_\mathcal{X}(y)^-$.
        \end{enumerate}
    \end{enumerate}
\end{definition}

For a formula $\varphi \in \Lbiltl$, define $\tau(\varphi)$ by recursively replacing 
instances of $\G\varphi$ by $\varphi\wedge{\X\G}\varphi$, and instances of $\E\varphi$ by $\varphi\vee\X\E\varphi$.
Formally, $\tau (\varphi)$ is definded inductively as follows.
\begin{center}
\begin{tabular}{l l l}
   $\tau (p)$ & $\coloneqq$ & $p$ for $p \in \Prop$ \\
   $\tau ({ \varphi \ast \psi})$ & $\coloneqq$ & $\tau (\varphi)\ast \tau (\psi)$ for $\ast \in \{\wedge, \vee, \rightarrow, \dimp\}$\\
   $ \tau({\X\varphi})$ & $\coloneqq$ & $\X\tau(\varphi)$ \\
   $\tau({\G\varphi})$ & $\coloneqq$ & $\tau(\varphi)\wedge \X {\G \tau (\varphi)}$\\
   $ \tau({\E\varphi})$ & $\coloneqq$ & $\tau(\varphi)\vee \X {\E\tau (\varphi)}$ \\
\end{tabular}
\end{center}

Given a set of formulas $\Sigma \subseteq \Lbiltl$, define $\tau(\Sigma)=\{\tau(\varphi) \mid \varphi\in \Sigma\}$.
Similarly, if $\Phi$ is a fleeting type then $\tau (\Phi)=(\tau ({\Phi^+}),\tau ({\Phi^-}))$, and if $\mathcal X$ is a fleetingly labelled poset then $\tau({\mathcal X})$ is the same as $\mathcal X$ but labelled by $\tau ({\ell_\mathcal X})$.
Finally, for a set of formulas $\Theta\subseteq\tau(\Sigma)$ define $\tau^{-1}(\Theta) = \{\varphi \in \Sigma \mid \tau(\varphi) \in \Theta\}$, and for a type $\Phi$ define $\tau^{-1} (\Phi) = (\tau^{-1} ({\Phi^+}),\tau^{-1}({\Phi^-}))$.

\begin{lemma}\label{lemWidehat}
For every formula $\varphi \in \Lbiltl$, $\vdash\varphi\leftrightarrow \tau(\varphi)$.
Moreover, if $\Phi$ is a fleeting $\tau(\Sigma)$-type, then $\tau^{-1}(\Phi)$ is a $\Sigma$-type.
\end{lemma}
\begin{proof}
    Recall that $\vdash \E \varphi \leftrightarrow \varphi \vee \X \E \varphi$ and $\vdash \G \varphi \leftrightarrow \varphi \wedge {\X \G} \varphi$. Therefore $\vdash \varphi \leftrightarrow \tau(\varphi)$ follows by a standard induction on the structure of $\varphi$. Now suppose that $\Phi = (\Phi^+, \Phi^-)$ is a fleeting $\tau(\Sigma)$-type. Then $\tau^{-1}(\Phi) \subseteq \Sigma \times \Sigma$. We check that $(\tau^{-1}(\Phi^+), \tau^{-1}(\Phi^-))$ is a $\Sigma$-type by checking all clauses of Definition \ref{d: Sigma type}. The conditions for the connectives $\wedge, \vee,\rightarrow, \dimp$ are straighforward. For example if $\varphi \rightarrow \psi \in \tau^{-1}(\Phi)^+$, then $\tau(\varphi \rightarrow \psi) \in \Phi^+$. By definition $\tau(\varphi \rightarrow \psi) = \tau(\varphi) \rightarrow \tau(\psi)$ and since $\Phi$ is a fleeting type, we have that $\tau(\varphi) \in \Phi^-$ or $\tau(\psi) \in \Phi^+$. Hence $\varphi \in \tau^{-1}(\Phi^-)$ or $\psi \in \tau^{-1}(\Phi^+)$. The interesting cases are for $\G$ and $\E$. First suppose $\G \varphi \in \tau^{-1}(\Phi^+)$. Then $\tau(\G \varphi) \in \Phi^+$. By definition $\tau(\G \varphi) = \tau(\varphi) \wedge {\X \G} \tau(\varphi)$. Since $\Phi$ is a fleeting type, we have $\tau(\varphi) \in \Phi^+$ and hence $\varphi \in \tau^{-1}(\Phi^+)$. Similarly if $\E \varphi \in \tau^{-1}(\Phi^-)$, then $\tau(\E \varphi) \in \Phi^-$. Since $\tau(\E \varphi) = \varphi \vee \X \E \tau(\varphi)$ we have by definition of a fleeting type that $\tau(\varphi) \in \Phi^-$ and hence that $\varphi \in \tau^{-1}(\Phi^-)$. We conclude that $\tau^{-1}(\Phi)$ is a $\Sigma$-type.
\end{proof}

It remains to define a suitable sensibility condition in order to apply Theorem \ref{thmISig}.
To this end, let $\Phi, \Psi$ be fleeting $\tau(\Sigma)$-types and define $\Phi \mathrel{S} \Psi$ if and only if $(\tau^{-1}(\Phi), \tau^{-1}(\Psi))$ is sensible.

\begin{lemma}
$S$ is a sensibility condition.
\end{lemma}
\begin{proof}
     Suppose $\Phi \mathrel{S} \Psi$ and $\Delta$ is a set of formulas closed under subformulas. Note that $(\tau^{-1}(\Phi) \upharpoonright_{\tau^{-1}(\Delta)}, \tau^{-1}(\Psi) \upharpoonright_{\tau^{-1}(\Delta)})$ is sensible. For example if $\X \varphi \in \tau^{-1}(\Delta)$ and $\X \varphi \in \tau^{-1}(\Phi^+)$, then $\X \tau(\varphi) \in \Delta$ and since $\Delta$ is subformula closed, $\tau(\varphi) \in \Delta$, implying that $\varphi \in \tau^{-1}(\Delta)$ and so, since $(\tau^{-1}(\Phi), \tau^{-1}(\Psi))$ is sensible, we have $\varphi \in \tau^{-1}(\Psi^+)$. The other cases are similar. Hence we obtain $\Phi \upharpoonright_\Delta \mathrel{S} \Psi \upharpoonright_\Delta$. Next, suppose $\Psi \subseteq \Psi'$. Then $\tau^{-1}(\Psi) \subseteq \tau^{-1}(\Psi')$ and since $(\tau^{-1}(\Phi), \tau^{-1}(\Psi))$ is sensible, so is $(\tau^{-1}(\Phi), \tau^{-1}(\Psi'))$ (recall that the conditions of `sensible' only go in one direction). Therefore $\Phi \mathrel{S} \Psi'$. We conclude that $S$ is a sensibility condition.
\end{proof}

Recall that the canonical model for the full language $\Lbiltl$ is denoted by $\fw_c$. 
 
\begin{proposition}\label{propSimMc}
Let $\Sigma \subseteq \Lbiltl$ be finite and closed under subformulas.
Then there exists a finite, acyclic $\Sigma$-labelled system $\sfrac{\mathcal M_\mathrm{c}}\Sigma $ and a strongly surjective dynamic simulation $E_*$ between $\sfrac{\mathcal M_\mathrm{c}}{\Sigma}$ and $\fw_c$.
Specifically, $E_*$ is the union of all simulations between the two structures.
\end{proposition}

\begin{proof}
By Theorem \ref{thmISig} there exists a finite, acyclic $\tau(\Sigma)$-labelled system $\mathbb U(\tau(\Sigma), S) = (U,\leq_{\mathbb{U}}, R_{\mathbb{U}},\ell_{\mathbb{U}} )$ and a strongly surjective dynamic simulation $E_\ast$ between $\mathbb U(\tau(\Sigma), S)$ and $\fw_c$. Define $J= dom(E_*)$ and set 
\begin{equation*}
    \sfrac{\mathcal M_\mathrm{c}}\Sigma =(J,\allowbreak{{\leq_\mathbb U}{\upharpoonright_J}},\allowbreak {{R_\mathbb U}{\upharpoonright_J}}, \allowbreak\tau^{-1}({\ell_{\mathbb{U}}}{\upharpoonright_J})).
\end{equation*}
In other words, $\sfrac{\mathcal M_\mathrm{c}}\Sigma$ is the same as $\mathbb U(\tau(\Sigma), S){\upharpoonright_{dom(E_*)}} $, except that $\ell_{\sfrac{\mathcal M_\mathrm{c}}\Sigma} (w) = \tau^{-1} {\ell_{\mathbb{U}}} (w)$ for every $w\in dom(E_*) $. By Lemma \ref{isquasi}, $\mathbb U(\tau(\Sigma), S){\upharpoonright_{dom(E_*)}} $ is a $\tau(\Sigma)$-labelled system, implying that $(J, \leq_{\sfrac{\mathcal M_\mathrm{c}}\Sigma})$ is a poset. We check the remaining conditions of a $\Sigma$-labelled poset. The labelling function $\ell_{\sfrac{\mathcal M_\mathrm{c}}\Sigma}$ assigns to each world $w \in J$ the label $\tau^{-1}(\ell_{\mathbb{U}}(w))$. Since $\ell_{\mathbb{U}}(w)$ is a fleeting $\tau(\Sigma)$-type, Lemma \ref{lemWidehat} implies that $\tau^{-1}(\ell_{\mathbb{U}}(w))$ is a $\Sigma$-type, implying that $\ell_{\sfrac{\mathcal M_\mathrm{c}}\Sigma}: J \longrightarrow \type_\Sigma$. For the defects suppose that $\varphi \rightarrow \psi \in \ell_{\sfrac{\mathcal M_\mathrm{c}}\Sigma}(w)^-$. Then $\tau(\varphi) \rightarrow \tau(\psi) \in \ell_{\mathbb{U}}(w)^-$. Since $\mathbb U(\tau(\Sigma), S){\upharpoonright_{dom(E_*)}}$ is a labelled system, there exists $v \geq_{\mathbb{U}} w$ with $\tau(\varphi) \in \ell_{\mathbb{U}}(v)^+$ and $\tau(\psi) \in \ell_{\mathbb{U}}(v)^+$ and $v \in dom(E_*)$. Hence, there exists $v \geq_{\sfrac{\mathcal M_\mathrm{c}}\Sigma} w$ with $\varphi \in \ell_{\sfrac{\mathcal M_\mathrm{c}}\Sigma}(v)^+$ and $\psi \in \ell_{\sfrac{\mathcal M_\mathrm{c}}\Sigma}(v)^-$. The case for $\dimp$ is similar and so we conclude that $(J, \leq_{\sfrac{\mathcal M_\mathrm{c}}\Sigma}, \ell_{\sfrac{\mathcal M_\mathrm{c}}\Sigma})$ is a $\Sigma$-labelled poset. Next, notice that $R_{\sfrac{\mathcal M_\mathrm{c}}\Sigma}$ is forth-up and forth-down confluent, since $R_{\sfrac{\mathcal M_\mathrm{c}}\Sigma} = R_{\mathbb{U}} \upharpoonright_{J}$. Moreover, whenever $w \Rel_{\sfrac{\mathcal M_\mathrm{c}}\Sigma} v$, then $\ell_{\mathbb{U}}(w) \mathrel{S} \ell_{\mathbb{U}}(v)$, implying that the pair $(\ell_{\sfrac{\mathcal M_\mathrm{c}}\Sigma}(w), \ell_{\sfrac{\mathcal M_\mathrm{c}}\Sigma}(v))$ is sensible. Hence $\sfrac{\mathcal M_\mathrm{c}}\Sigma$ is a $\Sigma$-labelled system. Since $\mathbb{U}(\tau(\Sigma), S)$ is finite and acyclic, so is $\sfrac{\mathcal M_\mathrm{c}}\Sigma$. \smallskip

It remains to show that $E_*$ is a strongly surjective dynamic simulation. By assumption, $E_*$ is a simulation between $\mathbb{U}(\tau(\Sigma), S)$, implying that $E_*$ is forth-up and forth-down confluent with respect to $(\leq_{\sfrac{\mathcal M_\mathrm{c}}\Sigma}, \leq_c)$ where $\leq_c$ is the intuitionistic order of $\fw_c$. Suppose $w \in J$ and $\Gamma \in W_c$ with $w \mathrel{E_*} \Gamma$. If $\varphi \in \ell_{\sfrac{\mathcal M_\mathrm{c}}\Sigma}(w)^+$, then $\tau(\varphi) \in \ell_{\mathbb{U}}(w)^+$ and so $\tau(\varphi) \in \Gamma$. Since $\vdash \varphi \leftrightarrow \tau(\varphi)$ by Lemma \ref{lemWidehat} and $\Gamma$ is a prime theory and hence deductively closed, we have $\varphi \in \Gamma$. Similarly, if $\varphi \in \ell_{\sfrac{\mathcal M_\mathrm{c}}\Sigma}(w)^-$, then $\varphi \not \in \Gamma$ by the same argument. Thus $E_*$ is a simulation. Now suppose $w \mathrel{E} \Gamma$ and $\Gamma \Rel_c \Delta$. Then there exists $v \in U$ with $w \Rel_{\mathbb{U}} v$ and $v E_* \Delta$, since $E_*$ is a dynamic simulation between $\mathbb{U}(\tau(\Sigma), S)$. Thus $v \in dom(E_*)$, implying that $v \in J$. Therefore $w \Rel_{\sfrac{\mathcal M_\mathrm{c}}\Sigma} v$ and $v E_* \Delta$, implying that $E_*$ is a dynamic simulation. Finally, let $w \in J$. Then $w \in U$ and since $E_*$ is strongly surjective there exists $\Gamma \in W_c$ with $w \mathrel{E_*} \Gamma$ and $\ell_{\mathbb{U}}(w) = \Gamma \cap \tau(\Sigma)$. Since $\vdash \varphi \leftrightarrow \tau(\varphi)$ we thus obtain that $\ell_{\sfrac{\mathcal M_\mathrm{c}}\Sigma}(w) = \Gamma \cap \Sigma$, implying that $E_*$ is strongly surjective.
\end{proof}

Therefore if a formula $\varphi \in \Lbiltl$ is falsifiable, we have $\not \vdash \varphi$ by soundness. By the Lindenbaum Lemma there exists a prime theory $\Gamma \in W_c$ with $\varphi \not \in \Gamma$. Thus there exists a world $w \in J$ with $\varphi  \in \ell_{\sfrac{\mathcal M_\mathrm{c}}\Sigma}(w)^-$. To obtain completeness, in the presence of Theorem \ref{theoQuasiToMod} it thus suffices to show that $\sfrac{\mathcal M_\mathrm{c}}\Sigma$ is not only a labelled system, but in fact a quasimodel. This is in no way trivial to establish, and will be done in the next sections using so called \emph{simulation formulas}.

\section{Simulation Formulas}\label{secSimForm}

As before, $\Sigma \subseteq \Lbiltl$ is assumed to be finite and closed under subformulas. Consider a $\Sigma$-labelled poset  $\mathcal{X}=(X, \leq_\mathcal{X}, \ell_\mathcal{X})$ and let $x, y \in X$. Recall that a \emph{non-repeating path} from $x$ to $y$ is a finite sequence $(\rho(i))_{i \leq n}$ of \emph{(pairwise) distinct} worlds, such that $\rho(0) = x$, $\rho(n) = y$, and for all $0 \leq i <n$ either $\rho(i)$ covers $\rho(i+1)$ or $\rho(i+1)$ covers $\rho(i)$. If $\rho = (\rho(i))_{i \leq n}$, the \emph{length} $\lvert \rho \rvert$ of $\rho$ is $n$. Let
\begin{equation*}
    \mathsf{ZZP}(x) \coloneqq \{ \rho \mid \rho \text{ is a non-repeating path starting at } x\}.
\end{equation*}

\begin{figure}\centering
    \begin{tikzpicture}
\draw   (-1.5,-0.5)-- (-1,-1) --(0,0) -- (1,1) -- (1.5,0.5)  ;
\draw (.5,.5) -- (1,0);
\draw (-.5,-.5) -- (-1,0);
\node at (-1.5,-0.5)[shape=circle,draw,fill=white,inner sep=1pt]  {\tiny 0};
\node at (-1,-1)[shape=circle,draw,fill=white,inner sep=1pt]  {\tiny 1};
\node at (-.5,-.5)[shape=circle,draw,fill=white,inner sep=1pt]  {\tiny 2};
\node at (0,0)[shape=circle,draw,fill=white,inner sep=1pt]  {\tiny 3};
\node at (.5,.5)[shape=circle,draw,fill=white,inner sep=1pt]  {\tiny 2};
\node at (1,1)[shape=circle,draw,fill=white,inner sep=1pt]  {\tiny 1};
\node at (1.5,.5)[shape=circle,draw,fill=white,inner sep=1pt]  {\tiny 0};
\node at (1,0)[shape=circle,draw,fill=white,inner sep=1pt]  {\tiny 0};
\node at (-1,0)[shape=circle,draw,fill=white,inner sep=1pt]  {\tiny 0};

\node at (0,0)[anchor=south east]{$x$};

    \end{tikzpicture}\caption{Example of an  $x$-induced tree $\mathsf T(x)$, with heights}\label{figure:tree}
\end{figure}
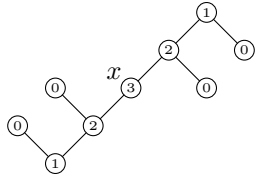

\begin{definition}\index{induced tree}
    Let $\mathcal{X}=(X, \leq_\mathcal{X}, \ell_\mathcal{X})$ be a finite, acyclic $\Sigma$-labelled poset and $x \in X$. The \emph{$x$-induced tree} is defined as $\mathsf{T}(x)\coloneqq (\mathsf{ZZP}(x), \sqsubset)$, where $\rho \sqsubset \rho'$ if and only if $\rho$ is a proper initial segment of $\rho'$ (see Figure \ref{figure:tree}).
\end{definition}

Observe that $\mathsf{T}(x)$ is a \emph{finite} tree with the path $(x)$ as \emph{root}. As before, Given $\rho \sqsubseteq \rho'$, we write $\rho' - \rho$ for the final segment of $\rho'$ after $\rho$. Moreover, we write ${\uparrow}(\rho' - \rho)$ if each element in $\rho' - \rho$ \emph{covers} its predecessor in $\rho'$, and ${\downarrow}(\rho' - \rho)$ if each element in $\rho' - \rho$ \emph{is covered} by its predecessor in $\rho'$. For $\rho \in \mathsf{ZZP}(x)$, the \emph{height} of $\rho$ is defined as $h(\rho) \coloneqq \max\{\lvert \rho' - \rho \rvert \mid \rho \sqsubseteq \rho'\}$. 

We now define, for $x$ in a finite, acyclic labelled poset, the simulation formulas $\chi^+(x)$ and $\chi^-(x)$, which together encode all worlds accessible from $x$ via a non-repeating path. Therefore, satisfying or falsifying $\chi^+(x)$ or $\chi^-(x)$ respectively at some world $y$ of a labelled poset is equivalent to the existence of a simulation involving $x$ and $y$; see Proposition \ref{propSimForm}.

We define $\chi^+(x)$ and $\chi^-(x)$ by working `outside-in' on $\mathsf{T}(x)$, i.e.~recursively from the leaves of $\mathsf T(x)$ to the root, exploiting the following.
\begin{enumerate}[label=(\roman*)]
\item 
By \emph{asserting} a formula $\varphi \dimp \psi$ we can express that there is a world \emph{below} where $\varphi$ holds and $\psi$ does not.
\item 
By \emph{denying} a formula $\varphi \rightarrow \psi$ we can express that there is a world \emph{above} where $\varphi$ holds and $\psi$ does not.
\end{enumerate}

We begin by defining for each path $\rho$ in $\mathsf T(x)$ different from $(x)$ a formula $\varphi_\rho$. The simulation formulas are then composed from these formulas $\varphi_\rho$. Recall that by convention $\bigwedge \emptyset \coloneqq \top$ and $\bigvee \emptyset \coloneqq \bot$.

\begin{definition}\label{def:simulation}\index{simulation!formula}
   Let $\mathcal{X}=(X, \leq_\mathcal{X}, \ell_\mathcal{X})$ be a finite, acyclic $\Sigma$-labelled poset, and $x \in X$. 
   For each $\rho = (\rho(0),\allowbreak \ldots,\allowbreak \rho(n)) \in \mathsf T(x)$ with $\lvert \rho \rvert > 0$ define the formula $\varphi_\rho$ by induction on $h(\rho)$. Suppose $\varphi_\rho'$ has been defined for each $\rho'$ with $h(\rho') < h(\rho)$.
   \begin{enumerate}
   \item  If $\rho(n-1) >_\mathcal{X} \rho(n)$, define
       \begin{equation*}
       \begin{split}
           \varphi_\rho \coloneqq &(\bigwedge \ell_\mathcal{X}(\rho(n))^+ \wedge \bigwedge_{\rho' \sqsupset \rho \colon  {\downarrow}(\rho' - \rho)} \varphi_{\rho'}) \dimp \\&(\bigvee \ell_\mathcal{X} (\rho(n))^- \vee \bigvee_{\rho' \sqsupset \rho \colon {\uparrow}(\rho' - \rho)} \varphi_{\rho'})
           \end{split}
       \end{equation*}
       \item  If $\rho(n-1) <_\mathcal{X} \rho(n)$, define
       \begin{equation*}
       \begin{split}
           \varphi_\rho \coloneqq & (\bigwedge \ell_\mathcal{X}(\rho(n))^+ \wedge \bigwedge_{\rho' \sqsupset \rho \colon {\downarrow}(\rho' - \rho)} \varphi_{\rho'}) \rightarrow \\& (\bigvee \ell_\mathcal{X} (\rho(n))^- \vee \bigvee_{\rho' \sqsupset \rho \colon {\uparrow}(\rho' - \rho)} \varphi_{\rho'})
           \end{split}
       \end{equation*}

   \end{enumerate}
   Then define $\chi^+(x)$ and $\chi^-(x)$ as follows. 
  \begin{itemize}
      \item[] \begin{equation*}
           \begin{split}
           \chi^+(x) \coloneqq &(\bigwedge \ell_\mathcal{X}(x)^+ \wedge \bigwedge_{\rho \sqsupset (x) \colon {\downarrow}(\rho - (x))} \varphi_{\rho}) \dimp \\&(\bigvee \ell_\mathcal{X} (x)^- \vee \bigvee_{\rho \sqsupset (x) \colon {\uparrow}(\rho - (x))}  \varphi_{\rho})
           \end{split}
       \end{equation*}

       \item[] \begin{equation*}
       \begin{split}
           \chi^-(x) \coloneqq &(\bigwedge \ell_\mathcal{X}(x)^+ \wedge \bigwedge_{\rho \sqsupset (x) \colon {\downarrow}(\rho - (x))} \varphi_{\rho}) \rightarrow \\&(\bigvee \ell_\mathcal{X}(x)^- \vee \bigvee_{\rho \sqsupset (x) \colon {\uparrow}(\rho -(x))} \varphi_{\rho})
           \end{split}
       \end{equation*}
  \end{itemize}
\end{definition}

For a path $\rho = (\rho(i))_{i\leq n}$ let $\final(\rho) \coloneqq \rho(n)$, i.e. $\final(\rho)$ denotes the final element of $\rho$. Let
\begin{equation*}
    \final[\mathsf{ZZP}(x)] = \{ \final(\rho) \mid \rho \in \mathsf{ZZP}(x)\}.
\end{equation*}

Recall that $\mathcal M_\mathrm{c} = (W_\mathrm{c},{\leq_\mathrm{c}},f_\mathrm{c},V_\mathrm{c})$ is the canonical model. Moreover, recall that by Lemma \ref{l: witnessing lemma for biltl} if $\varphi \rightarrow \psi \not \in \Gamma$ for $\Gamma \in W_c$, then there exists $\Delta \in W_c$ with $\Gamma \leq_c \Delta$ and $\varphi \in \Delta$ and $\psi \not \in \Delta$. Similarly, if $\varphi \dimp \psi \in \Gamma$, then there exists $\Delta \in W_c$ with $\Delta \leq_c \Gamma$ and $\varphi \in \Delta$ and $\psi \not \in \Delta$.

\begin{proposition}\label{propSimForm}
     Let $\sfrac{\mathcal M_\mathrm{c}}\Sigma = (U ,{\leq},R,\ell)$ and  $E_* \subseteq U\times W_\mathrm{c}$ be the strongly surjective dynamic simulation provided by Proposition \ref{propSimMc}. Let $x \in U$ and $\Gamma \in W_\mathrm c$. The following hold.
    \begin{enumerate}
        \item\label{condition+} $\chi^+(x) \in \Gamma$ if and only if there exists  $\Delta \in W_\mathrm c$ with $\Delta \leq_\mathrm{c} \Gamma$ such that $x \mathrel E_* \Delta$. 
         \item\label{condition-} $\chi^-(x) \in \Lbiltl \setminus \Gamma$ if and only if there exists  $\Delta \in W_\mathrm c$ with $\Gamma \leq_\mathrm c \Delta$ such that $x\mathrel E_* \Delta$. 
    \end{enumerate}
\end{proposition}
\begin{proof}
     We first prove the left-to-right direction of \ref{condition+} and \ref{condition-}. Given a path $\rho$, denote by $(\varphi_\rho)_\mathsf{L}$ the subformula on the left of $\varphi_\rho$'s principal connective and by $(\varphi_\rho)_\mathsf{R}$ the subformula on the right. Define a function $f: \final[\mathsf{ZZP}(x)] \longrightarrow W_c$ satisfying for all $\rho \in \mathsf{ZZP}(x)$ the property
    \begin{equation}\label{equation simulation formulas}
     (\varphi_\rho)_\mathsf{L} \in  f(\final(\rho)) \text{ and } (\varphi_\rho)_\mathsf{R} \in  f(\final(\rho))^c
    \end{equation}
    (where $\Delta^c = \Lbiltl \setminus \Delta$ for $\Delta \in W_c$) by induction on $\lvert \rho \rvert$ as follows, where we assume that for $(x)$ the formula $\varphi_{(x)} = \chi^+(x)$ or $\varphi_{(x)} = \chi^-(x)$, for \ref{condition+} and \ref{condition-}, respectively. For the base case observe that if $\rho \in \mathsf{ZZP}(x)$ and $\lvert \rho \rvert = 0$, then $\rho = (x)$. For \ref{condition+}, by assumption $\chi^+(x) \in \Gamma$, so there exists a world $\Delta \leq_c \Gamma$ such that $(\chi^+(x))_\mathsf{L} \in \Delta$ and $(\chi^+(x))_\mathsf{R} \in \Delta^c$. Define $ f(x) = \Delta$. For \ref{condition-}, by assumption $\chi^-(x) \not \in \Gamma$, so there exists a world $\Delta \geq_c \Gamma$ such that $(\chi^-(x))_{\mathsf{L}} \in \Delta$ and $(\chi^-(x))_{\mathsf{R}} \in \Delta^c$. Again, define $f(x) = \Delta$.

    For the induction step suppose that $f$ has been defined for all $\rho$ with $\lvert \rho \rvert \leq n$ and satisfies \eqref{equation simulation formulas}. Let $\rho \in \mathsf{ZZP}(x)$ with $\lvert \rho \rvert = n+1$. Let $\rho' \sqsubset \rho$ be the unique path such that $\lvert \rho - \rho'\rvert = 1$. We distinguish the following two cases.

    First, suppose $\final(\rho') < \final(\rho)$. Then the principal connective of $\varphi_\rho$ is an implication. Observe that $\varphi_\rho$ is a subformula of $\varphi_{\rho'}$ occurring on the right-hand side of the principal connective. By the induction hypothesis and the properties of a type, we have $\varphi_\rho \in(\final(\rho'))^c$. Hence there exists a world $\Delta \in W_c$ with $f(\final(\rho')) \leq_c \Delta$ and $(\varphi_\rho)_\mathsf{L} \in \Delta$ and $(\varphi_\rho)_\mathsf{R} \in \Delta^c$. Define $ f(\final(\rho)) \coloneqq \Delta$.
    
    Otherwise $\final
(\rho') > \final(\rho)$. Then the principal connective of $\varphi_\rho$ is a co-implication and $\varphi_\rho$ is a subformula of $\varphi_{\rho'}$ occurring on the left-hand side of the principal connective. By the induction hypothesis and the properties of a type, we have $\varphi_\rho \in f(\final(\rho')$. Hence there exists a world $\Delta \in W_c$ with $\Delta \leq_c f(\final(\rho'))$ and $(\varphi_\rho)_\mathsf{L} \in \Delta$ and $(\varphi_\rho)_\mathsf{R} \in \Delta^c$. Define $ f(\final(\rho)) \coloneqq \Delta$.
\smallskip

We claim that $f \subseteq U \times W_c$ is a simulation. Let $(u, f(u)) \in f$ for arbitrary $u \in \final[\mathsf{ZZP}(x)]$. Let $\rho$ be the unique path with $\final(\rho)=u$. We check that  all three conditions of a simulation are satisfied. \smallskip

\noindent \textsc{1.} By property \eqref{equation simulation formulas}, $(\varphi_{\rho})_{\mathsf{L}} \in f(u)$ and $(\varphi_{\rho})_{\mathsf{R}} \in f(u)^c$. Since $\bigwedge \ell_\mathcal{X}(u)^+$ is a subformula of $(\varphi_\rho)_L$ it follows from the properties of a type that $\ell_\mathcal{X}(u)^+ \subseteq f(u)$. Similarly, since $\bigvee \ell_\mathcal{X}(u)^-$ is a subformula of $(\varphi_\rho)_{\mathsf{R}}$ it follows that $\ell_\mathcal{X}(u)^- \subseteq f(u)^c$. Therefore $\ell_\mathcal{X}(u) \subseteq_\mathrm{T} (f(u), f(u)^c)$.
   
\noindent \textsc{2.} Suppose $u \leq u'$. First suppose that $u'$ covers $u$. Let $\rho' \in \mathsf{ZZP}(x)$ be the unique path with $\final(\rho') = u'$. There are two cases to consider: either $\rho \sqsubset \rho'$ or $\rho' \sqsubset \rho$. In both cases it follows immediately from construction that $f(u) \leq_c f(u')$, and hence that there exists $\Delta \in W_c$ with $f(u) \leq_c \Delta$ and $(u', \Delta) \in f$. In case $u \leq u'$ but $u'$ does not cover $u$, the result follows by induction on the number of worlds between $u$ and $u'$, using the above argument in the induction step (the base case is trivial).

\noindent \textsc{3.} Suppose $u' \leq_\mathcal{X} u$. The argument is then symmetrical to the previous case. \smallskip

Therefore, $f$ is a simulation relating $x$ with $\Delta$ for some $\Delta \in W_c$ with $\Delta \leq_c \Gamma$. Since $E_*$ is the union of all simulation between $\sfrac{\mathcal M_\mathrm{c}}\Sigma$ and $\fw_c$, we obtain that $x \mathrel{E_*} \Delta$.
\smallskip

Next, we prove the right-to-left direction of \ref{condition+} and \ref{condition-}. To that end we show by induction on the height of $\rho$ that whenever $\mathbf{f}(\rho) \mathrel{E_*} \Delta$, then
\begin{equation}\label{equation simulation formulas 2}
    (\varphi_\rho)_{\mathsf{L}} \in \Delta \text{ and } (\varphi_\rho)_{\mathsf{R}} \in \Delta^c.
\end{equation}

For the base suppose $h(\rho) = 0$. Then $\mathbf{f}(\rho) = u$ is a leaf of $\mathsf{ZZP}(x)$. Let $u \mathrel{E_*} \Delta$ for some $\Delta \in W_c$. By definition of a simulation $\ell(u)^+ \subseteq \Delta$ and $\ell(u)^- \subseteq \Delta^c$, implying that
\begin{equation}\label{equation simulation formulas 3}
    \bigwedge \ell(u)^+ \in \Delta \text{ and } \bigvee \ell(u)^- \in \Delta^c.
\end{equation}

The claim then follows from observing that $(\varphi_\rho)_{\mathsf{L}} = \bigwedge \ell(u)^+$ and $(\varphi_\rho)_{\mathsf{R}} = \bigvee \ell(u)^-$.

For the induction step suppose that (\ref{equation simulation formulas 2}) holds for all $\rho'$ with $h(\rho') \leq n$ and suppose $h(\rho) = n+1$. Let $\mathbf{f}(\rho) = u$ and $u \mathrel{E_*} \Delta$. First, note that (\ref{equation simulation formulas 3}) holds for $u$ by the same argument as before. Now suppose $\rho \sqsubset \rho'$ and ${\downarrow}(\rho' - \rho)$. Let $\mathbf{f}(\rho') = u'$. Then $u' < u$, and so by forth-down confluence there exists $\Delta' \in W_c$ with $\Delta' \leq_c \Delta$ and $u' \mathrel{E_*} \Delta'$. Note that $h(\rho') \leq n$. Thus by induction hypothesis we have $(\varphi_{\rho'})_\mathsf{L} \in \Delta'$ and $(\varphi_{\rho'})_\mathsf{R} \in (\Delta')^c$. Let $\rho' = (\rho'(i))_{i \leq k}$. Then $\rho'(k-1) > \rho'(k) = u'$, since $\rho \sqsubset \rho'$ and ${\downarrow}(\rho' - \rho)$. Therefore the main connective of $\varphi_{\rho'}$ is a co-implication. Thus by Lemma \ref{l: witnessing lemma for biltl} we have that $\varphi_{\rho'} \in \Delta$. Next, suppose $\rho \sqsubset \rho'$ and ${\uparrow}(\rho' - \rho)$. Then, by a similar argument, we find $u < \mathbf{f}(\rho') = u' \mathrel{E_*} \Delta'$ and $\Delta \leq_c \Delta'$. By induction hypothesis $(\varphi_{\rho'})_\mathsf{L} \in \Delta'$ and $(\varphi_{\rho'})_\mathsf{R} \in (\Delta')^c$. Since the main connective of $\varphi_{\rho'}$ is an implication, we conclude that $\varphi_{\rho'} \in \Delta^c$ by Lemma \ref{l: witnessing lemma for biltl}. Thus, by properties of a type, we have $(\varphi_\rho)_{\mathsf{L}} \in \Delta$ and $(\varphi_\rho)_\mathsf{R} \in \Delta^c$. \smallskip

Finally, suppose $\varphi_{(x)} = \chi^+(x)$ and there exists $\Gamma, \Delta \in W_c$ with $\Delta \leq_c \Gamma$ and $x \mathrel{E_*} \Delta$. By the above proof $\chi(x)^+_{\mathsf{L}} \in \Delta$ and $\chi(x)^+_{\mathsf{R}} \in \Delta^c$. By Lemma \ref{l: witnessing lemma for biltl} we conclude that $\chi^+(x) \in \Gamma$. Similarly if $\varphi_{(x)} = \chi^-(x)$ and there exists $\Gamma, \Delta \in W_c$ with $\Gamma \leq_c \Delta$ and $x \mathrel{E_*} \Delta$, then by the above proof $\chi(x)^-_{\mathsf{L}} \in \Delta$ and $\chi(x)^-_{\mathsf{R}} \in \Delta^c$. Thus Lemma \ref{l: witnessing lemma for biltl} implies that $\chi^-(x) \in \Gamma^c$. \qedhere
\end{proof}

Next we establish some $\biltl$-derivable properties of $\chi^+$ and $\chi^-$.
We begin with the former.
These properties are established by using Proposition \ref{propSimForm} to see that they are present in \emph{every} prime theory in the canonical model and so derivable.
As before, $\cqm = (U ,{\leq},R,\ell)$; the reflexive transitive closure of $R$ is denoted $R^*$.

\begin{proposition}\label{propsubplus}
Given $w \in U$ and $\psi\in \Sigma $, the following hold.
\begin{enumerate}
	\item\label{itPropsubplOne} If $\psi\in \ell({{w}})^-$, then
$\vdash \chi^+(w)\rightarrow (\chi^+(w) \dimp \psi ).$

	\item\label{itPropsubplOneb} If $\psi\in \ell({w})^+$, then
	$\vdash \chi^+(w) \rightarrow \psi.$

	

	\item\label{itPropsubplFive}
$\vdash\displaystyle \chi^+(w) \rightarrow \X\bigvee _{{{w}} \Rel {{{v}}}  } \chi^+(v).$
\end{enumerate}
\end{proposition}

\begin{proof}
\noindent  \ref{itPropsubplOne}.
Let $\Gamma\in W_\mathrm{c}$ and assume that $\psi\in \ell(w)^-$ and $\chi^+(w) \in \Gamma$. By properties of the canonical model, it suffices to show that $ \chi^+(w) \dimp \psi \in \Gamma $.
From $\chi^+(w) \in \Gamma$ and Proposition \ref{propSimForm}, case \ref{condition+}, we obtain $\Delta\leq_\mathrm{c} \Gamma$ such that $w \mathrel{E_*} \Delta$. Hence $\psi\in \Delta^c$ and, using Proposition \ref{propSimForm}, case \ref{condition+}, in the opposite direction, $\chi^+(w) \in \Delta$. These yield $\chi^+(w) \dimp \psi \in \Gamma$.
\smallskip

\noindent \ref{itPropsubplOneb}. If $\psi\in \ell ({w})^+$, as above, let $\Gamma\in W_\mathrm{c}$ be such that $\chi^+(w) \in \Gamma^+$, and $\Delta\leq_\mathrm{c} \Gamma$ with $ w \mathrel{E_*} \Delta$.
Then $\psi\in\Delta $, yielding $\psi\in \Gamma$.
\smallskip




\smallskip

\noindent \ref{itPropsubplFive}. Let $\Gamma$ be such that $\chi(w)^+ \in \Gamma$, so there exists $\Delta\leq_\mathrm{c} \Gamma$ with $w\mathrel E_* \Delta$. Since $E_*$ is dynamic, there is $v$ such that $w\mathrel R v$ and $v\mathrel E_* f_\mathrm{c}(\Delta)$.
By Proposition \ref{propSimForm}, we have $ \chi^+(v) \in  f_\mathrm{c} ( \Delta )$, which implies that $\X \chi^+(v) \in \Delta$.
Thus $\X \chi^+(v) \in \Gamma$ by definition of $\leq_\mathrm{c}$, so $\X \bigvee_{w\mathrel R  v} \chi^+(v) \in \Gamma^+$.
\end{proof}

The formula $\chi^-$ behaves `dually', as follows.

\begin{proposition}\label{propsubminus}
Given $w \in U $ and $\psi\in \Sigma $, the following hold.
\begin{enumerate}

\item\label{itPropsubOneb} If $\psi\in \ell({w})^+$, then
$\vdash  (\psi \rightarrow \chi^-(w)  )\rightarrow \chi^-(w) $.

	\item\label{itPropsubOne} If $\psi\in \ell({{w}})^-$, then
$\vdash \psi\rightarrow \chi^-(w)$.


	\item\label{itPropsubFive}
$\vdash\displaystyle \X\bigwedge _{{{w}} \mathrel R  {{{v}}}  } \chi^-(v) \rightarrow \chi^-(w).$

\end{enumerate}
\end{proposition}

\begin{proof}
\noindent \ref{itPropsubOneb}.
Suppose that $\psi\in \ell ({w})^+$. We prove the claim by contraposition.
If $\chi(w)^- \in \Gamma^c$ for some $\Gamma\in W_\mathrm{c}$, then there is $\Delta \geq _\mathrm{c} \Gamma$ such that $w \mathrel{E_*} \Delta$.
But then $\chi^-(w)\in \Delta^c$ and $\psi\in \Delta$, which implies that $ \psi\rightarrow\chi^-(w)  \in \Delta^c$; hence also $ \psi\rightarrow\chi^-(w)  \in \Gamma^c$, as required.
\smallskip

\noindent \ref{itPropsubOne}.
Assume that $\psi\in \ell ({w})^- \cap \Gamma$, and suppose that $\Delta \in W_c$ is such that $w \mathrel{ E_*} \Delta$.
Then $\psi\in \Delta^c$, which means we cannot have $\Delta \geq_c \Gamma$. Hence Proposition \ref{propSimForm} implies that $\chi^-(w)\notin \Gamma^c$, i.e.~$\chi^-(w)\in \Gamma$.
\smallskip


\noindent \ref{itPropsubFive}. 
Proceed by contraposition. If $\chi^-(w)\in \Gamma^c$ for some $\Gamma\in W_\mathrm{c}$, then there is $\Delta\geq _\mathrm{c}\Gamma$ such that $w \mathrel E_* \Delta$ by Proposition \ref{propSimForm}.
Since $E_*$ is dynamic, there is $v\in U$ such that $w\mathrel R v$ and $v\mathrel E_ * f_\mathrm{c}(\Delta)$.
Thus $\chi^- (v) \in  f_\mathrm{c}(\Delta)^c$; hence $\X \chi^-(v) \in   \Delta^c$, and by downward persistence, $\X \chi^-(v) \in   \Gamma^c$.
Hence $\X \bigwedge_{w\mathrel R v} \chi^- (v)  \in \Gamma^c$.
\end{proof}

\section{Completeness}\label{SecComp}\label{secComp}

The simulation formulas $\chi^+$ and $\chi^-$ are fundamental in the completeness proof. Specifically, they will be used to show that $\cqm$ is $\omega$-sensible and hence a quasimodel. Since validity over the class of quasimodels is equivalent to validity over the class of expanding models by Theorem \ref{theoQuasiToMod}, completeness will follow.
The following lemma is the first step towards establishing $\omega$-sensibility.
As above, we write $\cqm = (U,{\leq},R,\ell)$ and  $R^*$ for the reflexive transitive closure of $R$, and $E_* \subseteq U\times W_\mathrm{c}$ is the strongly surjective dynamic simulation between $\cqm$ and $\fw_c$, where $\fw_c$ is the canonical model. As in the completeness proof for $\mathrm{IM_{H}}$ in Chapter \ref{c: IM}, let $R^*(w)$ be the \emph{reachable component} of $w$, i.e. $R^*(w) = \{ v \in U \mid w \Rel^* v\}$. In order to show that $\cqm$ is $\omega$-sensible, we need to use formal induction over the reachable component of worlds. The following lemma will help us achieve this.

\begin{lemma}\label{syntactic}
If $\Sigma\subseteq \Lbiltl$ is finite and closed under subformulas, and ${{w}}\in U$, then:
\begin{enumerate}

\item $\vdash \bigvee _{w \mathrel R^* v}\chi^+({{{v}}}) \rightarrow \X \bigvee _{w \mathrel R^* v}\chi^+(v)  $, 

\item $\vdash \X \bigwedge _{w \mathrel R^* v}\chi^-(v)\rightarrow \bigwedge _{w \mathrel R^* v}\chi^-({{{v}}})$.

\end{enumerate}
\end{lemma}

\begin{proof}
The first item follows from Proposition \ref{propsubplus}, case \ref{itPropsubplFive}, as for any $w \mathrel{R^*}v$ we have that
\[\vdash \chi^+({{{v}}}) \rightarrow \X \bigvee _{v \mathrel R  u}\chi^+(u)  .\]
Since $v \mathrel R  u$ implies that $w \mathrel R^ *  u$,
\[\vdash \chi^+({{{v}}}) \rightarrow \X \bigvee _{w \mathrel R^*  u}\chi^+(u)  .\]
Since $v$ was arbitrary, we obtain
\[\vdash \bigvee _{w \mathrel R^* v}\chi^+({{{v}}}) \rightarrow \X \bigvee _{w \mathrel R^* u}\chi^+(u),  \]
which by a change of variables yields the original claim.

Item 2 is similar, but uses Proposition \ref{propsubminus}, case \ref{itPropsubFive}.
\end{proof}

In order to complete our proof that $\cqm$ is $\omega$-sensible, it suffices to apply the induction rules $\mathsf{Ind}_{\G}$ and $\mathsf{Ind}_{\E}$ of our calculus to the formulas of Lemma \ref{syntactic}.

\begin{proposition}[$\omega$-sensibility]\label{tempinc}\
\begin{enumerate}

\item\label{ittempincone} If ${{w}}\in U$ and $\E \psi\in \ell ({{w}})^+$, then there exists ${{{v}}}\in{R^*}({{w}})$ such that $\psi\in \ell ({{{v}}})^+$.

\item\label{ittempinctwo} If ${{w}}\in U$ and $\G \psi\in \ell ({{w}})^-$, then there exists ${{{v}}}\in{R^*}({{w}})$ such that $\psi\in \ell({{{v}}})^-$.

\end{enumerate}
\end{proposition}

\begin{proof} \noindent \ref{ittempincone} . Towards a contradiction, assume that ${{w}}\in U$ and $\E \psi\in \ell ({{w}})^+$, but for all ${{{v}}}\in{R^*}({{w}})$, we have $\psi \in \ell({{v}})^-$. 
By Lemma \ref{syntactic}, $\vdash \X \bigwedge \limits_{w \mathrel R^* v} \chi^-({{{v}}})\rightarrow \bigwedge\limits_{w \mathrel R^* v} \chi^-({{{v}}})$.
By the $\mathsf{Ind}_\E$ rule, ${\vdash \E \bigwedge \limits_{w \mathrel R^* v} \chi^-({{{v}}})}\rightarrow \bigwedge\limits_{w \mathrel R^* v} \chi^-({{{v}}})$;
in particular,
\begin{equation}\label{other}
\vdash \E \bigwedge _{w\mathrel R^* v} \chi^-({{{v}}})\rightarrow \chi^-({{w}}).
\end{equation}

Now let ${{{v}}}\in{R^*}({{w}})$.
By Proposition \ref{propsubminus}, case \ref{itPropsubOne}, and the assumption that $\psi \in \ell({{{v}}})^-$, we have that
$\vdash \psi \rightarrow \chi^-({{{v}}}) $,
and since ${{{v}}}$ was arbitrary,
$\vdash \psi \rightarrow \bigwedge_{w\mathrel R^* v}\chi^-({{{v}}}) $.
Using $\mathsf{Mon}_\E$, we further have that
$\vdash \E \psi \rightarrow \E \bigwedge_{w\mathrel R^* v}\chi^-({{{v}}})$.
This, along with (\ref{other}), shows that
$\vdash \E \psi \rightarrow \chi^-({{w}})$.
However, by Proposition \ref{propsubminus}, case \ref{itPropsubOneb}, and our assumption that $\E \psi\in \ell ({{w}})^+$, we have that
$\vdash  ( \E \psi \rightarrow \chi^-({{w}})  ) \rightarrow \chi^-(w)$.
Hence by $\mathsf{MP}$ we obtain $\vdash \chi^-({{w}}).$
Choosing $\Gamma\in W_\mathrm c$ such that $w\mathrel E_*\Gamma $, Proposition \ref{propSimForm}, case \ref{condition-}, yields $\chi^-({{w}}) \notin\Gamma$, but this contradicts $\vdash \chi^-({{w}})$.
We conclude that there is ${{{v}}}\in{R^*}({{w}})$ with $\psi \in \ell({{v}})^+$, as needed.
\smallskip

\noindent \ref{ittempinctwo}.
This is similar to the first item, but dualised.
Towards a contradiction, assume that ${{w}}\in U$ and $\G \psi\in \ell ({{w}})^-$ but, for all ${{{v}}}\in{R^*}({{w}})$, we have $\psi \in \ell({{w}})^+$.
By Lemma \ref{syntactic}, $\vdash \bigvee \limits_{w \mathrel R^* v} \chi^+({{{v}}}) \rightarrow \X \bigvee \limits_{w \mathrel R^* v} \chi^+({{{v}}})$.
By the $\mathsf{Ind}_{\G}$-rule, $\vdash \bigvee \limits_{w \mathrel R^* v} \chi^+({{{v}}})\rightarrow\Box \bigvee \limits_{w \mathrel R^* v} \chi^+({{{v}}})$;
in particular,
\begin{equation}\label{otherb}
\vdash\chi^+({{w}}) \rightarrow \G  \bigvee _{w\mathrel R^*v} \chi^+({{{v}}}) .
\end{equation}

Now let ${{{v}}}\in{R^*}({{w}})$.
By Proposition \ref{propsubplus}, case \ref{itPropsubplOneb}, and the assumption that $\psi \in \ell({{{v}}})^+$, we have that
$\vdash \chi^+({{{v}}}) \rightarrow \psi $,
and since ${{{v}}}$ was arbitrary,
$\vdash \bigvee_{w\mathrel R^* v}\chi^+({{{v}}})\rightarrow \psi $.
Using  $\mathsf{Mon}_{\G}$ we further have that
$\vdash  \G \bigvee_{w\mathrel R^* v}\chi^+({{{v}}}) \rightarrow \G \psi$.
This, along with (\ref{otherb}), shows that
\begin{equation}\label{eqPlusBox}
\vdash \chi^+({{w}}) \rightarrow \G \psi.
\end{equation}
By Proposition \ref{propsubplus}, case \ref{itPropsubplOne} and our assumption that $\G \psi\in \ell ({{w}})^-$, we have that
$\vdash \chi^+(w) \rightarrow  (  \chi^+({{w}}) \dimp \G\psi  )$.
Hence by \eqref{eqPlusBox} and Lemma \ref{Lemma exclusion property}, case \ref{itDimpMon}, we obtain $\vdash \chi^+({{w}}) \rightarrow (\G\psi\dimp\G\psi).$
Since $(\G\psi\dimp\G\psi) \equiv\bot$, this implies that $\chi^+({{w}})$ is contradictory.
Choosing $w\in U$ such that $w \mathrel E_* \Gamma$, Proposition \ref{propSimForm}, case \ref{condition+}, yields $\chi^+({{w}}) \in\Gamma$, which once again is impossible and we conclude that there is ${{{v}}}\in{R^*}({{w}})$ with $\psi \in \ell({{v}})^-$.
\end{proof}

\begin{corollary}\label{laststretch}
If $\Sigma\subseteq \Lbiltl$ is finite and closed under subformulas, then $\cqm$ is a quasimodel.
\end{corollary}

\begin{proof}
By Proposition \ref{propSimMc}, $ \cqm$ is a labelled system (and serial, since $\mathcal M_\mathrm c$ is), while by Proposition \ref{tempinc}, $R$ is $\omega$-sensible. So by Definition \ref{def:quasimodel}, $ \cqm$ is a quasimodel.
\end{proof}

 We are now ready to prove that our calculus is complete.

\begin{theorem}\label{theocomp}
Given $\varphi \in \Lbiltl$, the following are equivalent:
\begin{enumerate}
 \item\label{itOne}  $ \biltl \vdash\varphi$,  \item\label{itTwo} $\varphi$ is valid over the class of expanding models, \item\label{itThree} $\varphi$ is valid over the class of finite quasimodels.
\end{enumerate}

\end{theorem}

\begin{proof}
That \ref{itOne} implies \ref{itTwo} is Theorem \ref{t: soundness of biltl} and that \ref{itTwo} implies \ref{itThree} is Theorem \ref{theoQuasiToMod}.
We show that \ref{itThree} implies \ref{itOne} by contraposition.
Suppose $\varphi$ is an unprovable formula and let $\Sigma$ be the set of subformulas of $\varphi$.
Since $\varphi$ is unprovable, there exists $\Gamma\in W_\mathrm{c}$ with $\varphi\notin\Gamma$.
Since $E_*$ is strongly surjective, there is $w\in U$ such that $\varphi\in\ell(w)^-$ and $w\mathrel E_* \Gamma$.
Hence $w$ is a point in a finite quasimodel falsifying $\varphi$.
\end{proof}

\begin{corollary}
Derivability in $\biltl$ is decidable.
\end{corollary}

This follows from the fact that $\biltl$ is axiomatisable and has a finite quasimodel property (with a computable bound on the `finite').

\section{Conclusion}

The main contribution of this chapter is the presented sound and complete axiomatization for bi-intuitionistic linear temporal logic. This is the first finite axiomatization for a logic extending intuitionistic linear temporal logic over the full temporal language including `next', `eventually' and `henceforth' and thus solves a long standing open problem. Our method of proving completeness was based on the finite model construction from Chapter \ref{c: bi-int ml new} which was used to obtain a finite `filtrated' model which embeds into the canonical model via a dynamic simulation. The proof is thereby an interesting adaptation of classical techniques for LTL to the intuitionistic realm, illustrating in particular the significant increase in the difficulty of the mathematical theory. The crucial ingredient for obtaining a complete axiomatization is the co-implication, whose presence is exploited to prove $\iLTL$-validites such as the $\mathbf{RV}$-formula. As a consequence we noted that the axiomatization is not conservative over the $\dimp$-free fragment of the logic. In the completeness proof the co-implication is crucial in establishing that the modal accessibility relation of $\cqm$ is $\omega$-sensible for $\G$ case (while implication was used to prove the $\E$ case). This leads us to the first question left open by our work:

\begin{question}
    Can $\iLTL$ over the full language be finitely axiomatized without the co-implication?
\end{question}

We are optimistic that this is indeed possible, partly due to the fact that Men\'endez Turata managed to obtain a sound and complete cyclic proof system for $\iLTL$ over the full language~\cite{menendez_2024}, which might be used to extract a finite axiomatization. We leave this problem for future work. 

An interesting corollary of this chapter is that $\dimp$ cannot be extended to the class of topological models while validating bi-intuitionistic logic, since as verified in~\cite{Boudou_2021}, this would mean that the class of \emph{dynamic} topological models would validate $\biltl$, and hence validate $\mathbf{RV}$, which we know by Example \ref{ex: RV not valid over topological spaces} not to be the case.

\begin{corollary}\label{corNoDimp}
Suppose that $\dimp_\mathrm{top}$ assigns to each topological space  $(X,\tau)$ a binary operation $\tau \times \tau \to \tau$. (Here, $\tau$ is the collection of opens.) Consider the semantics that combines standard topological semantics for intuitionistic propositional logic with $\dimp_\mathrm{top}$ semantics for $\dimp$. Then the class of topological spaces does not validate propositional bi-intuitionistic logic.    
\end{corollary}

A natural next step to extend our work would be to turn to proof theory with the aim of developing an analytic sequent calculus for bi-intuitionistic linear temporal logic. This is a follow up problem to Question \ref{q: proof theory for biml}: an analytic sequent calculus for bi-intuitionistic modal logic could be extended by fixed point rules, admitting analytic non-wellfounded proofs. However, this would require to combine nested sequents with non-wellfounded branches, which is notoriously difficult to handle (see Question \ref{q: ICK proof theory}).


\backmatter

\printbibliography[heading=bibintoc]

\addcontentsline{toc}{chapter}{Index}
\printindex

\addcontentsline{toc}{chapter}{List of Figures}
\listoffigures

\addcontentsline{toc}{chapter}{List of Tables}
\listoftables

\chapter{English Summary}

This thesis develops the mathematical theory of intuitionistic dynamic logics - extensions of intuitionistic propositional logic with modalities and fixed point operators. Such systems provide formal tools for reasoning about change, such as encountered in mathematical systems evolving over time or in the knowledge state of an agent after an information update.\smallskip

 We investigate five intuitionistic dynamic logics: intuitionistic master modality, intuitionistic common knowledge logic, intuitionistic linear temporal logic, bi-intuitionistic modal logic and bi-intuitionistic linear temporal logic. On the proof theoretic side we develop sound and complete Hilbert-style axiomatizations as well as non-wellfounded and cyclic sequent calculi. On the semantic side we study these logics over various classes of dynamic models, which are birelational Kripke models satisfying confluence and frame conditions. We establish expressivity results, the finite model property, decidability, as well as complexity bounds. \smallskip

 The main contributions are threefold. First, we develop analytic cyclic sequent calculi for intuitionistic master modality and common knowledge logic, where completeness is obtained by a robust proof search argument. Second, we obtain the finite model property and decidability for bi-intuitionistic modal logic via an intricate combinatorial analysis of dynamic models. Third, we develop a sound and complete axiomatization for intuitionistic linear temporal logic featuring the temporal operators \emph{next}, \emph{eventually} and \emph{henceforth}, thereby providing a positive answer to the long-standing open question concerning the existence of a finite axiomatization.

\let\cleardoublepage\clearpage

\chapter{Nederlandstalige Samenvatting}

Dit proefschrift ontwikkelt de wiskundige theorie van intuïtionistische dynamische logica’s — uitbreidingen van de intuïtionistische propositionele logica met modaliteiten en fixpunt-operatoren. Dergelijke systemen bieden formele instrumenten om te redeneren over verandering, zoals die optreedt in wiskundige systemen die in de tijd evolueren of in de kennis­toestand van een agent na een informatie-update.\smallskip

We onderzoeken vijf intuïtionistische dynamische logica’s: intuïtionistische meester­modaliteit, intuïtionistische gemeenschappelijke-kennis­logica, intuïtionistische lineaire temporele logica, bi-intuïtionistische modale logica en bi-intuïtionistische lineaire temporele logica. Aan de bewijstheoretische kant ontwikkelen we correcte en volledige Hilbert-stijl-axiomatiseringen, evenals niet-welgegronde en cyclische sequent-calculi. Aan de semantische kant bestuderen we deze logica’s over verschillende klassen van dynamische modellen, dat wil zeggen birelationele Kripke-modellen die voldoen aan confluentie en framecondities. We bewijzen resultaten over expressiviteit, de eindig-model-eigenschap, beslisbaarheid en complexiteitsgrenzen. \smallskip

De belangrijkste bijdragen zijn drievoudig. Ten eerste ontwikkelen we analytische cyclische sequent-calculi voor intuïtionistische meester­modaliteit en gemeenschappelijke-kennis­logica, waarbij volledigheid wordt verkregen via een robuust bewijszoek-argument. Ten tweede verkrijgen we de eindig-model-eigenschap en beslisbaarheid voor de bi-intuïtionistische modale logica door middel van een verfijnde combinatorische analyse van dynamische modellen. Ten derde ontwikkelen we een correcte en volledige axiomatisering voor de intuïtionistische lineaire temporele logica met de temporele operatoren \emph{next}, \emph{eventually} en \emph{henceforth}, waarmee een positief antwoord wordt gegeven op de lang openstaande vraag naar het bestaan van een eindige axiomatisering.

\end{document}